\pdfoutput=1
\documentclass[11pt,oneside,openany]{uwthesis}
\usepackage[utf8]{inputenc}
\usepackage[T1]{fontenc}

\usepackage{amsmath, amssymb, mathtools, nccmath}
\usepackage{physics}
\usepackage{dsfont, bbm, bm}
\usepackage{cancel}
\usepackage{stackengine}

\usepackage{graphicx}
\usepackage{placeins}
\usepackage{tabularx, array, multirow, booktabs, diagbox}
\usepackage[dvipsnames]{xcolor}
\usepackage[normalem]{ulem}
\usepackage{soul}
\usepackage{xspace}

\usepackage{subcaption}
\usepackage{algorithm2e}
\usepackage{tikz}
\usetikzlibrary{quantikz2, circuits.logic.US}

\tikzset{
    gateT/.style={
        draw,
        not gate US,
        inner sep=5pt
    },
    gateO/.style={
        draw,
        circle,
        minimum width=1.67em,
        inner sep=2pt,
        append after command={
            \pgfextra {
                \fill (\tikzlastnode) circle[radius=1pt];
            }
        }
    }
}

\usepackage[square,numbers,comma,sort&compress]{natbib}

\graphicspath{{figures/SeqHT/}{figures/energy_loss/}{figures/gauge_fix/}}

\definecolor{lavenderindigo}{rgb}{0.58, 0.34, 0.92}
\definecolor{colorslice}{HTML}{000580}

\newcolumntype{C}[1]{>{\centering\let\newline\\\arraybackslash\hspace{0pt}}m{#1}}
\newcolumntype{Y}{>{\centering\arraybackslash}X}

\providecommand{\CE}{\mathcal{E}}

\providecommand{\CN}{\mathcal{N}}
\providecommand{\CO}{\mathcal{O}}
\providecommand{\CP}{\mathcal{P}}

\providecommand{\uone}{${\rm U}(1)$\,}
\providecommand{\sutwo}{${\rm SU}(2)$\,}

\providecommand{\nn}{\nonumber}
\providecommand{\htp}{H_{\square\square}}
\providecommand{\hathtp}{\hat{H}_{\square\square}}
\DeclareMathOperator*{\argmin}{arg\,min}
\newcommand{\SubFigRef}[2]{\ref{#1}{#2}}
\newcommand{\infiL}{{\mathcal{I}_L}}

\stackMath
\newcommand\tenq[2][1]{
 \def\useanchorwidth{T}%
  \ifnum#1>1%
    \stackunder[0pt]{\tenq[\numexpr#1-1\relax]{#2}}{\scriptscriptstyle\sim}%
  \else%
    \stackunder[1pt]{#2}{\scriptscriptstyle\sim}%
  \fi%
}

\usepackage{orcidlink}

\begin{document}
 
% ==========   Preliminary pages

\prelimpages
 
%
% ----- copyright and title pages
%
\Title{Digital Quantum Simulation of Dynamics in Interacting Fields}
\Author{Zhiyao Li}
\Year{2026}
\Program{Physics}

\Chair{Martin J. Savage}{Professor}{Department of Physics}
\Signature{Stephen R. Sharpe}
\Signature{Silas R. Beane}

\copyrightpage
\titlepage  

%
% ----- abstract
%

\abstract{
Matter under extreme conditions behaves as a non-equilibrium, strongly-interacting quantum system. One example is the quark-gluon plasma, which briefly filled the early universe and is now recreated in particle accelerators like the Large Hadron Collider. Such systems are in the non-perturbative regime of quantum chromodynamics (QCD), where their time evolution is classically intractable. Quantum computing offers a promising tool for simulating their real-time dynamics directly. Present-day noisy intermediate-scale quantum (NISQ) hardware is limited in qubit count and circuit depth, so developing efficient formulations and algorithms is key both to current demonstrations and to scaling toward experimentally relevant QCD simulations.

This dissertation spans different formulations of the \sutwo lattice gauge theory as well as the scalar field theory, constructing scalable quantum circuits for these formulations and introducing physics-aware truncation methods. All proposed simulations are validated on quantum hardware, with results in agreement with classical benchmarks. 

A scalable framework is established for simulating energy loss and hadronization in $1{+}1$-dimensional SU(2) lattice gauge theory, in which a heavy quark traverses a lattice of light quarks. Classical lattice QCD methods are translated into physics-informed quantum circuits to prepare the interacting ground state. The evolution is run on IBM's quantum processor {\tt ibm\_pittsburgh}, and observables including charge distribution and energy loss are measured on the hardware. A further study develops the quantum simulation of the fully gauge-fixed \sutwo Hamiltonian in a mixed basis, a formulation that is expected to remain efficient at all gauge couplings. The continuous gauge fields can be digitized with as few as three qubits per plaquette to reproduce low-energy observables with $10^{-3}$ precision.

Progress to larger systems on present-day hardware is limited by resource overhead, which motivates truncations that preserve the relevant observables. Sequency Hierarchy Truncation (SeqHT) exploits the hierarchy in the contributions of basis operators. Demonstrated on the adiabatic preparation of the $\lambda\phi^4$ ground state on \texttt{ibm\_sherbrooke}, SeqHT reduces circuit depth by ${\sim}30\%$, suppressing hardware noise in the extracted observables.

Together, these studies contribute toward quantum simulations of real-time QCD dynamics inaccessible to classical computation.
}
 
%
% ----- contents & etc.
%
\tableofcontents
\listoffigures
\listoftables
  
%
% ----- acknowledgments
%
\acknowledgments{
  I wish to express sincere appreciation to my advisor, Professor Martin J. Savage, for his guidance and support throughout my PhD journey. I am inspired by his passion for charting the unknown, his devotion to science, his confidence in his vision, and his ability to realize his vision that justifies the confidence. I thank him for shouldering the grants and administrative work so that we could focus entirely on research, for the eye-opening workshops and conferences, and for setting up beautiful offices for us with fancy standing desks, ergonomic chairs, and a view of the Olympic Mountains.
  
  I am deeply grateful to the InQubator for Quantum Simulation (IQuS) for providing a stimulating research environment, and to Katie Hennessey for her invaluable work that kept the group functioning. My time at IQuS was enriched by wonderful colleagues and collaborators; thank you to Marc Illa, Dorota Grabowska, Nikita Zemlevskiy, Henry Froland, Anthony Ciavarella, Roland Farrell, Saurabh Kadam, Sarah Powell, Hersh Singh, Xiaojun Yao, Jeremy Hartse, Ivan Chernyshev, Ramya Bhaskar, Sebastian Grieninger, Francesco Turro, Niklas Mueller, Jiunn-Wei Chen, Natalie Klco, Fabio Anza, Stephan Caspar, Ziyuan Li, Anne Lashbrook, and Roland de Putter. 

  I would like to thank the professors who served on my doctoral supervisory committee and reading committee for volunteering their precious time to guide me at pivotal times of my career. Thank you to Prof. Stephen Sharpe, Prof. Silas Beane, Prof. Sara Mouradian, Prof. Mark Rudner, Prof. Boris Blinov, and Prof. Kai-Mei Fu. I gratefully acknowledge the QuantumX Initiative at the University of Washington for expanding my understanding of the broader quantum ecosystem. 

  I appreciate the physics and fencing communities, as well as my professors and mentors, during my undergraduate years at Carnegie Mellon University, including Prof. Rachel Mandelbaum, Prof. Kunal Ghosh, Prof. Markus Deserno, Prof. Reinhard Schumacher, Prof. Stephen Garoff and more. I cherished my time interning at Riverlane working with Guen Prawiroatmodjo, Kallie Ferguson, Angela Burton, Christi Amend, Abe Asfaw, and Liz Durst, who created the best work environment one could hope for. I would like to thank Guen particularly for her very organized and supportive management style, and for continuing to look out for my growth afterward. I've also had the privilege of an illuminating experience at Quantum Computing bootcamp 2023 at Jefferson Lab. 
  
  During my years in Seattle, I was incredibly lucky to find vibrant communities that kept me grounded outside of work. I would like to thank Salle Auriol Seattle for being a stellar fencing club. My deepest gratitude goes to my coach Myles Luedecke, as well as coaches Marshall Hibnes, Russel Redding, and Paul Englund; our armorer Chris Aher; and managers Kundry Haberkern and Sarah Lippai. I also appreciate the numerous friends who constitute the great environment and made me a better fencer, including Ben Weldon, Vivian Rand, Steven Benack, Jack Witecki, Jenner Mandel, Katherine Sizov, Madeleine McCreary, Heather Chvany, David Westphal, Tzu-Chen Hsu, Shiou Tian Hsu, Juliya Reed, Daisy Ust, Andrew Kiluk, John Varney, Monica Morrison, Paul Riviere, Connor Dempsey, and Ernest Miu as well as the numerous others whom I've had the honor to fence with and talk to. 
  My appreciation also goes to the UW Women PhD Students group for exposing me to diverse knowledge, scintillating discussions and warm support. I have also had lots of fun at SeaUni, SeattleTKS, and the UW Wushu Club. 

  I've spent some fabulous time during the PhD with great Seattle friends. To name a few, Zixuan (Jasmine) Zhou, Zi Yang, Boxi Li, Kent Wilson, Gian Marco Visani, Artidoro Pagnoni, Jimmy Sinnis, Peijie He, Lingnan Shen, Beini (Rachel) Gu. I loved the holiday feasts, fun parties, ski trips, and volleyball games with Nikita, Murali Saravanan, Jesse Yue, Emily Meng, Yihui Liao, Jiahui Liao, and everyone else in this great friend group.  I feel fortunate to have enjoyed the beautiful nature Seattle has to offer. I appreciate the bright days when the mountain ranges come into view, as well as the rainy days that keep everything green and lush and sustain the ecosystem.
  
  My long-term friends are scattered around the world. Even though we are not always in close contact, their mere existence brings me serenity. Connecting with them always brings out a deep happiness in me: Jingyan (Amy) Xu, Wenyi Wang, Sarah Penfield, Swara Srinivasan, Tomasz Swierzewski, Teven Le Scao, Nick Rauen, Yizhe Liu, Sayan Mandal, Jingze Lu, Shibo Wang, Weiyi Sun, Shengming Li, Nick Speeney, Jiwoo Seo, Biqi Zhao, Lingbo Jin, and so many others. I would also like to thank my therapist Emily for providing a crucial space for clarity, balance, and personal growth; and my physical therapist Brendan for helping me take care of my body.

  Most importantly, I thank my parents for their unwavering love and support, and for their open-mindedness in always trusting my choices. I thank my father for his unparalleled wisdom and his devotion to family. I thank my mother for her intelligence and vision, and for fostering my ability to find happiness no matter the circumstances; I wish the same for her. I am grateful to my grandma and mother for being role models of strong, driven women. I thank my grandparents for all the delicious meals they cooked when I was growing up that made me so sturdy, and for the love and work that kept such a big family harmoniously together. I thank my aunts and uncles for all the love and care, and for their support of my parents in times of need. I thank my cousins Mingfei Jia and Yingfan Jia, who grew up with me and have always felt like siblings. I also appreciate my cousins Xiuli Li, Huanhuan Li, and Wenhui Zu, who watched me grow up and made me feel loved. I thank Ziwen Ye for sticking by my side all through graduate school, growing, learning, and exploring this world with me, loving (or tolerating) all the sides of me that are unknown to the world.

  Finally, I acknowledge luck, for growing up away from wars and conflicts and having the chance to build up who I am. How I wish it weren't a privilege. 

  The work in this dissertation was supported, in part, by U.S. Department of Energy, Office of Science, Office of Nuclear Physics, InQubator for Quantum Simulation (IQuS)\footnote{\url{https://iqus.uw.edu}} under Award Number DOE (NP) Award DE-SC0020970 via the program on Quantum Horizons: QIS Research and Innovation for Nuclear Science,\footnote{\url{https://science.osti.gov/np/Research/Quantum-Information-Science}}
  by the Department of Physics\footnote{\url{https://phys.washington.edu}}
  and the College of Arts and Sciences\footnote{\url{https://www.artsci.washington.edu}} at the University of Washington. 
  This research used resources of the Oak Ridge Leadership Computing Facility (OLCF), 
  which is a DOE Office of Science User Facility supported under Contract DE-AC05-00OR22725.
  We acknowledge the use of IBM Quantum services for this work. The views expressed are those of the authors, and do not reflect the official policy or position of IBM or the IBM Quantum team. This work was enabled, in part, by the use of advanced computational, storage and networking infrastructure provided by the Hyak supercomputer system at the University of Washington.\footnote{\url{https://itconnect.uw.edu/research/hpc}}
  We have made extensive use of Wolfram {\tt Mathematica}~\cite{Mathematica},
  {\tt python}~\cite{python3,Hunter:2007}, {\tt julia}~\cite{Julia-2017},
  {\tt jupyter} notebooks~\cite{PER-GRA:2007} in the {\tt Conda} environment~\cite{anaconda}, and IBM's quantum programming environment {\tt qiskit}~\cite{qiskit}.
  This research used resources of the National Energy Research Scientific Computing Center (NERSC), a Department of Energy Office of Science User Facility using NERSC award NP-ERCAP0032083.
}

%
% ----- dedication
%
\dedication{\begin{center}to my parents\end{center}}

%
% end of the preliminary pages

%
% ==========      Text pages
%

\textpages
 
% ========== Chapter 1
 
\chapter{Introduction}
Shortly after the Big Bang, our universe entered the quark epoch, during which it was filled with the quark-gluon plasma. This is one example of non-equilibrium systems of fundamental particles that are studied in nuclear and high-energy physics. Other examples include neutrinos in extreme astrophysical environments, 
e.g., Refs.~\cite{Malaney:1990aj,Savage:1990by,Pantaleone:1992eq,Pantaleone:1992xh,Loreti:1994ry,Qian:1994wh,Hoffman:1996aj,Liebendoerfer:2000cq,Liebendoerfer:2002xn,Balantekin:2004ug,Duan:2005cp,Balantekin:2006tg,Duan:2006an,Duan:2006jv,Janka:2006fh,Bruenn:2006oub,Duan:2010bg,Pehlivan:2011hp,Lentz:2012tel,Winteler:2012hu,Cherry:2013mv,Tamborra:2014aua,Tamborra:2014xia,Wanajo_2014,Mirizzi:2015eza,Izaguirre:2016gsx,Cervia:2019res,Muller:2019upo,Martin:2019dof,Johns:2019izj,Mezzacappa:2020pkk, Burrows:2020qrp,Nagakura:2021hyb,Hall:2021rbv,Martin:2021xyl,Gorda:2021znl,Johns:2021qby,Fuller:2022nbn,Patwardhan:2022mxg,Siwach:2022xhx,Foucart:2022bth,Illa:2022zgu,deGouvea:2022gut,Siwach:2023wzy,Balantekin:2023qvm,Martin:2023gbo,Martin:2023ljq,Bhaskar:2023sta,Balantekin:2023ayx,Chernyshev:2024kpu,DeLia:2024kjv},
matter in the interior of neutron stars, 
e.g., Refs.~\cite{Kurkela:2009gj,Lattimer:2012nd,Hebeler:2013nza,Kurkela:2014vha,Hell:2014xva,Holt:2016pjb,Baym:2017whm,Annala:2017llu,McLerran:2018hbz,Tews:2018iwm,Most:2018hfd,Annala:2019puf,Al-Mamun:2020vzu,Drischler:2021kxf,Altiparmak:2022bke,Alford:2022bpp,Lovato:2022vgq,Gorda:2022jvk,Mroczek:2023zxo,Annala:2023cwx},
and highly-inelastic collisions of hadrons created in the laboratory, 
e.g., Refs.~\cite{OPAL:1998arz,CMS:2011jlm,ALICE:2016fzo,ATLAS:2019rqw,ATLAS:2024nbm,Baker:2017wtt,Raychowdhury:2019iki,Dasgupta:2020itb,Kreshchuk:2020aiq,Beane:2021zvo,Qian:2021jxp,Atas:2021ext,Brambilla:2016wgg,Nijs:2023bok,Andronic:2024oxz,Biondini:2024aan}. The theoretical foundation underlying all of these systems is the Standard Model~\cite{Glashow:1961tr, Higgs:1964pj, Weinberg:1967tq, Salam:1968rm, Politzer:1973fx, Gross:1973id}, which classifies the known particles and describes the forces of nature. The dynamics of neutrinos are governed by the weak interaction, while
quantum chromodynamics (QCD) describes the strong interactions between quarks and gluons. Emergent phenomena such as confinement and hadronization lie in the non-perturbative regime of QCD, and first-principles simulations are important to understand their structure and dynamics. Quantum simulation gives direct access to the real-time dynamics of
these systems in regimes that classical methods cannot reach, and it
is the approach developed throughout this dissertation.

In this introduction, Sec.~\ref{sec:lgt} reviews lattice gauge theory,
the established framework for first-principles studies of QCD, and the
limitations of classical simulation methods. Sec.~\ref{sec:recipe} outlines the procedure for performing quantum simulations on quantum devices, Sec.~\ref{sec:hardware} reviews the current quantum hardware landscape along with the error-handling methods needed to extract useful results, Sec.~\ref{sec:formulation} introduces the Hamiltonian formulations relevant for this dissertation, and Sec.~\ref{sec:overview} lays out the road map of the chapters, summarizing main results and how they connect.

\section{From Gauge Theories to Quantum Simulations}
\label{sec:lgt}
Gauge symmetry is the organizing principle of the Standard Model. A global
symmetry applies the same transformation everywhere and relates physically
distinct states. A gauge, or local, symmetry is really a redundancy of the description: configurations
related by a gauge transformation are the same physical state. Requiring the theory to stay invariant under such local transformations introduces gauge fields, which carry the interactions between matter particles. The symmetry groups
describe the gauge fields and how they couple to matter. Each
symmetry group is a group of matrices. The U stands for unitary, so U(N) is the
group of $N\times N$ unitary matrices, and the $S$ stands for special,
restricting to those with determinant one. The simplest case is U(1), the
$1\times 1$ unitary matrices, which are just the complex phases $e^{i\theta}$.
An SU(N) group has $N^2-1$ independent generators, one for each gauge boson.

QCD, the theory of the strong interactions, is an SU(3) gauge theory,
where SU(3) rotates among the three color charges carried by quarks and its
eight generators correspond to the eight gluons. Much of this dissertation aims to study QCD, whose most important phenomena lie
beyond the reach of analytic methods. The strength of the interaction depends
on the energy scale. At shorter distances and high energies, the coupling becomes
weaker, displaying asymptotic freedom~\cite{Gross:1973id,
Politzer:1973fx}, and perturbation theory works well. At low energies, the coupling grows and perturbative
expansions no longer converge. This strongly coupled
regime is where confinement, the hadronization of quarks and gluons, and the generation of most of the mass of visible matter take
place. Studying these phenomena from first principles therefore calls for a
formulation of QCD that does not rely on perturbation theory.

Lattice gauge theory provides a way forward~\cite{Wilson:1974sk}. The continuum theory is placed on a discrete spacetime, with matter fields on
the sites and the gauge field carried by parallel transporters on the links
between them. Discretizing brings two benefits.
First, it defines the theory non-perturbatively, specifying its path integral,
the weighted sum over all field configurations, directly rather than as an
expansion in the coupling, so it applies in the strongly coupled regime where
perturbation theory fails. Second, it turns that path integral from an
infinite-dimensional object into a finite one that can be evaluated directly
on a computer. Physical results are then recovered from correlation
functions of gauge-invariant operators, extrapolated to the continuum by
taking the spacing to zero and increasing lattice points.

In its Euclidean formulation, the lattice path integral carries a real and
positive weight, so expectation values can be estimated by Monte Carlo sampling. This approach is a great tool in theoretical physics. Lattice QCD reproduces the masses of
the low-lying hadrons at the percent level~\cite{BMW:2008jgk}, supplies the
hadronic matrix elements that experiments use to test the Standard
Model~\cite{FlavourLatticeAveragingGroupFLAG:2024oxs}, and
determines the thermodynamics of QCD at zero baryon
density~\cite{Borsanyi:2013bia,
HotQCD:2014kol}. The same Euclidean Monte Carlo approach extends to nuclear physics as well, computing the properties of multi-baryon systems, nuclear forces, and the lightest nuclei directly from QCD, and supplying the nuclear
matrix elements needed to interpret experiments that probe fundamental
symmetries~\cite{Beane:2010em,
NPLQCD:2012mex, Savage:2016kon, Chang:2017eiq}.

Classical Monte Carlo simulations of QCD, though strong at extracting static properties, encounter a sign problem in real-time evolution, where the path-integral weight becomes an oscillating complex phase rather than a positive probability. Sign problems also arise when quantum Monte Carlo is applied to fermions. Such sign problems are NP-hard~\cite{Troyer:2004ge}, so no growth in classical resources is expected to overcome them. Consequently, the real-time observables central to these non-equilibrium systems, such as how quarks lose energy traversing a medium or how quarks and gluons hadronize, are currently accessible only through phenomenological models whose parameters are tuned to data and whose uncertainties cannot be systematically reduced~\cite{Bierlich:2022pfr}.

A Hamiltonian formulation of the lattice theory~\cite{Kogut:1974ag} gets past the sign problem by evolving quantum states directly in real time. The
cost is that the state must now be stored and propagated explicitly, and here classical computers meet a different limit. Describing the state requires a number of amplitudes equal to the dimension of its Hilbert space, which grows exponentially with system size. Additionally, the entanglement of the state challenges tensor-network methods, and its quantum magic (non-stabilizerness), together with entanglement, makes the system generally hard to compute classically~\cite{Robin:2026lqp, Jha:2026ror,Froland:2026aff}. Both entanglement and quantum magic are physical properties of the states that need to be captured.

 A quantum computer stores such states in the physical state of its qubits. A device of $n_q$ qubits spans a $2^{n_q}$-dimensional Hilbert space, so the qubit count scales linearly with the system size of the target theory. Since these quantum computers are fundamentally quantum in nature, they natively capture the entanglement and quantum magic of the complex quantum systems. In quantum simulations, complex quantum phenomena are modeled and studied on controllable quantum systems, so we can access the state of the system throughout its evolution, gaining valuable insight into the underlying dynamics. 
 
 In recent years, there has been rapid progress in the development of quantum simulations~\cite{Hall:2021rbv,Yeter-Aydeniz:2021olz,Amitrano:2022yyn,Illa:2022zgu,Illa:2022jqb,Chai:2023qpq,
Farrell:2024fit,
Farrell:2024mgu,
Davoudi:2024wyv,
Zemlevskiy:2024vxt, Illa:2025dou, Turco:2025jot,
Farrell:2025nkx,
Chai:2025qhf,
Davoudi:2025rdv, Chernyshev:2025lil, Than:2025gso, Chen:2025zeh, Jiang:2025ufg, Hartnett:2026abf,Cogburn:2026aqy, Chen:2026tvd,Zemlevskiy:2026kpc}. These advances are expected to lead to predictive capabilities for the structure and dynamics of systems of fundamental particles that are far from equilibrium (for recent reviews, see Refs.~\cite{Banuls:2019bmf,Klco:2021lap,Funcke:2021aps, Bauer:2022hpo,Bauer:2023qgm,Funcke:2023jbq,2023arXiv230300113B,DiMeglio:2023nsa}). 
Quantum hardware is developing rapidly, but present-day devices remain too noisy for quantum simulations to be comparable with experiment, or to outperform classical computing on a problem of physical interest. Preparing and validating the algorithms now means that when capable hardware arrives, the methods to use it are already in hand. This dissertation lays out simulation frameworks for using future quantum computers and tests proposed algorithms on state-of-the-art near term devices. Chapter~\ref{chap:energy_loss} targets energy loss and hadronization in strongly-interacting matter, the physics relevant to studying the quark-gluon plasma; Chapter~\ref{chap:gauge_fix} constructs the quantum simulation for a formulation that may be more efficient in approaching the continuum limit; and Chapter~\ref{chap:seqht} develops a truncation method that reduces the quantum resources required for simulations, particularly for systems exhibiting a hierarchy of length scales.

\section{Recipe of Quantum Simulations}
\label{sec:recipe}

 Quantum simulation of physical systems generally involves three stages: preparing the initial state, evolving the state in time, and measuring observables of interest. Each stage presents distinct challenges and opportunities on near-term hardware, which we describe in this section. We also discuss the entanglement and quantum magic that motivate the use of quantum hardware. The Hamiltonian formulations specific to the theories studied in this dissertation are detailed in Sec.~\ref{sec:formulation}.
 
\subsection{State Preparation}
Preparing a target initial state on a quantum computer is difficult. Generally, a set of quantum gates form a unitary that rotates the all-zero initial state of the quantum computer to the target state.
In a brute-force preparation of an arbitrary state, the number of gates grows exponentially with system size. There is good news and bad news here. The good news is that we are interested in preparing eigenstates of particular physical systems, which are not arbitrary. The bad news is that we often have access only to the Hamiltonian, which may be difficult to diagonalize, and we may not even know the target state in advance. The structure and symmetries of the Hamiltonian, however, can be exploited to prepare ground states more efficiently. The work in this dissertation mainly prepares ground states, and two key methods are used: adiabatic state preparation, and variational quantum eigensolver (VQE). 

In adiabatic state preparation~\cite{Farhi:2000ikn, Jordan:2012xnu}, the quantum system is initialized in the ground state of an easy-to-solve Hamiltonian. The Hamiltonian is then slowly varied to minimize energy injection into the system, so the state stays a ground state of the updated Hamiltonian. Through this process, the known ground state slowly evolves into the ground state of the target complex Hamiltonian. The difficulty is that slow variation requires many evolution time steps, making the procedure challenging on near-term devices.

Another state-preparation method is VQE~\cite{Peruzzo:2013bzg}. At each step, the initial state is evolved with a parametrized circuit, and the relevant observable of the resulting state is measured. This result, together with the current parameters, is then passed to a classical optimizer, which proposes the next set of parameters. For the purpose of preparing a ground state, energy is measured and minimized during the iterations, so the state is driven towards the ground state. VQE was originally designed to be a hybrid quantum-classical algorithm. However, with the noise on current quantum computers, the energy measurements do not have the resolution required for the classical optimizer to converge. In practice, both the optimization and the evolution are run on classical computers for the system sizes relevant in this dissertation. 

An extension of VQE is the ADAPT-VQE~\cite{Grimsley_2019}. Instead of using a fixed parametrized circuit, the state preparation circuit will be built step by step. At each step, an operator is picked from an operator pool and attached to the circuit. The selection criteria could be gradient of the cost function. If this operator is parametrized, VQE is then used to find the parameter. This can be further extended to SC-ADAPT-VQE~\cite{Farrell:2023fgd}. For systems with translational invariance, the variational circuits can be constructed to have translational symmetry. The parameters can then be determined for a range of smaller system sizes, then extrapolated to arbitrarily large systems with controllable errors. 

Other state preparation methods include mid-circuit measurements~\cite{Baumer:2023vrf,Farrell:2025nkx}, non-unitary variational approaches~\cite{Song:2025pwd}, sample-based diagonalization~\cite{Robledo-Moreno:2024pzz}, and more. Beyond ground-state and low-energy-state preparations, thermal state preparations for quantum simulations are also being developed~\cite{Motta:2019yya,Sagastizabal:2020wfq,Warren:2022evv,Holmes:2022fic,Consiglio:2023mev, Turro:2023nuk,Robertson:2026com}.

\subsection{Time Evolution}
Real-time evolution is most commonly implemented through Trotterization~\cite{Trotter:1959ytf, Suzuki:1976be, Hatano:2005gh}. After digitization, the Hamiltonian can be separated into non-commuting groups of Pauli operators. For a simple example, consider a Hamiltonian that can be divided into two parts:
\begin{equation}
    \hat{H} = \hat{H}_A + \hat{H}_B, \,\,\, [\hat{H}_A,\hat{H}_B] \neq 0
\end{equation}
On quantum hardware, we cannot directly evolve the system with $e^{-i\hat{H}t}$, but since all the operators within $\hat{H}_{A/B}$ commute, $e^{-i \hat{H}_At}e^{-i \hat{H}_Bt}$ can be directly applied. This is the foundational idea of Trotter evolution.
Trotter evolutions are not exact. According to the Baker-Campbell-Hausdorff formula, 
\begin{equation}
    e^{-i \hat{H}_At}e^{-i \hat{H}_Bt} = e^{-i (\hat{H}_A+\hat{H}_B)t-\frac{1}{2}[\hat{H}_A,\hat{H}_B]t^2 + \dots}
\end{equation}
Trotter errors can be systematically improved by taking multiple time steps with smaller step sizes, and higher order Trotterization. Generalizing to a Hamiltonian with $k$ commuting groups, and evolving with $n$ Trotter steps, the Lie-Trotter (first-order) and Suzuki-Trotter (second-order) evolutions are
\begin{align}
    e^{-it\sum_j \hat{H}_j} & \rightarrow \left(\prod_j e^{-i\hat{H}_j\frac{t}{n}} \right)^n\\
    & \rightarrow \left(\prod_{j=0}^{k-1} e^{-i\hat{H}_j\frac{t}{2n}} \prod_{j=k-1}^{0} e^{-i\hat{H}_j\frac{t}{2n}} \right)^n
\end{align}
The error of first-order Trotter evolution scales as $\mathcal{O}(t^2/n)$, and second-order $\mathcal{O}(t^3/n^2)$~\cite{Childs:2019hts}, with prefactors set by commutator norms that can be large. Higher-order Trotterizations do not necessarily give better results, because the prefactor of their more complicated commutators could be larger. The errors can generally be mitigated by taking more, smaller Trotter steps to reach the same time, which is limiting on near-term devices due to the increased circuit depth. 

\subsection{Measuring Observables}
Quantum measurement is projective: reading out a register collapses the
state and returns a single eigenvalue of the measured observable rather
than its expectation value. Expectation values are therefore reconstructed
statistically. The circuit is executed many times (shots), and the expectation value of an observable is estimated as the sample mean of the measured eigenvalues, with a statistical uncertainty (shot noise) that scales as $1/\sqrt{N}$ with the number of shots $N$. Most quantum hardware only measures in the computational basis, that is, the eigenbasis of the Pauli-$Z$ operator.
Basis rotations are applied before the $Z$ measurements to diagonalize the observables. Observables that are diagonal in the same basis can be measured simultaneously from the same set of shots. 

\subsection{Entanglement, Quantum Magic, and Hardness of Simulation}
Not every quantum system is expensive to simulate classically, so it is
worth identifying what makes a simulation hard and motivates the use of
quantum hardware.
Two measures are worth considering. The first is entanglement~\cite{Horodecki:2009zz,Zhao:2024xtl}, the quantum correlations between subsystems that cannot be described independently. Classical tensor network methods represent states efficiently when their entanglement is bounded, and become intractable when entanglement grows rapidly, as it generally does in real-time dynamics~\cite{Cirac:2008rbg}. The second is non-stabilizerness, or quantum magic~\cite{Leone:2021rzd}, which quantifies how far a state is from stabilizer states, the class of states generated by Clifford operations. Clifford gates, for example, the Hadamard, phase, and CNOT gates, map Pauli operators to other Pauli operators through conjugation. Quantum magic is related to the number of non-Clifford gates needed to prepare a state. 
The Gottesman-Knill theorem~\cite{Gottesman:1998hu,Aaronson:2004xuh} states that stabilizer circuits, which are circuits containing only Clifford gates, can be simulated efficiently on a classical computer, regardless of their entanglement. Universal quantum computation requires at least one non-Clifford gate. The T gate, a single-qubit phase rotation by $\pi/4$,  is one such example.
Magic reflects the amount of T gates needed to prepare a state, and is necessary (but not sufficient) for a quantum advantage.

On current devices, two-qubit gates such as CNOT are the leading source of error and therefore the primary resource constraint, even though CNOT is part of the Clifford gate group. Single-qubit T gates are comparatively cheap, with shorter gate times and higher fidelity. However, in the context of fault-tolerant quantum computing, this reverses. In fault-tolerant architectures, a logical qubit is encoded across many physical qubits, and quantum information is protected by redundancy. Some logical gates can be implemented transversially by acting the same physical gates on each individual physical qubit. So an error on one physical qubit cannot spread through the performance of this gate. Such gates are fault-tolerant and have constant depth. The Eastin-Knill theorem~\cite{Eastin:2009tem} forbids a universal transversal gate set, so at least one gate, which is usually the T gate, must be implemented in another way. A common approach distills noisy magic states into high fidelity ones~\cite{Bravyi:2004isx, Bravyi:2012lxn} in a T-gate factory, then consumes them to apply T gates to the data qubits. Additional methods to address the non-transversal gates are actively researched, for example, code switching~\cite{Anderson:2014jvy} and magic-state cultivation~\cite{Gidney:2024alh}. 

\section{Quantum Hardware}
\label{sec:hardware}
\subsection{Status of Current Hardware}
Several qubit modalities are currently being developed for quantum computing. Each modality has different strengths and weaknesses across the important metrics of qubit count, gate fidelity, gate time, and connectivity. Currently, no single platform is leading on all fronts. 

The foundations of superconducting qubits were laid out at UC Berkeley and recognized by the 2025 Nobel Prize in Physics~\cite{PhysRevLett.53.1260,PhysRevLett.55.1543, Devoret:1985yga}. The transmon design~\cite{Koch:2007hay} underlies most devices in use today, which are now built by companies such as IBM, Google, IQM, and Rigetti, with continued advances in academia such as fluxonium qubits at Maryland~\cite{Somoroff:2021elj} and large-scale processors at USTC~\cite{Gao:2024fik}. Superconducting quantum computers offer the fastest gates but are constrained by the limited connectivity of a fixed planar lattice. Trapped ions, developed at NIST and Innsbruck~\cite{PhysRevLett.74.4091,PhysRevLett.75.4714} and by Quantinuum and IonQ, provide all-to-all connectivity and the highest two-qubit gate fidelities~\cite{daSilva:2024tfc}, at the cost of slower and more expensive operations and fewer qubits. Neutral-atom platforms, developed by groups at Harvard, MIT, and Caltech~\cite{Jaksch:2000cus,Bluvstein:2023zmt,Manetsch:2024lwl,Chiu:2025uis}, and by companies such as QuEra, Pasqal, and Infleqtion, offer large qubit counts and reconfigurable geometries, but operate more slowly and rely on whole-array imaging for readout, which makes mid-circuit and partial measurements nontrivial. Other approaches under active development include silicon qubits~\cite{xue2022quantum, MembersoftheHRLQuantumTeam:2026slx}, photonic~\cite{Zhong:2020iql,Alexander:2024vys,AghaeeRad:2025lqu} and topological qubits~\cite{microsoft2025interferometric}, and qudit-based hardware~\cite{Kim:2025ywx}. The hardware demonstrations in this dissertation are carried out on IBM's superconducting processors.

\subsection{Error Handling}
Because quantum devices are noisy, handling errors is essential for extracting accurate physical observables from them. 
There are three aspects to quantum error handling.
Quantum error suppression and mitigation are important for running algorithms on present-day hardware and are used throughout this dissertation, while quantum error correction is essential for the longer-term goal of fault-tolerant quantum computation.

Error suppression lowers the error rate at the device and control level. One element we can control on the software level is dynamical decoupling (DD)~\cite{Viola:1998jx,Souza_2012,Ezzell:2022uat}, which applies pulses to idle qubits to suppress crosstalk from neighboring qubits. The pulse sequence and whether to enable DD are both design choices. If the circuit is dense, DD may hurt because of imperfect pulses.

Error mitigation uses large ensemble sizes and statistics to reduce the effect of errors in the observables of interest. There are a few established methods. Pauli Twirling~\cite{Wallman:2015uzh} converts coherence errors into incoherent errors. Readout Error Mitigation~\cite{Berg:2020ibi, Nation:2021kye} targets measurement noise. Operator Decoherence Renormalization~\cite{Urbanek:2021oej,ARahman:2022tkr,Farrell:2023fgd, Farrell:2024fit} implements a set of mitigation circuits to run alongside the physics circuits on hardware. These mitigation circuits share the structure of the physics circuits, but only contain Clifford gates, making them easy to simulate classically. The decoherence of the mitigation circuits can then be used to calibrate the hardware results of the physics circuits. Zero Noise Extrapolation~\cite{Li:2016vmf,Temme:2016vkz,PhysRevLett.120.210501,Klco:2018kyo,Kandala2019Error} amplifies the noise, for example  by inserting additional CNOT layers, and extrapolates back to the zero-noise limit. 

Quantum Error Correction encodes logical information redundantly across physical qubits and detects and corrects errors through repeated syndrome measurements. A key concept is the threshold: once the physical error rate falls below a code-dependent value, the logical error rate is suppressed exponentially in the code distance. There are a plethora of error correction and error detection codes, cataloged in the Error Correction Zoo~\cite{albert2026handbook}. The use of gauge redundancies and symmetries of physical systems as error correction codes is also explored~\cite{Rajput:2021trn, Carena:2024dzu, Spagnoli:2026qni}. Error-correction experiments on hardware so far center on quantum memory~\cite{Terhal:2013vbm}, the preservation of one or a few logical qubits. Below-threshold operation for quantum memory has recently been demonstrated on superconducting hardware~\cite{GoogleQuantumAIandCollaborators:2024efv}, though as a calibration-intensive demonstration on a single device rather than a generally available capability. Fault-tolerant logical operations, the gates required for general-purpose computation, have also been explored on trapped-ion and neutral-atom hardware~\cite{Bluvstein:2023zmt, daSilva:2024tfc, Reichardt:2024xfs, Bluvstein:2025ped}. Even though current demonstrations are at modest scales, the field is advancing rapidly, and useful fault-tolerant quantum computers are on the horizon. As hardware advances toward fault-tolerance, quantum simulations will need to incorporate error correction through codesign efforts, rather than waiting for general-purpose logical qubits.

\section{Hamiltonian Formulations}
\label{sec:formulation}
To simulate a field theory on a quantum computer, it must first be represented in a finite-dimensional Hamiltonian form and
digitized onto qubits. Quantum field theories have infinite-dimensional Hilbert spaces, both because space is continuous and because bosonic fields can have continuous values at each spatial point. The first is addressed by formulating the theory on a spatial lattice; the second by truncating and digitizing the local Hilbert space at each lattice site. Fermionic fields, on the other hand, are finite-dimensional and map directly to qubits. This section presents the formulations used throughout this dissertation from the digitization of scalar fields to gauge theories. 

\subsection{Digitizing the Scalar Fields}
For a scalar field, the field operator $\hat{\phi}_x$ at site $x$ can be represented in its own eigenbasis, sampled evenly at $2^{n_q}$ points for $n_q$ qubits across a finite domain $[-\phi_{\rm max}, \phi_{\rm max}]$, where $\phi_{\rm max}$ is the maximum field value allowed.
\begin{equation}
\hat{\phi}_x \ = \ \sum_{j=0}^{2^{n_q}-1} \phi_j \, \ket{\phi_j}\bra{\phi_j}_x
\ , \qquad
\phi_j \ = \ -\phi_{\rm max} \, + \, j \, \delta_\phi
\ ,
\label{eq:phi_digitization}
\end{equation}
with sampling interval $\delta_\phi = 2\phi_{\rm max}/(2^{n_q}-1)$.

Just as position and momentum are related by a Fourier transform in quantum mechanics, the conjugate-momentum operator $\hat{\Pi}^2$ is diagonal in the Fourier space. It is therefore implemented via a local quantum Fourier Transform, which are circuits that performs a discrete Fourier transform on the qubits. 
Digitizing the field and its conjugate momentum symmetrically
preserves the discrete symmetries of the Hamiltonian and improves the efficiency of the simulation~\cite{Klco:2018zqz}. While the discussion so far concerns the free field, interactions are straightforward to include. Any potential that is a function of the field, such as the $\lambda\phi^4$ term studied in Chapter~\ref{chap:seqht}, is diagonal in the field basis and can be implemented similarly to the field operator.

By the Nyquist-Shannon sampling theorem~\cite{Macridin:2018gdw,Macridin:2018oli}, if a continuous wavefunction is sampled across its full regions of support in both field space and conjugate-momentum space, it can be reconstructed with exponentially small errors introduced by the discretization. Since the free scalar field theory on one site is a harmonic oscillator with a Gaussian ground state, its localization allows for the determination of a field cutoff $\phi_{\rm max}$ that preserves the relevant information. There is a tradeoff with the choice of $\phi_{\rm max}$: for a fixed number of qubits, a smaller $\phi_{\rm max}$ allows a finer resolution in the central region, while a larger $\phi_{\rm max}$ contains the tails of the wavefunction better. 

Another feature of this digitization is the foundation of Chapter~\ref{chap:seqht}. Since we are working in the field basis, the field operators can be represented by a real-diagonal matrix, which can be decomposed into a sum of tensor products of single qubit operators $\hat{I}$s and $\hat{Z}$s. These tensor products form an orthogonal basis called the Walsh basis, which can be seen as the digital counterpart of a Fourier basis. Each term is labeled by a unique sequency, which is the digital analog of frequency. The decomposition of the low-energy wavefunctions are dominated by low-sequency terms, which is the structure that the truncation scheme in Chapter~\ref{chap:seqht} exploits.

\subsection{The Kogut-Susskind Formulation on the Lattice}

The Kogut-Susskind~\cite{Kogut:1974ag} Hamiltonian is one of the formulations that study the Yang-Mills lattice gauge theory. It adopts a staggered formalism, and represents fermionic matter fields on the sites and bosonic gauge fields on the links. 
The chromo-electric and chromo-magnetic contributions are
\begin{equation}
\hat{H}=\frac{g^2}{2 a^{d-2}} \sum_{b, \text { links }}\left|\hat{\mathbf{E}}^{(b)}\right|^2+\frac{1}{2 a^{4-d} g^2} \sum_{\text {plaquettes }}\left[2 N_c-\hat{\square}(\mathbf{x})-\hat{\square}^{\dagger}(\mathbf{x})\right]
\end{equation}
where $g$ is the coupling strength, $a$ is the lattice spacing between adjacent spatial sites, $d$ is the number of spatial dimensions, and $N_c$ is the number of colors. $\hat{\mathbf{E}}^{(b)}$ is the electric operator, and $\hat{\square}(\mathbf{x})$ is the plaquette operator. The full Hamiltonian also contains the staggered fermion mass term and the hopping term that couples the matter and the gauge fields.

In one spatial dimension, the gauge field carries no independent dynamics of its own, so it can be integrated out such that only the fermionic modes remain. Gauss's law can be enforced to constraints the electric field to the charges. With open boundary conditions in the axial gauge and one flavor, the Hamiltonian form becomes~\cite{Ciavarella:2023mfc}
\begin{equation}
\begin{aligned}
& \hat{H}=\hat{H}_{k}+\hat{H}_m+\hat{H}_g \\
& \hat{H}_{k}=\sum_{x, a} \frac{1}{2} \hat{\psi}_{x, a}^{\dagger} \hat{\psi}_{x+1, a}+\text { h.c. } \\
& \hat{H}_m=m \sum_{x, a}(-1)^x \hat{\psi}_{x, a}^{\dagger} \hat{\psi}_{x, a} \\
& \hat{H}_g=\sum_{x, c} \frac{g^2}{2}\left(\sum_{y\leq x} \hat{Q}_y^c\right)\left(\sum_{y\leq x} \hat{Q}_y^c\right),
\end{aligned}
\end{equation}
where $\hat{Q}_y^c$ is the chromo-electric charge at spatial site $y$, 
\begin{equation}
\hat{Q}_y^c=\sum_{a, b} \hat{\psi}_{y, a}^{\dagger} T_{a, b}^c \hat{\psi}_{y, b}.
\end{equation}
$T_{a, b}^c$ are proportional to the Gell-Mann matrices for SU(3), and Pauli matrices for SU(2). Labels $a$ and $b$ are the color indices. 

The fermion fields can be mapped to qubits via the Jordan-Wigner Transformation~\cite{Jordan:1928wi}, 
\begin{equation}
\begin{aligned}
\psi_i^{\dagger} & =\prod_{j<i}\left(-\hat{Z}_j\right) \sigma_i^{+} \\
\psi_i & =\prod_{j<i}\left(-\hat{Z}_j\right) \sigma_i^{-},
\end{aligned}
\end{equation}
with $\hat{\sigma}^\pm = (\hat{X} \pm i \hat{Y})/2$. Each fermionic mode is represented on one qubit, and the anti-commutation relations are implemented through strings of Pauli-$\hat{Z}$ operators. 

One of the most studied theories in one spatial dimension is the U(1) theory, also called the Schwinger Model. It is not directly used in this dissertation, but lays the theoretical foundation for the work in Chapter~\ref{chap:energy_loss}. The Schwinger model is a 1+1D quantum electrodynamics (QED) system, and it exhibits confinement. As an Abelian system, it is easier to study than non-Abelian systems, but can provide insight into more complicated non-Abelian theories. The work in Ref.~\cite{Farrell:2024mgu} is of particular importance to Chapter~\ref{chap:energy_loss}. There, the authors implemented background heavy charge in the Schwinger model via a discontinuity in Gauss's law, and studied how energy and momentum are transferred as a heavy hadron moves through a vacuum and a dense medium. Continuous motion was achieved by a superposition of the heavy charge location over two neighboring sites. Still, when the charge is moved too quickly, lattice artifacts appear and destroy the entanglement structure. Chapter~\ref{chap:energy_loss} extends this study to SU(2), where the discontinuity in Gauss's law and the superposition trick can no longer be used because of the non-commuting charge operators in SU(2). 

In two or more spatial dimensions, the gauge field cannot
be eliminated entirely, and its infinite-dimensional link Hilbert space must be truncated and digitized. A range of approaches is being explored to efficiently encode the continuous gauge groups on quantum computers, 
for example Loop-String-Hadron~\cite{Raychowdhury:2019iki,  Kadam:2022ipf,Kadam:2024ifg,Kadam:2025trs}, Large-$N_c$ truncations~\cite{Ciavarella:2024fzw,Ciavarella:2025bsg,Modi:2026syn}, or using an alternative formulation like the orbifold lattice~\cite{Buser:2020cvn,Bergner:2024qjl,Halimeh:2024bth}. A comprehensive comparison across all bases and formulations has yet to be performed. Chapter~\ref{chap:gauge_fix} works with one of the candidates that is designed to be efficient at all gauge couplings: the fully gauge-fixed Hamiltonian in the mixed basis. A maximal-tree gauge fixing
removes all redundant link degrees of freedom, and the remaining physical degrees of freedom are described in a
basis that interpolates between electric and magnetic descriptions, with a coupling-dependent digitization designed
to be efficient at all values of the gauge coupling.

\section{Dissertation Outline}
\label{sec:overview}
The goal of this dissertation is to develop and demonstrate resource-efficient methods for the digital quantum simulation of the real-time dynamics of quantum field theories relevant to nuclear and high-energy physics, and to validate them on present-day quantum hardware. 

Simulations of energy loss and hadronization are essential for understanding a range of phenomena in non-equilibrium strongly-interacting matter.
In Chapter~\ref{chap:energy_loss}, 
a framework for performing such simulations on a quantum computer is established and applied to a heavy quark moving across a 1+1D SU(2) lattice of light quarks.
Conceptual advances with regard to simulations of non-Abelian versus Abelian theories are developed. The evolution of the energy in light quarks, their local non-Abelian charge densities, and their multi-partite entanglement are computed.
Unlike in the Abelian theory, non-Abelian charge operators have non-trivial actions on arbitrary states, so, instead of implementing the background charges via discontinuities in Gauss's law, the heavy quarks are mapped to qubits alongside the light quarks. The heavy-quark motion is also limited to discrete steps among spatial lattice sites.

Further, scalable quantum circuits are used to prepare the interacting ground-state wavefunction in the presence of heavy quarks. Classical lattice QCD methods inspire the state preparation procedure. Hadronic interpolating operators implement the color entanglement among the heavy quarks and light quarks effectively, and Domain Decomposition addresses the heterogeneity of non-Abelian charge sectors across the lattice.
The discrete motion of heavy quarks between adjacent spatial sites is implemented using fermionic SWAP operations.
Quantum simulations of the dynamics of a system on $L=3$ spatial sites are performed using IBM's {\tt ibm\_pittsburgh} quantum computer using 18 qubits, for which the circuits for state preparation, motion, and one second-order Trotter step of time evolution have a two-qubit depth of 398 after transpilation.
A suite of error mitigation techniques  are used to extract the observables from the simulations, providing results that are in good agreement with classical simulations.
The framework presented in this chapter generalizes straightforwardly to other non-Abelian groups, including SU(3) for quantum chromodynamics.

The gauge field in Chapter~\ref{chap:energy_loss} could be integrated out because of the open boundary conditions in 1+1D. However, in higher-dimensions, the dynamical gauge fields cannot be eliminated and will be nontrivial to represent. Moreover, the electric basis is efficient only at strong coupling. To accurately approach the weak coupling limit, it is important to formulate systematically-improvable representations of lattice gauge theory Hamiltonians that are efficient at all values of the gauge coupling. Chapter~\ref{chap:gauge_fix} constructs the quantum simulation of an alternative Hamiltonian formulation: the fully gauge-fixed Hamiltonian defined in the mixed basis.  The smallest non-trivial system in this formulation, two plaquettes with open boundary conditions, is dissected. A mapping of the continuous gauge field degrees of freedom to qubit-based representations is developed. It is found that as few as three qubits per plaquette is sufficient to reach $10^{-3}$ precision on predictions for observables. Two distinct algorithms for implementing time evolution in the mixed basis are developed and analyzed in terms of quantum resource estimates. One algorithm has favorable scaling in circuit depth for large numbers of qubits, while the other is more practical when qubit count is limited. The latter algorithm is used in the measurement of a real-time observable on IBM's Heron superconducting quantum processor, {\tt ibm\_fez}. The quantum results match classical predictions with precision around $10^{-2}$. This work lays out a path forward for two- and three-dimensional simulations of larger systems, as well as demonstrating the viability of mixed-basis formulations for studying the properties of SU(2) gauge theories at all values of the gauge coupling.

Decoherence and noise limit the scale of simulations that can be run on present-day hardware, making it important to use the available quantum resources as efficiently as possible. In Chapter~\ref{chap:seqht}, we introduce the Sequency Hierarchy Truncation (SeqHT) scheme for compressing the resources required for state preparation and time evolution in quantum simulations, while preserving the relevant physics.
For the $\lambda\phi^4$ interaction in scalar field theory, or any interaction with a polynomial expansion, upper bounds on the contributions of operators of a given sequency are derived to confirm that truncation errors are bounded.
For the systems we have examined, observables computed in sequency-truncated wavefunctions, including quantum correlations as measured by magic, are found to step-wise converge to their exact values with increasing cutoff sequency, making this truncation scheme systematically improvable. 
The utility of SeqHT is demonstrated in the adiabatic state preparation of the $\lambda\phi^4$ anharmonic oscillator ground state 
using IBM's quantum computer {\tt ibm\_sherbrooke}. 
Using SeqHT, the depth of the required quantum circuits is reduced by $\sim 30\%$. Since less hardware noise is introduced in this shallower circuit, determinations of observables in the quantum simulations are significantly improved.
More generally,  SeqHT is expected to lead to a reduction in required resources for quantum simulations of systems with a hierarchy of length scales.
 
% ========== Chapter 2
 
\chapter{\texorpdfstring{A Framework for Simulating Energy-Loss and Hadronization\\in 1+1D SU(2) Lattice Gauge Theory}{A Framework for Simulating Energy-Loss and Hadronization in 1+1D SU(2) Lattice Gauge Theory}}
\label{chap:energy_loss}

\vspace{1em}
\begin{center}
    \itshape
    This chapter is associated with Ref.~\cite{Li:2025sgo}: 
    ``A Framework for Quantum Simulations of Energy-Loss and Hadronization in Non-Abelian Gauge Theories: SU(2) Lattice Gauge Theory in 1+1D''\\by Zhiyao Li, Marc Illa, and Martin J. Savage
\end{center}
\vspace{2em}

%%%%%%%%%%%%%%%%%%%%%%%%%%%%

\section{Introduction}
\label{sec:Intro}
\noindent
Extensive experimental campaigns in high-energy heavy-ion collisions are designed to reveal the structure and dynamics of matter under extreme conditions.  
The remarkable discovery of a near-ideal, low-viscosity quantum liquid of quarks and gluons produced in such collisions~\cite{PHENIX:2004vcz,BRAHMS:2004adc,PHOBOS:2004zne,STAR:2005gfr} has driven a program to measure and interpret its properties.
An important line of investigation is to quantify properties of transport in the liquid phase, where the identification of signatures and probes present a challenge.
One class of probes  is the production, motion and measurement of heavy quarks, including the production rates of quarkonia,  and processes such as dissociation and recombination.
Produced in hard collisions at early times, heavy quarks  witness the entire evolution of the exotic matter and its transition to hadronic final states. 
Therefore, the measured relative yields of heavy-quark systems reveal key attributes of the medium.
An excellent review of the state-of-the-art in this area can be found in Ref.~\cite{He_2023}.
An important element in extracting properties of this exotic matter is understanding how heavy quarks move and interact within the evolving system, including how they lose and gain energy as they move.
From a theoretical standpoint, first-principles simulations of these collisions present a monumental challenge.  
This challenge is not limited to evolution through exotic matter.   
Even simulations of high-energy probes impacting cold dense matter, such as a nucleus, or nuclear matter, suffer the same difficulties.  
While processes occurring over short-distances can be reliably determined in perturbation theory, the subsequent evolution (cascade) to longer length scales involves fragmentation and hadronization, both of which are intrinsically non-perturbative. 
Tuned to experimental data, modeling such processes is impressively successful, e.g., Geant4~\cite{GEANT4:2002zbu,Muskinja:2020ots,basaglia2024geant4gamechangerhigh,Bierlich:2022pfr},
but remains incomplete in describing fundamentally quantum observables, e.g., Refs.~\cite{Bauer:2019qxa,Klco:2021lap,Bauer:2022hpo,Gray:2022fou,Bauer:2023qgm,DiMeglio:2023nsa}.
While classical Euclidean-space
simulations of quantum chromodynamics (QCD) encounter a sign problem in summing complex phases in the time-evolution,
real-time evolution, starting from a specified initial state, is known to be efficient for quantum computers.

Constrained by available noisy intermediate-scale quantum (NISQ)~\cite{Preskill:2018jim} era quantum hardware, present-day quantum simulations of quantum field theories are focused on modest-sized, low-dimensional systems~\cite{Martinez:2016yna,
Klco:2018kyo,
Kokail:2018eiw,
Lu:2018pjk,
Klco:2019evd,
Mil:2019pbt,
Yang:2020yer,
Ciavarella:2021nmj,
Bauer:2021gup,
Atas:2021ext,
ARahman:2021ktn,
Mazzola:2021hma,
deJong:2021wsd,
Gong:2021bcp,
Zhou:2021kdl,
Riechert:2021ink,
Ciavarella:2021lel,
Nguyen:2021hyk,
Su:2022glk,
Illa:2022jqb,
Mildenberger:2022jqr,
ARahman:2022tkr,
Asaduzzaman:2022bpi,
Farrell:2022wyt,
Atas:2022dqm,
Farrell:2022vyh,
Mueller:2022xbg,
Pomarico:2023png,
Charles:2023zbl,
Zhang:2023hzr,
Ciavarella:2023mfc,
Farrell:2023fgd,
Meth:2023wzd,
Borzenkova:2023xaf,
Schuster:2023klj,
Angelides:2023noe,
Farrell:2024fit,
Kavaki:2024ijd,
Davoudi:2024wyv,
Turro:2024pxu,
Ciavarella:2024fzw,
Gyawali:2024hrz,
Gonzalez-Cuadra:2024xul,
Zemlevskiy:2024vxt,
Crippa:2024hso,
Ciavarella:2024lsp,
Than:2024zaj,
Turro:2025sec,
Alexandrou:2025vaj,
Farrell:2025nkx,
Davoudi:2025rdv,
Schuhmacher:2025ehh,
Chernyshev:2025lil,
Chen:2025zeh,
Cobos:2025krn,
Saner:2025nrq,
Than:2025gso},
with the Schwinger model~\cite{Schwinger:1962tp} (quantum electrodynamics (QED) in 1+1D) receiving significant attention.
Like QCD, the Schwinger model is a confining gauge theory, and remains a useful test-bed for exploring new algorithms and workflows and gaining improved understanding of infinite-volume and continuum limits.
This chapter is related to simulating transport, in a general sense, in non-equilibrium non-Abelian systems, 
and is on the path for robust simulations of jet physics and hadronization in QCD. 
There is a growing foundation for such simulations, e.g., 
Refs.~\cite{
Kuhn:2015zqa,
Kasper:2015cca,
Sala:2018dui,
Surace:2019dtp,
Magnifico:2019kyj,
Alexandru:2019nsa,
Li:2020uhl,
Bauer:2021gup,
Ji:2020kjk,
Barata:2021yri,
Li:2021zaw,
Yao:2022eqm,
Barata:2022wim,
Carena:2022kpg,
Florio:2023dke,
Barata:2023clv,
Florio:2024aix,
Alexandrou:2025vaj,
Barata:2025hgx}, 
along with progress in string breaking~\cite{
Hebenstreit:2013baa,
Lee:2023urk,
Belyansky:2023rgh,
Barata:2023jgd,
De:2024smi,
Surace:2024bht,
Ciavarella:2024lsp,
Gonzalez-Cuadra:2024xul,
grieninger2025st,
Borla:2025gfs,
Xu:2025abo,
Tian:2025mbv,
Cataldi:2025cyo}, 
and of the associated entanglement evolution~\cite{Chai:2023qpq,Papaefstathiou:2024zsu}.
Simulations of SU(2) and SU(3) theories including quarks and anti-quarks in 1+1D, with the gluon fields constrained by Gauss's law (either as dynamical fields or through long-range interactions), are now being performed using quantum computers~\cite{Klco:2019evd,
Ciavarella:2021nmj,
Atas:2021ext,
Farrell:2022wyt,
Farrell:2022vyh,
Atas:2022dqm,
Than:2024zaj,
Chernyshev:2025lil}.
With the Jordan-Wigner mapping of quarks, each spatial lattice site requires $2 n_f N_c$ qubits for $n_f$ flavors of light quarks with $N_c$ colors.
The inclusion of electroweak processes~\cite{Farrell:2022vyh,Chernyshev:2025lil} (using the low-energy effective field theory of the Standard Model) requires additional $4 n_g$ qubits encoding the charged-leptons and neutrinos, where $n_g$ is the number of generations included in the simulation.
Progress has also been made toward simulations using qudit systems, e.g., Refs.~\cite{Gustafson:2021qbt,
Meth:2023wzd,
Zache:2023cfj,
Gustafson:2023kvd,
Calajo:2024qrc,
Illa:2024kmf,
Gustafson:2024kym,
Crane:2024tlj,
Kurkcuoglu:2024cfv,
Gaz:2025ptu,
Jiang:2025ufg,
Huie:2025yzn,
Kurkcuoglu:2025gik,
Perez:2025cxl},
which have significant advantages in circuit depth and execution times.
Additionally, there have been a number of important quantum and classical algorithm developments for wavepacket preparation for scattering~\cite{Chai:2023qpq,
Farrell:2024fit,
Farrell:2024mgu,
Davoudi:2024wyv,
Zemlevskiy:2024vxt,
Turco:2025jot,
Farrell:2025nkx,
Chai:2025qhf,
Davoudi:2025rdv}, and for 
systematically-improvable truncation schemes~\cite{Li:2024lrl}.

Building on previous work that established a framework for quantum simulations of energy-loss and fragmentation in systems of electrically-charged particles in the context of Abelian lattice gauge theory (LGT)~\cite{Farrell:2024mgu}, this chapter addresses energy loss in non-Abelian lattice gauge theories.\footnote{This should be considered to be a specific case of quantum thermodynamics of a non-equilibrium process~\cite{YungerHalpern:2019mvj,Davoudi2024a}, in the context of a non-Abelian lattice gauge theory.}
In particular, we establish a framework to simulate the energy loss and collisions of particles in an SU(2) non-Abelian lattice gauge theory in 1+1D using quantum computers.
Static heavy quark fields are mapped to qubits in order to simulate hadron motion and collisions on the lattice (as well as to establish regions of dense matter). 
To prepare ground-state  wavefunctions in the presence of a heavy quark, SC-ADAPT-VQE~\cite{Grimsley_2019,Farrell:2023fgd} is combined with Domain Decomposition 
(DDec), an algorithm used extensively in traditional lattice QCD calculations~\cite{10.1093/oso/9780198501787.001.0001,saad2003iterative,Luscher:2003qa,Heybrock2014LatticeQW,frommer2014DD,Ce:2016idq}, leveraging the heterogeneity of charge sectors across different regions on the lattice. 
The heavy quarks are translated across the lattice via discrete fermionic-SWAP operators, introducing additional finite lattice-spacing artifacts that vanish in the continuum limit.
The states are then time-evolved to allow the light quarks to respond dynamically to the heavy quark motion. 
Scalable quantum circuits~\cite{Farrell:2023fgd} are developed for state preparation, discrete translation, and time evolution, enabling measurements of the time dependence of local non-Abelian charge fluctuations, entanglement, and the energy in the light quarks and gluons.
Complete simulations (state preparation, translation, time evolution and measurement, with associated mitigation circuits) of a lattice with $L=3$ spatial sites are performed, requiring a total of 18 qubits, using IBM's heron-processor quantum computers, with a transpiled CNOT depth of $\sim 400$. 
With multiple error-mitigation strategies applied, the results we obtain from the quantum simulations are in good agreement with corresponding classical simulations.

The simulations presented in this chapter were made possible by four conceptual advances.
One concerns the color entanglement between the heavy quark and the light degrees of freedom. 
Algorithms from traditional lattice QCD point toward working with color-singlet hadronic interpolating operators as an efficient way to handle color entanglement.
They provide a way to include the required symmetries of the wavefunction without qubit-wise manipulations.
A second development is appreciating the implications of mapping background static charges to the quantum register, and not including them through Gauss's law.
As the evolution operator for the light degrees of freedom does not commute with the heavy-quark fermionic SWAP operations, moving  heavy quarks to positions between lattice sites does not achieve what is intended, and as such, only translations between lattice sites are consistent.
Third, during the collision of two heavy quarks (when they are on adjacent sites), the SWAP operation interchanges them, and introduces an additional lattice artifact that can be addressed in two ways (or more).
Fourth, as the deviation of the wavefunction of the light degrees of freedom from the interacting vacuum state is localized around the heavy quark, it becomes an increasingly small part of total wavefunction infidelity in the variational process.
Therefore, partitioning the system in such a way that the heavy-quark modification is not sub-leading, DDec permits effective state preparation.

%%%%%%%%%%%%%%%%%%%%%%%%%%%%
\section{The SU(2) Lattice with Heavy Quarks}
\label{sec:what}
\begin{figure}[ht!]
    \centering
\includegraphics[width=\linewidth]{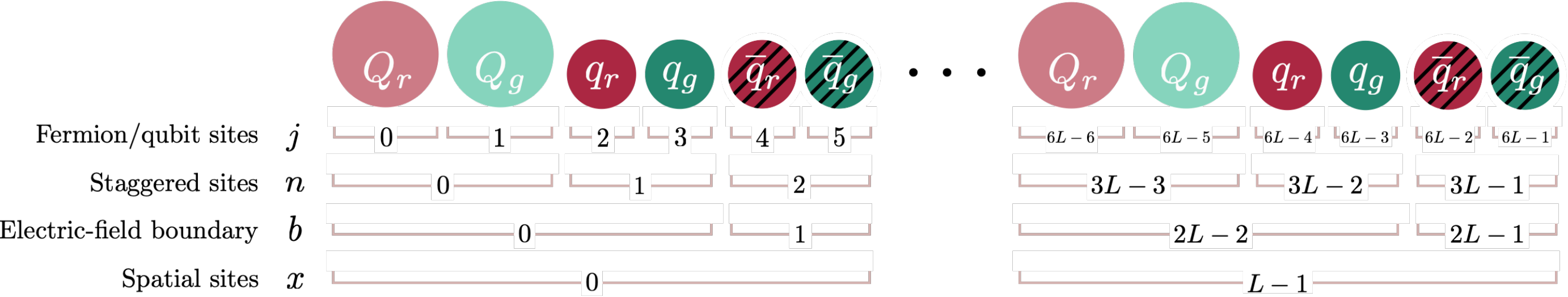}
\caption{
The lattice layout for heavy-quark, light-quark and light anti-quark fields on a lattice of spatial extent $L$.  
The spatial sites are labeled by $x$, which takes values from $x=0$ to $x=L-1$, the staggered sites are labeled by $n$ which takes values $n=0$ to $n=3L-1$, and the fermion (or qubit) sites are labeled by $j$ which takes values $n=0$ to $n=6L-1$.
The boundaries between quark and anti-quark sectors, denoted by $b$, correspond to the locations of chromo-electric field contributing to the Hamiltonian via the non-local charge operators. }
    \label{fig:layout_lattice}
\end{figure}

\noindent
The theory of focus in this chapter is the 1+1D SU(2) non-Abelian lattice gauge theory (LGT) with one flavor of light quark.
In the axial gauge with the Jordan-Wigner (JW) mapping of fermions to qubits, each spatial site of the lattice consists of non-dynamical heavy-quark fields, $Q_{r,g}$, light-quark fields, $q_{r,g}$, and light anti-quark fields, $\overline{q}_{r,g}$.
Heavy anti-quark fields are not included to reduce qubit overhead. 
The mapping of this system to qubits is shown in Fig.~\ref{fig:layout_lattice} for a lattice of spatial extent $L$.

The Hamiltonian describing these systems has three distinct contributions: 
kinetic energy, mass energy and chromo-electric interaction energy.
The kinetic energy operator  maps to spin operators acting on the spins associated with the 
light quark and anti-quark occupations.
In axial gauge with open boundary conditions (OBCs) it becomes,
\begin{equation}
\begin{aligned}
\hat H_k = 
& -\frac{1}{2}
\sum_{n\in q}\sum_{c=0,1}
\left(
\hat \sigma^+_{2n+c} \hat Z_{2n+1+c} \hat \sigma^-_{2n+2+c} 
 + \rm{h.c.} \right) \\
 & - \frac{1}{2}
\sum_{n \in \overline{q}}\sum_{c=0,1}
\left( 
\hat \sigma^+_{2n+c} \hat Z_{2n+1+c} \hat Z_{2n+2+c} \hat Z_{2n+3+c} \hat \sigma^-_{2n+4+c} + \rm{h.c.} \right)
\ ,
\label{eq:kinL}
\end{aligned}
\end{equation}
where $n$ is the staggered index, $c$ the color index, and $\hat{\sigma}^{\pm} = (\hat{X} \pm i \hat{Y})/2$. 
The first term corresponds to hopping between quarks and anti-quarks
within one spatial site, while the second term corresponds to hopping between adjacent sites.
For the heavy quark to remain static (localized), effectively ``integrating it out'' of the low-energy effective lattice theory,
its kinetic term is not included. 

The mass-energy contribution to the Hamiltonian receives contributions from the $Q$, $q$ and $\overline{q}$.
For arbitrary heavy-quark and light-quark masses,  
the mass term becomes 
\begin{equation}
\begin{aligned}
\hat H_m & =
m_Q \sum\limits_{n \in Q}
\frac{1}{2}\left(\hat I + \hat Z\right)_n
 +\ 
m_q \sum\limits_{n \in q}
\frac{1}{2}\left(\hat I + \hat Z\right)_n
 +\ 
m_q \sum\limits_{n \in \overline{q}}
\frac{1}{2}\left(\hat I - \hat Z\right)_n
\\ & \rightarrow \ 
\frac{m_q}{2}
\sum\limits_{n \in q, \overline{q}}\sum_{c=0,1}
(-1)^{\eta_n}\  \hat Z_{2n+c}
\ \ ,
\end{aligned}
\label{eq:massterm}
\end{equation}
where $\eta_q=0$ and $\eta_{\overline{q}} = 1$.
We have set $m_Q=0$ for simplicity of calculation, 
and removed contributions from identity operators.

Contributions to the Hamiltonian
from the chromo-electric fields are determined by combinations of the non-Abelian SU(2) charge operators~\cite{Atas:2021ext}, 
\begin{equation}
\begin{aligned}
\hat{\mathcal{Q}}_n^{(1)} & = 
\frac{1}{2}
\left(
\hat \sigma^+_{2n}\hat \sigma^-_{2n+1}+
\hat \sigma^-_{2n}\hat \sigma^+_{2n+1}
\right)
\ ,\ \\
\hat{\mathcal{Q}}_n^{(2)} & =  \frac{-i}{2}  
\left(
\hat \sigma^+_{2n}\hat \sigma^-_{2n+1}-
\hat \sigma^-_{2n}\hat \sigma^+_{2n+1}
\right)
\ ,\ \\
\hat{\mathcal{Q}}_n^{(3)} & =  \frac{1}{4}  \left(  \hat Z_{2n}-\hat Z_{2n+1} \right)
\ \ ,
\label{eq:Qadefs}
\end{aligned}
\end{equation}
that act on pairs of fermion sites.
With the boundary condition that the chromo-electric field for $x<0$ vanishes,
the field at a given quark-anti-quark boundary, $E_b^{(a)}$, 
(denoted by $b$, depicted in Fig.~\ref{fig:layout_lattice})
is proportional to the summed charges from $n=0$ to the boundary,
\begin{equation}
    E_b^{(a)}  \propto \sum_{n=0}^{b+1+\lfloor b/2 \rfloor } \hat{\mathcal{Q}}_n^{(a)}
    \ \ ,
\end{equation}
where $\lfloor y \rfloor$ denotes the floor of $y$.
With OBCs, and the requirement that the chromo-electric field outside of the lattice ($x<0$ and $x>L-1$) has to vanish, the summation is restricted to the electric-field boundaries in the interior of the lattice, $b \in \{0, 1, \ldots, 2L-2\}$.
Color neutrality of the lattice is enforced by including a large energy penalty term in the Hamiltonian for non-zero $E_b^{(a)}$ at $b=2L-1$~\cite{Farrell:2022wyt}.
The chromo-electric field energy contribution to the Hamiltonian becomes
\begin{equation}
\hat H_g = 
\frac{g^2}{2}
\sum\limits_{\substack{a=1,2,3}}
\sum\limits_{b=0}^{2L-2}
\left(E_b^{(a)}\right)^2
\ \ ,\ \ 
\hat H_\lambda =
\frac{\lambda^2}{2}
\sum\limits_{a=1,2,3}
\left(E_{b=2L-1}^{(a)}\right)^2
\ \ ,
\end{equation}
where the second contribution is the quadratic penalty for non color-singlet configurations of the complete lattice, and $\lambda\rightarrow\infty$ is, in principle, taken.
The complete Hamiltonian is given by
\begin{equation}
    \hat H = \hat H_k + \hat H_m + \hat H_g + \hat H_\lambda
    \ .
    \label{eq:fullHami}
\end{equation}

%%%%%%%%%%%%%%%%%%%%%%%%%%%%
\section{Conceptual Challenges and Advances}
\label{sec:Concepts}
\noindent
In order to study the energy loss of a heavy quark moving in the vacuum or through a region of finite density on a lattice, conceptual and algorithmic developments are required.
One challenge is to efficiently include heavy quarks into simulations, and understand the role of entanglement in color space across the lattice.   
A second is to develop quantum circuits to transport the heavy quark between adjacent spatial sites, and a third is to develop techniques to prepare the interacting ground-state in the presence of one or more heavy quarks.

%%%%%%%%%%%%%%%%%%%%%%%%%
\subsection{Color Entanglement}
\label{subsec:ColorEnt}
\noindent
In the case of U(1) gauge theory, heavy charges can be consistently included by discontinuities in Gauss's law and in values of the gauge field at each spatial location.
This is implemented by mapping the field to qubits (the occupation basis), where each basis state is an eigenstate of the electric charge operator.\footnote{When acted on by the electric charge operator, an electron remains an electron, and a positron remains a positron.}
In 1+1D QED, external charges are screened and charge-charge correlation functions fall off exponentially outside of a few confinement lengths.
The situation is more complex in the case of non-Abelian theories, even in 1+1D.
Specifically, because the non-Abelian charge operators act non-trivially on the fields, specifically, mixing states in the occupation basis, external charges cannot be efficiently included via discontinuities in Gauss's law and in the chromoelectric fields (see related works using tensor networks~\cite{Kuhn:2015zqa,Sala:2018dui}).\footnote{This leads to the use of a quadratic penalty for the energy in the electric field beyond the end of the lattice to enforce global color neutrality.}
Hence, heavy quarks are mapped to qubits in the same way as the light quarks, but a kinetic term is not included in the Hamiltonian to prevent its delocalization from a given lattice site.

To improve our intuition regarding the additional complexities of introducing  heavy quarks into a lattice of non-Abelian fields, we start by recalling the strong-coupling (SC) structure\footnote{The SC limit ($g\to\infty$) of the ground state is equivalent to ground state in the absence of the kinetic term in the Hamiltonian with arbitrary $g$.} of a single spatial site of the lattice Schwinger model with Kogut-Susskind (staggered) discretization of the light fermion fields, and with one-heavy charged particle.
In the absence of heavy charges, $n_Q=0$, the (anti-ferromagnetic) trivial-vacuum ground state is
\begin{equation}
    |\ Q^-\  e^- \ Q^+\ \ e^+ \rangle \rightarrow 
    |\downarrow \downarrow\ \uparrow \uparrow\rangle
    \ =\ |1100\rangle
    \ .
\label{eq:QEDSCQe}
\end{equation}
The minimum energy configuration when a static
negative electric charge is present, with vanishing global electric charge, is 
achieved by screening from a positron, giving the ($n_Q=1)$ state
\begin{equation}
    |\ Q^-\  e^- \ Q^+\ \ e^+ \rangle \rightarrow 
    |\uparrow \downarrow\ \uparrow \downarrow\rangle
    \ =\ |0101\rangle
    \ .
\label{eq:QEDSCQeQ1}
\end{equation}
The total energy of this state is higher than that of the trivial vacuum because of the single unit of electric flux absorbed by the classical $Q^-$ charge.
The structure of this state is somewhat trivial in the sense that it is unentangled, and the charge of the positron additively cancels that of the static charge (to ensure global charge neutrality).
This state can be thought of as the presence of a heavy meson, with quantum numbers of $Q^-e^+$, on top of the trivial vacuum.  
As a short-cut for this discussion, we could have considered the action of the simplest meson interpolating operator, $\hat O\sim \overline{Q}e$, on the trivial vacuum state.
As a point for the discussion that
follows, 
because the number of heavy charges in the system is fixed, 
if we consider only the presence or absence of $Q^-$'s,  the $Q^+$ field can be removed, reducing the state space to 
$|\ Q^-\  e^- \ e^+ \rangle$ 
(reducing the qubit footprint in quantum simulations).

The situation is more complex for non-Abelian theories.
We start by considering
one spatial site ($L=1$)
of a single flavor of color SU(2) fermions in the fundamental representation,
with the Kogut-Susskind discretization, along with a heavy-quark field, $Q$, (without heavy anti-quarks).
The lattice assignment of fields on a single spatial site, in terms of the two colors, red ($r$) and green ($g$), is $|\ Q_r\  Q_g\ q_r\  q_g\  \overline{q}_r\ \overline{q}_g\  \rangle$.
In the absence of a heavy quark, the SC vacuum state is
\begin{equation}
    |\ Q_r\  Q_g\ q_r\  q_g\  \overline{q}_r\ \overline{q}_g\  \rangle 
    \rightarrow 
    |\downarrow \downarrow \downarrow \downarrow\ \uparrow \uparrow\rangle
    \ =\ |111100\rangle
    \ .
\label{eq:SU2SCQ0}
\end{equation}
With a heavy-quark present, its non-Abelian charge can be screened in two ways.
The first is by the presence of light anti-quarks, corresponding to the presence of (color-singlet) heavy mesons (with interpolating operators of the form $\hat O \sim \sum_\alpha \overline{Q}{}^\alpha q_\alpha$). 
The second is by the presence of light quarks, corresponding to the presence of (color-singlet) heavy baryons (with interpolating operators of the form $\hat O \sim \sum_{\alpha\beta}  \epsilon_{\alpha\beta} \ \overline{Q}{}^\alpha \overline{q}{}^\beta$).\footnote{The Abelian theory does not have corresponding baryon operators.}
The latter is energetically favored (in the lattice theory) as there is no chromo-electric flux between the quark and anti-quark sectors.\footnote{As noted in Ref.~\cite{Atas:2021ext}, the baryon and meson masses converge in the continuum limit where lattice-spacing artifacts vanish.}
The SC ground-state wavefunction for this $n_Q=1$ system is of the form\footnote{The (entangled) wavefunction created from a meson interpolating operator is of the form
\begin{equation}
    |\ Q_r\ Q_g\ q_r\ q_g\ \overline{q}_r\ \overline{q}_g\  \rangle 
    \rightarrow 
\frac{1}{\sqrt{2}}\biggl[
|\uparrow \downarrow \downarrow \downarrow\ \downarrow \uparrow\rangle
\ +\ 
|\downarrow \uparrow \downarrow \downarrow\ \uparrow \downarrow\rangle
\biggl]
    \ =\ \frac{1}{\sqrt{2}}\biggl[
|011110\rangle \ +\ |101101\rangle
\biggr]
    \ .
\label{eq:SU2SCQ1mes}
\end{equation}
}
\begin{equation}
    |\ Q_r\  Q_g\ q_r\ q_g\  \overline{q}_r\ \overline{q}_g\  \rangle 
     \rightarrow 
\frac{1}{\sqrt{2}}\biggl[
|\uparrow \downarrow \downarrow \uparrow\ \uparrow \uparrow\rangle
\ -\ 
|\downarrow \uparrow \uparrow \downarrow\ \uparrow \uparrow\rangle
\biggr]
    \ =\ \frac{1}{\sqrt{2}}\biggl[
|011000\rangle \ -\ |100100\rangle
\biggr]
    \ ,
\label{eq:SU2SCQ1}
\end{equation}
where the colors of the heavy quark and the light quark
are entangled.
In the remainder of the manuscript, to set a convention, 
the ground state 
in the SC limit with $L$ spatial sites and $n_Q$ heavy quarks will be denoted by $|SC; L, n_Q\rangle$, while the true (interacting) ground state will be denoted by $|I; L, n_Q\rangle$.

%%%%%%%%%%%%%%%%%%%%
\subsection{Implementing Heavy-Quark Motion}
\label{subsec:HQmove}
\noindent
For simulations of the Schwinger model, continuous spatial translations of the classical background charge can be implemented by a position-dependent reweighting of the discontinuities in the Gauss's law constraint~\cite{Florio:2023dke,Florio:2024aix,Farrell:2024mgu}.  
For the reasons mentioned above, specifically the action of the non-Abelian charge operator, spatial translations in non-Abelian theories cannot be efficiently used in this way, and instead are implemented by fermionic-SWAP (FSWAP) operators~\cite{Verstraete:2008qpa,Cervera-Lierta:2018bia,Algaba:2023enr} acting among the heavy-quark lattice sites.
Acting on a two-qubit system $\{ |00\rangle , |01\rangle , |10\rangle, |11\rangle \}$, the continuous FSWAP operation is
\begin{align}
\hat F(\theta) \ = \ \hat {\rm FSWAP}(\theta) & =\  
\left(
\begin{array}{cccc}
1&0&0&0\\
0 & e^{i\theta/2}\cos{\frac{\theta}{2}} & -ie^{i\theta/2}\sin{\frac{\theta}{2}}&0\\
0 & -ie^{i\theta/2}\sin{\frac{\theta}{2}} & e^{i\theta/2}\cos{\frac{\theta}{2}}&0\\
0&0&0&e^{i\theta}\\
\end{array}
\right)
\\ & =\ 
e^{-i 
\frac{\theta}{4} 
\left( \hat Y\otimes\hat Y + \hat X\otimes\hat X + \hat I\otimes \hat Z + \hat Z\otimes \hat I - 2 \hat I \right)
}
\ ,
\nonumber\\
&
 \xrightarrow[\theta\rightarrow\pi]{}
\left(
\begin{array}{cccc}
1&0&0&0\\
0& 0&1&0\\
0&1&0&0\\
0&0&0&-1\\
\end{array}
\right)
    \ .
    \label{eq:FSWAP}
\end{align}
As explained in App.~\ref{app:fswap}, the FSWAP gate can only be applied with $\theta=\pi$.
In the quantum simulations that follow, the FSWAP gates will be implemented for the red and green heavy-quark sites independently.
Their fermionic nature will be retained in the transformation, by including $\hat Z$ operators acting on the light and heavy quarks in between,
\begin{equation}
\hat F_{r,g} = 
e^{-i \frac{\pi}{4} 
\left( \hat X\otimes (- \hat Z)^{\otimes 5} \otimes\hat X 
+ \hat Y\otimes (- \hat Z)^{\otimes 5} \otimes\hat Y
+\hat I^{\otimes 6}\otimes \hat Z  
+\hat Z\otimes \hat I^{\otimes 6} 
- 2 \hat I^{\otimes 7}
\right) }
    \ .
    \label{eq:FSWAP6sites}
\end{equation}

Simulating a system involving multiple heavy quarks with one or more of them in motion requires modeling collisions. In 1+1D, this means characterizing the behavior when one $Q$ moves across another $Q$ when the motion is implemented by FSWAP gates.
\begin{figure}[ht!]
    \centering
\includegraphics[width=0.6\linewidth]{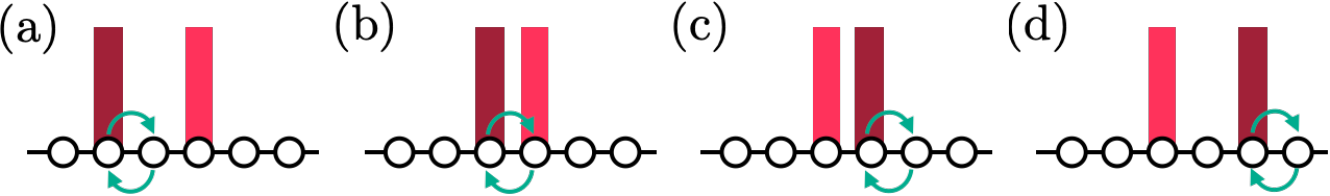}
\caption{A schematic of one heavy quark moving toward and passing a second heavy quark, implemented by FSWAP gates.
The circles denote spatial lattice sites, and the vertical colored bars denote heavy quarks.
Panels (a) through (d) show the discrete change of spatial site.  Panel (b) shows the 
action of the FSWAP gate when the heavy quarks are on adjacent site, in which the stationary heavy quark exchanges position with the moving heavy quark, displacing it by one lattice site.}
    \label{fig:QQswap}
\end{figure}
Consider the initial configuration shown in panel (a) of Fig.~\ref{fig:QQswap}. There are two heavy quarks, with one fixed in place and the other moving toward it, implemented by FSWAP gates.
As described in the caption of Fig.~\ref{fig:QQswap}, one important point to note is that when the heavy quarks are on adjacent spatial sites, they are exchanged by the action of 
FSWAP gates, displacing the stationary heavy quark by one site.
Subsequent motion involves displacing the moving heavy quark, but could also involve moving the stationary heavy quark back to its original location or not.
The difference between these possibilities is a lattice-spacing artifact that vanishes in the continuum limit.

%%%%%%%%%%%%%%%%%%%%%%%%%
\subsection{Ground-State Wavefunctions with Sources, and Domain Decomposition}
\label{subsec:GSDD}
\noindent
It is helpful to consider the structure of the ground-state wavefunction in the presence of a heavy quark.
Without the heavy quark, the ground-state wavefunction is invariant under discrete spatial translations, modified by effects from the boundaries.
In the case of OBCs, the boundaries modify the wavefunction by components that decrease exponentially away from the boundary, with a length scale set by the longest correlation length (the inverse mass of the lightest particle) in the theory.  
While local regions in the contained SU(2) charges fluctuate, the total charge of the lattice vanishes.
In the presence of a heavy quark, the light quarks are entangled with it in such a way to ensure global color neutrality.
Further, screening of the color charges means that far from the source the wavefunction approaches the ground state of the system without a heavy quark.   
A coherent sum of colored eigenstates arranges itself to generate these features.

In previous works, we developed scalable quantum circuits~\cite{Farrell:2023fgd,Farrell:2024fit} implemented by 
ADAPT-VQE~\cite{Grimsley_2019}, called SC-ADAPT-VQE (for a variant, see Ref.~\cite{Gustafson:2024bww}).
Classical computing was used to determine the operator sequence and their coefficients in creating a state approaching the ground state in modest-sized volumes, and then scaling to arbitrarily large systems on quantum computers in a controlled and converging way using the symmetries of the (lattice) system.
When wavepackets or heavy-quark sources are included, localized regions of the system are modified from the vacuum by operators with finite extent to approach the target wavefunction.

For the heavy-quark systems, it is the size of a heavy meson (or heavy baryon) that dictates the spatial extent of the region that needs to be modified away from the vacuum state.
This modified region is then combined with a region of the ground-state wavefunction without the heavy-quark source.
Both wavefunctions are classically optimized, and both suffer from boundary effects.
In such heterogeneous systems, DDec is often an efficient method to pursue.
We classically perform SC-ADAPT-VQE to determine the unitary operators to prepare the region with heavy-quarks, and combine it with a region of vacuum that is also prepared classically using SC-ADAPT-VQE.
This forms an initial wavefunction, which then undergoes another round of SC-ADAPT-VQE to optimize a sequence of unitary operators with support around the boundary.  
Specifically, operators that ``heal the wound'' caused by the boundary effects on both sides have a typical length-scale set by the confinement scale (or the lightest hadron).
For a system with multiple boundaries, as long as the boundaries are far from the heavy quark(s) and the edges, they can be healed with the same operators due to translational invariance.
This is made explicit in subsequent sections.

There are also practical reasons for using DDec in preparing these states.
For a large system with only one heavy quark, the volume of the wavefunction that is distorted away from the vacuum state is suppressed by the hadronic volume as opposed to the lattice volume. 
With increasing lattice volume, eventually this distortion becomes less than the tolerance of the optimizer, and the variational wavefunction becomes unresolvable between the interacting vacuum states with and without the heavy quark.
Therefore, a region around the heavy quark will necessarily need to be optimized separately from the rest of the lattice.

%%%%%%%%%%%%%%%%%%%%%%%%%%%%%%%%%%%%%%%%%
\section{Classical Simulation of SU(2) in 1+1D}
\label{sec:CSim}
\noindent
All the ingredients are in place to perform classical and quantum simulations of
heavy quarks moving through systems of light and heavy quarks in non-Abelian lattice gauge theories in 1+1D. 
It is helpful to consider the structure and dynamics of small systems in understanding how to perform simulations at scale. 
Small systems provide convenient environments to demonstrate the concepts and techniques discussed above, along with the subsequent implementation on quantum computers at scale.
As classical simulations are central to the framework of quantum simulations, we start by developing them for small-size systems, including for ``stitching them together'' in DDec, and for exploring their quantum complexity.
We also classically simulate time-evolution with discrete translation of the heavy quark, as discussed above.

%%%%%%%%%%%%%%%%%%%%%%%%%%%%
\subsection{Structure of a Single Spatial Site}
\label{sec:L1}
\noindent
We begin with a somewhat detailed discussion of the simplest system of a heavy quark, a light quark and a light anti-quark on one spatial site, $L=1$, with basis states denoted by $ |\ Q_r\  Q_g\ q_r\  q_g\  \overline{q}_r\ \overline{q}_g\  \rangle $.
As shown in Fig.~\ref{fig:layout_lattice}, the staggered sites are labeled by $n=\{0,1,2\}$, with one internal electric-field boundary, $b=0$ after $n=1$.
As discussed in Sec.~\ref{sec:what}, the Hamiltonian describing the system is 
\begin{equation}
\begin{aligned}
\hat H & =
-\frac{1}{2}
\sum_{c=0}^{1}
\left( 
\hat \sigma^+_{2+c} \hat Z_{3+c} \hat \sigma^-_{4+c}  + {\rm h.c.} \right)
 +
\frac{m_q}{2}
\sum\limits_{n=1}^{2} \sum_{c=0}^{1}
(-1)^{\eta_n}  \hat Z_{2n+c}\\
& +
\frac{g^2}{2}
\sum\limits_{a=1}^{3}
\left(\sum\limits_{n=0}^{1}\hat{\mathcal{Q}}_n^{(a)}\right)^2
 + 
\frac{\lambda^2}{2}
\sum\limits_{a=1}^{3}
\left(\sum\limits_{n=0}^{2}\hat{\mathcal{Q}}_n^{(a)}\right)^2
\ ,
\end{aligned}
\end{equation}
where $\eta_n$ is defined below Eq.~(\ref{eq:massterm}).

The SC ground state in the absence of heavy quarks is
\begin{equation}
    |SC; 1, 0\rangle = |\downarrow\downarrow\downarrow\downarrow\ \uparrow\uparrow\rangle 
     = |111100\rangle 
    \ ,
    \label{eq:SCGS10}
\end{equation}
and exact diagonalization of the Hamiltonian in the $2^6$ dimensional basis for $g=1$ and $m_q=0.1$ gives the interacting ground-state wavefunction,
\begin{equation}
|I; 1, 0\rangle = 
0.6663 |\downarrow\downarrow\downarrow\downarrow\ \uparrow\uparrow \rangle
\ +\ 
0.4383 |\downarrow\downarrow\downarrow\uparrow\ \uparrow\downarrow \rangle
\ -\ 
0.4383 |\downarrow\downarrow\uparrow\downarrow\ \downarrow\uparrow \rangle
\ +\ 
0.4143 |\downarrow\downarrow\uparrow\uparrow\ \downarrow\downarrow \rangle
\ \ ,
\label{eq:psiL1Q0}
\end{equation}
with an energy of $E_0=-0.6578$, and with spin expectation values $\langle \hat Z_j \rangle$ of 
\[\{ -1 , -1 , -0.2723 , -0.2723 , +0.2723 , +0.2723  \}.\]
In this state, the expectation values of the components of the Hamiltonian are
\begin{equation}
\langle \hat H_k \rangle = 
-0.9474
\ ,\ 
\langle \hat H_m \rangle  =
+0.14553
\ ,\ 
\langle \hat H_g \rangle =
+0.1440
\ ,\ 
\langle \hat H_\lambda\rangle = 0
\ \ .
\end{equation}

The operators required to create $|I; 1, 0\rangle$ from $|SC; 1, 0\rangle$ correspond to interpolating operators for color-singlet states in the spectrum, restricted to one-meson and one-baryon operators, defined to be
\begin{equation}
    \hat{O}_{M0} \equiv 
    \hat I^2 (\hat X \hat Z\hat Y - \hat Y \hat Z\hat X )   
    \hat I + \hat I^3  (\hat X \hat Z\hat Y - \hat Y \hat Z\hat X )
    \ ,\ 
    \hat O_{B0} \equiv   4i \hat I^2 (\hat\sigma^+\hat\sigma^+\hat\sigma^-\hat\sigma^- - {\rm h.c.}) 
     \ ,
     \label{eq:mesOpsL1}
\end{equation}
where the subscript $M$ denotes a meson operator, and $B$ denotes a baryon operator.
A variational wavefunction of the form
\begin{equation}
|\psi_{\rm var}\rangle = 
e^{-i \theta_2 \hat O_{B0}}\ 
e^{-i \theta_1  \hat{O}_{M0}}\ 
 |SC; 1, 0\rangle
\ ,\ \  
{\rm with}\ \ 
\theta_1 =  0.267215
\ ,\ \theta_2 = 0.05484
    \ ,
    \label{eq:L1nQ0VQE}
\end{equation}
recovers the wavefunction with an infidelity of ${\cal I} =  1 - |\langle\psi_{\rm var}|I; 1, 0\rangle|^2 \sim 10^{-6}$.
The infidelity can be reduced to zero with this complete operator set.

In systems with one heavy quark, $n_Q=1$, the SC vacuum is given in Eq.~\eqref{eq:SU2SCQ1}, and the exact wavefunction for $g=1$ and $m_q=0.1$
is
\begin{equation}
|I; 1, 1\rangle = 
0.6120 |\downarrow\uparrow\uparrow\downarrow\  \uparrow\uparrow \rangle
-
0.6120 |\uparrow\downarrow\downarrow\uparrow\ \uparrow\uparrow \rangle
+
0.3540 |\downarrow\uparrow\uparrow\uparrow\  \uparrow\downarrow \rangle
-
0.3540 |\uparrow\downarrow\uparrow\uparrow\  \downarrow\uparrow \rangle
\ \ ,
\end{equation}
giving expectation values
of the components of the Hamiltonian,
\begin{equation}
\langle \hat H_k \rangle  = 
-0.4334
\ ,\ 
\langle \hat H_m \rangle = 
+0.1501
\ ,\ 
\langle \hat H_g \rangle = 
+0.0940
\ ,\ 
\langle \hat H_{\rm ext.}\rangle = 0
\ \ .
\end{equation}
The energy of this state is 
$E_1=-0.1892$, 
giving a hadron mass (minus the mass of the heavy quark) of $\Lambda_Q=E_1 - E_0 = 0.4685$, with spin alignments of $\langle \hat Z_j \rangle = \{0,\allowbreak 0,\allowbreak 0.2507,\allowbreak 0.2507,\allowbreak 0.7492,\allowbreak 0.7492\}$.
The vanishing heavy-quark expectation values reflect the color entanglement with the light quarks.
Without the possibility of adding a baryon-antibaryon pair to the SC vacuum, due to the color entanglement with the heavy quark, only mesons can contribute to the wavefunction.
A variational wavefunction of the form
\begin{equation}
|\psi_{\rm var}\rangle = 
e^{-i \theta_1  \hat{O}_{M0}}\ 
 |SC; 1, 1\rangle
\ ,\ \  
{\rm with}\ \ 
\theta_1 =  0.26224
    \ ,
    \label{eq:L1nQ1VQE}
\end{equation}
recovers the exact wavefunction.

%%%%%%%%%%%%%%%%%%%%%%%%%%%%
\subsection{Hadronic Interpolating Operators for State Preparation}
\label{sec:L2}
\noindent
The $L=2$ system introduces a further layer of complexity, primarily in 
identification of the pool of variational operators.
Sec.~\ref{sec:what} contains the basic tools required to address the $L=2$ system,
and App.~\ref{app:L2} provides details of the exact diagonalization.
For notational clarity and specificity, site indices are added to the operators to indicate the spatial site that the first and last operator act on, i.e., $(x,y)$ in $\hat O_{Md}^{(x,y)}$ 
(if the operator acts only on one site, this is reduced to a single index, i.e., $\hat O_{M0}^{(x)}$).
The operators that we have identified for state preparation with sub-percent infidelity are:
\begin{equation}
\hat O_{M0}^{(0)}, \ 
\hat O_{M0}^{(1)}, \ 
\hat O_{M1}^{(0,1)}, \ 
\hat O_{B0}^{(0)}, \ 
\hat O_{B0}^{(1)}, \ 
\hat O_{B1}^{(0,1)} \ 
\ ,
\label{eq:OPxyL2}
\end{equation}
where
\begin{equation}
\hat{O}_{M1} \equiv (\hat Y \hat Z^3\hat X - \hat X \hat Z^3 \hat Y )  \hat I
+ \hat I   (\hat Y \hat Z^3\hat X - \hat X \hat Z^3 \hat Y )
\ ,\ 
    \hat O_{B1}  \equiv  -4i(\hat\sigma^+\hat\sigma^+ \hat{I}^2 \hat\sigma^-\hat\sigma^-
    - \rm{h.c.})
\ ,
\label{eq:OPxyL2def}
\end{equation}
where, as for the $L=1$ system, the $\hat X, \hat Y$ operators only act on the light quarks and anti-quarks.
These are both operators that ``straddle''  heavy $Q$ sites, i.e., the extra $\hat Z^2$ or $\hat I^2$ acts on  $Q_r, Q_g$ sites. 
A more detailed presentation of this operator pool can be found in Eq.~\eqref{eq:OPcomsL2}.
In general, for arbitrary systems, meson and baryon-antibaryon operators, or in fact any color-singlet configurations, with support across multiple sites, located at arbitrary positions on the lattice should be included.
Their relative contributions are determined by their (hadronic) form factors and the structure of the ground-state wavefunction.

In the variational process, as in previous works, the infidelity density, $\infiL$,  is chosen as the objective function,
\begin{equation}
\infiL = \frac{1}{L} 
\left( 1 - 
|\langle\psi_{\rm var}|\psi_{\rm target}\rangle|^2 \right)
\ ,
\label{eq:infidelitydens}
\end{equation} 
where $|\psi_{\rm target}\rangle$ is the target wavefunction (in this case  $|I; 2, n_Q\rangle$) and $|\psi_{\rm var}\rangle$ is the variational ansatz wavefunction that is being optimized to approach $|\psi_{\rm target}\rangle$, as discussed above in Sec.~\ref{sec:L1}.
This definition becomes increasingly useful as $\infiL\rightarrow 0$, as it gives comparable measures of the ``goodness of fit'' of the wavefunction per spatial site.

ADAPT-VQE is used, guided by physics input, to identify the optimal operator ordering and subsequent layer-by-layer parameter optimizations.  
To respect translational invariance and scalability in the $n_Q=0$ sector in the absence of OBCs, the meson operators are applied first to the SC vacuum in Eq.~\eqref{eq:psiSCL2Q0}, followed by the baryon operators, to give
\begin{equation}
\hat U^{n_Q=0}_{L=2} = 
e^{-i \theta_5 \hat O_{B1}^{(0,1)}}
e^{-i \theta_4 \hat O_{M1}^{(0,1)}}
e^{-i \theta_3 (\hat O_{B0}^{(0)}+\hat O_{B0}^{(1)})}
e^{-i \theta_2 (\hat O_{M0}^{(0)}+\hat O_{M0}^{(1)})}
e^{-i \theta_1 \hat O_{M1}^{(0,1)}}
\ ,
\label{eq:L2UNITQ0}
\end{equation}
with five variational parameters.
\begin{figure}[ht!]
    \centering
\includegraphics[width=0.48\linewidth]{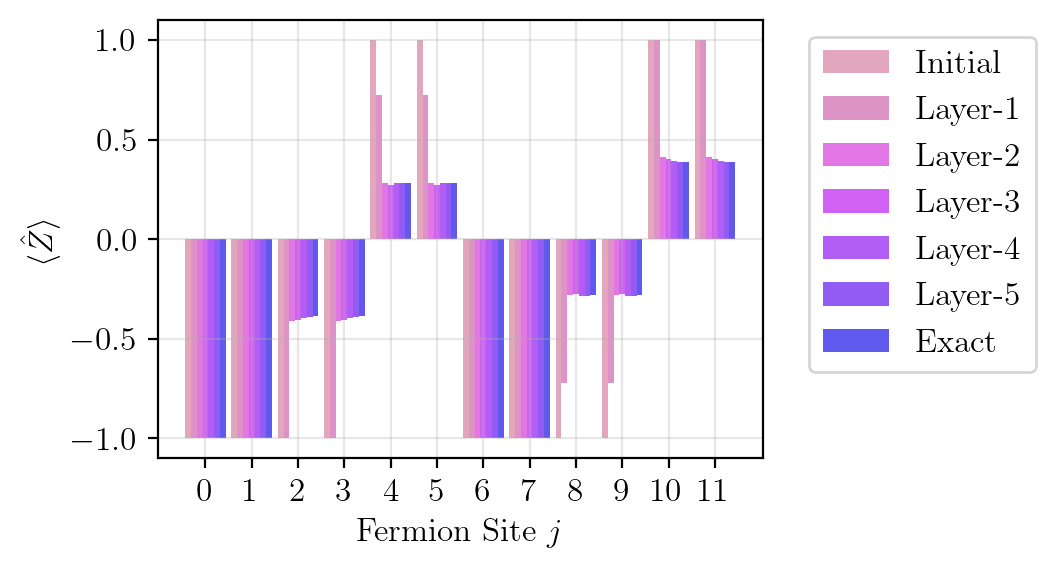}
\ \ 
\includegraphics[width=0.48\linewidth]{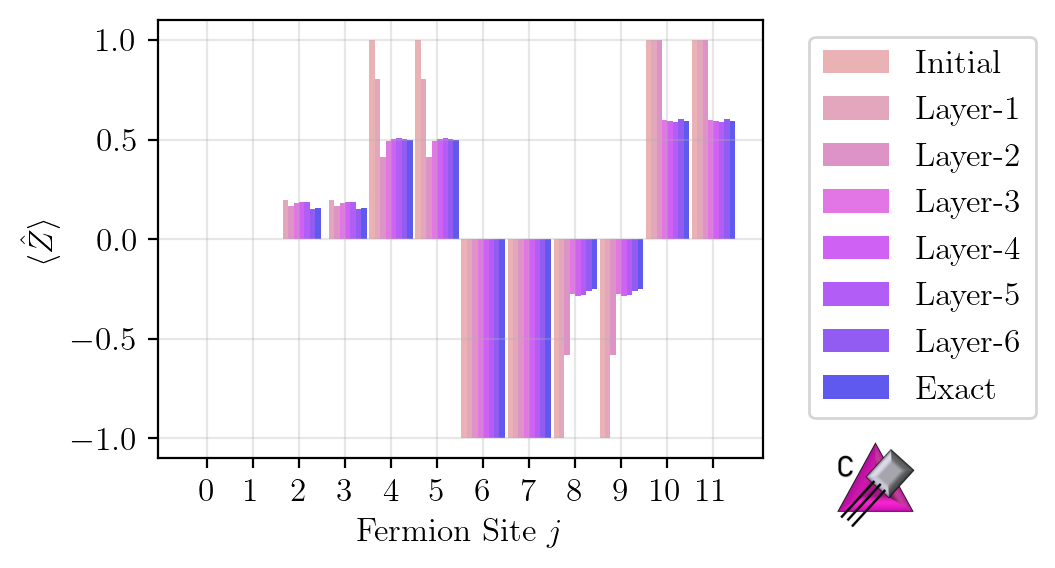}
\caption{The $\langle \hat Z_j \rangle$ as a function of the fermion site $j$, 
starting  from the SC vacuum, through the classically optimized layers, through to the exact values for $L=2$ with $m_q=0.1$ and $g=1.0$.  
The left panel shows the $\langle \hat Z_j \rangle$ in the $n_Q=0$ sector with up to five layers of optimization starting from $|SC; 2, 0\rangle$ toward $|I; 2, 0\rangle$.  
The right panel shows the analogous results in the $n_Q=1$ sector with up to six layers of optimization starting from $|SC; 2, 1\rangle$ toward $|I; 2, 1\rangle$.
The numerical values of the results are given in Table~\ref{tab:L2Q0Z} and Table~\ref{tab:L2Q1Z}.
}
    \label{fig:L2nQ01Z}
\end{figure}
Figure~\ref{fig:L2nQ01Z} shows the convergence of the $\langle \hat Z_j \rangle$ as a function of the fermion site $j$, starting from $|SC; 2, 0\rangle$ in the $n_Q=0$ sector. 
The small icons added to the figures throughout this manuscript represent wether they have been obtained using a classical simulator (purple triangle) or with quantum hardware (blue diamond)~\cite{Klco:2019xro}.
The meson layers are seen to converge the (single qubit) $\langle \hat Z_j \rangle$ to be close to their exact values, but the infidelity of the wavefunction is above $\sim 2\%$.
Including the baryon-antibaryon operators reduces the infidelity to below $\sim 0.1\%$.

The situation is more complicated in the $n_Q=1$ sector because of the lack of translational invariance.
The structure around the heavy quark emerges from the variational optimization of the operator pool, with an optimized wavefunction of the form

\begin{equation}
\hat U^{n_Q=1}_{L=2} = 
e^{-i \theta_6 \hat O_{M0}^{(0)}}\ 
e^{-i \theta_5 \hat O_{B1}^{(0,1)}}\
e^{-i \theta_4 \hat O_{B0}^{(1)}}\ 
e^{-i \theta_3 \hat O_{M0}^{(1)}}\ 
e^{-i \theta_2 \hat O_{M1}^{(0,1)}}\ 
e^{-i \theta_1 \hat O_{M0}^{(0)}}
\ ,
\label{eq:L2UNITQ1}
\end{equation}
applied to the SC state $|SC; 2, 1\rangle$.
The convergence of the fidelity of this six-parameter variational wavefunction is given in Table~\ref{tab:L2VQEmanuelQ1}, and of the associated single-qubit $\langle \hat Z_j \rangle$ in Table~\ref{tab:L2Q1Z}.
Figure~\ref{fig:L2nQ01Z} shows the convergence of the $\langle \hat Z_j \rangle$ as a function of the fermion site $j$, starting from $|SC; 2, 1\rangle$.
While the SC ground state has the light quarks screening the heavy quark color, it is the meson correlations that dominate the interacting ground state, as seen from the operator ordering.  This is discussed more in App.~\ref{app:L2}.

%%%%%%%%%%%%%%%%%%%%%%%%%%%%
\subsection{Implementing Domain Decomposition}
\label{sec:L3SP}
\noindent
The $L=3$ system, supported on 18 qubits, 
provides an opportunity to demonstrate a simple implementation of DDec by combining results and optimizations from the $L=1$ system, detailed in Sec.~\ref{sec:L1},
and the $L=2$ system, detailed in 
Sec.~\ref{sec:L2} and App.~\ref{app:L2}.
This is also the simplest system in which to implement heavy-quark motion.

Working with $g=1.0$ and $m_q=0.1$, the numerically exact wavefunctions for the $L=3$ system in the $n_Q=0,1$ systems are obtained using Lanczos algorithm, and are shown in App.~\ref{app:L3}.
Also given in that appendix are the energy densities associated with the components of the Hamiltonian and the associated $\langle \hat Z_j \rangle$.
The same methods used for $L=1$ and $L=2$ can be used to create the interacting ground state variationally with subpercent infidelity.  With $L=3$ there are additional operators that are added to the operator pool, 
which introduce meson and baryon-antibaryon correlations
between adjacent sites, 
\begin{equation}
\hat{O}_{M2} \equiv (\hat X \hat Z^7\hat Y - \hat Y \hat Z^7 \hat X )   \hat I+ \hat I   (\hat X \hat Z^7\hat Y - \hat Y \hat Z^7 \hat X )
\ ,\ 
    \hat O_{B2} \equiv  -4i(\hat\sigma^+\hat\sigma^+ \hat{I}^6 \hat\sigma^-\hat\sigma^-
    - {\rm h.c.})
\ \ ,
\label{eq:OM2B2def}
\end{equation}
in addition to the shorter-range correlations from the operators in Eq.~\eqref{eq:OPxyL2def} and Eq.~\eqref{eq:mesOpsL1}.
Thus, the operator pool is extended to
\begin{equation}
\hat O_{M0}^{(0)}, \ 
\hat O_{M0}^{(1)}, \ 
\hat O_{M0}^{(2)}, \ 
\hat O_{M1}^{(0,1)}, \ 
\hat O_{M1}^{(1,2)}, \ 
\hat O_{M2}^{(0,1)}, \ 
\hat O_{M2}^{(1,2)}, \ 
\hat O_{B0}^{(0)}, \ 
\hat O_{B0}^{(1)}, \ 
\hat O_{B0}^{(2)}, \ 
\hat O_{B1}^{(0,1)} \ 
\hat O_{B1}^{(1,2)} \ 
\hat O_{B2}^{(0,2)} \ 
\ .
\label{eq:OPxyL3}
\end{equation}
In $\hat{O}_{Md}$, $d$ indicates that there are $d^2 + d + 1$ number of $\hat Z$s in between the $\hat X$ and $\hat Y$; 
in $\hat{O}_{Bd}$, 
the number of $\hat I$s is $3d$ for even $d$ and $3d-1$ for odd $d$. 
As preparing the ground state in the $n_Q=0$ sector is not the focus of this work, 
we do not provide details or discussions (see App.~\ref{app:L3}), 
and focus on the $n_Q=1$ sector.

\begin{figure}[!ht]
    \centering \includegraphics[width=0.75\linewidth]{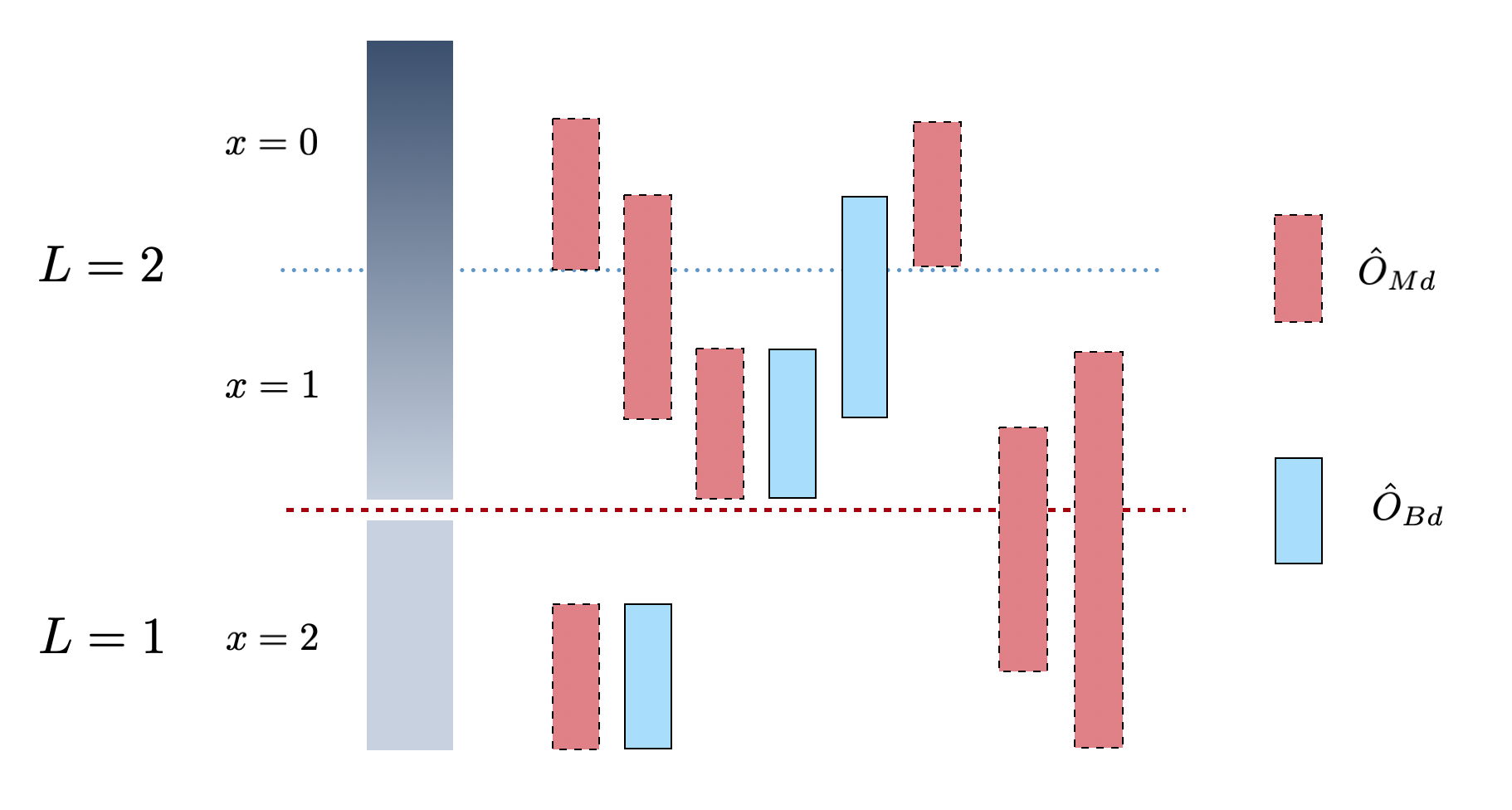}
    \caption{A block diagram for the DDec of the variational wavefunction in the $n_Q=1$ sector for $L=3$. 
    The $L=3$ wavefunction is initialized from the variationally optimized $n_Q=1, L=2$ (with the heavy quark located at $x=0$) and  $n_Q=0, L=1$ wavefunctions, as indicated by the operators acting either side of the dashed red line.  
    The re-optimized operators in each sector are implemented in parallel, followed by the application of boundary operators that entangle the two sectors, which are also optimized (with all of the operators globally optimized).
    }
    \label{fig:DDecL3}
\end{figure}

\begin{figure}[ht!]
    \centering
\includegraphics[width=0.8\linewidth]{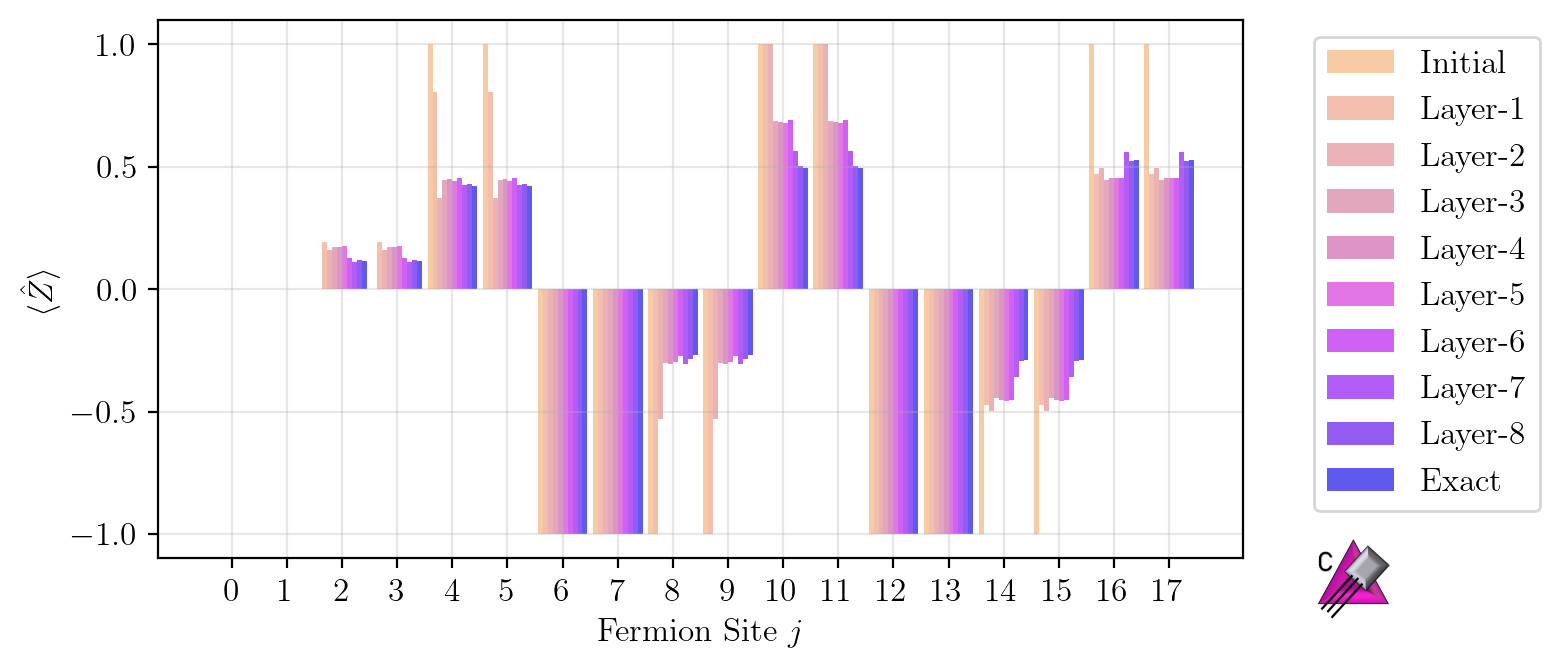}
\caption{The $\langle \hat Z_j \rangle$ as a function of fermion site $j$,  from $|SC; 3, 1, x=0\rangle$,  
through the 8 classically-optimized layers (corresponding to the application of the 10 operators in Eq.~\eqref{eq:L3U}, as shown in Fig.~\ref{fig:DDecL3}), 
through to the exact values from $|I; 3, 1, x=0\rangle$ for $m_q=0.1$ and $g=1.0$.  
The heavy quark is located at $x=0$ in sites $j=\{0,1\}$, and  $j=\{6,7,12,13\}$ correspond to unoccupied heavy-quark sites.
The numerical values of the results are given in Table~\ref{tab:L3Q1Z}.
}
    \label{fig:L3nQ1Z}
\end{figure}

To prepare the ground state of the $n_Q=1$ sector with a heavy quark located at $x=0$
using DDec,
the $L=3$ wavefunction is initialized with the operator sequences and angles
determined from the variationally-optimized 
$n_Q=1, L=2$, with the heavy quark located at $x=0$, 
and  $n_Q=0, L=1$ systems, as shown in the schematic diagram in Fig.~\ref{fig:DDecL3}.
As the $L=3$ target wavefunction differs from the tensor product of the 
$L=2$ and $L=1$ wavefunctions, 
with modifications by the boundary that
penetrate into both sectors determined by the confinement scale,
the fit angles in both sectors are re-optimized layer-by-layer.\footnote{Such re-optimizations in the bulk of the sectors
will become decreasingly significant with increasing lattice subvolumes, as the fit parameters asymptotically converge to fixed-point values in a confining theory~\cite{Klco:2019yrb,Klco:2020aud,Farrell:2023fgd,Farrell:2024fit}.}
This is followed by optimizing boundary operators that entangle the two sectors.
Through optimization, we have identified a variational wavefunction that is prepared by a unitary operator defined in terms of 10 angles,
\begin{equation}
\begin{aligned}
\hat U^{n_Q=1}_{L=3} = \ & 
e^{-i \theta_{10} \hat O_{M2}^{(1,2)} }\ 
e^{-i \theta_{9} \hat O_{M1}^{(1,2)} }\ 
e^{-i \theta_{8} \hat O_{M0}^{(0)} }\ 
e^{-i \theta_{7} \hat O_{B1}^{(0,1)} }\ 
e^{-i \theta_{6} \hat O_{B0}^{(1)} }\ 
\nonumber\\
&
\times e^{-i \theta_{5} \hat O_{M0}^{(1)} }\ 
e^{-i \theta_{4} \hat O_{B0}^{(2)} }\ 
e^{-i \theta_{3} \hat O_{M1}^{(0,1)} }\ 
e^{-i \theta_{2} \hat O_{M0}^{(2)} }\ 
e^{-i \theta_{1} \hat O_{M0}^{(0)} }
\ ,
\label{eq:L3U}
\end{aligned}
\end{equation}
acting on the $n_Q=1$ SC ground state, $|SC; 3, 1, x=0\rangle$,
given in Eq.~\eqref{eq:SCL3n1x0}, with a third index to denote the location of the heavy quark.
As shown in Fig.~\ref{fig:DDecL3}, the first two layers interleave operations on the $L=1$ and $L=2$ sections, followed by four layers acting on the $L=2$ section, and followed by two entangling layers at the boundary.
As each layer is applied, the fit angles from the individual sections provide starting points for further global optimizations.
The results of minimizing the infidelity of the  wavefunction prepared with the unitary transformation defined in Eq.~\eqref{eq:L3U}, with respect to the exact wavefunction, are shown in Table~\ref{tab:L3DoDeQ1}.
With the two layers of the $L=1$ applied in parallel with the four layers of the $L=2$ region, followed by two layers to heal the boundary, an infidelity density below $0.5\%$ is achieved.
The $\langle \hat Z_j\rangle$ obtained with each layer are  displayed in Fig.~\ref{fig:L3nQ1Z}, and given in Table~\ref{tab:L3Q1Z}.
The impact of optimization within each section is clear from the behavior of the $\langle \hat Z_j \rangle$, which converge rapidly within each sector, but slower in the boundary region $j=\{10,11\}$.

The main reason that DDec is effective in this small system when combining the ground states of the $L=2$ and $L=1$ sections as a first approximation to the $L=3$ wavefunction is that the coupling, 
$g=1$, is relatively strong, giving rise to short correlation lengths.
This implies that the boundary corrections are modest and easily captured. 
Approaching the continuum limit, where $g\rightarrow 0$, will require increasing spatial volumes to accommodate the increasing correlation length(s).

%%%%%%%%%%%%%%%%%%%%%%%%%%%%
\subsubsection{Color Complexity}
\label{sec:L3taus}
\noindent
Because of the key role of color entanglement in non-Abelian systems, we further explore its structure among the heavy quark and the light quarks, along with the quantum complexity measured by stabilizer R\'{e}nyi entropies (SRE)~\cite{Leone:2021rzd,Haug:2022vpg,Haug:2023hcs,Leone:2024lfr,Bittel:2025yhq}.
Of particular interest is the behavior of entanglement as a function of distance from the heavy quark to identify long-range quantum correlations, for which we chose to compute the mutual information $I^{(0,x)}_f$, which contains both classical and quantum correlations, and the 4-tangle~\cite{Wong:2000cmz}, $\tau_{4f}^{(0,x)}$, for $f=q,\overline{q}$ (light quarks and anti-quarks) separately.

The mutual information, $I^{(0,x)}_f$, between the heavy quark and light quarks or anti-quarks at spatial site $x$ is defined to be 
\begin{equation}
I^{(0,x)}_f = S(\hat\rho_0^{Q}) + S(\hat\rho_x^{f}) - S(\hat\rho_{0,x}^{Qf})
\ ,
\label{eq:I0x}
\end{equation}
where $\hat\rho_x^{f}$ is the ($4\times 4$) reduced density matrix of fermion-type $f$ at spatial site $x$, constructed by $\hat\rho_x^{f}={\rm Tr}_{\rm rest}[\hat\rho]$. 
$\hat\rho_{0,x}^{Qf}$ is the ($16\times 16$) reduced-density matrix from the pair of fermions, and $S(\hat\rho)$ is the von Neumann entropy, $S(\hat\rho)=-{\rm Tr}[\hat\rho\log_2\hat\rho]$.
The mutual information between the heavy quark and light anti-quarks is similarly defined.
The 4-tangle, $\tau_{4f}^{(0,x)}$, between the heavy quark at spatial site 0 and the light quarks (or anti-quarks) at spatial site $x$ is defined by matrix elements of the product $\hat Y_0^{Q_r} \hat Y_0^{Q_g} \hat Y_x^{f_r} \hat Y_x^{f_g}$,
\begin{equation}
\tau_{4f}^{(0,x)} = 
|\langle\psi| 
\hat Y_0^{Q_r} \hat Y_0^{Q_g} \hat Y_x^{f_r} \hat Y_x^{f_g}
| \psi^*\rangle |^2
\ .
\label{eq:tau40x}
\end{equation}
Typically, sub-tangles among the contributing sites, such as $\tau_{2q}^{(0,x)}$ are also evaluated, but these all vanish in this case.

\begin{figure}[ht!]
    \centering
\includegraphics[width=0.95\linewidth]{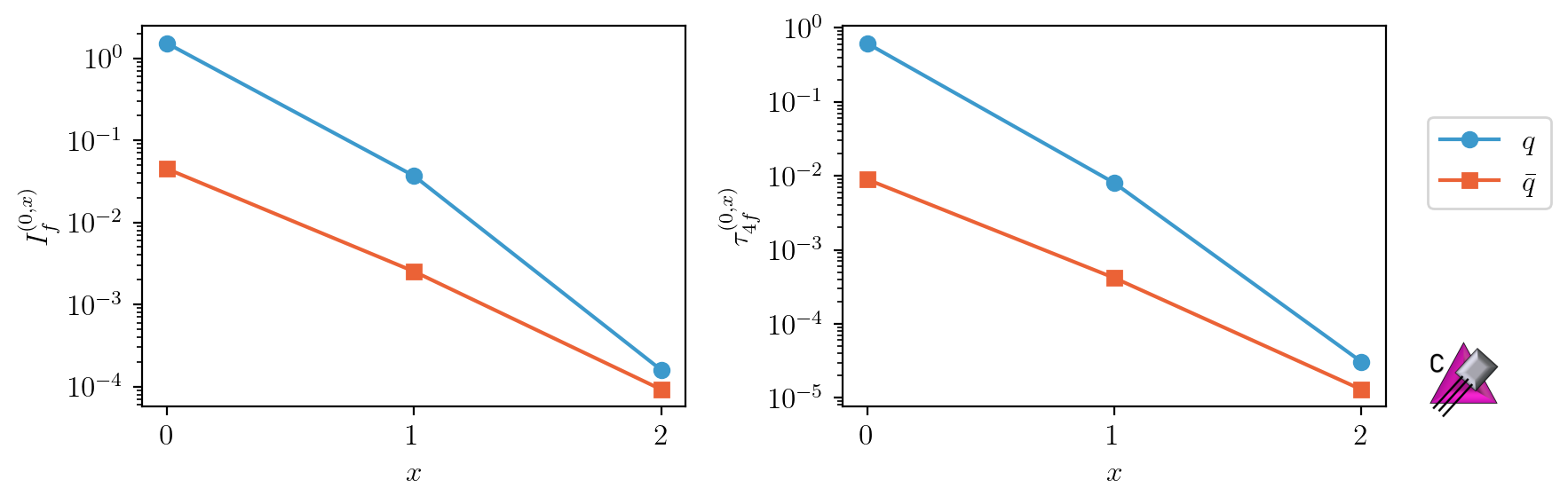}
\caption{The mutual information (left panel, defined in Eq.~\eqref{eq:I0x}) and 4-tangle (right panel, defined in Eq.~\eqref{eq:tau40x}) between the heavy quark at $x=0$ and the quarks and anti-quarks at $x=\{0,1,2\}$ in the $L=3$ system with $m_q=0.1$ and $g=1.0$.
Numerical values of the displayed points given in Table~\ref{tab:L3Q1Iijtau4}.
}
    \label{fig:L3nQ1Y4}
\end{figure}

Figure~\ref{fig:L3nQ1Y4} displays the mutual information and 4-tangles among the heavy and light quarks or anti-quarks as a function of spatial separation from the heavy quark, with numerical values given in Table~\ref{tab:L3Q1Iijtau4}.
As baryons preferentially screen the heavy-quark charges, the mutual information and 4-tangle with quarks are significantly larger than with anti-quarks.
The decrease of entanglement
with distance is consistent with exponential, as later confirmed in large systems in Sec.~\ref{sec:largeL}.
It is interesting to note that, in the variational wavefunction, the non-zero $x=2$ values, $\tau_{4q}^{(0,2)}$ and $\tau_{4\overline{q}}^{(0,2)}$, are only possible due to the boundary operators, and would vanish in their absence.

\begin{figure}[ht!]
    \centering
\includegraphics[width=0.4\linewidth]{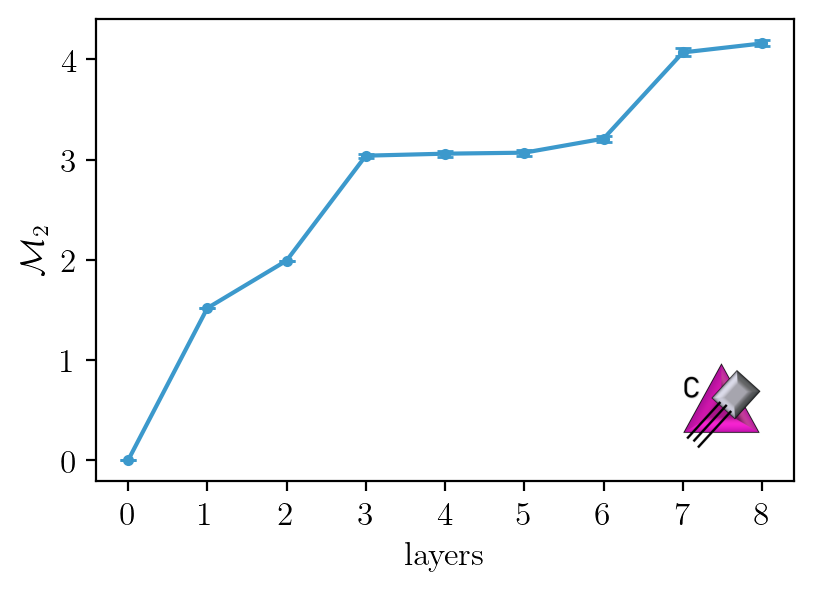}\ \ 
\caption{
The ${\cal M}_2$  non-stabilizerness in the $L=3$ wavefunction with $m_q=0.1$ and $g=1.0$ 
as a function of the number of variational layers for a  heavy quark at $x=0$.
Numerical values of the displayed points are given in Table~\ref{tab:L3DoDeQ1}.
}
    \label{fig:L3nQ1M2}
\end{figure}

A third quantity that we have examined is the quantum complexity, or quantum magic, measured by the SRE, 
${\cal M}_2$~\cite{Leone:2021rzd,Haug:2022vpg,Haug:2023hcs,Leone:2024lfr,Bittel:2025yhq}.
This provides a measure of the non-stabilizerness of a quantum state~\cite{Gottesman:1998hu,Aaronson:2004xuh,Bravyi_2016}, 
and how difficult it is for a classical computer to prepare it.  
A non-zero value of an SRE indicates that a state
lies outside of those
preparable 
by the Clifford gate set, 
and requires the inclusion of T-gates in quantum circuits.
It has recently been shown that the evolution of magic during 
Quantum Approximation Optimization Algorithm (QAOA) 
presents a complexity barrier, 
as the magic increases then decreases as the target state is approached with increasing numbers of unitary layers~\cite{Capecci:2025ull}.
To search for a complexity barrier in creating the ground states of the systems considered in this chapter, 
we compute the magic at each layer of operators creating the $L=3$, $n_Q=1$ ground state using DDec. 
For a system of $N$ qubits,
the SRE, ${\cal M}_\xi$, is defined as~\cite{Leone:2021rzd}
\begin{equation}
    {\cal M}_\xi = \frac{1}{1-\xi} 
    \log_2 \left( \sum_{\hat P\in \hat P_N} 
    \frac{\langle\psi| \hat P |\psi\rangle^{2\xi}}{2^N} 
    \right)
    \ \ ,
    \label{eq:SRE}
\end{equation}
where $\hat P$ is a string of Pauli operators, and the sum extends over all such Pauli strings for $N$ qubits, 
denoted by $\hat P_N$.\footnote{
For $\xi=2$, and writing the wavefunction in terms of eigenstates of the $\hat Z$ operator with eigenvalues 
$\pm 1$, $|\psi\rangle = \sum_{\bm\sigma} c_{\bm\sigma} |{\bm\sigma}\rangle$,
the expression in Eq.~\eqref{eq:SRE} 
reduces to a 4-replica summation~\cite{Tarabunga:2023xmv},
\begin{equation}
e^{-{\cal M}_2} = 
\sum_{ {\bm\sigma}^{(1)} , {\bm\sigma}^{(2)} , {\bm\sigma}^{(3)} , {\bm\sigma}^{(4)}}
c_{{\bm\sigma}^{(1)}}\ 
c_{{\bm\sigma}^{(2)}}\ 
c_{{\bm\sigma}^{(3)}}\ 
c_{   {\bm\sigma}^{(1)}   {\bm\sigma}^{(2)}   {\bm\sigma}^{(3)}   }\ 
c^*_{   {\bm\sigma}^{(1)}   {\bm\sigma}^{(2)}   {\bm\sigma}^{(4)}   }\ 
c^*_{   {\bm\sigma}^{(1)}   {\bm\sigma}^{(3)}   {\bm\sigma}^{(4)}   }\ 
c^*_{   {\bm\sigma}^{(2)}   {\bm\sigma}^{(3)}   {\bm\sigma}^{(4)}   }\ 
c^*_{{\bm\sigma}^{(4)}}
\ ,
\label{eq:Tara}
\end{equation}
where products of ${\bm\sigma}^{(s)} $ denote direct element by element multiplies.
This expression provides a parametric acceleration of evaluations in sparse wavefunctions, 
and is what we implemented in this chapter 
(via statistical sampling).
The summation in Eq.~\eqref{eq:SRE} scales as $ \sim 4^{n_Q}$ for $n_Q$ qubits, 
independent of the number of non-zero amplitudes contributing to $|\psi\rangle$.
The properties of the Pauli strings can be used to greatly reduce the number of evaluations.
In contrast, the summation in Eq.~(\ref{eq:Tara}) scales as $\sim {\cal N}_{\cal A}^4$, where 
${\cal N}_{\cal A}$ is the number of non-zero amplitudes.
}
The values of ${\cal M}_2$ for the $L=3$ system as a function of the variational layer are shown in Fig.~\ref{fig:L3nQ1M2} and given in Table~\ref{tab:L3DoDeQ1}.
One observes a monotonic increase (within sampling uncertainties) in magic with increasing numbers of layers.
The increase in ${\cal M}_2$ plateaus toward layer-6, but jumps when layer-7 is included, corresponding to the first boundary operator that joins the $L=2$ and $L=1$ regions.
In contrast to the QAOA observation, there is no evidence for a magic barrier in preparing the $L=3$ wavefunction.

%%%%%%%%%%%%%%%%%%%%%%%%%%%%
\subsection{Moving Heavy Quarks}
\label{sec:L3Qmotion}
\noindent
With the preparation of the ground state of the $L=3$ system with a heavy quark complete, we begin to explore the time evolution of the system after the heavy quark is moved from $x=0$ to $x=1$.  

%%%%%%%%%%%%%%%%%%%%%%%%%%%%
\subsubsection{Through the Lattice Vacuum}
\label{sec:L3QmotionVAC}
\begin{figure}[ht!]
    \centering
\includegraphics[width=0.99\linewidth]{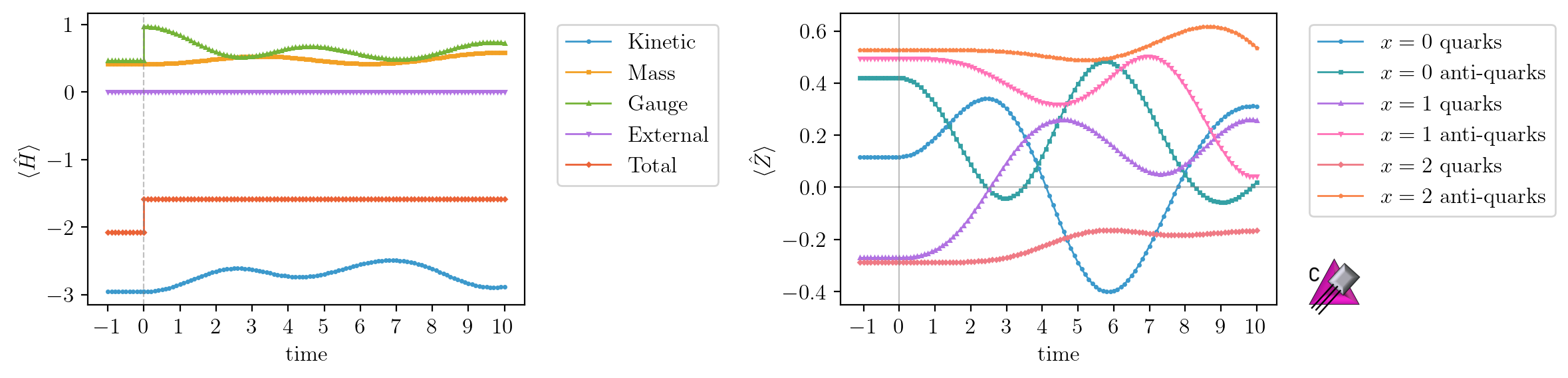}
\caption{The left panel shows the expectation values of the components of the Hamiltonian, and their total, as a function of time in the $L=3$ system.  
At $t=0$, the heavy quark is instantaneously moved from $x=0$ to $x=1$.
The right panels shows the associated time evolution of the $\langle \hat Z_j \rangle$. 
The right-justified dashed lines correspond to the values in the ground state with the heavy quark at $x=1$.
}
    \label{fig:L3Hi}
\end{figure}

\noindent
The exact (classically-determined) ground state of the $L=3$ system is prepared with a heavy-quark positioned at $x=0$.  
At $t=0$, the heavy-quark is instantaneously moved from $x=0$ to $x=1$ via the FSWAP operations defined in Eq.~(\ref{eq:FSWAP6sites}).
The system is then evolved under the exact Hamiltonian in discrete time steps, but without Trotterization with respect to the individual terms contributing to the Hamiltonian.
Using the same Hamiltonian parameters, $g=1.0$ and $m_q=0.1$, and a time step of $\delta t = 0.2$, the different contributions to $\hat H$, as well as their sum, are computed as a function of time and are displayed in the left panel of Fig.~\ref{fig:L3Hi}. 
The change of heavy-quark position at $t=0$ by one spatial site
injects energy into the system, by increasing the energy in the gauge field (at that instant).\footnote{In the infinite-volume and continuum limits, 
the magnitude of the discontinuity vanishes 
as $g\rightarrow 0$ ($a\rightarrow 0$, and the correlation length become 
infinitely large (in lattice units)), 
and with the boundaries infinitely far away from the heavy-quark position(s), 
the net energy injected vanishes by Lorentz invariance.
In contrast, as considered in our previous work~\cite{Farrell:2024mgu}, 
in the presence of matter, 
the net injected energy will not vanish, distributed into internal excitations and fragmented hadronic particles.
The extent to which the 
injected energy remains near the heavy quark in the form of stable hadronic excitation, or is radiated into propagating color-singlet states requires simulations of larger systems, 
and both effects are expected to vanish in the continuum and infinite-volume limits 
in the absence of a background medium. }
The average occupations of the quarks and anti-quarks,
$\langle \hat Z_j \rangle$,
are shown in the right panel of Fig.~\ref{fig:L3Hi}.
As expected, the responses of the quarks and anti-quarks closest to the heavy-quark's position are the most immediate and largest, while the responses at $x=2$ are delayed by the speed of light and suppressed
by distance.

The time evolution of the entanglement between the heavy quark and the light quarks or anti-quarks can also be explored.  
The 4-tangles of the initial state evolve with time, in such a way that in the continuum and infinite volume limits, the measures of entanglement will become Lorentz invariant.
That is to say, that the entanglement structure in the light quarks is expected to move with the heavy quark (at constant velocity).
\begin{figure}[ht!]
    \centering
 \includegraphics[width=0.9\linewidth]{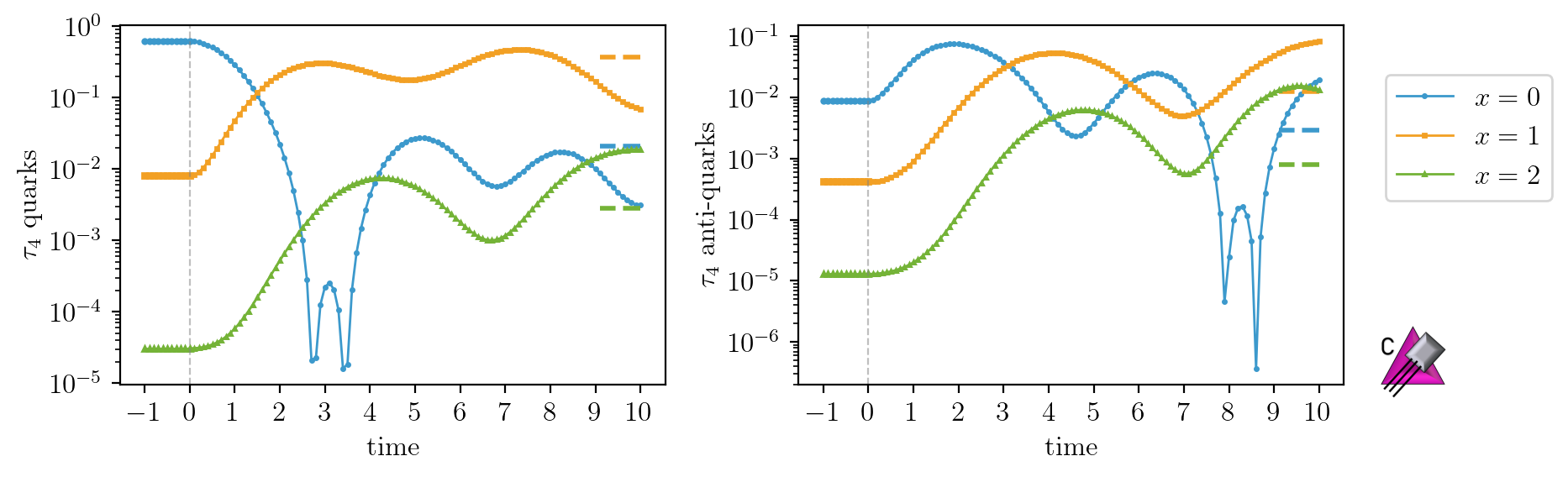}
\caption{The 4-tangles of the light quarks and anti-quarks with the heavy quark.
At $t=0$, the heavy quark is instantaneously moved from $x=0$ to $x=1$, and the definition of $\tau_4$ changes from
$\tau_{4f}^{(0,x)}$ to $\tau_{4f}^{(1,x)} $
The left panels shows the 4-tangle for the quarks, and the right panel for the anti-quarks.
The right-justified dashed lines correspond to the values in the ground state with the heavy quark at $x=1$, with numerical values given in Table~\ref{tab:L3Q1tau4x1}.
}
    \label{fig:tau4L3time}
\end{figure}
The 4-tangles, $\tau_{4f}^{(0,x)}(t)$,  defined in Eq.~\eqref{eq:tau40x} at $t=0$ as discussed above, becomes $\tau_{4f}^{(1,x)}(t)$ after the heavy quark is moved from $x=0$ to $x=1$.

Figure~\ref{fig:tau4L3time} shows the time evolution of the quark and anti-quark $\tau_4(t)$ for $L=3$ computed from the exact wavefunction(s).
Given the initial configuration, we anticipate that in the ideal, infinite-volume system, that the values for $x=0$ and $x=2$ should become approximately equal, up to internal excitations and reflections from the boundaries, due to the symmetry of the final state system.

%%%%%%%%%%%%%%%%%%%%%%%%%%%%
\subsubsection{In-Medium: Energy Loss and $\frac{dE}{dx}$}
\label{sec:L3QmotionIM}
\begin{figure}[ht!]
    \centering
\includegraphics[width=0.9\linewidth]{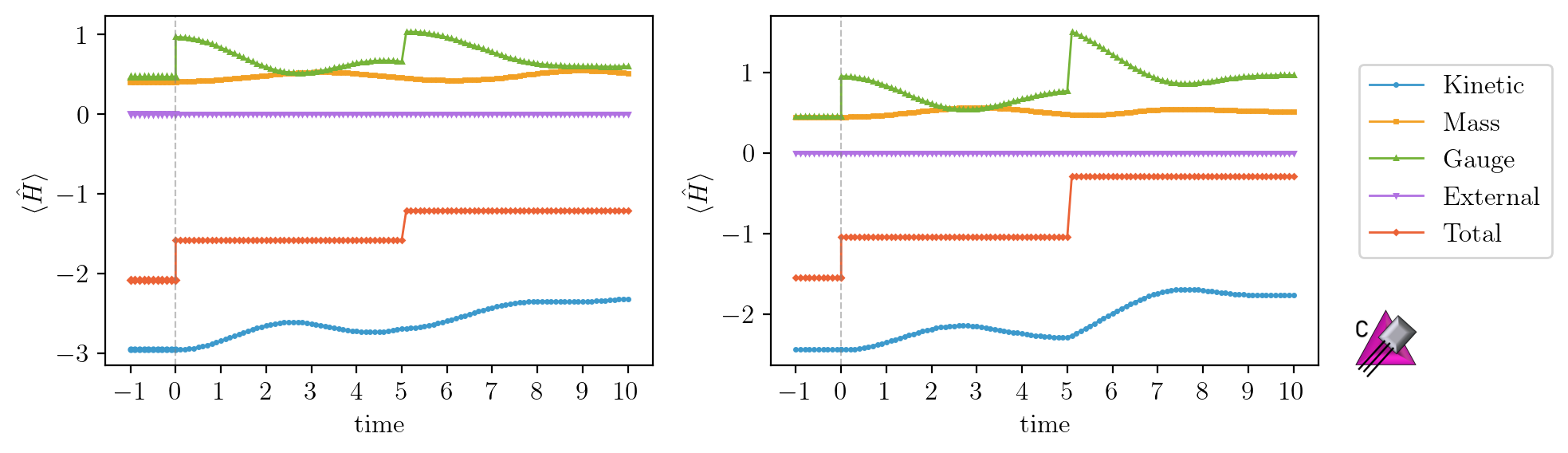}
\caption{
The contributions to the energy, and the total energy,
in the $L=3$ system, as the heavy quark initially at $x=0$ is moved to $x=1$ at $t=0$ and to $x=2$ at $t=5$.
The left panel shows the total energy in the motion across the vacuum, 
while the right panel shows total energy in the presence of a second heavy quark at $x=2$.
}
    \label{fig:Hix012}
\end{figure}

\noindent
The $L=3$ system, despite its limitations, permits a first exploration of the energy loss in the presence of matter.
To address this, and to estimate the instantaneous energy loss, $dE/dx$, in these systems with matter present, we perform two classical simulations, following the method presented in Ref.~\cite{Farrell:2024mgu}.  
Two systems are prepared: the ground state of the system with one heavy quark at $x=0$, and that with two heavy quarks, one at $x=0$ and the other at $x=2$.
In each, the heavy quark is moved from $x=0$ to $x=1$ at $t=0$ and then to $x=2$ and $t=5$, the latter time being arbitrarily selected.
The total energy of each system is determined in the three time intervals, ($t<0$, $0\le t < 5$ and $t\ge 5$) from which the difference between them can be translated into an estimator for $dE/dx$.
The contributions to the energy from components of the Hamiltonian, and the total energy as a function of time are shown in Fig.~\ref{fig:Hix012}.
The total energies in the three intervals are
\begin{equation}
    E_{\rm vac} = \{\ 0,\ 0.5002,\ 0.8721 \}   \ -\ 2.0806
    \ \ ,\ \ 
     E_{\rm med} = \{\ 0,\ 0.5030,\ 1.2595 \}\ -\ 1.5485
     \ .
\end{equation}
Estimates of $dE/dx$ are determined by the differences in total energy between adjacent plateaus, followed by a subtraction of the differences from moving in the lattice vacuum, giving,
\begin{align}
\frac{dE}{dx}^{({\rm vac})}\Big|_{x=(1,0)}^{(\Delta t=5)} & = 0.5002\ ,\ 
\frac{dE}{dx}^{({\rm vac})}\Big|_{x=(2,1)}^{(\Delta t=5)} \ =\ 0.3719\ \nonumber\\
\frac{dE}{dx}^{({\rm med})}\Big|_{x=(1,0)}^{(\Delta t=5)} & = 0.5030\ ,\ 
\frac{dE}{dx}^{({\rm med})}\Big|_{x=(2,1)}^{(\Delta t=5)} \ =\ 0.7564\ 
     \ .
\end{align}
Combining these values gives, 
\begin{align}
\langle \frac{dE}{dx}\rangle \Big|_{x=(1,0)}^{(\Delta t=5)} & = 
\frac{dE}{dx}^{({\rm med})}\Big|_{x=(1,0)}^{(\Delta t=5)}  -  \frac{dE}{dx}^{({\rm vac})}\Big|_{x=(1,0)}^{(\Delta t=5)} 
\ =\ 0.0028
\nonumber\\
\langle \frac{dE}{dx}\rangle \Big|_{x=(2,1)}^{(\Delta t=5)} & = 
\frac{dE}{dx}^{({\rm med})}\Big|_{x=(2,1)}^{(\Delta t=5)}  -  \frac{dE}{dx}^{({\rm vac})}\Big|_{x=(2,1)}^{(\Delta t=5)}
\ =\ 0.3846
     \ .
\end{align}
As the energy loss in the vacuum is entirely a lattice artifact, its subtraction from the in-medium energy loss removes this leading unphysical contribution.
These values show the impact of the matter (heavy quark) located at $x=2$ on the energy loss experienced by the moving heavy quark.
The truncations and approximations employed in this simple calculation are significant, and simulations are not expected to provide robust estimates of the true energy loss in the continuum theory.  
However, these results demonstrate the potential utility of this method.

%%%%%%%%%%%%%%%%%%
\subsection{Large-$L$ Tensor-Network Calculations}
\label{sec:largeL}
\noindent
In order to go beyond lattice sizes limited by state vector simulations, we use tensor network techniques to study the static and dynamical properties of systems with heavy quarks. 
We represent the states as matrix product states (MPS), and use the density matrix renormalization group (DMRG) algorithm to prepare them in the correct sector, and the time dependent variational principle (TDVP) algorithm to perform time evolution.
This is implemented via the {\tt ITensor}~\cite{fishman2022itensor} and {\tt ITensorMPOContruction}~\cite{Corbett:2025flm} packages. While for ground-state preparation and static property studies a maximum bond dimension of 200 was sufficient (the variance $\langle \psi | \hat{H}^2 | \psi \rangle - \langle \psi | \hat{H} | \psi \rangle^2$ was below $10^{-5}$), for time evolution it had to be increased to 600. 
The real-time dynamics was found to be considerably harder to simulate than the static properties.

\begin{figure}[ht!]
    \centering
\includegraphics[width=0.9\linewidth]{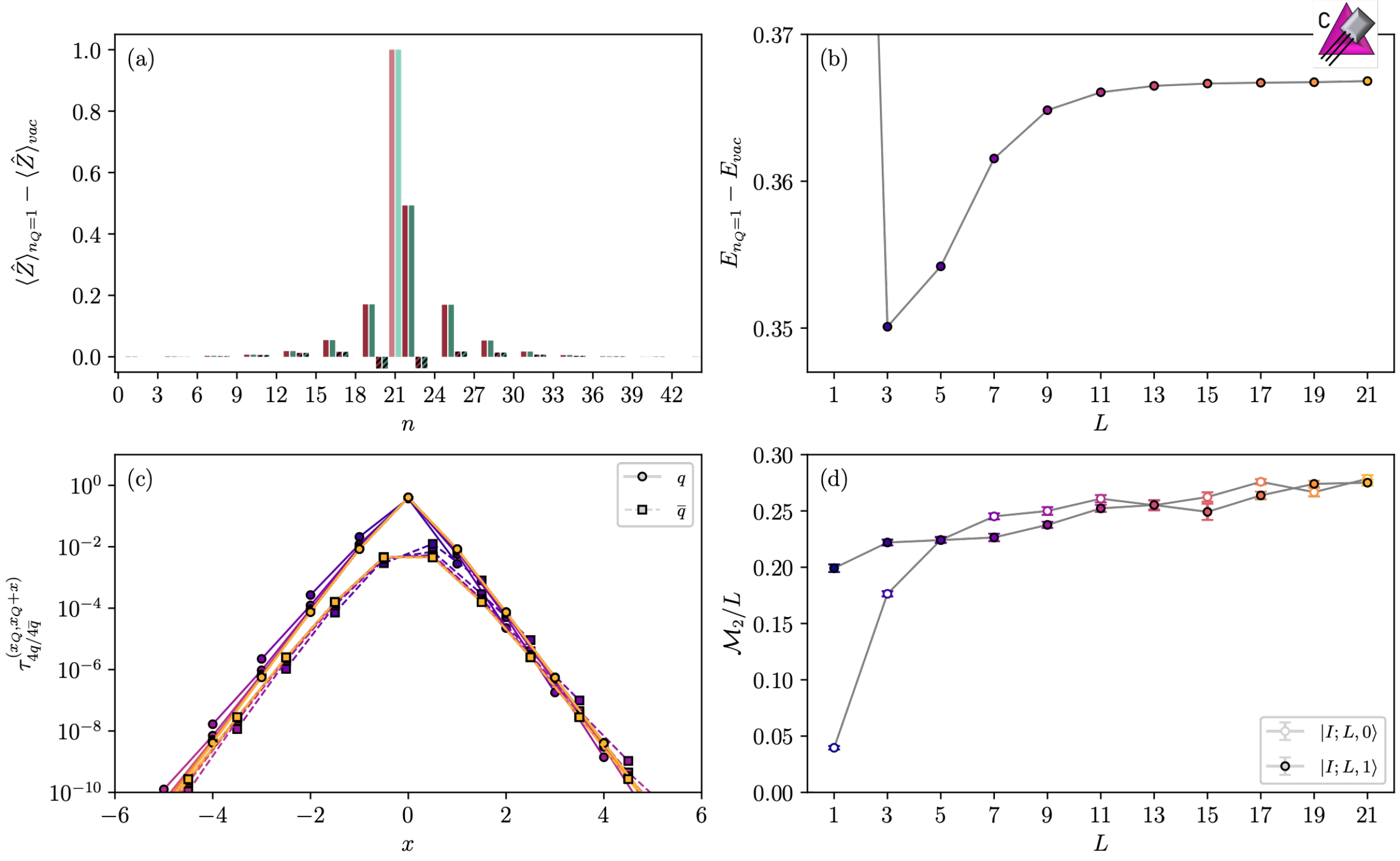}
\caption{Static properties of the system with and without a heavy quark at the center of the lattice, with $m_q=0.1$ and $g=1.0$ for large systems.
(a) The difference in the distribution of charges between the state with a heavy quark in the middle and the vacuum state, with $L=15$.  
(b) The energy difference with the vacuum as a function of the extent of the lattice $L$, reaching a plateau value of the infinite-volume gap. 
This is consistent with the expected exponential convergence of the mass of a single heavy meson with increasing lattice volume.
(c) 
The 4-tangles for quarks and anti-quarks, 
$\tau^{(x_Q,x_Q+x)}_{4q}$ and 
$\tau^{(x_Q,x_Q+x)}_{4\overline{q}}$,
for a series of $L$, where the position of the heavy quark $x_Q$ is at the middle of the lattice, displaying the exponential decay of correlations. The points for the anti-quarks have been shifted to the right for clarity. 
(d) 
The ${\cal M}_2/L$ SRE densities for both vacuum and heavy-quark states as a function of $L$.
The convergence is due to the heavy-quark wavefunction having support over a limited number of lattice sites.
}
    \label{fig:static_L}
\end{figure}
Figure~\ref{fig:static_L} shows the results obtained in a series of calculations performed on both the vacuum state, $|I,L,0\rangle$, and the state with a heavy-quark positioned at the mid-point of the lattice, $|I,L,1\rangle$, with $L\in\{1,3,\ldots,21\}$.
To observe how the light quarks screen and confine the heavy quark, in Fig.~\ref{fig:static_L}(a), 
the difference of the charges of each quark and color, measured by $\langle \hat Z \rangle$, 
between the state with and without a heavy quark, is shown.
Since the heavy quark is far away from the boundaries, a clean exponential decay in $\langle \hat Z \rangle$ is observed.

By positioning the heavy quark at the center of the lattice, the energy gap to the vacuum state can be extracted with exponential convergence in lattice size, as shown in Fig.~\ref{fig:static_L}(b).
Despite working with OBCs, where $1/L$ volume effects are present for typical quantities, this leading contribution cancels in the difference, as the heavy quark (and the screening light quarks) are sufficiently far from the boundaries.
This exponential decay of correlations is also seen in the mutual information and 4-tangles as a function of distance from the heavy quark. 
The 4-tangles $\tau^{(x_Q,x_Q+x)}_{4f}$ ($f = q, \bar q $), given in Eq.~\eqref{eq:tau40x}, are shown in Fig.~\ref{fig:static_L}(c) for a selection of (large) lattice volumes as a function of (spatial) distance to the heavy quark.
As lattice volumes increase, the correlation lengths converge.
Figure~\ref{fig:static_L}(d) shows the ${\cal M}_2/L$ SRE densities for both vacuum and heavy-quark states, computed via the perfect Pauli sampling method~\cite{Lami:2023naw,Haug:2023hcs} using 2000 samples. 
These densities are seen to converge with increasing volume, consistent with the heavy hadron having only limited spatial support on the lattice.
For the small $L$ values, as explained in Ref.~\cite{Hoshino:2025ine}, systems with OBCs should display a logarithmic dependence of $\mathcal{M}_2$ with $L$, and that is case, as seen in Fig.~\ref{fig:static_L}(d).

\begin{figure}[ht!]
    \centering
\includegraphics[width=\linewidth]{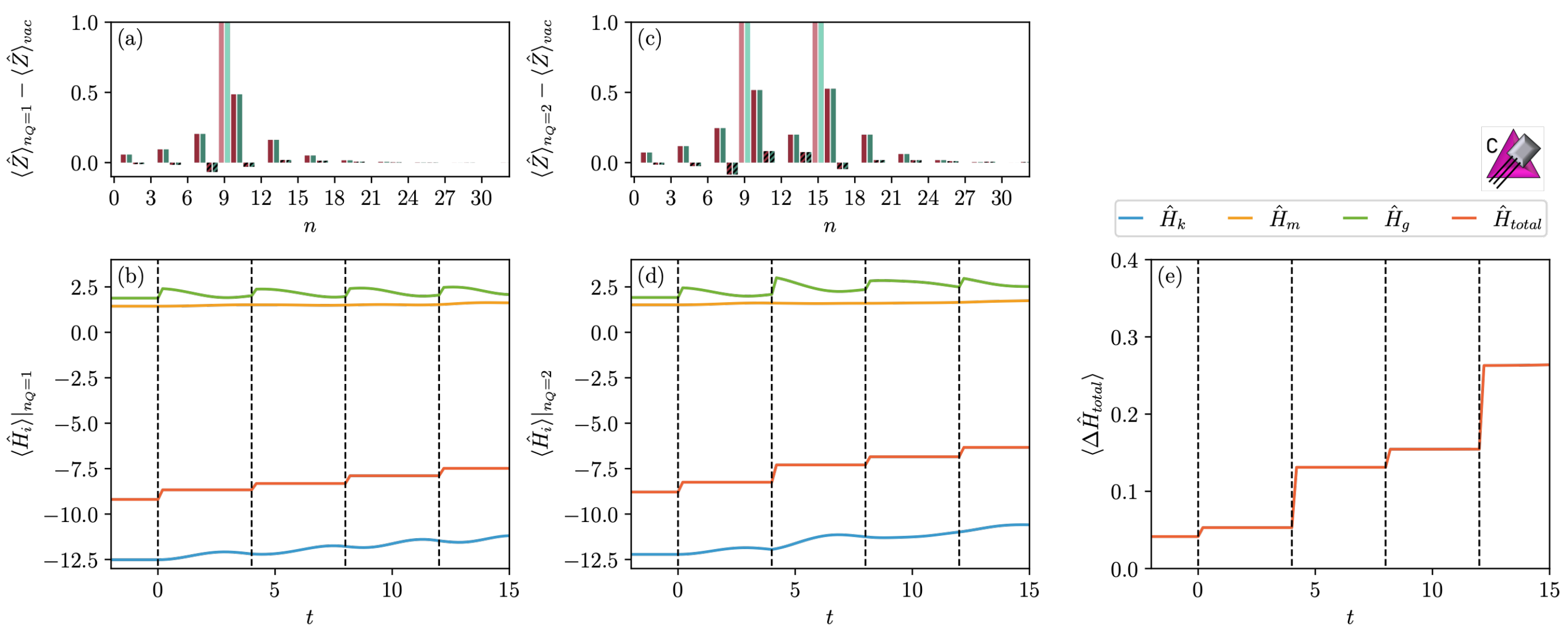}
\caption{Time evolution of the system with one and two heavy quarks, with $m_q=0.1$, $g=1.0$ and $L=11$. 
Panel (a) shows the initial charge distribution for a single heavy quark at $x=3$, and  
panel (b) shows the time evolution of the total energy and individual terms of the 
Hamiltonian as the charge is moved one site every $\Delta t = 4$ (dashed vertical lines). 
Panel (c) shows the initial charge distribution with two heavy quarks, at $x=3$ and $x=5$. 
(d) Same as (b), but moving the heavy quark initially at $x=3$, with another heavy quark static at $x = 5$.
(e) The difference in the energy contributions to the 
systems with and without a second heavy quark, defined in Eq.~\eqref{eq:diffH}.
}
    \label{fig:dynamic_L}
\end{figure}
Finally, we compute the associated energy differences when one heavy quark is moved across the lattice in the presence of a second heavy quark.
Figure~\ref{fig:dynamic_L} shows the initial charge distributions and time-dependent energies as one heavy quark is moved to adjacent spatial sites at regular time intervals, $\Delta t=4$, for an $L=11$ lattice.

Panels (a) and (b) show the results obtained without the second heavy quark, where any shift in energy is due to lattice artifacts (broken Lorentz symmetry).
Similarly, panels (c) and (d) show these quantities in the presence of a second heavy quark, where a difference in energy loss due to the collision is observed.
With vacuum and single-heavy quark contributions removed, panel (e) shows $\langle \Delta \hat{H}_{\rm total} \rangle$ for the two heavy-quark system as a function of time, defined as
\begin{equation}
    \langle \Delta \hat{H}_{\rm total} \rangle = 
    \langle \hat{H}_{\rm total} \rangle|_{n_Q=2} 
    - \langle \hat{H}_{\rm total} \rangle|_{n_Q=1,\ x(0)=3}
    - \langle \hat{H}_{\rm total} \rangle|_{n_Q=1,\ x=5}
    + \langle \hat{H}_{\rm total} \rangle|_{n_Q=0} 
    \ ,
\label{eq:diffH}
\end{equation}
where $\langle \hat{H}_{\rm total} \rangle|_{n_Q=2} $ refers to the total energy of the system with both heavy quarks, and $\langle \hat{H}_{\rm total} \rangle|_{n_Q=1,\ x(0)=3}$ refers to the energy with one moving quark starting at $x=3$.
Both $\langle \hat{H}_{\rm total} \rangle|_{n_Q=2} $ and $\langle \hat{H}_{\rm total} \rangle|_{n_Q=1,\ x(0)=3}$ are time dependent, depending upon the location and speed of the moving heavy quark. 
$\langle \hat{H}_{\rm total} \rangle|_{n_Q=1,\ x=5}$ refers to the energy with a static heavy quark at $x=5$, and $\langle \hat{H}_{\rm total} \rangle|_{n_Q=0}$ refer to the energy of the vacuum, both independent of time.
This difference between calculated energies removes the leading-order lattice artifacts, and would vanish if there were no interactions between the heavy mesons.

%%%%%%%%%%%%%%%%%%%%%%%%%%%%
\section{Quantum Simulation of SU(2) in 1+1D}
\label{sec:QSim}
\noindent
This section presents the methods, circuits and error-mitigation techniques that we have used to perform quantum simulations of dynamics in the system described above on three physical sites $L=3$ (which maps to 18 qubits), and the results we have obtained using IBM's {\tt heron} processor quantum computer {\tt ibm\_pittsburgh}.
The ground state of the system with a heavy quark present is first prepared via ADAPT-VQE and DDec, then the heavy quark is translated to an adjacent physical site by FSWAP operations. After translation, the state is time-evolved with the system Hamiltonian to simulate the dynamic response of the light quarks. 
The quantum circuits are designed with the objective of minimizing CNOT depth and count in the context of nearest-neighbor connectivity, as applicable to superconducting-qubit architectures. 
A resource analysis,  including the scaling of circuit resources with regard to the number of physical sites and time evolution steps,  is discussed. 
We measure the associated charge distributions and energy losses using IBM's {\tt ibm\_pittsburgh}, and compare with classically calculated results.

%%%%%%%%%%%%%%%%%%
\subsection{Algorithms, Techniques and Quantum Circuits}
\label{sec:ApTpC}
\noindent
There are two types of operators involved in the state preparation and time evolution: those with two $\hat \sigma^\pm$s, with $-\hat{Z}$s inserted in between to accommodate the fermion-exchange anti-symmetry, and those with four $\hat \sigma^\pm$s.
In the two $\hat \sigma^\pm$s case, the form $(\hat{\sigma}^{-}\hat Z^n\hat{\sigma}^{+} - {\rm h.c.})$ is used in the meson operators in state preparation (to furnish real wavefunctions), while $(\hat{\sigma}^{-}\hat Z^n\hat{\sigma}^{+}+ {\rm h.c.})$ is used in time evolution from the kinetic term. 
Operators with four $\hat \sigma^\pm$s can be implemented using GHZ transformations, as presented in Refs.~\cite{Stetina:2020abi,Farrell:2022wyt}. 
Baryon operators of the form $\hat\sigma^+\hat\sigma^+\hat\sigma^-\hat\sigma^- - {\rm h.c.}$ are used in state preparation, and those of the form $\hat\sigma^+\hat\sigma^-\hat\sigma^-\hat\sigma^+ + {\rm h.c.}$ are used in time evolution from the gauge Hamiltonian, implemented by GHZ transformation circuits.
The GHZ transformation used to implement the gauge Hamiltonian evolution is also used in measuring the energies.

%%%%%%%%%%%%
\subsubsection{R-box Constructions}
\label{sec:Rboxes}
\begin{figure}[htpb]
    \centering
    \includegraphics[width=0.75\textwidth]{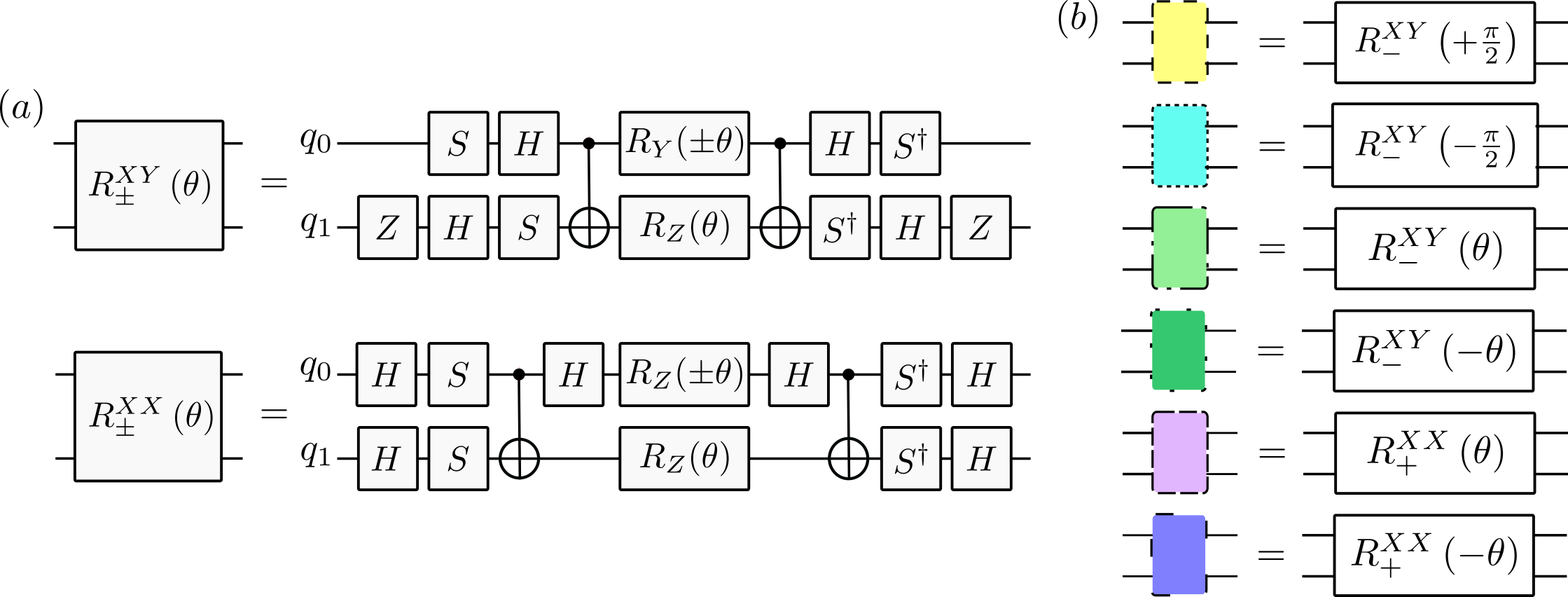}
    \caption{Panel (a) shows the circuits implementing $R_{\pm}^{XY}$ and $R_{\pm}^{XX}(\theta)$, and
    panel (b) shows the shorthand notation that will be used in subsequent (composite) circuits.}
    \label{fig:rbox}
\end{figure}

\noindent
Building upon the work in Refs.~\cite{Algaba:2023enr,Farrell:2023fgd,Farrell:2024fit}, the circuits for operators with two $\hat \sigma^\pm$s can be implemented for nearest-neighbor connectivity with two-qubit blocks, referred to as R-boxes, defined as
\begin{equation}
    R_{\pm}^{XY}(\theta) \equiv   e^{-i\frac{\theta}{2}(\hat{Y}\hat{X} \pm \hat{X} \hat{Y})} \ , \
    R_{\pm}^{XX}(\theta)  \equiv  e^{-i\frac{\theta}{2}(\hat{X}\hat{X} \pm \hat{Y} \hat{Y})} \ .
    \label{eq:RboxDef}
\end{equation}
Circuits for implementing these building blocks are shown in Fig.~\ref{fig:rbox}(a), and the shorthand notation we will use to represent them in subsequent (larger) circuits are given in Fig.~\ref{fig:rbox}(b).
These R-boxes can be sequentially layered to implement long-ranged operators in the form of $e^{-i\theta (\hat{\sigma}^{\pm}\hat Z^n\hat{\sigma}^{\mp} \pm {\rm h.c.})}$, examples of which are given in App.~\ref{appen:cir}.
%

%%%%%%%%%%
\subsubsection{State Preparation}
\label{sec:StatePrepCir}
\noindent 
A few steps are taken to prepare the ground state of the interacting theory in the presence of a heavy quark.
First, the  SC ground state is prepared, as given in Eq.~\eqref{eq:SU2SCQ1} for one spatial site, and Eqs.~\eqref{eq:nq2L2} and \eqref{eq:SCL3n1x0} for $L=2$ and $L=3$ sites, respectively.
The corresponding circuit to prepare the SC ground state for the spatial site with the heavy quark is given by
\begin{equation}
 \begin{tikzpicture}[baseline={([yshift=1.0 ex]current bounding box.center)}]
  \node[scale=0.8] {
\begin{quantikz}[row sep={0.9cm,between origins}]
    \lstick{$\ket{Q_r}$}      & \gate{X}\slice[style=colorslice]{} &          &           & \targ{}   &   \\
    \lstick{$\ket{Q_g}$}      & \gate{X}          & \gate{X} & \targ{}   & \ctrl{-1} &   \\   
    \lstick{$\ket{q_r}$}      & \gate{X}          & \gate{H} & \ctrl{-1} & \ctrl{1}  &   \\
    \lstick{$\ket{q_g}$}      & \gate{X}          &          &           & \targ{}   &   \\
    \lstick{$\ket{\bar q_r}$} &                   &          &           &           &   \\
    \lstick{$\ket{\bar q_g}$} &                   &          &           &           & 
\end{quantikz}
};
\end{tikzpicture}
\label{eq:stateprep1}
\end{equation}
with the remaining $L-1$ spatial sites are prepared with 
$\hat X\hat X\hat X\hat X\hat I\hat I$
acting on $|0\rangle^{\otimes 6 (L-1)}$.

DDec is used to prepare the interacting  ground state from the SC ground state,
as discussed in Sec.~\ref{sec:L3SP}.
For each lattice segment of length $l_i$, classical ADAPT-VQE is used to establish the sequence of unitary operators and their corresponding coefficients that prepare the ground state with fidelity greater than a prescribed value.
The lengths of these lattice segments should ideally be chosen to be larger than the longest correlation length in the theory.
After each lattice segment is optimized by minimizing the energy, the ground state of the full system is established by first applying the operators responsible for  preparing each segment, then addressing the interactions and boundary effects between segments by applying operators across each boundary. 
The coefficients of each of the operators are re-optimized at each stage of ``the build'' for the full ground state.
Because of the translation symmetry of regions of the state when far from the heavy quark, scalable operators can be implemented in these sections of the lattice.

Preparing the interacting heavy-quark ground state from the SC heavy-quark ground state differs from preparing it from the interacting vacuum.
That is because the segment containing the heavy quark is distinct from the vacuum, requiring a different operator sequence and set of angles to accommodate modifications to the light-quark wavefunction within correlation lengths of the heavy quark.
The same operator pool with the meson and baryon operators can be used to prepare the interacting vacuum and heavy-quark states from their corresponding SC states, but the operator sequences selected by ADAPT-VQE are different.

Collecting the results from Sec.~\ref{sec:CSim}, a general operator in the state preparation operator pool is a color singlet with significant overlap with states in the hadronic spectrum, i.e., a set of interpolating operators.   
This translates into a pool of relatively local meson and baryon operators (with size set by the confinement scale).  
These operators have their fermion operators separated by a modest number of sites, with the impact of the latter  suppressed by the confinement scale.
A set of interpolating meson operators that is found to be effective as circuit elements is (compiled from Eqs.~\eqref{eq:mesOpsL1}, \eqref{eq:OPxyL2def} and \eqref{eq:OM2B2def}):
\begin{align}
    \hat{O}_{M0} & \equiv (\hat X \hat Z\hat Y - \hat Y \hat Z\hat X )   \hat I + \hat I  (\hat X \hat Z\hat Y - \hat Y \hat Z\hat X )
    \ ,\nonumber\\
    \hat{O}_{M1} & \equiv (\hat Y \hat Z^3\hat X - \hat X \hat Z^3 \hat Y )  \hat I+ \hat I   (\hat Y \hat Z^3\hat X - \hat X \hat Z^3 \hat Y )\ ,\nonumber\\
    \hat{O}_{M2} & \equiv (\hat X \hat Z^7\hat Y - \hat Y \hat Z^7 \hat X )   \hat I+ \hat I   (\hat X \hat Z^7\hat Y - \hat Y \hat Z^7 \hat X )
     \ , 
     \label{eq:mesOps}
\end{align}
where it is understood that the $\hat X, \hat Y$ act on the light quarks and anti-quarks only.
$\hat{O}_{M0}$ acts on $r\bar r$ and $g\bar g$ on the same physical site, $\hat{O}_{M1}$ acts on $\bar r r$ and $\bar g g$ on two neighboring sites, and $\hat{O}_{M2}$ acts on $r\bar r$ and $g\bar g$ also on two neighboring sites. 
Note that $\hat O_{M1}$ has the opposite relative sign from $\hat O_{M0}$ and $\hat O_{M2}$ because of the number of $\hat Z$s it encounters when constructed from an operator commuting with the mass term.
The implementation of the $\hat U_{Md} = e^{-i\theta\hat{O}_{Md}}$ circuits can be found in App.~\ref{appen:cir}.
In terms of entangling gates, $\hat U_{M0}$ has depth 6 with a total number of 8 CNOT gates, $\hat U_{M1}$ has depth $14$ with a total number of 28 CNOT gates, and $\hat U_{M2}$ has depth 26 with a total number of 60 CNOT gates.

\begin{figure}[htpb]
    \centering \includegraphics[width=0.99\textwidth]{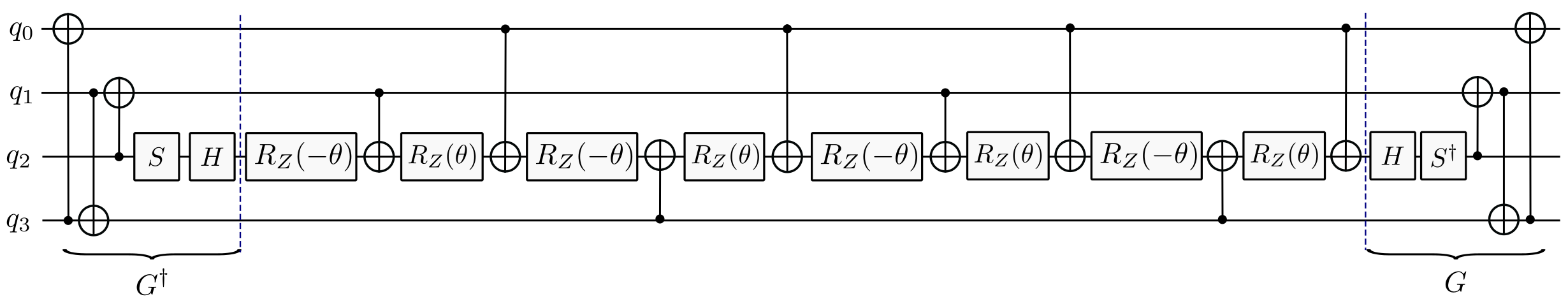}
    \caption{A circuit implementing $e^{-i\theta\hat O_{B0}}$, given
    in Eq.~\eqref{eq:B01}, where
    $G$ and $G^{\dagger}$ 
    implement a GHZ transformation. Additional details about constructing this circuit
    can be found in App.~\ref{appen:cir}.
    }
    \label{fig:baryon}
\end{figure}

The fundamental local baryon operators that create baryon-anti-baryon pairs on the same spatial site and  separated by one spatial site are,
\begin{equation}
    \hat O_{B0} = 4i(\hat\sigma^+\hat\sigma^+\hat\sigma^-\hat\sigma^- - {\rm h.c.}) 
    \ ,\ 
    \hat O_{B1} = -4i(\hat\sigma^+\hat\sigma^+ \hat{I}^2 \hat\sigma^-\hat\sigma^-
    - {\rm h.c.})
    \ \ .
    \label{eq:B01}
\end{equation}
The circuit for ${\exp}(-i\theta\hat O_{B0})$ is shown in Fig.~\ref{fig:baryon}, and ${\exp}(-i\theta\hat O_{B1})$ can be implemented with two additional qubits in the middle. Additional discussions can be found in App.~\ref{appen:cir}.
The CNOT depth of $\hat U_{Bd}$ on an all-to-all connected quantum device (all-to-all CNOT depth) is 14.

With the fundamental building blocks of the circuits established, the overall sequence of state preparation operators can be developed.
As described before, for notational clarity and specificity, an argument $(x)$ is added to $\hat O$ (for example, $O_{M0}^{(x)}$) to indicate the physical site the operator acts on (and $(x,y)$ if the operator acts on two sites).
The classical optimization of the sub-lattice operators and the implementation of DDec is detailed above Sec.~\ref{sec:L3SP}.
We repeat elements of that discussion for completeness.
As sketched in Fig.~\ref{fig:DDecL3}, preparing the $Q = 1$ ground state on the $x=0,1$ spatial sites of the $L=3$ lattice with sub-percent precision is accomplished by,
\begin{equation}
\begin{aligned}
|\psi_1\rangle = 
e^{-i (-0.2000) O_{M0}^{(0)}}&
e^{-i (0.0407) O_{B1}^{(0,1)}}
e^{-i (0.0196) O_{B0}^{(1)}}\\
& e^{-i (0.1820) O_{M0}^{(1)}}
e^{-i (0.2642) O_{M1}^{(0,1)}}
e^{-i (0.3802) O_{M0}^{(0)}}\ 
|SC,3,1\rangle
\ ,
\end{aligned}
\end{equation}
with the fit angles given in Table~\ref{tab:L2VQEmanuelQ1}.
The $Q = 0$ ground state on the $x=2$ site is accomplished by
\begin{equation}
    |\psi_2\rangle = e^{-i (0.0270) O_{B0}^{(2)}} e^{-i (0.2200) O_{M0}^{(2)}} \ 
    |\psi_1\rangle
    \ ,
\end{equation}
and sewing the two segments together to furnish the $L=3$ ground state with $n_Q=1$ is accomplished using
\begin{equation}
    \ket{\psi_{f}} = e^{-i (-0.0995) O_{M2}^{(1,2)}}e^{-i (0.2314) O_{M1}^{(1,2)}}\ket{\psi_{2}}
    \ .
\end{equation}

The all-to-all CNOT depth of the state-preparation circuit is depth 2 for the SC vacuum (with 3 CNOT gates) for a lattice of any size, is depth 50 for the DDec of the $L=2$ sublattice with cancellations and depth 20 for the $L=1$ sublattice (which can be performed in parallel).
The boundary operators require an additional depth of 38, which can be partially performed in parallel.
The total all-to-all CNOT depth for state preparation of the $L=3$ system with $n_Q=1$ is depth 72.

%%%%%%%%%%
\subsubsection{Heavy-Quark Translation and Time Evolution}
\noindent
To provide examples of the circuits that can implement the motion of heavy quarks, we move a heavy quark from the first physical site to the second using FSWAP operators with $-\hat Z$s inserted in the middle for fermion anti-symmetry.
Expression of the FSWAP operator for a single color is given in Eq.~\eqref{eq:FSWAP6sites}. 
The R-box constructions analogous to those used for state preparation can be used to implement this motion. 
For nearest-neighbor connectivity, the FSWAP circuit CNOT-depth is 22.
\begin{figure}[htpb]
    \centering
    \includegraphics{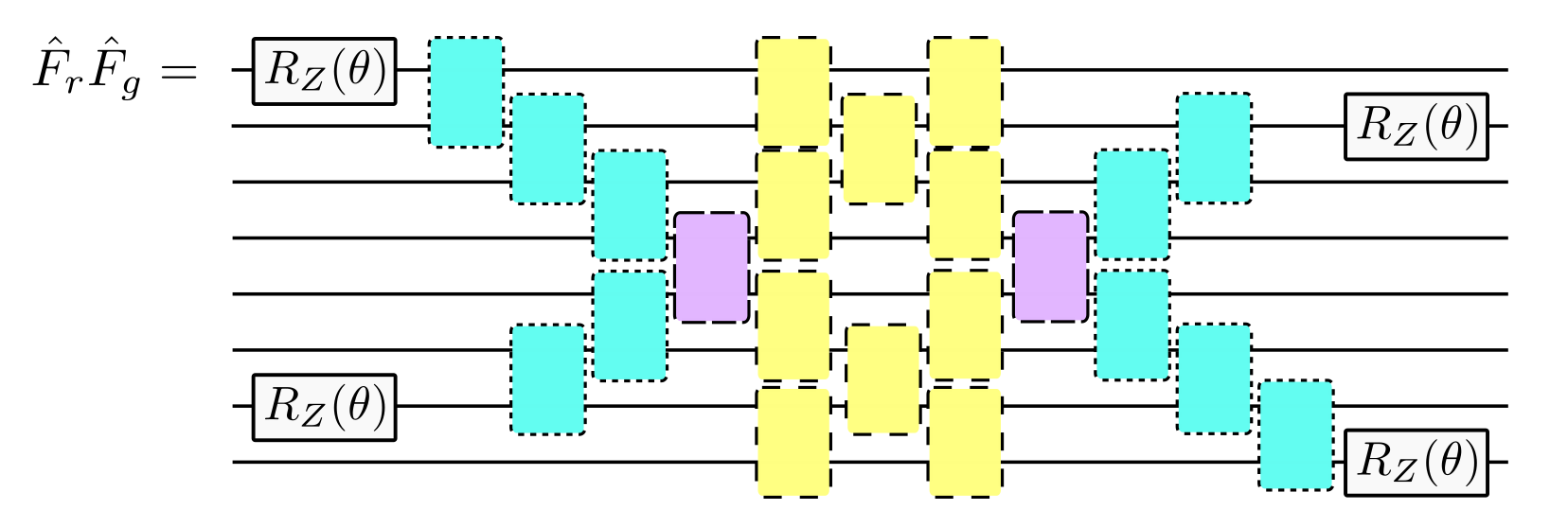}
    \caption{Circuits to implement the spatial translation of a heavy quark between adjacent spatial lattice sites using FSWAP circuits. 
    The $\theta$s appearing in circuit elements take the value of $-\pi/2$.
    }
    \label{fig:fswap}
\end{figure}
Figure~\ref{fig:fswap} shows the composite quantum circuit that implements motion of a heavy quark, where SU(2) symmetry is used to compress the depth of the circuit. Construction details can be found in App.~\ref{appen:cir}.

After translating the heavy quark by one spatial site, the state is time evolved with a Trotterized Hamiltonian. 
Implementation of the mass term is straight forward, as it is a product of single-qubit $Z$-rotations:
\begin{align}
    \hat U_{m,q}(t) & = e^{-i \frac{t m_q}{2} \hat Z_j} = R_Z(t m_q )\ \ :\ \ {\rm quarks} \ ,
    \nonumber\\
    \hat U_{m,\overline{q}} (t) & = e^{+i \frac{t m_q}{2} \hat Z_j} = R_Z(-t m_q )\ \ :\ \ {\rm antiquarks}    
    \ .
    \label{eq:Um}
\end{align}
For a given spatial site, these evolution operators are implemented by the circuit shown in Fig.~\ref{fig:h_full}.
As these are single-qubit gates, they do not contribute to the depth of the circuit.
We have not included the time evolution of mass Hamiltonian associated with the heavy quark sites. 
As their kinetic term is not included, their mass terms add only an overall phase to the wavefunction that is position independent.

The kinetic terms correspond to two types of operators, those with one $\hat Z$ in the middle (denoted $\hat{H}_{k0}$) and those with three (denoted $\hat{H}_{k1}$), as given in Eq.~\eqref{eq:UHK},
\begin{align}
        \hat{H}_{k0} & \equiv (\hat X \hat Z\hat X + \hat Y \hat Z\hat Y ) \otimes \hat I + \hat I \otimes (\hat X \hat Z\hat X + \hat Y \hat Z\hat Y )\ , \nonumber \\
        \hat{H}_{k1} & \equiv  (\hat X \hat Z ^3 \hat X + \hat Y \hat Z ^ 3 \hat Y ) \otimes \hat I + \hat I \otimes (\hat X \hat Z ^3 \hat X + \hat Y \hat Z ^ 3 \hat Y )
        \ .
        \label{eq:UHK}
\end{align}
$\hat{H}_{k0}$ acts on $r\bar r$ and $g\bar g$ on the same physical site, while $\hat{H}_{k1}$ acts on $\bar r r$ and $\bar g g$ on two neighboring sites.
When acting on the same spatial sites, $\hat{H}_{k0}$ and $\hat{H}_{k1}$ do not commute. 
An additional site label $(x)$ is included to indicate which site the operator acts on. 
The complete kinetic circuit is implemented using first order Trotterization, where a layer of $\hat{H}_{k0}$ evolution is first applied, followed by $\hat{H}_{k1}$ evolution, as given in Eq.~\eqref{eq:UHKevol},
\begin{equation}
        e^{-it\hat H_k} \rightarrow  \exp\left(i \frac{t}{4}\sum_{x = 0}^{L-1}\hat{H}_{k0}^{(x)}\right) \exp\left(i \frac{t}{4}\sum_{x = 0}^{L-2}\hat{H}_{k1}^{(x,x+1)}\right)\ .
        \label{eq:UHKevol}
\end{equation}
Time evolution by $\hat{H}_{k0}$ is implemented by a circuit with a two-qubit depth of 6, while evolution by $\hat{H}_{k1}$ is implemented by a circuit of depth 14.
The leading-order Trotterized kinetic-energy evolution operator acting on three physical sites is shown in Fig.~\ref{fig:h_full}.
This can be straightforwardly generalized to larger systems, as well as higher order Trotterizations.

\begin{figure}[htpb]
    \centering
    \includegraphics[width=\textwidth]{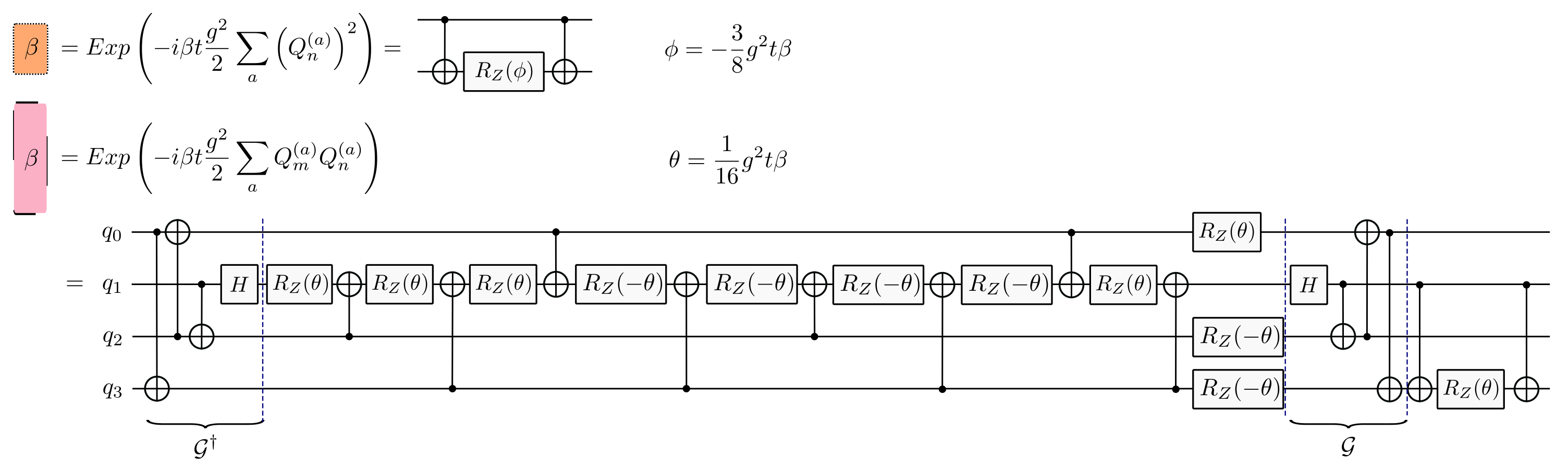}
    \caption{Circuit elements implementing evolution by the chromo-electric Hamiltonian,
    given in Eqs.~\eqref{eq:HgL3} and \eqref{eq:Qops}.
    The orange elements implement single-site contributions, while the  pink implements contributions from different spatial sites. 
}
\label{fig:hgauge_seg}
\end{figure}

\begin{figure}[htpb]
    \centering
    \includegraphics[width = \textwidth]{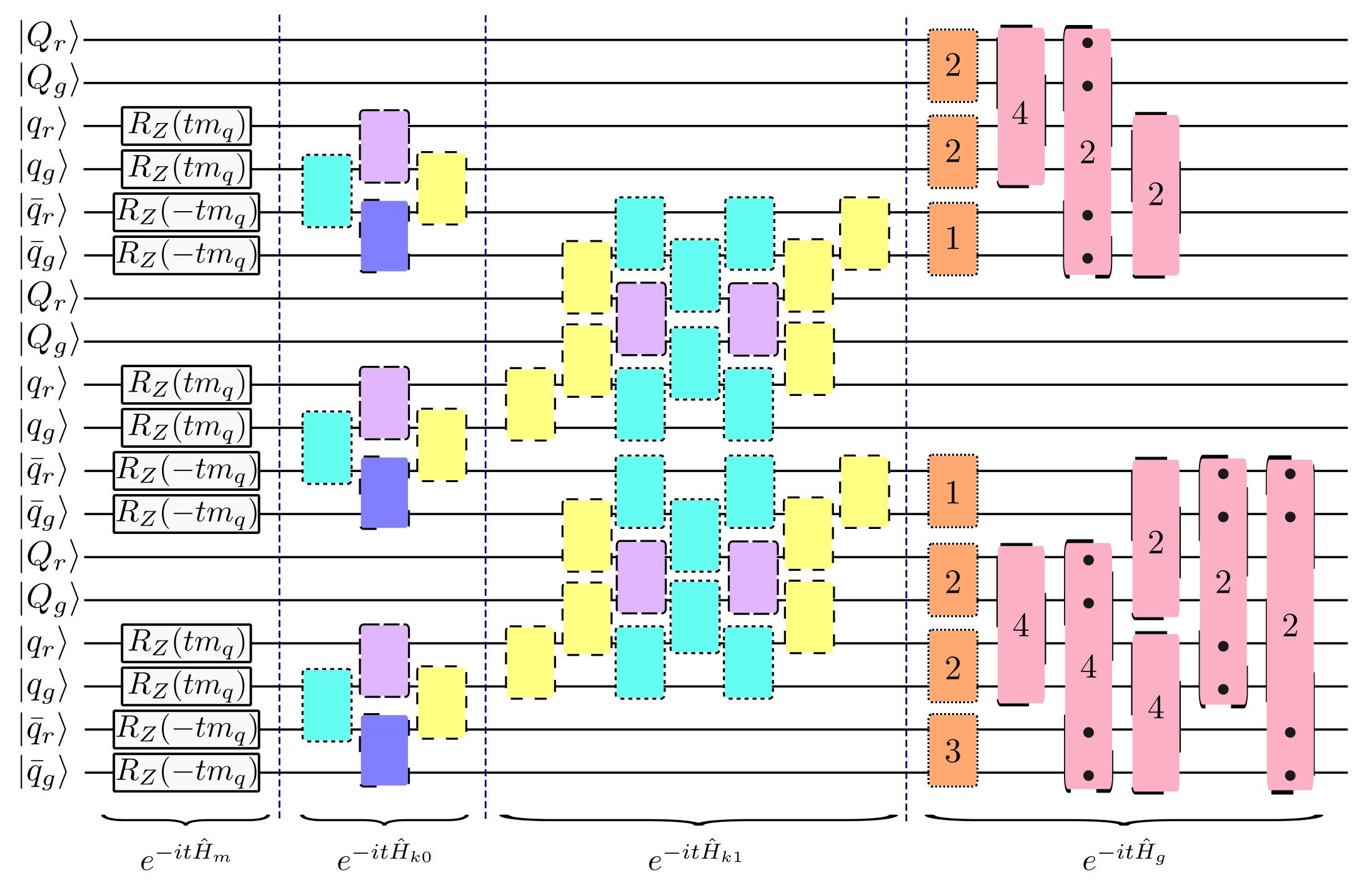}
    \caption{A quantum circuit implementing the terms involved in a Trotterized time evolution. The left-most block implements the mass terms,
    given in Eq.~\eqref{eq:Um}. 
    This is followed by
    the blocks related to the kinetic-energy operators, given in Eq.~\eqref{eq:UHKevol}.
    The angle defining the circuit elements, given in Fig.~\ref{fig:rbox}, is given by $\theta = t/2$. 
    Finally, the right-most block implements
    the complete chromo-electric contribution to the Hamiltonian (without optimization), given in Eqs.~\eqref{eq:HgL3} and \eqref{eq:Qops}.
    The orange and pink circuit elements are defined in Fig.~\ref{fig:hgauge_seg}, and the black dots on the pink elements indicate the engaged qubits.
    }
    \label{fig:h_full}
\end{figure}

The contribution from the energy in the chromo-electric field can be usefully expressed in its symmetric form. 
Making use of the lattice having overall vanishing color charge, the chromo-electric field at each of the sites can be obtained symmetrically from both ends of the lattice (similar to the construction in App.~\ref{app:L2} for $L=2$, and the general case in App.~\ref{app:explicitham}).
Using
\begin{align}
    E^{(a)}_0 & = \hat{\mathcal{Q}}^{(a)}_{0}+ \hat{\mathcal{Q}}^{(a)}_{1}\ ,\ E^{(a)}_1 =  E^{(a)}_0 + \hat{\mathcal{Q}}^{(a)}_{2}\ , \qquad {\rm starting\ from\ the}\ n=0\ {\rm site} \ ,\ \nonumber\\
    E^{(a)}_5 & = \hat{\mathcal{Q}}^{(a)}_{8}\ ,\ E^{(a)}_4 =  E^{(a)}_5 +\hat{\mathcal{Q}}^{(a)}_{6} + \hat{\mathcal{Q}}^{(a)}_{7}\ ,\  E^{(a)}_3 = E^{(a)}_4 + \hat{\mathcal{Q}}^{(a)}_{5}\ , 
    \nonumber \\
    & \qquad \qquad \qquad \qquad\qquad\qquad \qquad\qquad\qquad{\rm starting\ from\ the}\ n=8\ {\rm site}
\ , 
\end{align}
provides for the chromo-electric component of the Hamiltonian to be expressed as 
\begin{align}
H_g  =  \frac{g^2}{2} \sum_{a} & 
\left[ 2 \hat{\mathcal{Q}}_{0}^2 
+ 2 \hat{\mathcal{Q}}_{1}^2 
+ \hat{\mathcal{Q}}_{2}^2 
+ 4 \hat{\mathcal{Q}}_{0} \hat{\mathcal{Q}}_{1}
+ 2 \hat{\mathcal{Q}}_{0} \hat{\mathcal{Q}}_{2}
+ 2 \hat{\mathcal{Q}}_{1} \hat{\mathcal{Q}}_{2}
\right.\nonumber\\
& \ + 2 \hat{\mathcal{Q}}_{6}^2 + 2 \hat{\mathcal{Q}}_{7}^2 + 3 \hat{\mathcal{Q}}_{8}^2
+ 4 \hat{\mathcal{Q}}_{6} \hat{\mathcal{Q}}_{7} + 4 \hat{\mathcal{Q}}_{6} \hat{\mathcal{Q}}_{8}
+ 4\hat{\mathcal{Q}}_{7} \hat{\mathcal{Q}}_{8} \nonumber\\
& \ \left. + \hat{\mathcal{Q}}_{5}^2 + 2 \hat{\mathcal{Q}}_{5} \left( \hat{\mathcal{Q}}_{6} + \hat{\mathcal{Q}}_{7} + \hat{\mathcal{Q}}_{8} \right)
\right]
\ .
\label{eq:HgL3}
\end{align}
where the generator indices $a$ have been suppressed in Eq.~\eqref{eq:HgL3} for simplicity.
The distinct structures given in Eq.~\eqref{eq:HgL3}, acting on $r$-$g$ pairs, can be implemented by the two general operators 
\begin{align}
\sum_a \left(\hat{\mathcal{Q}}_n^{(a)}\right)^2 & = \frac{3}{8} \left( \hat I - \hat Z_{2n} \hat Z_{2n+1} \right) \ , \nonumber\\
\sum_a  \hat{\mathcal{Q}}_n^{(a)} \hat{\mathcal{Q}}_m^{(a)}  & = \frac{1}{16}\left(\hat Z_{2n} -\hat Z_{2n+1}\right)\left(\hat Z_{2m} -\hat Z_{2m+1}\right) + \frac{1}{2} \left(\hat\sigma_{2n}^+\hat\sigma_{2n+1}^- \hat\sigma_{2m}^-\hat\sigma_{2m+1}^+ \ +\ {\rm h.c.} \right) \ ,
\label{eq:Qops}
\end{align}
where $(\hat{\mathcal{Q}}_n^{(a)})^2$ corresponds to single-staggered-site contributions, 
while $\hat{\mathcal{Q}}_n^{(a)} \hat{\mathcal{Q}}_m^{(a)}$ corresponds to contributions from different staggered sites.
Same-site interactions contain only $\hat Z \hat Z$ operators and can be implemented straightforwardly. Interactions between different staggered sites can be implemented efficiently using a GHZ transformation~\cite{Farrell:2022vyh}.\footnote{The GHZ transformation $\mathcal{G}$ used here is to be distinguished from the GHZ transformation $G$ use earlier in the state preparation circuits. They facilitate the same goal of implementing a target operator by transforming the basis of efficient circuits in the computational basis. Since the target operators in state preparation and time evolution have different forms, the GHZ transformations are slightly different} The operator order from left to right correspond to circuit wires from top to bottom. 
\begin{align}
    \hat\sigma^+\hat\sigma^-\hat\sigma^-\hat\sigma^+ + {\rm h.c.} = 
    \frac{1}{8} & \left[
    \hat X\hat X\hat X\hat X
    +
    \hat X\hat X\hat Y\hat Y
    +
    \hat X\hat Y\hat X\hat Y
    -
    \hat X\hat Y\hat Y\hat X
    \right.
    \\
    & \left. -\hat Y\hat X\hat X\hat Y
    +
    \hat Y\hat X\hat Y\hat X
    +
    \hat Y\hat Y\hat X\hat X
    +
    \hat Y\hat Y\hat Y\hat Y
     \right] ,
    \nonumber\\
    \hat {\cal G}^\dagger \left(\hat\sigma^+\hat\sigma^-\hat\sigma^-\hat\sigma^+ + {\rm h.c.} 
    \right) \hat {\cal G} = 
        \frac{1}{8} & \left[
        -\hat Z\hat Z\hat Z\hat I - 
        \hat Z\hat Z\hat I\hat Z + 
        \hat I\hat Z\hat Z\hat Z + 
        \hat I\hat Z\hat I\hat I \right. \nonumber \\
        & \left. -
        \hat Z\hat Z\hat Z\hat Z - 
        \hat Z\hat Z\hat I\hat I + 
        \hat I\hat Z\hat Z\hat I + 
        \hat I\hat Z\hat I\hat Z  
     \right]\ .
    \label{eq:XYTOIZaa}
\end{align}
Additionally, the $\hat Z \hat Z$ interactions between staggered sites can also be transformed into a GHZ basis. 
The $\hat Z \hat Z$ operators in a GHZ basis on four qubits are\footnote{Note that one four-qubit GHZ transformation can only transform three $\hat Z \hat Z$ operators into a single $\hat Z$, so one of the four 
remains untransformed.
In the SU(3) case, all the $\hat Z \hat Z$ operators can be absorbed into the GHZ basis because there are $\binom{3}{2} = 3$ sets of 4-qubit GHZ transformations and $3 \times 3 = 9$ $\hat Z \hat Z$ operators.
The ratio between the number of $\hat Z \hat Z$ operators and GHZ transformations is $3$, the maximum number that can be absorbed. 
In the SU(2) case, there are $\binom{2}{2} = 1$ set of 4-qubit GHZ transformations and $2 \times 2 = 4$ $\hat Z \hat Z$ operators. 
The ratio between the number of $\hat Z \hat Z$ operators and GHZ transformations is $4$.}
\begin{equation}
\left(\hat Z_0 -\hat Z_{1}\right)\left(\hat Z_2 -\hat Z_{3}\right) = \hat Z_0\hat Z_2 - \hat Z_0\hat Z_3 -\hat Z_1\hat Z_2+\hat Z_1\hat Z_3 = \mathcal{G}(\hat Z_0 -\hat Z_3 - \hat Z_2)\mathcal{G}^{\dagger} + \hat Z_1\hat Z_3 \ .
\end{equation}
The associated quantum circuits for $\mathcal{G}$, $\mathcal{G}^{\dagger}$, and for the two types of interactions, are given in Fig.~\ref{fig:hgauge_seg}.
The two-qubit orange circuit element has depth 2, and each four-qubit pink circuit element has depth 16.
The total depth of the Trotterized time-evolution operator implementing the chromo-electric Hamiltonian (without optimization) is 78, as shown in Fig.~\ref{fig:h_full}.
This does not include truncations of the chromo-electric contributions made possible by confinement~\cite{Farrell:2024fit}, as the chromo-electric Hamiltonian is symmetrically implemented, and  interactions act on a maximum of two spatial sites.
We perform a second order Trotterization, where each step of the time evolution is
\begin{equation}
    U_2(t) = e^{-i\frac{t}{2}\hat{H}_{k1}}e^{-i\frac{t}{2}\hat{H}_{k0}}e^{-it\hat{H}_{m}}e^{-it\hat{H}_{g}}e^{-i\frac{t}{2}\hat{H}_{k0}} e^{-i\frac{t}{2}\hat{H}_{k1}}
    \ .
\end{equation}
Combining all of the contributions and without optimization, the total two-qubit gate depth of the circuit implementing one application of the complete second Trotterized time-evolution operator is 
\begin{equation}
{\rm depth} = 20 \ ({\rm Kinetic}) \times 2 + 0 \ ({\rm Mass}) + 78\  ({\rm Gauge})
\ =
 118
 \ .
\end{equation}
Since some parts of the circuits can be performed in parallel, the actual depth is 112. When performing multiple second-order Trotter steps, the $e^{-i\frac{t}{2}\hat{H}_{k1}}$ terms between consecutive steps can be combined. 
Therefore, the steps after the first one are shallower by the depth of $e^{-it\hat{H}_{k1}}$ (102 compared to 112). 
An analysis of the systematic errors in first- and second-order Trotterization can be found in App.~\ref{appen:resources}.

%%%%%%%%%%%%%%%%%%%%%%%%%%%%%%%%%%%%%%%%%
\subsection{Resource Projection and Mapping to Device}
\label{sec:resproj}
\noindent
The scaling of quantum resources is considered with respect to two simulation parameters, increasing the lattice size and increasing the number of Trotter steps in the time evolution (which permits a simulation to reach longer times with limited systematic errors).
On the axis of increasing the number of Trotter steps, state preparation is not impacted, and as each step of the time-evolution circuit has the same depth, the total depth of the time evolution grows as $\mathcal{O}(n)$ where $n$ is the number of steps.
For increasing the system size, the circuit depth for initial state preparation depends on the imposed constraints. 
It remains constant when the heavy quark is initialized 
near a boundary with a short  correlation length, but increases when higher precision or longer-range correlations are required.
The depth of each step of the Trotterized time evolution is affected differently by each term in the Hamiltonian. 
For the mass and kinetic terms, increasing the number of spatial lattice sites does not increase the depth of circuit elements required to implement one Trotter step.
However, the contribution from the chromo-electric term requires more attention because of its naive all-to-all connectivity.
To permit scalable simulation, the chromo-electric term needs to be truncated at a distance determined by the confinement scale, introducing an additional, but exponentially small, truncation error.

\begin{figure}[htpb]
    \centering
    \includegraphics[width=0.3\linewidth]{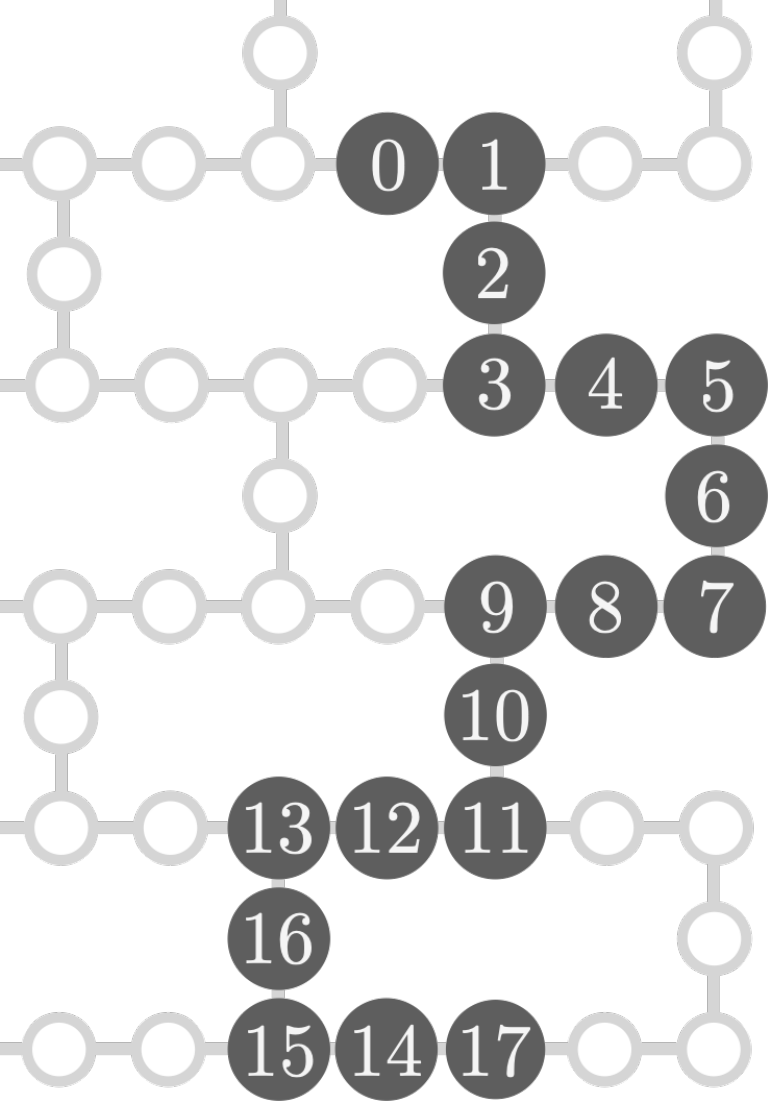}
    \caption{
    Qubit layout of the $L=3$ system on {\tt ibm\_pittsburgh}.
}
    \label{fig:layout}
\end{figure}

For transpilation to a target device, the circuits are adapted to the native gate set and the connectivity of the device.
Our target device is IBM's  {\tt ibm\_pittsburgh} with a heavy-hex connectivity.
Each CNOT gate is transpiled to a CZ gate plus single qubit rotations, so they are equivalent in the consideration of two-qubit-gate count. 
To optimize the depth of the transpiled time-evolution circuit, in particular the gauge term, we use the layout displayed in Fig.~\ref{fig:layout}, where the last qubits have been rearranged.
This is not necessarily the layout that gives the shallowest possible depth, since the goal is not to over-optimize this 18 qubit system. 
The qubits in Fig.~\ref{fig:layout} were selected by imposing a maximum error in the CZ gate of 0.005 on {\tt ibm\_pittsburgh} with a median CZ error of 0.0017. 
The resource requirements for each component of the simulation (state preparation and time evolution) are shown in Table~\ref{tab:component_resource}.
Further details, such as CZ depth distribution across the lattice, can be found in App.~\ref{appen:resources}.
\begin{table}[htpb]
\renewcommand{\arraystretch}{1.1}
    \centering
    \begin{tabular}{|r|c|c|}
        \hline
        Term & CZ All-to-All & CZ transpiled \\
        \hline
        \hline
        Strong Coupling & 2 & 2\\
        VQE Layers & 70 & 126\\
        Fermionic SWAP & 22 & 22 \\
        Time Evolution & 112 & 254\\\hline
        Total & 182 & 398 \\\hline
        Additional time step & 102 & 248 \\\hline\hline
    \end{tabular}
    \caption{
    Quantum resource requirements 
    (entangling gates)
    for preparing the initial states and for time evolution.
    The transpiled requirements are for the fixed qubit layout shown in Fig.~\ref{fig:layout}.
    The total resource requirements are less than the sum of the listed components 
    because some of the operations can be performed in parallel.
    }
    \label{tab:component_resource}
\end{table}
%

%%%%%%%%%%%%%%%%%%%%%%%%%%%%
\subsection{Results from Quantum Simulations}
\label{sec:QSresults}
\noindent
A suite of state preparation and time evolution circuits, along with the associated error-mitigation circuits, that have been detailed in previous sections, were implemented with IBM's {\tt qiskit}~\cite{qiskit} and executed on IBM's quantum computer {\tt ibm\_pittsburgh}.
The results of these simulations were analyzed classically with standard statistical techniques using {\tt python}, {\tt julia} and {\tt Mathematica}.
We focus on two observables related to heavy-quark motion: charge distribution and energy loss.

%%%%%%%%%%%%%%%%%%%%%%%%%%%%
\subsubsection{Charge Distribution}
\label{sec:chargeD}
\noindent
An initial state of the $L=3$ ($n_{\rm qubits} = 18$) system using Hamiltonian parameters $m_q=0.1$ and $g=1.0$ was prepared with a heavy quark residing on the  $x=0$ spatial site. 
The heavy quark is then moved to the $x=1$ spatial site and the system is time evolved forward to $t=1.0$ using a single second-order Trotter step.
Several error mitigation methods were applied to compensate for the noisy hardware. 
Pauli twirling~\cite{Wallman:2015uzh} was performed around the entangling gates, with 64 distinct twirled circuits for the production runs.
Dynamical decoupling~\cite{Viola:1998jx,Ezzell:2022uat} and measurement mitigation~\cite{Berg:2020ibi} were applied through the {\tt qiskit}~\cite{qiskit} Estimator primitive. 
To remove the remaining bias due to incoherence of the device, we apply operator-decoherence renormalization (ODR)~\cite{Urbanek:2021oej,ARahman:2022tkr,Farrell:2023fgd}. 
For that, we ran two types of circuits: the ``physics'' circuit that implements the process of interest, and two ``mitigation'' circuits.
The mitigation circuits are set up such that they share the same structure as the physics circuits, but are easy to simulate classically in that they only contain Clifford circuits. In the state preparation, the non-Clifford angles in the rotation gates\footnote{The gates $R_{\pm}^{XY}(\theta)$ and $R_{\pm}^{XX}(\theta)$ with $\theta = \pm \frac{\pi}{2}$ in Fig.~\ref{fig:rbox} do not get modified for the mitigation circuits.} are set to $\theta_{\rm mit}=0, \pi/2$.
In the time evolution, the mitigation circuits evolve the system forward then backward in time such that there is no change in an ideal simulation.\footnote{For one step of second-order Trotterized time evolution, the mitigation circuit becomes
\begin{equation}
    e^{i\frac{t}{2}\hat{H}_{k1}}e^{i\frac{t}{2}\hat{H}_{k0}}e^{-i0\hat{H}_{m}}e^{-i0\hat{H}_{g}}e^{-i\frac{t}{2}\hat{H}_{k0}} e^{-i\frac{t}{2}\hat{H}_{k1}} \ .
\end{equation}}
With these quantities, the rescaling factor to compensate for the decoherence can be computed:
\begin{equation}
    \eta_{\hat O} = \frac{\langle \hat O \rangle^{\rm mit}_{\rm meas}}{\langle \hat O \rangle^{\rm mit}_{\rm pred}} 
    \ \ \ \ \rightarrow \ \ \ \ 
    \langle \hat O \rangle^{\rm physics}_{\rm pred} = \frac{\langle \hat O \rangle^{\rm physics}_{\rm meas}}{\eta_{\hat O}} \ ,
\end{equation}
where the ``meas'' observable is the quantity extracted from the noisy device, and ``pred'' is the noiseless prediction.

\begin{figure}[ht!]
    \centering
\includegraphics[width=.45\linewidth]{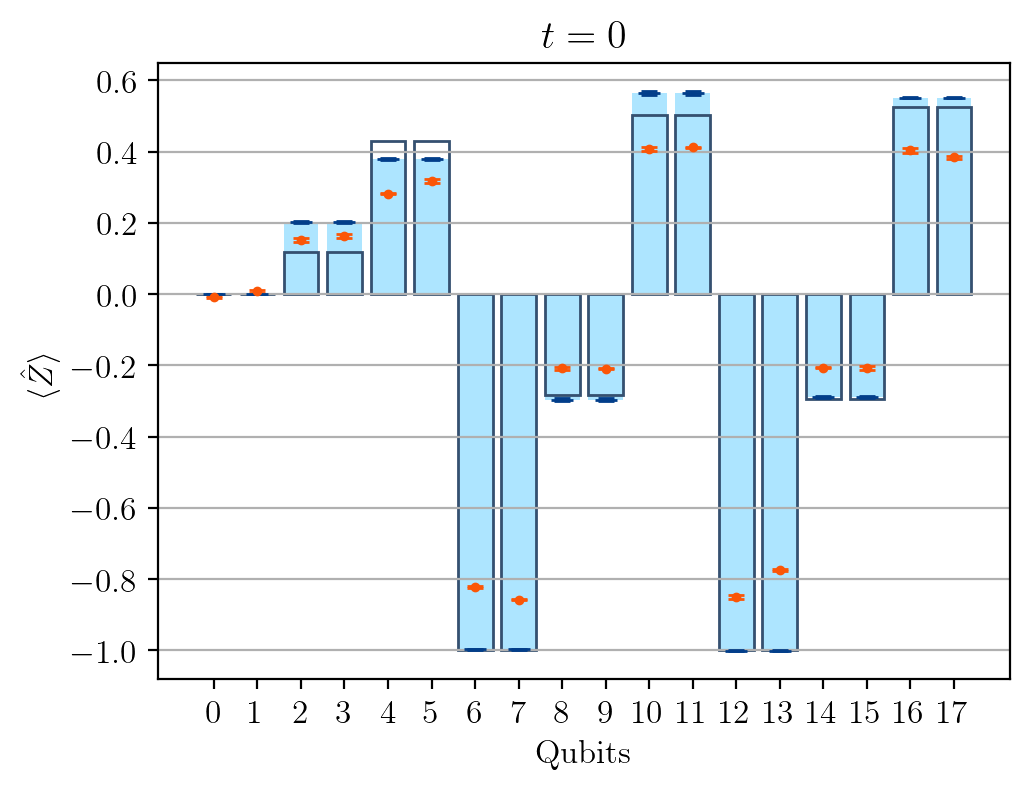}
\includegraphics[width=.45\linewidth]{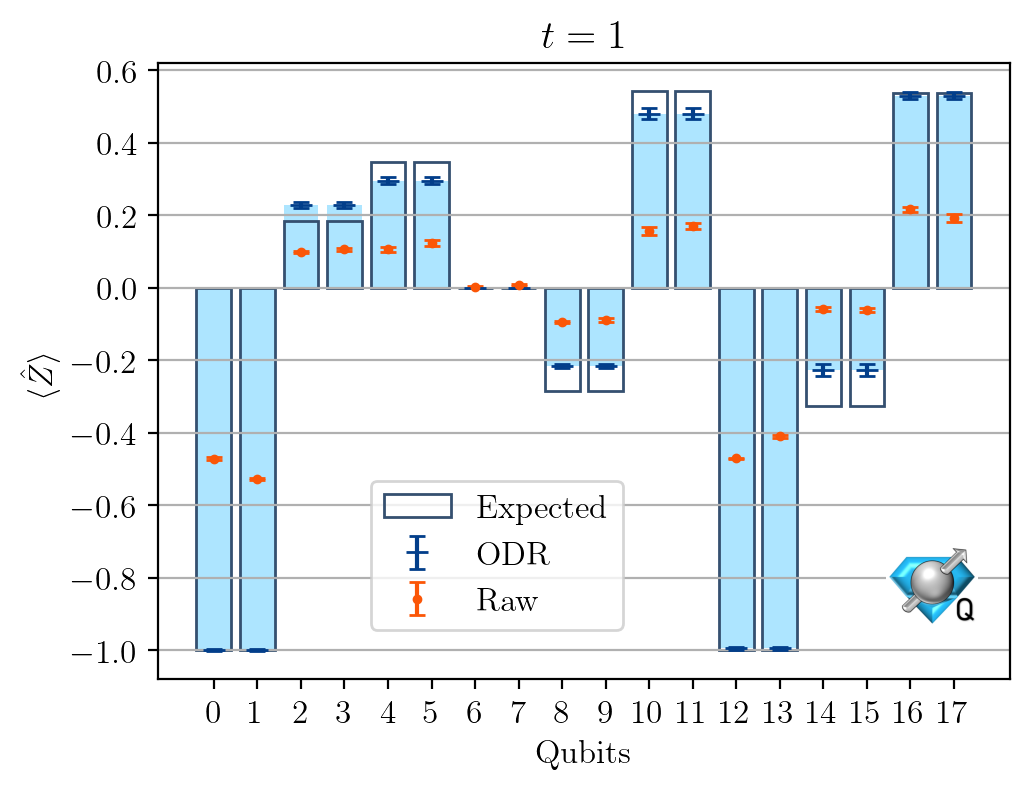}
\caption{
Results obtained from {\tt ibm\_pittsburgh} for $\langle \hat Z_j \rangle$ in the $L=3$ system.  
At $t=0$, the heavy quark is instantaneously moved from $x=0$ to $x=1$, then evolved forward in time to $t=1.0$ 
using one step of the second-order Trotterized time-evolution operator.
The SU(2) symmetry has been used to combine results for each red and green quark.
The left panel shows the expectation values of $\langle \hat Z_j \rangle$ before the heavy quark is moved; the right panel  shows the values
at $t=1$. The results are obtained using bootstrap resampling with the median and MAD.
The points with error bars correspond to the raw (orange circle) and mitigated (blue line) results obtained from {\tt ibm\_pittsburgh}, 
while the gray bar outlines are the expected results obtained using classical computing.
}
    \label{fig:IBMresultfig}
\end{figure}

The quantum circuits were transpiled to run on IBM hardware.
For each twirl of each type of circuit, 4096 measurements (shots) of $\langle \hat Z_j \rangle$ were performed in the computational basis. 
For each qubit, we obtain three results (one from the physics circuit, and two from the mitigation circuits), each for a total of 64 twirls.
For the mitigation circuits, $\langle \hat Z_j\rangle^{\rm mit}_{\rm pred}$ is expected to take three values, $\{-1,0,1\}$. 
The reason we chose to run two different mitigation circuits, is that with each of the mitigation circuits, some of the observables can take the value zero, incompatible with ODR.
With the two choices of $\theta_{\rm mit}$ selected, we have at least one non-zero value for $\langle \hat Z_j\rangle^{\rm mit}_{\rm pred}$.
Measurements with $|\langle \hat Z_j\rangle^{\rm mit}_{\rm meas}|<\lambda^{\rm mit}_{\rm cut}$ and $|\langle \hat Z_j\rangle^{\rm phys}_{\rm meas}|<\lambda^{\rm phys}_{\rm cut}$, are filtered out~\cite{Farrell:2024fit,Farrell:2025nkx}.
Additionally, the inherent SU(2) symmetry of the system, which requires that pairs of qubits within a staggered site to have the same value of $\langle \hat Z_j \rangle$, was used to combined results for the $r,g$ qubits.
The median and (re-scaled) median-average-deviation (MAD) were used as robust estimators of the predicted central value of $\langle \hat Z_j \rangle$ for each qubit, along with its $68\%$ confidence interval.
The value of $\lambda^{\rm mit}_{\rm cut}$ and $\lambda^{\rm phys}_{\rm cut}$ was set at $0.005$, data below which is considered unrecoverable for a circuit with $\sim 400$ two-qubit-gate depth. 
The results of this analysis are compiled in Table~\ref{tab:IBManalysis}, and displayed in Fig.~\ref{fig:IBMresultfig}.

The error-mitigated results obtained from {\tt ibm\_pittsburgh}
are in good agreement with expectations,
despite a transpiled two-qubit-gate circuit depth of $\sim 400$.  
The left panel of Fig.~\ref{fig:IBMresultfig} shows that there remain modest systematic errors that we have not identified, and that ODR does not remove all of the errors introduced by {\tt ibm\_pittsburgh} during circuit execution.
Some of these errors can be attributed to the time of execution of the circuits, where delay between device calibration and execution times can be 
significant.
The right panel displays the bootstrap distributions from 64 twirled samples for all the staggered sites from the raw data. While for some of the results their $\{r,g\}$-averaged values come from single peaked distributions, 
some come from non-trivial distributions, and in the worst case a double peaked distribution, making the use of robust statistics important.

%%%%%%%%%%%%%%%%%%%%%%%%%%%%
\subsubsection{Energy Loss}
\label{sec:QSresultsdEdx}
\noindent
Determining
the energy loss during the motion of the heavy quark can be accomplished by measuring the difference between the total energy immediately before and 
after 
translating the heavy quark between adjacent spatial sites. 
Given that the time evolution circuit is predominantly energy-preserving, 
and any small systematic errors are negligible,
the post-translation time-evolution step is unnecessary.
As considered in Sec.~\ref{sec:L3Qmotion}, it is only the energy in the gauge field that changes discontinuously during this process.  Therefore, only this component of the Hamiltonian  needs to be evaluated.
At the level of circuits, there are well-established ways to determine the energy in a state.
We choose to use a set of low-depth entangling operations to transform modest sets of commuting operators contributing to the Hamiltonian into the computational basis (with simultaneous transformations).
These sets of contributions are then individually  evaluated probabilistically from measurements. 
As only the chromo-electric energy is required to be evaluated, 
before and after translation of the heavy quark, 
the same GHZ transformation used in the time evolution of the gauge term can be applied to transform $\sigma^\pm\sigma^\mp\sigma^\mp\sigma^\pm$ into the computational basis for simpler measurements.
This transformation is implemented 
in quantum simulations 
by applying $\hat{\cal G}^\dagger$ defined in Fig.~\ref{fig:hgauge_seg} to the final state wavefunction on the quantum computer,
followed by measurements in the computational basis.
This circuit is applicable for any pair of quarks and anti-quarks or quarks.\footnote{In addition to this method, we have also explored the Hadamard test to assess its viability as an alternate technique, see App.~\ref{app:Energycircs}. 
We find it to be inefficient.}

To determine the difference in energy of the system when the heavy quark is moved from $x=0$ to $x=1$,
the only terms that need to be evaluated before and after are
\begin{align}
\Delta \hat H_{g} = &  
\ \frac{g^2}{2} \sum_a \ 
\left[\ 
2 ( \hat{\mathcal{Q}}^{(a)}_{0} )^2
\ +\ 
4 \hat{\mathcal{Q}}^{(a)}_{0}\hat{\mathcal{Q}}^{(a)}_{1}
\ +\ 
2 \hat{\mathcal{Q}}^{(a)}_{0} \hat{\mathcal{Q}}^{(a)}_{2}
\ \right]
\ ,
\nonumber\\
= & 
\ \frac{g^2}{2} \left[ 
\frac{3}{4}(1-\hat Z_0\hat Z_1)
+ \frac{1}{4} (\hat Z_0-\hat Z_1)(\hat Z_2-\hat Z_3) + \frac{1}{8} (\hat Z_0-\hat Z_1)(\hat Z_4-\hat Z_5)
\ \right. 
\nonumber\\
 &
 \quad \quad \left.  
  + \  2 \left(\hat\sigma^+_0\hat\sigma^-_1\hat\sigma^-_2\hat\sigma^+_3 + {\rm h.c.} 
    \right)
 + \left(\hat\sigma^+_0\hat\sigma^-_1\hat\sigma^-_4\hat\sigma^+_5 + {\rm h.c.} 
    \right)    
\right]\ ,
\nonumber\\
= & \  \sum_{i=1}^3\ 
\hat {\cal O}_i^\dagger \ ( C_i \cdot \hat P_i ) \ \hat {\cal O}_i
\ ,
\label{eq:HdiffE}
\end{align}
The expectation values of the operators in Eq.~\eqref{eq:HdiffE} vanish after the translation of the heavy quark to $x=1$ because $\hat{\mathcal{Q}}^{(a)}_{0}|\psi\rangle=0$.
Equation~\eqref{eq:estimator} gives the operators evaluated on the final state, the list of Pauli operators,
 $\hat P_i$, and the list of their coefficients, $C_i$.
\begin{align}
\hat {\cal O}_1\ =\ \hat I 
\qquad & :
\hat P_1 = \left[
\hat I,\ 
\hat Z_0\hat Z_1,\  \hat Z_0\hat Z_2,\   \hat Z_0\hat Z_3,\   \hat Z_1\hat Z_2 , \  
\hat Z_1\hat Z_3,\    \hat Z_0\hat Z_4,\    \hat Z_0\hat Z_5,\    \hat Z_1\hat Z_4,\   
\hat Z_1\hat Z_5
\right] 
\nonumber\\
& \qquad
C_1 = 
\frac{g^2}{2} \frac{1}{8}\left[ 
+6,-6, +2, -2, -2, +2, +1, -1,  -1, +1
\right] 
\nonumber\\
\hat {\cal O}_2\ =\ \hat{\cal G}_{01;23}^\dagger & :  
\hat P_2 = \left[
        -\hat Z_0\hat Z_1\hat Z_3, 
        \ \hat Z_1\hat Z_2 \hat Z_3,   
        \ -\hat Z_0\hat Z_1\hat Z_2,   
        \ \hat Z_1, \right. \\
        &\qquad \left. \ -\hat Z_0\hat Z_1\hat Z_2\hat Z_3,   
        \ \hat Z_1\hat Z_2,  
        \ \hat Z_1\hat Z_3,  
         \ -\hat Z_0 \hat Z_1  
\right]
\nonumber\\
& \qquad
C_2 = \frac{g^2}{2} \frac{1}{4}
\left[+1, +1, +1, +1, +1, +1, +1, +1\right]
\nonumber\\
 \hat {\cal O}_3\ =\ \hat{\cal G}_{01;45}^\dagger & : 
\hat P_3 = \left[
        -\hat Z_0\hat Z_1\hat Z_5, 
        \ \hat Z_1\hat Z_4 \hat Z_5,   
        \ -\hat Z_0\hat Z_1\hat Z_4,   
        \ \hat Z_1, \right. \\
        & \qquad \left.
        \ -\hat Z_0\hat Z_1\hat Z_4\hat Z_5,   
        \ \hat Z_1\hat Z_4,  
        \ \hat Z_1\hat Z_5,  
         \ -\hat Z_0 \hat Z_1  
\right]
\nonumber\\
& \qquad
C_3 = 
\frac{g^2}{2} \frac{1}{8}\left[+1, +1, +1, +1, +1, +1, +1, +1
\right]
\ .
\label{eq:estimator}
\end{align}
The implementation in {\tt qiskit} involves three different circuit evaluations.
Taking the difference between the summed estimators
before and after the heavy quark translation
provides the energy loss from the translation.
The operator structure in Eqs.~\eqref{eq:HdiffE} and \eqref{eq:estimator} will differ for computing the energy loss from heavy-quark translations between different adjacent spatial sites on lattices of arbitrary length.

The energy transfer to the system due to the heavy-quark translating from $x=0$ to $x=1$ was determined by running three different physics circuits, along with 
corresponding
mitigation circuits,\footnote{The mitigation circuits for the operators in $\hat{P}_1$ and $\hat{P}_2$ have the non-Clifford rotation angles set to 0, while no mitigation circuits were run for $\hat{P}_3$. This is because, for generating the $\hat{P}_3$ mitigation circuits, setting non-Clifford rotation angles to 0 or $\pi/2$ both result in 0 for noiseless predictions of observables, incompatible with ODR.} on the quantum computer.  
As the last operator in the state preparation circuit does not change the measurements,
a shallower circuit, with a CZ depth of 109, is used to measure the diagonal Pauli strings defined in Eq.~\eqref{eq:estimator}.

In addition to the error mitigation methods applied in the charge distribution measurements, zero-noise extrapolation (ZNE)~\cite{Li:2016vmf,Temme:2016vkz,PhysRevLett.120.210501,Klco:2018kyo,Kandala2019Error} is employed through 
{\tt qiskit} Runtime options~\cite{qiskit}. 
Noise factors-$(1,2,3)$ are executed on the device, where the different noise factors indicate the increase in circuit depth due to additional CZ gates (e.g., factor 1 is the original circuit, and factor 3 has three times the CZ count of the original circuit).
In post-processing, and before performing ZNE of the raw data, 
we perform ODR for each noise level using the 32 sets of Pauli-twirled physical and mitigation circuits.
These distributions are then linearly extrapolated in the noise-factor to the point of zero noise, providing an error-mitigated estimate of the quantity of interest.
Computing the three contributions to the energy difference given in Eq.~\eqref{eq:estimator}, both with and without ZNE for the sake of comparison, gives,
\begin{align}
\langle \Delta \hat H_{g}  \rangle^{\rm IBM}_{t<0} & =
0.3715(16) - 0.7243(30) - 0.0314(04) \ = \ -0.3841(30)
\ \ ({\rm ODR})
\, 
\\
\langle \Delta \hat H_{g}  \rangle^{\rm IBM}_{t<0} & = 
0.3491(19) - 0.7538(73) - 0.0390(07) \ = \ -0.4437(72)
\ \ ({\rm ODR+ZNE})
\ ,
\label{eq:estIBMvalues}
\end{align}
which are to be compared with the exact values, 
$\langle \Delta \hat H_{g}  \rangle^{\rm exact}_{t<0} = 0.3314-0.7951-0.0421=-0.5058$, 
obtained from a noiseless simulation of the quantum circuits.
Combined with $\langle \Delta \hat H_{g}  \rangle^{\rm IBM}_{0<t<1}=0$, this result implies an energy loss of 
$\delta \langle \Delta \hat H_{g}  \rangle^{\rm IBM}=+0.4437(72)$
due to the translation of the heavy quark from $x=0$ to $x=1$.
Further details about this determination can be found in App.~\ref{app:IBMruns}.
It is important to note the importance of combining ODR and ZNE in obtaining a result from {\tt ibm\_pittsburgh}
that is close to the expected value.
This is, in part, due to the modest cancellations among contributions to the sum, that magnify the impact of the individual errors.

%%%%%%%%%%%%%%%%%%%%%%%%%%%%
\section{Summary and Conclusions}
\label{sec:Summary}
\noindent
In this chapter, we have established a set of techniques that enable the real-time simulation of energy-loss and hadronization in non-Abelian lattice gauge theories in 1+1D.
As the action of the non-Abelian charge operators generally transform any given state, the methods used for such simulations in Abelian theories, including charges as discontinuities in Gauss's law, are inapplicable.  
To circumvent this problem, a non-dynamical  
heavy-quark field is included as an explicit degree of freedom.
With one light flavor of quark and anti-quark, and one flavor of heavy quark,
the 1+1D system with two colors can be mapped to a qubit register with six qubits per spatial site.
Motion of the heavy quark across the lattice is implemented by fermionic SWAP operators, with the condition that only complete translations between spatial sites are permitted, as anything different is incompatible 
with the 
time evolution of the light quarks.
This method introduces a further lattice-spacing artifact due to swapping in the presence of other heavy quarks in the lattice (that mimics matter), which can be mitigated at a subsequent time step.
Domain decomposition is introduced as a scalable way to 
prepare the (initial) ground state of a single heavy-quark positioned at one end of the lattice (and generalizes to all initial locations).
This permits the use to classical computing to provide an approximation (large) sections of the ground state, which can be ``stitched together'' on a quantum computer with localized boundary operators (due to confinement) found by variational methods, which we demonstrate for the (small) $L=3$ system.
The operator structures required for each step correspond to hadronic operators, both mesonic and baryonic, defined by position and form factors.  
The challenges that are naively faced, associated with color antisymmetry and entanglement, are naturally accommodated and rendered irrelevant by working with such hadronic operator structures.
For instance, a heavy quark can be created in a total color singlet lattice by acting with a 
heavy-baryon operator or heavy-meson operator
on the interacting vacuum state.   
This new understanding also allows for generalization to systems with larger number of colors,
or, as an example, with background charges in the adjoint representation.

Through extensive classical simulations of small systems, we have developed the necessary quantum circuits that can be scaled to arbitrarily large systems (made possible by truncations of the charge-charge operators enabled by confinement).
Using IBM's {\tt ibm\_pittsburgh},
these circuits (with a transpiled circuit depth $\sim 400$) have been used to perform simulations of single heavy-quark displaced by once spatial site and time evolved forward using one second-order Trotter step.
With a full complement of error-mitigation strategies, including Pauli-twirling, dynamic-decoupling, measurement twirling, ODR, and ZNE, results were obtained that are in good agreement with classical expectations.  

With all the elements in place to perform measurements of energy loss in non-Abelian systems in 1+1D, there are a number of near-term directions to pursue. 
One direction is to perform simulations on larger lattices, with the heavy quark initially located far from the boundaries.
This will permit a more robust implementation of domain decomposition for initial state preparation.
Further, it will allow for estimates of energy loss with reduced systematic uncertainties, and changing the Hamiltonian parameters to decrease the lattice spacing. 
Hadrons created by the motion of the heavy quark will have more time to propagate before encountering the boundaries and reflecting back to the heavy quark, providing an extended time interval to study the fundamental mechanisms without interference.
A second direction is to simulate energy loss and hadronization in the analogous SU(3) lattice gauge theory. 
Heavy quarks can now be included in existing frameworks of 1+1D SU(3) simulations~\cite{Ciavarella:2021nmj,Farrell:2022wyt,Farrell:2022vyh,Atas:2022dqm,Chernyshev:2025lil}. 
Such simulations with one flavor of light quarks require  9 qubits per spatial site ($rgb$ for the heavy and light quarks, and $\bar r \bar g \bar b$ for the light anti-quarks), and two-flavor simulations require 15 qubits per spatial site.
A third direction, which is less trivial and may require the development of additional techniques, is the extension to simulations in 2+1D and 3+1D.  
Significant progress is  being made in evolving quantum simulations of lattice field theories into 2+1D, with explorations in 3+1D in limited cases~\cite{Raychowdhury:2018osk,
Ciavarella:2021nmj,
Zache:2023dko,
Muller:2023nnk,
Kavaki:2024ijd,
Turro:2024pxu,
Gustafson:2023aai,
Kadam:2024ifg,
Fontana:2024rux,
Grabowska:2024emw,
Ciavarella:2024cyt,
Halimeh:2024bth,
Ciavarella:2024fzw,
Balaji:2025afl,
Illa:2025dou,
Illa:2025njz,
Jiang:2025ufg,
Yao:2025uxz}.
Typically, these involve simulating small volumes of the Kogut-Susskind Hamiltonian in the electric basis, with a significantly truncated gauge space, and without quantification of systematic errors.
However, full simulations with both quarks and gluons remain to be performed.  
Such simulations are a prerequisite for qualitative comparisons with experiment and observation, and for making robust predictions in regimes where experiment is not possible.

%%%%%%%%%%%%%%%%%%%%%%%%%%%%

% ========== Chapter 3

\chapter{\texorpdfstring{Digital Mapping and Simulation of\\the Fully Gauge-Fixed SU(2) Hamiltonian}{Digital Mapping and Simulation of the Fully Gauge-Fixed SU(2) Hamiltonian}}
\label{chap:gauge_fix}

\vspace{1em}
\begin{center}
    \itshape
    This chapter is associated with Ref.~\cite{Froland:2025bqf}: 
    ``Simulating Fully Gauge-Fixed SU(2) Hamiltonian Dynamics on Digital Quantum Computers''\\by Henry Froland, Dorota M. Grabowska, and Zhiyao Li
\end{center}
\vspace{2em}

%%%%%%%%%%%%%%%%%%%%%%%%%%%%
% Chapter: Digital Mapping and Simulation of the Fully Gauge-Fixed SU(2) Hamiltonian
%%%%%%%%%%%%%%%%%%%%%%%%%%%%

\section{Introduction}

A key step toward realizing large-scale lattice gauge theory simulations on quantum hardware is the development of efficient, physically-faithful representations of the formally infinite Hilbert space. The fundamental task is encoding the continuous gauge group into discrete degrees of freedom that can be implemented on a quantum device. There exist a variety of approaches that utilize group-theoretic constructions to address this task, including group irreducible representation (irrep) bases~\cite{Mathur:2004kr, mathur2006loop, mathur2007loop, Raychowdhury:2018osk, Raychowdhury:2019iki, Kadam:2022ipf, Kadam:2024ifg, Ciavarella:2021nmj, Klco:2019evd, Davoudi:2020yln, Byrnes:2005qx, Anishetty:2009nh, Banuls:2017ena,Davoudi:2022xmb}, group-element bases~\cite{PhysRevD.71.054503,Mathur:2015wba, DAndrea:2023qnr, Romiti:2023hbd,Grabowska:2024emw}, discrete subgroups~\cite{Alam:2021uuq, Alexandru:2021jpm, Gustafson:2022xdt, Gustafson:2023kvd, Gustafson:2024kym, Assi:2024pdn, Lamm:2024jnl, Muarari:2024dqx}, deformed gauge groups~\cite{Zache:2023dko,Hayata:2023bgh}, and dual bases~\cite{Kaplan:2018vnj, Bender:2020ztu, Haase:2020kaj,Miranda-Riaza:2025fus,Halimeh:2024bth,Bergner:2024qjl}. Each of these choices, however, presents its own particular limitations, ranging from inefficiency at weak coupling, complications with systematic improvement, and difficulties in making physically-accurate predictions. Therefore, it is of vital importance to compare these various formulations, with a particular focus on the resources necessary for extracting experimentally-relevant measurements. 

A recently introduced basis, the \textit{mixed basis}~\cite{DAndrea:2023qnr, Grabowska:2024emw}, serves as a middle ground between the irrep and group element representations of the gauge group. With the removal of gauge-redundant degrees of freedom, as well as the use of a coupling-dependent digitization scheme, this basis is believed to provide a resource-efficient formulation of \sutwo for all values of the gauge coupling. This is of particular importance when extracting physical parameters from numerical simulations, due to the necessary tunings of bare couplings to reach the appropriate large volume continuum limits. 

This chapter develops algorithms for simulating the real-time dynamics of the minimal non-trivial example of a fully gauge-fixed \sutwo Hamiltonian: two plaquettes with open boundary conditions. Sec.~\ref{sec:theory_section} provides a short review of gauge theory Hamiltonians and gauge fixing. It also details the group element, mixed and character irrep bases, each of which is optimized for distinct coupling regimes.  Sec.~\ref{sec:digitization} analyzes the mixed basis in detail, presenting the corresponding digitization scheme and mapping of the Hamiltonian into quantum circuits for time evolution. It also provides resource estimates for larger-scale simulations. Sec.~\ref{sec:MixedBasisCircuits} gives two different methods for translating the digitization into quantum circuits that can be implemented onto Noisy Intermediate-Scale Quantum (NISQ)-era devices. Finally, Sec.~\ref{sec:quantum_results} presents the results of a real-time simulation on {\tt ibm\_fez}, a 156 superconducting qubit chip. An in-depth explanation of the various error-mitigation strategies that are utilized in order to extract results from the noisy quantum simulation is also presented. Finally, Sec.~\ref{sec:Conclusions} discusses the implications of this chapter and proposes directions for future work and extensions to larger system sizes.
\begin{figure}[htbp]
    \centering
    \includegraphics[scale=.5]{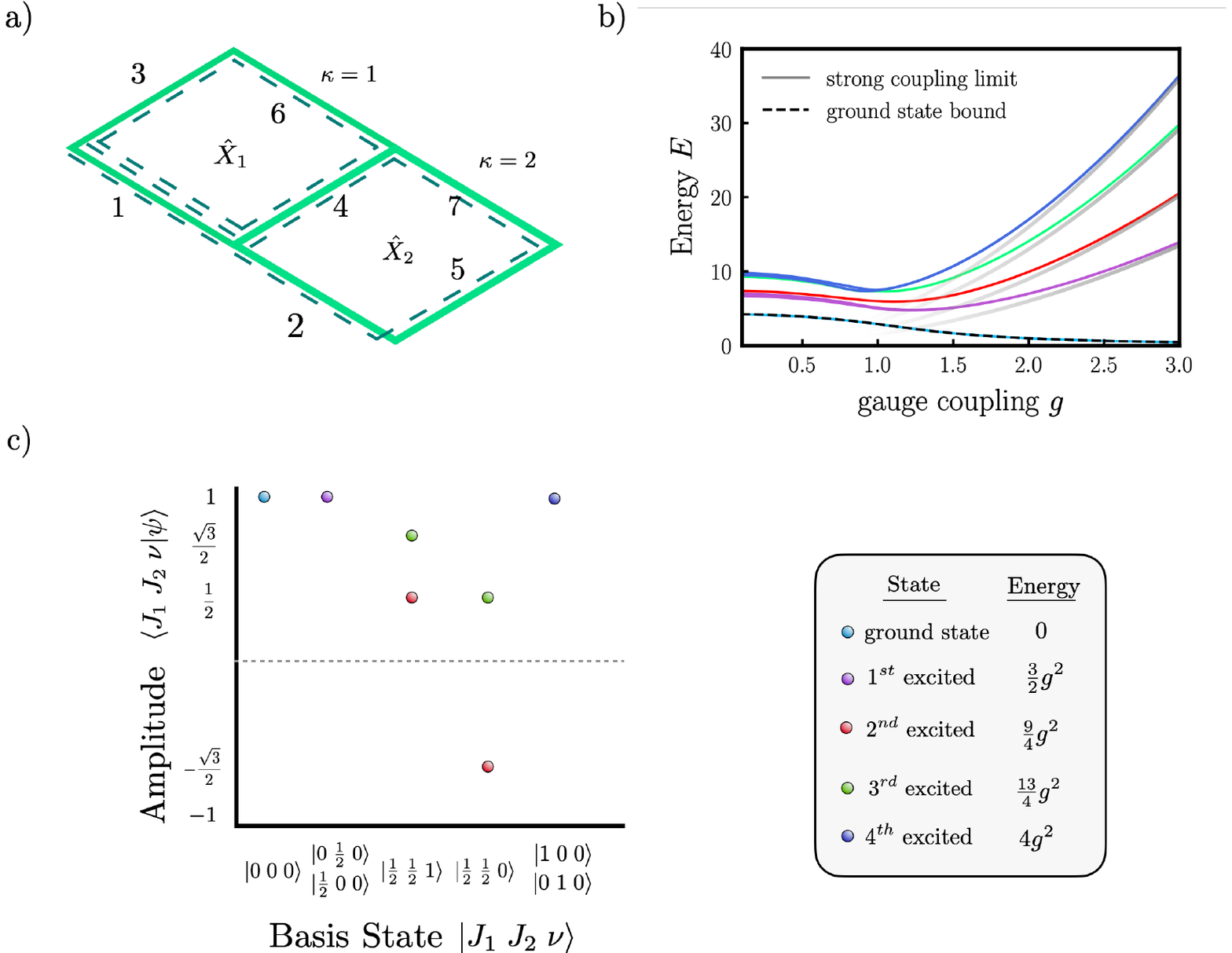}
    \caption{\textit{Spectrum of the Two Plaquette System} \textbf{a)} Two plaquette system with open boundary conditions. The system contains seven links, numbered $1$ through $7$, with the physical links denoted as $\kappa=1,2$. The magnetic degrees of freedom $\hat{X}_{1},\hat{X}_2$ are associated with each plaquette. The dashed lines that run around each plaquette denote the loops used in the max tree gauge fixing procedure. \textbf{b)} Spectrum of the Hamiltonian for different bare couplings. The solid lines show the solution obtained by FEM. The black dashed line on the bottom shows the energy bound of Eq.~\eqref{eq:bound}. The grey lines denote the strong coupling limit of the character irrep spectrum. This limit is found by ignoring the contribution of the magnetic Hamiltonian, which is only valid when the gauge coupling is large. \textbf{c)} Low-lying spectrum of the character irrep Hamiltonian in the strong coupling limit. The first and fourth excited state are two-fold degenerate and each state is a linear superposition of the stacked basis states. See Table~\ref{tab:CharIrrepStrong} for further details.}
    \label{fig:two_plaq_description}
\end{figure}
\section{Two Plaquette Hamiltonian}\label{sec:theory_section}
This section provides a concise overview of how to formulate lattice gauge theories using maximal-tree gauge fixing~\cite{PhysRevD.15.1128}, as well as how to remove the remaining global gauge transformation. It also introduces several different bases that can be used to derive analytic results for a wide range of (bare) coupling values. For a pedagogical review about formulating gauge theories on the lattice and understanding gauge symmetries in the max-tree construction, see Ref.~\cite{DAndrea:2023qnr}. Operators on the Hilbert space will generally be denoted with a hat, however when the meaning is unambiguous, the hat will be omitted for readability.

\subsection{Gauge Fixing Lattice Gauge Theories}
The Kogut-Susskind Hamiltonian~\cite{PhysRevD.11.395} is the starting point of this work. Written in terms of electric field link operators and magnetic plaquette operators, the general form of a lattice gauge Hamiltonian is
\begin{align}
\hat H = \frac{g^2}{2a}\sum_{\ell \in \text{links}}\hat E^2_\ell+ \frac{1}{2g^2a}\sum_{p \in \text{plaq.}}\Tr\left[2 - \hat P^{\phantom{\dagger}}_p-\hat P^\dagger_p\right]
\end{align}
where each plaquette is constructed out of gauge link operators, $g$ is the gauge coupling, and $a$ is the lattice spacing, usually set to the identity. 
Before introducing the commutation relations of the operators that appear in this Hamiltonian, it is beneficial to understand how these operators are related to the mathematics of Lie groups. 

Typically, a gauge link operator is defined as the path-ordered Wilson line of the gauge field,
\begin{align}
U_{rs}(\ell) \equiv \left[\CP \exp \left(i \int_{\boldsymbol{n}}^{\boldsymbol{n}+\boldsymbol{e}_i}dx^\mu A_\mu(x)\right)\right]_{rs}
\end{align}
where the gauge link $\ell(\boldsymbol{n}, \boldsymbol{e}_i)$ originates at point $\boldsymbol{n}$ and extends in the direction $\boldsymbol{e}_i$ and $r,s$ are color indices. The Hilbert space of this operator is the space of all square-integrable wavefunctions $L^2(G, \mathfrak{g})$ over the Lie group $G$ with respect to the Haar measure $d\mathfrak{g}$. Therefore, any gauge link operator can be written in the group element basis as
\begin{align}
\hat U_{rs}(\ell) \equiv \int d\mathfrak{g}_\ell \, u_{rs}(\mathfrak{g}_\ell)\ket{\mathfrak{g_\ell}}\bra{\mathfrak{g_\ell}}
\end{align}
where $u_{rs}$ are the matrices that define the group elements in the fundamental representation. The full Hilbert space is simply the tensor product of the local Hilbert spaces of each link.

The electric operators are also defined using the group element Hilbert space. In particular, the electric operators are responsible for translations in the space of group elements,
\begin{align}
\hat \Theta_{L\mathfrak{h}}= e^{i \phi^a(\mathfrak{h})\hat E^a_L}\qquad \hat \Theta_{R\mathfrak{h}}= e^{i \phi^a(\mathfrak{h})\hat E^a_R} 
\end{align}
where $\phi^a(\mathfrak{h})$ are parameters typically called normal coordinates, where $a\in\{1,\dots,\text{dim}(G)\}$. The operators $\hat\Theta$ act on the group element basis states as
\begin{align}
\hat \Theta_{L\mathfrak{h}}\ket{\mathfrak{g}}= \ket{\mathfrak{h}^{-1}\mathfrak{g}} \qquad \hat \Theta_{R\mathfrak{h}}\ket{\mathfrak{g}}= \ket{\mathfrak{g}\mathfrak{h}^{-1}} \, ;
\end{align}
since the left and right electric field operators act on states differently, both need to be present for non-Abelian gauge theories.

The electric field operators satisfy the commutation relations
\begin{gather}
\left[\hat E_L^a, \hat E_L^b\right]= - i f^{abc}\hat E_L^c \qquad \left[\hat E_R^a, \hat E_R^b\right]=  i f^{abc}\hat E_R^c \nonumber \\
\left[\hat E_L^a, \hat E_R^b\right] = 0 \, 
\end{gather}
and given that the left and right electric operators commute, they each furnish an independent Lie algebra. Additionally, each electric operator has a commutation relation with the gauge link operator
\begin{align}
\left[\hat E^a_{L}, \hat U^{(r)}_{mn}\right] = T^{(r)a}_{mm'}\hat U^{(r)}_{m'n} \quad \left[\hat E^a_{R}, \hat U^{(r)}_{mn}\right] = \hat U^{(r)}_{mn'}T^{(r)a}_{n'n}
\end{align}
where $r$ indicates the representation and is \textit{not} summed over. 

So far, there has been no mention of gauge invariance in this discussion, though all the ingredients are now in place. For a theory that is gauge invariant, there exist states that are equivalent to each other; these states are related via gauge transformations. In deriving the Kogut-Susskind Hamiltonian from the Wilson action, Weyl gauge is imposed. This is an incomplete gauge-fixing condition and purely spatial gauge transformations still remain. A gauge transformation $\Omega$ at the lattice site $\boldsymbol n$ is performed by applying the operator
\begin{align}
\hat \Theta_{\Omega}(\boldsymbol{n}) \equiv \exp\left(i \phi^a(\Omega)\hat G^a(\boldsymbol{n})\right)
\end{align}
where $\hat G^a$ is the Hermitian operator
\begin{align}
\hat G^a(\boldsymbol{n}) = \sum_{i=1}^d \left[\hat E^a_R(\boldsymbol{n}- \boldsymbol{e}_i, \boldsymbol{e}_i)-\hat E^a_L(\boldsymbol{n}, \boldsymbol{e}_i)\right]
\end{align}
and is commonly called the Gauss Law operator. When this operator acts on any state in the physical subspace,
\begin{align}
\hat G(\boldsymbol{n}) \ket{\Psi_\text{phys}} = 0 \quad \forall\, \boldsymbol{n} \, 
\end{align}
is satisfied. It is important that these equivalent states are not treated as distinct. One possible approach for avoiding this is gauge-fixing, a method for choosing a single representative state from each set of equivalent states.

Gauge fixing on a spatial lattice can be done systematically, without the need to solve the Gauss law equation at each site. Instead, the residual spatial gauge transformations that are still present after Weyl gauge-fixing are used to set certain gauge links to the identity. Every time a gauge transformation is used to fix a specific gauge link, neighboring links are also transformed. In two or more spatial dimensions, it is not possible to set all the gauge links to the identity, which is consistent with gauge bosons being dynamical in these dimensions. Ref.~\cite{PhysRevD.15.1128} showed that the largest set of gauge links that can be set to the identity are those that make up a maximal tree.\footnote{A maximal tree is a set of all gauge links such that if any additional link was added to the set, the lattice would now contain a closed loop. See Ref.~\cite{DAndrea:2023qnr} for more details} This convention removes all of the residual spatial gauge transformations except for those at a single lattice site, which will be taken to be the origin. This last spatial gauge transformation, which is a global gauge transformation, must be treated differently. Once all the spatial gauge redundancy is fully removed, the resultant Hamiltonian only spans the physical subspace and there are no more equivalent states.

This process has been completed for \sutwo lattice gauge Hamiltonians in two- and three-dimensions in Ref.~\cite{DAndrea:2023qnr,Grabowska:2024emw}. In the first work, the Kogut-Susskind Hamiltonian is gauge-fixed using the maximal tree gauge fixing procedure and rewritten in terms of canonically-conjugate operators, $\hat{\cal E}^a_{L,R}(\kappa)$ and $\hat X(\kappa)$. The operators $\hat{\cal E}^a_{L,R}(\kappa)$ are related to parallel-transported electric link operators and $\hat X(\kappa)$ are related to parallel-transported gauge links; both sets of operators are transported along a path $P(\kappa)$. In terms of these variables, the Hamiltonian is given by
\begin{align}
\label{eq:H_maxTree}
    \hat H_E &= \frac{g^2}{2 a} \sum_{\ell} \left( \sum_{\kappa \in t_+(\ell)} \hat{\mathcal{E}}^a_{L}(\kappa) - \sum_{\kappa \in t_{-}(\ell)} \hat{\mathcal{E}}^{a}_{R}(\kappa) \right)^2\nn\\
    \hat H_B &= \frac{1}{2g^2 a}\sum_p \Tr \left( \hat I - \prod_{\kappa\in p} \hat X(\kappa)^{\sigma(\kappa)} \right) + \mathrm{h.c.}
\end{align}
where $\sum_p$ and $\sum_\ell $ denotes sums over all plaquettes and all links of the lattice, $t_\pm(\ell)$ is the set of all physical links such that $\ell$ is contained in the path $P$ as a positive/negative link and $\sigma = \pm 1$, depending on whether the link is traversed in the positive or negative orientation. The precise form of the Hamiltonian depends on the chosen max-tree and path convention; physical results will be convention-independent.

In addition, Ref.~\cite{DAndrea:2023qnr} provides a group-element basis for constructing the Hilbert space of the Hamiltonian of Eq.~\eqref{eq:H_maxTree}. Using the axis-angle representation of rotations in three-dimensional space, each loop variable is represented by a two-dimensional matrix
\begin{align}
\label{eq:XOrig}
\hat X &=\left(
\begin{array}{cc}\cos \frac{\omega}{2}-i \cos \vartheta \sin \frac{\omega}{2} & -i e^{-i \varphi} \sin \vartheta \sin \frac{\omega}{2} \\
-i e^{i \varphi} \sin\vartheta \sin \frac{\omega}{2}  & \cos \frac{\omega}{2}+i \cos\vartheta \sin \frac{\omega}{2} \\
\end{array}\right) 
\end{align}
where the variables have ranges
\begin{align}
\omega \in [0, 2\pi],\;\;\theta \in [0, \pi],\;\;\phi \in [0, 2\pi] \,\, .
\end{align}
The corresponding canonically-conjugate variables are given by
\begin{align}
\hat{\boldsymbol{{\cal E}}}_{R/L} = \frac{1}{2}\left(\hat{\boldsymbol{\Sigma}} \pm \hat{\boldsymbol{L}}\right) \, .
\end{align}
In axis-angle coordinates, $\hat{\boldsymbol{L}}$ is given by
\begin{align}
\label{eq:Lop}
\hat L^x &= i \left( \sin\varphi \frac{\partial}{d \vartheta} + \cot\vartheta \cos\varphi \frac{\partial}{{\partial} \varphi}\right)\nonumber\\
\hat L^y &= i \left( -\cos\varphi \frac{\partial}{{\partial} \vartheta} + \cot\vartheta \sin\varphi \frac{\partial}{{\partial} \varphi}\right)\nonumber\\
\hat L^z &= -i \frac{\partial}{{\partial} \varphi} \, ,
\end{align}
which is simply the (conventional) angular momentum operator. The operator $\hat{\boldsymbol{\Sigma}}$ can be written in terms of $\hat{\boldsymbol{L}}$ and the position vector $\hat{\boldsymbol{\eta}}$ as
\begin{align}
\hat{\boldsymbol{\Sigma}} = 2 i \hat{\boldsymbol{\eta}} \pdv{}{\omega}+ \cot \frac{\omega}{2}\left(\hat{\boldsymbol{\eta}} \cross \hat{\boldsymbol{L}} \right) \, ,
\end{align}
where $\hat{\boldsymbol{\eta}}$ is the axis of rotation in three-dimensional space,
\begin{align}
\label{eq:EtaDef}
\hat{\boldsymbol{\eta}} \equiv \left(\sin \vartheta \cos \varphi,\sin \vartheta \sin \varphi, \cos \vartheta\right) \,\, .
\end{align}
The operator $\hat{\boldsymbol{\Sigma}}$ is analogous to the boost operator of the Lorentz group. The commutation relations of these two operators are 
\begin{align}
\left[\hat L^a, \hat L^b\right] &= i \epsilon^{abc}\hat L^c \nonumber \\
\left[\hat \Sigma^a, \hat \Sigma^b\right] &= i \epsilon^{abc}\hat L^c \nonumber \\
\left[\hat \Sigma^a, \hat L^b\right] &= i \epsilon^{abc}\hat \Sigma^c \, ,
\end{align}
whereas the commutation relations for the Lorentz group operators have a relative minus sign for the second and third commutation relation.
The Haar measure for each link in this basis is
\begin{align}
d\mathfrak{g}= 4 \sin^2\frac{\omega}{2}\sin \vartheta \, d \omega\, d\vartheta\, d\varphi
\end{align}

The fixing of the last (global) gauge transformation is carried out in Ref.~\cite{Grabowska:2024emw}. The key insight of that work is that the Hamiltonian can be conceptualized as a system of rods all connected together at the origin. The $\kappa^\text{th}$ rod has length $\omega_\kappa$ and its orientation is given by the polar angle $\vartheta_\kappa$ and azimuthal angle $\varphi_\kappa$. Such a system can be written either in a lab-frame coordinate system or in a body-frame coordinate system. An Euler rotation relates these two coordinate systems. In the lab-frame coordinate system, the location of each rod is parameterized as in Eq.~\eqref{eq:EtaDef}. This same configuration of rods can be written using body-frame variables, coupled with an Euler rotation. The definition of the body frame is arbitrary, but Ref.~\cite{Grabowska:2024emw} chooses to define the $\hat z$ axis of the body frame by aligning it with one rod and the $\hat x -\hat y$ plane via the location of another rod. In this coordinate system, the location of each rod is given by
\begin{align}
\boldsymbol{\eta}_1 &= \boldsymbol{R}\cdot \left(0, 0, 1\right) \\
\boldsymbol{\eta}_2 &= \boldsymbol{R}\cdot \left(\sin\Theta, 0, \cos \Theta\right) \nonumber \\
\boldsymbol{\eta}_\mu &= \boldsymbol{R}\cdot\left(\sin \theta_\mu \cos \phi_\mu,\sin \theta_\mu \sin \phi_\mu, \cos \theta_\mu\right) \qquad \mu \geq 3  \nonumber
\end{align}
where $\boldsymbol{R}$ is the rotation matrix written in terms of Euler angles. Rotation of the full system leaves the lengths of each rod invariant. In carrying out the change of basis to go from the lab-frame coordinate system to the body-frame coordinate system, the remaining global gauge transformation is made visible. The Hamiltonian can then be trivially fixed to the total gauge-singlet sector and only depends on the variables $\{\omega_\kappa, \Theta, \theta_\mu, \phi_\mu\}$; all dependence on the Euler angles is eliminated from the Hamiltonian. Ref.~\cite{Grabowska:2024emw} provides the explicit forms of all possible electric bilinears as well as magnetic loop operators in this new coordinate system. The new basis is called the `sequestered basis', as all the dependence on the total charge of the system is \textit{sequestered} into the Euler angles. The focus of this paper is to simulate the two plaquette version of this system, on a NISQ-era device.

\subsection{Two Plaquette Hamiltonian}
\label{sub_sec:two_plaquette_hamiltonian}
Using Eq.~\eqref{eq:H_maxTree} and a loop convention where each loop only encompasses one plaquette, the Hamiltonian for the two plaquette system with open boundary conditions, shown in Fig.~\SubFigRef{fig:two_plaq_description}{a}, is given by
\begin{align}
\label{eq:HTPOp}
\hathtp &=\frac{g^2}{2}\left(4\, \hat{{\cal E}}_1^2+6\, \hat{{\cal E}}_2^2 - 2\, \hat{\boldsymbol{{\cal E}}}_{R2}\cdot \hat{\boldsymbol{{\cal E}}}_{L2} \right.\nonumber \left.+ 2\,\hat{\boldsymbol{{\cal E}}}_{R1}\cdot \hat{\boldsymbol{{\cal E}}}_{R2} -4\,\hat{\boldsymbol{{\cal E}}}_{R1}\cdot \hat{\boldsymbol{{\cal E}}}_{L2}\right)\nonumber \\
&+\frac{1}{g^2}\left(4-\tr \hat X_1-\tr \hat X_2 \right) \, .
\end{align}
The lattice spacing has been set to the identity, as is conventionally done, and
\begin{align}
\hat{\boldsymbol{{\cal E}}}_{L \kappa} \cdot \hat{\boldsymbol{{\cal E}}}_{L \kappa}=\hat{\boldsymbol{{\cal E}}}_{R \kappa} \cdot \hat{\boldsymbol{{\cal E}}}_{R \kappa}\equiv  \hat{{\cal E}}_\kappa^2 \, ,
\end{align}
where $\kappa$ is not summed over.
A basis that spans the Hilbert space of $\hathtp$ must be defined to analyze the real-time behavior of this system. Additionally, both $\hat{\cal E}_\kappa$ and $\hat X$ are infinite-dimensional operators and the basis needs to be truncated. The calculation of any physical observable is basis-independent but truncation will affect this universality. The effects should disappear as the truncation is removed. This paper utilizes three different bases to measure both static and dynamic properties.

\subsubsection{Group Element Basis}
In the group element basis, the eigenstates $\ket{\mathfrak{g}}$ of the gauge link operators (or the loop operators) are the basis states. Any parameterization of $\ket{\mathfrak{g}}$ can be used for defining the action of the Hamiltonian on the Hilbert space but some parameterizations are advantageous from the perspective of digitization and global gauge fixing. For \sutwo, two natural parameterizations are Euler angles or axis-angle coordinates. The axis-angle parameterization is particularly convenient~\cite{DAndrea:2023qnr,Grabowska:2024emw}. Any wavefunction that spans only the physical Hilbert space \textit{i.e.} any gauge-singlet wavefunction is given by
\begin{equation}
\label{eq:RescalePsi}
\braket{\omega_1\, \omega_2\, \Theta}{\Psi} = \Psi(\omega_1,\omega_2, \Theta)\equiv \frac{u(\omega_1, \omega_2, \Theta)}{4\sin\frac{\omega_1}{2}\sin\frac{\omega_2}{2}}
\end{equation}
where the rescaled wavefunction $u(\omega_1, \omega_2, \Theta)$ is introduced to simplify later calculations. The magnetic component of the fully-gauged fixed  Hamiltonian is diagonal and only depends on the radial coordinates $\omega_{1}$ and $\omega_2$,
\begin{align}
\label{eq:HBGE}
 H^{(B)}_{\square\square}= \frac{2}{g^2}\left(2- \cos \frac{\omega_1}{2}-\cos \frac{\omega_2}{2}\right) \,.
\end{align}
The electric component of the Hamiltonian can be determined by evaluating each of the electric bilinears that appear in Eq.~\eqref{eq:HTPOp}. The explicit expressions for the electric bilinears are relegated to App.~\ref{app:GEB}. Combining the various elements together, the electric component of the Hamiltonian is given by
\begin{align}
\label{eq:HEGE}
\htp^{(E)}&=-\frac{g^2}{2}\left[\left\{4\left(\pdv[2]{}{\omega_1}+ \cot \frac{\omega_1}{2}\pdv{}{\omega_1}\right)-2 \csc^2\frac{\omega_1}{2}\CN\right.\right.\nonumber \\
&+\left.\left( \cot \frac{\omega_2}{2}\pdv{}{\omega_1}\right)\left(\sin \Theta\pdv{}{\Theta}\right)+\left(\omega_1 \leftrightarrow \omega_2\right)\right\}  \nonumber \\
&+ \left(\frac{1}{2}\cot \frac{\omega_1}{2}\cot \frac{\omega_2}{2}\right)\left(\sin \Theta\pdv{}{\Theta}\right)\nonumber \\
&-\left.\left(\frac{1}{2}\cos \Theta \cot\frac{\omega_1}{2}\cot\frac{\omega_2}{2}-\frac{1}{2}\right)\CN\right]
\end{align}
where $\CN$ is the differential operator,
\begin{align}
\label{eq:LegDifEq}
\CN = - \pdv[2]{}{\theta}- \cot \theta \pdv{}{\theta} \, .
\end{align}

In this basis, $\htp$ is a partial differential operator and its spectrum can be found numerically using Finite Element Methods (FEM). FEM is a general numerical approach for solving low-dimensional partial differential equations that utilizes a discrete mesh of the computational domain.\footnote{This is in contrast to Finite Difference Methods (FDM), which approximate differential operators as finite differences on discrete grid.} In order to ensure that the eigenfunctions of $\htp$ are normalizable and that the mesh does not encounter any singularities, FEM is applied to solve for the rescaled wavefunctions, $u(\omega_1, \omega_2, \Theta)$. In this case, the radial coordinates, $\omega_{1,2}$ have a range of $[0, 2\pi]$ with Dirichlet boundary conditions while the angular coordinate $\Theta$ has range $[0, \pi]$ with Neumann boundary conditions. Fig.~\SubFigRef{fig:two_plaq_description}{b} plots the low-energy spectrum as a function of the bare gauge coupling. These solutions provide a beneficial `exact' solution to which the mixed basis and character irrep basis can be compared. This is particularly helpful, as while the these bases can be used for larger system sizes, the FEM approach faces insurmountable computational barriers when applied to larger systems. Without fully gauge-fixing, FEM could not even be applied to the two-plaquette system as, for example, the (naively parameterized) two-plaquette version of the Kogut-Susskind Hamiltonian involves twenty-one parameters (three quantum numbers for each of the seven links). 

\subsubsection{Mixed Basis}\label{subsub_sec:mixed_basis}
In the mixed basis, the states that span the Hilbert space are labeled by the continuous variables $\omega_1, \omega_2$ and the discrete quantum number $\nu$. This discrete quantum number is related to the angular variable $\Theta$ via
\begin{align}
\braket{\Theta}{\nu} = P_\nu(\Theta)
\end{align}
with $P_\nu(\Theta)$ the (rescaled) Legendre polynomial of degree $\nu$.\footnote{The conventional normalization of the Legendre polynomials is 
\begin{align}
\int_0^\theta d(\cos \theta)\,P_{\nu'}(\theta)P_\nu(\theta) = \frac{2}{2\nu+1}\delta_{\nu'\nu}
\end{align}
but due to the requirement that the basis states of the Hilbert space be orthonormal, the Legendre polynomials in this work are rescaled.} Legendre polynomials are used to span the Hilbert space because they are the eigenfunctions of the operator $\CN$,  
\begin{align}
    \CN P_\nu(\theta) = \nu(\nu+1) P_\nu(\theta) \, .
\end{align}

The matrix elements in this basis are found by calculating
\begin{align}
\mel{\omega_1'\omega_2'\nu'}{\hathtp}{\omega_1\omega_2 \nu} & = \int d(\cos \Theta) d(\cos\Theta')  \braket{\nu'}{\Theta'}\times \mel{\omega_1'\omega_2'\nu'}{\hathtp}{\omega_1 \omega_2 n}\braket{\Theta}{\nu} \nonumber\\
&= \int d(\cos \Theta)  P^*_{\nu'}(\Theta)\htp P_\nu(\Theta)
\end{align}
where $\htp$ is the sum of the magnetic and electric Hamiltonians, written in the group element basis in Eq.~\eqref{eq:HBGE} and Eq.~\eqref{eq:HEGE}

To convert between the group element basis and the mixed basis, only four integrals need to be calculated:
\begin{align}
\int \,d(\cos\Theta)\, P_{\nu'}(\Theta)
\left\{\begin{array}{c}
1\\
\CN \\
\cos \Theta\\
\sin \Theta \pdv{}{\Theta}
\end{array}\right\}
P_{\nu}(\Theta) \, .
\end{align}
The results of these integrals, tabulated in App.~\ref{app:MB}, are used to recast expressions in $\htp$. The necessary digitization, particularly of the components that depend on $\omega_\kappa$, will be discussed below in Sec.~\ref{sec:digitization}.

Using the mixed basis, it is possible to derive a strong bound on the ground state energy of the two plaquette system. Recall that the ground state energy of a Hamiltonian, $E_0$, can be bounded from above by the expectation value of that Hamiltonian in any state $\Psi$,
\begin{align}
E_\Psi \equiv \mel{\Psi}{\hat H}{\Psi} \geq E_0 \, .
\end{align}
In general, the goodness of this bound depends on how well the wavefunction $\Psi$ matches the true ground state. A good ansatz for the two plaquette system is found by assuming that the ground state does not depend strongly on $\Theta$ and therefore the functional dependence of the ground state on $\omega_\kappa$ is encapsulated in the differential operator
\begin{align}
-2g^2\left(\pdv[2]{}{\omega_\kappa}+ \cot \frac{\omega_\kappa}{2}\pdv{}{\omega_\kappa}\right)+\frac{2}{g^2}\left(1-\cos \frac{\omega_\kappa}{2}\right) \, ,
\end{align}
which is the component of the Hamiltonian that is diagonal in $\nu$ for each $\omega$. This differential equation is related to that of the Mathieu equation. Therefore, the ansatz for the ground state is
\begin{align}
\label{eq:GSAnsatz}
u(\omega_1, \omega_2, \Theta) = \frac{\text{se}_2(-8/g^4, \omega_1/4)\text{se}_2(-8/g^4, \omega_2/4)}{\sqrt{2}} \,
\end{align}
where $\Psi$ and $u$ are related via Eq.~\eqref{eq:RescalePsi}. The parity odd Mathieu function, $\text{se}_n(q, x)$ has period  $\pi$ when $n$ is even or $2\pi$ when $n$ is odd. For this system, $n$ has to be even for $\psi$ to be finite on the full range of $\omega_\kappa$.  With this ansatz, $E_\Psi$ is 
\begin{equation}
\label{eq:bound}
E_\Psi = \frac{g^4 b_2\left(-8/g^4\right)-4 g^4+16}{4 g^2} =\begin{cases}
3\sqrt{2}&\qquad g\ll 1 \\
4/g^2 &\qquad g \gtrsim 1
\end{cases}
\end{equation}
where $b_n(q)$ is the characteristic number that guarantees that $\text{se}_n(q, x)$ has period $\pi$ or $2\pi$. Comparing this ansatz to the FEM solution demonstrates that there is very good agreement between the two for a wide range of gauge couplings.

\subsubsection{Character Irrep Basis}
\label{subsubsec:CharIrrep}
The Kogut-Susskind Hamiltonian is formulated in the group irrep basis, also called the electric basis. The eigenstates of the single link Hilbert space are $\ket{J M_L M_R}$ where $J$ is the total angular momentum and $M_L$ ($M_R$) is the angular momentum projected onto the $\hat z$ axis of the lab (body) frame; $J$ is also commonly referred to as the irrep of \sutwo group. In analogy, the basis of the Hilbert space of the fully-gauged fixed two-plaquette Hamiltonian can be chosen to be $\ket{J_1\, J_2\, \nu}$ where $J_{1,2}$ are again irreps of \sutwo. The relationship between the group element basis and this basis is
\begin{align}
\braket{\omega_1\,\omega_2\,\Theta}{J_1\,J_2\,\nu} = \chi^{J_1}_\nu(\omega_1)\chi^{J_2}_\nu(\omega_2)P_\nu(\Theta)
\end{align}
where $\chi^J_\nu$ are (rescaled) generalized character functions of the irreps of the rotation group. These functions satisfy the differential equation,
\begin{align}
\left[\pdv[2]{}{\omega}+\cot \frac{\omega}{2}+J(J+1)-\frac{\nu(\nu+1)}{4 \sin^2\frac{\omega}{2}}\right]\chi^J_\nu(\omega) = 0
\end{align}
where $J$ are integers and half-integers and $\nu$ are integers, bounded by $2J\geq \nu\geq 0$. An explicit construction of these character functions is done via Clebsch-Gordan coefficients,
\begin{align}
\chi_\nu^J(\omega) = \frac{i^\lambda}{2\sqrt{\pi}} \sum_{M}e^{-i M \omega}C^{JM}_{JM\nu 0}
\end{align}
where the prefactor of $1/(2\sqrt{\pi})$ arises due to the requirement that the basis states are orthonormal.

Relevant matrix elements can be calculated for generic $J_1, J_2$ and $\nu$ by making use of recurrence relations for the character functions~\cite{khersonskii1988quantum}. The complete expressions are presented in App.~\ref{app:CIB}. A key result is that the magnetic component of the Hamiltonian only changes the $J_\kappa$ quantum numbers by at most $\pm 1/2$. In contrast, the electric component of the Hamiltonian only changes the $\nu$ quantum number by at most $\pm 1$. With these matrix elements, it is possible to construct a finite-dimensional Hamiltonian, with cutoff $J_{\kappa}^\text{max}$. Note that $\nu$ is naturally truncated due to the requirement that 
\begin{align}
2\,\text{Min}(J_1, J_2)\geq \nu \geq 0
\end{align}
and since the Hamiltonian is symmetric under $J_1 \leftrightarrow J_2$ exchange, as is evident from the differential form of the Hamiltonian, both $J_1$ and $J_2$ have the same cut-off. 

Solving for the spectrum in the strong coupling limit is simple, as the electric component preserves the quantum numbers $J_1, J_2$ and the contributions from the magnetic component can be ignored. Additionally, since the quantum number $\nu$ is bounded from above by $2\,\text{Min}(J_1,J_2)$, for a given $J_1, J_2$, the relevant sub-block of the Hamiltonian is finite-dimensional. In short, in the strong coupling limit, at leading order, the Hamiltonian is block-diagonal. The results for the lowest states are shown in Fig.~\SubFigRef{fig:two_plaq_description}{c} and in Table~\ref{tab:CharIrrepStrong}.

This basis can be used for other values of the gauge coupling if the magnetic component of the Hamiltonian is included. In this case, the Hamiltonian is no longer diagonal in $J_{1,2}$ and its spectrum cannot be found analytically. Using exact diagonalization for different values of $J^\text{max}_\kappa$, it can be shown that at small values of the gauge coupling, the energy eigenstates diverge as $1/g^2$. However, using FEM, it is shown that the low-lying eigenstates become independent of $g$ in the limit of small $g$. These large truncation errors at weak coupling is why electric bases are disfavored for this regime.

\section{Digitization of the Mixed Basis}
\label{sec:digitization}
The Hilbert space degrees of freedom that may be represented by a register of qubits is finite and so a finite representation of the continuous gauge field must be determined. This is commonly referred to as a `digitization' (distinct from `discretization', which corresponds to representing a continuous space-time on a lattice). This section focuses on the manner in which the mixed basis can be digitized, in particular the derivatives with respect to $\omega$. Of key importance is how to efficiently implement derivative operators onto a quantum device, which will be the focus of the next section.
\subsection{Conventions}
In the following sections, various ideas and tools will appear in slightly different contexts. Therefore, to ease notation and presentation, the following conventions are made.

For a single qubit, one may define the set of Pauli operators 
\begin{align}
\sigma_{\alpha}=\{I,X,Y,Z\} \,\,.
\end{align}
For $N$-qubits, the set of all Pauli strings (modulo phase) is the Pauli group, and is defined as \begin{align}
\mathcal{P}_N=\{\otimes_{l=1}^N\sigma_{l}^{\alpha_l}\;\; | \;\; \alpha_{l}=0,1,2,3\} \,\,.
\end{align}
This group defines a complete basis for the set of all hermitian operators $O$ (e.g.\ a Hamiltonian or density matrix) on $N$ qubits, and is normalized as 
\begin{align}
TrP P'=2^N\,\delta_{P,P'} \quad \forall\, P,P'\in\mathcal{P} \,\, .
\end{align}
Given such a hermitian operator $O$ on $N$ qubits, its (normalized) Pauli decomposition is denoted as 
\begin{align}
\mathcal{D}(O)=\left\{(P,\frac{1}{2^N}TrP O)\;\; | \;\; P\in \mathcal{P}_N\right\} \,\, .
\end{align}
$\mathcal{D}(O)$ will be used to either denote the set of pairs as shown above or the sum itself, where the meaning will be clear based on context.

\subsection{Formulating Gauge Field Operators}
An important feature of a good digitization is that observables (e.g.\ expectation values of hermitian operators) converge quickly to the continuum limit as the resolution is increased. For fast convergence, the range of support of the wavefunctions of interest must be fully covered in both field and conjugate momentum space. In particular, if the support of the wave function is not sufficiently covered then predictions for observables will not converge to the correct solution. If the region of field space covered is much larger than the support of the wave function, then the digitization will sample regions of the wave function that are exponentially small and the convergence will be slow. The following digitization achieves good convergence properties by optimally spanning the regions of wave function support.

The continuous gauge field degrees of freedom represented by $\{\omega_1,\omega_2\}\in[0,2\pi)^{\otimes2}$ must be embedded into a register of $2n_q$ qubits with Hilbert space dimension $2^{2n_q}$. This is done by letting each $\omega_i$ take values on a grid of size $2^{n_q}$ with the range $[0,\omega_{\text{max}})$. The discretized variables are denoted as $\omega_i,$ for $i=1,2$ and assigned values 
\begin{align}
\omega_i &= \delta\omega(n+1/2) \qquad & n\in\{0,\dots,2^{n_q}-1\} \nonumber \\ \delta\omega &= \omega_{\text{max}}/2^{n_q} \, \, .
\end{align}
When performing classical simulations when the system is not embedded into qubits, the total grid size is denoted as $N_{\omega}$. In addition to the two $\omega$-registers, there is an additional quantum number $\nu$ that takes integer values and requires an additional $n_{\nu}$ qubits, giving a total simulation Hilbert space of size $2^{2n_q+n_{\nu}}$. This layout is graphically illustrated in Fig.~\SubFigRef{fig:digitization_details}{a}.

There are three types of terms that appear in the Hamiltonian, namely the first-order derivatives of $\omega_i$, the second-order derivates of $\omega_i$ and functions of $\omega_i$. Their action on the qubit register are treated separately. The field variable $\omega_i$ is defined by the operator
\begin{equation}
    \hat{\omega} = \frac{\delta\omega}{2}\left[n_qI-\sum_{j=0}^{n_q-1}2^j \hat Z_j\right]
\end{equation}
where $\hat Z_j$ is the Pauli $\hat Z$ operator acting on the $j$th qubit. Functions of $\hat{\omega}$ can then be constructed as $\mathcal{D}(f(\hat\omega))$, explicitly
\begin{align}
    f(\hat{\omega}) &= \sum_{P\in\{I,Z\}^{\otimes n_q}}f_{P} P\nonumber\\ 
    f_{P} &= \frac{1}{2^{n_q}}\text{Tr}[f(\hat{\omega})P]
\end{align}
where the sum above is over all strings of Pauli $Z$ operators. The second-order derivatives can be implemented using the exact lattice Laplacian with Dirichlet boundary conditions, which is given by
\begin{align}\label{eq:exact_laplacian}
\pdv[2]{}{\omega} &\mapsto \text{DST}^{-1}_{II}(-\hat{k}^2)\text{DST}_{II}\,\,
\end{align}
where $\text{DST}_{II}$ is a variation of the Discrete Sine Transform (DST), reflecting the boundary conditions in field space:
\begin{align}
    (\text{DST}_{II})_{kn} = 
    \begin{cases}
    \sqrt{\frac{2}{N}}\sin\frac{\pi(n+\frac{1}{2})(k+1)}{N}, & \text{for } k \neq N-1\\
    \sqrt{\frac{1}{N}}\sin\pi(n+\frac{1}{2}),  & \text{for } k = N-1
    \end{cases} \,\,.
\end{align}
Note that $\hat{k}$ is the lattice momentum operator, which is diagonal in Fourier space with eigenvalues $\pi k/\omega_{\text{max}}$ with $k\in\{1,\dots,N\}$. It has been shown that when using this exact formulation, the matrix elements of this term converge faster than polynomial to the continuum value as the grid spacing $\delta\omega$ goes to zero~\cite{Jordan:2011ci,Jordan:2017lea,Klco:2018zqz,Macridin:2018gdw}. Therefore it is advantageous to use compared to a finite difference approach.

The first-order derivative is, however, formulated using the finite central difference
\begin{align}\label{eq:finite_central_diff}
    \left(\pdv{}{\omega}\right)_{kn} = 
    \begin{cases}
    \frac{1}{2\delta\omega}, & \text{for } k-n=1,\\
    -\frac{1}{2\delta\omega}, & \text{for } k-n=-1,\\
    0 & \text{else}
    \end{cases}
\end{align}
The consequence of using the finite difference is that the digitized form of $\htp$ will only have polynomial convergence towards the continuum Hamiltonian as the spacing in field space is decreased. However, in the context of quantum simulation, even for  large $\delta\omega$ this error will generally be subleading to algorithmic errors stemming from approximations to continuous time evolution. While it is desirable to use spectral methods for the first-order derivative, as was done for the Laplacian, the Dirichlet boundary conditions complicate such an approach. This can be understood intuitively because the DST acts on data with odd symmetry, e.g.\ a sine wave. A first derivative  maps functions with odd symmetry to functions with even symmetry, and thus is incompatible with the use of a DST. Other types of spectral methods for first order derivatives with Dirichlet boundary conditions have been formulated, but often require non-uniform grid spacings, for example using Chebyshev nodes~\cite{Trefethen2013Approx}. This approach is undesirable as such grid choices have denser coverage towards the boundaries of the field region. This causes both oversampling of the singularities of the Hamiltonian as well as asymmetric sampling of the wavefunctions for low $g$.

\subsection{Optimal Gauge Field Truncation}

\begin{figure}[htbp]
    \centering
    \includegraphics[scale=.25]{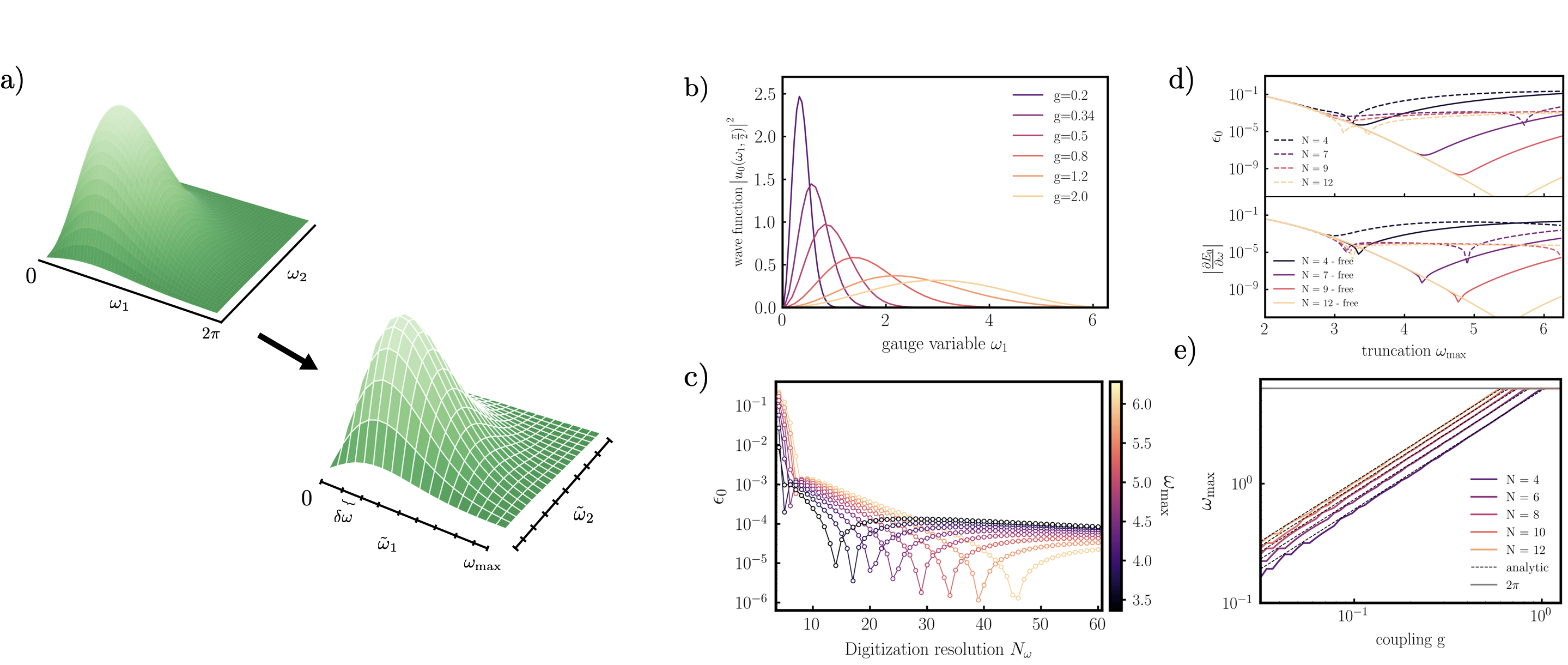}
    \caption{\textit{Digitization of Mixed Basis} \textbf{a)} The continuous range of $\omega$ is digitized on a discrete grid. As such the wave function will only be sampled at a discrete set of points. For the closest approximation to the undigitized theory, the maximum value of $\omega_{\text{max}}\leq 2\pi$ must be chosen. \textbf{b)} A slice of the ground state wave function for different values of the gauge coupling. As the gauge coupling decreases, the wave function becomes more concentrated around $\omega_1 = 0$, although it must go to $0$ at both boundaries due to the Dirichlet boundary conditions. \textbf{c)} Convergence of the ground state energy to the undigitized value as the digitization resolution $N_{\omega}$ is increased. The convergence is plotted for a variety of $\omega_{\text{max}}$. After quick convergence to $\epsilon_0\sim 10^{-3}$ a slower exponential approach occurs, followed by a plateau. \textbf{d)} \textit{(top)} Convergence to the ground state energy as $\omega_{\text{max}}$ is varied. The dashed lines correspond to the full two plaquette system while the solid lines correspond to the system without any first-derivative interactions. The effect of the first derivative reduces the precision that can be achieved. Even though this term limits precision, digitization errors are still below those due to time-evolution algorithms. \textit{(bottom)} The absolute value of the derivative of the energy w.r.t the truncation. It was heuristically found that the truncation at which this derivative takes a minimum corresponds to the optimal truncation for the analytically solvable case. This method is still applicable even when the exact solution is not known. \textbf{e)} The solution of the stationary condition for a range of gauge couplings. The agreement is very good with the analytical solution Eq.~\eqref{eq:OmegaMax}.}
    \label{fig:digitization_details}
\end{figure}

In analogy with digitizations of scalar fields, the gauge field is truncated at some value $\omega_{\text{max}}\leq 2\pi$. Fig.~\SubFigRef{fig:digitization_details}{b} shows the amplitude of the ground state wave function
\begin{align}
u_0(\omega_1,\omega_2) = \left|\braket{\omega_1,\omega_2}{\psi_{\text{gs}}}\right|^2
\end{align}
along the $\omega_1$ axis ($\omega_2=\pi/2$) for different coupling strengths. As the gauge coupling is decreased, $u_0$ becomes increasingly localized at small $\omega_1$, decaying exponentially for large values. Therefore, the optimal $\omega_{\text{max}}$ must be characterized as a function of gauge coupling. This is done by considering the precision 
\begin{equation}\label{eq:precision}
    \epsilon_0=\left|1-\frac{E'_{0}}{E_0}\right|
\end{equation}
where $E'_0$ is the ground state energy predicted by the digitization and $E_0$ is the `true' ground state energy. Similar results are observed for low-lying eigenstates. The solutions for the digitization are compared to those found via FEM methods in Fig.~\SubFigRef{fig:two_plaq_description}{b}, the strong coupling solutions of Sec.~\ref{subsubsec:CharIrrep} and the ansatz of Eq.~\eqref{eq:GSAnsatz}, and are in good agreement. The solutions found via FEM are of particular use, and they have a relative precision of approximately $10^{-4}$.

Fig.~\SubFigRef{fig:digitization_details}{c} shows the precision for different choices of $\omega_{\text{max}}$ truncation as a function of digitization resolution $N_\omega$. The convergence is characterized by three stages. The first stage is rapid convergence to $\epsilon_0\sim 10^{-3}$ level for small $N_{\omega}$. Following this, the precision decays more slowly, though still exponentially, as $N_{\omega}$ is increased. Eventually, a plateau is reached. The precision value of this plateau decreases as $\omega_{\text{max}}$ increases. There are sharp spikes observed before the plateau is reached, which are attributed to the digitized solution crossing the FEM solution, as opposed to approaching it. Such crossings are purely incidental and should not be interpreted as signs of optimality in representation of the wavefunction. Since the precision of the FEM solution is limited, it is not possible to study convergence below precision levels of $10^{-4}$. Despite this it is clear that even for grid sizes as low as $N_{\omega}=8\;(n_q=3)$ it is possible to reach $\epsilon_0\sim 10^{-3}$ levels of precision. This is at least an order of magnitude below algorithmic error due to time-evolution and on par with device noise-levels for near-term devices.

The optimal truncation is found by scanning over a range of $\omega_{\text{max}}$ values for a fixed $g,N_{\omega}$ and calculating the corresponding $\epsilon_0$. The top plot of Fig.~\SubFigRef{fig:digitization_details}{d} shows the precision $\epsilon_0$ for $\htp$. Also considered is the `free' effective theory of $\htp$, where terms containing a first derivative are subtracted off. The theory is not completely free, as there is still a collective $\nu$ quantum number, but the interactions due to the first derivatives are no longer present. This theory is studied to understand how $\epsilon_0$ behaves without the error induced by first derivatives. The ground state of the latter Hamiltonian can be found exactly and is given by the ansatz in Eq.~\eqref{eq:GSAnsatz}. The corresponding energy is Eq.~\eqref{eq:bound}. For the system without first derivatives, the precision decays exponentially in $\omega_{\text{max}}$ until the optimal value is reached, and then increases due to over-sampling regions of field space where the wave function has no support. However, once first derivatives are included, the precision plateaus at a value orders of magnitude higher, and can only be increased by increasing $N_{\omega}$. Thus there is a range of choices of $\omega_{\text{max}}$. The sharp dips are again due to crossings and should be ignored when considering the optimality of $\omega_{\text{max}}$.

The gauge field truncation can be optimized without knowing the exact value of the ground state energy by looking at the derivative of the energy with respect to $\omega_{\text{max}}$. It is empirically observed that the optimal $\omega_{\text{max}}$ can equivalently be defined by the stationary condition
\begin{equation}
\label{eq:stationary_condition}
    \omega_{\text{max}} = \argmin_{\omega}\left|\frac{\partial E_i(\omega)}{\partial \omega}\right|\,\,.
\end{equation}
This condition can also be observed for the digitization of the scalar field. Fig.~\SubFigRef{fig:digitization_details}{d} (bottom) shows the derivative of the ground state energy predicted by the digitization. For the system without first derivatives, the minimum of these curves align with the points where the ground state energy most closely approaches the expression Eq.~\eqref{eq:bound}. When first derivatives are included, the precision plateaus at low values for a range of $\omega_{\text{max}}$. The optimal truncation in the absence of first derivatives is found by scanning over a range of gauge couplings and finding the solution of stationary condition, Eq.~\eqref{eq:stationary_condition}. Fig.~\SubFigRef{fig:digitization_details}{e} shows that the numerical solution found by the scan closely matches the analytical solution~\cite{DAndrea:2023qnr}
\begin{equation}
    \omega_{\text{max}} = \text{min}\left(g(N-1)\sqrt{\frac{\sqrt{8}\pi}{N}},2\pi\right)
    \label{eq:OmegaMax}
\end{equation}
which was found by minimizing $\langle[\partial/\partial\omega,\omega]-1\rangle$ for the ground state of a single plaquette.

\begin{figure}[htbp]
    \centering
    \includegraphics[scale=.35]{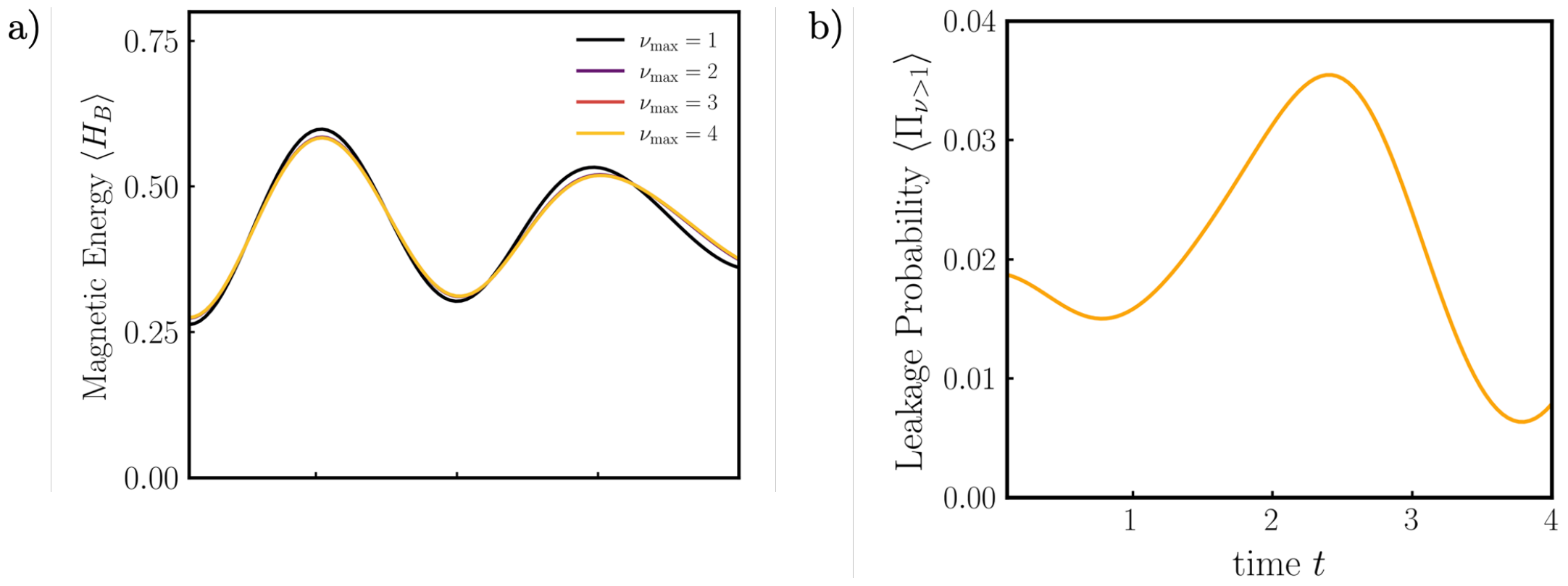}
    \caption{\textit{Accuracy of the $\nu$ Truncation} \textbf{a)} Calculation of the time dependent magnetic energy for the low-energy state $\ket{5}$, defined in Eq.~\eqref{eq:low_energy_state}, as the maximum $\nu$ value is increased for $g=0.5$ and $N_{\omega}=10$. This choice of digitization resolution ensures that any errors due to the $\omega_1,\omega_2$ truncation will be of order $10^{-2}$. \textbf{b)} The probability $\langle\Pi_{\nu>1}\rangle$ of leaking out of the $\nu=0,1$ sector as a function of time for $g=0.5, N_{\omega}=10$ and $\nu_\text{max} = 5$.}
    \label{fig:nu_truncation}
\end{figure}

In addition to the truncation of the gauge field, the discrete quantum number $\nu$ must also be truncated. An advantage of the mixed basis is that the low-lying eigenstates are supported solely in the $\nu=0,1$ sector, and so only a single qubit is required to represent this quantum number. Furthermore, $\htp$ only connects states that have $|\nu-\nu'|\leq 1$, and so for low-energy simulations heavy truncations on this degree of freedom will have little effect on the precision. To study the effect of different $\nu$ truncations on the time evolution of observables, the system is initialized in the state
\begin{align}\label{eq:low_energy_state}
\ket{5}=\frac{1}{\sqrt{5}}(\ket{E_0}+\ket{E_1}+\ket{E_2}+\ket{E_3}+\ket{E_4}) \, \,
\end{align}
where $\ket{E_n}$ is the $n^{\rm th}$ eigenstate of the digitized Hamiltonian and then time evolved according to different maximum $\nu$ values $\nu_{\text{max}}$. Fig.~\SubFigRef{fig:nu_truncation}{a} shows the time dependence of the expectation value for the magnetic part of the Hamiltonian $H_B$. As $\nu_{\text{max}}$ is increased, the values of $\langle H_B\rangle$ rapidly converge to the same curve, suggesting that $\nu_{\text{max}}=2$ is already sufficient for extracting high-precision estimates of observables.

The leakage out of the $\nu=0,1$ sectors is also studied. Defining the projector onto states with $\nu > 1$ as
\begin{equation}
    \Pi_{\nu>1} = \mathbb{I} - \sum_{J_1,J_2}\left(\ket{J_1J_20}\bra{J_1J_20} + \ket{J_1J_21}\bra{J_1J_21}\right),
\end{equation}
the expectation value $\langle \Pi_{\nu > 1}\rangle$ encodes the probability that the state has leaked out of the low-energy $\nu=0,1$ subspace. Fig.~\SubFigRef{fig:nu_truncation}{b} shows the leakage probability as a function of time. As the state evolves, the leakage does not grow noticeably over time, showing that a low-energy state with little support on values of $\nu > 1$ will not evolve to states with larger support over a range of times.

\section{Circuits for Time Evolution}
\label{sec:MixedBasisCircuits}

In this section the digitization presented above is translated into quantum circuits that can be implemented on current quantum devices. The circuit architecture for implementing time evolution is not unique, and two distinct approaches are developed that have different depths based on the level of digitization that is required for the simulation task. 

In the first approach to circuit design, a single Trotter step is formed from component sub-circuits. This has the advantage of avoiding any pre-computation of circuits as  the form of the components are already known. Additionally, most of these components have an asymptotic scaling that is no more than quadratic in $n_q$. For the component implementing a diagonal unitary that encodes various transcendental functions, exact implementations scale exponentially in the number of elementary gates with $n_q$, the number of qubits used to digitize $\omega_1$ and $\omega_2$. While this is seen as an unavoidable feature of this system, implementing such gates up to a desired level of precision scales more favorably than the worst-case exponential~\cite{Li:2024lrl}. 

This approach gives the greatest benefit when high resolution in field space is required. When the number of qubits per plaquette is of order one, it is advantageous to employ a different approach. This begins by directly decomposing the Hamiltonian into a sum of Pauli strings and implementing the time evolution operator by a series of rotations. This approach scales exponentially in $n_q$, although dropping strings with weights below some threshold is found to give circuit-depth reductions that outweigh the losses in accuracy. The second approach is more suited to simulation tasks where only low resolution is field space is necessary. The following sections give explicit circuit decompositions for each of these approaches and studies the accuracy and different approximations that go into each. The relative resource counts are then compared.

\subsection{Circuit from Differential Operators}
\label{subsec:DifOpCircuits}

\begin{figure}[htbp]
    \centering
\includegraphics[width=\textwidth]{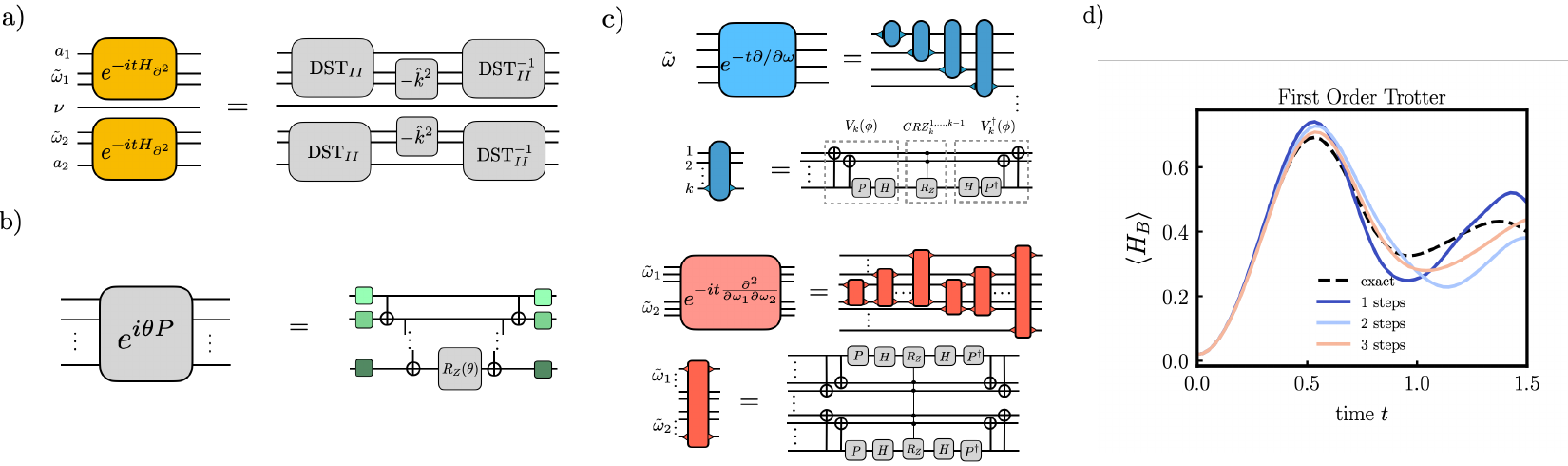}
    \caption{\textit{Circuits for the  Digitization Presented in Sec.~\ref{subsec:DifOpCircuits}, which constructs sub-components for each class of differential operators} \textbf{a)} Implementation of the exact exponential of the second derivative, making use of Discrete Sine Transforms. This subcircuit only entangles qubits within each $\omega$ register and does not touch the $\nu$ register. The ancillas $a_1,a_2$ associated with each $\omega$ register are the outermost qubits. \textbf{b)} An important circuit identity used in this work: The exponential of an $N-$body Pauli string $P$ is implemented by the staircase of $CX$s connecting all the qubits participating in the gate, followed by an $R_Z(\theta)$ rotation on one of the qubits and then another $CX$ staircase. The green single qubit gates denote a local change of bases to change $P$ into a string of $Z$s on all qubits e.g.\ if there is an $X$ on qubit $i$ then this will be a Hadamard $H$ and if it is a $Y$ then it will be a Hadamard followed by a phase gate $HS$. The different shades denote that the local basis change can depend on the particular Pauli string $P$ being implemented.
   \textbf{c)} Circuits for implementing first order derivatives and mixed derivatives using the finite difference representation. \textit{(top)} Implements the exponential of a single first derivative. Each element in this circuit implements one of the terms in the sum Eq.~\eqref{eq:partial_MPO}. The bottom qubit of each step is the target qubit for both the $V_k$ and $CRZ$ gates. \textit{bottom} An example of one of circuit elements $e^{-it\frac{\partial}{\partial\omega_1}\frac{\partial}{\partial\omega_2}}$ that interleaves the top circuit. \textbf{d)} Noiseless simulator results for the expectation value of the magnetic Hamiltonian at a different number of Trotter steps.}
    \label{fig:fine_dig_circ}
\end{figure}

This approach proceeds by organizing the terms in the
two-plaquette Hamiltonian according to the highest order
differential operator each term contains. In particular, the terms in the Hamiltonian are grouped in the following way
\begin{equation}
    \htp = H_{\partial^2}+H_{\partial}+H_{\omega}
\end{equation}
where $H_{\partial^2}$ collects all the terms that have only a second derivative, $H_{\partial}$ contains any terms that have a first order derivative, and $H_{\omega}$ is all remaining terms, \textit{i.e.} those that are only functions of $\omega$ with $\nu$-dependent coefficients. The total time evolution operator is Trotterized according to these groupings, $e^{-it\htp}\approx e^{-itH_{\partial^2}}e^{-itH_{\partial}}e^{-itH_{\omega}}$.

The circuits for $e^{-itH_{\partial^2}}$ and $e^{-itH_{\omega}}$ are straightforward, and mirror the circuits involved in simulating the scalar field. The Laplacian term is implemented by switching to the Fourier basis with a DST$_{II}$, which is based off of the Quantum Fourier Transform (QFT)~\cite{klappenecker2001discrete}. In this basis each second derivative is diagonal and proportional to the square of the conjugate momentum operator $\hat{k}^2$,
\begin{equation}
    \hat{k}^2= \frac{\pi}{\omega_{\text{max}}}\left(\frac{2^{n_q}-1}{2}I-\sum_j^{n_q-1}2^{j-1}Z_j\right)^2
\end{equation}
The circuit implementing a DST$_{II}$ is shown in Fig~\SubFigRef{fig:fine_dig_circ}{a}. While using the QFT as an algorithmic primitive, extra elements (not shown) are required to enforce the boundary conditions. The DST circuit requires an additional ancilla qubit changing the qubit count per gauge variable to be $n_{q}+1$. $H_{\omega}$ is completely diagonal in the mixed basis and so it can be represented as sum of Pauli $Z$ strings. It is implemented using the primitive shown in Fig.~\SubFigRef{fig:fine_dig_circ}{b} with $\nu$-dependent rotation angles.

The construction of the circuits for $e^{-itH_{\partial}}$ is more involved. Recalling the results from Sec.~\ref{sub_sec:two_plaquette_hamiltonian}, the matrix representation of the Hamiltonian is organized into $(\nu,\nu')$ sectors, where nontrivial sectors satisfy the selection rule $|\nu-\nu'|\leq 1$. The $\nu=\nu'$ case only involves the circuits already discussed. For $\nu\neq \nu'$, the Hamiltonian $H_{\partial}$ is expressed as
\begin{equation}
    H_{\partial} = \sum_{\nu}H^{+}_{\nu}\ket{\nu}\bra{\nu+1}+\sum_{\nu}H^{-}_{\nu}\ket{\nu}\bra{\nu-1}
\end{equation}
where the expressions for $H^\pm_\nu$ are given in App.~\ref{app:MB}.  

The coupling of different $\nu$ sectors is reminiscent of a finite difference matrix in $\nu$ space. As already noted, implementing such an operator exactly requires a prohibitive number of gates. This is where the expressive power of the mixed basis shines through. Truncating to $\nu_{\rm max}=1$ gives the off-diagonal Hamiltonian the form
\begin{align}
H_{\partial}=\sigma^{+}H^{+}_{0}+\sigma^{-}H^{-}_{1}
\end{align}
where $\sigma^{\pm}$, defined  as
\begin{align}
\sigma^{\pm}=\frac{1}{2}(X\pm iY) \, ,
\end{align}
acts on the single-qubit $\nu$-register. Applying $HS^{\dagger}$ to the $\nu$-register turns the projectors into 
\begin{align}
\sigma^{\pm}\mapsto \frac{1}{2}\left(Z\pm iX\right) \,\, ,
\end{align} 
where $H$ is a Hadamard and $S$ is a $\pi/2$ phase gate. Thus under this transformation, $H_{\partial}$ maps to
\begin{align}
H_{\partial}\mapsto \frac{1}{2}\left(Z\otimes H_{\text{sym}}-iX\otimes H_{\text{anti}}\right)
\end{align}
where $H_{\text{sym}}, H_{\text{anti}}$ are the symmetric and anti-symmetric parts of $H^{+}_0$,\footnote{$H_0^{+}$ and $H_1^{-}$ are hermitian conjugates}
\begin{align}
H_{\text{sym}}&=H_0^{+}+H_1^{-}\nonumber \\
H_{\text{anti}}&=H_0^{+}-H_1^{-} \, .
\end{align}
Explicitly, these components are given by
\begin{align}
    H_{\text{sym}} &= \frac{1}{\sqrt{3}}\pdv{}{\omega_1}\pdv{}{\omega_2}+\frac{1}{4\sqrt{3}}\cot \frac{\omega_1}{2}\cot \frac{\omega_2}{2}\nonumber\\
    H_{\text{anti}} &= -\frac{1}{2\sqrt{3}}\left(\cot \frac{\omega_1}{2}\pdv{}{\omega_2}+\cot \frac{\omega_2}{2}\pdv{}{\omega_1}\right)\nonumber \,\,.
\end{align}
The complete time evolution induced by the first derivatives may be approximated by
\begin{align}\label{eq:h_partial_evo}
    e^{-itH_{\partial}} &\approx e^{-itg^2Z\otimes H_{\text{sym}}}e^{-itg^2X\otimes H_{\text{anti}}}\nonumber\\
    &\approx e^{-i\frac{g^2t}{\sqrt{3}}Z\otimes\pdv{}{\omega_1}\pdv{}{\omega_2}}e^{-i\frac{g^2t}{4\sqrt{3}}Z\otimes \cot \frac{\omega_1}{2}\cot \frac{\omega_2}{2}}\nonumber\\
    &\hspace{30pt}e^{-\frac{g^2t}{2\sqrt{3}}X\otimes \cot \frac{\omega_1}{2}\pdv{}{\omega_2}}e^{-\frac{g^2t}{2\sqrt{3}}X\otimes \cot \frac{\omega_2}{2}\pdv{}{\omega_1}}
\end{align}
where the tensor product separates the $\nu$ register from the $\omega$ registers and the dependence of gauge coupling has been reinserted.

Therefore, the problem of implementing the time evolution due to the $H_{\partial}$ terms is reduced to exponentiation of the finite difference operator. It has been recognized that this operator can be recast as a sum of Matrix Product Operators (MPO)~\cite{sato2024hamiltonian} with an induced Trotter error of $t^2(n_q-1)/2$ . To see this, one starts from the identity
\begin{align}\label{eq:partial_MPO}
\pdv{}{\omega} &\mapsto\frac{-i}{2\delta\omega}\sum_k I^{\otimes n_{q}-k}\otimes \\
    &V_k\left(-\frac{\pi}{2}\right)\left(Z_k\otimes\ket{1}\bra{1}^{\otimes k-1}\right)V_k^{\dagger}\left(-\frac{\pi}{2}\right) \,. \nonumber
\end{align}
where $\hat Z_k$ is the Pauli $\hat Z$ operator on the $k^{\rm th}$ qubit in the corresponding register and $V_k(\phi)$ is a unitary change of basis defined by
\begin{equation}
    V_k(\phi) = \left(\prod_m^{k-1}\text{CX}_k^m\right)P_k(\phi)H_k \,
\end{equation}
where $\text{CX}_k^m$ is a CNOT gate with control on qubit $m$ and target on qubit $k$, $P_k(\phi)=\text{diag}(1,e^{i\phi})$ is a phase gate on qubit $k$, and $H_k$ is a Hadamard on qubit $k$, not to be confused with a Hamiltonian. After taking the exponential and Trotterizing the sum over $k$ in Eq.~\eqref{eq:partial_MPO}, the operator $e^{-t\partial/\partial\omega}$ can be implemented by a sequence of controlled $CRZ_k^{1,\dots,k-1}(t/\delta\omega)$ rotations for each qubit in the register $k$, where $CRZ_k^{1,\dots,k-1}(\theta)$ is a controlled $R_Z(\theta)$ with target $k$ and controls $1,\dots,k-1$.

Inspection of Eq.~\eqref{eq:h_partial_evo} shows that it is insufficient to implement terms just of the form $e^{-t\partial/\partial\omega}$, as there are entangling operations between the two gauge variable registers and the $\nu$ register. Therefore, it is necessary to modify the above circuits by interleaving the extra operations between each set of $V_k$ and $V_k^{\dagger}$. As an example, the first term will need to implement an operator which is the product of a first derivative on each $\omega$. Expanding the MPO expression for the first derivative gives
\begin{align*}
    \pdv{}{\omega_1}\pdv{}{\omega_2} &\mapsto -\frac{1}{4\delta\omega^2}\sum_{k,k'}I^{\otimes n_q-k}I'^{\otimes n_q-k'}\otimes\\
    &\hspace{10pt}V_kV'_k\left(Z_kZ_k'\otimes \ket{1}\bra{1}^{\otimes k-1}\ket{1'}\bra{1'}^{\otimes k'-1}\right)V_k^{\dagger}V_k'^{\dagger}
\end{align*}
where the (un)primed operators denote operators acting on the ($\omega_1$)$\omega_2$ register and the argument of the $V_k,V'_k$ have been suppressed. A similar interleaving can be made for the remaining terms in Eq.~\eqref{eq:h_partial_evo}, and for concreteness, the circuits structures for the case just discussed is shown in Fig.~\SubFigRef{fig:fine_dig_circ}{c}.

Fig.~\SubFigRef{fig:fine_dig_circ}{d} shows different numbers of first-order Trotter steps of these circuits at various times $t$. If the noise profile of the device is partially or completely characterized, there exists an optimal step size for a target total simulation time~\cite{Clinton:2021pdy,Zemlevskiy:2023eyw}. For even a single first-order step, this method remains accurate at up to a step size of approximately $t=0.5$, at which point dephasing is observed and deviations from the exact curve begin to grow. Modifying these circuits to perform a second-order step, with error that falls off as $\mathcal{O}(t^3)$ is straightforward, and requires modifications of the $V_k$ circuits as outlined in~\cite{sato2024hamiltonian}.

\subsection{Circuits by Pauli Truncation}
\label{subsec:coarse}

Simulating time evolution requires converting the digitized time evolution operator $U=e^{-i t \htp}$ into a sequence of two-qubit gates directly implementable on a quantum device. The most straightforward way to do this, while not the most efficient, is to calculate $\mathcal{D}(\htp)$. Once this decomposition is performed, the sum in the exponent of $U$ can be split into a product over Paulis using a Trotter-Suzuki formula, which approximates the exact time evolution operator. This time evolution operator $U_{T}(t)=\prod_Pe^{-i t c_P P}$, where $(P,c_p)\in \mathcal{D}(\htp)$, can be turned into circuit-level operations using the identity displayed in Fig.~\SubFigRef{fig:fine_dig_circ}{b}. Generally this procedure will have a two-qubit gate count that scales as $\mathcal{O}(4^{2n_q+n_{\nu}})$. However, many of the rotations, particularly those corresponding to high weight strings, will have small angles that do not greatly impact the expectation value of observables over some range of evolution times. The rotations with small angles less than $\theta_{\text{min}}$ can be dropped, which gives a compressed approximation of the exact time evolution operator.

It is necessary to understand the impact of dropping small angles for different levels of gauge field digitization and at different values of the gauge coupling. By removing these terms, the system is evolved instead by the Hamiltonian 
\begin{align}
H_{\delta}&=\htp + \theta_{\text{min}}V \nonumber \\
V&=-\sum_{c_P\leq \theta{\text{min}}}\left(\frac{c_P}{\theta_{\text{min}}}\right)P
\end{align}
where $\delta$ corresponds to the percentage of terms in the Pauli decomposition that have coefficient below $\theta_{\text{min}}$. The time evolution implemented on the device is given by the unitary 
\begin{align}
U_{\delta}(t)=UU_{\theta}(t)
\end{align}
where the $U_\theta$ (and its component) matrix, given by
\begin{align}
U_{\theta}(t)&=e^{-i\theta_{\text{min}}\int_0^td\tau V_I(\tau)} \nonumber \\
V_I(\tau)&=e^{-i\tau\htp} Ve^{i\tau\htp} \,\, ,
\end{align}
is the error term in the interaction picture. Therefore the extent to which the perturbation $V$ grows under the dynamics of the two plaquette system directly controls the appropriate level of truncation. The error in evolution on some observable $O$ is quantified by
\begin{align}
\langle O(t) \rangle_{\delta}=\langle U^{\dagger}_{\theta} O(t)U^{\phantom{\dagger}}_{\theta} \rangle \, .
\end{align}
which is the time-evolved value of the observable. This is compared to the untruncated value $\langle O(t) \rangle$ ($\delta=0$).

\begin{figure}[htbp]
    \centering
    \includegraphics[width=\textwidth]{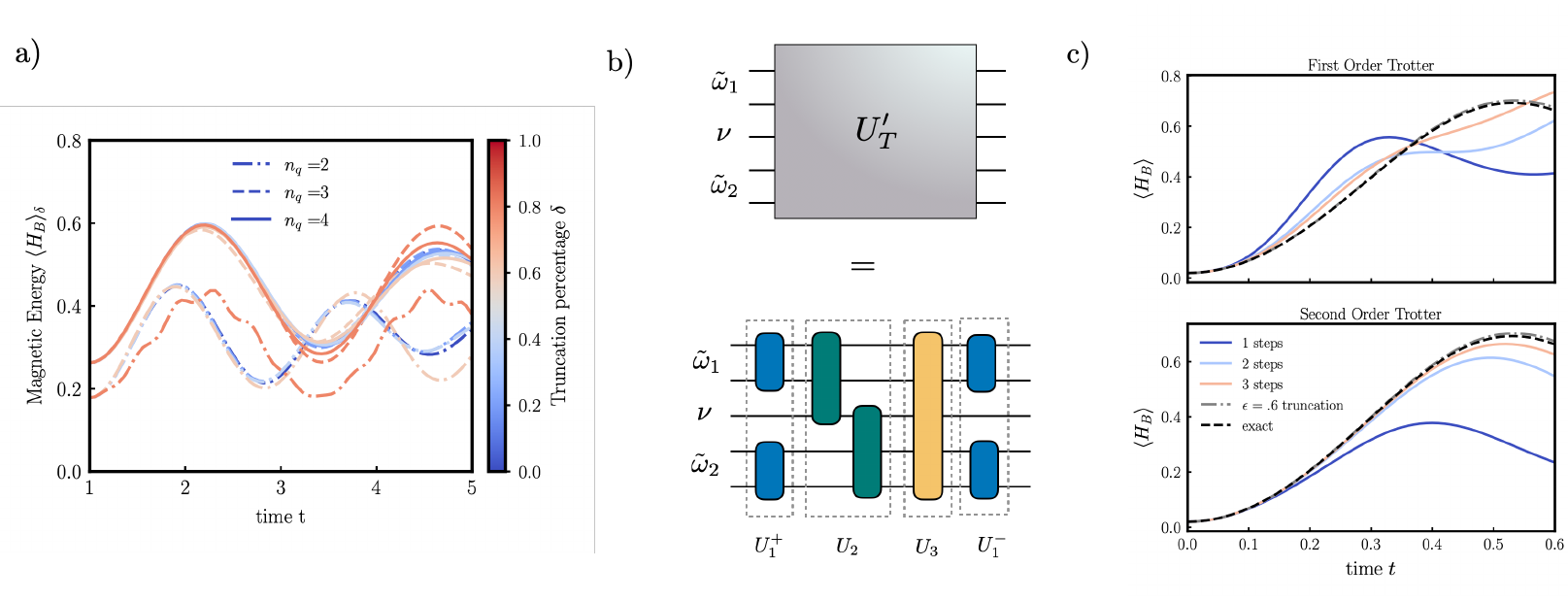}
    \caption{\textit{Approximation for Time Evolution Circuits Presented in Sec.~\ref{subsec:coarse}} \textbf{a)} The expectation value of the magnetic part of the Hamiltonian for differing digitization resolutions and percentage of terms dropped. Rapid convergence is observed for $n_q\geq 3$ and for truncation levels $\delta<.6$. \textbf{b)} Different stages of the time evolution operator for a second-order Trotter step. The silver circuit denotes the entire Trotter step. The blue sub-circuits acts only on the $\omega$ registers, typically involving the largest angles, and is identical for each. The green sub-circuits couple each $\omega$ register to the $\nu$ register and the circuit structure is again identical for both $\omega$. The yellow sub-circuit couple everything and generally contain the smallest rotations, therefore being the elements most affected by small-angle truncations. It also involves the highest weight terms, and so focusing on truncating these elements allows the circuit to be compressed at only a small cost in algorithmic error. \textbf{c)} The accuracy of first- and second-order Trotter on magnetic observables. Repeating the blue sub-circuit in the second step allows the circuit to take much larger time steps while only needing a small amount of extra gates to implement.}
    \label{fig:circuit_stuff}
\end{figure}

To understand the effect of Pauli truncation error, the system is initialized in the state $\ket{5}$ defined in Eq.~\eqref{eq:low_energy_state} and the expectation value of the magnetic part of the Hamiltonian $H_B$ is calculated. Fig~\SubFigRef{fig:circuit_stuff}{a} shows this expectation value for various levels of truncation and gauge field digitization. For a coarse field digitization of $n_q=2$, the expectation value is generally more sensitive to truncations. As the digitization resolution is increased, a larger percentage of the terms in the Pauli decomposition are able to be dropped, showing that an increasingly small fraction of the terms in the decomposition are responsible for capturing the relevant dynamics. For diagonal matrices, this can be understood via the Sequency Hierarchy  Truncation~\cite{Li:2024lrl}. The extension to non-diagonal matrices has not yet been developed, but similar arguments should hold. 

Having truncated the Hamiltonian to retain only the Pauli strings that are most relevant for implementing time evolution, the time evolution operator $U_{\delta}$ must be converted into a quantum circuit. To minimize the number of CNOTs each Trotter step contains, the Pauli rotations are grouped to model the terms that appear in the Hamiltonian. This structure is presented in Fig.~\SubFigRef{fig:circuit_stuff}{b}. The first group includes strings that only act on individual $\omega$ registers, the second group involves strings that couple a single $\omega$ register to the $\nu$ register and have a $Z$ acting on the $\nu$ register, and the third group includes strings that act on all of the qubits. The reason for classifying the rotations in this way are two-fold. First, as already mentioned, this grouping respects how the digitized Hamiltonian couples different quantum numbers, and so this structure establishes a hierarchy in the significance of how each term contributes to time evolution. The terms in group three generally have much smaller rotation angles and are the ones to be dropped first. Second, clustering terms that acts on smaller subsets of qubits makes compilation easier, as only CNOT cancellations over a subset of qubits need to be considered. For compilation the open source software package $\textbf{bqskit}$ is used~\cite{bqskit}. The circuit that implements the $i^{\rm th}$ group is defined to be $U_i$, and so the first-order product formula is $U_T=U_3U_2U_1$. As group one is restricted to weight-$n_q$ Paulis, it will generally involve less CNOTs to implement, and therefore it is advantageous to use the second-order Trotter formula $U'_T=U^-_1U_3U_2U^+_1$, where $U_1^{\pm}$ involve the same set of operations as $U_1$ with the rotation angle $c_P t$ replaced by $c_P t/2$ and $\pm$ denotes the same/reversed ordering of operations as compared to $U_1$. The explicit rotations for each term $U_1,U_2,U_3$ are recorded in App.~\ref{sec:appendixA}.

Fig.~\SubFigRef{fig:circuit_stuff}{c} shows the evolution of the magnetic energy using both first- and second-order formulas for a variable number of Trotter steps. The second-order method allows for longer steps, giving similar levels of accuracy as three steps of first-order Trotter but a notably smaller amount of CNOTs. Also shown for comparison is the exact evolution at $\delta=0.66$ levels of truncation. Even with over half of the Pauli strings dropped, for long enough times, the error will be subleading to algorithmic error due to Trotterization.

\subsection{Resource Estimates}
To count the gates needed to implement a Trotter step using the differential operator method, the three stages are considered individually. For the Laplacian term, the quantum DST circuits can be implemented using $\mathcal{O}(n_q^2)$ elementary quantum gates, where the quadratic scaling is a direct result of the use of the Quantum Fourier Transform. The diagonal momentum operator requires an additional $n_q^2$ gates, so this entire stage requires only $\mathcal{O}(n_q^2)$.

Implementing the time evolution $e^{-t\partial/\partial\omega}$ requires at most $9n_q^2-33n_q+34$ CNOT gates for $n_q\geq 3$~\cite{sato2024hamiltonian}. Using the interleaving method previously discussed, the two qubit gate count is the product of the gate counts for the two types of terms that are interleaved, e.g.\ $e^{-it\pdv{}{\omega_1}\pdv{}{\omega_2}}$ will require $(9n_q^2-33n_q+34)^2\sim\mathcal{O}(n_q^4)$. The decomposition of the other terms involving functions of $\omega_1,\omega_2$ in Eq.~\eqref{eq:h_partial_evo} will asymptotically require $\mathcal{O}(2^{2n_q+1})$ $Z$ strings that must be classically pre-computed. This situation mirrors the counts estimated for compact \uone gauge theories~\cite{Kane:2022ejm}. In this case it was found that using the Walsh basis to encode functions of the gauge variables and truncating interactions with small angles was sufficient to break the exponential scaling to polynomial, suggesting an efficient path forward to completely reducing the scaling to polynomial in $n_q$. In absence of these more specialized methods the gate count for a single Trotter step using the differential operator method will scale as $\mathcal{O}(2^{2n_q+1})$. It is once more emphasized that the only place where the exponential scaling enters is in exactly implementing the diagonal functions of $\omega_1,\omega_2$.

Turning now to the gate counts for the Pauli truncation method, even though a large percentage of the gates can be dropped to calculate time-evolved quantities up to some specified level of error, the scaling will generally be exponential in the number of qubits per plaquette. However, for large-scale simulations it may be more efficient to utilize the qubit budget for increasing the lattice volumes as opposed to increasing the resolution in field space. For instance, on a two-dimensional $(L+1)\times(L+1)$ lattice, the dominant interaction is between pairs of plaquettes. The number of such interactions is $\binom{L^2}{2}$, which is quartic in the linear dimension $L$. This suggests that the only exponential scaling is due to resolving the Hilbert space at each physical link. As already seen in the previous section, ground state properties can be resolved with precision of $10^{-3}$ already at $n_q=3$. Understanding the resource requirements for the Pauli truncation circuits up to per-mille precision is done by considering the number of two-qubit gates needed for time-evolution.

\begin{figure}[htbp]
    \centering
    \includegraphics[scale=0.35]{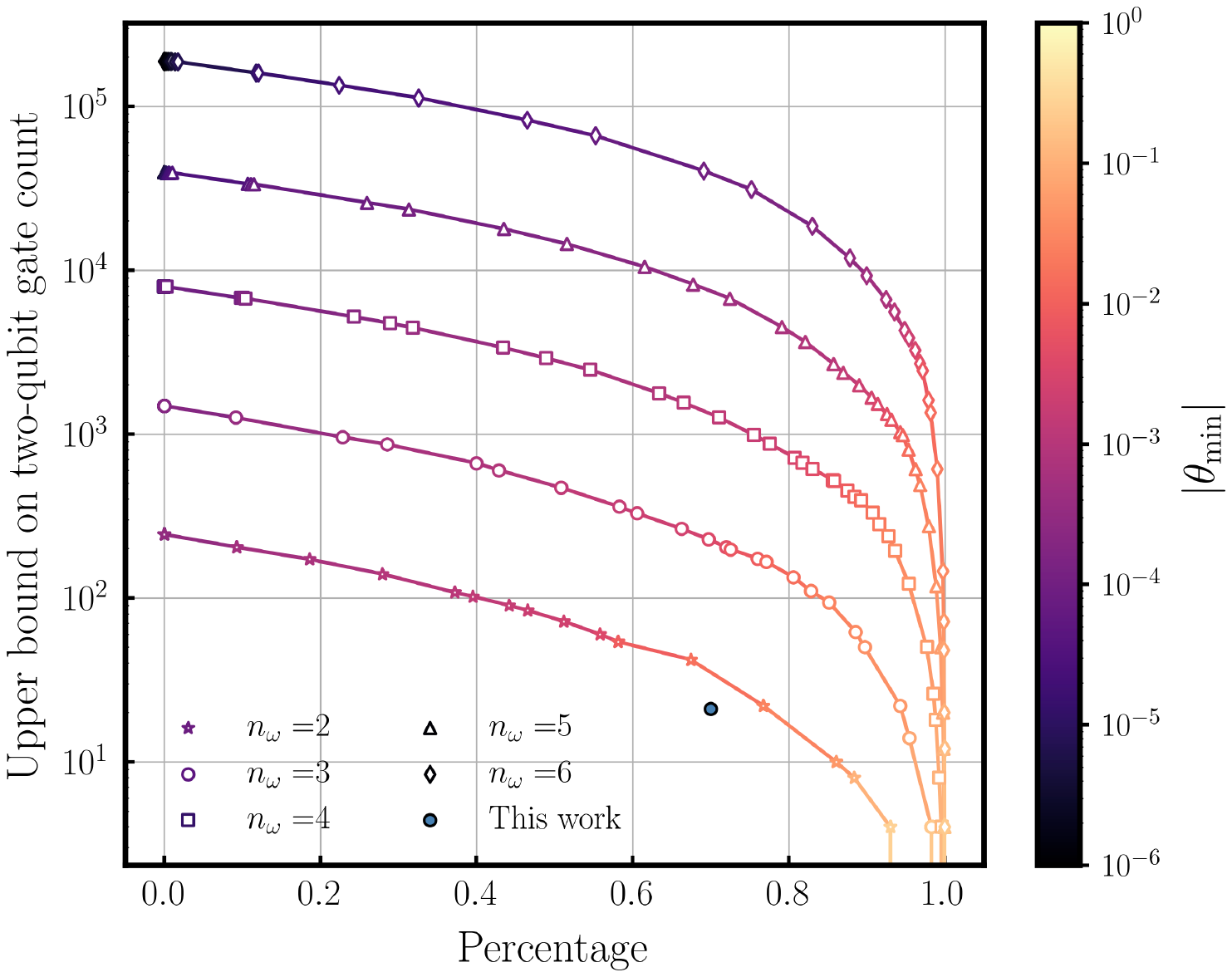}
    \caption{\textit{Upper Bounds on Resource Estimates for the Pauli Truncation Circuits} The cumulative weight of the Pauli decomposition of $\htp$ is shown as a function of the truncation percentage $\delta$. The color of the curves denotes the magnitude of the coefficient in the decomposition. The blue point demonstrates the truncation level and actual gate depth used in the quantum simulation in this work, highlighting the effect of circuit compilation.}
    \label{fig:gate_count}
\end{figure}

Analytically decomposing $\htp$ into Pauli strings for arbitrary $n_q$ is challenging, and so an upper-bound on the number of two-qubit gates is estimated by observing that every Pauli string of weight $w\geq 2$ in the Hamiltonian will require $2(w-1)$ CNOTs, as shown in Fig.~\SubFigRef{fig:fine_dig_circ}{b}. The estimate
\begin{equation}
    \#(\text{Two Qubit}) \leq \sum_{P\in\mathcal{D}(\htp)}2(w(P) - 1)
\end{equation}
is numerically calculated for digitization resolutions of $\htp$ up to $n_q=6$. Fig.~\ref{fig:gate_count} shows the upper bound as a function of the percentage of truncated strings from the Hamiltonian. This upper bound still scales exponentially in the number of qubits per plaquette, although this scaling can be improved by a factor of $2-3$ with appropriate truncation. In addition, careful compilation of the circuits can bring this number down by another factor of $\sim 2$. However, compilation is also generally a hard problem, so to compare the raw estimate to the number of post-compilation gates, the number of gates used on the quantum device in this work is also shown.

\section{Real-Time Dynamics in Two Plaquettes}
\label{sec:quantum_results}
\begin{figure}[htbp]
     \centering
     \includegraphics[scale=.4]{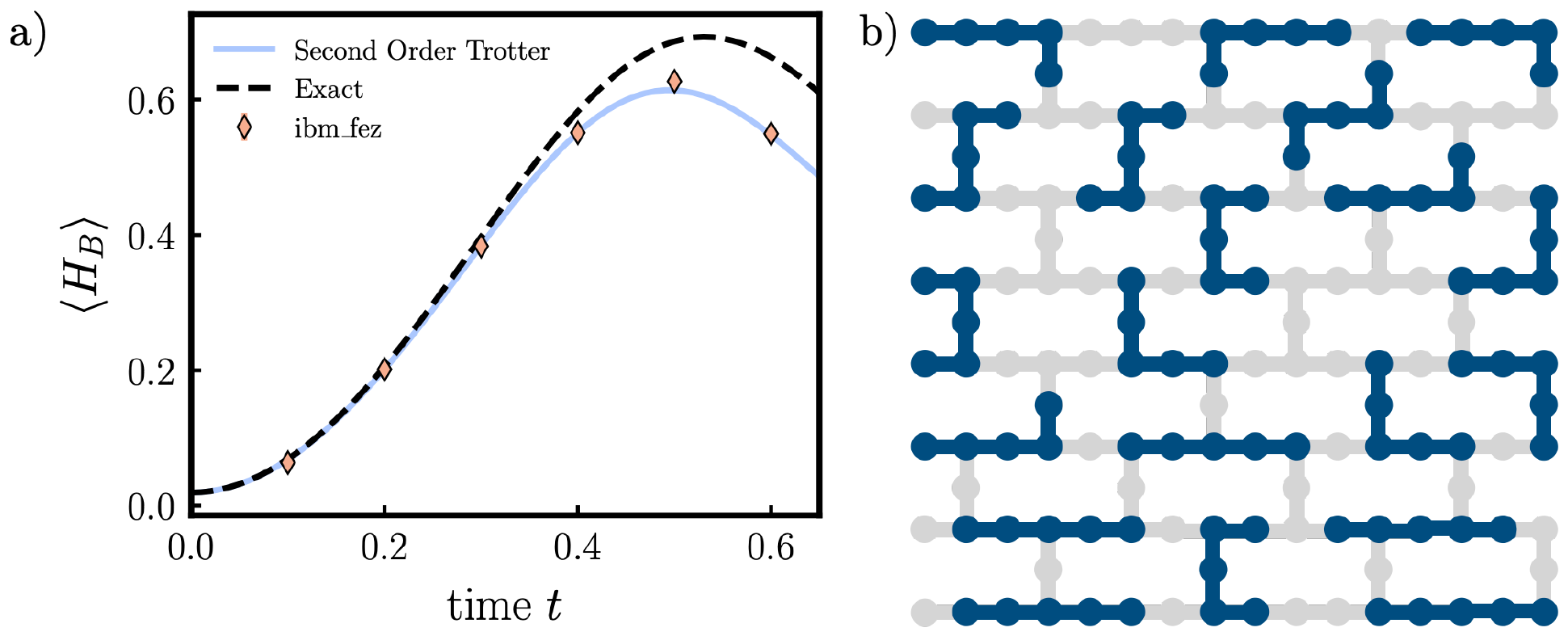}
     \caption{\textit{Quantum Simulation of $\langle H_B\rangle$ Dynamics} \textbf{a)} The time evolution due to $\htp$ at gauge coupling $g=0.5$. The dashed line corresponds to the solution obtained by directly taking the matrix exponential of the digitized Hamiltonian and the solid line is the Trotterized evolution. The points are the mitigated results from {\tt ibm\_fez}. \textbf{b)} The physical layout of the Heron chip. The highlighted regions are an example of the parallel layouts used to generate the data point at $t=0.6$.}\label{fig:quantum_results}
 \end{figure}

This section presents the results of a real-time simulation on the 156 qubit {\tt ibm\_fez} superconducting quantum chip. The circuit architecture used is the Pauli truncation approach presented in Sec.~\ref{subsec:coarse} at a truncation level of $\delta=0.66$ on five qubits ($n_q=2,n_{\nu}=1$). Preparing the example low-energy state $\ket{5}$ in Eq.~\eqref{eq:low_energy_state} requires additional circuits for state preparation. To isolate the performance of the time evolution circuits, the system is initialized in the product state $\ket{00000}$. Note that all of the states in the Hilbert space that $\htp$ acts upon satisfy Gauss' law, and so this product state is still a valid physical state. The system is then time evolved at gauge coupling $g=0.5$ and the expectation values of the magnetic part of the Hamiltonian $\langle H_B\rangle$ are measured. For the gauge coupling considered here, $\mathcal{D}(H_B)$ has the following explicit decomposition into Pauli observables:
\begin{align}
H_B = {} & 0.386\, I\otimes I\otimes I\otimes I\otimes I -0.0701\, I\otimes I\otimes I\otimes I\otimes Z \nonumber \\
     & -0.1445\, I\otimes I\otimes I\otimes Z\otimes I
       +0.0326\, I\otimes I\otimes I\otimes Z\otimes Z \nonumber \\
     & -0.0701\, I\otimes I\otimes Z\otimes I\otimes I
       -0.1448\, I\otimes Z\otimes I\otimes I\otimes I \nonumber \\
     & +0.0325\, I\otimes Z\otimes Z\otimes I\otimes I \, .
\end{align}
Fig.~\SubFigRef{fig:quantum_results}{a} shows the evolution of this observable from time $t = 0.1$ to $t=0.6$, in increments of $0.1$. To make maximum use of the device, up to 20 copies of the circuit were run in parallel at different locations on the chip, with 100 active qubits for every job. To limit the effect of cross-talk error between different parallel circuits, the layout routing was chosen such that at least one inactive qubit separated any active qubits in different subcircuits. An example of the layout used to generate the $t=0.6$ data point is shown in Fig.~\SubFigRef{fig:quantum_results}{b}.

Due to the inherent noisiness of current machines, utilizing various layers of error mitigation was crucial in obtaining these results. Once the bare circuits are created and layouts chosen, Dynamical Decoupling (DD)~\cite{Viola:1998jx,Ezzell:2022uat} was added to suppress idle errors and further mitigate the effects of cross-talk between qubits. In addition, Pauli Twirling (PT)~\cite{Wallman:2015uzh} and measurement twirling~\cite{Berg:2020ibi} are applied to each circuit. In many cases, these methods allows the dominant noise channels of the device to be approximated as a depolarizing channel. The benefit is that the impact of a depolarizing channel on observable estimates is effectively dealt with using the final mitigation technique, Operator Decoherence Renormalization (ODR)~\cite{Farrell:2023fgd,Farrell:2024fit,Urbanek:2021oej,ARahman:2022tkr}.

ODR is a very powerful error mitigation technique that allows
the raw data from the device to be `renormalized' in order to extract accurate physical results. For this method, two types of circuits are run on the quantum device. The first is the circuit for the physical process of interest, referred to as the physics circuit. The second, called the mitigation circuit, is a circuit that has the same pattern of single- and two-qubit gates, but is classically efficient to simulate. Classically simulating the mitigation circuit provides a noise-free and therefore numerically exact result to which the quantum simulation can be compared.  Assuming the dominant error channel can be approximated by sufficiently weak depolarizing noise, the quantum estimate of the observable $O$ will deviate from the true value by a multiplicative factor $p_O$, \textit{i.e.} $\langle O\rangle_{\text{meas}} = p_O \langle O\rangle_{\text{true}}$. If the physics circuit is affected by the noise in the same way as the mitigation circuit, the factors of $p_O$ between both are the same. Therefore, the renormalized estimate of the physical observable is given by
 \begin{equation}\label{eq:odr}
     \langle O_{\text{phys}}\rangle_{\text{true}} = \langle O_{\text{phys}}\rangle_{\text{meas}}\frac{\langle O_{\text{mit}}\rangle_{\text{true}}}{\langle O_{\text{mit}}\rangle_{\text{meas}}}
 \end{equation}
 where the subscripts $\textit{meas} /\textit{true}$ denote the estimate using the raw/renormalized quantum data and the subscripts $\textit{phys} /\textit{mit}$ denote the physics/mitigation circuits.

 For this technique to work, the choice of mitigation circuit is of central importance. For the smaller time steps, the physics circuits were `Cliffordized', meaning that the rotation angles of all $R_Z$ gates were rounded to the nearest $\pi / 2$. This transforms the mitigation circuit into a Clifford circuit with the same gate instructions as the physics circuit. When the time step grows large enough, the Cliffordization procedure can return a mitgation circuit that has vanishing expectation values, causing Eq~\ref{eq:odr} to become inapplicable. In this case the Trotter structure of the circuit is utilized for designing the mitigation circuit. For all but $t=0.1$, two steps of the second order formula were used, each with time step $t/2$, to reach time $t$. The gate sequence of this circuit can be rearranged so that when the second step has time step $-t/2$, the first step will be undone and the resultant state will be the initial state. As such all expectation values of Pauli $Z$ strings will be $1$ and so ODR can be used at a minimal addition to circuit depth.

 To generate the points in Fig.~\SubFigRef{fig:quantum_results}{a}, between $20$ to $60$ Pauli twirls are used with eight measurement twirls per Pauli twirl and $100$ shots per circuit. The measurement twirls are averaged together and then ODR is applied to each Pauli twirl. The resultant renormalized Pauli twirls are then bootstrap resampled to get the mean and standard error for each run. These values are reported in Table~\ref{tab:reported_values}, which also includes the depth of the circuit for each time step as well as the exact value of the Trotterized evolution. The exact value is calculated by Exact Diagonalization.
 \begin{table}[h]
     \centering
     \begin{tabularx}{\textwidth}{X X X X}
     \toprule
     Time $t$ & depth & {\tt ibm\_fez} & Target \\
     \midrule
     0.1 & 25 & 0.0637~(16) & 0.0683 \\
     0.2 & 48 & 0.2013~(41) & 0.2015 \\
     0.3 & 48 & 0.3834~(41) & 0.3846 \\
     0.4 & 51 & 0.5509~(54) & 0.5510 \\
     0.5 & 51 & 0.6269~(82) & 0.6137 \\
     0.6 & 51 & 0.549~(11) & 0.5471 \\
     \bottomrule
     \end{tabularx}
     \caption{Estimated and target (noiseless) values of $\langle H_B\rangle$ as a function of time.}
     \label{tab:reported_values}
 \end{table}

 \section{Discussions and Conclusion}
\label{sec:Conclusions}
Gauge theories are ubiquitous in descriptions of nature, appearing in the Standard Model, many condensed matter systems and theories of gravity. $\textit{Ab initio}$ calculations of their properties are vital for comparing theoretical predictions to experimental results. This is particularly true for QCD, whose strongly-coupled nature at low energies gives rise to a complex array of emergent phenomena as well as important contributions to high-energy processes. While massive theoretical and algorithmic developments in classical computing have allowed us to probe many aspects of gauge theories, there remain a plethora of open questions that do not seem amenable to these methods. For example, the determination of the phase diagram of finite-density QCD and direct, nonperturbative calculations of real-time dynamics of hadrons appear inaccessible via classical techniques due to sign and signal-to-noise problems. Even our knowledge of whether the electroweak gauge interactions in the Standard Model can be given a precise, nonperturbatively regularized definition using lattice field theory faces similar difficulties, as sign problems obstruct the verification of candidate regularizations of chiral gauge theories. 

Quantum simulations of Hamiltonian formulations of lattice gauge theories provide a powerful alternative framework for exploring strongly correlated quantum systems. There is an increased need for theoretical tools that make target systems amenable for such quantum simulations. For lattice gauge theories, a necessary step is constructing a finite-dimensional representation of the Hamiltonian, so that the continuous gauge groups of the Standard Model can be encoded into a discrete register of qubits. Ideally, this encoding can be efficiently simulated at all values of the (bare) gauge coupling and is also systematically improvable. In the last decade, many Hamiltonians for both Abelian and non-Abelian gauge theories have been constructed. The proposal put forth in Refs.~\cite{DAndrea:2023qnr,Grabowska:2024emw}, combining  gauge fixing and a basis inspired by the axis-angle representation of \sutwo, is one such promising approach for simulating real-time dynamics that have hitherto been inaccessible on classical devices.

The focus of this work was constructing the full algorithmic pipeline for simulating the real-time dynamics of the fully gauge-fixed two plaquette system with open boundary conditions, working in the mixed basis. While the explicit focus was on the two plaquette system, many of the developments are applicable to larger systems in two- and three-dimensions. The starting point was from the Hamiltonian written in the group element basis and ended with the measurement of an observable on a quantum processor. Throughout the development of this workflow, several key points were addressed.

First, this work validated that previously introduced mixed-basis truncation schemes work well for larger system sizes; an avenue for additional study is whether this scheme, particularly the optimal value of $\omega_\text{max}$ given in Eq.~\eqref{eq:OmegaMax}, will continue to provide small truncation errors as system size is increased further. An important development in this work was demonstrating how the inclusion of first-order derivatives, only present in systems with two or more plaquettes, inhibits the exponential convergence predicted by the Nyquist-Shannon theorem. In particular, approximating first derivatives with finite-differences introduces digitization errors that scale polynomially with the truncation scale. This work demonstrated that with a qubit allowance of three qubits per plaquette, the low-energy properties of the two-plaquette system could be recovered with a precision of $10^{-3}$. However, it is not clear how this precision depends on the lattice volume and therefore an important question that must be addressed in future work is whether the inclusion of first-derivative terms dramatically limits the precision achievable in this formulation. Such first derivative terms appear in many other quantum many-body systems beyond \sutwo, and developing formulations of such terms compatible with quantum simulation is an area where more developments are needed.

Second, this work presented two distinct algorithms for implementing time-evolution on a digital quantum computer. The two algorithms had different quantum resource scalings and thus had different regimes of applicability. The first, described in Sec.~\ref{subsec:DifOpCircuits}, sorted the Hamiltonian according to the highest order differential operator that each term contained. A sub-circuit was constructed for each type of term. The benefit of this algorithm is that it scales more favorably in circuit depth for a large number of qubits and gives a Trotterization scheme with less error. Additionally, it can be easily adapted for systems with a larger number of plaquettes. However, it could not be implemented on currently-available quantum processors due to its depth. The second algorithm, presented in Sec.~\ref{subsec:coarse}, decomposed the Hamiltonian directly into Pauli strings, truncating the Pauli strings that had a coefficient below some \textit{a posteriori} determined value. This resulted in a quantum circuit that was sufficiently small as to be simulated on {\tt ibm\_fez}.

Lastly, this work determined an error mitigation strategy for simulating the real-time dynamics of the two-plaquette system on NISQ-era hardware. Using previously-developed methods, namely Dynamical Decoupling, Pauli Twirling and Operator Decoherence Renormalization, were all necessary to achieve percent-level agreement between the quantum and classical results. It is not yet clear how the necessary quantum resource will scale with lattice volume and gauge coupling. Furthermore, different low-energy observables may require different error mitigation strategies.

The fully-gauged fixed two-plaquette \sutwo Hamiltonian is a very fruitful system for understanding some of the hurdles that lie ahead on the path to simulating QCD on quantum processors. It is a relatively simple system, exactly solvable using FEM, but contains both non-Abelian gauge dynamics and non-trivial interacting degrees of freedom. This work charts a course towards simulations of larger two- and three-dimensional systems while also demonstrating the viability of the mixed-basis formulation for studying the properties of \sutwo gauge theories at all values of the gauge coupling.

% ========== Chapter 4

\chapter{Sequency Hierarchy Truncation 
(SeqHT) for Adiabatic State Preparation and Time Evolution}
\label{chap:seqht}

\vspace{1em}
\begin{center}
    \itshape
    This chapter is associated with Ref.~\cite{Li:2024lrl}: 
    ``Sequency Hierarchy Truncation (SeqHT) for Adiabatic State Preparation and Time Evolution in Quantum Simulations''\\by Zhiyao Li, Dorota M. Grabowska, and Martin J. Savage
\end{center}
\vspace{2em}

%%%%%%%%%%%%%%%%
\section{Introduction}
\label{sec:intro}
\noindent
While the previous chapters contribute toward large-scale, real-time simulations of quantum chromodynamics (QCD), the scale of the simulations that can be achieved on present-day quantum hardware is constrained by the quantum resource overhead, such as the CNOT count and entangling gate depth. To mitigate these limits, it is essential to develop systematic truncation schemes that compress the resource requirement while preserving relevant physical observables. Rather than trying to build this optimization directly within the complex constraints of gauge fields, it is practical to establish our method in a simpler interacting system. This chapter uses the scalar $\lambda\phi^4$ model for that purpose, introducing a resource-efficient truncation scheme that target specific energy sectors, enabling resource savings while preserving an accurate representation of the underlying physics.

Despite recent demonstrations of systems with a small number of error-corrected logical qubits~\cite{Bravyi:2023qpn,Bluvstein:2023zmt,daSilva:2024tfc} and promising results from simulations that used more than 100 qubits~\cite{Kim:2023bwr,Yu:2022ivm,Chen:2023tfg,Shtanko:2023tjn,Baumer:2023vrf,Liao:2023eug,Farrell:2023fgd,Farrell:2024fit}, NISQ-era digital quantum computers remain limited in the quality of their entangling gates and the circuit depths that can be executed before the quantumness of the device is effectively lost~\cite{ibmquantumweb,quantinuumweb,ionqweb,queraweb}.
While current quantum computers are not yet practical for most real-world applications, research and development are progressing rapidly.
Significant effort is being placed on creating efficient and effective quantum circuits that provide results for the target observables within specified error tolerances. 
This can include simulating systems that are nearby in theory space 
(for example, a leading order Hamiltonian in a perturbative expansion about a target Hamiltonian)~\cite{Cohen:2021imf,Davoudi:2020yln,Davoudi:2022xmb}, 
developing perturbative schemes~\cite{Politzer:1973fx,Sun:2021mdd,Mitarai:2022via,Li:2022zbj,Li:2024uzc},
and identifying truncations of the Hamiltonian based upon the emergent properties of the systems such as confinement and gaps~\cite{Zohar:2012ay,Buyens:2015tea,Magnifico:2019kyj,Honda:2021aum,Mildenberger:2022jqr}.
One major recent advance in the simulation of lattice gauge theories is the use of confinement to truncate operator contributions based on spatial separation combined with operators that are scalable to arbitrarily large numbers of spatial sites~\cite{Farrell:2023fgd,Farrell:2024fit}. Error mitigation is essential for recovering meaningful results from quantum computers due to decoherence especially for deeper circuits~\cite{Bennett:1996gf,Viola:1998jx,Viola:1998gg,Dur:2005obk,Emerson2007Symmetrized,Dankert2009Exact,Souza_2012,Li:2016vmf,Temme:2016vkz,Martinez:2016yna,RevModPhys.88.041001,Endo2018Practical,Klco:2019evd,Kandala2019Error,He:2020udd,Tran:2020azk,Zhang:2021lzr,Nguyen:2021hyk,Urbanek:2021oej,ARahman:2022tkr,Leyton-Ortega:2022xfu}.
Results obtained from analog devices also require careful error-mitigation analyses~\cite{Childs:2019rty,Childs:2019hts,Georgopoulos:2021fyi,Flannigan:2022bwi,Cai:2022rnq,Zemlevskiy:2023eyw}.

The formal aspects of using (ideal) quantum computers to simulate $\lambda\phi^4$ theory have been established in the pioneering papers by Jordan, Lee and Preskill (JLP)~\cite{Jordan:2012xnu,Jordan:2011ci,Jordan:2014tma,Jordan:2017lea}. %
These papers developed all of the necessary tools for determining S-matrix elements, and have shown that the problem is BQP-complete (when classical background sources are included).
This last point means that any system that can be simulated efficiently on a quantum computer can be mapped (with polynomial-scaling resources) to this particular scalar field theory. 
A number of subsequent works~\cite{Somma:2015bcw,Klco:2018zqz,Macridin:2018oli,Barata:2020jtq,Macridin:2021uwn,YeterAydeniz2020Imaginary}
further examined the theory and identified further digitization possibilities, including using a harmonic oscillator basis to encode the scalar field as opposed to the eigenstates of the field operator. 
JLP showed that a judicious choice of conjugate-momentum operator implemented via a local quantum Fourier transform (QFT) exponentially suppresses digitization errors, with corrections to the continuum suppressed by factors of $\sim e^{-\alpha/\delta_\phi^2}$ where $\alpha$ is a digitization-independent factor and $\delta_\phi$ is the sampling interval in field space. 
Furthermore, JLP proposed using adiabatic state preparation (ASP)  to prepare the vacuum and wavepackets in the interacting theory; this method requires slowly evolving from the ground state of the free theory, with $\lambda = 0$, to non-zero $\lambda$ along a Trotterized trajectory in $\lambda(s)$. 
Using currently available devices, implementing this algorithm (even) on small systems results in states with relatively large errors compared to the target state. 
While methods continue to  become available for preparing the ground state and optimizing time evolution circuits, including variational quantum eigensolver (VQE)~\cite{Peruzzo:2013bzg,McClean2016Theory,Kandala:2017vok,Ferguson:2020qyf,Atas:2021ext,Farrell:2022wyt,Farrell:2022vyh,Cao:2024yew},
ADAPT-VQE and more (for example Refs.~\cite{Tang:2019tpm,Grimsley_2019,Ciavarella:2021lel,Farrell:2024fit,Farrell:2023fgd, Kaplan:2017ccd, Kokail:2018eiw, Roggero:2019myu,Ciavarella:2022qdx,Kane:2023jdo,Keever:2023gqo,Smith:2024nbh}), state preparation remains a generically challenging problem that nominally lies outside of the BQP complexity class. % 
This means that algorithms that can efficiently prepare states (within a specified fidelity) are important to identify, and will likely be of general utility.

In this chapter, we build upon these advances to identify a new convergent truncation that reduces quantum resources required for state preparation and time evolution of smooth and bounded wavefunctions.  
This truncation is guided by a hierarchy in the contributions to low-energy observables from basis operators based upon their sequency~\cite{Klco:2021jxl}.
We call this scheme  Sequency Hierarchy Truncation (SeqHT), and demonstrate its utility in preparing the ground-state wavefunction of $\lambda\phi^4$ interacting scalar-field theory. 
Because the implementation of SeqHT will be iterative in many setting, for example systematically increasing the cutoff sequency until changes in observables fall below a predetermined threshold, it should be considered a hybrid quantum-classical algorithm.
We derive rigorous upper bounds on the contributions of operators with regard to their sequency, and study the convergence of observables and quantum computational complexity (magic) as a function of increasing cutoff sequency. 
We construct the quantum circuits for adiabatically preparing the ground state of $\lambda\phi^4$, utilizing sequency truncation and JLP's method of starting from the non-interacting theory. 
An important element of the circuit development is that 
sequency-ordered operators allow for (maximum) CNOT-gate cancellations in 
implementing diagonal unitaries in preparing the non-interacting ground state~\cite{Welch_2014,Klco:2021jxl, Kane:2022ejm}.\footnote{For the time evolution circuit, 
overlapping $ZZ$ gates requires fewer CNOT gates than arranging them in sequency order.}
These circuits are then executed, along with their partner mitigation circuits, using IBM's superconducting-qubit quantum computer {\tt ibm\_sherbrooke}~\cite{ibmquantumweb}. 
The results of these simulations 
clearly demonstrate the utility of SeqHT  for improving the fidelity of adiabatically-prepared states (and evolution) because of the reduced quantum resource requirements. 
While the results we present here are compared with the analogous results obtained using classical computing methods, the actual utility of the method is on preparing states that cannot be accessed using these classical approaches.

While it is important to advance quantum simulations of scalar field theories, 
a more important objective for our work is to reduce the quantum resources required to simulate lattice gauge theories, both Abelian and non-Abelian. 
The wavefunction of the gauge-field is expected to be smooth, and localized near the scale of the 
physics of interest (after renormalization). 
As such, we anticipate that SeqHT will also have utility in quantum simulations of (2+1)\,D and (3+1)\,D quantum chromodynamics (QCD)~\cite{MILC:2009mpl,Wiese:2014rla,Lamm:2019bik,Yamamoto:2020eqi,Ciavarella:2021nmj,Homeier:2022mkg,Muller:2023nnk,Ciavarella:2023mfc,Turro:2024pxu}.

%%%%%%%%%%%%%%%%%%%%%%%%%%%%%%%%%%%%%%%%%%%
\section{Sequency Hierarchy of Operators and Wavefunctions}
\label{sec:seqops}
\noindent
Hermitian operators in their own eigenbasis are represented by real diagonal matrices 
which can be decomposed into an orthonormal basis of tensor products of the $2\times 2$ identity matrix, $\hat{I}$, 
and Pauli matrix $\hat{Z}$, in the form of 
\begin{equation}
\hat{\CO}=\sum_{\nu=0}^{2^{n_q}-1} \beta_\nu \hat{\CO}_\nu \quad, \quad \hat{\CO}_\nu=\bigotimes_{j=0}^{n_q-1} \hat{\sigma}_{[j]} = \hat{\sigma}_{[0]} \otimes \hat{\sigma}_{[1]} \otimes \dots \otimes \hat{\sigma}_{[n_q-1]}   
\quad, \quad
\label{eq:decomp}
\end{equation}
where $n_q$ is the number of qubits and each $\hat{\sigma}_{[j]}$ is either $\hat{I}$ or $\hat{Z}$ acting on qubit $j$.
The subscript $\nu$ is the sequency index that uniquely labels each basis operator, and $\beta_\nu$ are the sequency coefficients 
(sequency and sequency indexing will be explained later in this section). 
Since diagonal matrices commute, 
the time-evolution operator $e^{-i\hat{\CO}t}$ can be decomposed as
\begin{align}
e^{-i\hat{\CO}t} = \prod\limits_{\nu=0}^{2^{n_q}-1} e^{-i\beta_\nu\hat{\CO}_{\nu}t} \, .
\end{align}
Similarly, digital wavefunctions can be decomposed as 
\begin{equation}
\label{eq:wavedecomp}
|\psi\rangle=\sum_{\nu=0}^{2^{n_q}-1} \alpha_{\nu}|\psi_{\nu}\rangle \quad , \quad |\psi_{\nu}\rangle = \bigotimes_{j=0}^{n_q-1} |\ell_{[j]}\rangle = |\ell_{[0]}\rangle \otimes |\ell_{[1]}\rangle \otimes \dots \otimes |\ell_{[n_q-1]}\rangle  \quad ,
\end{equation}
where the normalization constant for basis states $|\psi_{\nu}\rangle$ is omitted, $\alpha_{\nu}$ are the sequency coefficients and $|\ell_{[j]}\rangle$ is either the column vector $(1,1)$ or $(1,-1)$ (the diagonal of $\hat{I}$  or $\hat{Z}$) for the $j$th qubit. 

The diagonals of the basis operators $\hat{\CO}_\nu$ in Eq.~\eqref{eq:decomp} and the basis vectors $|\psi_{\nu}\rangle$ in Eq.~\eqref{eq:wavedecomp} are Walsh functions \cite{a0fadc48-fe74-34f3-99da-fc82ca467d96}, which form a complete and orthogonal set of functions; the diagonal of $\hat{\CO}_\nu$ corresponds to the $\nu^\text{th}$ row of a sequency-ordered Walsh-Hadamard matrix. 
This matrix can be constructed by rearranging the rows of the natural-ordered Walsh-Hadamard matrix, which can be easily constructed via a tensor product:
\begin{equation}
\label{eq:WHT}
\mathrm{H}_{n_q}=\mathrm{H}^{\otimes n_q} \quad, \quad \mathrm{H}=\left(\begin{array}{cc}
1 & 1 \\
1 & -1
\end{array}\right)
\ .
\end{equation}
This Walsh series representation of discrete functions is the digital counterpart of the Fourier series and have an important role in digital signal processing~\cite{Shukla:2023oub}. 
The harmonics constituting a Fourier series can each be indexed by an integer $n$ that uniquely corresponds to a frequency given by $\omega_n = n$/Period. 
Similarly, Walsh basis operators $\CO_\nu$ and basis states $|\psi_{\nu}\rangle$ have a bijection to sequency, the digital analog of frequency, denoted as $\nu$. 
Sequencies of a digital function can be obtained by counting the 
number of zero crossings (the number of times the sign of a function changes) on their diagonals, so $\nu \in \mathbb{N}$. 
Since there is a parallel between the sequency decomposition of $\hat{\CO}_\nu$ and $|\psi_{\nu}\rangle$, we will focus the rest of the discussions in this section on $\hat{\CO}_\nu$, but we note that the discussed properties apply to both (with the exception of circuit constructions for unitary evolutions $e^{i\beta\hat{\CO}_\nu}$). %

In such a sequency expansion, the physics of low-energy configurations lead to a finite range of support in frequency (sequency) space, resulting in a hierarchy in the sequency coefficients that can be leveraged in quantum simulation. 
Basis operators with coefficients below a certain threshold, $\Lambda_{\rm cut}$, can be truncated to reduce the circuit size for implementing a target unitary.
In earlier approaches, either classical resources were utilized to  calculate all Walsh coefficients to determine which were below the desired $\Lambda_{\rm cut}$~\cite{Kane:2022ejm}, or a mix of quantum and classical resources were used to iteratively tune target observables to ascertain convergence below the desired error threshold~\cite{Klco:2021jxl}.
As the classical computing resources required for decomposing target operators scale exponentially with system size, large-system simulations remained impossible without a effective method to select operators for truncation. 
Sequency hierarchy provides insight to this problem. 
As will be shown later in this chapter, the value of the coefficient $\beta_\nu$ can be bounded from above when considering its contribution to any polynomial operator that appears in the Hamiltonian; this also then bounds from above $\beta_\nu$ for any operator that can be written as a power expansion of polynomial operators. This upper bound decreases for operators with increasing sequency, exhibiting a sequency hierarchy.
This allows the determination of a cutoff in sequency space $\nu_{\rm cut}$ that respect the threshold $\Lambda_{\rm cut}$ without calculating the whole decomposition. 
Thus, for large systems, it is viable to construct only a select set of low-sequency operators, obtain their coefficients (by projecting the target operator onto this basis), and implement a reduced pool of quantum circuits.

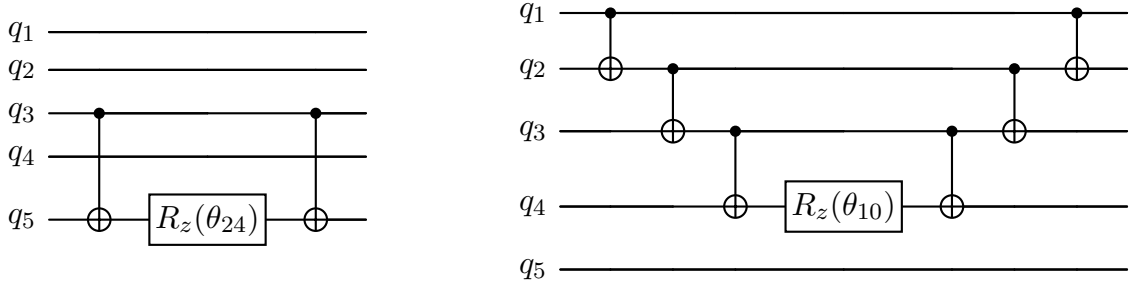
\begin{figure}[htpb]
\centering
\[
\begin{quantikz}[scale=1.1, transform shape]
    \lstick{$q_1$} & \qw & \qw & \qw & \qw \\
    \lstick{$q_2$} & \qw & \qw & \qw & \qw \\
    \lstick{$q_3$} & \ctrl{2} & \qw & \ctrl{2} & \qw \\
    \lstick{$q_4$} & \qw & \qw & \qw & \qw \\
    \lstick{$q_5$} & \targ{} & \gate{R_z(\theta_{24})} & \targ{} & \qw 
\end{quantikz}
\hspace{50pt}
\begin{quantikz}[scale=1.1, transform shape]
    \lstick{$q_1$} & \ctrl{1} & \qw & \qw & \qw & \qw & \qw & \ctrl{1} & \qw \\
    \lstick{$q_2$} & \targ{} & \ctrl{1} & \qw & \qw & \qw & \ctrl{1} & \targ{} & \qw \\
    \lstick{$q_3$} & \qw & \targ{} & \ctrl{1} & \qw & \ctrl{1} & \targ{} & \qw & \qw \\
    \lstick{$q_4$} & \qw & \qw & \targ{} & \gate{R_z(\theta_{10})} & \targ{} & \qw & \qw & \qw \\
    \lstick{$q_5$} & \qw & \qw & \qw & \qw & \qw & \qw & \qw & \qw
\end{quantikz}
\]
\caption{Example circuits for Walsh basis operators $\hat{\CO}_{24}$ (left) and $\hat{\CO}_{10}$ (right) where $24$ and $10$ are the sequency indices. 
Note that $q_1$ is the most significant qubit and $q_5$ is the least; $R_z$ is a single-qubit rotation gate about the Z axis.
}
\label{fig:cir}
\end{figure}

Quantum circuits implementing $e^{i\beta t \hat{\CO}_\nu}$ are required for 
time evolution induced by sequency operators. 
Two sequency-adjacent basis operators $\hat{\CO}_\nu$ and $\hat{\CO}_{\nu+1}$ differ in only one $\hat{\sigma}_{[j]}$. 
One can be transformed into another by flipping the $\hat{\sigma}_{[j]}$ on the most significant qubit that was not updated in the transformation from $\hat{\CO}_{\nu-1}$ to $\hat{\CO}_{\nu}$.
This resembles the Gray Code, a binary encoding protocol where two adjacent values are only one bit flip away. 
In fact, the locations of the Pauli $\hat{Z}$ operators in $\hat{\CO}_\nu$ correspond to the $1$'s in the bit-reversed Gray binary encoding for its sequency. 
A given sequency's corresponding tensor-product operator can be constructed by reversing the bits of the Gray code representation of the sequency and mapping $0 \rightarrow \hat{I}$ and $1 \rightarrow \hat{Z}$. 
Note that the more significant digits in the Gray code correspond to the less significant qubits in the computational basis (the rightmost bit in the big-Endian notation). 
This correlation can be leveraged to construct operators of any desired sequency. 
For example, the gray code for the integer 24 is $10100$, and the corresponding Hermitian sequency operator $\hat{\CO}_{24}$ is $\hat{I}\otimes \hat{I}\otimes \hat{Z}\otimes \hat{I}\otimes \hat{Z}$; the corresponding unitary time-evolution operator for this observable is $e^{-i\beta_{24} t\hat{Z}_3\hat{Z}_5}$. Similarly, for the operator $\hat{\CO}_{10}$ implemented on five qubits, the 5-digit gray code is $01111$, which corresponds to $\hat{Z} \otimes \hat{Z} \otimes \hat{Z} \otimes \hat{Z} \otimes \hat{I}$.
The circuits for these examples are shown in Fig.~\ref{fig:cir}, where $R_z(\theta) = exp(-i\theta \hat{Z}/2)$ and $\theta = 2\beta t$. 
Note that in this convention, $q_1$ is the most significant qubit and $q_5$ is the least. %

Sequency is independent of the number of qubits from the perspective of constructing a sequency operator for systems of different sizes. % 
When the number of qubits is increased, an operator with the same sequency is acquired by adding $I$s to the less significant qubits of the operator, a procedure analogous to adding $0$s to the bigger digits of Gray code. 
Since they are merely equivalent operators in a more precise system, the few lowest sequency operators still captures the bulk of low frequency information when the system size increases, and higher sequency operators can be neglected when studying low energy behaviors. % 

%%%%%%%%%%%%%%%%%
\section{\texorpdfstring{Truncated Adiabatic State Preparation for $\lambda\phi^4$ Scalar Field Theory}{Truncated Adiabatic State Preparation for lambda phi4 Scalar Field Theory}}
\label{sec:TASP}
\noindent
This section introduces the SeqHT scheme, which utilizes sequency and its associated hierarchy to reduce resource requirements for quantum simulations. 
To demonstrate the utility
of SeqHT, it is applied to adiabatic state preparation of $\lambda \phi^4$ theory.
The Hamiltonian of this target theory, a real massive scalar field with quartic self interactions ($\phi^4$ theory) in one spatial dimension, is given by 
\begin{equation}
\hat H^{\text {latt.}}=\sum_{k} \frac{1}{2} \hat{\Pi}_k^2 + \frac{1}{2} m^2 \hat{\phi}_k^2 - \frac{1}{2}\hat{\phi}_k\nabla^2\hat{\phi}_k+\frac{\lambda}{4 !} \hat{\phi}_k^4
\ ,
\label{eq:Hami}
\end{equation}
where $\hat{\phi}$ is the field operator, $\hat{\Pi}$ the conjugate momentum with a commutation relation $\left[\hat{\phi}_j, \hat{\Pi}_k\right]=i \delta_{j k}$, $k$ denotes the lattice site, and $\lambda$ the coupling constant. 
The potential energy term $\hat{\phi}^4$ generates self-interactions. 

As first discussed in the papers by Jordan, Lee and Preskill~\cite{Jordan:2012xnu,Jordan:2011ci,Jordan:2014tma,Jordan:2017lea}, state preparation of this theory on quantum devices can be accomplished by first initializing the ground state of the exactly solvable non-interacting theory ($\lambda = 0)$ and then adiabatically turning on interactions through unitary evolutions, with $\lambda$ slowly increasing until it reaches the desired value. 
This process will evolve the free theory ground state into the ground state of the interaction theory, providing that the theory has a non-vanishing mass gap throughout the evolution. 

For the purposes of demonstration, SeqHT is applied to adiabatic state preparation of $\phi^{4}$ theory on one spatial site and $m$ is taken to be $1$. 
Effective truncation and digitization are required to map this theory to the registers of a digital quantum computer. 
In this chapter, both $\hat{\phi}$ and $\hat{\Pi}$ are symmetrically digitized~\cite{Klco:2018zqz}, with a field cutoff $\phi_{\rm max}=4$, with further details on these choices explained in App.~\ref{appen:dig}. 
$H_{\hat{\phi}}$ and $H_{\hat{\Pi}}$ are both diagonal matrices in their respective eigenbasis and can both be decomposed into Walsh basis operators. 
For this system, SeqHT becomes effective  when the system size 
is increased to five qubits per site, where the $\hat{\phi}^4$ term can be truncated based on sequency hierarchy while preserving proper representation of the system. 
For fewer than five qubits, the simulation is too coarse to be truncated further. 

There is a finite range of $\lambda$ for which the theory 
is suitable to be simulated in this manner. 
When $\lambda$ becomes large, the wavefunction becomes increasingly sharply peaked. 
Eventually it approaches a delta function, which cannot be effectively represented. 
This could potentially be resolved by dynamically rescaling the field cutoff and qubit mapping for better representation of the features, in other words, zooming in on the nonzero region of the  evolved state. 
This paper works with $\lambda = 10$, which generates a ground-state wavefunction in the interacting theory that can be reasonably represented without rescaling $\phi_{\rm max}$. 
%
%%%%%%%%%%%%%%%%%
%
\begin{figure}[htpb]
    \centering
    \includegraphics[width=0.5\linewidth]{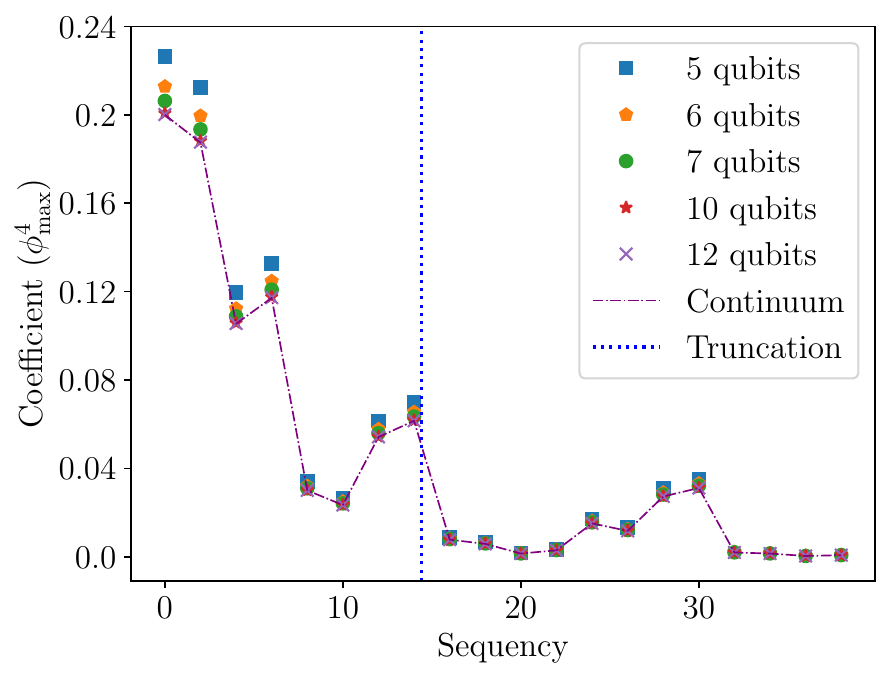}
    \caption{
The coefficients of Pauli strings contributing to the $\hat{\phi}^4$ operator, which is digitized with increasing  $n_q$.
Each Pauli string has a well-defined sequency, 
and operators with sequency below $40$ are displayed. 
(Note that for $n_q=5$, $\hat{\phi}^4$ only decomposes 
into operators with sequency below $32$.) 
The coefficients are calculated by $Tr(\hat{\phi}^{4\dag} \hat{\CO}_\nu)$. 
Values for the continuum limit are connected by a dot-dashed line for display purposes. 
The  numerical values used in the subsequent analysis implement a $\nu_{\rm cut}= 14$ sequency truncation, and the values of the results displayed in this figure can be found in Table~\ref{tab:coefficients}. }
    \label{fig:coeffs}
\end{figure}

On five qubits or more, SeqHT can be performed for the 
$\lambda\hat{\phi}^4$ interaction term. 
Since $\hat{\phi}^4$ is symmetric, it is decomposed into symmetric Walsh basis operators, which are associated with even sequencies because symmetric functions have an even number of zero crossings. 
The even sequency terms correspond to Pauli strings with an even number of $\hat{Z}$. Moreover, since $\hat{\phi}^4$ is constructed by multiplication of digitized field operator $\hat{\phi}$, it only consists of two-body operators and four-body operators other than identity, where an $n$-body operator refers to an $\hat{\CO}_\nu$ with $n$ appearances of $\hat{Z}$. This operator structure, where only $n$-body operators are involved with $n$ not greater than the order of the interaction, is a feature of the JLP basis and extends to larger number of qubits. 
After SeqHT, the remaining set of operators still construct a symmetric Hamiltonian. 
Because of this, the symmetries of the ground states prepared with adiabatic evolution under the SeqHT Hamiltonian are also preserved despite errors in adiabaticity and truncation.  

\begin{figure*}[htpb]
\centering
\begin{subfigure}[b]{0.45\textwidth}
            \centering
            \includegraphics[width=\textwidth]{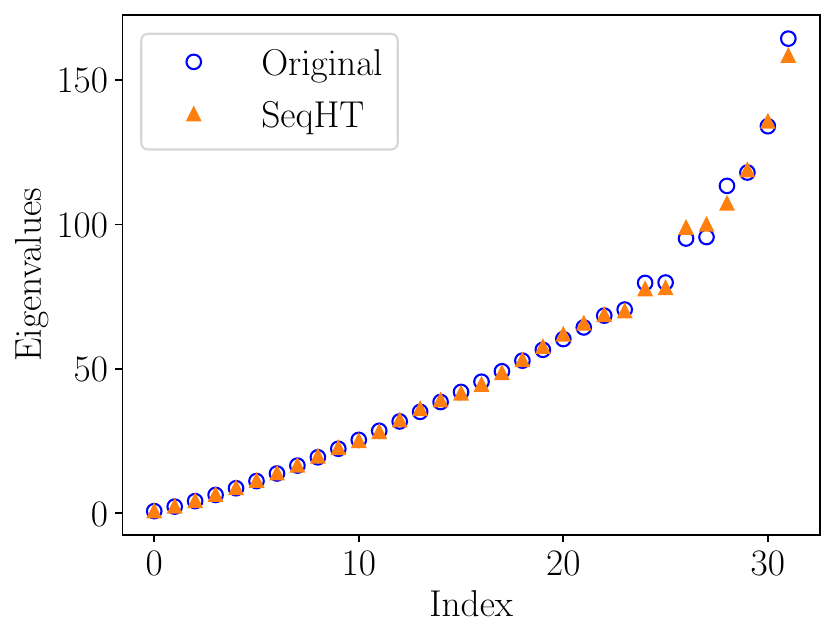}
\end{subfigure}
\begin{subfigure}[b]{0.45\textwidth}  
            \centering 
            \includegraphics[width=\textwidth]{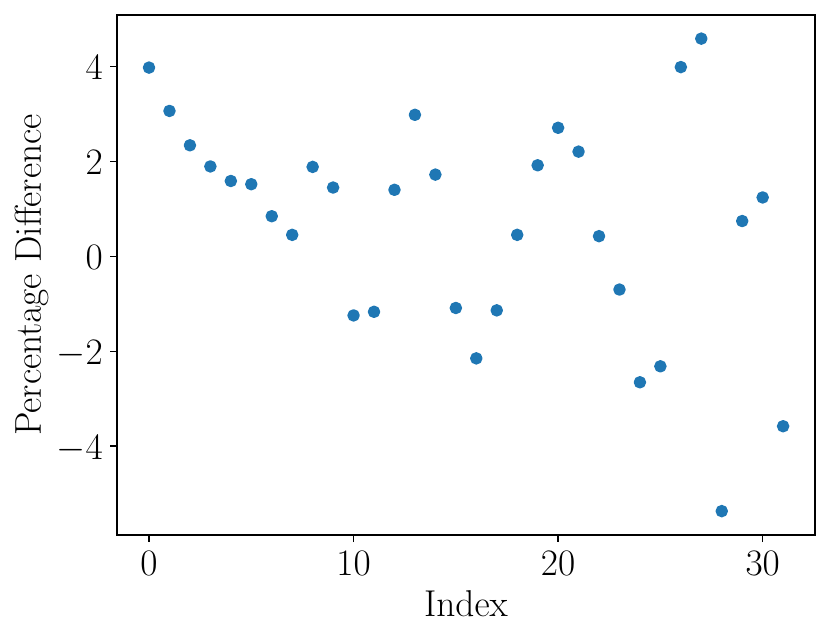}
\end{subfigure}
\caption{The left panel shows the eigenvalues of full $\phi^4$ theory Hamiltonian and the SeqHT Hamiltonian, both of which are digitized on $n_q = 5$ with $\phi_{\rm max} = 4$ and $\lambda = 10$.
The right panel shows the percentage differences between the two sets of eigenvalues. 
Numerical values for the results displayed in this figure can be found in Table~\ref{tab:eigenvalues}.}
\label{fig: eigens}
\end{figure*}

\begin{figure*}[htpb]
        \centering
        \begin{subfigure}[b]{0.45\textwidth}
            \centering
            \includegraphics[width=\textwidth]{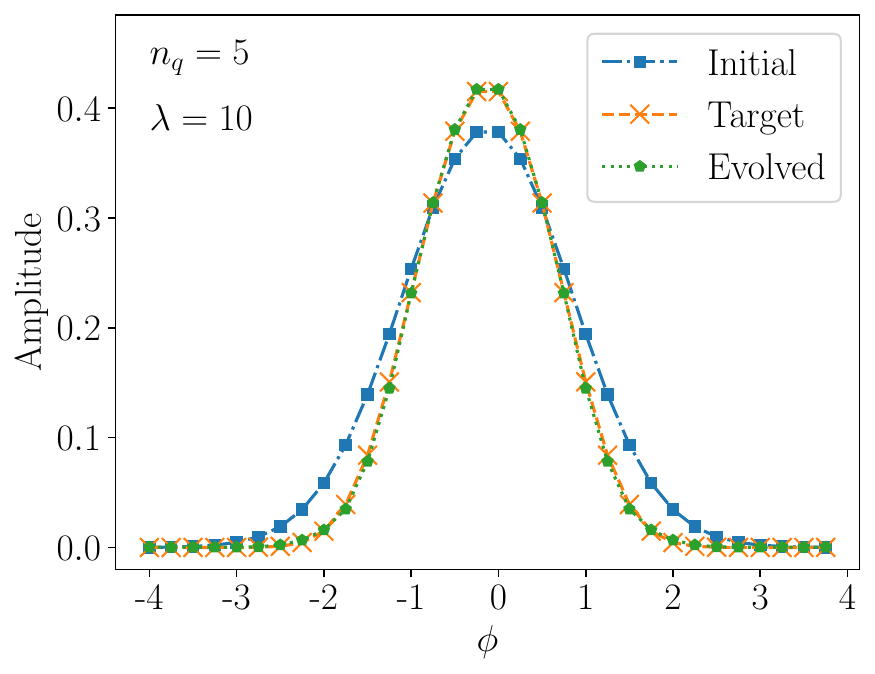}
        \end{subfigure}
        \begin{subfigure}[b]{0.45\textwidth}  
            \centering 
            \includegraphics[width=\textwidth]{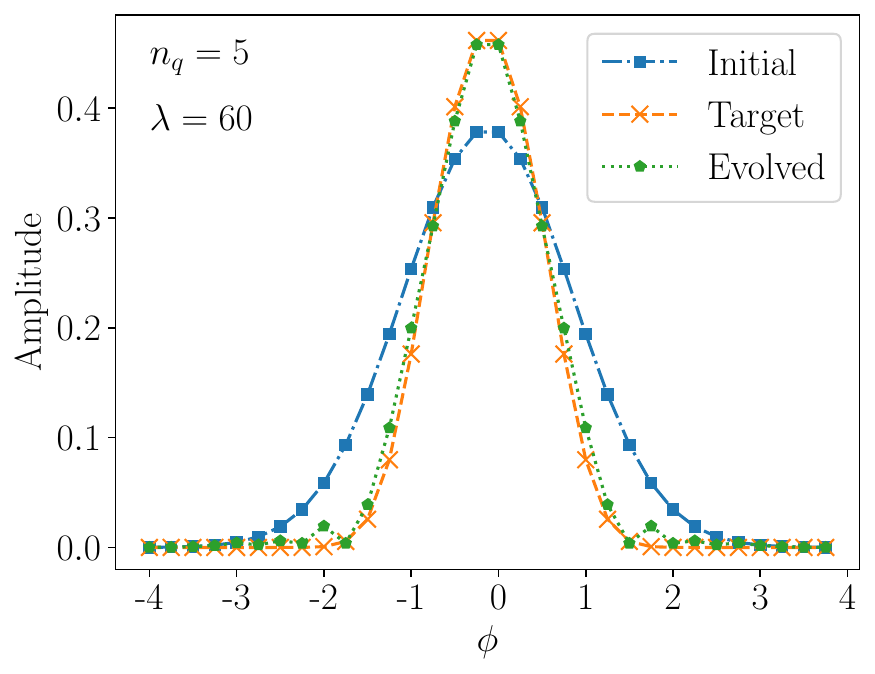}   
        \end{subfigure}
\caption{Amplitudes of the initial ground state, the target ground state of the interacting theory, and the state prepared via the SeqHT ASP procedure which uses five adiabatic steps (not trotterized), with total time $t=5$. 
In the left panel, the model uses $\lambda = 10$ and achieves a fidelity of $0.9999$. The prepared state is sufficiently close to the target state that the curves essentially coincide. In the right panel, $\lambda = 60$ and fidelity $0.9978$. 
The $\phi^4$ theory Hamiltonian is digitized onto five qubits 
($n_q=5$)
with a $\phi_{\rm max} = 4$. 
Numerical values for the results displayed in this figure can be found in Table~\ref{tab:gs_5qubits}.} 
        \label{fig:amp}
    \end{figure*}

\begin{figure}[htpb]
    \centering
    \includegraphics[width=0.45\linewidth]{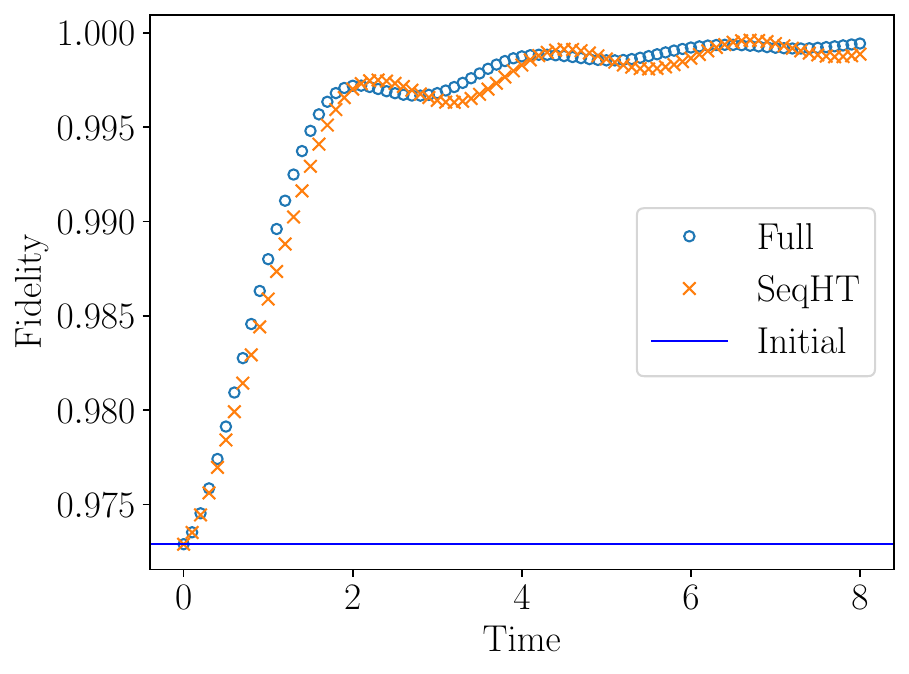}
    \caption{The fidelity of the $\lambda\phi^4 $ ground-state wavefunction prepared with complete adiabatic evolution (blue circles) and with SeqHT procedure (orange crosses). 
The Hamiltonian is digitized on five qubits with $\phi_{\rm max} = 4$ and $\lambda = 10$. 
Size of each time step is fixed at $\delta t = 0.1$ and number of steps is taken from $0$ to $80$ for a total time from $0$ to $8$. 
The horizontal line indicates the initial overlap ($0.9729$) of the free theory and the interacting theory ground states. 
Numerical values in this plot can be found in Table~\ref{tab:fidelity_time}.}
    \label{fig:time env}
\end{figure}

Figure~\ref{fig:coeffs} presents  the coefficients of basis operators, in sequency order, obtained from decomposing $\hat{\phi}^4$. 
For larger numbers of qubits, the coefficients of sequency operators start converging, and a general trend of decreasing coefficients for operators of increasing sequencies is observed; further analysis of this observation will be discussed in Sec.~\ref{sec:largenQ}. 
The cutoff sequency index $\nu_{\rm cut}$ is defined such that basis operators with sequency greater than $\nu_{\rm cut}$ are truncated. Subsequent analysis employs $\nu_{\rm cut} = 14$ for $\hat{\phi}^4$. In quantum simulation, the identity term with $\nu=0$ can also be ignored as it only impacts the global phase. 
For a sufficiently small system, the digitized Hamiltonian can be
exactly diagonalized, with numerically determined eigenvalues and eigenvectors that can be used to check the validity of the truncation scheme.
In the five-qubit case, when $\nu_{\rm cut}$ is set as indicated in Fig.~\ref{fig:coeffs}, the eigenvalues of the SeqHT Hamiltonian are found to deviate from those of the 
full Hamiltonian at the $5\%$-level, shown in Fig.~\ref{fig: eigens}.
Figure~\ref{fig:amp} shows the ground-state wavefunction of the free theory (the initial state), the state prepared via the SeqHT ASP procedure with total time $t=5$ and five adiabatic steps, and the exact ground state (determined by numerical diagonalization) of the interacting theory. 
For larger $\lambda$, the same SeqHT parameters achieve a slightly worse overlap, which can be compensated for by increasing the total time and number of steps.
The fidelity of the state preparation procedure is determined by the overlap of the adiabatically-prepared wavefunction with the target ground-state wavefunction. 
The fidelity of states prepared with full adiabatic time evolution and SeqHT evolution are presented as a function of time in Fig.~\ref{fig:time env}.

\begin{figure*}[htpb]
        \centering
        \begin{subfigure}[b]{0.45\textwidth}
            \centering
            \includegraphics[width=\textwidth]{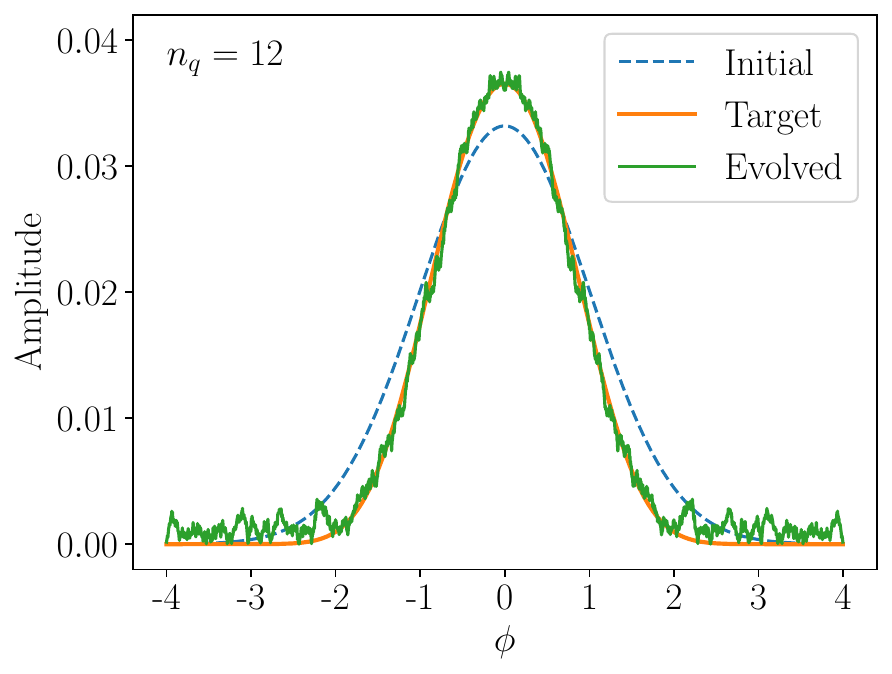}
        \end{subfigure}
        \begin{subfigure}[b]{0.45\textwidth}  
            \centering 
            \includegraphics[width=\textwidth]{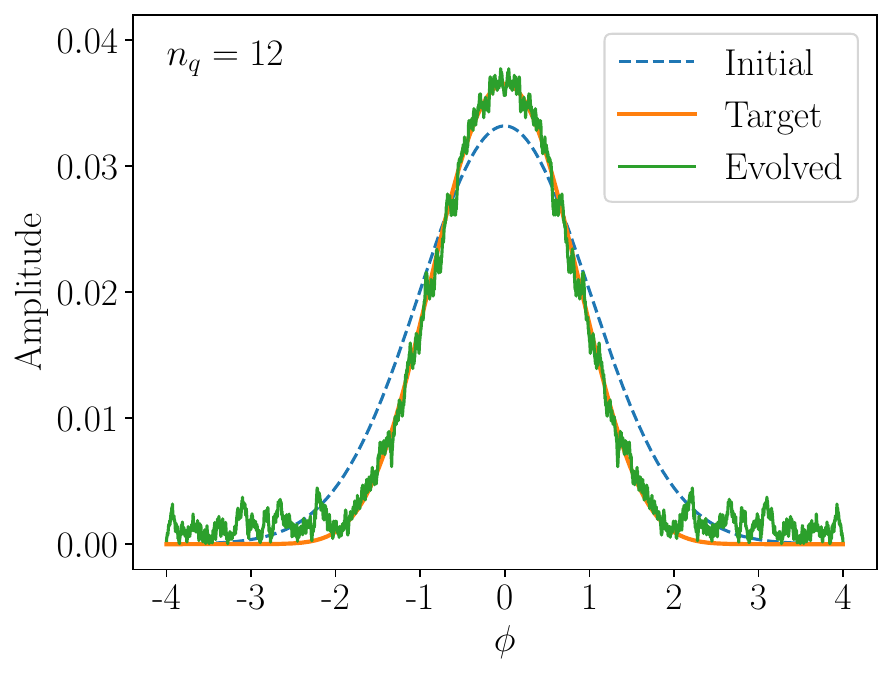}
        \end{subfigure}
\caption{Trotterized SeqHT ASP on twelve qubits ($n_q=12$). 
The Hamiltonian is digitized in the basis of the field operator, using $\phi_{\rm max}=4$ and $\lambda = 10$. 
Each adiabatic step consists of one second-order Trotter step implemented over a total evolution time of 1.6, and the system is evolved with 4 adiabatic steps. 
The left panel employs $\nu_{\rm cut}=14$ for the $\hat{\phi}^4$ operator, and achieves a fidelity of $0.9923$.
In the right panel,  $\hat{\phi}^2$ term is also truncated, to $\nu_{\rm cut}= 30 $. The adiabatically prepared wavefunction fidelity is $0.9897$.} 
\label{fig:12qubitamp}
\end{figure*}

For larger systems, SeqHT on other terms in the Hamiltonian
can be explored. 
For example, the $\hat{\phi}^2$ term can also be drastically truncated in a 12-qubit system. 
Wavefunctions prepared for a 12-qubit system by Trotterized ASP is shown in Fig.~\ref{fig:12qubitamp}. 
The time scan in App.~\ref{appen:timescan} guided what ASP 
parameters to use. 
SeqHT for the $\hat{\phi}^4$ term is implemented with $\nu_{\rm cut}=14$, leaving the same number of operators as for five qubits. 
In the left panel of Fig.~\ref{fig:12qubitamp}, no other truncations are implemented, while in the right panel the $\hat{\phi}^2$ term is also truncated, with $\nu_{\rm cut} = 30$, leaving ten two-body operators. 
For a n-qubit system with $\nu_{\rm cut}=14$, the number of four-body operators required to simulate the $\hat{\phi}^4$ term is reduced from $\binom{n_q}{4}$ to one. 
Two-body operator resource requirements are also reduced when truncation of the $\hat{\phi}^2$ term is implemented. 
In the case of $n_q=12$, the 
$66$ two-qubit operators are reduced down to just $10$ for $\nu_{\rm cut}=14$.

While we have not implemented a sequency truncation in conjugate-momentum space, we expect that such a truncation will also converge with increasing $\nu^{(\Pi)}_{\rm cut}$. 
The value of $\nu^{(\Pi)}_{\rm cut}$ may or may not coincide with its partner in $\phi$-space, as intrinsically they are unrelated. 
However, such an truncation will disturb the utility of the Nyquist-Shannon sampling theorem~\cite{Macridin:2018gdw,Macridin:2018oli} which
eliminates all power-law digitization effects.
This is based upon uniform sampling in conjugate-momentum space, and the use of the exact dispersion relation following the local quantum Fourier transform, and not the lattice relation from the finite-difference operator in $\phi$-space. 
Therefore, we anticipate larger than naively expected errors introduced by a sequency truncation in conjugate-momentum space. 
This is verified through exact calculations. %

%%%%%%%%%%%%%%%%%%%%%%%%%%%%%%%%%%%%%%%%%%%%%%%%%%%%%%%%%%
\section{Analysis of S\MakeLowercase{eq}HT in the Limit of a Large Number of Qubits}
\label{sec:largenQ}
\noindent
This section provides an upper bound on the sequence coefficient $\beta_\nu$, for an arbitrary polynomial function. This result can be used to determine an appropriate $\nu_{\rm cut}$ for a given polynomial operator without calculating all $\beta_\nu$.
This upper bound is found by utilizing the patterns in the locations of level crossings for sequence operators of dimension $2^{n_q} \times 2^{n_q}$, with $n_q$ arbitrary; these patterns can then be used to extrapolate to a large $n_q$ limit. 
This result extends to any well-behaved function which 
can be carried out by utilizing an appropriate Taylor expansion(s).

Consider the digitization of a polynomial function $F(x) = x^p$,
where $p$ is a positive integer. 
In the basis of eigenstates of the $\hat x$ operator ($\hat x |x\rangle = x |x\rangle$),
this function maps to a diagonal operator that can be decomposed into sequency operators $\hat\CO_\nu$, defined in Eq.~(\ref{eq:decomp}). 
There exist a hierarchy in sequency such that,
\begin{align}
\beta_\nu \gtrsim \beta_{\nu'} \quad \text{for} \quad \lfloor\log_2\nu'\rfloor > \lfloor\log_2\nu\rfloor
\end{align}
and therefore, in order to achieve a desired precision, 
it is sufficient to only include operators with sequency up to $\nu_{\rm cut}$ with the upper bound $ B_{\nu_{\rm cut}} < \Lambda_{\rm cut}$, the coefficient threshold.
This also limits the size of operators to only $n_\text{cut}$-body operators with
$n_\text{cut}\leq \lceil \log_2 \nu_\text{cut} \rceil$,
hence reducing the depth of quantum circuits required for implementation. 

For sufficiently large $n_q$, $\beta_\nu$ approaches
\begin{align}
\lim_{n_q\rightarrow \infty} \beta_\nu \rightarrow \int_{-x_M}^{x_M}dx 
\ F(x) \ \Theta_\nu(x; x_M)
\ ,
\end{align}
where $\Theta(x)_\nu$ are Heaviside step functions with $\nu$ level crossings that occur in the same patterns as sequency operators $\hat{\CO}_\nu$, in the range $x = [-x_M, x_M]$ (see App.~\ref{app:HW}); $x_M$ is the maximum value of $x$ used in the digitization. In the case of the digitized $\lambda \phi^4$ theory, $x$ is $\phi$ and $x_M = \phi_{\rm max}$.
In order to better compare various sequency coefficients, it is useful to define a normalized version of $\beta_\nu$ in the limit of large $n_q$,~\footnote{The normalization coefficient is chosen to be the integral of the polynomial function on the interval $x_M>x>-x_M$.}
\begin{align}
\tilde \beta_\nu = \frac{p+1}{2 x_M^{p+1}} \int_{-x_M}^{x_M}dx 
\ F(x)\  \Theta_\nu(x; x_M) 
\ ,
\end{align}
with $\tilde \beta_0 = 1$, and $p$ the positive integral order of the polynomial. 
Since the integrand has maximal support around $x\sim x_M$,
$\tilde\beta_\nu$ can be bounded from above by $\tilde B_\nu$, 
\begin{align}
\tilde B_\nu \equiv \CN_p \left( \int_{x_\nu}^{x_M}dx\, x^p+ \int^{-x_\nu}_{-x_M}dx\, x^p\right) 
\ \ ,\ \  \CN_p = \frac{p+1}{2 x^{p+1}_M} 
\, ,
\label{eq:BoundBetaEven}
\end{align}
for  $\nu \ne 0$,
where $x_\nu$ is the location of the level crossing closest to $x_M$ for $\hat{\CO}_\nu$. 
For $F(x)$ an even function, $\tilde B_\nu = 0$ for odd sequency $\nu$,
and for even $\nu$, 
\begin{align}
\tilde B_0 = 1 \qquad\text{and}\qquad  \tilde B_\nu \equiv 1-\left(\frac{x_\nu}{x_M}\right)^{1+p} \qquad \nu \in \text{positive even integers}.
\end{align}
It is important to note that this is an upper bound on the value of $\tilde\beta_\nu$ because $\Theta_\nu(x;x_M)$ becomes negative at $x<x_\nu$; while it can become positive at even smaller values of $x$, the next largest (and negative) contribution to the integrand comes from $x_\nu > x > x'_\nu$, with $x'_\nu$ being the location of the second level crossing.

The expression for $x_\nu$ can be derived by utilizing the aforementioned pattern in the level crossings for sequency operators $\hat{\CO}_\mu$ of dimension $n_q$. This is performed explicitly in App.~\ref{app:HW}, with the result that
\begin{align}
x_\nu = x_M \left(1 - \frac{1}{2^{\lfloor \log_2 \nu\rfloor}}\right)
\ ,
\end{align}
Therefore, an upper bound on the (normalized) sequency coefficient is
\begin{align}
\tilde B_\nu = 1 - \left(1 - \frac{1}{2^{\lfloor \log_2 \nu\rfloor}}\right)^{1+p}
\ ,
\label{eq:tildeBnu}
\end{align}
for positive, even sequency index. 
This upper bound  significantly overestimates $\beta_\nu$ for $\nu$ not close to an integer power of two, but it gives an intuitive argument for the number of operators to include in the truncated sum. 
The bound could be further refined by including additional contributions to the integral. 
%
%%% %%% %%% %%% %%%
\begin{figure}[htpb]
\subfloat{%
  \includegraphics[width=0.45\columnwidth]{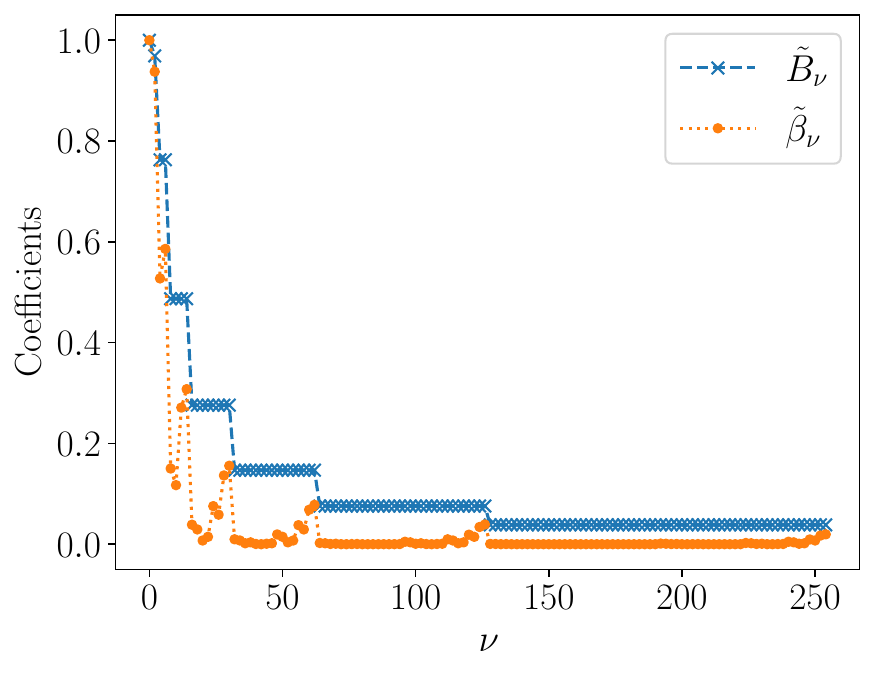}
}
\subfloat{%
  \includegraphics[width=0.45\columnwidth]{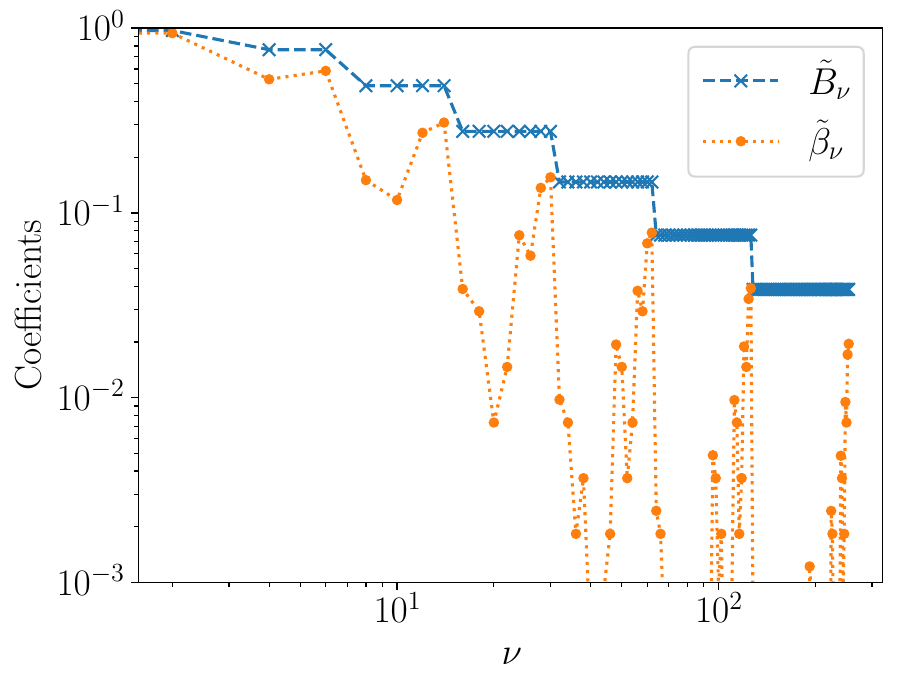}
}
\caption{Comparison between calculated coefficients,
$\tilde\beta_\nu$,
and upper bounds, $\tilde B_\nu$,
for a $x^4$ potential. 
The blue dashed lines are the $\tilde B_\nu$, and the orange dotted lines are the $\tilde\beta_\nu$. 
The left panel shows the results for eight qubits linearly, while right panel displays the results in a log-log scale. 
The vertical axis is the value of the normalized coefficients while the horizontal axis is $\nu$. 
Numerical values of the results displayed in this figure are included in Table~\ref{tab:bound}.
}
\label{fig:UpperBound}
\end{figure}
%%% %%% %%% %%% %%%

%
Figure~\ref{fig:UpperBound} shows the values of the coefficients $\tilde \beta_\nu$, along with the upper bound, $\tilde B_\nu$ for digitization using eight qubits. The upper bounds are respected by the computed $\tilde \beta_\nu$, as expected. 
With an increasing number of qubits, and associated digitization of a smooth function in the Hilbert space, the range of values for the various coefficients  can be substantial, with the largest values approaching the bound. 
An important aspect of this is that the bound decreases with increasing sequency, allowing for a sequency truncation, even in the ``worst'' case function.
This analysis can be repeated for odd polynomial functions, which result in an analogous result. Namely for $F(x)$ an odd polynomial, $\tilde B_\nu = 0$ for even sequency and $\tilde B_\nu$ is as given in Eq.~(\ref{eq:tildeBnu}), except with $\nu$ being positive and odd.

The approach developed above can be generalized to arbitrary functions.
In the case of an even function with support near $x_M$, 
Eq.~(\ref{eq:BoundBetaEven}) can be explicitly evaluated.
For example, 
for $F(x) = 1-\cos x$, the upper bound on the (normalized) sequency coefficients,
for $x_M < \pi/2$,  is given by
\begin{eqnarray}
\tilde B_\nu^{(F)} & = & 
1-\frac{x_\nu - \sin x_\nu}{x_M - \sin x_M}.
\end{eqnarray}
Lastly, the results 
for polynomials of definite parity can be 
combined to derive an upper bound on sequency coefficients
of any well-behaved 
function. 
In particular, for a function $F(x)$ with a polynomial expansion, 
an upper bound for the sequency coefficients is
\begin{eqnarray}
B_\nu^{(F)} & = & \sum^\infty_{p=0} |a_p| B_\nu^{(p)}\ \ {\rm for} \ \ 
F(x) \ =\  \sum^\infty_{p=0} a_p x^p 
\, .
\end{eqnarray}
Note that $B_\nu^{(p)}$ is the non-normalized upper bound $B_\nu^{(p)} = \tilde B_\nu^{[p]}/\CN_p$.

%%%%%%%%%%%%%%%%%%%%%%%%%%%%%%%%%%%%%%%%%%%%%%
\section{Computational Complexity: The Behavior of Magic with S\MakeLowercase{eq}HT}
\label{sec:magic}
\noindent
It is interesting to understand the behavior of the computational complexity with increasing levels of sequency truncation. 
In particular, we focus on the quantum computational complexity required to establish a given wavefunction as measured by its magic (non-stabilizerness)~\cite{Aaronson:2004xuh,Bravyi:2004isx,Stahlke_2014,Pashayan_2015,Bravyi_2016,Leone:2021rzd,Leone:2022lrn}, reflecting the number of T-gates in the quantum circuit. 
For a given wavefunction on a register of qubits, the magic in the state can be determined by the matrix elements of the complete set of $n$-qubit Pauli strings,
\begin{eqnarray}
    c_{P} & = & \langle\psi |  \hat P | \psi \rangle
    \ \ ,
    \label{eq:cP}
\end{eqnarray}
where $\hat P \in  \{  \hat{P}_1 \otimes \hat{P}_2  \otimes ... \otimes \hat{P}_n  \}$ and $\hat{P}_i \in \{ \hat I, \hat X, \hat Y, \hat Z  \}$. 
For a stabilizer state 
(a state that can be prepared efficiently using classical resources with a Clifford gate set), 
only $d=2^{n_q}$ of the $d^2=4^{n_q}$ matrix elements are non-zero, 
and take the values of $\pm 1$.
It is useful to define $\Xi_P \equiv c_P^2/d$ which can be identified as probabilities that satisfy $\sum\limits_P \Xi_P = 1$. 
Using $\Xi_P$, there are a number of ways that are used to quantify the  non-stabilizerness of the state, including the Renyi-entropy and linear magic. 
We use the latter, defined by 
\begin{eqnarray}
    {\cal M}_{\rm lin}(|\psi\rangle) & = & 1-d \sum_P \Xi_P^2
    \ \ ,
    \label{eq:magic}
\end{eqnarray}
which vanishes for stabilizer states.

We consider the convergence properties of magic in the digitization and  sequency truncation of a one-dimensional Gaussian wavefunction centered in the middle of the Hilbert space. 
Specifically, we calculate the magic of a digitized  Gaussian wavepacket of width $\sigma=1/\sqrt{2}$, with $\phi_{\rm max}=4.0$, scanning over a range of $n_q$ for which the wavefunction itself is well-contained on the register. 
As described above, the wavefunction $|\psi\rangle$ is digitized onto the $n_q$ qubits, to give $|\psi\rangle_d$ and renormalized such that ${}_d\langle\psi | \psi\rangle_d=1$. 

\begin{figure}[!ht]
\includegraphics[width=0.45\columnwidth]
{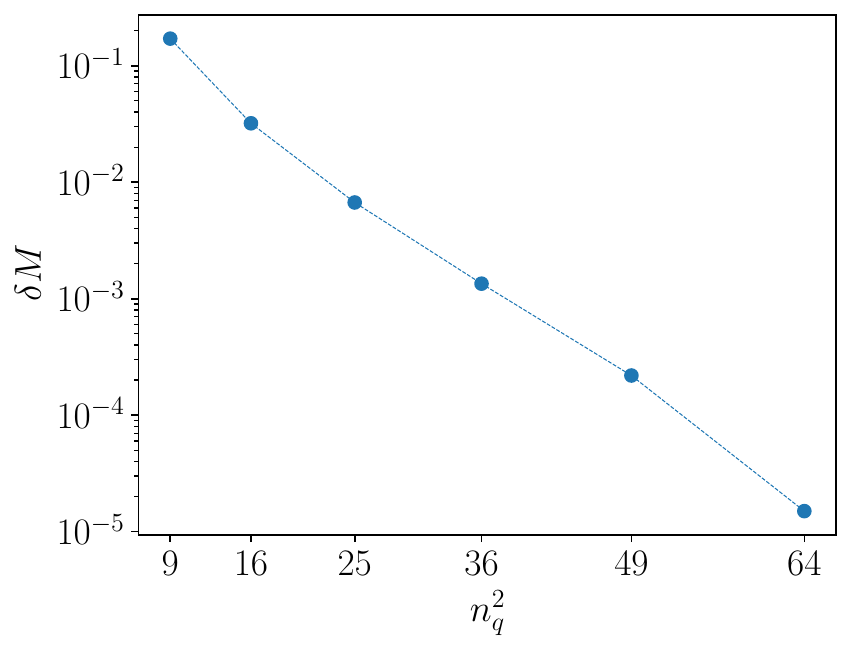}
\includegraphics[width=0.45\columnwidth]
{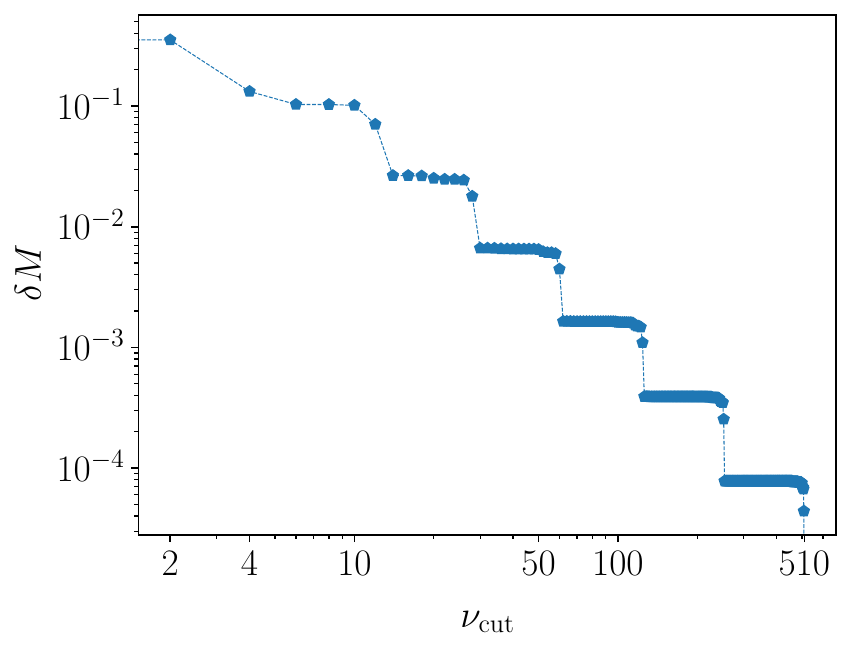}
\caption{
The convergence of magic in a Gaussian wavefunction with $\sigma=1/\sqrt{2}$ 
centered in the middle of the $n$-qubit Hilbert spaces with $\phi_{ max}=4$. 
The left panel shows the deviation from the asymptotic value as a function of the number of qubits, while the right panel shows the deviation from the asymptotic value as a function of $\nu_{\rm cut}$ for $n_q=9$ using a log-log scale. 
Numerical values for the results displayed in this figure can be found in Table~\ref{tab:magicLam9} and Table~\ref{tab:magicn}.}
\label{fig:GaussnQ}
\end{figure}

One interesting property to investigate is the magic in the digitized Gaussian wavefunctions $|\psi\rangle_d$ as a function of the square of the number of qubits used for digitization with fixed $\phi_{ max}$. 
Figure~\ref{fig:GaussnQ} shows the convergence of magic to its asymptotic value in the Gaussian wavefunction as a function of the number of qubits on a log-linear scale.
%%% %%% %%% %%% %%%
%%% %%% %%% %%% %%%
The magic is found to converge to a fixed value,
${\cal M}_{\rm lin} = 0.362007$ (for these parameters),
with an error, $\epsilon$, scaling in a way that is consistent with 
$\log \frac{1}{|\epsilon|} \sim {\cal O}(n_q^2)$.
This result demonstrates that, for the Gaussian wavefunction, and more generally continuous functions with bounded support, its magic can be defined by the limit of the digitized wavefunction. %

Another property to study is the magic in the digitized Gaussian wavefunctions $|\psi\rangle_d$ as a function of the cutoff sequency $\nu_{\rm cut}$. The digitized state is projected onto Hadamard-Walsh functions, which are then used to construct a truncated wavefunction including sequencies up to $\nu_{\rm cut}$, $|\psi\rangle_d^{(\nu_{\rm cut})}$, which are also renormalized to give $\prescript{(\nu_{\rm cut})}{d}\langle\psi | \psi\rangle_d^{(\nu_{\rm cut})}=1$. 
As shown in the right panel of Fig.~\ref{fig:GaussnQ}, the magic of the Gaussian wavefunction prepared on $n_q=9$ qubits approaches its asymptotic value with increasing precision as $\nu_{\rm cut}$ is increased, but in a step-wise fashion. 
It is observed to be scaling approximately linearly on the log-log plot. 
A cutoff sequency can be selected such that SeqHT retains the asymptotic quantum computational complexity of the system within a given error threshold. Any cutoff at or above this sequency yields results that lie within this threshold. With more quantum computational resources available, the cutoff sequency can be systematically increased to include higher sequency terms thus refining precision.

%%%%%%%%%%%%%%%%%%%%%%%%%%%%%%%%%%%%%%%%%%%%%%
\section{Quantum Simulations}
\label{sec:Qsims}
\noindent
Generally, 
implementing adiabatic time evolution using a digital quantum computer requires that
the evolution operator is Trotterized.  
Each adiabatic step is realized using a fixed number of Trotter steps, and each Trotter step consists of the same set of gates with different rotation angles to capture the time dependence of the Hamiltonian. 
Due to limitations of NISQ-era quantum devices, such as imperfect gate operations and limited coherence time, resources need to be optimally distributed to minimize the overall error in a simulation.

From a physics perspective, the fidelity of wavefunctions prepared using ASP improves with increasing time intervals 
over which the Hamiltonian is evolved. 
On the other hand, the fidelity of a Trotterized evolved wavefunction decreases with increasing time intervals. 
Therefere, there is an intrinsic tension in using Trotterized evolution to implement ASP; while this is of no consequence for ideal quantum computers, it is important for realistic quantum computers, particularly NISQ-era devices.
In practice, there is a tuning, or optimization, that has to be performed to use these combined methods on a real-world device. 

\begin{figure*}[h!t]
        \centering
        \begin{subfigure}[b]{0.44\textwidth}
            \centering
            \includegraphics[width=\textwidth]{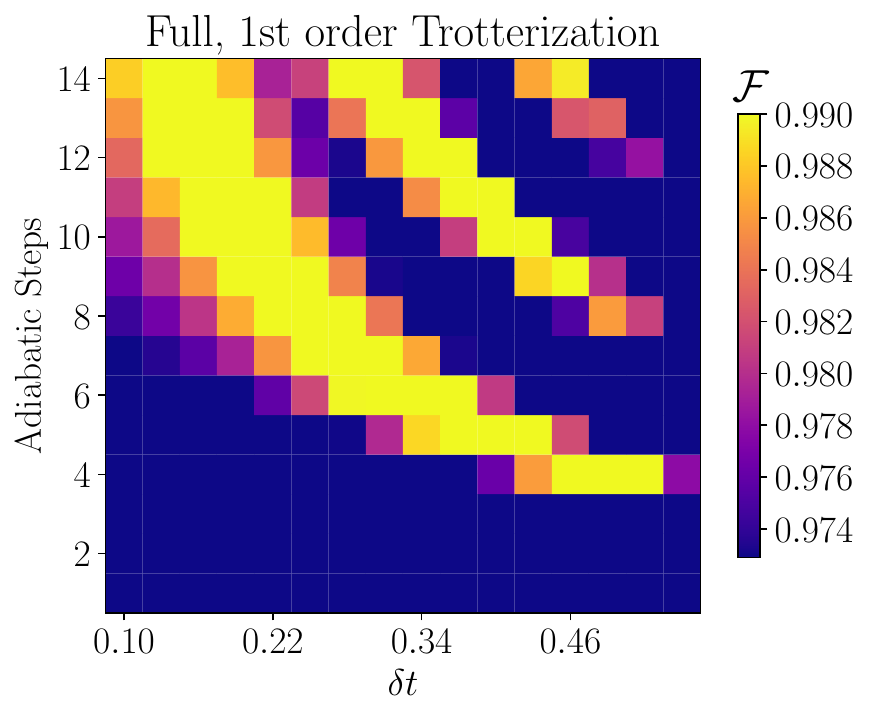}
        \end{subfigure}
        \begin{subfigure}[b]{0.44\textwidth}  
            \centering 
            \includegraphics[width=\textwidth]{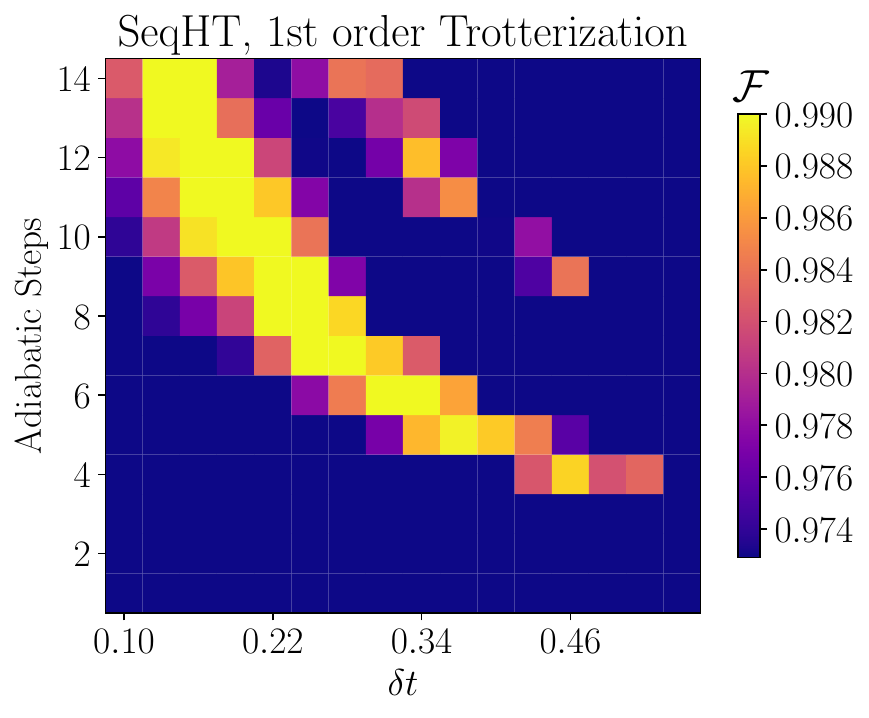}
        \end{subfigure}
        
        \begin{subfigure}[b]{0.44\textwidth}
            \centering
            \includegraphics[width=\textwidth]{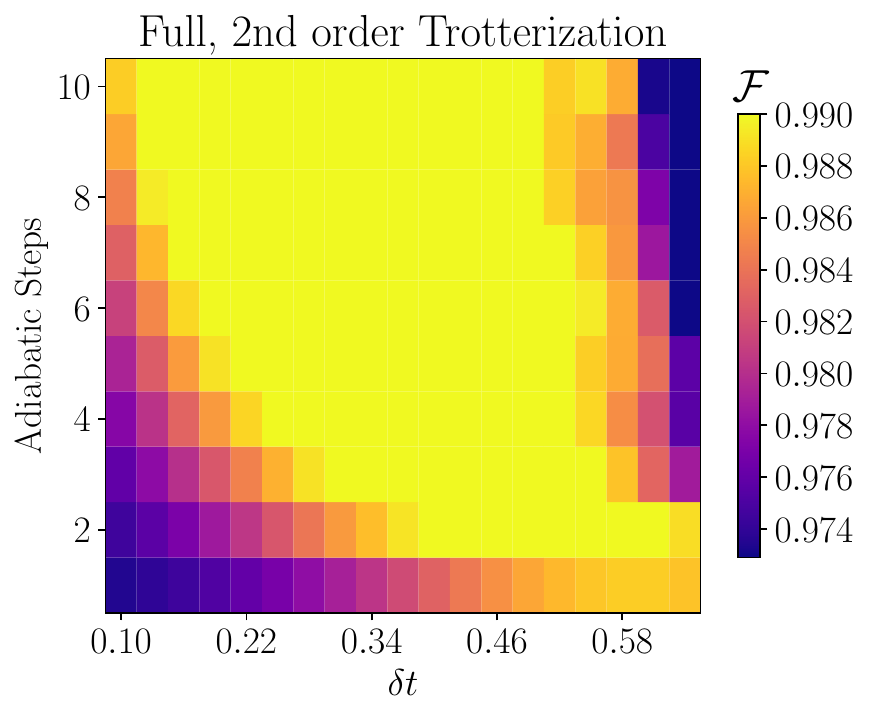}
        \end{subfigure}
        \begin{subfigure}[b]{0.44\textwidth}  
            \centering 
            \includegraphics[width=\textwidth]{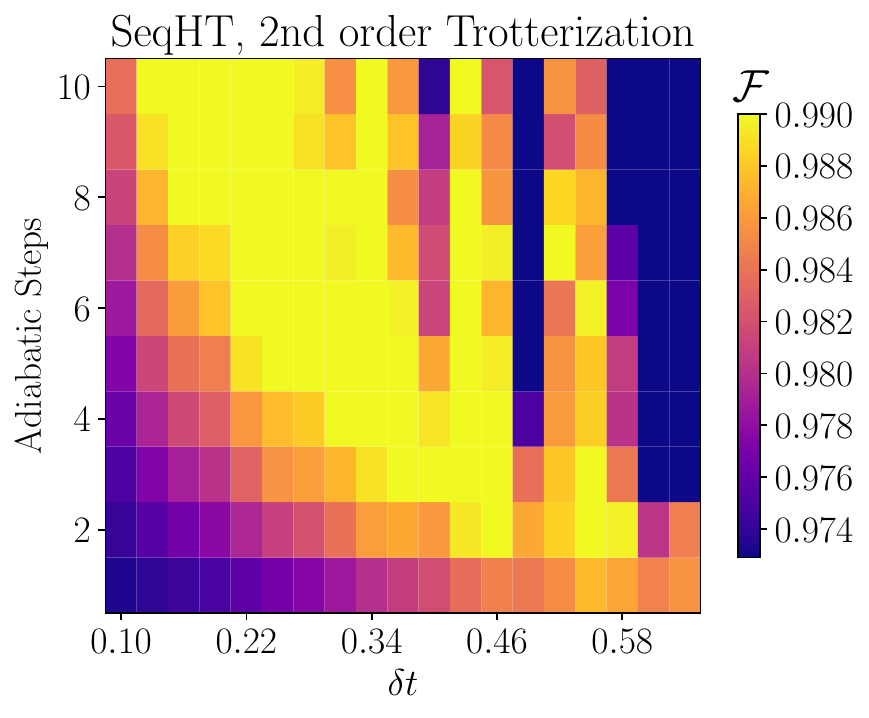}
        \end{subfigure}
\caption{The fidelity $\mathcal{F}$ of the $\lambda\phi^4$ ground state adiabatically prepared using a noiseless quantum simulator as a function of time-step size $\delta t$ and the number of adiabatic steps, for a five-qubit system ($n_q=5$) with $\phi_{\rm max}=4$ and $\lambda = 10$. 
Each full (left) or SeqHT (right) adiabatic step is comprised of one first-order (upper) or one second-order (lower) Trotter step. 
The initial overlap of the free-theory ground state ($\lambda=0$) and target ground state is $0.9729$. 
Basis operators of the interacting term with $\nu>14$ are truncated. 
Numerical values for the results displayed in this figure can be found in Table~\ref{tab:5qubit_scan_1st} and Table~\ref{tab:5qubit_scan_2nd}.
 }
\label{fig: time scans}
\end{figure*}

To focus on the study of ASP, we use one Trotter step for each adiabatic step. 
Figure~\ref{fig: time scans} presents (tuning) scans of state fidelity of the ground state of the $\lambda\phi^4$ theory as a function of time-step size $\delta t$ and the number of adiabatic steps; the ground state has been adiabatically prepared on a classical noiseless quantum simulator.
Both the full and SeqHT adiabatic evolution with $\nu_{\rm cut}=14$ for both first-order and second-order Trotterization are displayed. 
SeqHT evolution performs comparably well to the full evolution in both cases. 
Second-order Trotterization has a larger region of parameters that generate effective results compared to  first-order Trotterization. % 
Therefore, we choose to work with two adiabatic steps, each with one second-order Trotterized evolution, in the analysis that follows. 
 
%%%%%%%%%%%%%%%%%%%%%%%%%%%%%%%%%%%%%%%%%%%%%%
\subsection{Quantum Circuits and Resource Requirements}
\label{subsec:Qcircs}
\noindent
The first step of the adiabatic evolution to $\lambda\phi^4$ from the non-interacting theory  is to prepare the ground state of the non-interacting theory on the quantum register. 
The circuit that prepares a general unitary transformation on $n_q$ qubits with all-to-all connectivity is presented in Ref.~\cite{M_tt_nen_2004}. 
Since the ground state of the non-interacting theory can be defined to have positive and real amplitudes, all rotations in the circuit are 
about the $y$-axis.
The rotation angles for this sequence can either be directly calculated~\cite{Kitaev:2008vci,M_tt_nen_2004,Klco:2019xro} or obtained via a variational minimization procedure, for instance via a Variational Quantum Eigensolver (VQE).
Symmetry of the free theory ground state can be exploited by preparing half of the state on all but the most significant qubit, then reflecting it to form the full state~\cite{Klco:2019xro}. 
In the circuit shown in Fig.~\ref{fig:freeprep}, $F$ denotes the operators used to reflect about the midpoint of the Hilbert space, and the other circuit elements constitute a general state preparation circuit. 
This initialization requires $2^{n_q-1}+n_q-3$ CNOT gates for $n_q$ qubits with all-to-all connectivity. 
Figure~\ref{fig:reflect} shows the general reflection circuit $F$ 
(left), and a nearest-neighbor connectivity adaptation (right) that is based on the discussions in Ref.~\cite{Park:2023goh}. \\
\begin{figure}[!ht]
    \centering
\[
\begin{quantikz}[row sep=0.1cm, column sep=0.05cm]
    \lstick{$q_1$} & \qw & \qw & \qw & \qw & \qw & \qw & \qw & \qw & \qw & \qw & \qw & \qw & \qw & \qw & \qw & \qw & \qw & \qw & \qw & \qw & \qw & \qw & \qw & \qw & \qw & \qw & \qw & \gate[5]{F} & \qw \\
    \lstick{$q_2$} & \gate{R} & \ctrl{1} & \qw & \ctrl{1} & \qw & \qw & \ctrl{2} & \qw & \qw & \qw & \ctrl{2} & \qw & \qw & \qw & \qw & \qw & \qw & \ctrl{3} & \qw & \qw & \qw & \qw & \qw & \qw & \qw & \ctrl{3} & \qw & \qw \\
    \lstick{$q_3$} & \gate{R} & \targ{} & \gate{R} & \targ{} & \ctrl{1} & \qw & \qw & \qw & \ctrl{1} & \qw & \qw & \qw & \qw & \ctrl{2} & \qw & \qw & \qw & \qw & \qw & \qw & \qw & \ctrl{2} & \qw & \qw & \qw & \qw & \qw & \qw \\
    \lstick{$q_4$} & \qw & \qw & \qw & \gate{R} & \targ{} & \gate{R} & \targ{} & \gate{R} & \targ{} & \gate{R} & \targ{} & \ctrl{1} & \qw & \qw & \qw & \ctrl{1} & \qw & \qw & \qw & \ctrl{1} & \qw & \qw & \qw & \ctrl{1} & \qw & \qw & \qw & \qw \\
    \lstick{$q_5$} & \qw & \qw & \qw & \qw & \qw & \qw & \qw & \qw & \qw & \qw & \gate{R} & \targ{} & \gate{R} & \targ{} & \gate{R} & \targ{} & \gate{R} & \targ{} & \gate{R} & \targ{} & \gate{R} & \targ{} & \gate{R} & \targ{} & \gate{R} & \targ{} & \qw & \qw
\end{quantikz}
\]
\caption{Quantum circuit for preparing the ground state 
of the non-interacting theory on a quantum computer adapted from Ref.~\cite{Klco:2019xro}. 
For a positive and real wavefunction, all of the rotations are about the $y$-axis. 
$F$ denotes the circuitry to reflect the wavefunction about the mid-point of the Hilbert space, and is given in Fig.~\ref{fig:reflect}. }
\label{fig:freeprep}
\end{figure}
\begin{figure}[!ht]
\centering
\[
\begin{quantikz}
    \lstick{$q_1$} & \gate{H} & \ctrl{4} & \ctrl{3} & \ctrl{2} & \ctrl{1} & \qw \\
    \lstick{$q_2$} & \qw      & \qw      & \qw      & \qw & \targ{}  & \qw \\
    \lstick{$q_3$} &  &     &          & \targ{} &  &     \\
    \lstick{$q_4$} & \qw      & \qw      & \targ{}  & \qw & \qw      & \qw \\
    \lstick{$q_5$} & \qw      & \targ{}  & \qw      & \qw & \qw      & \qw 
\end{quantikz}
\hspace{50pt}
\begin{quantikz}
    \lstick{$q_1$} & \gate{H} & \qw      & \qw      & \ctrl{1} & \qw      & \qw      & \qw      & \qw \\
    \lstick{$q_2$} & \qw      & \qw      & \ctrl{1} & \targ{}  & \ctrl{1} & \qw      & \qw      & \qw \\
    \lstick{$q_3$} & \qw      & \ctrl{1} & \targ{}  & \qw      & \targ{}  & \ctrl{1} & \qw      & \qw \\
    \lstick{$q_4$} & \ctrl{1} & \targ{}  & \qw      & \qw      & \qw      & \targ{}  & \ctrl{1} & \qw \\
    \lstick{$q_5$} & \targ{}  & \qw      & \qw      & \qw      & \qw      & \qw      & \targ{}  & \qw 
\end{quantikz}
\]
\caption{Quantum circuits for reflecting a wavefunction 
about the mid-point of the Hilbert space
on a device with all-to-all connectivity (left) as discussed in Ref.~\cite{Klco:2019xro}, 
and on a device with nearest-neighbor connectivity (right).}
\label{fig:reflect}
\end{figure}
 
For the time evolution generated by the Hamiltonian given in Eq.~(\ref{eq:Hami}), since the mass term  and the interaction term  commute, the circuit can be Trotterized into $\Tilde{\Phi}(s,t)$ and $\Tilde{\Pi}(t)$ contributions; these operators are defined to be
\begin{equation}
\Tilde{\Phi}(s,t) \equiv e^{-i (\frac{1}{2}\hat{\phi}^2+\frac{\lambda(s)}{4!}\hat{\phi}^4) t}
\qquad \text{and} \qquad
\Tilde{\Pi}(t) \equiv e^{-i \frac{1}{2} \hat{\Pi}^2 t}. 
\end{equation}
Detailed construction of the quantum circuits for 
$\Tilde{\Phi}(s,t)$ and $\Tilde{\Pi}(t)$ can be found later in this section and in App.~\ref{appen:seqht_cir}. Each adiabatic step in the evolution from the non-interacting ground state to the $\lambda\phi^4$ ground state contains one Trotter step. 
In a second-order Trotter step, the circuit implements the unitary operator,
\begin{equation}
\hat U_{\rm 2nd}^{(1)} \ =\ \Tilde{\Phi}\left(s,\delta t/2\right)\,  \Tilde{\Pi}(\delta t) \, \Tilde{\Phi}\left(s,\delta t/2\right)\ .
\end{equation}
Between adjacent  second-order Trotterized adiabatic steps, 
the last term of the previous step and the first term of the current step are combined (because $\phi^2$ and $\phi^4$ commute) and therefore
\begin{equation}
\hat U_{\rm 2nd} \ =\ \Tilde{\Phi}(s_1,\delta t/2)\,  \Tilde{\Pi}(\delta t) \, \Tilde{\Phi}(s_1+s_2,\delta t/2)\,  \Tilde{\Pi}(\delta t) \, \Tilde{\Phi}(s_2,\delta t/2)\ . 
\label{eqn:physicscir}
\end{equation}
The quantum circuits employed to implement the error-mitigation strategy of decoherence renormalization (DR) and operator-DR (ODR) are designed such that they share similar structure with the physics circuit, but can be efficiently simulated  classically. The corresponding circuits are implemented as 
\begin{equation}
\hat U_{\rm 2nd}^{(2,\text{DR})} \ =\ \Tilde{\Phi}(s_1,\delta t/2)\,  \Tilde{\Pi}(\delta t) \, \Tilde{\Phi}(s_1+s_2,0)\,  \Tilde{\Pi}(-\delta t) \, \Tilde{\Phi}(s_2,-\delta t/2)\ \ ,
\label{eqn:miticir}
\end{equation}
corresponding to forward evolution for half of the time and backward evolution for the other half.

Nearest-neighbor connectivity is considered as the following quantum simulations are carried out on IBM's superconducting-qubit quantum computers. 
Two-body operators $R_{Z_aZ_b} = e^{i\theta \hat{Z}_a\hat{Z}_b}$ with the same most significant qubit can be overlaid to improve efficiency~\cite{Farrell:2024fit}, shown in Fig.~\ref{fig:phi2}. 
Additional CNOT cancellations from neighboring blocks further reduces the depth. 
As derived in Ref.~\cite{Farrell:2024fit}, 
the total number of CNOTs $N$ and circuit depth $D$ for the construction of all $R_{ZZ}$ gates with regard to the number of qubits $n_q$ is
\begin{equation}
N=2\binom{n_q}{2}, \quad D=n_q\left(n_q-2\right)+3.
\end{equation}
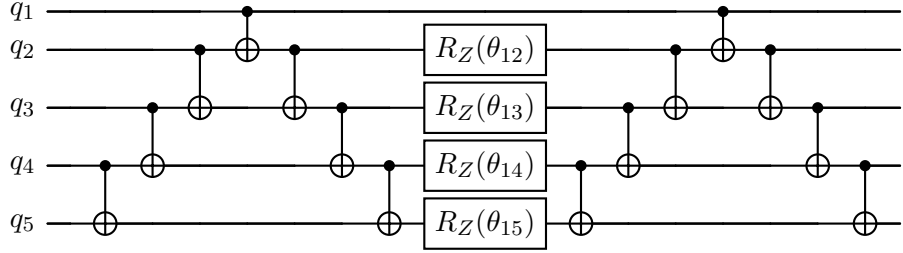
\begin{figure}[htpb]
\[
\begin{quantikz}[row sep=0.1cm, column sep=0.3cm]
    \lstick{$q_1$} & \qw & \qw & \qw & \qw & \ctrl{1} & \qw & \qw & \qw & \qw & \qw & \qw & \qw & \ctrl{1} & \qw & \qw & \qw & \qw \\
    \lstick{$q_2$} & \qw & \qw & \qw & \ctrl{1} & \targ{} & \ctrl{1} & \qw & \qw & \gate{R_Z(\theta_{12})} & \qw & \qw & \ctrl{1} & \targ{} & \ctrl{1} & \qw & \qw & \qw \\
    \lstick{$q_3$} & \qw & \qw & \ctrl{1} & \targ{} & \qw & \targ{} & \ctrl{1} & \qw & \gate{R_Z(\theta_{13})} & \qw & \ctrl{1} & \targ{} & \qw & \targ{} & \ctrl{1} & \qw & \qw \\
    \lstick{$q_4$} & \qw & \ctrl{1} & \targ{} & \qw & \qw & \qw & \targ{} & \ctrl{1} & \gate{R_Z(\theta_{14})} & \ctrl{1} & \targ{} & \qw & \qw & \qw & \targ{} & \ctrl{1} & \qw \\
    \lstick{$q_5$} & \qw & \targ{} & \qw & \qw & \qw & \qw & \qw & \targ{} & \gate{R_Z(\theta_{15})} & \targ{} & \qw & \qw & \qw & \qw & \qw & \targ{} & \qw
\end{quantikz}
\]
\caption{Quantum circuit for implementing multiple two-body $R_{ZZ}$ operators on a quantum computer from Ref.~\cite{Farrell:2024fit}.}
\label{fig:phi2}
\end{figure}
In conjugate-momentum space, $\hat \Pi^2$ can be implemented with the same set of basis operators as $\hat{\phi}^2$, but with different coefficients~\cite{Jordan:2012xnu}.
The Quantum Fourier Transform (QFT) is used to take these basis operators to the $\phi$-space, with a CNOT number and gate depth that scales as $n_q^{2}+n_q-4$~\cite{Park:2023goh} (with nearest neighbor connectivity for each QFT circuit). 
The resource requirements are calculated in the case where the controlled-phase gate is not a native gate of the quantum device, and is implemented using CNOTs and single-qubit rotations. % 

For a n-qubit system, when the interacting term $\hat{\phi}^4$ is implemented with $\nu_{\rm cut} = 14$, the number of four-body operators required to simulate the $\hat{\phi}^4$ term is reduced from $\binom{n_q}{4}$ to one. 
Each four-body operator is implemented with six CNOTs and total CNOT depth six, shown in Fig.~\ref{fig:cir}. 
Two-body operators in $\hat{\phi}^4$ can be absorbed into the two body operators of $\hat{\phi}^2$ and pose no additional cost. 

Table~\ref{tab:resourc} presents the two-qubit gate counts and depths for implementing a single adiabatic step (left panel) and the entire simulation with free theory initialization (right panel) before and after SeqHT. 
We find  SeqHT results in a $\sim 35\%$ reduction in depth for a single step and $\sim 29 \%$ reduction for the complete simulation. % 
Note that performing QFT and inverse QFT wth nearest-neighbor connectivity constitutes more than half of the two-qubit gate depth required for a single adiabatic step, which could potentially be improved by running on a quantum computer with all-to-all connectivity, such as a trapped-ion system; in such systems, the two-qubit gate depth of QFT scales as $4n_q-6$ instead.
 
\begin{table*}[htpb]
    \centering
        \begin{subtable}{0.33\textwidth}
        \centering
        \begin{tabular}{|c|c|c|}
            \hline
             & Depth & Count \\ \hline
             Full & 156 & 173 \\ \hline
            Truncated & 101 & 117 \\ \hline
        \end{tabular}
         \end{subtable}
        \begin{subtable}{0.25\textwidth}
        \centering
        \begin{tabular}{|c|c|}
            \hline
            Depth & Count \\ \hline
            291 & 321 \\ \hline
             208 & 237 \\ \hline
        \end{tabular}
    \end{subtable}
    \caption{The entangling-gate resources required to perform a single adiabatic step time evolution not including free theory initialization (left panel) and the complete simulation with two adiabatic steps each containing one second-order Trotter step (right panel), before and after SeqHT with $\nu_{\rm cut}=14$ using {\tt ibm\_sherbrooke}.
}
    \label{tab:resourc}
\end{table*}

%%%%%%%%%%%%%%%%%%%%%%%%%%%%%%%%%%%%%%%%%%%%%%
\subsection{Quantum Simulations using IBM's Quantum Computers}
\label{subsec:QsimIBM}
\noindent
The  ASP algorithms,
with and without SeqHT,
were run on IBM's superconducting qubit quantum computer {\tt ibm\_sherbrooke} 
with an Eagle r3 processor and ECR gates as native two-qubit operations. 
For both simulations, qubits with longer coherence time are selected, but this approach is only necessary for the full evolution as the significantly longer circuit approaches the decoherence boundary. 
The error mitigation techniques used  were read-out error mitigation~\cite{Nation:2021kye} through the {\tt qiskit}~\cite{qiskit} Runtime Sampler primitive, Dynamical Decoupling~\cite{Viola:1998jx,Souza_2012,Ezzell:2022uat} via {\tt qiskit} Transformation Pass~\cite{Javadi-Abhari:2024kbf}, Pauli Twirling~\cite{Wallman:2015uzh}, and Operator Decoherence Renormalization (ODR)~\cite{Farrell:2023fgd, Farrell:2024fit}. 
Employing Pauli-twirling in both sets of circuits converts coherent errors into incoherent errors, which can be suppressed by large ensemble sizes. 
This approach substantially reduces the error introduced into observables due to the quantum device's decoherence. 
In ODR, results from the physical circuits are renormalized by results from the mitigation circuits, as described in Eq.~(\ref{eqn:miticir}). 
Both the physical and mitigation circuits employ 80 Pauli-twirled instances with 8000 shots per twirl, with corresponding physical circuit and mitigation circuit twirled the same way. 
The central values and uncertainties are obtained from bootstrap resampling over the 80 twirls with 1000 resamples. 

\begin{figure*}[ht!]
        \centering
        \begin{subfigure}[b]{0.45\textwidth}
            \centering
            \includegraphics [width=\textwidth]{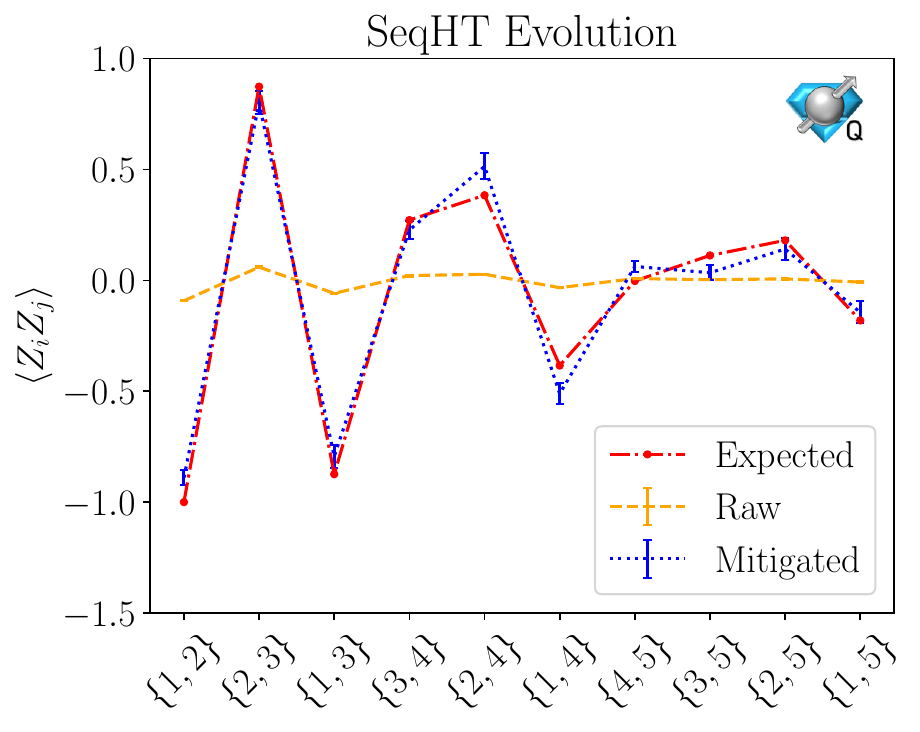}  
            \label{fig:devicetrun2}
        \end{subfigure}
        \begin{subfigure}[b]{0.45\textwidth}  
            \centering 
            \includegraphics[width=\textwidth]{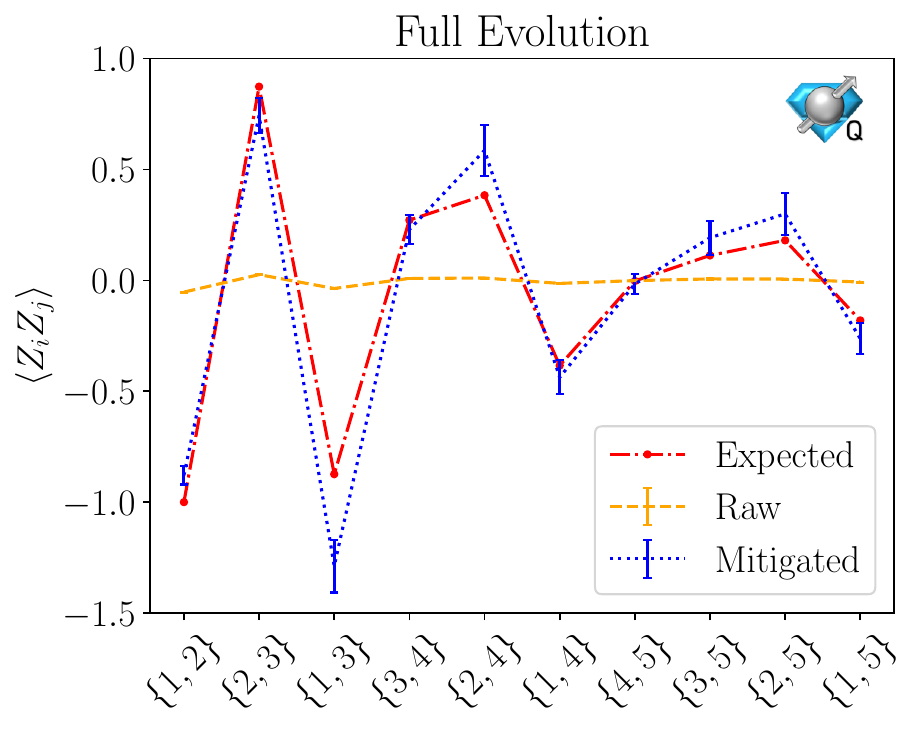}
            \label{fig:untrundevice2}
        \end{subfigure}
\caption{Expectation values of $\hat Z_i \hat Z_j$ operators
in the ground state of $\lambda\phi^4$ theory with $\phi_{\rm max}=4$ and $\lambda = 10$ measured from quantum simulations using IBM's {\tt ibm\_sherbrooke}. % 
The left panel computes the expectation values in the ground state prepared using SeqHT adiabatic evolution with $\nu_{\rm cut}=14$, while the right panel uses the full adiabatic evolution. 
Both evolutions involve two adiabatic steps, each consisting of one second-order Trotter step. 
Raw results and error-mitigated results are displayed, along the corresponding results from classical computations (from the numerically solved target state). 
The $x$-axis is $\{i,j\}$.
Both the physical and mitigation circuits employ 80 Pauli twirls with 8000 shots per twirl. 
The SeqHT circuit employs 237 two-qubit gates with depth 208 while the full circuit uses 321 two-qubit gates with depth 291.
The points are joined for display purposes. 
Numerical values for the results displayed in this figure can be found in Table~\ref{tab:devicenum}. 
} 
\label{fig:device2steps}
\end{figure*}

Fig.~\ref{fig:device2steps} shows a comparison between the expectation values $\langle \hat{Z}\hat{Z} \rangle$ obtained using the SeqHT adiabatic evolution and using the full adiabatic evolution. 
Encouragingly, the results from the SeqHT evolution are found to be superior to those of the full evolution, with smaller errors introduced by fewer gates in the shallower circuits and a reduced decoherence due to a shorter circuit run-time.
The full circuit implementations display larger deviations of central values of observables 
from the known results, and also substantially larger uncertainties in some instances. 
The truncated circuit is able to reproduce $\langle \hat{Z}\hat{Z}\rangle$s close to the values calculated from the exactly diagonalized interacting wavefunctions. 
As anticipated, the raw results obtained from the full circuit 
exhibit more sensitivity to device decoherence
than those from the truncated circuit, 
as can be seen 
in Fig.~\ref{fig:device2steps} and  
in the numerical values displayed in Table~\ref{tab:devicenum}. 
Note that we studied the $\langle \hat{Z}\hat{Z}\rangle$ observables instead of state fidelity or amplitudes since recovering states and state fidelities from a quantum computer requires methods such as quantum-state tomography~\cite{Ohliger:2013efs, Aaronson:2017qlb} or classical shadows~\cite{Kokail:2020opl,Huang:2020tih, Koh:2020qak, Struchalin_2021}; additionally, ODR has not yet been shown to be effective for these methods. % 

%%%%%%%%%%%%%%%%%%%%%%%%%%%%%%%
\section{Summary and Outlook}
\label{sec:concdisc}
\noindent
We have introduced a systematically-improvable 
truncation scheme that makes use of the sequency hierarchies present in many physical systems.
The Sequency Hierarchy Truncation (SeqHT) scheme reduces the quantum resources needed to
prepare and time-evolve states within a target fidelity for a given simulation. 
We have derived upper bounds on the sequency coefficients of polynomial interactions,  
allowing for a convergent sequency truncation; 
these results are generalized to provide upper bounds on the sequency coefficients of any well-behaved function. 
SeqHT is expected to be particularly effective in improving results obtained in NISQ-era quantum simulations, where device decoherence  remains a limitation.

As a demonstration of the potential of SeqHT,
we have performed quantum simulations of the adiabatic preparation of the ground state of 
$\lambda\phi^4$ from that of the harmonic oscillator
using a time-ordered Trotterization evolution on IBM's quantum computer  {\tt ibm\_sherbrooke}.
Specifically, we have computed the expectation values of Pauli strings 
of two $\hat Z$ operators in these prepared state, 
and compared with exact results from classical computations.
These results, 
with cutoff sequency 
tuned to remain within overall error tolerances,
generally have smaller uncertainties than those obtained with the full evolution.
Further, they are in better agreement with the classically-computed results.

As part of the study of the convergence of SeqHT, we considered the convergence of the measurement of quantum computational complexity using magic
(non-stabilizerness) in a single-site Gaussian wavefunction with sequency truncation.
The magic in a digitized Gaussian wavefunction is found to converge to 
a fixed value, with an error $\epsilon$ scaling 
as $\log \frac{1}{|\epsilon|} \sim {\cal O}(n_q^2)$
with increasing number of qubits used in its digitization.
Further, the magic is found to step-wise 
converge with increasing sequency cutoff, $\nu_{\rm cut}$, 
towards the exact value. 

When extending to multiple spatial sites in a scalar field theory, the hopping term must also be included. SeqHT can be applied here by treating each site individually, and then combining the truncated representation.
In more general systems with non-diagonal operators, SeqHT can become feasible through suitable basis changes or transformations, for example, by applying the Quantum Fourier Transform or by using circuits to transform into the GHZ basis. Because SeqHT is intrinsically associated with a qubit hierarchy, any encoding that preserves differential structure allows for practical application.

SeqHT is an organizational method based on the structure of wavefunctions and probes.
We have demonstrated its utility in the context of adiabatic state preparation of the $\lambda\phi^4$ ground state using Trotterized evolution.
However, it has more general applicability as it does not depend on the choice of methods or algorithms.
Consequently, we anticipate that it can be used in 
simulations of systems exhibiting hierarchies of scales,
which encompass many systems of physical interest. 
In the context of nuclear physics applications, 
SeqHT may,
for example,
accelerate quantum simulations of 
mean-field descriptions of nuclear matter and 
relativistic hydrodynamics describing heavy-ion collisions.
Further, because its efficacy is related to the structure and dynamics of physical systems, 
we anticipate that SeqHT can be 
fruitfully implemented in digital, analog and hybrid
quantum simulations.

% ========== Conclusion

\chapter{Conclusion}

The aim of this dissertation has been to advance the digital quantum simulation of real-time dynamics in interacting quantum field theories, targeting problems in nuclear and high-energy physics that lie beyond the reach of classical computation. The work spans the full pipeline, from theoretical formulation through algorithm and circuit design to demonstration on present-day superconducting hardware, extracting observables from noisy devices and validating them against classical benchmarks. The long-term goal is the first-principles simulation of real-time dynamics in quantum chromodynamics (QCD) and, ultimately, the Standard Model. 

Chapter~\ref{chap:energy_loss} proposed a framework to simulate energy loss and hadronization in a 1+1D \sutwo lattice gauge theory. Heavy-quark fields were included as additional degrees of freedom to capture the color entanglement between the light quarks and the dense background. Classical lattice QCD methods, such as hadronic interpolating operators and domain decomposition, were used to prepare the initial state on the quantum computer together with Adaptive Variational Quantum Eigensolver (ADAPT-VQE), and the motion of heavy quarks was implemented via fermionic swap operators. This process was simulated on {\tt ibm\_pittsburgh} using 18 qubits with a two-qubit gate depth of around 400, and the charge distribution and energy loss were measured. This framework extends straightforwardly to 1+1D SU(3), at the cost of additional qubits and longer-range circuits. 

In search of formulations that are efficient at all gauge couplings and extensible to higher dimensions, Chapter~\ref{chap:gauge_fix} developed the quantum simulation of an alternative Hamiltonian formulation: the fully gauge-fixed Hamiltonian in the mixed basis. The focus of this chapter was to examine the viability of simulating this continuous theory on a digital quantum computer. The continuous gauge fields were mapped onto qubit registers, reaching $10^{-3}$ precision in low-energy observables with as few as three qubits per plaquette. The dynamics were then simulated on {\tt ibm\_fez} using five qubits, and the measured magnetic energy was in agreement with classical benchmarks with around $10^{-2}$ precision. This established a complete pipeline from the continuous formulation to its hardware simulation, while also finding that the first-derivative terms limit the achievable precision, which will require careful considerations when extending to larger systems.

To use resources more efficiently on present-day hardware, Chapter~\ref{chap:seqht} introduced a systematically-improvable truncation method that exploits the hierarchy in the contributions of basis operators to low-energy observables. Sequency Hierarchy Truncation (SeqHT) was demonstrated in the adiabatic state preparation of the interacting ground state of $\lambda\phi^4$ theory on {\tt ibm\_sherbrooke}, reducing the two-qubit gate depth by around $30\%$. Because the shallower truncated circuit accumulates less hardware noise and decoherence, it outperformed the untruncated circuit on the device. Both the target observables and the quantum complexity, as measured by magic, were found to converge step-wise as the cutoff sequency is raised. Beyond circuit truncations, SeqHT serves as a broader organizational tool for systems exhibiting hierarchies of scales.

As one-dimensional studies of QCD mature and quantum hardware capabilities improve, a natural next step is to explore simulations in two and higher dimensions with dynamical gauge fields. This will require developing formulations and encoding schemes that balance gauge symmetry preservation, resource efficiency, and compatibility with near-term devices. Another direction is to incorporate quantum error correction in quantum simulations using codesign principles, exploring how physical redundancies and symmetries can act as stabilizers to sustain coherent dynamics over longer time scales. 

With continued progress in quantum hardware, algorithms, and lattice formulations, quantum simulation can move beyond validating against classical calculations towards its broader aim: 
a first-principles description of real-time dynamics of far-from-equilibrium matter. These
regimes are inaccessible to the Euclidean formulation of lattice QCD, where Monte Carlo
sampling is obstructed by the sign problem. Dynamical properties such as the energy loss and
hadronization of probes traversing the quark-gluon plasma are at present described through phenomenological modeling rather than
computed from the underlying theory with controlled, quantifiable uncertainties. Quantum
simulation offers direct access to these observables in real time, with systematically improvable errors. Its reach extends beyond comparison with experiment, which observes only late-time states. 
Quantum simulation can also make controlled Standard Model predictions,
with quantified uncertainties, against which searches for physics beyond the Standard Model
can be measured.
The author hopes that the frameworks, algorithms, and demonstrations presented here will be of use to others working toward this goal. 

% ==========   Appendices
%
\appendix

\raggedbottom\sloppy

% ========== Appendix A

\chapter{Energy Loss}
\section{Details on the Fermionic SWAP}
\label{app:fswap}
\noindent
In this appendix, 
we show how the value of the rotation in the FSWAP gate, Eq.~\eqref{eq:FSWAP}, cannot be arbitrary in these simulations.
To understand this limitation, it is helpful to consider a simple model system comprised of two sites $x=\{0,1\}$, defined in the ket (of one of the colors) $|\psi\rangle=|Q_0 Q_1\rangle$ in the occupation basis, with an initial condition of having the heavy quark residing at $x=0$, $|\psi\rangle_0=|\uparrow \downarrow \rangle$. 
In an attempt to move the heavy quark to $x = 1$ using continuous spatial translations, a FSWAP gate is applied, $\hat F(\theta)$, followed by time evolution, followed by a second FSWAP gate, $\hat F(\phi)$. 

Working in the reduced space with states $\{ |\uparrow\downarrow\rangle, |\downarrow\uparrow\rangle \}$, imagine that the evolution operator has the form 
\begin{equation}
\hat U = e^{i{\frac{\alpha}{2}}} e^{i  {\frac{\eta}{2}}\hat Z} 
\ ,
\end{equation}
where $\eta$ and $\alpha$ are phases resulting from the time evolution of the system between the applications of the two FSWAP gates.  
These phase shifts 
model the effect of the time evolution of the light quarks on the wavefunction including the heavy-quark.
The resulting state is 
\begin{align}
|\tilde\psi\rangle \  & =\ 
\hat F(\phi) \cdot \hat U \cdot \hat F(\theta) |\psi\rangle_0
\nonumber\\
& = e^{i\alpha/2}\ e^{i\theta/2}\ e^{i\phi/2}\
\left[\ 
|\uparrow\downarrow\rangle 
\left(
\cos{\frac{\eta}{2}} \cos{\frac{\theta+\phi}{2}} + i \sin{\frac{\eta}{2}} \cos{\frac{\theta-\phi}{2}}
\right)
\right.\nonumber\\
& \left.
-i|\downarrow\uparrow\rangle 
\left(
\cos{\frac{\eta}{2}} \sin{\frac{\theta+\phi}{2}} - i \sin{\frac{\eta}{2}} \sin{\frac{\theta-\phi}{2}}
\right)
\ \right]
    \ ,
    \label{eq:QQmove}
\end{align}
In particular, when the system with a heavy-quark at $x=0$ is in its ground state, when moved instantaneously to $x=1$ it is in a combination of excited states.
In the case where 
$\theta\ne\phi\ne\pi$ but
$\theta+\phi=\pi$, it would have been ideal for the heavy-quark to have moved from the $x=0$ site to the $x=1$ site, but this is not the situation when $\eta\ne 0$.
Forming the location of the heavy quark from the expectation value, from matrix elements of the operator
\begin{equation}
\hat R = \frac{1}{4} \left(\  2\hat I +\hat Z_2-\hat Z_1\ \right)
    \label{eq:Rop}
\end{equation}
in $|\tilde\psi\rangle$ gives
\begin{equation}
\langle\tilde\psi| \ \hat R \ |\tilde\psi\rangle = 
\frac{1}{2}
\left(
1 - \cos^2{\frac{\eta}{2}} \cos ({\theta+\phi})
 - \sin^2{\frac{\eta}{2}} \cos ({\theta-\phi})
\right)
\ .
\label{eq:RinPsitilde}
\end{equation}
Attempting to move the heavy-quark from $x=0$ to $x=1$ via an intermediate $x=\frac{1}{2}$ location, $\theta=\phi=\pi/2$, leads to the location of the heavy quark being
$\langle \hat R \rangle = 1-\sin^2{\frac{\eta}{2}}$.  
In contrast, making one move to the adjacent lattice site, 
with $\theta=\pi$, $\phi=0$, 
or with $\theta=0$, $\phi=\pi$, 
gives $\langle \hat R \rangle = 1$.

%%%%%%%%%%%%%%%%%%%%%%%%%%%%
\section{\texorpdfstring{Details of the $L=2$ System}{Details of the L=2 System}}
\label{app:L2}
\noindent
Details of calculations in the $L=2$ system that are summarized in Sec.~\ref{sec:L2} are presented in this appendix.
While this system follows relatively straightforwardly from the $L=1$ system, 
there are aspects relevant for simulations in larger systems, 
as discussed in the main text.
Following the same path as in Sec.~\ref{sec:L1}, we present results for the eigenstates, energies, spin-observables, and the variational operators and wavefunctions in the 
$n_Q=\{0,1\}$ sectors.

The wavefunction for a state describing  an $L=2$ system
is formed from a linear combination of occupation basis states, 
$|\ Q_r\  Q_g\ q_r\  q_g\  \overline{q}_r\ \overline{q}_g
\ \ Q_r\  Q_g\ q_r\  q_g\  \overline{q}_r\ \overline{q}_g\  \rangle$,
with staggered site labels $n=\{0,1,\ldots,5\}$.
The kinetic energy operator receives contributions from the light quarks, 
\begin{align}
\hat H_k = &
-\frac{1}{2}
\sum_{n=1,4}\sum_{c=0,1}
\left(\ 
\hat \sigma^+_{2n+c} \hat Z_{2n+1+c} \hat \sigma^-_{2n+2+c} 
\ +\ {\rm h.c.}
\right)
\nonumber\\
&
-\frac{1}{2}
\sum_{n=2}\sum_{c=0,1}
\left(\ 
\hat \sigma^+_{2n+c} \hat Z_{2n+1+c} \hat Z_{2n+2+c} \hat Z_{2n+3+c} \hat \sigma^-_{2n+4+c} 
\ +\ {\rm h.c.}
\right)
\ \ ,
\end{align}
where the second term comes for hopping between anti-quarks at site $x=0$ and quarks at $x=1$.
The mass term, using the same arguments as before, takes the form
\begin{equation}
\hat H_m = 
\frac{m_q}{2}
\sum\limits_{n=1,2,4,5}\sum_{c=0,1}
(-)^{\eta_n}\  \hat Z_{2n+c}
\ \ ,
\end{equation}
where $\eta_{1,4}=0$ and $\eta_{2,5} = 1$.
The gauge-field contribution to the Hamiltonian is,
\begin{align}
\hat H_g & = 
\frac{g^2}{2}
\sum\limits_{\substack{a=1,2,3}}
\left[\ 
\left(\hat{\mathcal{Q}}_0^{(a)} +  \hat{\mathcal{Q}}_1^{(a)}\right)^2
\ +\ 
\left(\hat{\mathcal{Q}}_0^{(a)} + \hat{\mathcal{Q}}_1^{(a)}+ \hat{\mathcal{Q}}_2^{(a)}\right)^2 \right.
\nonumber\\
\ &+\ 
\left. \left( \hat{\mathcal{Q}}_0^{(a)} + \hat{\mathcal{Q}}_1^{(a)} + \hat{\mathcal{Q}}_2^{(a)} + \hat{\mathcal{Q}}_3^{(a)} + \hat{\mathcal{Q}}_4^{(a)}\right)^2
\ \right]
\ \ , 
\nonumber\\
\hat H_\lambda & = 
\frac{\lambda^2}{2}
\sum\limits_{a=1,2,3}
\left( \hat{\mathcal{Q}}_0^{(a)} + \hat{\mathcal{Q}}_1^{(a)} + \hat{\mathcal{Q}}_2^{(a)} + \hat{\mathcal{Q}}_3^{(a)} + \hat{\mathcal{Q}}_4^{(a)} + \hat{\mathcal{Q}}_5^{(a)} \right)^2
\ \ ,
\end{align}
with the charge operators defined in Eq.~\eqref{eq:Qadefs},
which can be replaced with the more compact expression (see App.~\ref{app:explicitham})
\begin{equation}
\hat H_g = 
\frac{g^2}{2}
\sum\limits_{\substack{a=1,2,3}}
\left[\ 
\left( \hat{\mathcal{Q}}_0^{(a)} + \hat{\mathcal{Q}}_1^{(a)}\right)^2
\ +\ 
\left( \hat{\mathcal{Q}}_0^{(a)} + \hat{\mathcal{Q}}_1^{(a)} + \hat{\mathcal{Q}}_2^{(a)}\right)^2
\ +\ 
\left( \hat{\mathcal{Q}}_5^{(a)}\right)^2
\ \right]
\ \ ,
\end{equation}
with $\lambda\rightarrow \infty$ to enforce the color-neutrality constraint.

We use the Lanczos algorithm~\cite{Lanczos:1950zz} to generate the ground-state energies and wavefunctions.
Working with $g=1.0$ and $m_q=0.1$ in the $n_Q=0$ sector,
our starting state is the SC vacuum is
\begin{equation}
|SC; 2, 0\rangle = 
| \downarrow\downarrow\downarrow\downarrow\ \uparrow\uparrow\ 
\downarrow\downarrow\downarrow\downarrow\ \uparrow\uparrow\ \rangle
\ \ .
\label{eq:psiSCL2Q0}
\end{equation}
Iterating the Lanczos algorithm to convergence with the system Hamiltonian gives
\begin{align}
|I; 2, 0\rangle  = & \ 
0.47806 | \downarrow\downarrow\downarrow\downarrow\ \uparrow\uparrow\ 
\downarrow\downarrow\downarrow\downarrow\ \uparrow\uparrow\ \rangle
\nonumber\\
& +  
0.27134 | \downarrow\downarrow\downarrow\downarrow\ \uparrow\uparrow\ 
\downarrow\downarrow\downarrow\uparrow\ \uparrow\downarrow\ \rangle
\ +\ 
0.27134 | \downarrow\downarrow\downarrow\uparrow\ \uparrow\downarrow\ 
\downarrow\downarrow\downarrow\downarrow\ \uparrow\uparrow\ \rangle
\nonumber\\
& -
0.27134 
| \downarrow\downarrow\downarrow\downarrow\ \uparrow\uparrow\ 
\downarrow\downarrow\uparrow\downarrow\ \downarrow\uparrow\ \rangle
\ -\ 
0.27134 
| \downarrow\downarrow\uparrow\downarrow\ \downarrow\uparrow\ 
\downarrow\downarrow\downarrow\downarrow\ \uparrow\uparrow\ \rangle
\nonumber\\
& + 
0.22182 
| \downarrow\downarrow\uparrow\uparrow\ \downarrow\downarrow\ 
\downarrow\downarrow\downarrow\downarrow\ \uparrow\uparrow\ \rangle
\ +\ 
0.22182 
| \downarrow\downarrow\downarrow\downarrow\ \uparrow\uparrow\ 
\downarrow\downarrow\uparrow\uparrow\ \downarrow\downarrow\ \rangle
\ +\ \cdots
\ \ ,
\end{align}
which exhibits the structure of mesons and baryon-antibaryon-pair excitations above the SC vacuum.
The ground-state energy of 
$E_0=-1.5179$, 
with corresponding energy density of 
$E_0/L\equiv \epsilon_0=-0.7589$.
The matrix elements of the components of the Hamiltonian density are
\begin{equation}
\langle \hat H_k \rangle/L \ =\ 
-1.04757
\ ,\ 
\langle \hat H_m \rangle/L \ =\ 
+0.13346
\ ,\ 
\langle \hat H_g \rangle/L = 
+0.15512
\ ,\ 
\langle \hat H_{\rm ext.}\rangle/L \ =\ 0
\ \ ,
\end{equation}
and the resulting spin alignments are 
\begin{equation}
\begin{aligned}
\langle \hat Z_j \rangle = &
\{  -1, -1, -0.38569, -0.38569, 0.279695, 0.279695,
\\ & -1, -1,
-0.279695, -0.279695, 0.38569, 0.38569 \}
\ \ .
\end{aligned}
\end{equation}

There are three meson operators (two acting on a single spatial site, $\hat O_{M0}^{(0)}$ and $\hat O_{M0}^{(1)}$, and one between sites, $\hat O_{M1}^{(0,1)}$), and three baryon operators with analogous actions ($\hat O_{B0}^{(0)}$, $\hat O_{B0}^{(1)}$, and $\hat O_{B1}^{(0,1)}$) that can be used to prepare the interacting ground state with sub-percent precision,
\begin{align}
    \hat O_{M0}^{(0)} & \equiv
    \hat I^2 \left(
\hat X \hat Z \hat Y \hat I 
- \hat Y\hat Z \hat X \hat I
+ \hat I \hat X \hat Z \hat Y
- \hat I \hat Y \hat Z \hat X \right) \hat I^6 \ ,
    \nonumber\\
\hat O_{M0}^{(1)} & \equiv
\hat I^8 \left(
    \hat X\hat Z \hat Y \hat I 
    - \hat Y\hat Z \hat X \hat I
    + \hat I \hat X\hat Z \hat Y  
    - \hat I\hat Y\hat Z \hat X \right) \ ,
        \nonumber\\
    \hat O_{M1}^{(0,1)} & \equiv
    -\hat I^4
    \left(
    \hat X\hat Z^3 \hat Y \hat I
    -\hat Y\hat Z^3 \hat X \hat I
+
    \hat I\hat X   \hat Z^3 \hat Y
    -\hat I \hat Y  \hat Z^3 \hat X
\right)  \hat I^2 \ ,
        \nonumber\\
    \hat O_{B0}^{(0)} & \equiv 4 i \hat I^2\left(
    \hat\sigma^+\hat\sigma^+\hat\sigma^-\hat\sigma^-
    -
    \hat\sigma^-\hat\sigma^-\hat\sigma^+\hat\sigma^+ \right)
    \hat I^6 \ ,
        \nonumber\\
    \hat O_{B0}^{(1)}  & \equiv 4 i \hat I^8\left(
    \hat\sigma^+\hat\sigma^+\hat\sigma^-\hat\sigma^-
    -
    \hat\sigma^-\hat\sigma^-\hat\sigma^+\hat\sigma^+ \right) \ ,
        \nonumber\\
    \hat O_{B1}^{(0,1)}  & \equiv -4 i \hat I^4\left(
     \hat\sigma^+\hat\sigma^+ \hat{I^2}\hat\sigma^-\hat\sigma^-
    -
    \hat\sigma^-\hat\sigma^-\hat{I^2} \hat\sigma^+\hat\sigma^+ \right)\hat I^2
\ .
\label{eq:OPcomsL2}
\end{align}
To identify the optimal operator ordering and subsequent layer-by-layer parameter optimizations, the meson operators are applied first
to the SC vacuum in Eq.~\eqref{eq:psiSCL2Q0},
followed by the baryon operators, to give Eq.~\eqref{eq:L2UNITQ0},
\begin{equation}
\hat U = 
e^{-i \theta_5 \hat O_{B1}^{(0,1)}}
e^{-i \theta_4 \hat O_{M1}^{(0,1)}}
e^{-i \theta_3 (\hat O_{B0}^{(0)}+\hat O_{B0}^{(1)})}
e^{-i \theta_2 (\hat O_{M0}^{(0)}+\hat O_{M0}^{(1)})}
e^{-i \theta_1 \hat O_{M1}^{(0,1)}}
\ .
\label{eq:L2UNITQ0app}
\end{equation}
The form of this variational ansatz is motivated by the symmetries of the system - the translational invariance (modulo OBCs) and color neutrality.
Table~\ref{tab:L2VQEmanuelQ0} shows the results of tuning the parameters defining the wavefunction ansatz in Eq.~\eqref{eq:L2UNITQ0} to the exact $L=2$, $n_Q=0$ wavefunction.
\begin{table}[htbp]
\renewcommand{\arraystretch}{1.1}
\small
\begin{tabularx}{\textwidth}{| Y || Y | Y | Y | Y | Y |  Y |}
 \hline
 {\rm Quantity}
 & $\hat I$
 & $\hat O_{M1}^{(0,1)}$
 & $\hat O_{M0}^{(0)}+\hat O_{M0}^{(1)}$
 & $\hat O_{B0}^{(0)}+\hat O_{B0}^{(1)}$
 & $\hat O_{M1}^{(0,1)}$
 & $\hat O_{B1}^{(0,1)}$
 \\
\hline\hline 
 $\infiL$
& 0.3857 
& 0.3479 & 0.0202 & 0.0133 & 0.0020  & 0.0007\\
\hline
\hline
$\theta_i$ 
& - 
& 0.1907 & & & &\\
 \hline
& - 
& 0.1291 & 0.2967 & & &\\
 \hline
& - 
& 0.1291 & 0.2543 & 0.0469 & &\\
 \hline
& - 
& 0.2264 & 0.2802 & 0.0588 & -0.1498 &\\
 \hline
& - 
& 0.2316 & 0.2790 & 0.0637 & -0.1691 &  0.0289 \\
 \hline
 \hline
\end{tabularx}
\caption{
Results from  fitting the variational wavefunction  in Eq.~\eqref{eq:L2UNITQ0} with $m_q=0.1$ and $g=1.0$ through five layers, optimized at each level.  
Shown are the infidelity density $\infiL$,  defined in Eq.~\eqref{eq:infidelitydens}, and fit angles $\theta_i$, for the $n_Q=0$ state with $L=2$.
}
\label{tab:L2VQEmanuelQ0}
\end{table}
The expectation values $\langle \hat Z_j \rangle$ are given in Table~\ref{tab:L2Q0Z} and displayed in Fig.~\ref{fig:L2nQ01Z}.
They show the convergence with the addition of each layer of the state preparation in Eq.~\eqref{eq:L2UNITQ0}, and subsequent re-optimization.
\begin{table}[htbp]
\renewcommand{\arraystretch}{1.2}
\setlength{\tabcolsep}{.5pt}
\resizebox{\textwidth}{!}{
\begin{tabularx}{1.3\textwidth}{|c || Y || Y | Y | Y | Y | Y |  Y | Y | Y |  Y | Y | Y |}
 \hline
 state
 & 0
 & 1
 & 2
 & 3
 & 4
 & 5
 & 6
 & 7
 & 8
 & 9
 & 10
 & 11
 \\
\hline\hline 
$|SC; 2, 0\rangle$ 
&  -1 & -1 & -1 & -1 & 1 & 1 & -1 & -1 & -1 & -1 & 1 & 1 \\
layer-1 
&  -1 & -1 & -1 & -1 & 0.7220 & 0.7220 & -1 & -1 & -0.7220 & -0.7220
   & 1 & 1 \\
layer-2 &  -1 & -1 & -0.4127 & -0.4127 & 0.2810 & 0.2810 & -1 & -1 & -0.2810
   & -0.2810 & 0.4127 & 0.4127 \\
layer-3 &  -1 & -1 & -0.4048 & -0.4048 & 0.2731 & 0.2731 & -1 & -1 & -0.2731 &
   -0.2731 & 0.4048 & 0.4048 \\
layer-4 &  -1 & -1 & -0.3936 & -0.3936 & 0.2841 & 0.2841 & -1 & -1 & -0.2841
   & -0.2841 & 0.3936 & 0.3936 \\
layer-5 &  -1 & -1 & -0.3890 & -0.3890 & 0.2842 & 0.2842 & -1 & -1 & -0.2842
   & -0.2842 & 0.3890 & 0.3890 \\
$|I; 2, 0\rangle$ &  -1 & -1 & -0.3856 & -0.3856 & 0.2796 & 0.2796 & -1 & -1 & -0.2796
   & -0.2796 & 0.3856 & 0.3856 \\
 \hline
 \hline
\end{tabularx}}
\renewcommand{\arraystretch}{1}
\caption{
The $\langle \hat Z_j \rangle$ as a function of the fermion site $j$, starting from the SC vacuum, through the classically optimized layers, 
and comparing to the exact values,
for the $n_Q=0$ state with $L=2$, $m_q=0.1$ and $g=1.0$.
These results are displayed in Fig.~\ref{fig:L2nQ01Z}.
}
\label{tab:L2Q0Z}
\end{table}

In the $n_Q=1$ sector with the heavy quark located at $x=0$, 
the SC vacuum is
\begin{equation}
|SC; 2, 1\rangle = 
\frac{1}{\sqrt{2}}
\biggl[\ 
| \uparrow\downarrow\downarrow\uparrow\ \uparrow\uparrow\ 
\downarrow\downarrow\downarrow\downarrow\ \uparrow\uparrow\ \rangle
- 
| \downarrow\uparrow\uparrow\downarrow\ \uparrow\uparrow\ 
\downarrow\downarrow\downarrow\downarrow\ \uparrow\uparrow\ \rangle
\biggr]
\ \ ,
\label{eq:nq2L2}
\end{equation}
and is no longer spatially symmetric because of the heavy quark at $x=0$. 
However,  as with the $n_Q=0$ sector, 
to identify the optimal operator ordering and subsequent layer-by-layer parameter optimizations, 
we first implement the meson operators followed by the baryon operators.
The meson operators are sequentially implemented with distance from the heavy quark, 
followed by two baryon operators, and then followed by the closest meson operator,
\begin{equation}
\hat U  = 
e^{-i \theta_6 \hat O_{M0}^{(0)}}\ 
e^{-i \theta_5 \hat O_{B1}^{(0,1)}}\ 
e^{-i \theta_4 \hat O_{B0}^{(1)}}\ 
e^{-i \theta_3 \hat O_{M0}^{(1)}}\ 
e^{-i \theta_2 \hat O_{M1}^{(0,1)}}\ 
e^{-i \theta_1 \hat O_{M0}^{(0)}}
\ ,
\label{eq:L2UNITQ1app}
\end{equation}
applied to $|SC; 2, 1\rangle$, reproducing Eq.~\eqref{eq:L2UNITQ1} in the main text. 
The operator sequence has the mesons implemented sequentially away from the heavy quark, one at the boundary, baryon-antibaryon pairs sequentially are applied moving closer to the heavy quark, with a meson operator at the heavy quark.
The fit angles and infidelities are given in Table~\ref{tab:L2VQEmanuelQ1},
\begin{table}[htbp]
\begin{tabularx}{\textwidth}{| Y || Y | Y | Y | Y | Y |  Y | Y | Y | Y |}
 \hline
 {\rm Quantity}
 & $\hat I$
 & $\hat O_{M0}^{(0)}$
 & $\hat  O_{M1}^{(0,1)}$
 & $\hat O_{M0}^{(1)}$
 & $\hat O_{B0}^{(1)}$
 & $\hat O_{B1}^{(0,1)}$
 & $\hat O_{M0}^{(0)}$
 \\
\hline\hline 
 $\infiL$
& 0.336982 
& 0.2965 & 0.1855 & 0.02256 & 0.02048 & 0.01959 & 0.003619\\
\hline
\hline
$\theta_i$ 
& - 
& 0.2309 & & & & &\\
& - 
& 0.2096 & 0.2473 & & & &\\
& - 
& 0.2231 & 0.2152 & 0.2563 & & &\\
& - 
& 0.2231 & 0.2149 & 0.2318 & 0.03233 & &\\
& - 
& 0.2223 & 0.2025 & 0.2283 & 0.03208 &  0.02261 &\\
& - 
& 0.3862 & 0.2358 & 0.2282 & 0.03233 &  0.02613 & -0.1837\\
 \hline
 \hline
\end{tabularx}
\caption{
Results from  fitting the variational wavefunction  in Eq.~\eqref{eq:L2UNITQ1}
with $m_q=0.1$ and $g=1.0$ through six layers, optimized at each level.  
Shown are the infidelity density, $\infiL$, 
defined in Eq.~\eqref{eq:infidelitydens},
and fit angles, $\theta_i$, for the $n_Q=1$ state with $L=2$.
}
\label{tab:L2VQEmanuelQ1}
\end{table}
and the expectation values $\langle \hat Z_j \rangle$ are given in Table~\ref{tab:L2Q1Z} and displayed in Fig.~\ref{fig:L2nQ01Z}.  
They show the pattern of convergence with the addition of each layer of state preparation
in Eq.~\eqref{eq:L2UNITQ1}, and subsequent re-optimizations.
\begin{table}[htbp]
\renewcommand{\arraystretch}{1}
\setlength{\tabcolsep}{.5pt}
\resizebox{\textwidth}{!}{
\begin{tabularx}{1.3\textwidth}{|c || Y | Y | Y | Y | Y | Y |  Y | Y | Y |  Y | Y | Y |}
 \hline
 State
 & 0
 & 1
 & 2
 & 3
 & 4
 & 5
 & 6
 & 7
 & 8
 & 9
 & 10
 & 11
 \\
\hline\hline 
$|SC; 2, 1\rangle$
&  0 & 0 & 0 & 0 & 1 & 1 & -1 & -1 & -1 & -1 & 1 & 1 \\
layer-1 
&  0 & 0 & 0.1971 & 0.1971 & 0.8028 & 0.8028 & -1 & -1 & -1 & -1 & 1
   & 1. \\
layer-2 &  0 & 0 & 0.1660 & 0.1660 & 0.4148 & 0.4148 & -1 & -1 & -0.5809 &
   -0.5809 & 1 & 1 \\
layer-3 &  0 & 0 & 0.1839 & 0.1839 & 0.4941 & 0.4941 & -1 & -1 & -0.2752 &
   -0.2752 & 0.5971 & 0.5971 \\
layer-4 &  0 & 0 & 0.1844 & 0.1844 & 0.5049 & 0.5049 & -1 & -1 & -0.2843 &
   -0.2843 & 0.5948 & 0.5948 \\
layer-5 &  0 & 0 & 0.1854 & 0.1854 & 0.5061 & 0.5061 & -1 & -1 & -0.2796 &
   -0.2796 & 0.5880 & 0.5880 \\
   layer-6 & 0 & 0 & 0.1533 & 0.1533 & 0.5034 & 0.5034 & -1 & -1 & -0.2581 &
   -0.2581 & 0.6013 & 0.6013 \\ \hline
$|I; 2, 1\rangle$ &  0 & 0 & 0.1554 & 0.1554 & 0.4989 & 0.4989 & -1 & -1 & -0.2498 & -0.2498 & 0.5955 & 0.5955 \\
 \hline
 \hline
\end{tabularx}}
\caption{
The $\langle \hat Z_j \rangle$ as a function of fermion site $j$, 
from the SC vacuum, $|SC; 2, 1\rangle$,  
through the classically optimized layers, 
through to the exact values from $|I; 2, 1\rangle$
for the $n_Q=0$ state with $L=2$, $m_q=0.1$ and $g=1.0$.
These results are displayed in Fig.~\ref{fig:L2nQ01Z}.
}
\label{tab:L2Q1Z}
\end{table}
The energy of the ground state in this sector is 
$E_1=-1.162$, giving a hadron mass of 
$\Lambda_Q=E_1 - E_0 =0.3557$.

%%%%%%%%%%%%%%%%%%%%%%%%%%%%
\section{\texorpdfstring{Details of the $L=3$ System}{Details of the L=3 System}}
\label{app:L3}
\noindent
Further details related to the $L=3$ results presented in Sec.~\ref{sec:L3SP} are presented in this appendix.
Specifically, we
present results for the eigenstates, energies, spin-observables, color-entanglement, and variational operators and wavefunctions in the $n_Q=0,1$ sectors.

Working with $g=1.0$ and $m_q=0.1$ (for the sake of comparison), 
in the light-sector ($n_Q=0$),
the ground-state wavefunction is determined using the Lanczos algorithm
applied to the SC vacuum, 
$|SC; 3, 0\rangle$, requiring more than 50 iterations to converge,
\begin{align}
|I; 3, 0\rangle = & \
0.345173 | \downarrow\downarrow\downarrow\downarrow\ \uparrow\uparrow\ 
\downarrow\downarrow\downarrow\downarrow\ \uparrow\uparrow\ 
\downarrow\downarrow\downarrow\downarrow\ \uparrow\uparrow \rangle
\nonumber\\
& \ +\ 
0.19079 | \downarrow\downarrow\downarrow\uparrow\ \uparrow\downarrow\ 
\downarrow\downarrow\downarrow\downarrow\ \uparrow\uparrow\
\downarrow\downarrow\downarrow\downarrow\ \uparrow\uparrow \rangle
\nonumber\\
&
\ -\ 
0.19079 | \downarrow\downarrow\downarrow\downarrow\ \uparrow\uparrow\ 
\downarrow\downarrow\downarrow\downarrow\ \uparrow\uparrow\ 
\downarrow\downarrow\uparrow\downarrow\ \downarrow\uparrow \rangle
\nonumber\\
& \ +\ 
0.19079 
| \downarrow\downarrow\downarrow\downarrow\ \uparrow\uparrow\ 
\downarrow\downarrow\downarrow\downarrow\ \uparrow\uparrow\
\downarrow\downarrow\downarrow\uparrow\ \uparrow\downarrow \rangle
\nonumber\\
&
\ -\ 
0.19079 
| \downarrow\downarrow\uparrow\downarrow\ \downarrow\uparrow\ 
\downarrow\downarrow\downarrow\downarrow\ \uparrow\uparrow\
\downarrow\downarrow\downarrow\downarrow\ \uparrow\uparrow \rangle
\ +\ \cdots
\ \ .
\label{eq:L3exact}
\end{align}
It has an energy of $E_0=-2.3994$
(with a corresponding energy density of $\epsilon_0=-0.7998$),
and matrix elements of the components of the Hamiltonian density,
\begin{equation}
\langle \hat H_k \rangle/L \ =\ 
-1.0846
\ ,\ 
\langle \hat H_m \rangle/L \ =\ 
+0.1264
\ ,\ 
\langle \hat H_g \rangle/L = 
+0.1583
\ ,\ 
\langle \hat H_{\rm ext.}\rangle \ =\ 0
\ \ .
\label{eq:HdesL3}
\end{equation}
The $\langle \hat Z_j \rangle$,
exhibiting the expected reflection symmetry,
are 
\begin{align}
\langle \hat Z_j \rangle = &  \ 
\big\{-1, -1, -0.4411, -0.4411, 0.2877, 0.2877,
\nonumber  \\ 
& \ \ -1, -1,
-0.3744, -0.3744, 0.3744, 0.3744,
\nonumber \\
& \ \   -1, -1, -0.2877,
-0.2877, 0.4411, 0.4411\  
\big\}
\ \ .
\label{eq:ZnL3n0}
\end{align}

In the $n_Q=1$ sector with the heavy quark located at $x=0$, 
the SC wavefunction is
\begin{equation}
|I; 3, 1, x=0\rangle  = 
\frac{1}{\sqrt{2}}
\biggl[
| \uparrow\downarrow\downarrow\uparrow\ \uparrow\uparrow\ 
\downarrow\downarrow\downarrow\downarrow\ \uparrow\uparrow\
\downarrow\downarrow\downarrow\downarrow\ \uparrow\uparrow \rangle
\ -\  
| \downarrow\uparrow\uparrow\downarrow\ \uparrow\uparrow\ 
\downarrow\downarrow\downarrow\downarrow\ \uparrow\uparrow\
\downarrow\downarrow\downarrow\downarrow\ \uparrow\uparrow \rangle
\biggl]
\ .
\label{eq:SCL3n1x0}
\end{equation}
and the ``exact'' wavefunction, determined by the Lanczos algorithm, is 
\begin{align}
|I; 3, 1, x=0\rangle & = \
0.273228| \uparrow\downarrow\downarrow\uparrow\ \uparrow\uparrow\ 
\downarrow\downarrow\downarrow\downarrow\ \uparrow\uparrow\
\downarrow\downarrow\downarrow\downarrow\ \uparrow\uparrow \rangle
\nonumber\\
&
\ -\  
0.273228| \downarrow\uparrow\uparrow\downarrow\ \uparrow\uparrow\ 
\downarrow\downarrow\downarrow\downarrow\ \uparrow\uparrow\
\downarrow\downarrow\downarrow\downarrow\ \uparrow\uparrow \rangle
\nonumber\\
&
\ -\ 
0.161826 | \uparrow\downarrow\downarrow\uparrow\ \uparrow\downarrow\ 
\downarrow\downarrow\downarrow\uparrow\ \uparrow\uparrow\
\downarrow\downarrow\downarrow\downarrow\ \uparrow\uparrow\ \rangle
\nonumber\\
&
\ -\ 
0.161826 | \downarrow\uparrow\uparrow\downarrow\ \downarrow\uparrow\ 
\downarrow\downarrow\uparrow\downarrow\ \uparrow\uparrow\
\downarrow\downarrow\downarrow\downarrow\ \uparrow\uparrow\ \rangle
\ +\ \cdots
\ \ ,
\label{eq:L3exactn1}
\end{align}
with an energy of $E_1=-2.0806$.
This gives rise to a hadron mass of $\Lambda_Q = 0.3188$. 
The resulting spin alignments are 
\begin{align}
\langle \hat Z_j \rangle  = &  \
\big\{0, \ \ \ 0,\ \  0.1169, \ \ \ 0.1169, 0.4204, 0.4204, 
\nonumber\\
& \ -1, -1, -0.2705, 
-0.2705, 0.4939, 0.4939, 
\nonumber \\
& \ -1, -1, -0.2878, -0.2878,  0.5270, 0.5270
\big\}
\ \ .
\end{align}
\begin{table}[htbp]
\renewcommand{\arraystretch}{1}
\setlength{\tabcolsep}{.5pt}
\resizebox{\textwidth}{!}{
\begin{tabularx}{1.2\textwidth}{| Y || Y | Y | Y | Y | Y |  Y | Y | Y |Y | Y |  Y | Y | Y |}
 \hline
 {\rm Quantity}
 & $\hat I$
 & $\hat O_{M0}^{(0)} $
 & $\hat O_{M0}^{(2)} $
 & $\hat O_{M1}^{(0,1)} $
 & $\hat O_{B0}^{(2)} $
 & $\hat O_{M0}^{(1)}$
 & $\hat O_{B0}^{(1)}$
 & $\hat O_{B1}^{(0,1)}$
 & $\hat O_{M0}^{(0)} $
 & $\hat O_{M1}^{(1,2)} $
 & $\hat O_{M2}^{(1,2)}$
 \\
\hline\hline 
 $\infiL$
& 0.2835 
& & 0.2258 
& & 0.1557 
& 0.0889 & 0.0881  & 0.0875
& 0.0765 & 0.0171 & 0.0043
 \\
\hline
\hline
 ${\cal M}_2$ 
 & 0 
 & & 1.519(0)
 & & 1.993(0)
 & 3.04(2)
 & 3.06(3)
 & 3.07(3)
 & 3.21(3)
 & 4.07(4)
 & 4.16(3)
\\
\hline
\hline
 $\theta_i$ 
& - & 0.2270 & 0.2687 & & & &&&&&\\
 & - & 0.2024 & 0.2363 & 0.2644 & 0.0328 &&&&&&\\
 & - & 0.2141 & 0.2411 & 0.2369 & 0.0358 & 0.2280&&&&& \\
 & - & 0.2137 & 0.2375 & 0.2360 & 0.0380 & 0.2092&0.0268&&&&\\
 & - & 0.2165 & 0.2386 & 0.2203 & 0.0386 & 0.2072& 0.0261& 0.0247 &&&\\
 & - & 0.3706 & 0.2401 & 0.2591 & 0.0370 & 0.2102& 0.0235& 0.0278 & -0.1878 &&\\
 & - & 0.3753 & 0.2139 & 0.2675 & 0.0325 & 0.1831& 0.0187& 0.0401 & -0.1997 & 0.2002&\\
 & - & 0.3802 & 0.2200 & 0.2642 & 0.0270 & 0.1820& 0.0196& 0.0407 & -0.2000 & 0.2314& -0.0995\\
 \hline
 \hline
\end{tabularx}}
\caption{
Results from  fitting the variational wavefunction in Eq.~\eqref{eq:L3U}
with $m_q=0.1$ and $g=1.0$ through eight layers, optimized at each level.  
DDec of the $L=3$ system is used to determine the sequence of operators from the
$L=1$ and $L=2$ systems, along with their initial angles in optimization.
Shown are the infidelity density, $\infiL$, 
defined in Eq.~\eqref{eq:infidelitydens},
the stabilizer Renyi entropy ${\cal M}_2$
and fit angles $\theta_i$, for the $n_Q=1$ state with $L=3$.
}
\label{tab:L3DoDeQ1}
\end{table}
Table~\ref{tab:L3DoDeQ1} shows the results of a layer-by-layer optimization of  parameters in the variational wavefunction in Eq.~\eqref{eq:L3U}.
Table~\ref{tab:L3Q1Z} provides the numerical values of 
$\langle \hat Z_j\rangle$ obtained 
from optimizing each layer preparing 
$|I; 3, 1, x=0\rangle$ state with $m_q=0.1$ and $g=1.0$,
which are displayed in Fig.~\ref{fig:L3nQ1Z}.
\begin{table}[htbp]
\renewcommand{\arraystretch}{1.2}
\setlength{\tabcolsep}{.5pt}
\resizebox{\textwidth}{!}{
\begin{tabularx}{1.2\textwidth}{| c || Y | Y | Y | Y || Y | Y |  Y | Y || Y |  Y | Y | Y |}
 \hline
 State
 & 2
 & 3
 & 4
 & 5
 & 8
 & 9
 & 10
 & 11
 & 14
 & 15
 & 16
 & 17
 \\
\hline\hline 
$|SC; 3, 1\rangle$
 & 0 & 0 & 1 & 1  & -1 & -1 & 1 & 1  & -1 & -1 & 1 & 1
   \\
layer-1 &  
 0.1935 & 0.1935 & 0.8064 & 0.8064  & -1 & -1 & 1 & 1  & $-0.4727$ & $-0.4727$ & 0.4727 & 0.4727 
   \\
layer-2 &  
 0.1593 & 0.1593 & 0.3711 & 0.3711  & -0.5305 &
   -0.5305 & 1 & 1  & -0.4961 & -0.4961 & 0.4961 & 0.4961
   \\
layer-3  & 0.1705 & 0.1705 & 0.4447 & 0.4447  & -0.3033 &
   -0.3033 & 0.6880 & 0.6880 &  -0.4444 & -0.4444 & 0.4444 &
   0.4444 
   \\
layer-4  &   0.1728 & 0.1728 & 0.4501 & 0.4501  & -0.3069 &
   -0.3069 & 0.6839 & 0.6839  & -0.4533 & -0.4533 &
   0.4533 & 0.4533 \\
layer-5  & 0.1763 & 0.1763 & 0.4437 & 0.4437 & -0.2980 &
   -0.2980 & 0.6779 & 0.6779 &  -0.4558 & -0.4558 &
   0.4558 & 0.4558 \\
layer-6  &  0.1264 & 0.1264 & 0.4557 & 0.4557 &  -0.2747 &
   -0.2747 & 0.6926 & 0.6926 &  -0.4536 & -0.4536 &
   0.4536 & 0.4536 \\
layer-7  &  0.1115 & 0.1115 & 0.4253 & 0.4253 &  -0.3049 &
   -0.3049 & 0.5661 & 0.5661 &  -0.3581 & -0.3581 &
   0.5600 & 0.5600 \\
layer-8  &   0.1177 & 0.1177 & 0.4310 & 0.4310 &  -0.2834 &
   -0.2834 & 0.5035 & 0.5035 &  -0.2929 & -0.2929 & 0.5240
   & 0.5240 \\ \hline
$|I; 3, 1\rangle$
   & 0.1169 & 0.1169 & 0.4204 & 0.4204 &  -0.2705 &
   -0.2705 & 0.4940 & 0.4940 &  -0.2878 & -0.2878 &
   0.5270 & 0.5270 \\
\hline
\hline
\end{tabularx}}
\caption{
The $\langle \hat Z_j \rangle$ as a function of fermion site number $j$, starting from the $L=3$ SC ground state, 
$|SC; 3, 1, x=0\rangle$, 
through the classically optimized layers,  
and comparing to the exact values $|I; 3, 1, x=0\rangle$ with $m_q=0.1$ and $g=1.0$.
The values for the heavy-quark sites are
$\langle \hat Z_j \rangle=0$ for $j=\{0,1\}$, and
$\langle \hat Z_j \rangle=-1$ for $j=\{6,7,12,13\}$.
These results are displayed in Fig.~\ref{fig:L3nQ1Z}.
}
\label{tab:L3Q1Z}
\end{table}

Table~\ref{tab:L3Q1Iijtau4} displays the values of the mutual information and 4-tangles among the heavy quark and light quarks and anti-quarks for $L=3$.
\begin{table}[htbp]
\begin{tabularx}{\textwidth}{| c || Y | Y | Y ||}
 \hline
Quantity
& $x=0$
& $x=1$
& $x=2$
 \\
\hline\hline 
$I^{(0,x)}_q$ & 1.5182 & 0.0367 & $1.59\times 10^{-4}$\\
$I^{(0,x)}_{\bar{q}}$ & 0.0448 & $2.52\times 10^{-3}$ & $9.20\times 10^{-5}$\\
\hline
$\tau^{(0,x)}_{4q}$ & 0.618 & $8.03 \times 10^{-3}$ & $3.07\times 10^{-5}$\\
$\tau^{(0,x)}_{4\bar{q}}$ & $8.95 \times 10^{-3}$ & $4.17 \times 10^{-4}$ 
& $1.30\times 10^{-5}$ \\
\hline
\hline
\end{tabularx}
\caption{ The mutual information and 4-tangles among the heavy quark 
at $x=0$
and light quarks and anti-quarks for $L=3$ with $m_q=0.1$ and $g=1.0$.}
\label{tab:L3Q1Iijtau4}
\end{table}
%

%%%%%%%%%%%%%%%%%%%%%%%%%%%%
\section{\texorpdfstring{Details of the Heavy-Quark Motion for $L=3$}{Details of the Heavy-Quark Motion for L=3}}
\label{app:L3motion}
\begin{figure}[htbp]
    \centering
    \includegraphics[width=0.8\linewidth]{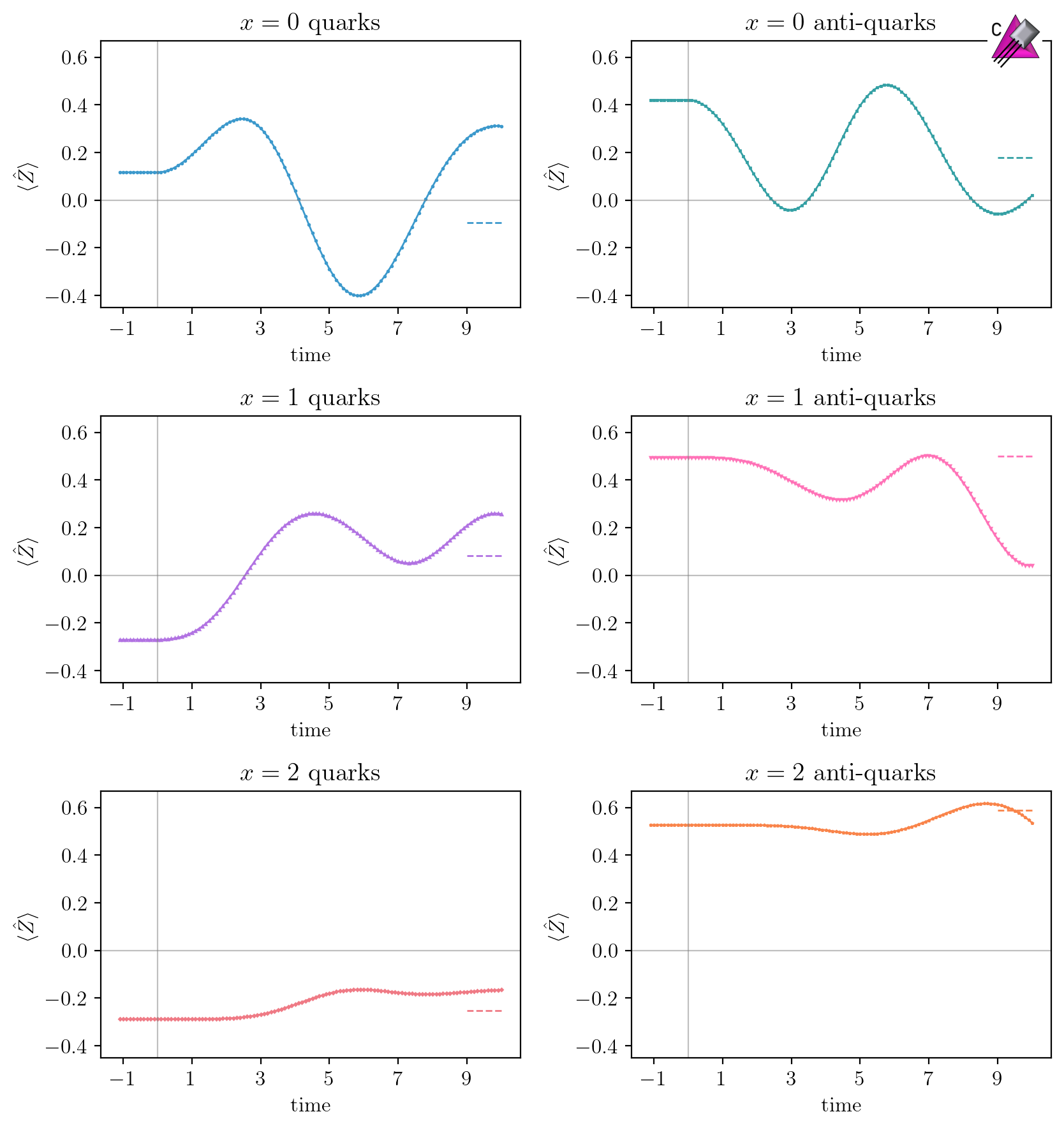}
\caption{The individual expectation values of the $\hat Z_j$ as a function of time in the $L=3$ system.  
At $t=0$, the heavy quark is instantaneously moved form $x=0$ to $x=1$.
}
    \label{fig:L3Ziind}
\end{figure}

\noindent
Figure.~\ref{fig:L3Hi} of the main text displays the time evolution of the 
$\langle \hat Z_j \rangle$ subsequent to moving the heavy quark from $x=0$ to $x=1$.
It is helpful to display the time dependencies individually, as shown in 
Fig.~\ref{fig:L3Ziind}.
\begin{table}[htbp]
\renewcommand{\arraystretch}{1.0}
\begin{tabularx}{\textwidth}{| c || Y | Y | Y ||}
 \hline
Quantity
& $x=0$
& $x=1$
& $x=2$
 \\
\hline\hline 
$\tau^{(1,x)}_{4q}$ & 0.0209 & $0.3704$ & $2.800\times 10^{-3}$\\
$\tau^{(1,x)}_{4\bar{q}}$ & $2.924 \times 10^{-3}$ 
& $1.238 \times 10^{-2}$ 
& $8.016\times 10^{-4}$ \\
\hline
\hline
\end{tabularx}
\caption{ The mutual information and 4-tangles among the heavy quark 
at $x=1$
and light quarks and anti-quarks for $L=3$ with $m_q=0.1$ and $g=1.0$.}
\label{tab:L3Q1tau4x1}
\end{table}
%

%%%%%%%%%%%%%%%%%%%%%%%%%%%%
\section{Circuit Construction}
\label{appen:cir}
\noindent 
This appendix is a supplement to
Sec.~\ref{sec:ApTpC} and gives further details 
about the construction of the quantum circuits used
in our simulations.

\begin{figure}[htbp]
    \centering
    \includegraphics[width=\textwidth]{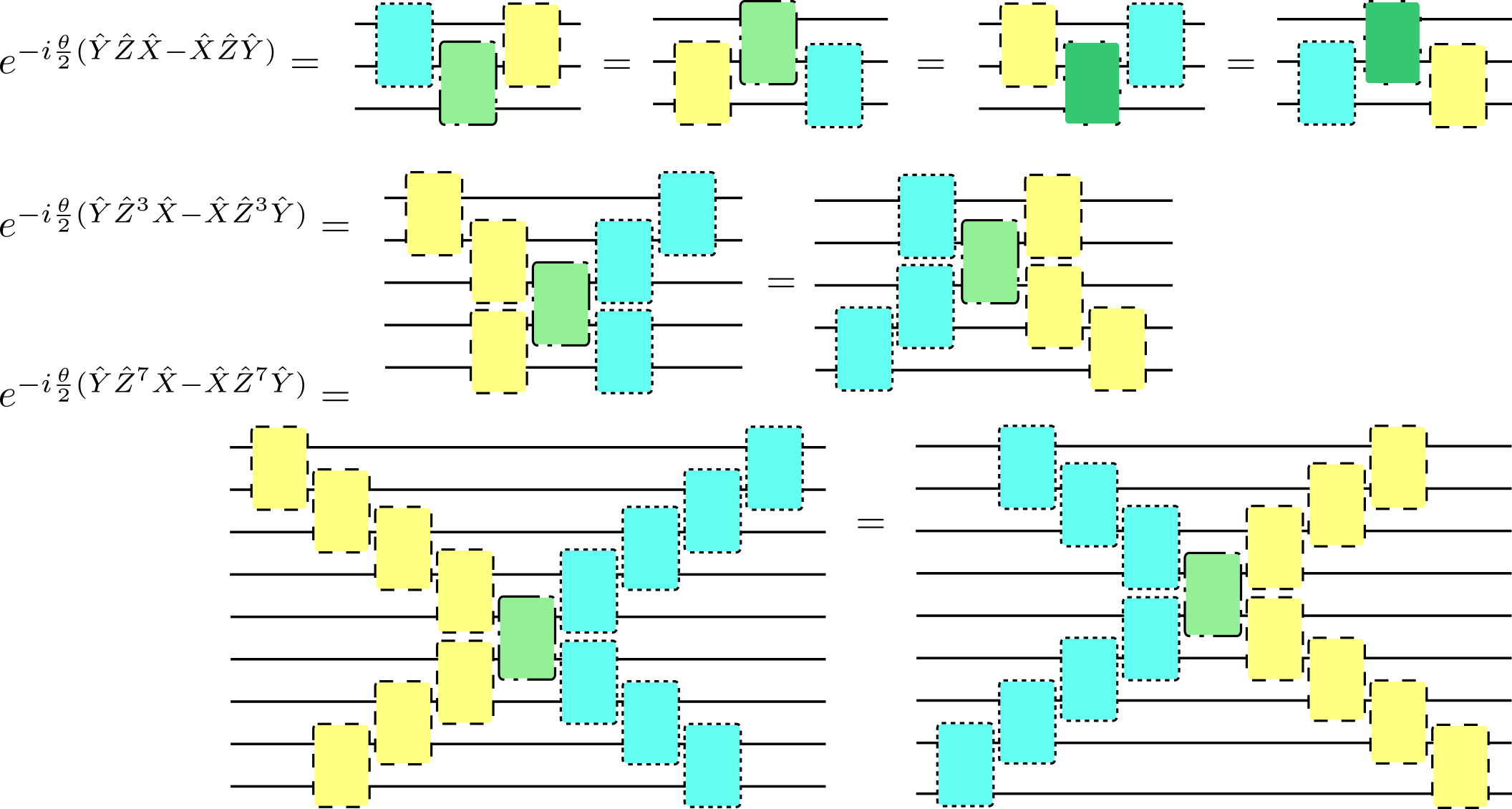}
    \caption{Different circuits implementing $e^{-i \frac{\theta}{2} (\hat Y \hat Z^n \hat X - \hat X \hat Z^n\hat Y )}$, the building blocks for implementing the meson operators defined in Eq.~\eqref{eq:mesOps}.
    The R-box composite elements in the circuit are defined in Fig.~\ref{fig:rbox}.
    }
    \label{fig:rboxbig}
\end{figure}

\begin{figure}[htbp]
    \centering
    \includegraphics[width=\textwidth]{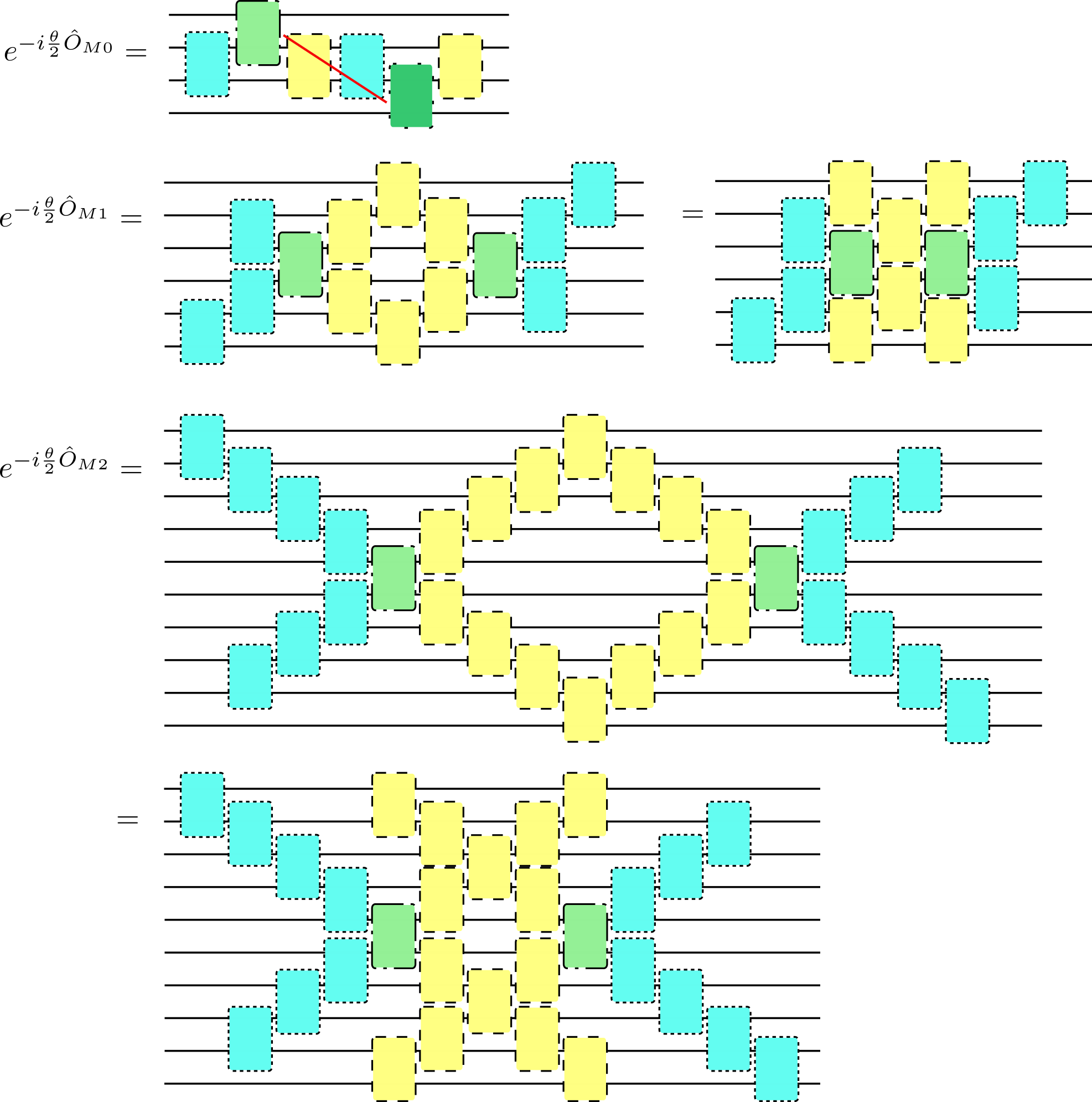}
    \caption{Circuits that implement unitary transformations
    from the meson operators.  
    The red line in first circuit denotes a cancellation between block operators.
    The CNOT-depth of these circuits are: 8, 28, 60, respectively. The R-box composite elements in the circuit are defined in  Fig.~\ref{fig:rbox}.
    }
    \label{fig:ohn}
\end{figure}

In the operator pool for state preparation, there are two types of operators, meson operators and baryon operators, as explained in Sec.~\ref{sec:L3SP} and Sec.~\ref{sec:StatePrepCir}. 
Meson operators are in the form of $e^{-i \frac{\theta}{2} \left((\hat Y \hat Z^n \hat X - \hat X \hat Z^n\hat Y )\hat I+\hat I(\hat Y \hat Z^n \hat X - \hat X \hat Z^n\hat Y )\right)}$.
Since $(\hat Y \hat Z^n \hat X - \hat X \hat Z^n\hat Y )\hat I$ commutes with $\hat I(\hat Y \hat Z^n \hat X - \hat X \hat Z^n\hat Y )$, they can be implemented exactly by subsequently applying the segments shown in Fig~\ref{fig:rboxbig}.
Figure~\ref{fig:rboxbig} also
shows examples of how unitary operators of the form
$e^{-i \frac{\theta}{2} (\hat Y \hat Z^n \hat X - \hat X \hat Z^n\hat Y )}$
can be implemented in multiple ways with different ``dressings'' of $R_{+}^{XY}$ or $R_{-}^{XY}$. 
This flexibility allows us to select a
variation to use in 
subsequent applications, 
and further reduce the depth of the circuit by taking advantage of the SU(2) symmetry.
Figure~\ref{fig:ohn} implements such products, along with circuit compression from identified cancellations. In particular, the variations chosen form two ``V''-shaped (or ``$\Lambda$''-shaped) operator structures in the middle that can be replaced with an ``X''-shape that reduces the circuit depth. 

\begin{figure}[htpb]
    \centering 
    \includegraphics[width = 0.6\textwidth]{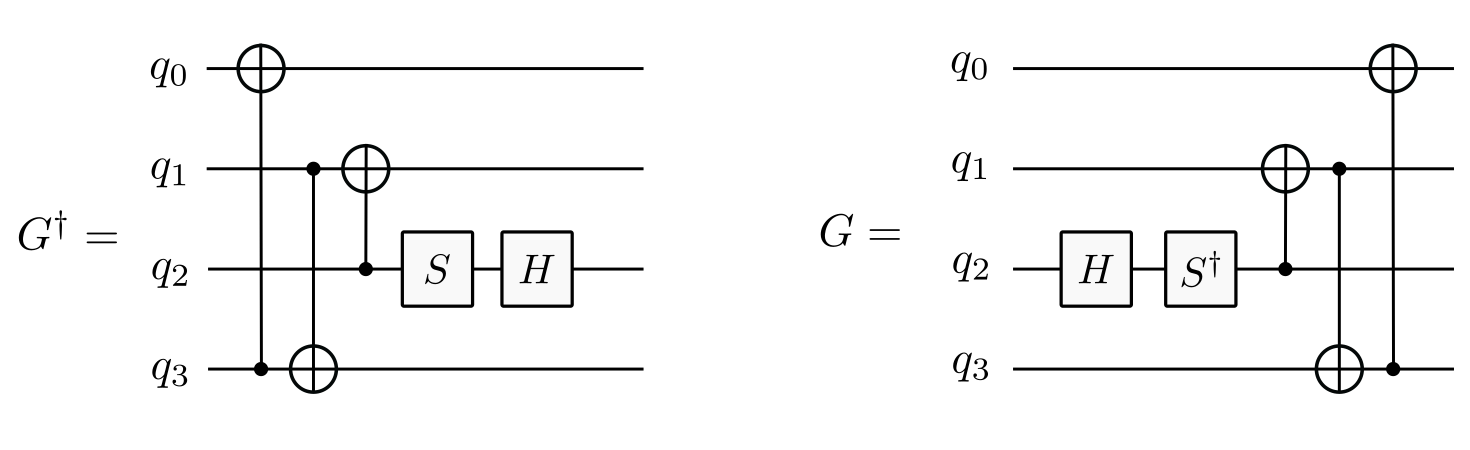}
    \caption{Circuits that furnish the GHZ transformation for baryon operators in state preparation.
    }
    \label{fig:ghz_state_prep}
\end{figure}

For baryon operators, GHZ transformations are used to transform the operator into the computational basis
\begin{align}
\label{eqn:stateprepGHZ}
\hat O_{B0} & \equiv 4i(\hat\sigma^+\hat\sigma^+\hat\sigma^-\hat\sigma^- - {\rm h.c.}) \nonumber \\
& = \frac{1}{2}(-\hat{X}\hat{X}\hat{Y}\hat{X}
+\hat{X}\hat{Y}\hat{X}\hat{X}
+\hat{Y}\hat{Y}\hat{X}\hat{Y}
+\hat{Y}\hat{X}\hat{X}\hat{X} \nonumber \\
& -\hat{X}\hat{X}\hat{X}\hat{Y} -\hat{X}\hat{Y}\hat{Y}\hat{Y}+\hat{Y}\hat{Y}\hat{Y}\hat{X}-\hat{Y}\hat{X}\hat{Y}\hat{Y}) \nonumber \\
& = \frac{1}{2} G(-\hat{I}\hat{I}\hat{Z}\hat{I}+\hat{I}\hat{Z}\hat{Z}\hat{I}-\hat{Z}\hat{Z}\hat{Z}\hat{I}+\hat{Z}\hat{Z}\hat{Z}\hat{Z} \nonumber \\
& -\hat{I}\hat{Z}\hat{Z}\hat{Z}+\hat{I}\hat{I}\hat{Z}\hat{Z}-\hat{Z}\hat{I}\hat{Z}\hat{Z}+\hat{Z}\hat{I}\hat{Z}\hat{I})G^{\dagger} \ ,
\end{align}
where $G$ is defined in Fig.~\ref{fig:ghz_state_prep}. The operators in the diagonal basis commute, so they can be arranged in a fashion that minimizes the number of CNOT gates required.
In Eq.~\eqref{eqn:stateprepGHZ}, the diagonal operators are listed in the order in which they are applied in the quantum circuit.

\begin{figure}[htbp]
    \centering
    \includegraphics[width=\textwidth]{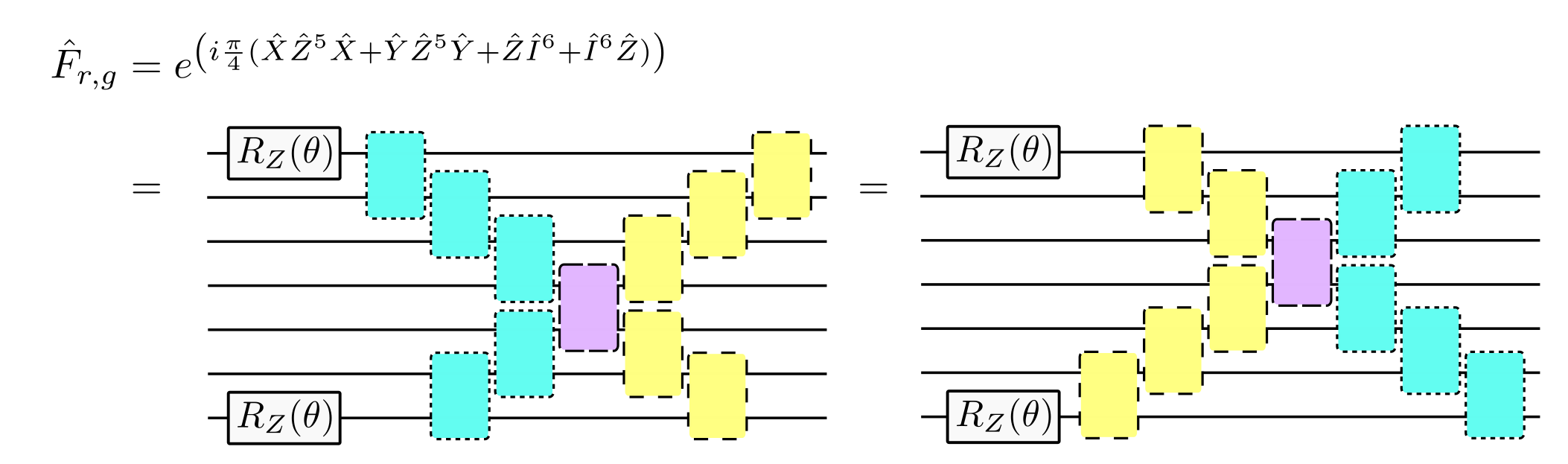}
    \caption{Circuits to implement the spatial translation of one color of SU(2) heavy quark between adjacent spatial lattice sites using an FSWAP circuit. 
    The $\theta$s appearing in circuit elements take the value of $-\pi/2$.
    }
    \label{fig:fswap_seg}
\end{figure}

The FSWAP gate for one color is shown in Fig.~\ref{fig:fswap_seg}, which is then combined in similar ways as for the meson operators in state preparation to perform the full Fermionic swap.

\begin{figure}[htbp]
    \centering
    \includegraphics[width=\textwidth]{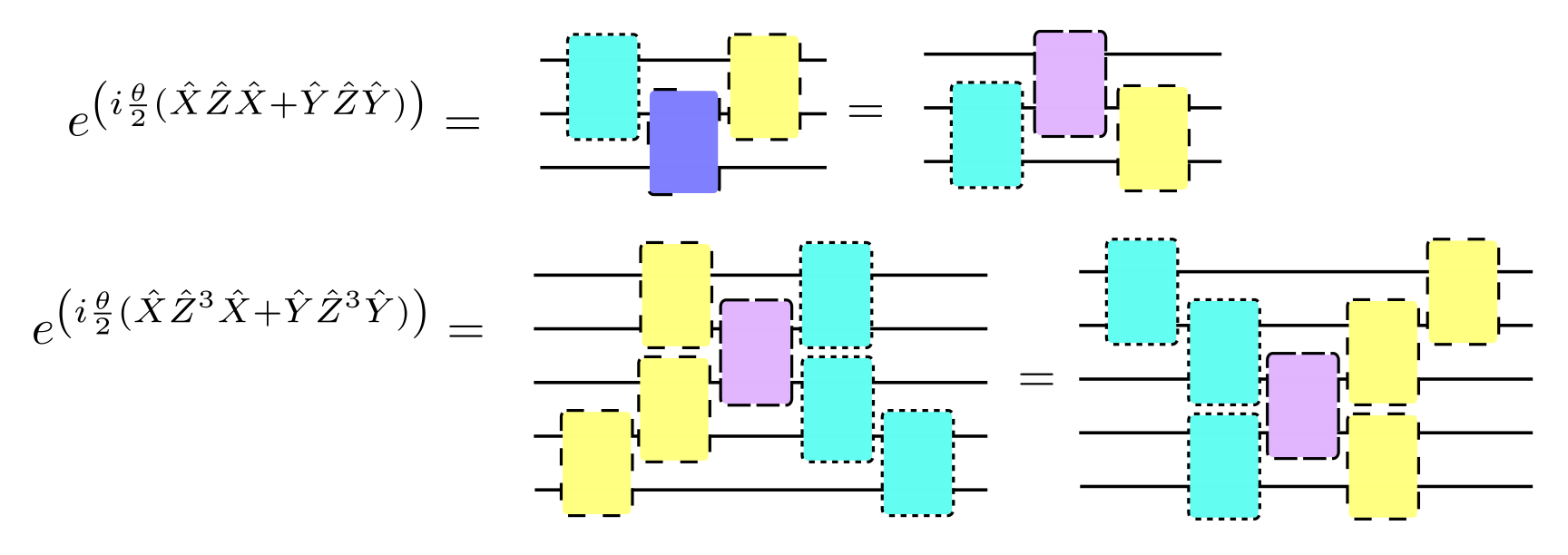}
    \caption{
    Elements of the quantum circuits used to implement the kinetic-energy operator.
    The different circuit elements are given in Fig.~\ref{fig:rbox}.}
    \label{fig:hkin}
\end{figure}

The single-color components used to construct the kinetic energy term in the Hamiltonian
are shown in Fig.~\ref{fig:hkin}. 
%

%%%%%%%%%%%%%%%%%%%%%%%%%%%%%%%%%%%%%%%%%%%%%%%%%%%
\section{Additional Circuit Resource Analysis}
\label{appen:resources}
\noindent
This appendix presents additional analysis of quantum circuit resources, including the CNOT depth of circuit segments, CNOT distributions over the lattice, and analysis of Trotter errors. 
The depth of each component in state preparation and time evolution is shown in Table~\ref{tab:time_evo_resource}. Note that evolution of the mass term has only single qubit rotations, and are not included in the table. 
\begin{table}[htpb]
    \centering
    \renewcommand{\arraystretch}{1}
    \begin{tabular}{|c || c | c | c | }
        \hline
        State prep. & All-to-All & Transpiled   \\
        \hline \hline
        $\hat O_{M0}$ & 6 & 6  \\
        $\hat O_{M1}$ & 14 & 14 \\
        $\hat O_{M2}$ & 26 & 27 \\
        $\hat O_{B0}$ & 14 & 35 \\
        $\hat O_{B1}$ & 14 & 43 \\
         \hline \hline
    \end{tabular}
    \begin{tabular}{|c || c | c | c |}
        \hline
        Time ev. & All-to-All & Transpiled  \\
        \hline  \hline 
        $\hat{H}_{k0}$ & 6 & 11\\
        $\hat{H}_{k1}$ & 14  & 19\\
        $\hat{H}_{g}$ (total) & 78 & 202\\ \hline
        $\hat{\mathcal{Q}}_n\hat{\mathcal{Q}}_{n+1}$ & 16 & 41 \\
        $\hat{\mathcal{Q}}_n\hat{\mathcal{Q}}_{n+2}$ & 16 & 43\\
        $\hat{\mathcal{Q}}_n\hat{\mathcal{Q}}_{n+3}$ & 16 & 49\\
         \hline \hline
    \end{tabular}
    \caption{
    The two-qubit-gate depth
    of each component in the state preparation (left table) and time evolution (right table). The ``Transpiled'' column correspond to the CZ depth after transpiling onto devices with heavy-hex connectivity. This number could vary between different layouts and transpilation optimization.
    }
\label{tab:time_evo_resource}
\end{table}

\begin{figure}[htbp]
    \centering
    \includegraphics[width=0.9\linewidth]{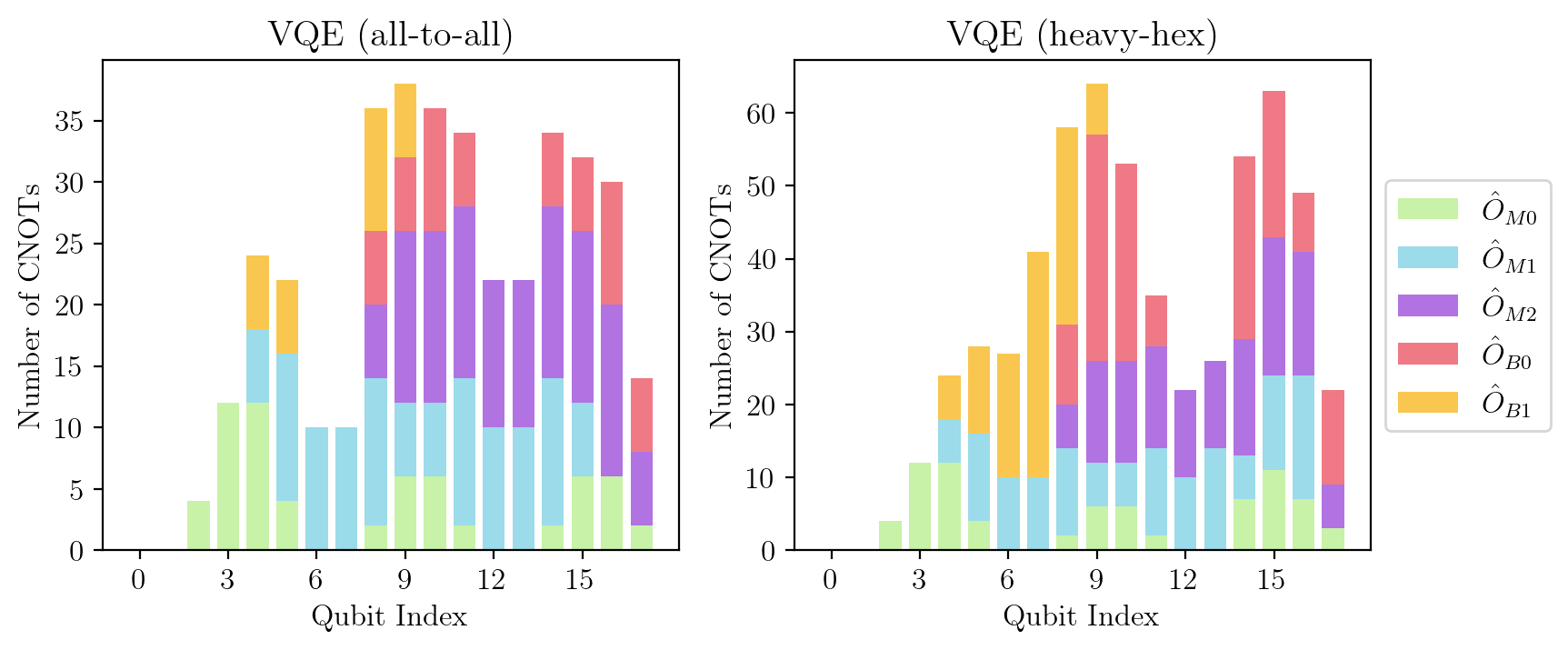}
    \caption{The CNOT depth distribution of components in the state preparation. 
    The left panel corresponds to  all-to-all connectivity,
    while the right panel is after transpiling to {\tt ibm\_pittsburgh}. 
    The number of CNOTs on each qubit is smaller than the total CNOT depth because some operators can't be executed in parallel.}
    \label{fig:vqe_cnot_dist}
\end{figure}

\begin{figure}[htbp]
    \centering
    \includegraphics[width=0.9\linewidth]{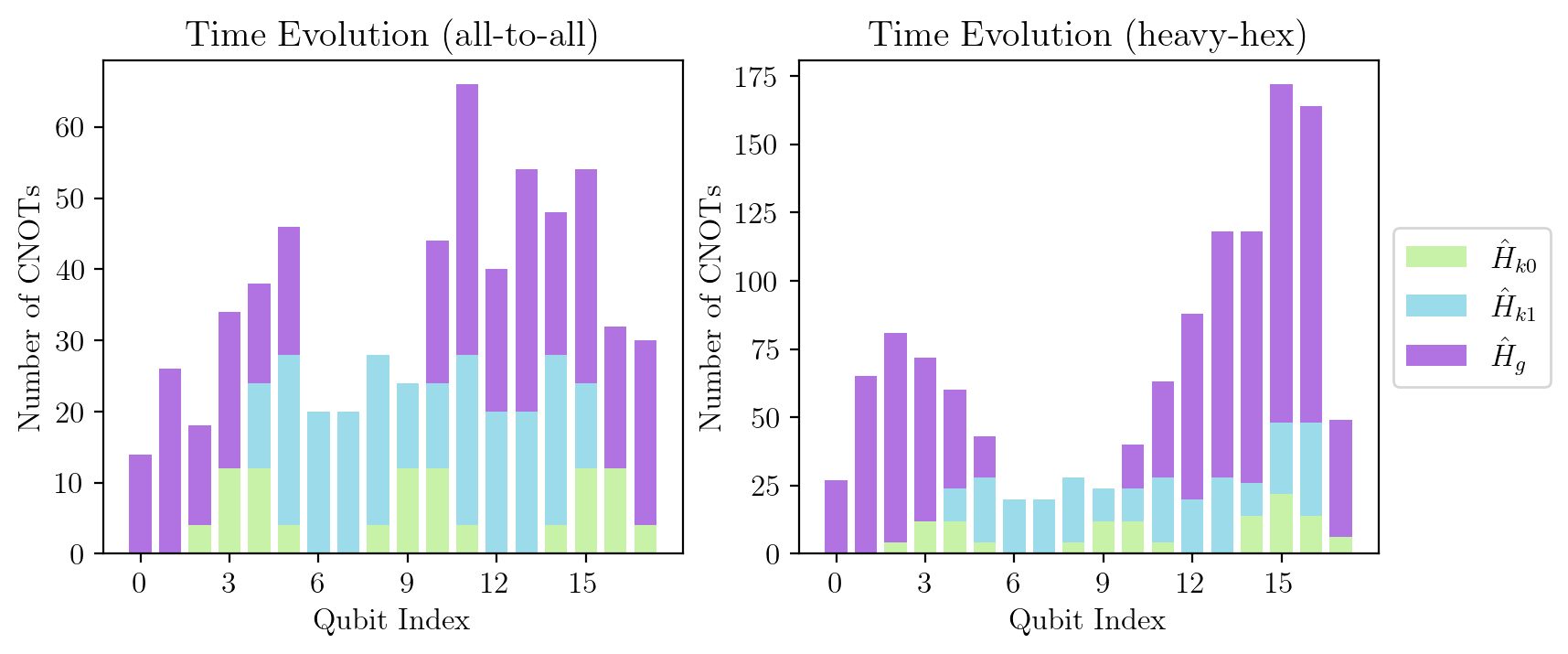}
    \caption{The CNOT depth distribution of components in the time evolution. 
    The left panel corresponds to  all-to-all connectivity, while the right panel is after transpiling to {\tt ibm\_pittsburgh}. 
    Because of the use of second-order Trotterized time evolution, the gate-counts associated with $\hat{H}_{k0}$ and $\hat{H}_{k1}$ count are doubled.
    The number of CNOTs on each qubit is smaller than the total CNOT depth because some operators can't be executed in parallel.    }
    \label{fig:time_evo_cnot_dist}
\end{figure}

\begin{figure}[htpb]
    \centering
    \includegraphics[width=0.9\linewidth]{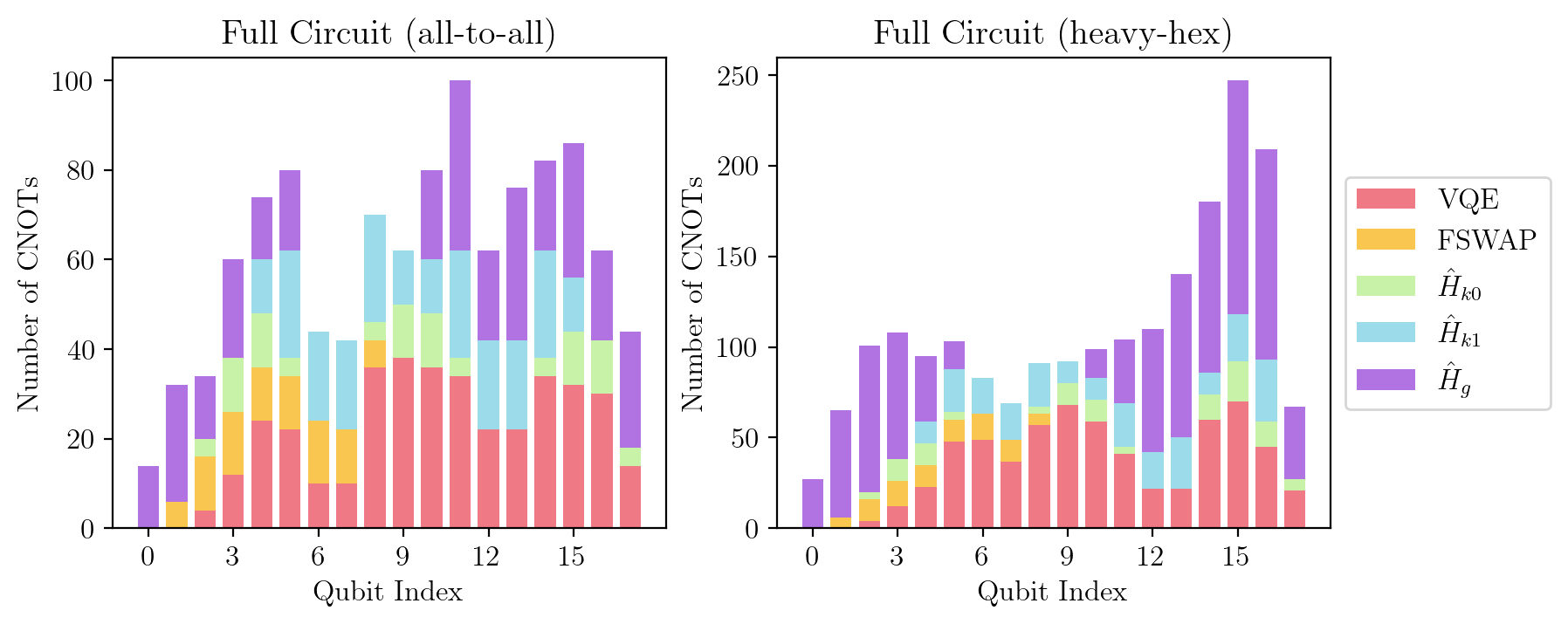}
    \caption{CNOT depth distribution of components in the complete process. 
    The left panel corresponds to  all-to-all connectivity, while the right panel is after transpiling to {\tt ibm\_pittsburgh}.
    }
    \label{fig:full_cnot_dist}
\end{figure}

CNOT depth distributions over qubits for state preparation (Fig.~\ref{fig:vqe_cnot_dist}), time evolution (Fig.~\ref{fig:time_evo_cnot_dist}) and for the full process (Fig.~\ref{fig:full_cnot_dist})  are analyzed, both for all-to-all connectivity and for the layout we use on {\tt ibm\_pittsburgh}
(with heavy-hex connectivity).
This allows for better understanding of the contributions to depth from each component over sectors of the lattice.
Upon transpilation to nearest-neighbor connectivity, the CNOT depth of $\hat O_{B1}$ and $e^{-it\hat H_g}$ increase significantly, while the other terms remain nearly unchanged, as they are already constructed for nearest-neighbor connectivity. 

\begin{figure}[tpb]
    \centering
    \includegraphics[width=0.85\linewidth]{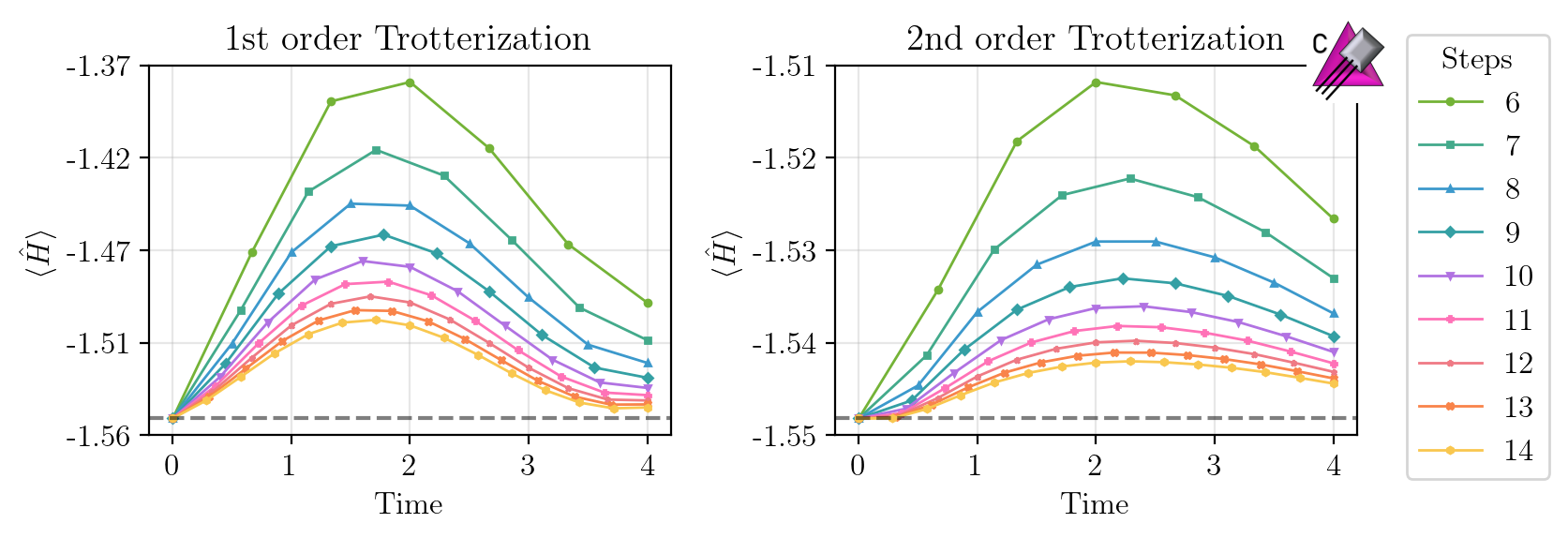}
    \caption{
    Energy conservation in the presence of Trotter errors.
    The left panel shows $\langle \hat H \rangle$ as a function of time for first-order Trotterization, 
    while the right panel shows the same for second-order Trotterization,
    using a number of Trotter steps between 6 to 19.
    }
    \label{fig:energy_scans}
\end{figure}

Violations of energy conservation
caused by Trotter errors over various step sizes and number of steps 
are shown in Fig.~\ref{fig:energy_scans} for both first- and second-order Trotterization. Second-order Trotterization preserves energy for longer times with comparable circuit depth.
%

%%%%%%%%%%%%%%%%%%%%%%%%%%%%
\section{The Hadamard Test for Computing Energies}
\label{app:Energycircs}
\begin{figure}[b!]
    \centering
       \begin{tikzpicture}
        \node[scale=1.1] {
            \begin{quantikz}[row sep={0.9cm,between origins}]
    \lstick{$|\psi\rangle$} &          &  & \gate{\hat U(t)}     &  & \\
    \lstick{$|0\rangle$}    & \gate{H} & \gate{S}  &  \ctrl{-1} & \gate{H} &  \meter{}
           \end{quantikz}
        };
    \end{tikzpicture}
    \caption{The circuit implementing the evolution operator via a 
    Hadamard test, 
    from which part or all of the total energy of the system can be extracted.}
  \label{eq:HadamardTestHcirc}
\end{figure}
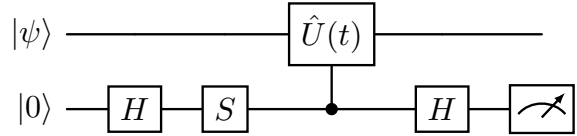

\noindent 
While we have measured energy losses using the method discussed in the main text, with results presented in Sec.~\ref{sec:QSresultsdEdx},
we have also assessed the implementation of the Hadamard test for determining the energy in the system
using classical simulations.
We have found it to be effective but inefficient,
requiring a suite of measurements and  subsequent extrapolation.
Including an ancilla with the register containing a state $|\psi\rangle$,
and acting with the controlled-evolution operator $C\hat U(t)$ 
(controlled by the ancilla qubit),
\begin{align}
    &|\phi\rangle =  |0\rangle |\psi\rangle
    \ ,\ 
     \hat S\hat H\otimes\hat I|\phi\rangle\ =\ 
     \frac{1}{\sqrt{2}} \left( |0\rangle + i|1\rangle\right)|\psi\rangle
     \ ,
     \nonumber\\
&(\hat H \otimes \hat I) \cdot C\hat U(t) \cdot (\hat S\hat H \otimes \hat I)|\phi\rangle\
     =
          \frac{1}{2}
     \left[
     |0\rangle \left(\hat I+i\hat U \right)|\psi\rangle
     \ +\ 
          |1\rangle \left(\hat I-i\hat U\right)|\psi\rangle
     \right]
     \ ,
\end{align}
implemented with the circuit shown in Fig.~\ref{eq:HadamardTestHcirc}.

The probabilities of the ancilla qubit measurements are 
\begin{equation}
{\rm Prob}(0) = \frac{1}{4} \langle \psi | 2 \hat I - i \left(\hat U^\dagger - \hat U\right) |\psi\rangle
\ ,\ 
{\rm Prob}(1) =  \frac{1}{4} \langle \psi | 2 \hat I + i \left(\hat U^\dagger - \hat U\right) |\psi\rangle
\ \ .
\end{equation}
Using the exact evolution operator $\hat U (t) = e^{-i t \hat H}$
leads to
\begin{equation}
{\rm Prob}(0) - {\rm Prob}(1) = t  \langle \psi | \hat H |\psi\rangle\ +\ {\cal O}(t^3)
\ \ .
\label{eq:fromprobstoH}
\end{equation}
By evaluating this quantity over a range of short times, 
$\langle\hat H\rangle$ can be determined.

For a Trotterized evolution operator, the extracted energy corresponds to that of an effective Hamiltonian, modified by the finite time step.  
A leading-order Trotterization will give rise to a correction  to Eq.~\eqref{eq:fromprobstoH} that is quadratic in time, giving a relative linear correction.
To mitigate this systematic error, a series of simulations should be performed over a range of time steps, with a polynomial extrapolation of the results to estimate the energy for a vanishing time step size.

\begin{figure}[htpb]
    \centering
    \includegraphics[width=0.4\linewidth]{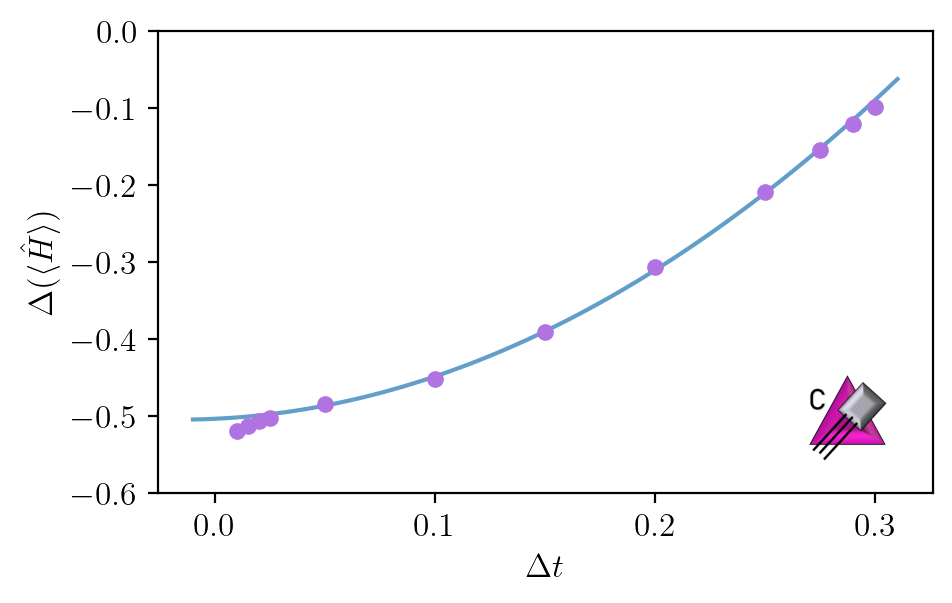}
    \caption{Results obtained from the  {\tt qiskit} simulator for the energy injected into the light quarks by moving the heavy quark from $x=0$ to $x=1$ in the $L=3$ system (without other matter present) (purple points), 
    with Hamiltonian parameters $m_q=0.1$ and $g=1.0$. 
    The blue curve corresponds to a (representative) quadratic fit to the results from $\Delta t= 0.05-0.275$.
    }
    \label{fig:L3x0x1SIM}
\end{figure}

Using the {\tt qiskit} simulator to measure the energy difference induced by the $x=0$ to $x=1$ displacement of the heavy quark without matter present,
we have implemented this technique over a range of time steps, obtaining the result shown in Fig.~\ref{fig:L3x0x1SIM}.
The results for small $\Delta t$ are subject to precision issues, and are not included in the fit. 
From these (classical) simulations, the central value of the 
extrapolated energy is found to be 
$\delta \langle \Delta \hat H_{g}  \rangle=-0.50(2)$ 
which is to be compared with the exact result of $0.5058$,
and the result that we have obtained from {\tt ibm\_pittsburgh} using the other method 
$\delta \langle \Delta \hat H_{g}  \rangle^{\rm IBM}=-0.5301(48)$. 
The quoted error is estimated from the range of values determined from varying the order of fit polynomials and selection of fit data.
While encouraged by the results and the fit displayed in Fig.~\ref{fig:L3x0x1SIM}, 
this makes clear that multiple sets of measurements are required, 
and an understanding of their systematic errors are needed for the method to provide reliable results.
Further, the required extrapolation adds additional errors that lead to a result that is inferior to that from the method used in this chapter, presented in Sec.~\ref{sec:QSresultsdEdx}.

%%%%%%%%%%%%%%%%%%%%%%%%%%%%
\section{Explicit Form of the Hamiltonian}
\label{app:explicitham}
\noindent 
The Hamiltonian for a single flavor of light and heavy quarks, with an arbitrary number of colors $N_c$ and sites $L$, is given by:
\begin{align}
    \hat{H}_{k} = & \ \frac{1}{2}\sum_{x=0}^{L-1}\sum_{c=0}^{N_c-1} \left[ \sigma_{i(3x+1,c)}^+ \left ( \bigotimes_{j=1}^{N_c-1}[-Z_{i(3x+1,c)+j} ] \right ) \sigma_{i(3x+1,c) + N_c}^- +\rm{h.c.} \right] \nonumber\\
    & \ + \frac{1}{2}\sum_{x=0}^{L-2}\sum_{c=0}^{N_c-1} \left[ \sigma_{i(3x+2,c)}^+ \left ( \bigotimes_{j=1}^{2 N_c-1}[-Z_{i(3x+2,c)+j} ] \right ) \sigma_{i(3x+2,c) + 2 N_c}^- +\rm{h.c.} \right] \ , \\
    \hat{H}_m  = & \ \frac{1}{2} \sum_{x=0}^{L-1} \sum_{c=0}^{N_c-1} \left(m_Q \left[  Z_{i(3x,c)} + 1 \right]+m_q \left[  Z_{i(3x+1,c)} + 1 \right]+m_q \left[  -Z_{i(3x+2,c)} + 1 \right]\right) \ , \\
    \hat{H}_g = & \ \frac{g^2}{2} \sum_{n=0}^{3L-2} \left( 2L-\lceil \tfrac{2n+1}{3} \rceil \right) \hat{\mathcal{Q}}^{(a)}_n \hat{\mathcal{Q}}^{(a)}_n + g^2 \sum_{n=0}^{3L-3} \sum_{m=n+1}^{3L-2} \left( 2L-\lceil \tfrac{2m+1}{3} \rceil \right) \hat{\mathcal{Q}}^{(a)}_n \hat{\mathcal{Q}}^{(a)}_m \ , \\
    \hat{H}_\lambda = & \ \frac{\lambda^2}{2} \sum_{n=0}^{3L-2} \hat{\mathcal{Q}}^{(a)}_n \hat{\mathcal{Q}}^{(a)}_n + \lambda^2 \sum_{n=0}^{3L-3} \sum_{m=n+1}^{3L-2} \hat{\mathcal{Q}}^{(a)}_n \hat{\mathcal{Q}}^{(a)}_m \ ,
\end{align}
with $i(n,c) = (N_c n+ c)$, and the products of the charges defined as~\cite{Farrell:2022wyt}
\begin{align}
    4 \hat{\mathcal{Q}}_{n}^{(a)} \, \hat{\mathcal{Q}}_{n}^{(a)} =& \ \frac{N_c^2-1}{2} - \left (1+\frac{1}{N_c} \right )\sum_{c=0}^{N_c-2}\sum_{c' = c+1}^{N_c-1} Z_{i(n,c)} Z_{i(n,c')} \ ,  \nonumber \\[4pt]
    8 \hat{\mathcal{Q}}_{n}^{(a)} \, \hat{\mathcal{Q}}_{m}^{(a)} =& \ 4 \sum_{c=0}^{N_c-2}\sum_{c'=c+1}^{N_c-1} \left[ \sigma^+_{i(n,c)} \ \left(\otimes Z \right)_{(n,c,c')} \ \sigma^-_{i(n,c')} \sigma^-_{i(m,c)} \ \left(\otimes Z \right)_{(m,c,c')} \ \sigma^+_{i(m,c')} + {\rm h.c.}
    \right] \  \nonumber \\
    &+ \sum_{c=0}^{N_c-1} \sum_{c'=0}^{N_c-1}\left (\delta_{cc'} - \frac{1}{N_c} \right)Z_{i(n,c)} Z_{i(m,c')}\ , \nonumber \\
    \left(\otimes Z \right)_{(n,c,c')}  \equiv & \ \bigotimes_{k=1}^{c'-c-1} Z_{i(n,c)+k}
    \ .
    \label{eq:QnfQmfpN}
\end{align}
As mentioned in Refs.~\cite{Farrell:2023fgd,Farrell:2024fit}, in the $Q=0$ sector, 
there is a freedom in the way Gauss' law can be imposed.
Instead of starting with ${\bf E}^{(a)}=0$ at one end of the lattice and working toward the other side of the lattice,
the contributions can be determined by starting from ${\bf E}^{(a)}=0$ at both ends, and working toward the middle.
In our case, it takes the following form,
\begin{equation}
\begin{aligned}
    \hat{H}_g = \sum_{b=0}^{L-2}E_b^2+\sum_{b=L}^{2L-1}E_b^2 & =\  \frac{g^2}{2} \sum _{n=0}^{\lfloor \tfrac{3L-1}{2} \rfloor } \left[\left(L- \lceil \tfrac{2n+1}{3}\rceil \right) \hat{\mathcal{Q}}^{(a)}_n \hat{\mathcal{Q}}^{(a)}_n \right. \nonumber \\
    & + \left. \left(L-\lceil \tfrac{2n-1}{3}\rceil \right) \hat{\mathcal{Q}}^{(a)}_{3L-n-1}\hat{\mathcal{Q}}^{(a)}_{3L-n-1} \right] \nonumber\\
    & + g^2 \sum _{n=0}^{\lfloor \tfrac{3 L}{2}\rfloor } \left[\sum_{m=n+1}^{\lfloor \tfrac{3 L-2}{2} \rfloor } \left(L-\lceil \tfrac{2m+1}{3} \rceil \right) \hat{\mathcal{Q}}^{(a)}_n \hat{\mathcal{Q}}^{(a)}_m \right. \nonumber \\
    & \left. + \sum_{m=1}^{n-1} \left(L+1-\lceil \tfrac{2n}{3}\rceil \right) \hat{\mathcal{Q}}^{(a)}_{3L-n} \hat{\mathcal{Q}}^{(a)}_{3L-m}\right]
\end{aligned}
\end{equation}
\begin{equation}
\begin{aligned}
    \hat{H}_g = \sum_{b=0}^{L-1}E_b^2+\sum_{b=L+1}^{2L-1}E_b^2 & =\  \frac{g^2}{2} \sum _{n=0}^{\lfloor \tfrac{3L}{2} \rfloor } \left[\left(L + 1 - \lceil \tfrac{2n+1}{3}\rceil \right) \hat{\mathcal{Q}}^{(a)}_n \hat{\mathcal{Q}}^{(a)}_n \right. \\
    & +\left. (1-\delta_{n,0})\left(L-\lceil \tfrac{2n}{3}\rceil \right) \hat{\mathcal{Q}}^{(a)}_{3L-n}\hat{\mathcal{Q}}^{(a)}_{3L-n} \right] \nonumber\\
    & + g^2 \sum _{n=0}^{\lfloor \tfrac{3 L}{2}\rfloor } \left[\sum_{m=n+1}^{\lfloor \tfrac{3 L}{2} \rfloor } \left(L-\lceil \tfrac{2m-2}{3} \rceil \right) \hat{\mathcal{Q}}^{(a)}_n \hat{\mathcal{Q}}^{(a)}_m \right. \\
    & + \left. \sum_{m=1}^{n-1} \left(L-\lceil \tfrac{2n}{3}\rceil \right) \hat{\mathcal{Q}}^{(a)}_{3L-n} \hat{\mathcal{Q}}^{(a)}_{3L-m}\right]
\end{aligned}
\end{equation}
%

%%%%%%%%%%%%%%%%%%%%%%%%%%%%
\section{Results from Simulations using IBM's Quantum Computers}
\label{app:IBMruns}
\noindent 
The results obtained from simulations performed using {\tt ibm\_pittsburgh} 
discussed in Sec.~\ref{sec:QSresults}
and displayed in Fig.~\ref{fig:IBMresultfig}
are shown in Table~\ref{tab:IBManalysis}. 
\begin{table}[htbp]
\renewcommand{\arraystretch}{1}
\setlength{\tabcolsep}{.5pt}
\resizebox{\textwidth}{!}{
\begin{tabularx}{1.5\textwidth}{|c|| Y | Y | Y | Y | Y | Y | Y | Y | Y |}
 \hline
\diagbox{Set up}{Qubit} & 0,1 & 2,3 & 4,5 & 6,7 & 8,9 & 10,11 & 12,13 & 14,15 & 16,17 \\
\hline\hline
\multirow{2}{*}{Raw} & -0.008(3) & 0.152(5) & 0.282(2) & -0.822(2) & -0.208(4) & 0.408(5) & -0.850(7) & -0.206(3) & 0.404(6) \\
 & 0.008(4) & 0.163(7) & 0.318(6) & -0.859(1) & -0.210(1) & 0.413(2) & -0.776(3) & -0.208(5) & 0.385(4) \\
\hline
ODR & 0.000(0) & 0.203(4) & 0.380(4) & -0.997(2) & -0.297(4) & 0.564(5) & -1.002(2) & -0.289(3) & 0.552(3) \\
\hline
Expected & 0.000 & 0.118 & 0.431 & -1.000 & -0.282 & 0.503 & -1.000 & -0.294 & 0.524 \\
\hline
\end{tabularx}}
\\
\resizebox{\textwidth}{!}{
\begin{tabularx}{1.5\textwidth}{|c|| Y | Y | Y | Y | Y | Y | Y | Y | Y |}
 \hline
\diagbox{Set up}{Qubit} & 0,1 & 2,3 & 4,5 & 6,7 & 8,9 & 10,11 & 12,13 & 14,15 & 16,17 \\
\hline\hline
\multirow{2}{*}{Raw} & -0.472(4) & 0.099(3) & 0.106(7) & 0.002(4) & -0.095(3) & 0.156(11) & -0.470(1) & -0.059(5) & 0.216(7) \\
 & -0.527(3) & 0.106(5) & 0.123(8) & 0.006(3) & -0.089(5) & 0.170(8) & -0.410(4) & -0.062(6) & 0.193(11) \\
\hline
ODR & -1.000(3) & 0.228(7) & 0.295(9) & 0.000(0) & -0.216(5) & 0.480(15) & -0.994(4) & -0.227(16) & 0.530(10) \\
\hline
Expected & -1.000 & 0.183 & 0.348 & 0.000 & -0.285 & 0.541 & -1.000 & -0.325 & 0.537 \\
\hline
\end{tabularx}}
\caption{
The values of $\langle \hat Z_j \rangle$ measured using  {\tt ibm\_pittsburgh} for the $L=3$ systems using Hamiltonian parameters $m_q=0.1$ and $g=1.0$.  An initial state was prepared with a heavy quark residing on the  $x=0$ spatial site (upper table). The heavy quark
is then moved to the $x=1$ spatial site, then the system is
time evolved forward to $t=1.0$ using one second-order Trotter step (lower table), as discussed in the main text. 
As discussed in Sec.~\ref{sec:chargeD}, the applied cuts are
$\lambda^{\rm mit}_{\rm cut} = \lambda^{\rm phys}_{\rm cut} = 0.005$ for the ODR points. 
}
\label{tab:IBManalysis}
\end{table}

\begin{figure}[htbp]
    \centering
\includegraphics[width=\linewidth]{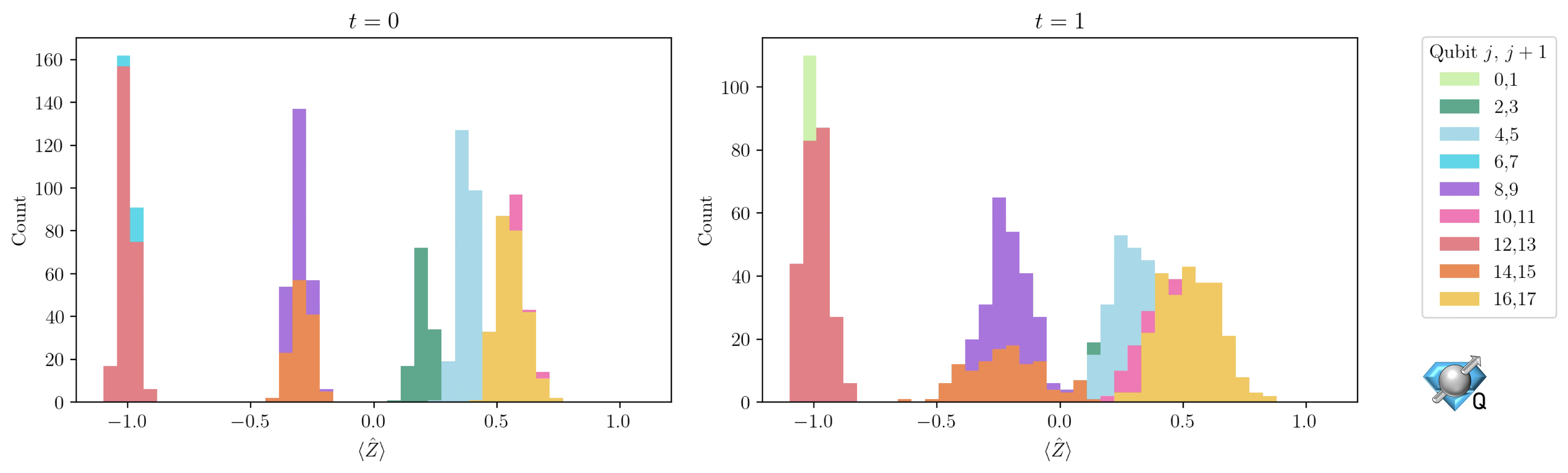}
\caption{
Results obtained from {\tt ibm\_pittsburgh} for $\langle \hat Z_j \rangle$ in the $L=3$ system.  
At $t=0$, the heavy quark is instantaneously moved from $x=0$ to $x=1$, then evolved forward in time to $t=1.0$ 
using one step of the second-order Trotterized time-evolution operator.
The SU(2) symmetry has been used to combine results for each red and green quark.
The left panel shows the distribution of each $\langle \hat Z_j \rangle$ before the heavy quark is moved; the right panel  shows the distribution after time evolution.
}
    \label{fig:IBMresultdist}
\end{figure}

Fig.~\ref{fig:IBMresultdist} shows histograms of results for $\langle \hat Z_j \rangle$ for each pair of qubits. Each color corresponds to a different staggered site.

The evaluation of the energy of the system at times immediately before and after translating the heavy quark by one spatial lattice site
is accomplished by running physics and mitigation circuits (described in the text) 
with different levels of noise for ZNE on IBM's quantum computers.
For each of the three ensemble measurements, matrix elements of operators diagonal in the computational basis are evaluated in post-processing directly from measurements, as described in the main text. 
\begin{figure}[htbp]
    \centering
    \includegraphics[width=0.32\linewidth]{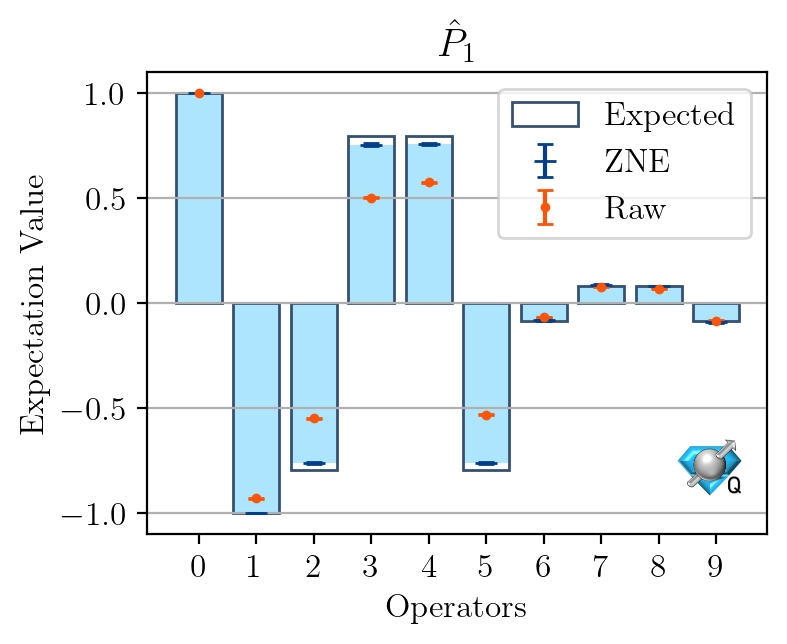}
    \includegraphics[width=0.32\linewidth]{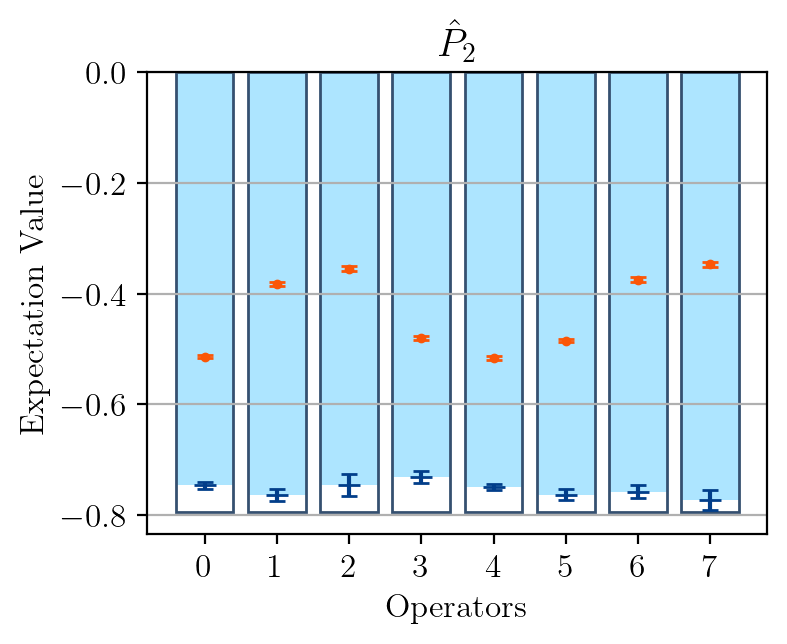}
    \includegraphics[width=0.325\linewidth]{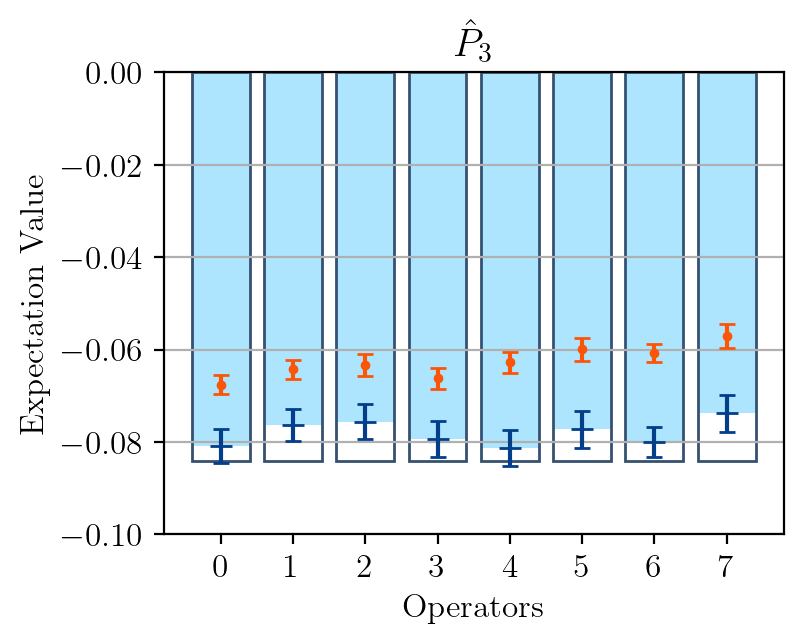}
    \caption{Results for matrix elements of Pauli strings from the three quantum circuits used to evaluate the change in energy due to the heavy quark being moved from $x=0$ to $x=1$.
    The left panel shows the raw and mitigate results, along with expected noiseless values from classical computation, corresponding to $\hat {P}_1 $ given in Eq.~\eqref{eq:estimator}, while the middle and right panels correspond to $\hat {P}_2$ and $\hat {P}_3$,   respectively.
    The points with error bars correspond to the raw and mitigated results obtained from {\tt ibm\_pittsburgh}, 
while the gray bar outlines are the expected results obtained using classical computing.
 The numerical values of the displayed results are given in Tables~\ref{tab:MEest1}, \ref{tab:MEest2} and \ref{tab:MEest3}.
 }
    \label{fig:est123}
\end{figure}
\begin{figure}[htbp]
    \centering
    \includegraphics[width=0.9\linewidth]{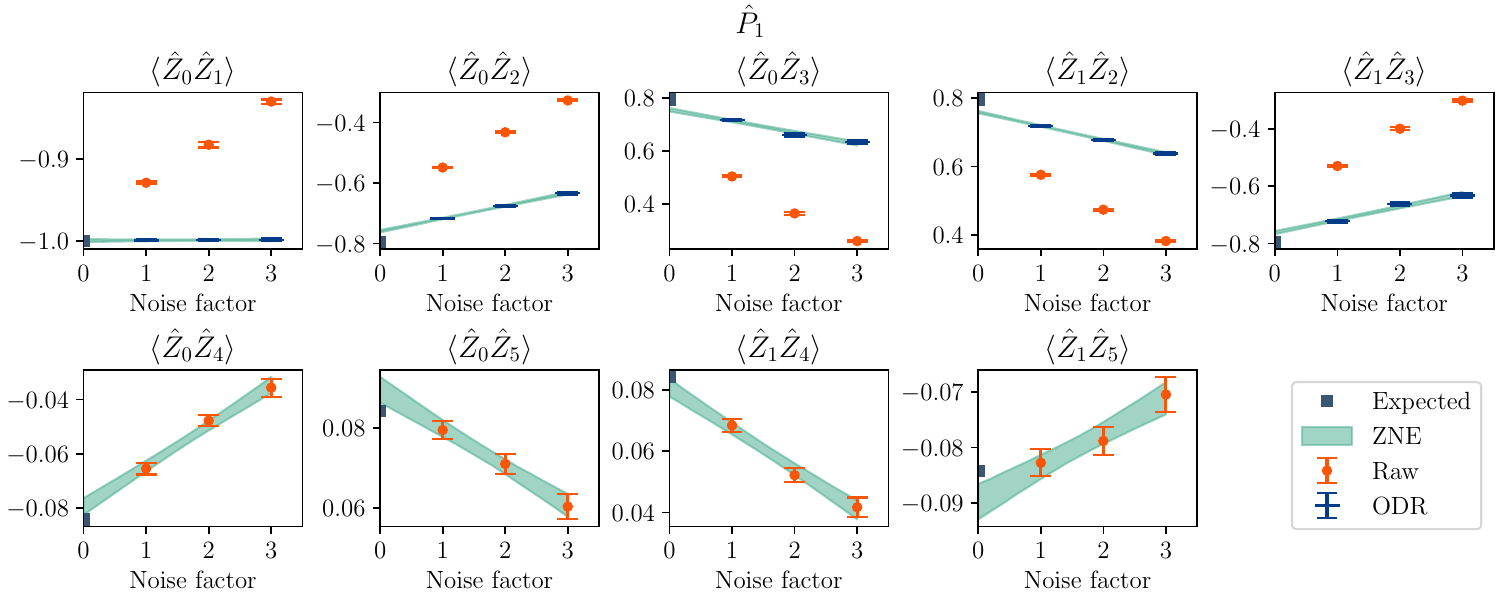}
    \includegraphics[width=0.7\linewidth]{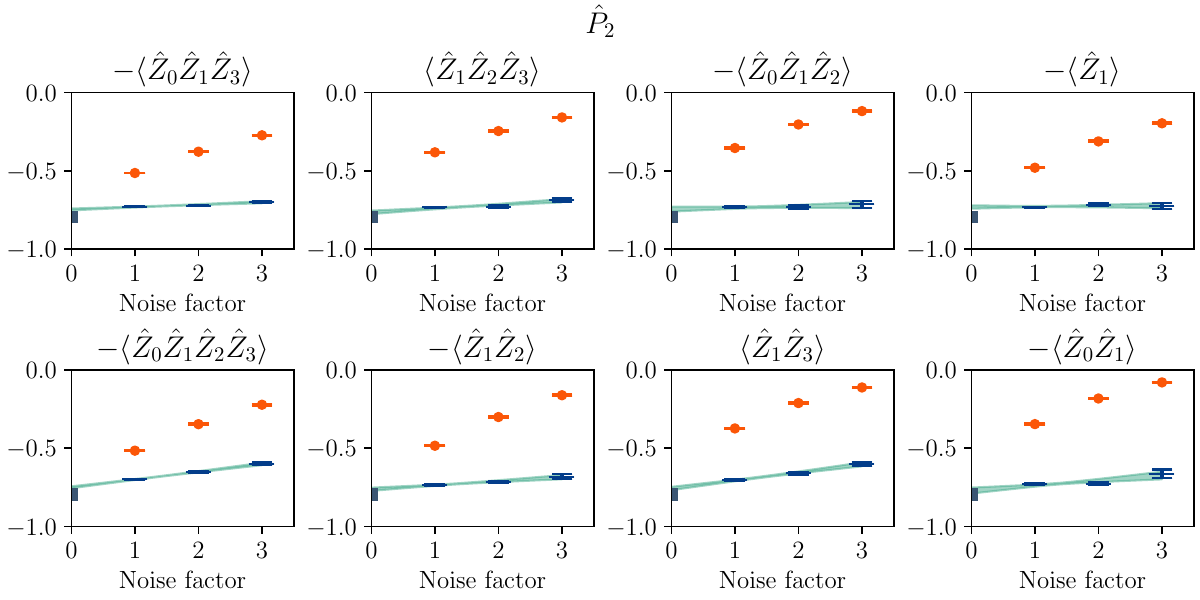}
    \includegraphics[width=0.72\linewidth]{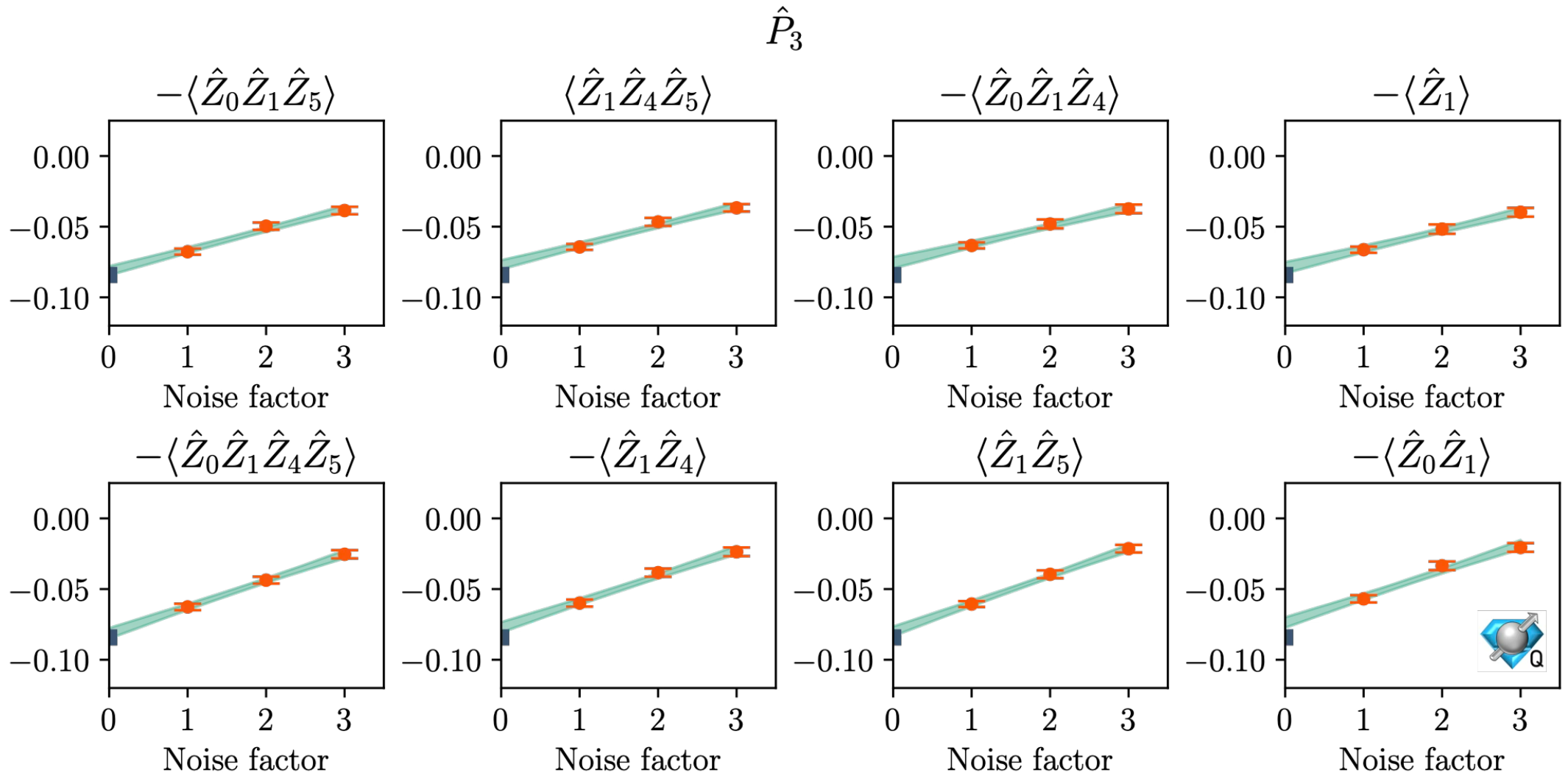}
    \caption{ZNE results for the matrix elements of Pauli strings from the three quantum circuits used to evaluate the change in energy due to the heavy quark being moved from $x=0$ to $x=1$.
    In each panel ($\hat {\cal O}_1$ top, $\hat {\cal O}_2$ middle, $\hat {\cal O}_3$ bottom), each subpanel shows the raw and ODR-mitigated results (when available) for each Pauli string, along with expected noiseless values from classical computation, and the linear ZNE band fit.
    }
    \label{fig:est123_zne}
\end{figure}
The final results for the matrix elements of the Pauli strings given in Eq.~\eqref{eq:estimator}
obtained from the three different circuits run on {\tt ibm\_pittsburgh} are displayed in Fig.~\ref{fig:est123} and 
given in Tables~\ref{tab:MEest1}, \ref{tab:MEest2} and \ref{tab:MEest3}. 
The ZNE fits are displayed in Fig.~\ref{fig:est123_zne}.
\begin{table}[htbp]
\renewcommand{\arraystretch}{1.1}
\begin{tabularx}{\textwidth}{|c || Y | Y | Y |Y| Y| }
 \hline
\multicolumn{1 }{|c||}{ }
&  \multicolumn{3}{ c |}{Matrix Elements $\hat P_1 $ from {\tt ibm\_pittsburgh}} & \multicolumn{2}{c |}{Classical }\\
 \hline
Operator & Raw & ODR & ODR $+$ ZNE & Expected & Exact \\
\hline\hline 
$\hat I$ & 1.0000(0) & 1.0000(0) & 1.0000(0) & 1.0 & 1.0\\
$\hat Z_0 \hat Z_1$ & -0.9288(15) & -0.9991(11) & -0.9994(20) & -1.0 & -1.0\\
$\hat Z_0 \hat Z_2$ & -0.5488(25) & -0.7174(28) & -0.7597(53) & -0.7951  & -0.7862\\
$\hat Z_0 \hat Z_3$ & 0.5039(38) & 0.7185(45) & 0.7561(85) & 0.7951 & 0.7862\\
$\hat Z_1 \hat Z_2$ & 0.5761(22) & 0.7189(24) & 0.7581(49) & 0.7951 & 0.7862\\
$\hat Z_1 \hat Z_3$ & -0.5302(37) & -0.7224(45) & -0.7613(81) & -0.7951 & -0.7862\\
$\hat Z_0 \hat Z_4$ & -0.0653(22) & - & -0.0792(37) & -0.0842 & -0.0946\\
$\hat Z_0 \hat Z_5$ & 0.0794(25) & - & 0.0894(40) & 0.0842 & 0.0946\\
$\hat Z_1 \hat Z_4$ & 0.0683(22) & - & 0.0807(36) & 0.0842 & 0.0946\\
$\hat Z_1 \hat Z_5$ & -0.0826(25) & - & -0.0895(39) & -0.0842 & -0.0946\\
 \hline
 \hline
\end{tabularx}
\caption{
Matrix elements of diagonal Pauli strings in the computational basis of the initial state ($t<0$) prepared on  {\tt ibm\_pittsburgh}, 
corresponding to $\hat {\cal O}_1 $ 
given in Eq.~\eqref{eq:estimator}.
These results are displayed in Fig.~\ref{fig:est123}.
The expected results show the value from a noiseless circuit applied to the ground state prepared on the quantum computer, while the value in parentheses corresponds to the result from the exact ground state.
Results for the last few operators were not mitigated via ODR because setting non-Clifford rotation angles to 0 or $\pi/2$ both result in 0 for noiseless predictions of these observables, incompatible with ODR.}
\label{tab:MEest1}
\end{table}
\begin{table}[htbp]
\renewcommand{\arraystretch}{1.1}
\resizebox{\textwidth}{!}{
\begin{tabularx}{1.1\textwidth}{|c || Y | Y | Y |Y| Y| }
 \hline
\multicolumn{1 }{|c||}{ } 
&  \multicolumn{3}{ c |}{Matrix Elements $\hat P_2 $ from {\tt ibm\_pittsburgh}} & \multicolumn{2}{c |}{ Classical}\\
 \hline
Operator & Raw & ODR &  ODR $+$ ZNE & Expected & Exact  \\
\hline\hline 
$-\hat Z_0\hat Z_1\hat Z_3$ & -0.5141(29) & -0.7286(27) & -0.746(06) & -0.7951 & -0.7862 \\
$\hat Z_1\hat Z_2 \hat Z_3$ & -0.3820(41) & -0.7334(46) & -0.764(11) & -0.7951 & -0.7862 \\
$-\hat Z_0\hat Z_1\hat Z_2$ & -0.3545(48) & -0.7328(50) & -0.746(20) & -0.7951 & -0.7862 \\
$\hat Z_1$ & -0.4804(36) & -0.7323(37) & -0.731(10) & -0.7951 & -0.7862 \\
$-\hat Z_0\hat Z_1\hat Z_2\hat Z_3$ & -0.5160(33) & -0.6996(27) & -0.749(06) & -0.7951 & -0.7862 \\
$\hat Z_1\hat Z_2$ & -0.4846(33) & -0.7344(40) & -0.763(09) & -0.7951 & -0.7862 \\
$\hat Z_1\hat Z_3$ & -0.3742(47) & -0.7039(44) & -0.758(12) & -0.7951 & -0.7862 \\
$-\hat Z_0 \hat Z_1$ & -0.3466(44) & -0.7283(56) & -0.772(18) & -0.7951 & -0.7862 \\
\hline
\hline
\end{tabularx}}
\caption{Matrix elements of diagonal Pauli strings in the computational basis of the initial state ($t<0$) prepared on  {\tt ibm\_pittsburgh}, corresponding to $\hat {\cal O}_2$ given in Eq.~\eqref{eq:estimator}.
These results are displayed in Fig.~\ref{fig:est123}. The expected results show the value from a noiseless circuit applied to the ground state prepared on the quantum computer, while the value in parentheses corresponds to the result from the exact ground state.
}
\label{tab:MEest2}
\end{table}
\begin{table}[htbp]
\renewcommand{\arraystretch}{1}
\resizebox{\textwidth}{!}{
\begin{tabularx}{1.2\textwidth}{|c || Y | Y | Y | Y |}
 \hline
\multicolumn{1}{|c||}{} & \multicolumn{2}{c|}{Matrix Elements $\hat P_3$ from {\tt ibm\_pittsburgh}} & \multicolumn{2}{c|}{Classical}\\
 \hline
Operator & Raw & ZNE & Expected & Exact \\
\hline\hline
$-\hat Z_0\hat Z_1\hat Z_5$ & -0.0676(22) & -0.0808(37) & -0.0842 & -0.0946 \\
$\hat Z_1\hat Z_4 \hat Z_5$ & -0.0643(21) & -0.0763(35) & -0.0842 & -0.0946 \\
$-\hat Z_0\hat Z_1\hat Z_4$ & -0.0633(24) & -0.0756(37) & -0.0842 & -0.0946 \\
$\hat Z_1$ & -0.0662(23) & -0.0794(39) & -0.0842 & -0.0946 \\
$-\hat Z_0\hat Z_1\hat Z_4\hat Z_5$ & -0.0628(22) & -0.0813(39) & -0.0842 & -0.0946 \\
$\hat Z_1\hat Z_4$ & -0.0600(25) & -0.0773(40) & -0.0842 & -0.0946 \\
$\hat Z_1\hat Z_5$ & -0.0607(20) & -0.0800(32) & -0.0842 & -0.0946 \\
$-\hat Z_0 \hat Z_1$ & -0.0570(26) & -0.0738(40) & -0.0842 & -0.0946 \\
 \hline
 \hline
\end{tabularx}}
\caption{
Matrix elements of diagonal Pauli strings in the computational basis of the initial state ($t<0$) prepared on  {\tt ibm\_pittsburgh}, 
corresponding to $\hat {\cal O}_3$ 
given in Eq.~\eqref{eq:estimator}.
These results are displayed in Fig.~\ref{fig:est123}.
The expected results show the value from a noiseless circuit applied to the ground state prepared on the quantum computer, while the value in parentheses corresponds to the result from the exact ground state. ODR mitigation circuits were not run for $\hat{P}_3$ because setting non-Clifford rotation angles to 0 or $\pi/2$ both result in 0 for the noiseless predictions of observables, 
incompatible with implementing ODR.
}
\label{tab:MEest3}
\end{table}
%

%%%%%%%%%%%%%%%%%%%%%%%%%%%%%%%%%%%%%%%%%%%%%%%%%%%%%%%%

% ========== Appendix B

\chapter{Gauge-Fixed Hamiltonian Dynamics}
%%%%%%%%%%%%%%%%%%%%%%%%%%%%
% Appendix: Gauge-Fixed Hamiltonian Dynamics
% Appendix of Ref. Froland:2025bqf (body lives in chapters/gauge_fix.tex)
%%%%%%%%%%%%%%%%%%%%%%%%%%%%

\section{Explicit Form of Quantum Circuits}
\label{sec:appendixA}
This section gives explicit forms for the circuits that are constructed in Sec.~\ref{subsec:coarse} by specifying the Pauli decomposition $\mathcal{D}(\htp)$. The circuit used to calculate Fig.~\SubFigRef{fig:quantum_results}{a} is formed by exponentiating the terms in this decomposition. Table~\ref{tab:delta_high} gives the terms in the decomposition that were retained for the circuits implemented on the quantum device, whereas Table~\ref{tab:delta_low} gives the terms in the decomposition that were dropped. 

The \textbf{subcircuit} column specifies the subcircuits $U_1,U_2$ or $U_3$, represented in accordance with their graphical depiction in Fig.~\SubFigRef{fig:circuit_stuff}{b}. The \textbf{Pauli String} column denotes a specific Pauli string, where the left-most index corresponds to the top wire in the circuit element and the rightmost corresponds to the bottom wire. The $\delta$ column denotes at which truncation level the corresponding string would be dropped. For Fig.~\SubFigRef{fig:quantum_results}{a}, the truncation level of $0.66$ is chosen.  As such, all entries in Table~\ref{tab:delta_high} have values $\delta>0.66$ and all entries in Table~\ref{tab:delta_low} have values $\delta<0.66$.

\section{Explicit Matrix Elements of Hamiltonian Operators}
This appendix contains the explicit expression for various electric bilinear operators that appear in the Hamiltonian. The matrix elements of these operators are presented in three different bases: group element, mixed and character irrep. 

\subsection{Group Element Basis}
\label{app:GEB}
In the group element basis, the Hilbert space basis states are spanned by the three continuous variables, $\omega, \theta, \phi$. Using this representation, the electric bilinear matrix elements are given by
\begin{align}\label{eq:electric_terms}
\CE_1^2&= - \left(\pdv[2]{}{\omega_1}+ \cot \frac{\omega_1}{2}\pdv{}{\omega_1}\right)+\frac{1}{4}\csc^2\frac{\omega_1}{2} \CN \nonumber \\
\CE_2^2&= - \left(\pdv[2]{}{\omega_2}+ \cot \frac{\omega_2}{2}\pdv{}{\omega_2}\right)+\frac{1}{4}\csc^2\frac{\omega_2}{2} \CN \nonumber \\
\boldsymbol{{\cal E}}_{R2}\cdot \boldsymbol{{\cal E}}_{L2} &= \boldsymbol{\CE}_2^2-\frac{1}{2}\CN \nonumber \\
\boldsymbol\CE_{1R}\cdot \boldsymbol \CE_{2R} &=-\cos \Theta \pdv{}{\omega_1}{ \omega_2}\nonumber\\
&+\frac{1}{2}\sin \Theta\left(\cot \frac{\omega_2}{2}\pdv{}{\omega_1}+\cot \frac{\omega_1}{2}\pdv{}{\omega_2}\right.\nonumber\\
&+\left.\frac{1}{2}\cot \frac{\omega_1}{2}\cot \frac{\omega_2}{2}\pdv{}{\Theta}\right) \nonumber \\
&-\frac{1}{4}\left(1+\cos \Theta \cot \frac{\omega_1}{2}\cot \frac{\omega_2}{2} \right) \CN \nonumber \\
\boldsymbol\CE_{1R}\cdot \boldsymbol \CE_{2L}&= \boldsymbol\CE_{1R}\cdot \boldsymbol \CE_{2R}+\frac{1}{2}\CN
\end{align}
where $\CN$ is the operator for which the Legendre polynomials are eigenfunctions, given in Eq.~\eqref{eq:LegDifEq}

\subsection{Mixed Basis}
\label{app:MB}
The mixed basis and the group element basis are related via
\begin{align}
\braket{\omega_1' \, \omega_2'\, \nu}{\omega_1 \, \omega_2\, \Theta} = P_\nu (\Theta)\delta(\omega_1-\omega_1')\delta(\omega_2-\omega_2')
\end{align}
To convert the Hamiltonian from the group element basis to the mixed basis, these four integrals
\begin{align}
\int d(\cos\Theta)\, P_{\nu'}(\Theta)P_{\nu}(\Theta)&= \delta_{\nu'\nu} \\
\int d(\cos\Theta)\,P_{\nu'}(\Theta)\CN P_{\nu}(\Theta)&= \nu(\nu+1)\delta_{\nu'\nu} \nonumber \\
\int d(\cos\Theta)\, P_{\nu'}(\Theta)\cos \Theta P_{\nu}(\Theta)&=\frac{(\nu+1)\, \delta_{\nu'\nu+1}}{\sqrt{(2\nu+1)(2\nu+3)}} \nonumber\\
&+ \frac{\nu\, \delta_{\nu'\nu-1}}{\sqrt{(2\nu+1)(2\nu-1)}}\nonumber \\
\int d(\cos\Theta) P_{\nu'}(\Theta)\sin \Theta \pdv{P_{\nu}(\Theta)}{\Theta}&= \frac{\nu\left(\nu+1\right) \,\delta_{\nu'\nu+1}}{\sqrt{(2\nu+1)(2\nu+3)}}\nonumber\\
&-\frac{\nu\left(\nu+1\right)\,\delta_{\nu'\nu-1}}{\sqrt{(2\nu+1)(2\nu-1)}} \nonumber
\end{align}
are used. Note that the unconventional normalization on $P_\nu(\Theta)$ is due to the requirement that the Hilbert space basis be orthonormal.

\begin{table}[htpb!]
\centering
\renewcommand{\arraystretch}{0.9}

\setlength{\tabcolsep}{3pt}
\resizebox{!}{0.3\textheight}{
\begin{tabular}{|c||c|c|c|}
\hline
\textbf{Subcircuit}
& \textbf{Pauli String}
&\textbf{Coefficient}
&\textbf{Truncation Percentage} $\boldsymbol{\delta}$\\
\hline

\multirow{9}{*}{%
\parbox{0.3\textwidth}{%
\centering
\includegraphics[width=0.06\textwidth]{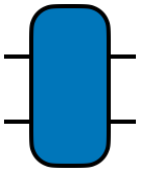}
}}
&&&\\
& I$\,\otimes\,$X & -4.358185 & 0.9857 \\
& Z$\,\otimes\,$Z & 3.049102 & 0.9429 \\
& X$\,\otimes\,$X & -2.468268 & 0.9143 \\
& X$\,\otimes\,$I & 1.565659 & 0.8571 \\
& Y$\,\otimes\,$Y & -1.396263 & 0.7857 \\
& I$\,\otimes\,$Z & 0.884597 & 0.7429 \\
& Z$\,\otimes\,$I & 0.519954 & 0.7143 \\
&&&\\
\hline

\multirow{6}{*}{%
\parbox{0.3\textwidth}{%
\centering
\includegraphics[width=0.06\textwidth]{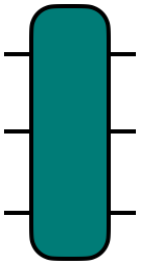}
}}
&&&\\
& Z$\,\otimes\,$I$\,\otimes\,$I & -3.778734 & 0.9571 \\
& Z$\,\otimes\,$Z$\,\otimes\,$I & -1.678690 & 0.8857 \\
& Z$\,\otimes\,$I$\,\otimes\,$Z & -1.452504 & 0.8143 \\
& Z$\,\otimes\,$Z$\,\otimes\,$Z & -1.392269 & 0.7571 \\
&&&\\
\hline

\multirow{6}{*}{%
\parbox{0.3\textwidth}{%
\centering
\includegraphics[width=0.06\textwidth]{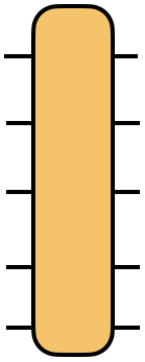}
}}
&&&\\
&&&\\
& X$\,\otimes\,$I$\,\otimes\,$I$\,\otimes\,$I$\,\otimes\,$I & 0.283820 & 0.6857 \\
& Y$\,\otimes\,$I$\,\otimes\,$Y$\,\otimes\,$I$\,\otimes\,$I & -0.226330 & 0.6714 \\
& & & \\
&&&\\
\hline
\end{tabular}
}
\caption{Terms in the Pauli string decomposition of $\htp$ that are retained in the circuits. The colored circuit elements correspond to those in the main text. The Pauli string column denote which the rotations $e^{-ic_p P}$ in the circuit for the string $P$ and corresponding coefficient $c_p$. The truncation percentage $\delta$ denotes the fraction of Pauli strings that would need to be truncated for this term to have the smallest coefficient. As an example, all terms in this table were included in the circuit which was run at $\delta=.66$ and so all $\delta$ values in this table are greater than $.66$.}
\label{tab:delta_high}
\end{table}

\begin{table}[t]
\centering
\renewcommand{\arraystretch}{0.8}

\setlength{\tabcolsep}{3pt}
\resizebox{!}{0.36\textheight}{
\begin{tabular}{|c||c|c|c|}
\hline
\textbf{Subcircuit}
& \textbf{Pauli String}
& \textbf{Coefficient}
& \textbf{Truncation Percentage} $\delta$ \\
\hline
\multirow{45}{*}{%
\parbox{0.3\textwidth}{%
\centering
\includegraphics[width=0.1\textwidth]{subcirc3.pdf}
}}
& Y$\,\otimes\,$I$\,\otimes\,$I$\,\otimes\,$I$\,\otimes\,$Y & -0.226330 & 0.6571 \\
& X$\,\otimes\,$I$\,\otimes\,$I$\,\otimes\,$Z$\,\otimes\,$I & 0.189503 & 0.6429 \\
& X$\,\otimes\,$Z$\,\otimes\,$I$\,\otimes\,$I$\,\otimes\,$I & 0.189503 & 0.6286 \\
& X$\,\otimes\,$I$\,\otimes\,$Y$\,\otimes\,$I$\,\otimes\,$Y & -0.180485 & 0.6143 \\
& Y$\,\otimes\,$I$\,\otimes\,$Y$\,\otimes\,$Z$\,\otimes\,$I & -0.151118 & 0.6000 \\
& Y$\,\otimes\,$Z$\,\otimes\,$I$\,\otimes\,$I$\,\otimes\,$Y & -0.151118 & 0.5857 \\
& X$\,\otimes\,$I$\,\otimes\,$I$\,\otimes\,$I$\,\otimes\,$Z & 0.136188 & 0.5714 \\
& X$\,\otimes\,$I$\,\otimes\,$Z$\,\otimes\,$I$\,\otimes\,$I & 0.136188 & 0.5571 \\
& X$\,\otimes\,$Z$\,\otimes\,$I$\,\otimes\,$Z$\,\otimes\,$I & 0.126529 & 0.5429 \\
& Y$\,\otimes\,$X$\,\otimes\,$Y$\,\otimes\,$I$\,\otimes\,$I & 0.113165 & 0.5286 \\
& Y$\,\otimes\,$Y$\,\otimes\,$X$\,\otimes\,$I$\,\otimes\,$I & -0.113165 & 0.5143 \\
& Y$\,\otimes\,$I$\,\otimes\,$I$\,\otimes\,$X$\,\otimes\,$Y & 0.113165 & 0.5000 \\
& Y$\,\otimes\,$I$\,\otimes\,$I$\,\otimes\,$Y$\,\otimes\,$X & -0.113165 & 0.4857 \\
& X$\,\otimes\,$I$\,\otimes\,$I$\,\otimes\,$Z$\,\otimes\,$Z & 0.110025 & 0.4714 \\
& X$\,\otimes\,$Z$\,\otimes\,$Z$\,\otimes\,$I$\,\otimes\,$I & 0.110025 & 0.4571 \\
& Y$\,\otimes\,$I$\,\otimes\,$Y$\,\otimes\,$I$\,\otimes\,$Z & -0.108602 & 0.4429 \\
& Y$\,\otimes\,$I$\,\otimes\,$Z$\,\otimes\,$I$\,\otimes\,$Y & -0.108602 & 0.4286 \\
& X$\,\otimes\,$I$\,\otimes\,$Z$\,\otimes\,$Z$\,\otimes\,$I & 0.090931 & 0.4143 \\
& X$\,\otimes\,$Z$\,\otimes\,$I$\,\otimes\,$I$\,\otimes\,$Z & 0.090931 & 0.4000 \\
& X$\,\otimes\,$I$\,\otimes\,$Y$\,\otimes\,$X$\,\otimes\,$Y & 0.090243 & 0.3857 \\
& X$\,\otimes\,$I$\,\otimes\,$Y$\,\otimes\,$Y$\,\otimes\,$X & -0.090243 & 0.3714 \\
& X$\,\otimes\,$X$\,\otimes\,$Y$\,\otimes\,$I$\,\otimes\,$Y & 0.090243 & 0.3571 \\
& X$\,\otimes\,$Y$\,\otimes\,$X$\,\otimes\,$I$\,\otimes\,$Y & -0.090243 & 0.3429 \\
& Y$\,\otimes\,$I$\,\otimes\,$Y$\,\otimes\,$Z$\,\otimes\,$Z & -0.087739 & 0.3286 \\
& Y$\,\otimes\,$Z$\,\otimes\,$Z$\,\otimes\,$I$\,\otimes\,$Y & -0.087739 & 0.3143 \\
& Y$\,\otimes\,$X$\,\otimes\,$Y$\,\otimes\,$Z$\,\otimes\,$I & 0.075559 & 0.3000 \\
& Y$\,\otimes\,$Y$\,\otimes\,$X$\,\otimes\,$Z$\,\otimes\,$I & -0.075559 & 0.2857 \\
& Y$\,\otimes\,$Z$\,\otimes\,$I$\,\otimes\,$X$\,\otimes\,$Y & 0.075559 & 0.2714 \\
& Y$\,\otimes\,$Z$\,\otimes\,$I$\,\otimes\,$Y$\,\otimes\,$X & -0.075559 & 0.2571 \\
& X$\,\otimes\,$Z$\,\otimes\,$I$\,\otimes\,$Z$\,\otimes\,$Z & 0.073463 & 0.2429 \\
& X$\,\otimes\,$Z$\,\otimes\,$Z$\,\otimes\,$Z$\,\otimes\,$I & 0.073463 & 0.2286 \\
& X$\,\otimes\,$I$\,\otimes\,$Z$\,\otimes\,$I$\,\otimes\,$Z & 0.065348 & 0.2143 \\
& Y$\,\otimes\,$I$\,\otimes\,$Z$\,\otimes\,$X$\,\otimes\,$Y & 0.054301 & 0.2000 \\
& Y$\,\otimes\,$I$\,\otimes\,$Z$\,\otimes\,$Y$\,\otimes\,$X & -0.054301 & 0.1857 \\
& Y$\,\otimes\,$X$\,\otimes\,$Y$\,\otimes\,$I$\,\otimes\,$Z & 0.054301 & 0.1714 \\
& Y$\,\otimes\,$Y$\,\otimes\,$X$\,\otimes\,$I$\,\otimes\,$Z & -0.054301 & 0.1571 \\
& X$\,\otimes\,$I$\,\otimes\,$Z$\,\otimes\,$Z$\,\otimes\,$Z & 0.052795 & 0.1429 \\
& X$\,\otimes\,$Z$\,\otimes\,$Z$\,\otimes\,$I$\,\otimes\,$Z & 0.052795 & 0.1286 \\
& X$\,\otimes\,$X$\,\otimes\,$Y$\,\otimes\,$X$\,\otimes\,$Y & -0.045121 & 0.1143 \\
& X$\,\otimes\,$X$\,\otimes\,$Y$\,\otimes\,$Y$\,\otimes\,$X & 0.045121 & 0.1000 \\
& X$\,\otimes\,$Y$\,\otimes\,$X$\,\otimes\,$X$\,\otimes\,$Y & 0.045121 & 0.0857 \\
& X$\,\otimes\,$Y$\,\otimes\,$X$\,\otimes\,$Y$\,\otimes\,$X & -0.045121 & 0.0714 \\
& Y$\,\otimes\,$X$\,\otimes\,$Y$\,\otimes\,$Z$\,\otimes\,$Z & 0.043870 & 0.0571 \\
& Y$\,\otimes\,$Y$\,\otimes\,$X$\,\otimes\,$Z$\,\otimes\,$Z & -0.043870 & 0.0429 \\
& Y$\,\otimes\,$Z$\,\otimes\,$Z$\,\otimes\,$X$\,\otimes\,$Y & 0.043870 & 0.0286 \\
& Y$\,\otimes\,$Z$\,\otimes\,$Z$\,\otimes\,$Y$\,\otimes\,$X & -0.043870 & 0.0143 \\
& X$\,\otimes\,$Z$\,\otimes\,$Z$\,\otimes\,$Z$\,\otimes\,$Z & 0.042652 & 0.0000 \\
\hline
\end{tabular}
}
\caption{The columns are the same as in the Table~\ref{tab:delta_high}. This table contains all of the Pauli string in the decomposition $\mathcal{D}(\htp)$ that were dropped in the quantum simulations. As such all entries in the $\delta$ column are less than $\delta=.66$. Note that all of the terms dropped are included in the yellow circuit element. They all have either an $X$ or a $Y$ located on the $\nu$ register. Additionally, they are generally much higher weight and have support on a greater number of qubits.}
\label{tab:delta_low}
\end{table}

\subsection{Character Irrep Basis}
\label{app:CIB}
The characer irrep basis and the group element basis are related via
\begin{align}
\braket{J_1 J_2 \nu}{\omega_1\omega_2\Theta} = \chi^{J_1}_\nu(\omega_1) \chi^{J_2}_\nu (\omega_2) P_\nu (\Theta)
\end{align}
where $\chi^J_\nu(\omega)$ are generalized character functions of the irreducible representations of the rotation group. These functions are normalized such that
\begin{align}
\int_0^{2\pi} 4 \sin^2\frac{\omega}{2}\chi^{J'}_{\nu}(\omega) \chi^{J}_{\nu}(\omega) = \delta_{JJ'}
\end{align}
where the factor of four is due to the explicit form of the Haar measure for \sutwo parameterized using the axis-angle coordinates.

To convert between the group element basis and the character irrep basis, several matrix elements need to be computed. For the magnetic Hamiltonian, the only relevant matrix element is
\begin{align}
&\mel{J'_1\,J'_2\,\nu'}{\Tr \hat X_1}{J_1\,J_2\,\nu}= \frac{1}{ \sqrt{2}}\,\delta_{J_2'J^{\phantom{'}}_2}\delta_{\nu'\nu}  \\
&\hspace{15pt}\times \left(\sqrt{\frac{(2J_1-\nu +1) (2 J_1+\nu +2)}{(J_1+1) (2 J_1+1)}}\,\delta_{J_1' J^{\phantom{'}}_1+\frac{1}{2}} \right. \nonumber \\
&\hspace{15pt}\left.+\sqrt{\frac{(2 J_1-\nu ) (2J_1+\nu +1)}{J_1 (2 J_1+1)}}\,\delta_{J_1' J^{\phantom{'}}_1-\frac{1}{2}}\right) \, ,
\end{align}
with the matrix element for $\Tr \hat X_2$ the same, except with the exchange of $1\leftrightarrow 2$. For the electric component of the Hamiltonian, the relevant matrix elements are
\begin{align}
&\mel{J'_1\,J'_2\,\nu'}{\hat\CE_\kappa^2}{J_1\,J_2\,\nu}= J_\kappa(J_\kappa+1)\delta_{J_1'J^{\phantom{'}}_1}\delta_{J_2'J^{\phantom{'}}_2}\delta_{\nu'\nu}\nonumber \\
&\mel{J'_1\,J'_2\,\nu'}{\hat{\boldsymbol{\CE}}_{R2}\cdot \hat{\boldsymbol{\CE}}_{R2}}{J_1\,J_2\,\nu}=\delta_{J_1'J^{\phantom{'}}_1}\delta_{J_2'J^{\phantom{'}}_2}\delta_{\nu'\nu}\nonumber \\
&\hspace{30pt}\times\left(J_2(J_2+1)-\frac{\nu(\nu+1}{2}\right) \nonumber \\
&\mel{J'_1\,J'_2\,\nu'}{\hat{\boldsymbol{\CE}}_{R1}\cdot \hat{\boldsymbol{\CE}}_{R2}} {J_1\,J_2\,\nu}=-\frac{1}{4}\,\delta_{J_1'J^{\phantom{'}}_1}\delta_{J_2'J^{\phantom{'}}_2}\nonumber \\
&\hspace{30pt}\times\left(\nu(\nu+1)\delta_{\nu'\nu}+\delta_{\nu'\nu+1}\frac{c^+_{J_1 \nu}c^+_{J_2 \nu}(\nu+1)}{\sqrt{(2\nu+1)(2\nu+3)}}\right.\nonumber \\
&\hspace{30pt}\left.+\delta_{\nu'\nu-1}\frac{c^-_{J_1 \nu}c^-_{J_2 \nu}\,\nu }{\sqrt{(2\nu+1)(2\nu-1)}}\right) \nonumber \\
&\mel{J'_1\,J'_2\,\nu'}{\hat{\boldsymbol{\CE}}_{R1}\cdot \hat{\boldsymbol{\CE}}_{L2}}{J_1\,J_2\,\nu}=\frac{1}{4}\,\delta_{J_1'J^{\phantom{'}}_1}\delta_{J_2'J^{\phantom{'}}_2}\nonumber \\
&\hspace{30pt}\times\left(\nu(\nu+1)\delta_{\nu'\nu}-\delta_{\nu'\nu+1}\frac{c^+_{J_1 \nu}c^+_{J_2 \nu}(\nu+1)}{\sqrt{(2\nu+1)(2\nu+3)}}\right.\nonumber \\
&\hspace{30pt}\left.-\delta_{\nu'\nu-1}\frac{c^-_{J_1 \nu}c^-_{J_2 \nu}\,\nu }{\sqrt{(2\nu+1)(2\nu-1)}}\right)
\end{align}
where in the first expression $\kappa = 1,2$ and the coefficients $c^\pm_{J\nu}$ are given by
\begin{align}
c^+_{J \nu}&= \sqrt{(2 J+\nu +2) (2 J-\nu)} \nonumber \\
c^-_{J \nu}&= \sqrt{(2 J+\nu+1 ) (2 J-\nu+1)}
\end{align}

The full Hamiltonian will not be given here, though key components will. Of particular interest are the components that contain only first derivatives with respect to the $\omega$ variables, grouped into $H_\partial$. This Hamiltonian is given by
\begin{equation}
    H_{\partial} = \sum_{\nu}\ket{\nu}\bra{\nu+1}H^{+}_{\nu}+\sum_{\nu}H^{-}_{\nu}\ket{\nu}\bra{\nu-1}
\end{equation}
where $H^\pm$ are 
\begin{align}
H^{+}_{\nu} &= A_{\nu}\pdv{}{\omega_1}\pdv{}{\omega_2}+B_{\nu}\left(\cot \frac{\omega_1}{2}\pdv{}{\omega_2}+\cot \frac{\omega_2}{2}\pdv{}{\omega_1}\right)\nonumber\\
&+\frac{1}{4}\left(\nu(\nu+1)A_{\nu}-2B_{\nu}\right)\cot \frac{\omega_1}{2}\cot \frac{\omega_2}{2}\nonumber\\
H^{-}_{\nu} &= D_{\nu}\pdv{}{\omega_1}\pdv{}{\omega_2}+C_{\nu}\left(\cot \frac{\omega_1}{2}\pdv{}{\omega_2}+\cot \frac{\omega_2}{2}\pdv{}{\omega_1}\right)\nonumber\\
&+\frac{1}{4}\left(\nu(\nu+1)D_{\nu}-2C_{\nu}\right)\cot \frac{\omega_1}{2}\cot \frac{\omega_2}{2}
\end{align}
with the coefficients 
\begin{align}
    A_{\nu} &= \frac{(\nu+1)}{\sqrt{(2\nu+1)(2\nu+3)}}\nonumber\\
    B_{\nu} &= -\frac{(\nu+1)^2}{2\sqrt{(2\nu+1)(2\nu+3)}}\nonumber\\
    C_{\nu} &= \frac{\nu^2}{2\sqrt{(2\nu-1)(2\nu+1)}}\nonumber\\
    D_{\nu} &= \frac{\nu}{\sqrt{(2\nu-1)(2\nu+1)}} \, .
\end{align}
The relation that $H_{\nu}^{+}=(H_{\nu+1}^-)^{\dagger}$ ensures hermiticity of the overall expression.

For the quantum simulations presented in this work, only $H_0^+$ and $H_1^-$ is needed and so they are written out explicitly here:
\begin{align}
H^+_0&=\frac{1}{\sqrt{3}}\pdv{}{\omega_1}\pdv{}{\omega_2}+\frac{1}{4\sqrt{3}}\cot \frac{\omega_1}{2}\cot \frac{\omega_2}{2}\nonumber \\
&-\frac{1}{2\sqrt{3}}\left(\cot \frac{\omega_1}{2}\pdv{}{\omega_2}+\cot \frac{\omega_2}{2}\pdv{}{\omega_1}\right) \nonumber \\
&= (H^-_1)^\dagger \, \, .
\end{align}

\section{Table of Lowest Energy Eigenstates in the Character Irrep Basis}
The energy spectrum of the two plaquette Hamiltonian in the strong coupling limit can be found exactly when working in the character irrep basis. This is due to the electric component of the  Hamiltonian being block diagonal, with each block of finite dimension $2\, \text{Min}(J_1, J_2)+1$; the magnetic component is negligible and can be ignored. The energies and expressions for the several lowest lying eigenstates are showin in Table~\ref{tab:CharIrrepStrong}.

\begin{table}[t]
    \centering
\renewcommand{\arraystretch}{0.9}
\resizebox{!}{0.4\textheight}{$
\begin{array}{|l|c|c|c|} \hline
\text{\textbf{Level of State}}& \text{\textbf{Degeneracy}} & \text{\textbf{In Irrep Basis:}}\ket{J_1, J_2, \nu} & \text{\textbf{Energy}}\\\hline
 &&&\\
\text{Ground} & \text{None} & \ket{0,\,0,\,0}& 0 \\
 &&&\\ \hline
&&&\\
&  & \ket{\frac{1}{2},\,0,\,0} &  \\
\text{First Excited}&\text{Two-Fold}&&\frac{3}{2}g^2\\
 &  & \ket{0,\,\frac{1}{2},\,0} &\\
 &&&\\ \hline
 &&&\\
\text{Second Excited} & \text{None} & \frac{1}{2}\ket{\frac{1}{2},\,\frac{1}{2},\,0}-\frac{\sqrt{3}}{2}\ket{\frac{1}{2},\,\frac{1}{2},\,1} & \frac{9}{4}g^2 \\
&&&\\ \hline
&&&\\
\text{Third Excited} & \text{None} & \frac{\sqrt{3}}{2}\ket{\frac{1}{2},\,\frac{1}{2},\,0}+\frac{1}{2}\ket{\frac{1}{2},\,\frac{1}{2},\,1} & \frac{13}{4}g^2 \\
&&&\\ \hline
&&&\\
& & \ket{1,\,0,\,0} &  \\
\text{Fourth Excited} & \text{Two-Fold}&&4g^2\\
 &  & \ket{0,\,1,\,0} & \\
 &&&\\ \hline
 &&&\\
& &\frac{1}{\sqrt{3}} \ket{1,\,\tfrac{1}{2},\,0}-\sqrt{\frac{2}{3}} \ket{1,\,\tfrac{1}{2},\,1} &  \\
\text{Fifth Excited} & \text{Two-Fold}&&\frac{9}{2}g^2\\
&&\frac{1}{\sqrt{3}} \ket{\tfrac{1}{2},\,1,\,0}-\sqrt{\frac{2}{3}} \ket{\tfrac{1}{2},\,1,\,1}&\\
 &&&\\ \hline
&&&\\
& &{\sqrt\frac{2}{3}} \ket{1,\,\tfrac{1}{2},\,0}+\frac{1}{\sqrt{3}} \ket{1,\,\tfrac{1}{2},\,1} &  \\
&&&\\
\text{Sixth Excited} & \text{Three-Fold}& \sqrt{\frac{2}{3}} \ket{\tfrac{1}{2},\,1,\,0}+ \frac{1}{\sqrt{3}}\ket{\tfrac{1}{2},\,1,\,1}&6g^2\\
 &&&\\
 &&\frac{1}{3}\ket{1,\,1,\,0}-\frac{1}{\sqrt{3}}\ket{1,\, 1,\, 1}+\frac{\sqrt{5}}{3}\ket{1,\,1,\,2}&\\
 &&&\\ \hline
\end{array}
$}
\caption{Energies and explicit eigenstate constructions for the seven lowest-lying eigenstates of the two plaquette Hamiltonian in the strong coupling limit. For the states that are degenerate in this limit, the magnetic Hamiltonian breaks the degeneracy and splits the states.}
\label{tab:CharIrrepStrong}
\end{table}

% ========== Appendix C
 
\chapter{Sequency Hierarchy Truncation}
\section{Digitization}
\label{appen:dig}
\noindent
The scalar field theory is digitized following Ref.~\cite{PhysRevA.65.042323,Macridin:2018gdw,Klco:2018zqz,Macridin:2018oli} which is rooted in the JLP Formalism~\cite{Jordan:2012xnu,Jordan:2011ci,Jordan:2014tma,Jordan:2017lea}. 
A scalar field can be mapped to the $2^{n_q}$ states in the Hilbert space, in uniform intervals, 
defined by a selected maximum value of the field, Specifically,
\begin{eqnarray}
    \phi & \rightarrow & \{ \ -\phi_{\rm max} + j \delta\phi\ \ \}
    \ \ ,\ \ 
    \delta\phi \ =\ {2\phi_{\rm max}\over 2^{n_q}-1}
    \ \ ,\ \ 
j\ \in \ [0,2^{n_q}-1 ]
    \ .
    \label{eq:fieldvalues}
\end{eqnarray}

The $\hat{\Pi}^2$ term contributing to the Hamiltonian describing the dynamics of the scalar field (Eq.~\eqref{eq:Hami}) can be constructed with twisted boundary condition~\cite{Lin_2001,Sachrajda:2004mi,Bedaque:2004kc,Briceno:2013hya} to preserve its symmetry. 
Symmetrically digitizing both $\phi$ and $\Pi$ preserves the discrete symmetries for the Hamiltonian and improves efficiency of quantum simulation~\cite{Klco:2018zqz}. 
For the sake of having enough support for the evolved interacting theory wavefunctions in our digitization, the maximum field value $\phi_{\rm max}$ is taken as $4$ instead of its optimal value in relation to the number of qubits $n_q$ defined in \cite{Bauer:2021gek, Kane:2022ejm}, 
\begin{equation}
\label{eqn:optimal}
\phi_{\max }=\frac{2^{n_q}}{2} \sqrt{\frac{\sqrt{8} \pi}{2^{n_q}}}. %
\end{equation}

The non-interacting scalar field theory (free theory) on one site can be viewed as a quantum harmonic oscillator (QHO). As is well known, analytic solutions to the eigenstates of the QHO in the continuum are the Hermite polynomials. 
With mass $\hbar = m = 1$, $n$ being the $n^{\rm th}$ energy level, $H_n$ the Hermite Polynomial, these eigenstates and associated energy eigenvalues are
\begin{equation}
\begin{aligned}
E_n & =n+\frac{1}{2}
\ \ \ ,\ \ \ 
\left\langle x \mid \psi_n\right\rangle 
& =\sqrt{\frac{1}{\sqrt{\pi} 2^n n !}} e^{-\frac{x^2}{2}} H_n\left[x\right]
\ \ \ ,\ \ \ 
n = 0,1,2,...
\end{aligned}
\end{equation}

This analytical wavefunction in the continuum is digitized by first sampling on the selected set of field values, and then appropriately renormalizing,
\begin{eqnarray}
\psi & \rightarrow & \{ \, \psi(\phi_j)\, \}\ \equiv\ \{\, \tilde \psi_d(\phi_j)\, \}
    \quad , \quad
\sum_j |A \tilde \psi_d (\phi_j) |^2  =  1
\ ,
\end{eqnarray}
which defines the digitized wavefunction,
\begin{equation}
    \{\,\psi_d (\phi_j)\,\} \equiv \{\,A \tilde \psi_d (\phi_j)\,\}
    \ .
\end{equation}
For a wavefunction with support within the register, which becomes increasingly densely sampled with increasing numbers of qubits, the digitized wavefunction amplitudes $\psi_d (\phi_j)$ approaches $\psi(\phi_j) \sqrt{\delta\phi}$.
A quantum simulation involving such a scalar field will requiring using appropriate $\phi_{\rm max}$ to optimize the use of quantum resources.

\begin{figure*}[htpb]
        \centering
        \begin{subfigure}[b]{0.45\textwidth}  
            \centering 
            \includegraphics[width=\textwidth]{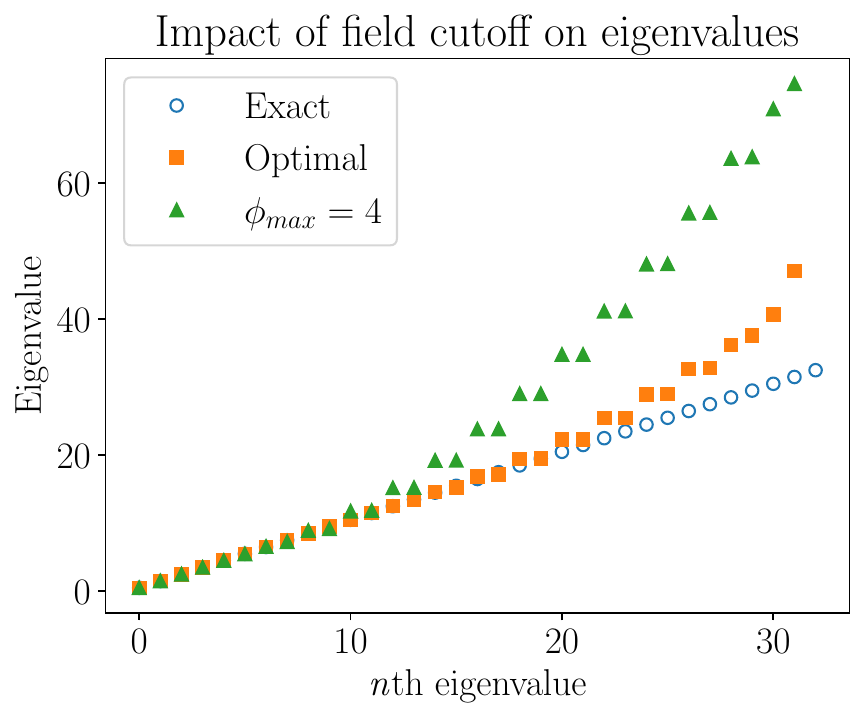}
        \end{subfigure}
        \begin{subfigure}[b]{0.45\textwidth}
            \centering
            \includegraphics[width=\textwidth]{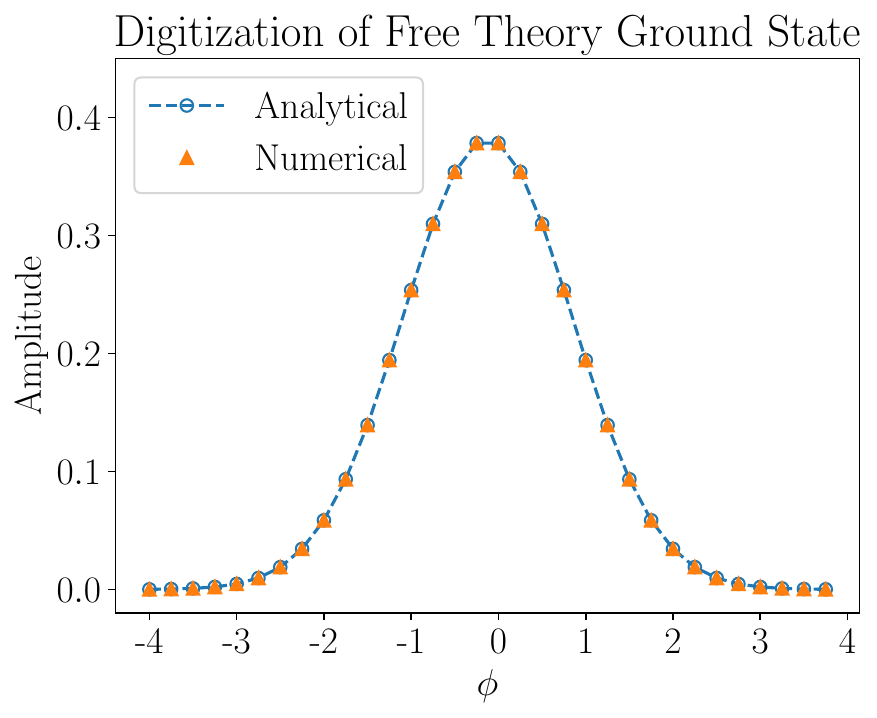} 
        \end{subfigure}
\caption{Comparison of analytic and digitized eigenvalues and wavefunctions for the non-interacting theory. 
Five qubits, $n_q=5$, are used for the digitization. 
In the left plot, the blue circular points are the analytical eigenvalues; orange square points are the eigenvalues of a digitized QHO Hamiltonian with optimal maximum field value defined in Eq.~(\ref{eqn:optimal}), while  the green triangular points are the corresponding eigenvalues with $\phi_{\rm max} = 4$.
The right plot shows the ground-state wavefunctions. 
Blue circular points are the ground state obtained from normalizing the analytical solution, connected with dashed line for better visual representation. % 
Orange triangles are the numerically solved ground state from the digitized Hamiltonian with $\phi_{\rm max} = 4$ using $n_q=5$.
Numerical values for the results displayed in this figure can be found in Table~\ref{tab:gs_5qubits} and Table~\ref{tab:eigenvalues}. } 
        \label{fig: analytical}
    \end{figure*}

To check that the one-site free theory Hamiltonian is digitized appropriately, we compare the eigenvalues and eigenstates that are numerically solved from the digitized Hamiltonian with the analytical solutions. 
The analytical eigenvalues, the eigenvalues of a digitized Hamiltonian with optimal max field value defined in \ref{eqn:optimal}, and with $\phi_{\rm max} = 4$ are shown in the left panel of Fig.~\ref{fig: analytical}. 
The eigenvalues start to deviate at higher energies but faithfully reproduce low-energy behaviors. 
The ground states acquired from directly digitizing the analytical solution and from numerically solving the digitized Hamiltonian are shown in the right panel of Fig.~\ref{fig: analytical}. It can be observed that they have good agreement.

%%%%%%%%%%%%%%%%%%%%
\section{Details about the Sequency Hierarchy and the Hadamard-Walsh Basis}
\label{app:HW}
\noindent
The diagonals of the basis operator $\hat\CO_\nu$, defined in Eq.~(\ref{eq:decomp}), are given by the entries in the $\nu^\text{th}$ row of the sequency-ordered Walsh-Hadamard matrix, $H_{n_q}$, where the entries of $H_{n_q}$ are given by
\begin{align}
H^{[n_q]}_{ij} &=\left(-1\right)^{\nu^{[n_q]}_{ij}} 
\ ,
\end{align}
with $i, j$ indexed from 1. Here,  the integer $\nu$ is given by
\begin{align}
\nu^{[n_q]}_{ij} &= \sum_{k =0}^{n_q-1} b_{k, i -1}p_{k, j-1} 
\, ,
\end{align}
where $b_{i,j}$ is the $i^\text{th}$ bit of the binary representation of the integer $j$,
\begin{align}
p_{i,j} = \begin{cases}
b_{n_q-1, j} \quad & i = 0 \\
b_{n_q-i, j}+b_{n_q-i-1, j} & i = 1, \dots n_q
\end{cases} \, .
\end{align} 
Note that the sequency-ordered Walsh-Hadamard matrix is a reordering of the recursively-defined natural-ordered Hadamard matrix in Eq.~\eqref{eq:WHT}. 
Using the orthogonality of the sequency operators, the value of sequency coefficients $\beta_\nu$ is given by
\begin{align}
\beta_\nu = \text{Tr} f(x_j) \hat{\CO}_\nu
\ ,
\end{align}
where $f(x_j)$ is the digitization of the function $F(x)$,
written as a diagonal matrix. 
Assuming an equipartitioning of the variable $x$ 
(see App.~\ref{appen:dig}), 
$x_j$ is given by 
\begin{align}
x_j = -x_M+ j \, \delta x  
\ \ ,\ \ 
 \delta x = \frac{2 x_M}{2^{n_q}-1} \ \ {\rm with}\ \  x_M>0
 \ ,
\end{align}
where $n_q$ is the number of qubits used to digitize $x$ and $x_M$ is the maximum value of $x$ used in the simulation. 

If $n_q$ is sufficiently large, the trace can be approximated by an integral with computable corrections using the Poisson resummation formula, and $\beta_\nu$ becomes
\begin{align}
\lim_{n_q\rightarrow \infty} \beta_\nu \rightarrow \int_{-x_M}^{x_M}dx 
\ F(x) \ \Theta_\nu(x; x_M)
\ ,
\end{align}
The Heaviside step functions with $\nu$ level crossings are given by
\begin{align}
\Theta_0(x; x_M) &= 1  , \nonumber \\
\Theta_1(x;  x_M) &= 2 H(x)-1 ,  \nonumber\\
\Theta_2(x;  x_M) &= 2 \left[H(x- x_M/2)-H(x+ x_M/2)\right]+1 , \nonumber \\
\Theta_3(x;  x_M) &= 2\left[H(x) - H(x- x_M/2) - H(x+ x_M/2)\right] +1 
\, ,
\end{align}
where $H(x)$ is the Heaviside step function. %

In order to derive an explicit form of $x_\nu$, it is helpful to understand how to generate sequency operators recursively. 
Let us assume that the sequency operators $\hat{\CO}^{[n_q]}_\nu$, which are matrices of dimension $2^{n_q} \times 2^{n_q}$, have already been constructed, and for this derivation, a superscript is added to the operators in order to  denote their dimension. 
The sequency operators $\hat{\CO}^{[n_q+1]}_\nu$ are then given by
\begin{align}
\hat{\CO}^{[n_q+1]}_{\nu} &= \hat{\CO}^{[n_q]}_{\nu} \otimes \hat{I}  ,  \nonumber \\
\hat{\CO}^{[n_q+1]}_{2^{n_q+1}-1-\nu} &= \hat{\CO}^{[n_q]}_{\nu} \otimes \hat{Z} \qquad  2^{n_q}-1 \geq \nu \geq 0 
\, .
\end{align}
This is already sufficient to derive an explicit expression for $x_\nu$. 
Starting with $n_q=2$, the Walsh-Hadamard matrix is given by
\begin{align}
H_2 &= \left(
\begin{array}{@{}cc!{\color{blue}\vline width 0.6pt}cc@{}}
 1 & 1 & 1 & 1 \\
  1 & 1 & -1 & -1 \\
 1 & -1 & -1 & 1 \\
 1 & -1 & 1 & -1
\end{array}
\right) \qquad \text{with} \qquad 
\nu = \left(\begin{array}{cccc}
0\\
1\\
2\\
3
\end{array}\right) \, ,
\end{align}
where the rows have been labeled by sequency index and the vertical line between column two and three marks the location of the origin when converting the discrete entries in each row to the continuous function $\Theta_\nu(x; x_M)$.
In particular, note that for $\nu = 0$, both entries to the left of the line are $+1$, while for $\nu = 2$, one entry is $+1$ while the other is $-1$. 
Therefore,
\begin{align}
n^D_0 = 2\qquad n^D_1 = 2 \qquad n_2^D =1 =\frac{1}{2}n^D_0 \qquad x_3^D = 1 =\frac{1}{2}n^D_1 \,
\end{align}
where $n^D_\nu$ is defined to be the number of entries, reading left to right, before a level crossing occurs OR the `origin' is reached; in this convention $n^D_0 = 2^{n-1}$. 
To construct the sequency operators for $n_q = 3$, each entry of $H_2$ is expanded into a 2-component vector, either $\{1\}\rightarrow \{1, 1\}$, or $\{1\}\rightarrow \{1, -1\}$, depending on whether the desired sequency index is less than eight or greater. 
In particular
\begin{align}
H_3 &= \left(
\begin{array}{cccc!{\color{blue}\vline width 0.6pt}cccc}
1&1 & 1&1 & 1&1 & 1&1 \\
1& -1 & 1&-1 & 1&-1 & 1&-1 \\ \hline
1&1 & 1&1 & -1&-1 & -1&-1 \\
1&-1 & 1&-1 & -1&1 & -1&1 \\ \hline
1&1 & -1&-1 & -1&-1 & 1&1 \\
1&-1 & -1&1 & -1&1 & 1&-1 \\ \hline
1&1 & -1&-1 & 1&1 & -1&1 \\
1&-1 & -1&1 & 1&-1 & -1&-1 
\end{array}
\right) \qquad \text{with} \qquad 
\nu = \left(\begin{array}{cccc}
0\\
7\\ \hline
1\\
6\\ \hline
2\\
5 \\ \hline
3 \\
4
\end{array}\right) \, ,
\end{align}
where the vertical line still marks the origin and the horizontal lines denote sequency operators $\hat{\CO}^{[n_q+1]}_\nu$ that are constructed out of the same $\hat{\CO}^{[n_q+1]}_\nu$. 
For example, $\hat{\CO}^{[3]}_{0}$ and $\hat{\CO}^{[3]}_{7}$ are created by taking the tensor product of $\hat{\CO}^{[2]}_{0}$ with $\hat{I}$ and $\hat{Z}$, respectively. 
The key thing to notice is that all the sequency operators with $\nu \geq 4$ are constructed by taking $\{1\} \rightarrow \{1, -1\}$ and therefore for these four operators, there is only a single $+1$ entry before a level crossing to $-1$. 
Therefore, in this case,
\begin{align}
n^D_0 &= 4 \ ,\  
n^D_2 = 2 = \frac{1}{2}n^D_0 \ ,\  
n^D_{4,6} =\frac{1}{2}n^D_2= \frac{1}{4}n^D_0 , \nonumber \\
n^D_1 &= 4 \ ,\  
n^D_3 = 2 = \frac{1}{2}n^D_1 \ ,\  
n^D_{5,7} =\frac{1}{2}n^D_3= \frac{1}{4}n^D_1
\ . %
\end{align}
From this, a clear pattern emerges and it is simple enough to convince oneself, using recursive arguments, that
\begin{align}
n^D_\nu &= \frac{1}{2}n^D_{\lfloor \log_2 \nu\rfloor-1} \ =\ 
\frac{1}{2^2}n^D_{\lfloor \log_2 \nu\rfloor-2} \ =\ 
\frac{1}{2^{\lfloor \log_2 \nu\rfloor}}n_0^D 
\ ,\  \nu > 0 
\, .
\end{align}
Recalling that $x = x_\nu$ is the position of the last level crossing before $x = x_M$, the parameter $n_\nu^D$ is related to $x_\nu$ via
\begin{align}
x_\nu = x_M \left(1 - \frac{1}{2^{\lfloor \log_2 \nu\rfloor}}\right)
\ ,
\end{align}
and inserting this into Eq.~\eqref{eq:BoundBetaEven}, the upper bound on the sequency coefficient is given by
\begin{align}
\tilde B_\nu = 1 - \left(1 - \frac{1}{2^{\lfloor \log_2 \nu\rfloor}}\right)^{1+p}
\ ,
\end{align}
for positive, even sequency index.

Because $x_\nu$ is known and the integral that gives $\tilde B_\nu$ can be evaluated, the derivation of $\tilde B_\nu$ for functions that are odd polynomial powers is straightforward. 
Thus, the upper bound for any polynomial functions can be shown to be
\begin{align}
\tilde \beta_\nu^\text{B} = \begin{cases} 1 - \left(1- \frac{1}{2^{\lfloor \text{Log}_2 \nu \rfloor}}\right)^{p+1} &\qquad \nu, p \in \text{Evens}\quad \text{OR}\quad \nu, p \in \text{Odds} \\
0 &\qquad \text{all other cases}
\end{cases}
\ . 
\end{align}
While this upper bound does not capture the finer structure of the coefficients $\tilde \beta_\nu$, as is clear from Fig.~\ref{fig:UpperBound}, it is useful in estimating a reasonable truncation scale in achieving a target precision. %

%%%%%%%%%%%%%%%%%%%%
\section{Additional Details about the Quantum Circuits Used for Simulation}
\label{appen:seqht_cir}
\noindent
For the time evolution discussed around Eq.~(\ref{eqn:physicscir}), the circuit contains two segments $\Tilde{\Phi}(s,t)$ and $\Tilde{\Pi}(t)$, with
\begin{equation}
\Tilde{\Phi}(s,t) \equiv e^{-i (\frac{1}{2}\hat{\phi}^2+\frac{\lambda(s)}{4!}\hat{\phi}^4) t}
\ \ ,\ \  
\Tilde{\Pi}(t) \equiv e^{-i \frac{1}{2} \hat{\Pi}^2 t}
\ .
\end{equation}
In their own eigenbasis, 
\begin{equation}
     \hat{\Pi}^2 = \left(\frac{(2^{n_q}-1)\pi}{2^{n_q}\phi_{\rm max}}\right)^2 \hat{\CO}^{(2,n_q)} \ , \quad \hat{\phi}^2 = \left(\frac{2\phi_{\rm max}}{2^{n_q}-1}\right)^2 \hat{\CO}^{(2,n_q)} \ , \quad \hat{\phi}^4 = \left(\frac{2\phi_{\rm max}}{2^{n_q}-1}\right)^4 \hat{\CO}^{(4,n_q)}
     \ ,
\end{equation}
where
\begin{equation}
\begin{aligned}
\hat{\CO}^{(2,n_q = 5)} = & \ \sum\beta_\mu\hat{\CO}_\mu \\ = & \  64\, \hat{Z} \otimes \hat{Z} \otimes \hat{I} \otimes \hat{I} \otimes \hat{I}+32\, \hat{Z} \otimes \hat{I} \otimes \hat{Z} \otimes \hat{I} \otimes \hat{I}+16\, \hat{Z} \otimes \hat{I} \otimes \hat{I} \otimes \hat{Z} \otimes \hat{I} \\
& +8\, \hat{Z} \otimes \hat{I} \otimes \hat{I} \otimes \hat{I} \otimes \hat{Z}+16\, \hat{I} \otimes \hat{Z} \otimes \hat{Z} \otimes \hat{I} \otimes \hat{I}+8\, \hat{I} \otimes \hat{Z} \otimes \hat{I} \otimes \hat{Z} \otimes \hat{I} 
\\
& +4\, \hat{I} \otimes \hat{Z} \otimes \hat{I} \otimes \hat{I} \otimes \hat{Z}+4\, \hat{I} \otimes \hat{I} \otimes \hat{Z} \otimes \hat{Z} \otimes \hat{I}+2\, \hat{I} \otimes \hat{I} \otimes \hat{Z} \otimes \hat{I} \otimes \hat{Z} \\
& +\hat{I} \otimes \hat{I} \otimes \hat{I} \otimes \hat{Z} \otimes \hat{Z}+\frac{341}{4} \hat{\mathds{I}} .
\end{aligned}
\end{equation}
The decomposition of the $\lambda \hat{\phi}^4$ interaction term for $n_q = 5$ is presented in Table~\ref{tab:coefficients}. 

\begin{figure}[htpb]
\centering
\[
\resizebox{\textwidth}{!}{
\begin{quantikz}[row sep=0.15cm, column sep=0.2cm]
    & \qw & \qw & \qw & \ctrl{1} & \qw & \qw & \qw & \qw & \qw & \qw & \qw & \ctrl{1} & \qw & \qw & \qw & \qw & \qw & \qw & \qw & \qw & \qw & \qw \\
    & \qw & \qw & \ctrl{1} & \targ{} & \ctrl{1} & \qw & \qw & \gate{R_Z(\theta_{12})} & \qw & \qw & \ctrl{1} & \targ{} & \qw & \qw & \qw & \qw & \ctrl{1} & \qw & \qw & \qw & \qw & \qw \\
    & \qw & \ctrl{1} & \targ{} & \qw & \targ{} & \ctrl{1} & \qw & \gate{R_Z(\theta_{13})} & \qw & \ctrl{1} & \targ{} & \ctrl{1} & \qw & \gate{R_Z(\theta_{23})} & \qw & \ctrl{1} & \targ{} & \qw & \qw & \ctrl{1} & \qw & \qw \\
    & \ctrl{1} & \targ{} & \qw & \qw & \qw & \targ{} & \ctrl{1} & \gate{R_Z(\theta_{14})} & \ctrl{1} & \targ{} & \qw & \targ{} & \ctrl{1} & \gate{R_Z(\theta_{24})} & \ctrl{1} & \targ{} & \ctrl{1} & \gate{R_Z(\theta_{34})} & \ctrl{1} & \targ{} & \ctrl{1} & \qw \\
    & \targ{} & \qw & \qw & \qw & \qw & \qw & \targ{} & \gate{R_Z(\theta_{15})} & \targ{} & \qw & \qw & \qw & \targ{} & \gate{R_Z(\theta_{25})} & \targ{} & \qw & \targ{} & \gate{R_Z(\theta_{35})} & \targ{} & \gate{R_Z(\theta_{45})} & \targ{} & \qw 
\end{quantikz}}
\]
\caption{A quantum circuit for implementing $\Phi^2_5(\theta)$ which consists of all two-body $R_{ZZ}$ operators on five qubits. }
\label{fig:phi2complete}
\end{figure}
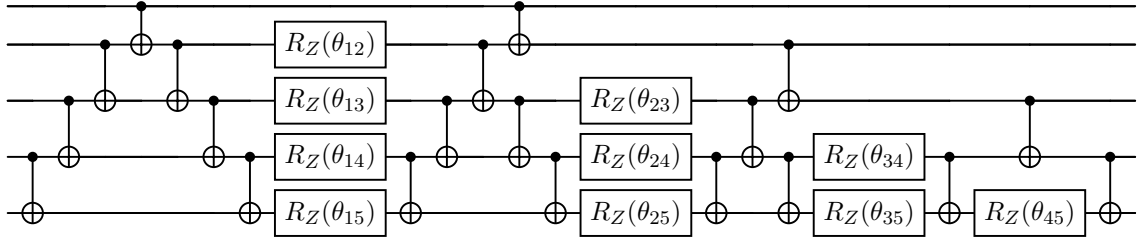

Since $\hat{\CO}^{(2,n_q)}$ consists of all combinations of $\hat{Z}_a\hat{Z}_b$ pairs and all these pairs commute, $\Phi^2_{n_q}(\theta) = e^{i\theta \hat{\CO}^{(2,n_q)}}$ can be implemented by a circuit of a complete set of two-qubit rotation gates $e^{i\theta_{ab} \hat{Z}_a\hat{Z}_b}$. 
Ref.~\cite{Farrell:2024fit} presents an efficient  way to apply these rotations. 
A circuit for $\Phi^2_{n_q = 5}(\theta)$ is shown in Fig.~\ref{fig:phi2complete}. 
For the complete implementation of $\Tilde{\Phi}(s,t)$, four-body operators $R_{ZZZZ}$ are also included, and  a circuit for implementing $R_{ZZZZ}$ is shown in Fig.~\ref{fig:cir}.

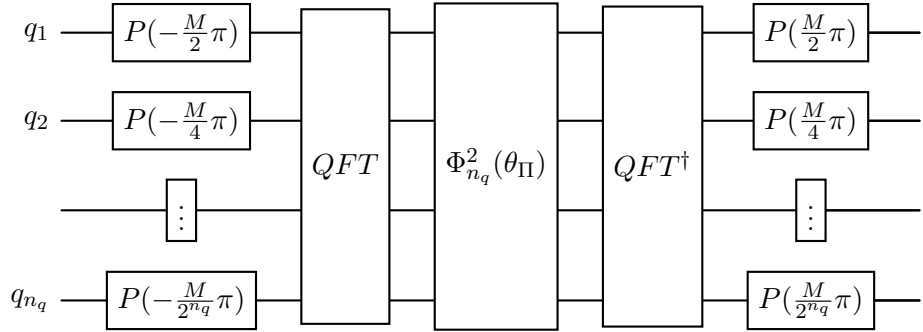
\begin{figure}[htpb]
    \centering
\[
\begin{quantikz}[row sep=0.4cm, column sep=0.6cm]
    \lstick{$q_1$} & \gate{P(-\frac{M}{2}\pi)} & \gate[4]{QFT} & \gate[4]{\Phi^2_{n_q}(\theta_\Pi)} & \gate[4]{QFT^\dag} & \gate{P(\frac{M}{2}\pi)} & \qw \\
    \lstick{$q_2$} & \gate{P(-\frac{M}{4}\pi)} & & & & \gate{P(\frac{M}{4}\pi)} & \qw \\
    & \gate{\vdots} & & & & \gate{\vdots} & \qw \\
    \lstick{$q_{n_q}$} & \gate{P(-\frac{M}{2^{n_q}}\pi)} & & & & \gate{P(\frac{M}{2^{n_q}}\pi)} & \qw 
\end{quantikz}
\]
    \caption{Circuit for implementing $\Tilde{\Pi}(t)$ in $\phi$-space with Symmetric QFT, where 
    $M = \sum\limits_{j=0}^{n-1} 2^j$. }
    \label{fig:pi2}
\end{figure}

A circuit for implementing $\Tilde{\Pi}(t)$ in field space using the Symmetric QFT~\cite{Klco:2018zqz} is shown in Fig.~\ref{fig:pi2}, 
where the phase gate is defined as
$P(\theta) = \left(\begin{array}{cc}
1 & 0 \\
0 & e^{i\theta}
\end{array}\right)
\ $ 
and $M = \sum\limits_{j=0}^{n-1} 2^j$. 
An efficient circuit to implement the QFT with nearest-neighbor connectivity, as is relevant for implementing using IBM's quantum computers, can be found in Fig.~6 of Ref.~\cite{Park:2023goh}. 

%%%%%%%%%%%%%%%
\section{Additional Fidelity Scans}
\label{appen:timescan}
\begin{figure*}[htpb]
    \centering
     \begin{subfigure}[b]{0.45\textwidth}
            \centering
            \includegraphics[width=\textwidth]{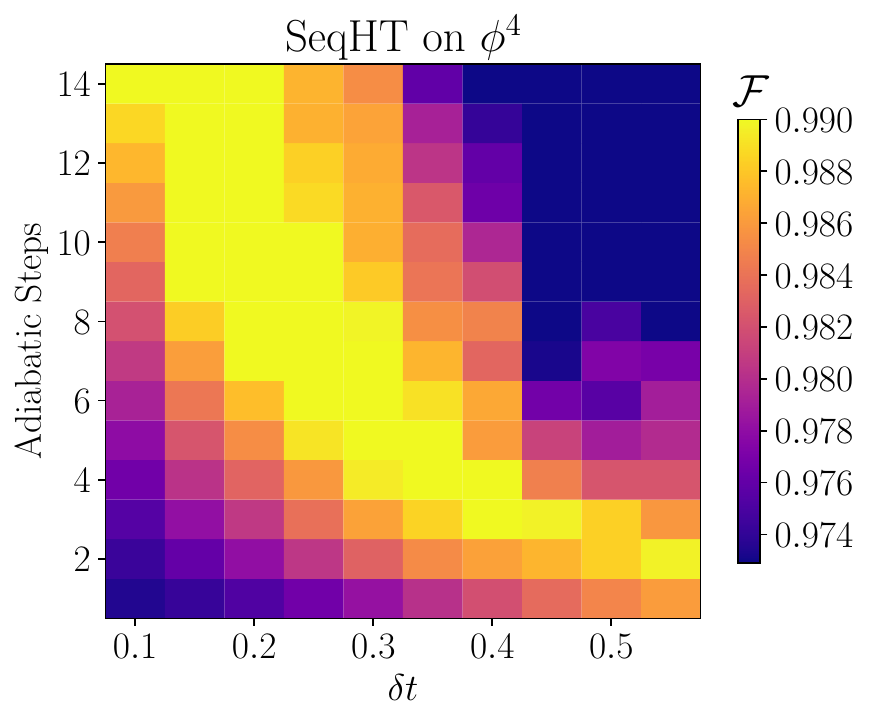}
        \end{subfigure}
        \begin{subfigure}[b]{0.45\textwidth}  
            \centering 
            \includegraphics[width=\textwidth]{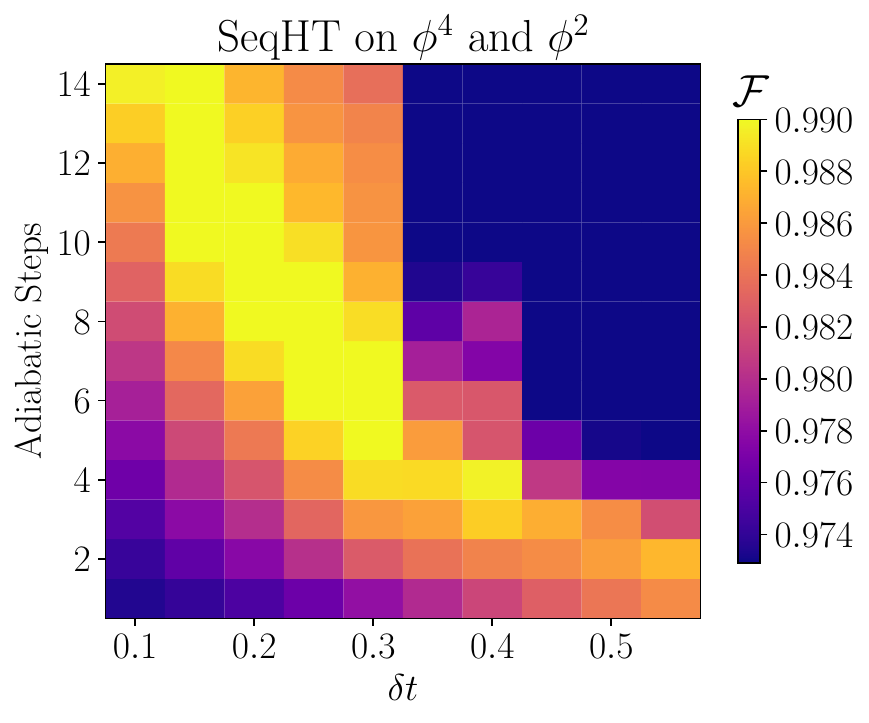}
        \end{subfigure}
\caption{ The fidelity of the $\lambda \phi^4$ ground state prepared using a noiselss quantum simulator as a function of time-step size and the number of adiabatic steps. The left panel employs $\nu_{\rm cut}=14$ for the $\hat{\phi}^4$ operator while in the right panel,  $\hat{\phi}^2$ term is also truncated, to $\nu_{\rm cut}= 30 $. 
The Hamiltonian is digitized on 12 qubits with a maximum field value $\phi_{\rm max}=4$ and $\lambda = 10$. 
Each adiabatic step consists of one second-order Trotter step. Numerical data used to plot this scan can be found in Table~\ref{tab:12_scan}.} 
        \label{fig:12_scan}
    \end{figure*}

\noindent
Fig.~\ref{fig:12_scan} presents scans of state fidelity of the $\lambda \phi^4$ ground state prepared using a noiseless quantum simulator as a function of time-step size $\delta t$ and the number of adiabatic steps for a 12-qubit system. 
Adiabatic evolutions with $\hat{\phi}^4$ truncation and additional $\hat{\phi}^2$ truncation are studied with second-order Trotterization.
The evolution with additional $\hat{\phi}^2$ truncation performs comparably well.
Scans of full adiabatic evolution (without SeqHT) is not performed for twelve qubits because completely decomposing $\hat{\phi}^4$ and executing all the four-body operators is too costly.

%%%%%%%%%%%%%%%%%%%%
\section{Magic Results}
\label{app:magic}
\noindent
In this appendix, we present tables with the results shown in Fig.~\ref{fig:GaussnQ} in Sec.~\ref{sec:magic}. 
The (undigitized) Gaussian wavefunction that we examine is given by
\begin{eqnarray}
 \psi(\phi) & = &   \frac{1}{\sqrt{\sigma\sqrt{2\pi}}}\ 
 e^{-(\phi-\phi_0)^2/(4\sigma^2)}
 \ ,
\end{eqnarray}
where  $\phi_0=0$ and $\sigma=1/\sqrt{2}$ has been used in our numerical analysis. The magic in the digitized Gaussian wavefunction as a function of the number of qubits are 
given in Table~\ref{tab:magicn}.
\begin{table}[htbp]
\centering
\begin{tabular}{l|c}
   $n_q$   &  $M_{\rm lin}$  \\
   \toprule
 3 & 0.19103 \\
 4 & 0.329949 \\
 5 & 0.355307 \\
 6 & 0.360661 \\
 7 & 0.361788 \\
 8 & 0.361992 \\
 9 & 0.362007 \\
 \end{tabular}
\caption{Linear magic in the digitized Gaussian wavefunction described in the main text 
as a function of the number of qubits, $n_q$.
The wavefunction is centered in the Hilbert space with a width of $\sigma=1/\sqrt{2}$. 
These results, after subtraction of the asymptotic value,
are shown in the left panel of Fig.~\ref{fig:GaussnQ}.}
\label{tab:magicn}
\end{table}
Table~\ref{tab:magicLam9}  shows the results for the magic in the reconstructed  sequency-truncated digital wavefunction described in the main text using
$n_q=9$ qubits as a function of $\nu_{\rm cut}$. 

\begin{table}[htbp]
\renewcommand{\arraystretch}{1}
\resizebox{\textwidth}{!}{
\centering
\scalebox{0.95}{
\begin{tabular}{c|c|| c|c || c|c|| c|c  || c|c || c|c || c|c|| c|c}
   $\nu_{\rm cut}$   &  $M_{\rm lin}$ & 
   $\nu_{\rm cut}$   &  $M_{\rm lin}$ &    
   $\nu_{\rm cut}$   &  $M_{\rm lin}$ & 
   $\nu_{\rm cut}$   &  $M_{\rm lin}$ & 
   $\nu_{\rm cut}$   &  $M_{\rm lin}$ & 
   $\nu_{\rm cut}$   &  $M_{\rm lin}$ &    
   $\nu_{\rm cut}$   &  $M_{\rm lin}$ & 
   $\nu_{\rm cut}$   &  $M_{\rm lin}$  
      \\
   \toprule
0 & 0 & 64 & 0.360366 & 128 & 0.361616 & 192 & 0.361617 & 256 & 0.361929 &320 & 0.361929 & 384 & 0.361929 & 448 & 0.361929 \\
2 & 0.00877133 & 66 & 0.360366 & 130 & 0.361616 & 194 & 0.361617 & 258 &0.361929 & 322 & 0.361929 & 386 & 0.361929 & 450 & 0.361929 \\
4 & 0.229904 & 68 & 0.36037 & 132 & 0.361617 & 196 & 0.361618 & 260 & 0.361929 & 324 & 0.361929 & 388 & 0.361929 & 452 & 0.361929 \\
6 & 0.25894 & 70 & 0.360372 & 134 & 0.361617 & 198 & 0.361618 & 262 & 0.361929 & 326 & 0.361929 & 390 & 0.361929 & 454 & 0.361929 \\
8 & 0.259058 & 72 & 0.360372 & 136 & 0.361617 & 200 & 0.361618 & 264 & 0.361929 & 328 & 0.361929 & 392 & 0.361929 & 456 & 0.361929 \\
10 & 0.260592 & 74 & 0.360372 & 138 & 0.361617 & 202 & 0.361618 & 266 & 0.361929 & 330 & 0.361929 & 394 & 0.361929 & 458 & 0.361929 \\
12 & 0.291475 & 76 & 0.360373 & 140 & 0.361617 & 204 & 0.361618 & 268 & 0.361929 & 332 & 0.361929 & 396 & 0.361929 & 460 & 0.36193 \\
14 & 0.335475 & 78 & 0.360373 & 142 & 0.361617 & 206 & 0.361618 & 270 & 0.361929 & 334 & 0.361929 & 398 & 0.361929 & 462 & 0.36193 \\
16 & 0.335475 & 80 & 0.360373 & 144 & 0.361617 & 208 & 0.361618 & 272 & 0.361929 & 336 & 0.361929 & 400 & 0.361929 & 464 & 0.36193 \\
18 & 0.335676 & 82 & 0.360373 & 146 & 0.361617 & 210 & 0.361618 & 274 & 0.361929 & 338 & 0.361929 & 402 & 0.361929 & 466 & 0.36193 \\
20 & 0.336793 & 84 & 0.360373 & 148 & 0.361617 & 212 & 0.361618 & 276 & 0.361929 & 340 & 0.361929 & 404 & 0.361929 & 468 & 0.36193 \\
22 & 0.337285 & 86 & 0.360373 & 150 & 0.361617 & 214 & 0.361618 & 278 & 0.361929 & 342 & 0.361929 & 406 & 0.361929 & 470 & 0.36193 \\
24 & 0.337302 & 88 & 0.360373 & 152 & 0.361617 & 216 & 0.361618 & 280 & 0.361929 & 344 & 0.361929 & 408 & 0.361929 & 472 & 0.36193 \\
26 & 0.337705 & 90 & 0.360373 & 154 & 0.361617 & 218 & 0.361618 & 282 & 0.361929 & 346 & 0.361929 & 410 & 0.361929 & 474 & 0.36193 \\
28 & 0.344123 & 92 & 0.360373 & 156 & 0.361617 & 220 & 0.361619 & 284 & 0.361929 & 348 & 0.361929 & 412 & 0.361929 & 476 & 0.36193 \\ 
30 & 0.355368 & 94 & 0.360373 & 158 & 0.361617 & 222 & 0.361619 & 286 & 0.361929 & 350 & 0.361929 & 414 & 0.361929 & 478 & 0.36193 \\
32 & 0.355368 & 96 & 0.360373 & 160 & 0.361617 & 224 & 0.361619 & 288 & 0.361929 & 352 & 0.361929 & 416 & 0.361929 & 480 & 0.36193 \\
34 & 0.355381 & 98 & 0.360376 & 162 & 0.361617 & 226 & 0.361619 & 290 & 0.361929 & 354 & 0.361929 & 418 & 0.361929 & 482 & 0.36193 \\
36 & 0.355442 & 100 & 0.360391 & 164 & 0.361617 & 228 & 0.361623 & 292 & 0.361929 & 356 & 0.361929 & 420 & 0.361929 & 484 & 0.361931 \\
38 & 0.355467 & 102 & 0.360397 & 166 & 0.361617 & 230 & 0.361625 & 294 & 0.361929 & 358 & 0.361929 & 422 & 0.361929 & 486 & 0.361931 \\
40 & 0.355468 & 104 & 0.360398 & 168 & 0.361617 & 232 & 0.361625 & 296 & 0.361929 & 360 & 0.361929 & 424 & 0.361929 & 488 & 0.361931 \\
42 & 0.355471 & 106 & 0.360398 & 170 & 0.361617 & 234 & 0.361625 & 298 & 0.361929 & 362 & 0.361929 & 426 & 0.361929 & 490 & 0.361931 \\
44 & 0.355482 & 108 & 0.360401 & 172 & 0.361617 & 236 & 0.361625 & 300 & 0.361929 & 364 & 0.361929 & 428 & 0.361929 & 492 & 0.361931 \\
46 & 0.355485 & 110 & 0.360402 & 174 & 0.361617 & 238 & 0.361626 & 302 & 0.361929 & 366 & 0.361929 & 430 & 0.361929 & 494 & 0.361931 \\
48 & 0.355485 & 112 & 0.360402 & 176 & 0.361617 & 240 & 0.361626 & 304 & 0.361929 & 368 & 0.361929 & 432 & 0.361929 & 496 & 0.361931 \\
50 & 0.355534 & 114 & 0.360414 & 178 & 0.361617 & 242 & 0.361629 & 306 & 0.361929 & 370 & 0.361929 & 434 & 0.361929 & 498 & 0.361932 \\
52 & 0.355798 & 116 & 0.360479 & 180 & 0.361617 & 244 & 0.361645 & 308 & 0.361929 & 372 & 0.361929 & 436 & 0.361929 & 500 & 0.361936 \\
54 & 0.355914 & 118 & 0.360508 & 182 & 0.361617 & 246 & 0.361652 & 310 & 0.361929 & 374 & 0.361929 & 438 & 0.361929 & 502 & 0.361938 \\
56 & 0.355917 & 120 & 0.360508 & 184 & 0.361617 & 248 & 0.361652 & 312 & 0.361929 & 376 & 0.361929 & 440 & 0.361929 & 504 & 0.361938 \\
58 & 0.356018 & 122 & 0.360534 & 186 & 0.361617 & 250 & 0.361659 & 314 & 0.361929 & 378 & 0.361929 & 442 & 0.361929 & 506 & 0.36194 \\
60 & 0.35755 & 124 & 0.360912 & 188 & 0.361617 & 252 & 0.361753 & 316 & 0.361929 & 380 & 0.361929 & 444 & 0.361929 & 508 & 0.361963 \\
62 & 0.360366 & 126 & 0.361616 & 190 & 0.361617 & 254 & 0.361929 & 318 & 0.361929 & 382 & 0.361929 & 446 & 0.361929 & 510 & 0.362007 \\
\end{tabular}  }}
\caption{Linear magic as a function of $\nu_{\rm cut}$ for the Gaussian wavefunction described in the main text with $n_q=9$ qubits. %   
Results are shown in the right panel of Fig.~\ref{fig:GaussnQ}. %  
As the odd-sequency coefficients vanish for this wavefunction, corresponding results for the magic are not shown.
}
\label{tab:magicLam9}
\end{table}
%

%%%%%%%%%%%%%%%%%%%%
\section{Tables of Results}
\label{app:deviceresults}
\noindent
The numerical data used to plot the figures in the text is provided in this appendix. 
Table~\ref{tab:devicenum} corresponds to Fig.~\ref{fig:device2steps} in the main text. 
It gives the expectation values of the Pauli strings in the $\lambda\phi^4$ ground-state wavefunction prepared on {\tt ibm\_sherbrooke} using the full ASP and SeqHT ASP, both error-mitigated and unmitigated, along with exact results. 
The uncertainties are obtained form bootstrap resampling using 100 resamples. 
Table~\ref{tab:coefficients} gives the coefficients of the Pauli strings contributing to the $\hat{\phi}^4$ operator as a function of $\nu$ for increasing $n_q$ and in the continuum limit, as shown in Fig~\ref{fig:coeffs}. 
Table~\ref{tab:eigenvalues} gives the $\hat{\phi}^4$ eigenvalues from exactly solving a digitized interacting Hamiltonian with and without SeqHT, as shown in Fig.~\ref{fig: eigens}, and the analytical eigenvalues of the free theory and the eigenvalues from a QHO Hamiltonian digitized with optimal maximum field value defined in Eq.~(\ref{eqn:optimal}) and with $\phi_{\rm max} = 4$, as shown in the left panel of Fig.~\ref{fig: analytical}. 
Table~\ref{tab:gs_5qubits} gives the amplitudes of the SeqHT evolved and target ground states of the interacting theory with $\lambda = 10$ and $\lambda = 60$, and the ground state wavefunction of the free theory obtained analytically and numerically, shown in Fig.~\ref{fig:amp} and the right panel of Fig.~\ref{fig: analytical}.
Table~\ref{tab:fidelity_time} shows the fidelity of the $\lambda\phi^4 $ ground-state wavefunction prepared with complete adiabatic evolution and with the SeqHT procedure as a function of time, corresponding to Fig.~\ref{fig:time env}.
Table~\ref{tab:bound} shows the calculated sequency coefficients,
$\tilde\beta_\nu$,
and upper bounds, $\tilde B_\nu$,
for a $x^4$ potential, as displayed in Fig.~\ref{fig:UpperBound}.
Table~\ref{tab:5qubit_scan_1st}, Table~\ref{tab:5qubit_scan_2nd}, and Table~\ref{tab:12_scan} shows the fidelity of the adiabatically prepared $\lambda\phi^4$ ground state as a function of $\delta t$ and the number of adiabatic steps with $\phi_{\rm max}=4$ and $\lambda = 10$, for $n_q=5$ and $n_q=12$; each adiabatic step consists of one first-order or second-order Trotter step. Results are shown in Fig.~\ref{fig: time scans} and Fig.~\ref{fig:12_scan}. Data for Fig.~\ref{fig:12qubitamp} can be provided upon request.

\begin{table}[htbp]
\centering
\resizebox{\textwidth}{!}{
\begin{tabular}{c|c|c||c|c||c|c}
&&& \multicolumn{2}{c||}{Truncated} & \multicolumn{2}{c}{Full} \\
\hline
 Sequency & Operator & Ideal & Raw & Mitigated & Raw & Mitigated \\
\hline
2 & $Z_1Z_2$ & -0.9999 & -0.092 (03) & -0.890 (35) & -0.053 (02) & -0.879 (42) \\
4 & $Z_2Z_3$ & 0.8737 & 0.060 (03) & 0.800 (52) & 0.026 (03) & 0.743 (80) \\
6 & $Z_1Z_3$ & -0.8738 & -0.058 (03) & -0.794 (53) & -0.036 (02) & -1.29 (12) \\
8& $Z_3Z_4$ & 0.2718 & 0.022 (04) & 0.227 (42) & 0.009 (02) & 0.229 (65) \\
12 & $Z_2Z_4$ & 0.3839 & 0.028 (03) & 0.514 (59) & 0.011 (02) & 0.59 (12) \\
14 & $Z_1Z_4$ & -0.3839 & -0.033 (03) & -0.511 (48) & -0.014 (02) & -0.437 (77) \\
16 & $Z_4Z_5$ & -0.0027 & 0.008 (03) & 0.063 (27) & -0.001 (02) & -0.016 (44) \\
24 & $Z_3Z_5$ & 0.1126 & 0.003 (03) & 0.035 (34) & 0.007 (03) & 0.192 (76) \\
28 & $Z_2Z_5$ & 0.1804 & 0.007 (03) & 0.142 (51) & 0.006 (02) & 0.300 (95) \\
30 & $Z_1Z_5$ & -0.1805 & -0.008 (03) & -0.145 (50) & -0.007 (02) & -0.263 (70) \\
\end{tabular}}
\caption{ 
The results obtained for strings of Pauli operators evaluated in the $\lambda\phi^4$ ground-state wavefunction prepared on {\tt ibm\_sherbrooke} using SeqHT ASP, both error-mitigated and unmitigated, along with exact results, as displayed in Fig.~\ref{fig:device2steps}. 
The uncertainties are obtained from bootstrap resampling using 1000 samples.
}
\label{tab:devicenum}
\end{table}
\begin{table}[htbp]
\renewcommand{\arraystretch}{0.8}
\resizebox{\textwidth}{!}{
\centering
\begin{tabular}{c|c||c|c|c|c|c|c|c|c}
\multicolumn{1}{c}{}& & \multicolumn{8}{c}{Coefficients in $\hat{\phi}^4$ Expansion} \\
\hline
Sequency & Operator & $n_q=5$ & $n_q=6$ & $n_q=7$ & $n_q=8$ & $n_q=10$ & $n_q=11$ & $n_q=12$ & Continuum \\
\hline
0 & I & 57.94 & 54.48 & 52.82 & 52.01 & 51.40 & 51.30 & 51.25 & 51.20 \\
2 & $Z_1Z_2$ & 54.36 & 51.09 & 49.52 & 48.76 & 48.19 & 48.09 & 48.05 & 48.00 \\
4 & $Z_2Z_3$ & 30.62 & 28.75 & 27.86 & 27.43 & 27.11 & 27.05 & 27.03 & 27.00 \\
6 & $Z_1Z_3$ & 33.99 & 31.93 & 30.95 & 30.47 & 30.12 & 30.06 & 30.03 & 30.00 \\
8 & $Z_3Z_4$ & 8.720 & 8.185 & 7.932 & 7.809 & 7.718 & 7.703 & 7.695 & 7.687 \\
10& $Z_1Z_2Z_3Z_4$ & 6.812 & 6.390 & 6.191 & 6.095 & 6.023 & 6.012 & 6.006 & 6.000 \\
12& $Z_2Z_4$ & 15.74 & 14.77 & 14.32 & 14.09 & 13.93 & 13.90 & 13.89 & 13.88 \\
14& $Z_1Z_4$ & 17.85 & 16.77 & 16.25 & 16.00 & 15.81 & 15.78 & 15.77 & 15.75 \\
16& $Z_4Z_5$ & 2.246 & 2.109 & 2.043 & 2.012 & 1.988 & 1.984 & 1.982 & 1.980 \\
18 & $Z_1Z_2Z_4Z_5$ & 1.703 & 1.598 & 1.548 & 1.524 & 1.506 & 1.503 & 1.501 & 1.500 \\
20& $Z_2Z_3Z_4Z_5$ & 0.4258 & 0.3994 & 0.3870 & 0.3809 & 0.3765 & 0.3757 & 0.3754 & 0.3750 \\
22 & $Z_1Z_3Z_4Z_5$ & 0.8516 & 0.7988 & 0.7739 & 0.7618 & 0.7529 & 0.7515 & 0.7507 & 0.7500 \\
24 & $Z_3Z_5$ & 4.386 & 4.118 & 3.990 & 3.928 & 3.882 & 3.875 & 3.871 & 3.867 \\
26& $Z_1Z_2Z_3Z_5$ & 3.406 & 3.195 & 3.096 & 3.047 & 3.012 & 3.006 & 3.003 & 3.000 \\
28 & $Z_2Z_5$ & 7.921 & 7.436 & 7.206 & 7.094 & 7.012 & 6.998 & 6.991 & 6.984 \\
30& $Z_1Z_5$ & 9.030 & 8.483 & 8.222 & 8.094 & 8.000 & 7.984 & 7.977 & 7.969 \\
32& $Z_5Z_6$ & -- & 0.5311 & 0.5146 & 0.5066 & 0.5007 & 0.4998 & 0.4993 & 0.4988 \\
34& $Z_1Z_2Z_5Z_6$ & -- & 0.3994 & 0.3870 & 0.3809 & 0.3765 & 0.3757 & 0.3754 & 0.3750 \\
36 & $Z_2Z_3Z_5Z_6$& -- & 0.09985 & 0.09674 & 0.09523 & 0.09412 & 0.09393 & 0.09384 & 0.09375 \\
38& $Z_1Z_3Z_5Z_5$ & -- & 0.1997 & 0.1935 & 0.1905 & 0.1882 & 0.1879 & 0.1877 & 0.1875 \\
\end{tabular}}
\caption{The coefficients of the Pauli strings contributing to the $\hat{\phi}^4$ operator as a function of $\nu$ for increasing $n_q$ and in the continuum limit, as shown in Fig~\ref{fig:coeffs}. 
The points shown in Fig.~\ref{fig:coeffs} are normalized by $\phi_{\rm max}^4 = 256$. 
}
\label{tab:coefficients}
\end{table}

\begin{table}[h!]
\centering
\renewcommand{\arraystretch}{0.8}
\resizebox{!}{0.4\textheight}{
\begin{tabular}{c||c|c|c|c|c}
& \multicolumn{2}{c|}{$\hat{\phi}^4$ theory} & \multicolumn{3}{c}{QHO} \\
\hline
Index & Truncated & Original & $\phi_{\rm max} = 4$ & Optimal $\phi_{\rm max}$ & Analytical \\
\hline
0  & 0.7003 & 0.6735 & 0.500 & 0.500 & 1.5 \\
1  & 2.304  & 2.236  & 1.500 & 1.500 & 2.5 \\
2  & 4.239  & 4.142  & 2.500 & 2.500 & 3.5 \\
3  & 6.399  & 6.279  & 3.499 & 3.500 & 4.5 \\
4  & 8.739  & 8.603  & 4.505 & 4.500 & 5.5 \\
5  & 11.25  & 11.08  & 5.472 & 5.500 & 6.5 \\
6  & 13.82  & 13.70  & 6.573 & 6.500 & 7.5 \\
7  & 16.53  & 16.45  & 7.276 & 7.500 & 8.5 \\
8  & 19.68  & 19.31  & 8.916 & 8.500 & 9.5 \\
9  & 22.60  & 22.28  & 9.188 & 9.500 & 10.5 \\
10 & 25.02  & 25.34  & 11.76 & 10.50 & 11.5 \\
11 & 28.16  & 28.50  & 11.85 & 11.49 & 12.5 \\
12 & 32.18  & 31.74  & 15.19 & 12.52 & 13.5 \\
13 & 36.11  & 35.06  & 15.23 & 13.44 & 14.5 \\
14 & 39.13  & 38.46  & 19.22 & 14.60 & 15.5 \\
15 & 41.49  & 41.94  & 19.23 & 15.25 & 16.5 \\
16 & 44.51  & 45.49  & 23.83 & 16.87 & 17.5 \\
17 & 48.55  & 49.11  & 23.83 & 17.15 & 18.5 \\
18 & 53.03  & 52.79  & 29.02 & 19.41 & 19.5 \\
19 & 57.64  & 56.55  & 29.03 & 19.52 & 20.5 \\
20 & 61.90  & 60.27  & 34.80 & 22.28 & 21.5 \\
21 & 65.71  & 64.29  & 34.81 & 22.32 & 22.5 \\
22 & 68.68  & 68.39  & 41.15 & 25.47 & 23.5 \\
23 & 70.00  & 70.49  & 41.17 & 25.47 & 24.5 \\
24 & 77.60  & 79.72  & 48.08 & 28.95 & 25.5 \\
25 & 78.01  & 79.86  & 48.12 & 28.97 & 26.5 \\
26 & 98.89  & 95.09  & 55.58 & 32.68 & 27.5 \\
27 & 99.99  & 95.60  & 55.66 & 32.83 & 28.5 \\
28 & 107.2  & 113.3  & 63.61 & 36.19 & 29.5 \\
29 & 118.8  & 118.9  & 63.83 & 37.63 & 30.5 \\
30 & 135.6  & 133.0  & 70.90 & 40.70 & 31.5 \\
31 & 158.5  & 164.3  & 74.60 & 47.08 & 32.5 \\
\end{tabular}}
\caption{The $\hat{\phi}^4$ columns show the eigenvalues from exactly solving a digitized interacting Hamiltonian with and without SeqHT, as shown in Fig.~\ref{fig: eigens}. The QHO columns show the analytical eigenvalues of the free theory and the eigenvalues from a QHO Hamiltonian digitized with optimal maximum field value defined in Eq.~(\ref{eqn:optimal}) and with $\phi_{\rm max} = 4$, as shown in the left panel of Fig.~\ref{fig: analytical}. }
\label{tab:eigenvalues}
\end{table}
\begin{table}[h!]
\centering
\renewcommand{\arraystretch}{0.8}
\resizebox{!}{0.4\textheight}{
\begin{tabular}{c|c||c|c||c|c}
\multicolumn{2}{c||}{$\lambda = 10$} & \multicolumn{2}{c||}{$\lambda = 60$} & \multicolumn{2}{c}{Free } \\
\hline
Evolved & Target & Evolved & Target & Digitized & Analytical \\
\hline
\rule{0pt}{3ex} 3.572$\times 10^{-4}$ & 9.630$\times 10^{-11}$ & 1.191$\times 10^{-4}$ & 3.370$\times 10^{-10}$ & 8.155$\times 10^{-5}$ & 1.280$\times 10^{-4}$ \\
1.053$\times 10^{-4}$ & 4.029$\times 10^{-9}$ & 1.662$\times 10^{-4}$ & -4.556$\times 10^{-10}$ & 3.292$\times 10^{-4}$ & 3.476$\times 10^{-4}$ \\
2.473$\times 10^{-4}$ & 1.058$\times 10^{-7}$ & 6.891$\times 10^{-4}$ & 6.488$\times 10^{-10}$ & 8.752$\times 10^{-4}$ & 8.830$\times 10^{-4}$ \\
1.886$\times 10^{-4}$ & 1.824$\times 10^{-6}$ & 1.740$\times 10^{-3}$ & -9.599$\times 10^{-10}$ & 2.095$\times 10^{-3}$ & 2.099$\times 10^{-3}$ \\
2.277$\times 10^{-4}$ & 2.127$\times 10^{-5}$ & 4.153$\times 10^{-3}$ & 1.759$\times 10^{-9}$ & 4.665$\times 10^{-3}$ & 4.667$\times 10^{-3}$ \\
5.275$\times 10^{-4}$ & 1.733$\times 10^{-4}$ & 2.375$\times 10^{-3}$ & 3.498$\times 10^{-8}$ & 9.709$\times 10^{-3}$ & 9.710$\times 10^{-3}$ \\
2.305$\times 10^{-3}$ & 1.017$\times 10^{-3}$ & 6.063$\times 10^{-3}$ & 1.997$\times 10^{-6}$ & 1.890$\times 10^{-2}$ & 1.890$\times 10^{-2}$ \\
6.672$\times 10^{-3}$ & 4.436$\times 10^{-3}$ & 3.768$\times 10^{-3}$ & 5.189$\times 10^{-5}$ & 3.441$\times 10^{-2}$ & 3.441$\times 10^{-2}$ \\
1.593$\times 10^{-2}$ & 1.483$\times 10^{-2}$ & 1.950$\times 10^{-2}$ & 7.073$\times 10^{-4}$ & 5.863$\times 10^{-2}$ & 5.863$\times 10^{-2}$ \\
3.498$\times 10^{-2}$ & 3.914$\times 10^{-2}$ & 3.802$\times 10^{-3}$ & 5.454$\times 10^{-3}$ & 9.345$\times 10^{-2}$ & 9.345$\times 10^{-2}$ \\
7.829$\times 10^{-2}$ & 8.402$\times 10^{-2}$ & 3.901$\times 10^{-2}$ & 2.570$\times 10^{-2}$ & 1.394$\times 10^{-1}$ & 1.394$\times 10^{-1}$ \\
1.450$\times 10^{-1}$ & 1.508$\times 10^{-1}$ & 1.089$\times 10^{-1}$ & 7.987$\times 10^{-2}$ & 1.944$\times 10^{-1}$ & 1.944$\times 10^{-1}$ \\
2.315$\times 10^{-1}$ & 2.321$\times 10^{-1}$ & 1.999$\times 10^{-1}$ & 1.762$\times 10^{-1}$ & 2.538$\times 10^{-1}$ & 2.538$\times 10^{-1}$ \\
3.141$\times 10^{-1}$ & 3.137$\times 10^{-1}$ & 2.927$\times 10^{-1}$ & 2.957$\times 10^{-1}$ & 3.099$\times 10^{-1}$ & 3.099$\times 10^{-1}$ \\
3.807$\times 10^{-1}$ & 3.791$\times 10^{-1}$ & 3.883$\times 10^{-1}$ & 4.015$\times 10^{-1}$ & 3.540$\times 10^{-1}$ & 3.540$\times 10^{-1}$ \\
4.173$\times 10^{-1}$ & 4.153$\times 10^{-1}$ & 4.580$\times 10^{-1}$ & 4.618$\times 10^{-1}$ & 3.784$\times 10^{-1}$ & 3.784$\times 10^{-1}$ \\
4.173$\times 10^{-1}$ & 4.153$\times 10^{-1}$ & 4.580$\times 10^{-1}$ & 4.618$\times 10^{-1}$ & 3.784$\times 10^{-1}$ & 3.784$\times 10^{-1}$ \\
3.807$\times 10^{-1}$ & 3.791$\times 10^{-1}$ & 3.883$\times 10^{-1}$ & 4.015$\times 10^{-1}$ & 3.540$\times 10^{-1}$ & 3.540$\times 10^{-1}$ \\
3.141$\times 10^{-1}$ & 3.137$\times 10^{-1}$ & 2.927$\times 10^{-1}$ & 2.957$\times 10^{-1}$ & 3.099$\times 10^{-1}$ & 3.099$\times 10^{-1}$ \\
2.315$\times 10^{-1}$ & 2.321$\times 10^{-1}$ & 1.999$\times 10^{-1}$ & 1.762$\times 10^{-1}$ & 2.538$\times 10^{-1}$ & 2.538$\times 10^{-1}$ \\
1.450$\times 10^{-1}$ & 1.508$\times 10^{-1}$ & 1.089$\times 10^{-1}$ & 7.987$\times 10^{-2}$ & 1.944$\times 10^{-1}$ & 1.944$\times 10^{-1}$ \\
7.829$\times 10^{-2}$ & 8.402$\times 10^{-2}$ & 3.901$\times 10^{-2}$ & 2.570$\times 10^{-2}$ & 1.394$\times 10^{-1}$ & 1.394$\times 10^{-1}$ \\
3.498$\times 10^{-2}$ & 3.914$\times 10^{-2}$ & 3.802$\times 10^{-3}$ & 5.454$\times 10^{-3}$ & 9.345$\times 10^{-2}$ & 9.345$\times 10^{-2}$ \\
1.593$\times 10^{-2}$ & 1.483$\times 10^{-2}$ & 1.950$\times 10^{-2}$ & 7.073$\times 10^{-4}$ & 5.863$\times 10^{-2}$ & 5.863$\times 10^{-2}$ \\
6.672$\times 10^{-3}$ & 4.436$\times 10^{-3}$ & 3.768$\times 10^{-3}$ & 5.189$\times 10^{-5}$ & 3.441$\times 10^{-2}$ & 3.441$\times 10^{-2}$ \\
2.305$\times 10^{-3}$ & 1.017$\times 10^{-3}$ & 6.063$\times 10^{-3}$ & 1.997$\times 10^{-6}$ & 1.890$\times 10^{-2}$ & 1.890$\times 10^{-2}$ \\
5.275$\times 10^{-4}$ & 1.733$\times 10^{-4}$ & 2.375$\times 10^{-3}$ & 3.498$\times 10^{-8}$ & 9.709$\times 10^{-3}$ & 9.710$\times 10^{-3}$ \\
2.277$\times 10^{-4}$ & 2.127$\times 10^{-5}$ & 4.153$\times 10^{-3}$ & 1.759$\times 10^{-9}$ & 4.665$\times 10^{-3}$ & 4.667$\times 10^{-3}$ \\
1.886$\times 10^{-4}$ & 1.824$\times 10^{-6}$ & 1.740$\times 10^{-3}$ & -9.599$\times 10^{-10}$ & 2.095$\times 10^{-3}$ & 2.099$\times 10^{-3}$ \\
2.473$\times 10^{-4}$ & 1.058$\times 10^{-7}$ & 6.891$\times 10^{-4}$ & 6.488$\times 10^{-10}$ & 8.752$\times 10^{-4}$ & 8.830$\times 10^{-4}$ \\
1.053$\times 10^{-4}$ & 4.029$\times 10^{-9}$ & 1.662$\times 10^{-4}$ & -4.556$\times 10^{-10}$ & 3.292$\times 10^{-4}$ & 3.476$\times 10^{-4}$ \\
3.572$\times 10^{-4}$ & 9.630$\times 10^{-11}$ & 1.191$\times 10^{-4}$ & 3.370$\times 10^{-10}$ & 8.155$\times 10^{-5}$ & 1.280$\times 10^{-4}$ \\
\end{tabular}}
\caption{Amplitudes of the SeqHT evolved and target ground states of the interacting theory with $\lambda = 10$ and $\lambda = 60$, and the ground state wavefunction of the free theory obtained analytically and numerically, shown in Fig.~\ref{fig:amp} and the right panel of Fig.~\ref{fig: analytical}. }
\label{tab:gs_5qubits}
\end{table}
\begin{table}
\centering
\renewcommand{\arraystretch}{0.8}
\resizebox{!}{0.4\textheight}{
\begin{tabular}{c|c|c||c|c|c||c|c|c}
     $t$ &   Full &   Trunc &     $t$ &   Full &   Trunc &     $t$ &   Full &   Trunc \\
\hline
 0.0 &  0.973 &   0.973 & 2.7 &  0.997 &   0.997 & 5.4 &  0.999 &   0.998 \\
 0.1 &  0.974 &   0.974 & 2.8 &  0.997 &   0.997 & 5.5 &  0.999 &   0.998 \\
 0.2 &  0.975 &   0.974 & 2.9 &  0.997 &   0.997 & 5.6 &  0.999 &   0.998 \\
 0.3 &  0.976 &   0.976 & 3.0 &  0.997 &   0.996 & 5.7 &  0.999 &   0.998 \\
 0.4 &  0.977 &   0.977 & 3.1 &  0.997 &   0.996 & 5.8 &  0.999 &   0.998 \\
 0.5 &  0.979 &   0.978 & 3.2 &  0.997 &   0.996 & 5.9 &  0.999 &   0.998 \\
 0.6 &  0.981 &   0.980 & 3.3 &  0.997 &   0.996 & 6.0 &  0.999 &   0.999 \\
 0.7 &  0.983 &   0.981 & 3.4 &  0.998 &   0.997 & 6.1 &  0.999 &   0.999 \\
 0.8 &  0.985 &   0.983 & 3.5 &  0.998 &   0.997 & 6.2 &  0.999 &   0.999 \\
 0.9 &  0.986 &   0.984 & 3.6 &  0.998 &   0.997 & 6.3 &  0.999 &   0.999 \\
 1.0 &  0.988 &   0.986 & 3.7 &  0.998 &   0.997 & 6.4 &  0.999 &   0.999 \\
 1.1 &  0.990 &   0.987 & 3.8 &  0.999 &   0.998 & 6.5 &  0.999 &   1.000 \\
 1.2 &  0.991 &   0.989 & 3.9 &  0.999 &   0.998 & 6.6 &  0.999 &   1.000 \\
 1.3 &  0.992 &   0.990 & 4.0 &  0.999 &   0.998 & 6.7 &  0.999 &   1.000 \\
 1.4 &  0.994 &   0.992 & 4.1 &  0.999 &   0.999 & 6.8 &  0.999 &   1.000 \\
 1.5 &  0.995 &   0.993 & 4.2 &  0.999 &   0.999 & 6.9 &  0.999 &   1.000 \\
 1.6 &  0.996 &   0.994 & 4.3 &  0.999 &   0.999 & 7.0 &  0.999 &   0.999 \\
 1.7 &  0.996 &   0.995 & 4.4 &  0.999 &   0.999 & 7.1 &  0.999 &   0.999 \\
 1.8 &  0.997 &   0.996 & 4.5 &  0.999 &   0.999 & 7.2 &  0.999 &   0.999 \\
 1.9 &  0.997 &   0.997 & 4.6 &  0.999 &   0.999 & 7.3 &  0.999 &   0.999 \\
 2.0 &  0.997 &   0.997 & 4.7 &  0.999 &   0.999 & 7.4 &  0.999 &   0.999 \\
 2.1 &  0.997 &   0.997 & 4.8 &  0.999 &   0.999 & 7.5 &  0.999 &   0.999 \\
 2.2 &  0.997 &   0.997 & 4.9 &  0.999 &   0.999 & 7.6 &  0.999 &   0.999 \\
 2.3 &  0.997 &   0.998 & 5.0 &  0.999 &   0.999 & 7.7 &  0.999 &   0.999 \\
 2.4 &  0.997 &   0.997 & 5.1 &  0.999 &   0.998 & 7.8 &  0.999 &   0.999 \\
 2.5 &  0.997 &   0.997 & 5.2 &  0.999 &   0.998 & 7.9 &  0.999 &   0.999 \\
 2.6 &  0.997 &   0.997 & 5.3 &  0.999 &   0.998 & 8.0 &  0.999 &   0.999 \\
\end{tabular}}
\caption{The fidelity of the $\lambda\phi^4 $ ground-state wavefunction prepared with complete adiabatic evolution and with SeqHT procedure as a function of time $t$, as described in the main text and shown in Fig.~\ref{fig:time env}.}
\label{tab:fidelity_time}
\end{table}

\begin{table}[htbp]
\renewcommand{\arraystretch}{0.8}
\resizebox{!}{0.4\textheight}{
    \centering
    \begin{tabular}{r|r|r||r|r|r||r|r|r||r|r|r}
 $\nu$ &   Coeff &   Bound &   $\nu$ &   Coeff &   Bound &   $\nu$ &   Coeff &   Bound &   $\nu$&   Coeff &   Bound  \\
\hline
  0 & 1.0000 &  1.0000 &      64 &  0.0024 &  0.0757 &     128 &  0.0006 &  0.0385 &     192 &  0.0012 &  0.0385 \\
  2 & 0.9375 &  0.9688 &      66 &  0.0018 &  0.0757 &     130 &  0.0005 &  0.0385 &     194 &  0.0009 &  0.0385 \\
  4 & 0.5274 &  0.7627 &      68 &  0.0005 &  0.0757 &     132 &  0.0001 &  0.0385 &     196 &  0.0002 &  0.0385 \\
  6 & 0.5859 &  0.7627 &      70 &  0.0009 &  0.0757 &     134 &  0.0002 &  0.0385 &     198 &  0.0005 &  0.0385 \\
  8 & 0.1502 &  0.4871 &      72 &  0.0001 &  0.0757 &     136 &  0.0000 &  0.0385 &     200 &  0.0001 &  0.0385 \\
 10 & 0.1172 &  0.4871 &      74 &  0.0000 &  0.0757 &     138 &  0.0000 &  0.0385 &     202 &  0.0000 &  0.0385 \\
 12 & 0.2710 &  0.4871 &      76 &  0.0002 &  0.0757 &     140 &  0.0001 &  0.0385 &     204 &  0.0001 &  0.0385 \\
 14 & 0.3076 &  0.4871 &      78 &  0.0005 &  0.0757 &     142 &  0.0001 &  0.0385 &     206 &  0.0002 &  0.0385 \\
 16 & 0.0387 &  0.2758 &      80 &  0.0000 &  0.0757 &     144 &  0.0000 &  0.0385 &     208 &  0.0000 &  0.0385 \\
 18 & 0.0293 &  0.2758 &      82 &  0.0000 &  0.0757 &     146 &  0.0000 &  0.0385 &     210 &  0.0000 &  0.0385 \\
 20 & 0.0073 &  0.2758 &      84 &  0.0000 &  0.0757 &     148 &  0.0000 &  0.0385 &     212 &  0.0000 &  0.0385 \\
 22 & 0.0146 &  0.2758 &      86 &  0.0000 &  0.0757 &     150 &  0.0000 &  0.0385 &     214 &  0.0000 &  0.0385 \\
 24 & 0.0755 &  0.2758 &      88 &  0.0001 &  0.0757 &     152 &  0.0000 &  0.0385 &     216 &  0.0000 &  0.0385 \\
 26 & 0.0586 &  0.2758 &      90 &  0.0000 &  0.0757 &     154 &  0.0000 &  0.0385 &     218 &  0.0000 &  0.0385 \\
 28 & 0.1364 &  0.2758 &      92 &  0.0001 &  0.0757 &     156 &  0.0000 &  0.0385 &     220 &  0.0001 &  0.0385 \\
 30 & 0.1556 &  0.2758 &      94 &  0.0002 &  0.0757 &     158 &  0.0001 &  0.0385 &     222 &  0.0001 &  0.0385 \\
 32 & 0.0097 &  0.1468 &      96 &  0.0049 &  0.0757 &     160 &  0.0000 &  0.0385 &     224 &  0.0024 &  0.0385 \\
 34 & 0.0073 &  0.1468 &      98 &  0.0037 &  0.0757 &     162 &  0.0000 &  0.0385 &     226 &  0.0018 &  0.0385 \\
 36 & 0.0018 &  0.1468 &     100 &  0.0009 &  0.0757 &     164 &  0.0000 &  0.0385 &     228 &  0.0005 &  0.0385 \\
 38 & 0.0037 &  0.1468 &     102 &  0.0018 &  0.0757 &     166 &  0.0000 &  0.0385 &     230 &  0.0009 &  0.0385 \\
 40 & 0.0005 &  0.1468 &     104 &  0.0002 &  0.0757 &     168 &  0.0000 &  0.0385 &     232 &  0.0001 &  0.0385 \\
 42 & 0.0000 &  0.1468 &     106 &  0.0000 &  0.0757 &     170 &  0.0000 &  0.0385 &     234 &  0.0000 &  0.0385 \\
 44 & 0.0009 &  0.1468 &     108 &  0.0005 &  0.0757 &     172 &  0.0000 &  0.0385 &     236 &  0.0002 &  0.0385 \\
 46 & 0.0018 &  0.1468 &     110 &  0.0009 &  0.0757 &     174 &  0.0000 &  0.0385 &     238 &  0.0005 &  0.0385 \\
 48 & 0.0194 &  0.1468 &     112 &  0.0097 &  0.0757 &     176 &  0.0000 &  0.0385 &     240 &  0.0048 &  0.0385 \\
 50 & 0.0146 &  0.1468 &     114 &  0.0073 &  0.0757 &     178 &  0.0000 &  0.0385 &     242 &  0.0037 &  0.0385 \\
 52 & 0.0037 &  0.1468 &     116 &  0.0018 &  0.0757 &     180 &  0.0000 &  0.0385 &     244 &  0.0009 &  0.0385 \\
 54 & 0.0073 &  0.1468 &     118 &  0.0037 &  0.0757 &     182 &  0.0000 &  0.0385 &     246 &  0.0018 &  0.0385 \\
 56 & 0.0378 &  0.1468 &     120 &  0.0189 &  0.0757 &     184 &  0.0000 &  0.0385 &     248 &  0.0095 &  0.0385 \\
 58 & 0.0293 &  0.1468 &     122 &  0.0146 &  0.0757 &     186 &  0.0000 &  0.0385 &     250 &  0.0073 &  0.0385 \\
 60 & 0.0683 &  0.1468 &     124 &  0.0342 &  0.0757 &     188 &  0.0000 &  0.0385 &     252 &  0.0171 &  0.0385 \\
 62 & 0.0781 &  0.1468 &     126 &  0.0391 &  0.0757 &     190 &  0.0000 &  0.0385 &     254 &  0.0195 &  0.0385 \\
\end{tabular}}
\caption{Calculated sequency coefficients, $\tilde\beta_\nu$,and upper bounds, $\tilde B_\nu$, for a $x^4$ potential, as shown in Fig.~\ref{fig:UpperBound}.}
    \label{tab:bound}
\end{table}

\begin{table}[htpb]

    \centering
    \renewcommand{\arraystretch}{0.8}
\resizebox{\textwidth}{!}{
    \begin{tabular}{r|r|r|r|r|r|r|r|r|r|r|r|r|r|r|r|r|r}
    \hline
  &\backslashbox{\#}{$\delta t$} & 0.1 &   0.13 &   0.16 &   0.19 &   0.22 &   0.25 &   0.28 &   0.31 &   0.34 &   0.37 &   0.4 &   0.43 &   0.46 &   0.49 &   0.52 &   0.55 \\
\hline
 \multirow{14}{3em}{Full} & 1 &  0.971 &  0.970 &  0.968 &  0.966 &  0.964 &  0.961 &  0.958 &  0.954 &  0.950 &  0.945 & 0.940 &  0.934 &  0.928 &  0.921 &  0.914 &  0.906 \\
& 2& 0.970 &  0.968 &  0.965 &  0.962 &  0.959 &  0.955 &  0.950 &  0.946 &  0.941 &  0.936 & 0.931 &  0.926 &  0.921 &  0.916 &  0.912 &  0.907 \\
& 3& 0.969 &  0.967 &  0.964 &  0.961 &  0.958 &  0.955 &  0.952 &  0.950 &  0.947 &  0.946 & 0.944 &  0.944 &  0.945 &  0.951 &  0.960 &  0.971 \\
&  4& 0.969 &  0.967 &  0.965 &  0.963 &  0.962 &  0.961 &  0.961 &  0.961 &  0.963 &  0.968 & 0.976 &  0.986 &  0.994 &  0.997 &  0.992 &  0.978 \\
& 5& 0.970 &  0.969 &  0.968 &  0.968 &  0.968 &  0.970 &  0.973 &  0.980 &  0.989 &  0.996 & 0.998 &  0.993 &  0.982 &  0.967 &  0.950 &  0.932 \\
& 6&0.971 &  0.971 &  0.971 &  0.973 &  0.976 &  0.982 &  0.990 &  0.997 &  0.998 &  0.992 & 0.981 &  0.966 &  0.951 &  0.934 &  0.923 &  0.922 \\
& 7& 0.972 &  0.974 &  0.976 &  0.979 &  0.986 &  0.994 &  0.999 &  0.996 &  0.987 &  0.973 & 0.960 &  0.946 &  0.937 &  0.947 &  0.966 &  0.972 \\
& 8 & 0.974 &  0.977 &  0.980 &  0.987 &  0.995 &  0.999 &  0.995 &  0.984 &  0.972 &  0.959 & 0.949 &  0.955 &  0.975 &  0.986 &  0.981 &  0.964 \\
& 9 & 0.976 &  0.980 &  0.986 &  0.994 &  0.999 &  0.995 &  0.985 &  0.973 &  0.962 &  0.959 & 0.971 &  0.989 &  0.992 &  0.980 &  0.949 &  0.918 \\
& 10 & 0.979 &  0.984 &  0.991 &  0.998 &  0.997 &  0.988 &  0.977 &  0.967 &  0.966 &  0.981 & 0.994 &  0.991 &  0.975 &  0.941 &  0.926 &  0.936 \\
& 11 & 0.981 &  0.987 &  0.996 &  0.998 &  0.991 &  0.981 &  0.971 &  0.971 &  0.985 &  0.996 & 0.990 &  0.971 &  0.942 &  0.942 &  0.960 &  0.972 \\
& 12 & 0.983 &  0.991 &  0.998 &  0.996 &  0.986 &  0.976 &  0.973 &  0.986 &  0.997 &  0.991 & 0.972 &  0.948 &  0.955 &  0.975 &  0.978 &  0.942 \\
& 13 & 0.986 &  0.995 &  0.998 &  0.991 &  0.982 &  0.975 &  0.984 &  0.997 &  0.993 &  0.976 & 0.954 &  0.964 &  0.982 &  0.983 &  0.943 &  0.914 \\
& 14 & 0.988 &  0.997 &  0.997 &  0.988 &  0.979 &  0.981 &  0.995 &  0.996 &  0.982 &  0.960 & 0.968 &  0.987 &  0.989 &  0.947 &  0.932 &  0.948 \\
\hline
\end{tabular}}
\renewcommand{\arraystretch}{0.8}
\resizebox{\textwidth}{!}{
    \begin{tabular}{r|r|r|r|r|r|r|r|r|r|r|r|r|r|r|r|r|r}
 &  \backslashbox{\#}{$\delta t$} &   0.1 &   0.13 &   0.16 &   0.19 &   0.22 &   0.25 &   0.28 &   0.31 &   0.34 &   0.37 &   0.4 &   0.43 &   0.46 &   0.49 &   0.52 &   0.55 \\
\hline
  \multirow{14}{3em}{SeqHT} & 1 & 0.971 &  0.969 &  0.968 &  0.965 &  0.963 &  0.960 &  0.956 &  0.952 &  0.948 &  0.943 & 0.937 &  0.931 &  0.924 &  0.917 &  0.909 &  0.901 \\
 & 2 & 0.969 &  0.967 &  0.964 &  0.960 &  0.957 &  0.952 &  0.947 &  0.942 &  0.937 &  0.931 & 0.926 &  0.922 &  0.914 &  0.906 &  0.906 &  0.906 \\
 & 3 & 0.968 &  0.966 &  0.963 &  0.959 &  0.956 &  0.951 &  0.947 &  0.945 &  0.943 &  0.940 & 0.936 &  0.938 &  0.941 &  0.941 &  0.957 &  0.965 \\
 & 4 & 0.968 &  0.966 &  0.963 &  0.959 &  0.959 &  0.958 &  0.956 &  0.955 &  0.956 &  0.963 & 0.971 &  0.982 &  0.989 &  0.982 &  0.983 &  0.968 \\
 & 5 & 0.968 &  0.967 &  0.965 &  0.963 &  0.964 &  0.965 &  0.967 &  0.977 &  0.987 &  0.990 & 0.988 &  0.985 &  0.976 &  0.945 &  0.935 &  0.925 \\
 & 6 & 0.968 &  0.969 &  0.968 &  0.967 &  0.972 &  0.978 &  0.985 &  0.990 &  0.993 &  0.986 & 0.968 &  0.960 &  0.940 &  0.908 &  0.916 &  0.916 \\
 & 7 & 0.969 &  0.971 &  0.972 &  0.974 &  0.983 &  0.991 &  0.992 &  0.988 &  0.983 &  0.965 & 0.938 &  0.944 &  0.931 &  0.912 &  0.955 &  0.962 \\
 & 8 & 0.971 &  0.974 &  0.977 &  0.981 &  0.992 &  0.998 &  0.989 &  0.972 &  0.962 &  0.947 & 0.928 &  0.949 &  0.968 &  0.945 &  0.963 &  0.963 \\
 & 9 & 0.972 &  0.977 &  0.983 &  0.988 &  0.996 &  0.992 &  0.977 &  0.963 &  0.954 &  0.945 & 0.951 &  0.975 &  0.984 &  0.926 &  0.938 &  0.923 \\
 & 10 & 0.974 &  0.981 &  0.989 &  0.994 &  0.993 &  0.984 &  0.967 &  0.958 &  0.961 &  0.971 & 0.973 &  0.978 &  0.962 &  0.877 &  0.939 &  0.932 \\
 & 11 & 0.976 &  0.985 &  0.994 &  0.993 &  0.988 &  0.977 &  0.964 &  0.961 &  0.980 &  0.985 & 0.966 &  0.954 &  0.932 &  0.884 &  0.970 &  0.956 \\
 & 12 & 0.978 &  0.989 &  0.997 &  0.990 &  0.981 &  0.973 &  0.962 &  0.977 &  0.988 &  0.977 & 0.937 &  0.935 &  0.943 &  0.902 &  0.950 &  0.944 \\
& 13 & 0.980 &  0.993 &  0.997 &  0.984 &  0.976 &  0.972 &  0.975 &  0.980 &  0.982 &  0.954 & 0.929 &  0.955 &  0.969 &  0.889 &  0.928 &  0.905 \\
& 14 & 0.983 &  0.996 &  0.996 &  0.979 &  0.973 &  0.978 &  0.984 &  0.984 &  0.972 &  0.946 & 0.950 &  0.972 &  0.964 &  0.865 &  0.929 &  0.938 \\
\end{tabular}}
    \caption{The fidelity of the adiabatically prepared $\lambda\phi^4$ ground state as a function of $\delta t$ and the number of adiabatic steps, for $n_q=5$ with $\phi_{\rm max}=4$ and $\lambda = 10$. Each adiabatic step consists of one first-order Trotter step. Results are shown in the top panels of Fig.~\ref{fig: time scans}.}
    \label{tab:5qubit_scan_1st}
\end{table}

\begin{table}[htpb]
\centering
\renewcommand{\arraystretch}{0.8}
\resizebox{\textwidth}{!}{
    \begin{tabular}{r|r|r|r|r|r|r|r|r|r|r|r|r|r|r|r|r|r|r|r|r}
 \multicolumn{2}{c|}{\backslashbox{\#}{$\delta t$}} & 0.1 &   0.13 &   0.16 &   0.19 &   0.22 &   0.25 &   0.28 &   0.31 &   0.34 &   0.37 &   0.4 &   0.43 &   0.46 &   0.49 &   0.52 &   0.55 &   0.58 &   0.61 &   0.64 \\
\hline
 \multirow{10}{.6em}{F} & 1 & 0.973 &  0.974 &  0.975 &  0.975 &  0.976 &  0.977 &  0.978 &  0.979 &  0.980 &  0.982 & 0.983 &  0.984 &  0.985 &  0.987 &  0.987 &  0.988 &  0.988 &  0.988 &  0.988 \\
& 2 & 0.975 &  0.976 &  0.977 &  0.979 &  0.980 &  0.982 &  0.984 &  0.986 &  0.988 &  0.989 & 0.991 &  0.992 &  0.993 &  0.994 &  0.994 &  0.995 &  0.994 &  0.992 &  0.989 \\
& 3 &  0.976 &  0.978 &  0.980 &  0.982 &  0.985 &  0.987 &  0.989 &  0.991 &  0.993 &  0.995 & 0.996 &  0.997 &  0.998 &  0.997 &  0.995 &  0.992 &  0.988 &  0.983 &  0.979 \\
& 4 & 0.978 &  0.980 &  0.983 &  0.986 &  0.989 &  0.991 &  0.993 &  0.995 &  0.997 &  0.998 & 0.998 &  0.996 &  0.995 &  0.992 &  0.991 &  0.989 &  0.985 &  0.982 &  0.976 \\
& 5 & 0.979 &  0.983 &  0.986 &  0.989 &  0.992 &  0.995 &  0.997 &  0.998 &  0.998 &  0.997 & 0.996 &  0.995 &  0.994 &  0.992 &  0.990 &  0.988 &  0.987 &  0.984 &  0.976 \\
& 6 & 0.981 &  0.985 &  0.989 &  0.992 &  0.995 &  0.997 &  0.998 &  0.997 &  0.996 &  0.996 & 0.996 &  0.995 &  0.994 &  0.993 &  0.992 &  0.989 &  0.987 &  0.983 &  0.968 \\
& 7 & 0.983 &  0.987 &  0.991 &  0.995 &  0.997 &  0.997 &  0.997 &  0.996 &  0.996 &  0.996 & 0.996 &  0.996 &  0.996 &  0.994 &  0.991 &  0.988 &  0.986 &  0.979 &  0.961 \\
& 8 & 0.985 &  0.989 &  0.993 &  0.996 &  0.997 &  0.997 &  0.996 &  0.996 &  0.996 &  0.997 & 0.998 &  0.998 &  0.995 &  0.992 &  0.988 &  0.986 &  0.986 &  0.977 &  0.961 \\
& 9 & 0.987 &  0.991 &  0.995 &  0.997 &  0.997 &  0.997 &  0.996 &  0.997 &  0.997 &  0.999 & 0.999 &  0.997 &  0.994 &  0.990 &  0.988 &  0.987 &  0.984 &  0.975 &  0.952 \\
& 10 & 0.988 &  0.993 &  0.997 &  0.997 &  0.997 &  0.997 &  0.997 &  0.998 &  0.999 &  0.999 & 0.998 &  0.996 &  0.992 &  0.990 &  0.988 &  0.989 &  0.987 &  0.973 &  0.948 \\
\end{tabular}}
\renewcommand{\arraystretch}{0.8}
\resizebox{\textwidth}{!}{
    \begin{tabular}{r|r|r|r|r|r|r|r|r|r|r|r|r|r|r|r|r|r|r|r|r}
    \hline
 \multicolumn{2}{c|}{\backslashbox{\#}{$\delta t$}}&   0.1 &   0.13 &   0.16 &   0.19 &   0.22 &   0.25 &   0.28 &   0.31 &   0.34 &   0.37 &   0.4 &   0.43 &   0.46 &   0.49 &   0.52 &   0.55 &   0.58 &   0.61 &   0.64 \\
\hline
  \multirow{10}{.6em}{T} & 1 & 0.973 &  0.974 &  0.974 &  0.975 &  0.976 &  0.977 &  0.978 &  0.979 &  0.980 &  0.981 & 0.982 &  0.984 &  0.985 &  0.984 &  0.985 &  0.987 &  0.986 &  0.985 &  0.986 \\
 &2 &0.974 &  0.975 &  0.977 &  0.978 &  0.980 &  0.981 &  0.982 &  0.984 &  0.986 &  0.987 & 0.986 &  0.989 &  0.991 &  0.987 &  0.988 &  0.993 &  0.990 &  0.980 &  0.985 \\
& 3 & 0.975 &  0.977 &  0.979 &  0.980 &  0.983 &  0.986 &  0.986 &  0.987 &  0.989 &  0.990 & 0.990 &  0.995 &  0.994 &  0.984 &  0.988 &  0.992 &  0.984 &  0.959 &  0.971 \\
& 4 & 0.976 &  0.979 &  0.982 &  0.983 &  0.986 &  0.987 &  0.988 &  0.992 &  0.996 &  0.993 & 0.989 &  0.994 &  0.991 &  0.975 &  0.986 &  0.988 &  0.980 &  0.939 &  0.961 \\
&5 & 0.977 &  0.981 &  0.984 &  0.985 &  0.989 &  0.992 &  0.991 &  0.992 &  0.992 &  0.991 & 0.987 &  0.991 &  0.989 &  0.967 &  0.986 &  0.988 &  0.981 &  0.918 &  0.952 \\
&6 & 0.979 &  0.983 &  0.986 &  0.988 &  0.992 &  0.994 &  0.992 &  0.992 &  0.994 &  0.990 & 0.981 &  0.990 &  0.987 &  0.957 &  0.984 &  0.990 &  0.977 &  0.894 &  0.942 \\
&7 & 0.980 &  0.985 &  0.988 &  0.989 &  0.994 &  0.996 &  0.993 &  0.990 &  0.991 &  0.987 & 0.982 &  0.991 &  0.990 &  0.958 &  0.990 &  0.986 &  0.976 &  0.863 &  0.933 \\
&8 & 0.981 &  0.987 &  0.990 &  0.991 &  0.995 &  0.994 &  0.991 &  0.991 &  0.993 &  0.985 & 0.981 &  0.990 &  0.986 &  0.944 &  0.989 &  0.987 &  0.972 &  0.836 &  0.916 \\
&9 & 0.982 &  0.989 &  0.993 &  0.993 &  0.994 &  0.995 &  0.989 &  0.988 &  0.992 &  0.988 & 0.979 &  0.988 &  0.985 &  0.937 &  0.982 &  0.985 &  0.970 &  0.827 &  0.898 \\
&10 & 0.984 &  0.991 &  0.994 &  0.991 &  0.995 &  0.995 &  0.990 &  0.985 &  0.991 &  0.986 & 0.974 &  0.990 &  0.982 &  0.924 &  0.986 &  0.983 &  0.968 &  0.781 &  0.879 \\
\end{tabular}}
    \caption{The fidelity of the adiabatically prepared $\lambda\phi^4$ ground state as a function of $\delta t$ and the number of adiabatic steps, for $n_q=5$ with $\phi_{\rm max}=4$ and $\lambda = 10$. Each adiabatic step consists of one second-order Trotter step. Results are shown in the bottom panels of Fig.~\ref{fig: time scans}. }
    \label{tab:5qubit_scan_2nd}
\end{table}

\begin{table}[htpb]
    \centering
    \renewcommand{\arraystretch}{0.8}
\resizebox{!}{0.2\textheight}{
    \begin{tabular}{c|c|c|c|c|c|c|c|c|c|c|c}
    \hline
& \backslashbox{\#}{$\delta t$} & 0.1 & 0.15& 0.2 & 0.25& 0.3 & 0.35& 0.4 & 0.45& 0.5 & 0.55\\
\hline
\multirow{14}{3em}{Full} & 1 & 0.973 & 0.974 & 0.975 & 0.977 & 0.978 & 0.980 & 0.982 & 0.984 & 0.985 & 0.986 \\
& 2 & 0.974 & 0.976 & 0.978 & 0.980 & 0.983 & 0.985 & 0.986 & 0.987 & 0.988 & 0.990 \\
& 3 & 0.975 & 0.978 & 0.981 & 0.984 & 0.986 & 0.989 & 0.990 & 0.990 & 0.988 & 0.986 \\
&4 & 0.977 & 0.980 & 0.983 & 0.986 & 0.989 & 0.992 & 0.992 & 0.985 & 0.982 & 0.982 \\
& 5 & 0.978 & 0.982 & 0.985 & 0.989 & 0.992 & 0.991 & 0.986 & 0.981 & 0.979 & 0.980 \\
& 6 & 0.979 & 0.984 & 0.988 & 0.992 & 0.992 & 0.989 & 0.987 & 0.977 & 0.976 & 0.979 \\
& 7 & 0.981 & 0.986 & 0.990 & 0.992 & 0.991 & 0.987 & 0.983 & 0.973 & 0.977 & 0.977 \\
& 8 & 0.982 & 0.988 & 0.992 & 0.992 & 0.990 & 0.985 & 0.985 & 0.971 & 0.975 & 0.968 \\
& 9 & 0.983 & 0.990 & 0.993 & 0.992 & 0.988 & 0.984 & 0.982 & 0.968 & 0.968 & 0.972 \\
& 10 & 0.985 & 0.992 & 0.993 & 0.990 & 0.987 & 0.984 & 0.980 & 0.962 & 0.971 & 0.968 \\
& 11 & 0.986 & 0.994 & 0.993 & 0.989 & 0.987 & 0.982 & 0.976 & 0.956 & 0.970 & 0.962 \\
& 12 & 0.987 & 0.995 & 0.992 & 0.988 & 0.987 & 0.980 & 0.976 & 0.960 & 0.963 & 0.960 \\
& 13 & 0.989 & 0.995 & 0.991 & 0.987 & 0.986 & 0.979 & 0.974 & 0.950 & 0.965 & 0.963 \\
& 14 & 0.990 & 0.995 & 0.990 & 0.987 & 0.985 & 0.976 & 0.972 & 0.949 & 0.962 & 0.955 \\
    \end{tabular}}
    \renewcommand{\arraystretch}{0.8}
\resizebox{!}{0.2\textheight}{
    \begin{tabular}{c|c|c|c|c|c|c|c|c|c|c|c}
    \hline
& \backslashbox{\#}{$\delta t$} & 0.1 & 0.15& 0.2 & 0.25& 0.3 & 0.35& 0.4 & 0.45& 0.5 & 0.55\\
\hline
\multirow{14}{3em}{SeqHT} & 1 & 0.973 & 0.974 & 0.975 & 0.976 & 0.978 & 0.980 & 0.981 & 0.983 & 0.984 & 0.985 \\
& 2 &  0.974 & 0.976 & 0.978 & 0.980 & 0.983 & 0.984 & 0.985 & 0.985 & 0.986 & 0.987 \\
& 3 & 0.975 & 0.978 & 0.980 & 0.983 & 0.986 & 0.986 & 0.988 & 0.987 & 0.985 & 0.982 \\
& 4 & 0.977 & 0.980 & 0.982 & 0.985 & 0.989 & 0.989 & 0.990 & 0.981 & 0.978 & 0.977 \\
& 5 & 0.978 & 0.982 & 0.984 & 0.988 & 0.991 & 0.986 & 0.982 & 0.976 & 0.973 & 0.973 \\
& 6 & 0.979 & 0.983 & 0.986 & 0.991 & 0.991 & 0.983 & 0.982 & 0.972 & 0.968 & 0.971 \\
& 7 &  0.980 & 0.985 & 0.989 & 0.991 & 0.990 & 0.979 & 0.977 & 0.967 & 0.973 & 0.968 \\
& 8 & 0.982 & 0.987 & 0.990 & 0.991 & 0.989 & 0.976 & 0.979 & 0.963 & 0.965 & 0.957 \\
& 9 & 0.983 & 0.989 & 0.991 & 0.990 & 0.987 & 0.973 & 0.974 & 0.960 & 0.957 & 0.960 \\
& 10 & 0.984 & 0.991 & 0.991 & 0.989 & 0.986 & 0.972 & 0.971 & 0.953 & 0.962 & 0.953 \\
& 11 & 0.986 & 0.992 & 0.990 & 0.987 & 0.986 & 0.969 & 0.969 & 0.944 & 0.961 & 0.948 \\
& 12 & 0.987 & 0.993 & 0.989 & 0.987 & 0.985 & 0.967 & 0.965 & 0.951 & 0.949 & 0.942 \\
& 13 & 0.988 & 0.993 & 0.988 & 0.986 & 0.985 & 0.964 & 0.964 & 0.935 & 0.953 & 0.944 \\
& 14 & 0.990 & 0.993 & 0.987 & 0.985 & 0.984 & 0.961 & 0.961 & 0.938 & 0.949 & 0.935 \\
    \end{tabular}}
    \caption{The fidelity of the adiabatically prepared $\lambda\phi^4$ ground state as a function of $\delta t$ and the number of adiabatic steps, for $n_q=12$ with $\phi_{\rm max}=4$ and $\lambda = 10$. Each adiabatic step consists of one second-order Trotter step. Results are shown in Fig.~\ref{fig:12_scan}.}
    \label{tab:12_scan}
\end{table}

%%%%%%%%%%%%%%%%%%%%%%%%%%%%%%%%%%%%%%%

% ==========   Bibliography
%
\bibliographystyle{unsrt}
\bibliography{uwthesis}

@article{Jordan:1928wi,
    author = "Jordan, Pascual and Wigner, Eugene P.",
    title = "{About the Pauli exclusion principle}",
    doi = "10.1007/BF01331938",
    journal = "Z. Phys.",
    volume = "47",
    pages = "631--651",
    year = "1928"
}

@article{Robledo-Moreno:2024pzz,
    author = "Robledo-Moreno, Javier and others",
    title = "{Chemistry beyond the scale of exact diagonalization on a quantum-centric supercomputer}",
    eprint = "2405.05068",
    archivePrefix = "arXiv",
    primaryClass = "quant-ph",
    reportNumber = "RIKEN-iTHEMS-Report-24",
    doi = "10.1126/sciadv.adu9991",
    journal = "Sci. Adv.",
    volume = "11",
    number = "25",
    pages = "adu9991",
    year = "2025"
}

@article{Zhao:2024xtl,
    author = "Zhao, Qi and Zhou, You and Childs, Andrew M.",
    title = "{Entanglement accelerates quantum simulation}",
    eprint = "2406.02379",
    archivePrefix = "arXiv",
    primaryClass = "quant-ph",
    doi = "10.1038/s41567-025-02945-2",
    journal = "Nature Phys.",
    volume = "21",
    number = "8",
    pages = "1338--1345",
    year = "2025"
}

@article{Motta:2019yya,
    author = "Motta, Mario and Sun, Chong and Tan, Adrian Teck Keng and Rourke, Matthew J. O' and Ye, Erika and Minnich, Austin J. and Brand{\~a}o, Fernando G. S. L. and Chan, Garnet Kin-Lic",
    title = "{Determining eigenstates and thermal states on a quantum computer using quantum imaginary time evolution}",
    eprint = "1901.07653",
    archivePrefix = "arXiv",
    primaryClass = "quant-ph",
    doi = "10.1038/s41567-019-0704-4",
    journal = "Nature Phys.",
    volume = "16",
    pages = "205--210",
    year = "2019"
}

@article{Sagastizabal:2020wfq,
    author = "Sagastizabal, R. and others",
    title = "{Variational preparation of finite-temperature states on a quantum computer}",
    eprint = "2012.03895",
    archivePrefix = "arXiv",
    primaryClass = "quant-ph",
    doi = "10.1038/s41534-021-00468-1",
    journal = "npj Quantum Inf.",
    volume = "7",
    pages = "130",
    year = "2021"
}

@article{Warren:2022evv,
    author = "Warren, Ada and Zhu, Linghua and Mayhall, Nicholas J. and Barnes, Edwin and Economou, Sophia E.",
    title = "{Adaptive variational algorithms for quantum Gibbs state preparation}",
    eprint = "2203.12757",
    archivePrefix = "arXiv",
    primaryClass = "quant-ph",
    month = "3",
    year = "2022"
}

@article{Holmes:2022fic,
    author = "Holmes, Zoe and Muraleedharan, Gopikrishnan and Somma, Rolando D. and Subasi, Yigit and {\c{S}}ahino{\u{g}}lu, Burak",
    title = "{Quantum algorithms from fluctuation theorems: Thermal-state preparation}",
    eprint = "2203.08882",
    archivePrefix = "arXiv",
    primaryClass = "quant-ph",
    reportNumber = "LA-UR-22-21364",
    doi = "10.22331/q-2022-10-06-825",
    journal = "Quantum",
    volume = "6",
    pages = "825",
    year = "2022"
}

@article{Robertson:2026com,
    author = "Robertson, Reece and Consiglio, Mirko and Stevens, Josey and Doucet, Emery and Apollaro, Tony J. G. and Deffner, Sebastian",
    title = "{Variational Gibbs State Preparation on Trapped-Ion Devices}",
    eprint = "2603.03801",
    archivePrefix = "arXiv",
    primaryClass = "quant-ph",
    month = "3",
    year = "2026"
}

@article{Consiglio:2023mev,
    author = "Consiglio, Mirko and Settino, Jacopo and Giordano, Andrea and Mastroianni, Carlo and Plastina, Francesco and Lorenzo, Salvatore and Maniscalco, Sabrina and Goold, John and Apollaro, Tony J. G.",
    title = "{Variational Gibbs state preparation on noisy intermediate-scale quantum devices}",
    eprint = "2303.11276",
    archivePrefix = "arXiv",
    primaryClass = "quant-ph",
    doi = "10.1103/PhysRevA.110.012445",
    journal = "Phys. Rev. A",
    volume = "110",
    number = "1",
    pages = "012445",
    year = "2024"
}

@article{Turro:2023nuk,
    author = "Turro, Francesco",
    title = "{Quantum Imaginary Time Propagation algorithm for preparing thermal states}",
    eprint = "2306.16580",
    archivePrefix = "arXiv",
    primaryClass = "quant-ph",
    reportNumber = "IQuS@UW-21-054",
    month = "6",
    year = "2023"
}

@article{Cirac:2008rbg,
    author = "Cirac, J. Ignacio and Vollbrecht, Karl Gerd H. and Wolf, Michael M. and Schuch, Norbert",
    title = "{On entropy growth and the hardness of simulating time evolution}",
    eprint = "0801.2078",
    archivePrefix = "arXiv",
    primaryClass = "quant-ph",
    doi = "10.1088/1367-2630/10/3/033032",
    journal = "New J. Phys.",
    volume = "10",
    number = "3",
    pages = "033032",
    year = "2008"
}

@article{Horodecki:2009zz,
    author = "Horodecki, Ryszard and Horodecki, Pawel and Horodecki, Michal and Horodecki, Karol",
    title = "{Quantum entanglement}",
    eprint = "quant-ph/0702225",
    archivePrefix = "arXiv",
    doi = "10.1103/RevModPhys.81.865",
    journal = "Rev. Mod. Phys.",
    volume = "81",
    pages = "865--942",
    year = "2009"
}

@article{Song:2025pwd,
    author = "Song, Jeffrey Z. and Kishony, Gilad and Berg, Erez and Rudner, Mark S.",
    title = "{Vari-Cool: a non-unitary quantum variational protocol for simulated cooling}",
    eprint = "2510.09749",
    archivePrefix = "arXiv",
    primaryClass = "quant-ph",
    month = "10",
    year = "2025"
}

@article{Baumer:2023vrf,
    author = {B{\"a}umer, Elisa and Tripathi, Vinay and Wang, Derek S. and Rall, Patrick and Chen, Edward H. and Majumder, Swarnadeep and Seif, Alireza and Minev, Zlatko K.},
    title = "{Efficient Long-Range Entanglement Using Dynamic Circuits}",
    eprint = "2308.13065",
    archivePrefix = "arXiv",
    primaryClass = "quant-ph",
    doi = "10.1103/PRXQuantum.5.030339",
    journal = "PRX Quantum",
    volume = "5",
    number = "3",
    pages = "030339",
    year = "2024"
}

@inproceedings{Gottesman:1998hu,
    author = "Gottesman, Daniel",
    title = "{The Heisenberg representation of quantum computers}",
    booktitle = "{22nd International Colloquium on Group Theoretical Methods in Physics}",
    eprint = "quant-ph/9807006",
    archivePrefix = "arXiv",
    reportNumber = "LAUR-98-2848, LA-UR-98-2848",
    pages = "32--43",
    month = "7",
    year = "1998"
}

@article{Aaronson:2004xuh,
    author = "Aaronson, Scott and Gottesman, Daniel",
    title = "{Improved simulation of stabilizer circuits}",
    eprint = "quant-ph/0406196",
    archivePrefix = "arXiv",
    doi = "10.1103/PhysRevA.70.052328",
    journal = "Phys. Rev. A",
    volume = "70",
    number = "5",
    pages = "052328",
    year = "2004"
}

@article{Hartnett:2026abf,
    author = "Hartnett, Gavin S. and Najafi, Khadijeh Sona and Khindanov, Aleksei and Liao, Haoran and Schutzman, Michael and Hush, Michael R. and Biercuk, Michael J. and Baum, Yuval",
    title = "{Fast, accurate, high-resolution simulation of large-scale Fermi-Hubbard models on a digital quantum processor}",
    eprint = "2605.04025",
    archivePrefix = "arXiv",
    primaryClass = "quant-ph",
    month = "5",
    year = "2026"
}

@article{Cogburn:2026aqy,
    author = "Cogburn, Cameron V. and Grieninger, Sebastian and Kharzeev, Dmitri E.",
    title = "{Quantum Simulation of Nucleon-Antinucleon Interaction in Large-$N$ QCD$_2$ on an IBM Quantum Nighthawk Processor}",
    eprint = "2606.02574",
    archivePrefix = "arXiv",
    primaryClass = "quant-ph",
    reportNumber = "IQuS@UW-21-127, NT@UW-26-11",
    month = "6",
    year = "2026"
}

@article{Chen:2026tvd,
    author = {Chen, Jiunn-Wei and Chen, Yu-Ting and Meher, Ghanashyam and M{\"u}ller, Berndt and Sch{\"a}fer, Andreas and Yao, Xiaojun},
    title = "{Thermalization of SU(2) Lattice Gauge Fields on Quantum Computers}",
    eprint = "2603.23948",
    archivePrefix = "arXiv",
    primaryClass = "hep-lat",
    reportNumber = "IQuS@UW-21-122",
    month = "3",
    year = "2026"
}

@article{Zemlevskiy:2026kpc,
    author = "Zemlevskiy, Nikita A.",
    title = "{Exclusive Scattering Channels from Entanglement Structure in Real-Time Simulations}",
    eprint = "2603.15621",
    archivePrefix = "arXiv",
    primaryClass = "quant-ph",
    reportNumber = "IQuS@UW-21-121",
    month = "3",
    year = "2026"
}

@article{Than:2025gso,
    author = "Than, Anton T. and Kadam, Saurabh V. and Vikramaditya, Vinay and Nguyen, Nhung H. and Liu, Xingxin and Davoudi, Zohreh and Green, Alaina M. and Linke, Norbert M.",
    title = "{Observation of quantum-field-theory dynamics on a spin-phonon quantum computer}",
    eprint = "2509.11477",
    archivePrefix = "arXiv",
    primaryClass = "quant-ph",
    reportNumber = "UMD-PP-025-05, IQuS@UW-21-107",
    month = "9",
    year = "2025"
}

@article{Chen:2025zeh,
    author = "Chen, Jiunn-Wei and Chen, Yu-Ting and Meher, Ghanashyam",
    title = "{Parton Distributions on a Quantum Computer}",
    eprint = "2506.16829",
    archivePrefix = "arXiv",
    primaryClass = "hep-lat",
    month = "6",
    year = "2025"
}

@article{Kadam:2025trs,
    author = "Kadam, Saurabh V. and Naskar, Aahiri and Raychowdhury, Indrakshi and Stryker, Jesse R.",
    title = "{Loop-string-hadron approach to SU(3) lattice Yang-Mills theory, II: Operator representation for the trivalent vertex}",
    eprint = "2512.11796",
    archivePrefix = "arXiv",
    primaryClass = "hep-lat",
    reportNumber = "IQuS@UW-21-116, IQuS@UW-21-116",
    month = "12",
    year = "2025"
}

@article{Peruzzo:2013bzg,
    author = "Peruzzo, Alberto and McClean, Jarrod and Shadbolt, Peter and Yung, Man-Hong and Zhou, Xiao-Qi and Love, Peter J. and Aspuru-Guzik, Al{\'a}n and O'Brien, Jeremy L.",
    title = "{A variational eigenvalue solver on a photonic quantum processor}",
    eprint = "1304.3061",
    archivePrefix = "arXiv",
    primaryClass = "quant-ph",
    doi = "10.1038/ncomms5213",
    journal = "Nature Commun.",
    volume = "5",
    number = "1",
    pages = "4213",
    year = "2014"
}

@article{Ciavarella:2025bsg,
    author = "Ciavarella, Anthony N. and Burbano, Ivan M. and Bauer, Christian W.",
    title = "{Efficient truncations of SU(Nc) lattice gauge theory for quantum simulation}",
    eprint = "2503.11888",
    archivePrefix = "arXiv",
    primaryClass = "hep-lat",
    doi = "10.1103/ylqb-phv5",
    journal = "Phys. Rev. D",
    volume = "112",
    number = "5",
    pages = "054514",
    year = "2025"
}

@article{Modi:2026syn,
    author = "Modi, Neel S. and Ciavarella, Anthony N. and Halimeh, Jad C. and Bauer, Christian W.",
    title = "{Large Nc Truncations for SU(Nc) Lattice Yang-Mills Theory with Fermions}",
    eprint = "2602.02344",
    archivePrefix = "arXiv",
    primaryClass = "hep-lat",
    month = "2",
    year = "2026"
}

@article{Trotter:1959ytf,
    author = "Trotter, H. F.",
    title = "{On the product of semi-groups of operators}",
    doi = "10.1090/s0002-9939-1959-0108732-6",
    journal = "Proc. Am. Math. Soc.",
    volume = "10",
    number = "4",
    pages = "545--551",
    year = "1959"
}

@article{Hatano:2005gh,
    author = "Hatano, Naomichi and Suzuki, Masuo",
    title = "{Finding Exponential Product Formulas of Higher Orders}",
    eprint = "math-ph/0506007",
    archivePrefix = "arXiv",
    doi = "10.1007/11526216_2",
    journal = "Lect. Notes Phys.",
    volume = "679",
    pages = "37",
    year = "2005"
}

@article{Suzuki:1976be,
    author = "Suzuki, M.",
    title = "{Generalized Trotter's Formula and Systematic Approximants of Exponential Operators and Inner Derivations with Applications to Many Body Problems}",
    doi = "10.1007/BF01609348",
    journal = "Commun. Math. Phys.",
    volume = "51",
    pages = "183--190",
    year = "1976"
}

@article{Farhi:2000ikn,
    author = "Farhi, Edward and Goldstone, Jeffrey and Gutmann, Sam and Sipser, Michael",
    title = "{Quantum Computation by Adiabatic Evolution}",
    eprint = "quant-ph/0001106",
    archivePrefix = "arXiv",
    reportNumber = "MIT-CTP-2936, MIT-CTP-2936",
    month = "1",
    year = "2000"
}

@article{Bravyi:2004isx,
    author = "Bravyi, Sergei and Kitaev, Alexei",
    title = "{Universal quantum computation with ideal Clifford gates and noisy ancillas}",
    eprint = "quant-ph/0403025",
    archivePrefix = "arXiv",
    doi = "10.1103/PhysRevA.71.022316",
    journal = "Phys. Rev. A",
    volume = "71",
    number = "2",
    pages = "022316",
    year = "2005"}

@article{Anderson:2014jvy,
    author = "Anderson, Jonas T. and Duclos-Cianci, Guillaume and Poulin, David",
    title = "{Fault-Tolerant Conversion between the Steane and Reed-Muller Quantum Codes}",
    eprint = "1403.2734",
    archivePrefix = "arXiv",
    primaryClass = "quant-ph",
    doi = "10.1103/PhysRevLett.113.080501",
    journal = "Phys. Rev. Lett.",
    volume = "113",
    number = "8",
    pages = "080501",
    year = "2014"
}

@article{Gidney:2024alh,
    author = "Gidney, Craig and Shutty, Noah and Jones, Cody",
    title = "{Magic state cultivation: growing T states as cheap as CNOT gates}",
    eprint = "2409.17595",
    archivePrefix = "arXiv",
    primaryClass = "quant-ph",
    month = "9",
    year = "2024"
}

@article{Eastin:2009tem,
    author = "Eastin, Bryan and Knill, Emanuel",
    title = "{Restrictions on Transversal Encoded Quantum Gate Sets}",
    eprint = "0811.4262",
    archivePrefix = "arXiv",
    primaryClass = "quant-ph",
    doi = "10.1103/PhysRevLett.102.110502",
    journal = "Phys. Rev. Lett.",
    volume = "102",
    number = "11",
    pages = "110502",
    year = "2009"
}

@article{Bravyi:2012lxn,
    author = "Bravyi, Sergey and Haah, Jeongwan",
    title = "{Magic-state distillation with low overhead}",
    eprint = "1209.2426",
    archivePrefix = "arXiv",
    primaryClass = "quant-ph",
    doi = "10.1103/PhysRevA.86.052329",
    journal = "Phys. Rev. A",
    volume = "86",
    number = "5",
    pages = "052329",
    year = "2012"
}

@article{Rajput:2021trn,
    author = "Rajput, Abhishek and Roggero, Alessandro and Wiebe, Nathan",
    title = "{Quantum error correction with gauge symmetries}",
    eprint = "2112.05186",
    archivePrefix = "arXiv",
    primaryClass = "quant-ph",
    reportNumber = "IQuS@UW-21-014",
    doi = "10.1038/s41534-023-00706-8",
    journal = "npj Quantum Inf.",
    volume = "9",
    number = "1",
    pages = "41",
    year = "2023"
}

@article{Spagnoli:2026qni,
    author = "Spagnoli, Luca and Roggero, Alessandro and Wiebe, Nathan",
    title = "{Qudit stabiliser codes for $\mathbb{Z}_N$ lattice gauge theories with matter}",
    eprint = "2602.20661",
    archivePrefix = "arXiv",
    primaryClass = "quant-ph",
    month = "2",
    year = "2026"
}

@article{albert2026handbook,
  title={Handbook of Error-Correcting Codes},
  author={Albert, Victor V and Faist, Philippe},
  journal={arXiv preprint arXiv:2606.11484},
  year={2026}
}

@article{GoogleQuantumAIandCollaborators:2024efv,
    author = "Acharya, Rajeev and others",
    collaboration = "Google Quantum AI and Collaborators",
    title = "{Quantum error correction below the surface code threshold}",
    eprint = "2408.13687",
    archivePrefix = "arXiv",
    primaryClass = "quant-ph",
    doi = "10.1038/s41586-024-08449-y",
    journal = "Nature",
    volume = "638",
    number = "8052",
    pages = "920--926",
    year = "2025",
    note = "[Erratum: Nature 653, E5 (2026)]"
}

@article{Terhal:2013vbm,
    author = "Terhal, Barbara M.",
    title = "{Quantum error correction for quantum memories}",
    eprint = "1302.3428",
    archivePrefix = "arXiv",
    primaryClass = "quant-ph",
    doi = "10.1103/RevModPhys.87.307",
    journal = "Rev. Mod. Phys.",
    volume = "87",
    number = "2",
    pages = "307",
    year = "2015"
}

@article{Reichardt:2024xfs,
    author = "Reichardt, Ben W. and others",
    title = "{Fault-tolerant quantum computation with a neutral atom processor}",
    eprint = "2411.11822",
    archivePrefix = "arXiv",
    primaryClass = "quant-ph",
    month = "11",
    year = "2024"
}

@article{Bluvstein:2025ped,
    author = "Bluvstein, Dolev and others",
    title = "{A fault-tolerant neutral-atom architecture for universal quantum computation}",
    eprint = "2506.20661",
    archivePrefix = "arXiv",
    primaryClass = "quant-ph",
    doi = "10.1038/s41586-025-09848-5",
    journal = "Nature",
    volume = "649",
    number = "8095",
    pages = "39--46",
    year = "2026",
    note = "[Erratum: Nature 650, E3 (2026)]"
}

@article{MembersoftheHRLQuantumTeam:2026slx,
    author = "Abraham, Michael and others",
    collaboration = "Members of the HRL Quantum Team, Collaborators",
    title = "{A digitally controlled silicon quantum processing unit}",
    eprint = "2604.16216",
    archivePrefix = "arXiv",
    primaryClass = "quant-ph",
    month = "4",
    year = "2026"
}

@article{Kim:2025ywx,
    author = "Kim, Taeyoon and others",
    title = "{Ultracoherent superconducting cavity-based multiqudit platform with error-resilient control}",
    eprint = "2506.03286",
    archivePrefix = "arXiv",
    primaryClass = "quant-ph",
    reportNumber = "FERMILAB-PUB-25-0379-SQMS",
    month = "6",
    year = "2025"
}

@article{Gao:2024fik,
    author = "Gao, Dongxin and others",
    title = "{Establishing a New Benchmark in Quantum Computational Advantage with 105-qubit Zuchongzhi 3.0 Processor}",
    eprint = "2412.11924",
    archivePrefix = "arXiv",
    primaryClass = "quant-ph",
    doi = "10.1103/PhysRevLett.134.090601",
    journal = "Phys. Rev. Lett.",
    volume = "134",
    number = "9",
    pages = "090601",
    year = "2025"
}

@article{Somoroff:2021elj,
    author = "Somoroff, Aaron and Ficheux, Quentin and Mencia, Raymond A. and Xiong, Haonan and Kuzmin, Roman V. and Manucharyan, Vladimir E.",
    title = "{Millisecond Coherence in a Superconducting Qubit}",
    eprint = "2103.08578",
    archivePrefix = "arXiv",
    primaryClass = "quant-ph",
    doi = "10.1103/PhysRevLett.130.267001",
    journal = "Phys. Rev. Lett.",
    volume = "130",
    number = "26",
    pages = "267001",
    year = "2023"
}

@article{microsoft2025interferometric,
  title={Interferometric single-shot parity measurement in InAs--Al hybrid devices},
  author={Microsoft Azure Quantum and Aghaee, Morteza and Alcaraz Ramirez, Alejandro and Alam, Zulfi and Ali, Rizwan and Andrzejczuk, Mariusz and Antipov, Andrey and Astafev, Mikhail and Barzegar, Amin and Bauer, Bela and others},
  journal={Nature},
  volume={638},
  number={8051},
  pages={651--655},
  year={2025},
  publisher={Nature Publishing Group UK London}
}

@article{xue2022quantum,
  title={Quantum logic with spin qubits crossing the surface code threshold},
  author={Xue, Xiao and Russ, Maximilian and Samkharadze, Nodar and Undseth, Brennan and Sammak, Amir and Scappucci, Giordano and Vandersypen, Lieven MK},
  journal={Nature},
  volume={601},
  number={7893},
  pages={343--347},
  year={2022},
  publisher={Nature Publishing Group UK London}
}

@article{Zhong:2020iql,
    author = "Zhong, Han-Sen and others",
    title = "{Quantum computational advantage using photons}",
    eprint = "2012.01625",
    archivePrefix = "arXiv",
    primaryClass = "quant-ph",
    doi = "10.1126/science.abe8770",
    journal = "Science",
    volume = "370",
    number = "6523",
    pages = "1460--1463",
    year = "2020"
}

@article{Alexander:2024vys,
    author = "Alexander, Koen and others",
    title = "{A manufacturable platform for photonic quantum computing}",
    eprint = "2404.17570",
    archivePrefix = "arXiv",
    primaryClass = "quant-ph",
    doi = "10.1038/s41586-025-08820-7",
    journal = "Nature",
    volume = "641",
    number = "8064",
    pages = "876--883",
    year = "2025"
}

@article{AghaeeRad:2025lqu,
    author = "Aghaee Rad, H. and others",
    title = "{Scaling and networking a modular photonic quantum computer}",
    doi = "10.1038/s41586-024-08406-9",
    journal = "Nature",
    volume = "638",
    number = "8052",
    pages = "912--919",
    year = "2025"
}

@article{PhysRevLett.53.1260,
  title = {Resonant Activation from the Zero-Voltage State of a Current-Biased Josephson Junction},
  author = {Devoret, Michel H. and Martinis, John M. and Esteve, Daniel and Clarke, John},
  journal = {Phys. Rev. Lett.},
  volume = {53},
  issue = {13},
  pages = {1260--1263},
  numpages = {0},
  year = {1984},
  month = {Sep},
  publisher = {American Physical Society},
  doi = {10.1103/PhysRevLett.53.1260},
  url = {https://link.aps.org/doi/10.1103/PhysRevLett.53.1260}
}

@article{PhysRevLett.55.1543,
  title = {Energy-Level Quantization in the Zero-Voltage State of a Current-Biased Josephson Junction},
  author = {Martinis, John M. and Devoret, Michel H. and Clarke, John},
  journal = {Phys. Rev. Lett.},
  volume = {55},
  issue = {15},
  pages = {1543--1546},
  numpages = {0},
  year = {1985},
  month = {Oct},
  publisher = {American Physical Society},
  doi = {10.1103/PhysRevLett.55.1543},
  url = {https://link.aps.org/doi/10.1103/PhysRevLett.55.1543}
}

@article{Devoret:1985yga,
    author = "Devoret, Michel H. and Martinis, John M. and Clarke, John",
    title = "{Measurements of Macroscopic Quantum Tunneling out of the Zero-Voltage State of a Current-Biased Josephson Junction}",
    doi = "10.1103/PhysRevLett.55.1908",
    journal = "Phys. Rev. Lett.",
    volume = "55",
    number = "18",
    pages = "1908",
    year = "1985"
}

@article{Koch:2007hay,
    author = "Koch, Jens and Yu, Terri M. and Gambetta, Jay and Houck, A. A. and Schuster, D. I. and Majer, J. and Blais, Alexandre and Devoret, M. H. and Girvin, S. M. and Schoelkopf, R. J.",
    title = "{Charge-insensitive qubit design derived from the Cooper pair box}",
    eprint = "cond-mat/0703002",
    archivePrefix = "arXiv",
    doi = "10.1103/physreva.76.042319",
    journal = "Phys. Rev. A",
    volume = "76",
    number = "4",
    pages = "042319",
    year = "2007"
}

@article{PhysRevLett.74.4091,
  title = {Quantum Computations with Cold Trapped Ions},
  author = {Cirac, J. I. and Zoller, P.},
  journal = {Phys. Rev. Lett.},
  volume = {74},
  issue = {20},
  pages = {4091--4094},
  numpages = {0},
  year = {1995},
  month = {May},
  publisher = {American Physical Society},
  doi = {10.1103/PhysRevLett.74.4091},
  url = {https://link.aps.org/doi/10.1103/PhysRevLett.74.4091}
}

@article{PhysRevLett.75.4714,
  title = {Demonstration of a Fundamental Quantum Logic Gate},
  author = {Monroe, C. and Meekhof, D. M. and King, B. E. and Itano, W. M. and Wineland, D. J.},
  journal = {Phys. Rev. Lett.},
  volume = {75},
  issue = {25},
  pages = {4714--4717},
  numpages = {0},
  year = {1995},
  month = {Dec},
  publisher = {American Physical Society},
  doi = {10.1103/PhysRevLett.75.4714},
  url = {https://link.aps.org/doi/10.1103/PhysRevLett.75.4714}
}

@article{Jaksch:2000cus,
    author = "Jaksch, D. and Cirac, J. I. and Zoller, P. and Rolston, S. L. and C{\^o}t{\'e}, R. and Lukin, M. D.",
    title = "{Fast Quantum Gates for Neutral Atoms}",
    eprint = "quant-ph/0004038",
    archivePrefix = "arXiv",
    doi = "10.1103/PhysRevLett.85.2208",
    journal = "Phys. Rev. Lett.",
    volume = "85",
    number = "10",
    pages = "2208",
    year = "2000"
}

@article{Manetsch:2024lwl,
    author = "Manetsch, Hannah J. and Nomura, Gyohei and Bataille, Elie and Lv, Xudong and Leung, Kon H. and Endres, Manuel",
    title = "{A tweezer array with 6,100 highly coherent atomic qubits}",
    eprint = "2403.12021",
    archivePrefix = "arXiv",
    primaryClass = "quant-ph",
    doi = "10.1038/s41586-025-09641-4",
    journal = "Nature",
    volume = "647",
    number = "8088",
    pages = "60--67",
    year = "2025"
}

@article{Li:2025sgo,
    author = "Li, Zhiyao and Illa, Marc and Savage, Martin J.",
    title = "{A Framework for Quantum Simulations of Energy-Loss and Hadronization in Non-Abelian Gauge Theories: SU(2) Lattice Gauge Theory in 1+1D}",
    eprint = "2512.05210",
    archivePrefix = "arXiv",
    primaryClass = "quant-ph",
    reportNumber = "IQuS@UW-21-115",
    month = "12",
    year = "2025"
}

@article{Froland:2025bqf,
    author = "Froland, Henry and Grabowska, Dorota M. and Li, Zhiyao",
    title = "{Simulating Fully Gauge-Fixed SU(2) Hamiltonian Dynamics on Digital Quantum Computers}",
    eprint = "2512.22782",
    archivePrefix = "arXiv",
    primaryClass = "quant-ph",
    month = "12",
    year = "2025"
}

@misc{Perez:2025cxl,
    author = "Perez, Sebastian Osorio and Murairi, Edison M. and Gustafson, Erik J. and Lamm, Henry",
    title = "{Primitive Quantum Gates for an SU(3) Discrete Subgroup: {\ensuremath{\Sigma}}(72{\texttimes}3)}",
    eprint = "2511.17437",
    archivePrefix = "arXiv",
    primaryClass = "hep-lat",
    reportNumber = "FERMILAB-PUB-25-0674-SQMS-STUDENT-T",
    month = "11",
    year = "2025"
}

@article{Gustafson:2024kym,
    author = "Gustafson, Erik J. and Ji, Yao and Lamm, Henry and Murairi, Edison M. and Perez, Sebastian Osorio and Zhu, Shuchen",
    title = "{Primitive quantum gates for an SU(3) discrete subgroup: {\ensuremath{\Sigma}}(36{\texttimes}3)}",
    eprint = "2405.05973",
    archivePrefix = "arXiv",
    primaryClass = "hep-lat",
    reportNumber = "FERMILAB-PUB-24-0132-SQMS-T, TUM-HEP-1506/24",
    doi = "10.1103/PhysRevD.110.034515",
    journal = "Phys. Rev. D",
    volume = "110",
    number = "3",
    pages = "034515",
    year = "2024"
}

@article{Gustafson:2023kvd,
    author = "Gustafson, Erik J. and Lamm, Henry and Lovelace, Felicity",
    title = "{Primitive quantum gates for an SU(2) discrete subgroup: Binary octahedral}",
    eprint = "2312.10285",
    archivePrefix = "arXiv",
    primaryClass = "hep-lat",
    reportNumber = "FERMILAB-PUB-23-753-SQMS-T",
    doi = "10.1103/PhysRevD.109.054503",
    journal = "Phys. Rev. D",
    volume = "109",
    number = "5",
    pages = "054503",
    year = "2024"
}

@article{Gustafson:2024bww,
    author = "Gustafson, Erik and others",
    title = "{Surrogate Constructed Scalable Circuits ADAPT-VQE in the Schwinger model}",
    eprint = "2408.12641",
    archivePrefix = "arXiv",
    primaryClass = "quant-ph",
    reportNumber = "FERMILAB-PUB-24-0456-SQMS-T",
    doi = "10.1103/PhysRevApplied.23.064002",
    journal = "Phys. Rev. Applied",
    volume = "23",
    number = "6",
    pages = "064002",
    year = "2025"
}

@article{Robin:2026lqp,
    author = "Robin, Caroline E. P. and Savage, Martin J.",
    title = "{Quantum Complexity and New Directions in Nuclear Physics and High-Energy Physics Phenomenology}",
    eprint = "2604.26376",
    archivePrefix = "arXiv",
    primaryClass = "quant-ph",
    reportNumber = "IQuS@UW-21-124",
    month = "4",
    year = "2026"
}

@article{Jha:2026ror,
    author = "Jha, Raghav G. and Toga, Goksu C. and Taher, Jaber I. and Bakalov, Bojko N. and Kemper, Alexander F.",
    title = "{Magic and entanglement in 1+1-dimensional SU(2) lattice gauge theory}",
    eprint = "2606.09971",
    archivePrefix = "arXiv",
    primaryClass = "quant-ph",
    month = "6",
    year = "2026"
}

@article{Froland:2026aff,
    author = "Froland, Henry and Grabowska, Dorota M.",
    title = "{Measuring Non-Stabilizerness in an SU(2) Lattice Gauge Theory}",
    eprint = "2606.14842",
    archivePrefix = "arXiv",
    primaryClass = "quant-ph",
    reportNumber = "IQuS@UW-21-129",
    month = "6",
    year = "2026"
}

@article{Bravyi_2016,
   author = "Bravyi, Sergey and Gosset, David",
    title = "{Improved Classical Simulation of Quantum Circuits Dominated by Clifford Gates}",
    eprint = "1601.07601",
    archivePrefix = "arXiv",
    primaryClass = "quant-ph",
    doi = "10.1103/PhysRevLett.116.250501",
    journal = "Phys. Rev. Lett.",
    volume = "116",
    number = "25",
    pages = "250501",
    year = "2016"}

@article{Haug:2022vpg,
    author = "Haug, Tobias and Piroli, Lorenzo",
    title = "{Quantifying nonstabilizerness of matrix product states}",
    eprint = "2207.13076",
    archivePrefix = "arXiv",
    primaryClass = "quant-ph",
    doi = "10.1103/PhysRevB.107.035148",
    journal = "Phys. Rev. B",
    volume = "107",
    number = "3",
    pages = "035148",
    year = "2023"
}

@article{Haug:2023hcs,
    author = "Haug, Tobias and Piroli, Lorenzo",
    title = "{Stabilizer entropies and nonstabilizerness monotones}",
    eprint = "2303.10152",
    archivePrefix = "arXiv",
    primaryClass = "quant-ph",
    doi = "10.22331/q-2023-08-28-1092",
    journal = "Quantum",
    volume = "7",
    pages = "1092",
    year = "2023"
}

@article{Leone:2024lfr,
    author = "Leone, Lorenzo and Bittel, Lennart",
    title = "{Stabilizer entropies are monotones for magic-state resource theory}",
    eprint = "2404.11652",
    archivePrefix = "arXiv",
    primaryClass = "quant-ph",
    doi = "10.1103/PhysRevA.110.L040403",
    journal = "Phys. Rev. A",
    volume = "110",
    number = "4",
    pages = "L040403",
    year = "2024"
}

@misc{Bittel:2025yhq,
    author = "Bittel, Lennart and Leone, Lorenzo",
    title = "{Operational interpretation of the Stabilizer Entropy}",
    eprint = "2507.22883",
    archivePrefix = "arXiv",
    primaryClass = "quant-ph",
    month = "7",
    year = "2025"
}

@book{10.1093/oso/9780198501787.001.0001,
    author = {Quarteroni, Alfio and Valli, Alberto},
    title = {Domain Decomposition Methods for Partial Differential Equations},
    publisher = {Oxford University Press},
    year = {1999},
    month = {05},
    isbn = {9780198501787},
    doi = {10.1093/oso/9780198501787.001.0001},
    url = {https://doi.org/10.1093/oso/9780198501787.001.0001},
}

@book{saad2003iterative,
  title={Iterative Methods for Sparse Linear Systems: Second Edition},
  author={Saad, Y.},
  isbn={9780898715347},
  lccn={2002044644},
  series={Other Titles in Applied Mathematics},
  url={https://books.google.com/books?id=qtzmkzzqFmcC},
  year={2003},
  publisher={Society for Industrial and Applied Mathematics}
}

@article{Luscher:2003qa,
    author = "Luscher, Martin",
    title = "{Solution of the Dirac equation in lattice QCD using a domain decomposition method}",
    eprint = "hep-lat/0310048",
    archivePrefix = "arXiv",
    reportNumber = "CERN-TH-2003-250",
    doi = "10.1016/S0010-4655(03)00486-7",
    journal = "Comput. Phys. Commun.",
    volume = "156",
    pages = "209--220",
    year = "2004"
}

@article{Ce:2016idq,
    author = "C{\`e}, Marco and Giusti, Leonardo and Schaefer, Stefan",
    title = "{Domain decomposition, multi-level integration and exponential noise reduction in lattice QCD}",
    eprint = "1601.04587",
    archivePrefix = "arXiv",
    primaryClass = "hep-lat",
    reportNumber = "DESY-16-007",
    doi = "10.1103/PhysRevD.93.094507",
    journal = "Phys. Rev. D",
    volume = "93",
    number = "9",
    pages = "094507",
    year = "2016"
}

@article{frommer2014DD,
    author = {Frommer, Andreas and Kahl, Karsten and Krieg, Stefan and Leder, Bj{\"o}rn and Rottmann, Matthias},
    title = "{Adaptive Aggregation-Based Domain Decomposition Multigrid for the Lattice Wilson--Dirac Operator}",
    eprint = "1303.1377",
    archivePrefix = "arXiv",
    primaryClass = "hep-lat",
    doi = "10.1137/130919507",
    journal = "SIAM J. Sci. Comput.",
    volume = "36",
    number = "4",
    pages = "A1581--A1608",
    year = "2014"
}

@article{Heybrock2014LatticeQW,
  title={Lattice QCD with Domain Decomposition on Intel{\textregistered} Xeon Phi Co-Processors},
  author={Simon Heybrock and B{\'a}lint Jo{\'o} and Dhiraj D. Kalamkar and Mikhail Smelyanskiy and Karthikeyan Vaidyanathan and Tilo Wettig and Pradeep K. Dubey},
  journal={SC14: International Conference for High Performance Computing, Networking, Storage and Analysis},
  year={2014},
  pages={69-80},
  url={https://api.semanticscholar.org/CorpusID:16604039}
}

@article{Turro:2025sec,
    author = "Turro, Francesco and Yao, Xiaojun",
    title = "{Emergent hydrodynamic mode on SU(2) plaquette chains and quantum simulation}",
    eprint = "2502.17551",
    archivePrefix = "arXiv",
    primaryClass = "hep-ph",
    reportNumber = "IQuS@UW-21-095",
    doi = "10.1103/PhysRevD.111.094502",
    journal = "Phys. Rev. D",
    volume = "111",
    number = "9",
    pages = "094502",
    year = "2025"
}

@article{Stetina:2020abi,
    author = "Stetina, Torin F. and Ciavarella, Anthony and Li, Xiaosong and Wiebe, Nathan",
    title = "{Simulating Effective QED on Quantum Computers}",
    eprint = "2101.00111",
    archivePrefix = "arXiv",
    primaryClass = "quant-ph",
    doi = "10.22331/q-2022-01-18-622",
    journal = "Quantum",
    volume = "6",
    pages = "622",
    year = "2022"
}

@article{Wong:2000cmz,
    author = "Wong, Alexander and Christensen, Nelson",
    title = "{Potential multiparticle entanglement measure}",
    eprint = "quant-ph/0010052",
    archivePrefix = "arXiv",
    doi = "10.1103/PhysRevA.63.044301",
    journal = "Phys. Rev. A",
    volume = "63",
    number = "4",
    pages = "044301",
    year = "2001"
}

@article{Viola:1998jx,
    author = "Viola, Lorenza and Lloyd, Seth",
    title = "{Dynamical suppression of decoherence in two state quantum systems}",
    eprint = "quant-ph/9803057",
    archivePrefix = "arXiv",
    doi = "10.1103/PhysRevA.58.2733",
    journal = "Phys. Rev. A",
    volume = "58",
    pages = "2733",
    year = "1998"
}

@article{Ezzell:2022uat,
    author = "Ezzell, Nic and Pokharel, Bibek and Tewala, Lina and Quiroz, Gregory and Lidar, Daniel A.",
    title = "{Dynamical decoupling for superconducting qubits: A performance survey}",
    eprint = "2207.03670",
    archivePrefix = "arXiv",
    primaryClass = "quant-ph",
    doi = "10.1103/PhysRevApplied.20.064027",
    journal = "Phys. Rev. Applied",
    volume = "20",
    number = "6",
    pages = "064027",
    year = "2023"
}

@article{Berg:2020ibi,
    author = "Berg, Ewout van den and Minev, Zlatko K. and Temme, Kristan",
    title = "{Model-free readout-error mitigation for quantum expectation values}",
    eprint = "2012.09738",
    archivePrefix = "arXiv",
    primaryClass = "quant-ph",
    doi = "10.1103/PhysRevA.105.032620",
    journal = "Phys. Rev. A",
    volume = "105",
    number = "3",
    pages = "032620",
    year = "2022"
}

@article{Wallman:2015uzh,
    author = "Wallman, Joel J. and Emerson, Joseph",
    title = "{Noise tailoring for scalable quantum computation via randomized compiling}",
    eprint = "1512.01098",
    archivePrefix = "arXiv",
    primaryClass = "quant-ph",
    doi = "10.1103/PhysRevA.94.052325",
    journal = "Phys. Rev. A",
    volume = "94",
    number = "5",
    pages = "052325",
    year = "2016"
}

@article{Urbanek:2021oej,
    author = "Urbanek, Miroslav and Nachman, Benjamin and Pascuzzi, Vincent R. and He, Andre and Bauer, Christian W. and de Jong, Wibe A.",
    title = "{Mitigating depolarizing noise on quantum computers with noise-estimation circuits}",
    eprint = "2103.08591",
    archivePrefix = "arXiv",
    primaryClass = "quant-ph",
    doi = "10.1103/PhysRevLett.127.270502",
    journal = "Phys. Rev. Lett.",
    volume = "127",
    number = "27",
    pages = "270502",
    year = "2021"
}

@misc{Yao:2025uxz,
    author = "Yao, Xiaojun",
    title = "{On Quantum Simulation of QED in Coulomb Gauge}",
    eprint = "2507.01089",
    archivePrefix = "arXiv",
    primaryClass = "quant-ph",
    reportNumber = "IQuS@UW-21-106",
    month = "7",
    year = "2025"
}

@misc{Halimeh:2024bth,
    author = {Halimeh, Jad C. and Hanada, Masanori and Matsuura, Shunji and Nori, Franco and Rinaldi, Enrico and Sch{\"a}fer, Andreas},
    title = "{A universal framework for the quantum simulation of Yang-Mills theory}",
    eprint = "2411.13161",
    archivePrefix = "arXiv",
    primaryClass = "quant-ph",
    reportNumber = "RIKEN-iTHEMS-Report-24",
    month = "11",
    year = "2024"
}

@article{Fontana:2024rux,
    author = "Fontana, Pierpaolo and Riaza, Marc Miranda and Celi, Alessio",
    title = "{Efficient Finite-Resource Formulation of Non-Abelian Lattice Gauge Theories beyond One Dimension}",
    eprint = "2409.04441",
    archivePrefix = "arXiv",
    primaryClass = "quant-ph",
    doi = "10.1103/k9p6-c649",
    journal = "Phys. Rev. X",
    volume = "15",
    number = "3",
    pages = "031065",
    year = "2025"
}

@article{Grabowska:2024emw,
    author = "Grabowska, Dorota M. and Kane, Christopher F. and Bauer, Christian W.",
    title = "{Fully gauge-fixed SU(2) Hamiltonian for quantum simulations}",
    eprint = "2409.10610",
    archivePrefix = "arXiv",
    primaryClass = "quant-ph",
    reportNumber = "IQuS@UW-21-087",
    doi = "10.1103/PhysRevD.111.114516",
    journal = "Phys. Rev. D",
    volume = "111",
    number = "11",
    pages = "114516",
    year = "2025"
}

@article{Kadam:2024ifg,
    author = "Kadam, Saurabh V. and Naskar, Aahiri and Raychowdhury, Indrakshi and Stryker, Jesse R.",
    title = "{Loop-string-hadron approach to SU(3) lattice Yang-Mills theory: Hilbert space of a trivalent vertex}",
    eprint = "2407.19181",
    archivePrefix = "arXiv",
    primaryClass = "hep-lat",
    reportNumber = "IQuS@UW-21-086",
    doi = "10.1103/PhysRevD.111.074516",
    journal = "Phys. Rev. D",
    volume = "111",
    number = "7",
    pages = "074516",
    year = "2025"
}

@article{Zache:2023dko,
   author = "Zache, Torsten V. and Gonz{\'a}lez-Cuadra, Daniel and Zoller, Peter",
    title = "{Quantum and Classical Spin-Network Algorithms for q-Deformed Kogut-Susskind Gauge Theories}",
    eprint = "2304.02527",
    archivePrefix = "arXiv",
    primaryClass = "quant-ph",
    doi = "10.1103/PhysRevLett.131.171902",
    journal = "Phys. Rev. Lett.",
    volume = "131",
    number = "17",
    pages = "171902",
    year = "2023"
}

@article{Raychowdhury:2018osk,
    author = "Raychowdhury, Indrakshi and Stryker, Jesse R.",
    title = "{Solving Gauss's Law on Digital Quantum Computers with Loop-String-Hadron Digitization}",
    eprint = "1812.07554",
    archivePrefix = "arXiv",
    primaryClass = "hep-lat",
    reportNumber = "INT-PUB-18-062",
    doi = "10.1103/PhysRevResearch.2.033039",
    journal = "Phys. Rev. Res.",
    volume = "2",
    number = "3",
    pages = "033039",
    year = "2020"
}

@misc{Kurkcuoglu:2024cfv,
    author = {Kurkcuoglu, Doga Murat and Lamm, Henry and Maestri, Andrea},
    title = "{Qudit Gate Decomposition Dependence for Lattice Gauge Theories}",
    eprint = "2410.16414",
    archivePrefix = "arXiv",
    primaryClass = "quant-ph",
    reportNumber = "FERMILAB-PUB-24-0612-SQMS-T",
    month = "10",
    year = "2024"
}

@article{Ji:2020kjk,
    author = "Ji, Yao and Lamm, Henry and Zhu, Shuchen",
    title = "{Gluon Field Digitization via Group Space Decimation for Quantum Computers}",
    eprint = "2005.14221",
    archivePrefix = "arXiv",
    primaryClass = "hep-lat",
    reportNumber = "FERMILAB-PUB-20-198-T",
    doi = "10.1103/PhysRevD.102.114513",
    journal = "Phys. Rev. D",
    volume = "102",
    number = "11",
    pages = "114513",
    year = "2020"
}

@article{Alexandru:2019nsa,
    author = "Alexandru, Andrei and Bedaque, Paulo F. and Harmalkar, Siddhartha and Lamm, Henry and Lawrence, Scott and Warrington, Neill C.",
    title = "{Gluon Field Digitization for Quantum Computers}",
    eprint = "1906.11213",
    archivePrefix = "arXiv",
    primaryClass = "hep-lat",
    doi = "10.1103/PhysRevD.100.114501",
    journal = "Phys. Rev. D",
    volume = "100",
    number = "11",
    pages = "114501",
    year = "2019"
}

@article{Carena:2022kpg,
    author = "Carena, Marcela and Lamm, Henry and Li, Ying-Ying and Liu, Wanqiang",
    title = "{Improved Hamiltonians for Quantum Simulations of Gauge Theories}",
    eprint = "2203.02823",
    archivePrefix = "arXiv",
    primaryClass = "hep-lat",
    reportNumber = "FERMILAB-PUB-21-674-T",
    doi = "10.1103/PhysRevLett.129.051601",
    journal = "Phys. Rev. Lett.",
    volume = "129",
    number = "5",
    pages = "051601",
    year = "2022"
}

@article{PhysRevLett.120.210501,
  author = "Dumitrescu, E. F. and McCaskey, A. J. and Hagen, G. and Jansen, G. R. and Morris, T. D. and Papenbrock, T. and Pooser, R. C. and Dean, D. J. and Lougovski, P.",
    title = "{Cloud Quantum Computing of an Atomic Nucleus}",
    eprint = "1801.03897",
    archivePrefix = "arXiv",
    primaryClass = "quant-ph",
    doi = "10.1103/PhysRevLett.120.210501",
    journal = "Phys. Rev. Lett.",
    volume = "120",
    pages = "210501",
    year = "2018"
}

@article{Calajo:2024qrc,
    author = "Calaj{\`o}, Giuseppe and Magnifico, Giuseppe and Edmunds, Claire and Ringbauer, Martin and Montangero, Simone and Silvi, Pietro",
    title = "{Digital Quantum Simulation of a (1+1)D SU(2) Lattice Gauge Theory with Ion Qudits}",
    eprint = "2402.07987",
    archivePrefix = "arXiv",
    primaryClass = "quant-ph",
    doi = "10.1103/PRXQuantum.5.040309",
    journal = "PRX Quantum",
    volume = "5",
    number = "4",
    pages = "040309",
    year = "2024"
}

@article{Than:2024zaj,
    author = "Than, Anton T. and others",
    title = "{The phase diagram of quantum chromodynamics in one dimension on a quantum computer}",
    eprint = "2501.00579",
    archivePrefix = "arXiv",
    primaryClass = "quant-ph",
    doi = "10.1038/s41467-025-65198-w",
    journal = "Nat. Commun.",
    volume = "16",
    number = "1",
    pages = "10288",
    year = "2025"
}

@article{Gustafson:2021qbt,
    author = "Gustafson, Erik",
    title = "{Prospects for Simulating a Qudit Based Model of (1+1)d Scalar QED}",
    eprint = "2104.10136",
    archivePrefix = "arXiv",
    primaryClass = "quant-ph",
    doi = "10.1103/PhysRevD.103.114505",
    journal = "Phys. Rev. D",
    volume = "103",
    number = "11",
    pages = "114505",
    year = "2021"
}

@misc{Crane:2024tlj,
    author = "Crane, Eleanor and others",
    title = "{Hybrid Oscillator-Qubit Quantum Processors: Simulating Fermions, Bosons, and Gauge Fields}",
    eprint = "2409.03747",
    archivePrefix = "arXiv",
    primaryClass = "quant-ph",
    month = "9",
    year = "2024"
}

@misc{Chai:2025qhf,
    author = {Chai, Yahui and Guo, Yibin and K{\"u}hn, Stefan},
    title = "{Scalable Quantum Algorithm for Meson Scattering in a Lattice Gauge Theory}",
    eprint = "2505.21240",
    archivePrefix = "arXiv",
    primaryClass = "quant-ph",
    month = "5",
    year = "2025"
}

@article{Turco:2025jot,
    author = "Turco, Matteo and Quinta, Gon{\c{c}}alo M. and Seixas, Jo{\~a}o and Omar, Yasser",
    title = "{Creation of wave packets for quantum chromodynamics on quantum computers}",
    eprint = "2501.13069",
    archivePrefix = "arXiv",
    primaryClass = "quant-ph",
    doi = "10.1103/2nb4-171l",
    journal = "Phys. Rev. D",
    volume = "112",
    number = "3",
    pages = "034506",
    year = "2025"
}

@article{Zache:2023cfj,
    author = "Zache, Torsten V. and Gonz{\'a}lez-Cuadra, Daniel and Zoller, Peter",
    title = "{Fermion-qudit quantum processors for simulating lattice gauge theories with matter}",
    eprint = "2303.08683",
    archivePrefix = "arXiv",
    primaryClass = "quant-ph",
    doi = "10.22331/q-2023-10-16-1140",
    journal = "Quantum",
    volume = "7",
    pages = "1140",
    year = "2023"
}

@misc{Huie:2025yzn,
    author = "Huie, William and Conefrey-Shinozaki, Cianan and Jia, Zhubing and Draper, Patrick and Covey, Jacob P.",
    title = "{Three-qubit encoding in ytterbium-171 atoms for simulating 1+1D QCD}",
    eprint = "2507.18426",
    archivePrefix = "arXiv",
    primaryClass = "quant-ph",
    month = "7",
    year = "2025"
}

@misc{Kurkcuoglu:2025gik,
    author = {K{\"u}rk{\c{c}}{\"u}oglu, Doga Murat and Lamm, Henry and Ogunkoya, Oluwadara and Pierattelli, Leonardo},
    title = "{The 2T-quoctit: a two-mode bosonic qudit for high energy physics}",
    eprint = "2509.21941",
    archivePrefix = "arXiv",
    primaryClass = "quant-ph",
    reportNumber = "FERMILAB-PUB-25-0143-SQMS-STUDENT-T",
    month = "9",
    year = "2025"
}

@misc{Barata:2025hgx,
    author = "Barata, Jo{\~a}o and Rico, Enrique",
    title = "{Real-time simulation of jet energy loss and entropy production in high-energy scattering with matter}",
    eprint = "2502.17558",
    archivePrefix = "arXiv",
    primaryClass = "hep-ph",
    reportNumber = "CERN-TH-2025-019",
    month = "2",
    year = "2025"
}

@misc{Saner:2025nrq,
    author = "Saner, S. and B{\u{a}}z{\u{a}}van, O. and Webb, D. J. and Araneda, G. and Ballance, C. J. and Srinivas, R. and Lucas, D. M. and Berm{\'u}dez, A.",
    title = "{Real-Time Observation of Aharonov-Bohm Interference in a $\mathbb{Z}_2$ Lattice Gauge Theory on a Hybrid Qubit-Oscillator Quantum Computer}",
    eprint = "2507.19588",
    archivePrefix = "arXiv",
    primaryClass = "quant-ph",
    month = "7",
    year = "2025"
}

@misc{Schuhmacher:2025ehh,
    author = "Schuhmacher, Julian and Su, Guo-Xian and Osborne, Jesse J. and Gandon, Anthony and Halimeh, Jad C. and Tavernelli, Ivano",
    title = "{Observation of hadron scattering in a lattice gauge theory on a quantum computer}",
    eprint = "2505.20387",
    archivePrefix = "arXiv",
    primaryClass = "quant-ph",
    month = "5",
    year = "2025"
}

@misc{Cobos:2025krn,
    author = {Cobos, Jes{\'u}s and Fraxanet, Joana and Benito, C{\'e}sar and di Marcantonio, Francesco and Rivero, Pedro and Kap{\'a}s, Korn{\'e}l and Werner, Mikl{\'o}s Antal and Legeza, {\"O}rs and Bermudez, Alejandro and Rico, Enrique},
    title = "{Real-Time Dynamics in a (2+1)-D Gauge Theory: The Stringy Nature on a Superconducting Quantum Simulator}",
    eprint = "2507.08088",
    archivePrefix = "arXiv",
    primaryClass = "quant-ph",
    reportNumber = "CERN-TH-2025-111",
    month = "7",
    year = "2025"
}

@misc{Farrell:2025nkx,
    author = "Farrell, Roland C. and Zemlevskiy, Nikita A. and Illa, Marc and Preskill, John",
    title = "{Digital quantum simulations of scattering in quantum field theories using W states}",
    eprint = "2505.03111",
    archivePrefix = "arXiv",
    primaryClass = "quant-ph",
    reportNumber = "IQuS@UW-21-099",
    month = "5",
    year = "2025"
}

@article{Alexandrou:2025vaj,
    author = {Alexandrou, Constantia and Athenodorou, Andreas and Blekos, Kostas and Polykratis, Georgios and K{\"u}hn, Stefan},
    title = "{Realizing string breaking dynamics in a Z2 lattice gauge theory on quantum hardware}",
    eprint = "2504.13760",
    archivePrefix = "arXiv",
    primaryClass = "hep-lat",
    doi = "10.1103/r6sr-dv13",
    journal = "Phys. Rev. D",
    volume = "112",
    number = "11",
    pages = "114506",
    year = "2025"
}

@misc{Crippa:2024hso,
    author = "Crippa, Arianna and Jansen, Karl and Rinaldi, Enrico",
    title = "{Analysis of the confinement string in (2 + 1)-dimensional Quantum Electrodynamics with a trapped-ion quantum computer}",
    eprint = "2411.05628",
    archivePrefix = "arXiv",
    primaryClass = "hep-lat",
    month = "11",
    year = "2024"
}

@article{Gonzalez-Cuadra:2024xul,
    author = "Gonzalez-Cuadra, Daniel and others",
    title = "{Observation of string breaking on a (2 + 1)D Rydberg quantum simulator}",
    eprint = "2410.16558",
    archivePrefix = "arXiv",
    primaryClass = "quant-ph",
    doi = "10.1038/s41586-025-09051-6",
    journal = "Nature",
    volume = "642",
    number = "8067",
    pages = "321--326",
    year = "2025"
}

@misc{Gyawali:2024hrz,
    author = "Gyawali, Gaurav and others",
    title = "{Observation of disorder-free localization using a (2+1)D lattice gauge theory on a quantum processor}",
    eprint = "2410.06557",
    archivePrefix = "arXiv",
    primaryClass = "quant-ph",
    month = "10",
    year = "2024"
}

@article{Balaji:2025afl,
    author = "Balaji, Praveen and Conefrey-Shinozaki, Cian{\'a}n and Draper, Patrick and Elhaderi, Jason K. and Gupta, Drishti and Hidalgo, Luis and Lytle, Andrew and Rinaldi, Enrico",
    title = "{Quantum circuits for SU(3) lattice gauge theory}",
    eprint = "2503.08866",
    archivePrefix = "arXiv",
    primaryClass = "hep-lat",
    doi = "10.1103/k8f6-yft8",
    journal = "Phys. Rev. D",
    volume = "112",
    number = "5",
    pages = "054511",
    year = "2025"
}

@article{Ciavarella:2024cyt,
    author = "Ciavarella, Anthony N. and Bauer, Christian W.",
    title = "{Quantum Simulation of Large N Lattice Gauge Theories}",
    eprint = "2411.16704",
    archivePrefix = "arXiv",
    primaryClass = "hep-lat",
    doi = "10.22323/1.466.0206",
    journal = "PoS",
    volume = "LATTICE2024",
    pages = "206",
    year = "2025"
}

@article{Ciavarella:2024fzw,
    author = "Ciavarella, Anthony N. and Bauer, Christian W.",
    title = "{Quantum Simulation of SU(3) Lattice Yang-Mills Theory at Leading Order in Large-Nc Expansion}",
    eprint = "2402.10265",
    archivePrefix = "arXiv",
    primaryClass = "hep-ph",
    doi = "10.1103/PhysRevLett.133.111901",
    journal = "Phys. Rev. Lett.",
    volume = "133",
    number = "11",
    pages = "111901",
    year = "2024"
}

@article{Jiang:2025ufg,
    author = "Jiang, Jacky and Klco, Natalie and Di Matteo, Olivia",
    title = "{Non-Abelian dynamics on a cube: Improving quantum compilation through qudit-based simulations}",
    eprint = "2506.10945",
    archivePrefix = "arXiv",
    primaryClass = "quant-ph",
    doi = "10.1103/rrmd-fqx6",
    journal = "Phys. Rev. D",
    volume = "112",
    number = "7",
    pages = "074512",
    year = "2025"
}

@article{YungerHalpern:2019mvj,
    author = "Yunger Halpern, Nicole and Beverland, Michael E. and Kalev, Amir",
    title = "{Noncommuting conserved charges in quantum many-body thermalization}",
    eprint = "1906.09227",
    archivePrefix = "arXiv",
    primaryClass = "quant-ph",
    doi = "10.1103/PhysRevE.101.042117",
    journal = "Phys. Rev. E",
    volume = "101",
    number = "4",
    pages = "042117",
    year = "2020"
}

@article{Li:2024lrl,
    author = "Li, Zhiyao and Grabowska, Dorota M. and Savage, Martin J.",
    title = "{Sequency Hierarchy Truncation (SeqHT) for Adiabatic State Preparation and Time Evolution in Quantum Simulations}",
    eprint = "2407.13835",
    archivePrefix = "arXiv",
    primaryClass = "quant-ph",
    reportNumber = "UW-21-084",
    doi = "10.22331/q-2025-09-29-1865",
    journal = "Quantum",
    volume = "9",
    pages = "1865",
    year = "2025"
}

@article{Zemlevskiy:2024vxt,
    author = "Zemlevskiy, Nikita A.",
    title = "{Scalable quantum simulations of scattering in scalar field theory on 120 qubits}",
    eprint = "2411.02486",
    archivePrefix = "arXiv",
    primaryClass = "quant-ph",
    reportNumber = "IQuS@UW-21-090",
    doi = "10.1103/qr72-51v1",
    journal = "Phys. Rev. D",
    volume = "112",
    number = "3",
    pages = "034502",
    year = "2025"
}

@article{Gaz:2025ptu,
    author = "Gaz, Emanuele and Popov, Pavel P. and Pardo, Guy and Lewenstein, Maciej and Hauke, Philipp and Zohar, Erez",
    title = "{Quantum simulation of non-Abelian lattice gauge theories: A variational approach to D8 with dynamical matter}",
    eprint = "2501.17863",
    archivePrefix = "arXiv",
    primaryClass = "hep-lat",
    doi = "10.1103/l8b6-h5s5",
    journal = "Phys. Rev. Res.",
    volume = "7",
    number = "3",
    pages = "033012",
    year = "2025"
}

@article{Chai:2023qpq,
    author = {Chai, Yahui and Crippa, Arianna and Jansen, Karl and K{\"u}hn, Stefan and Pascuzzi, Vincent R. and Tacchino, Francesco and Tavernelli, Ivano},
    title = "{Fermionic wave packet scattering: a quantum computing approach}",
    eprint = "2312.02272",
    archivePrefix = "arXiv",
    primaryClass = "quant-ph",
    doi = "10.22331/q-2025-02-19-1638",
    journal = "Quantum",
    volume = "9",
    pages = "1638",
    year = "2025"
}

@misc{Davoudi:2025rdv,
    author = "Davoudi, Zohreh and Hsieh, Chung-Chun and Kadam, Saurabh V.",
    title = "{Quantum computation of hadron scattering in a lattice gauge theory}",
    eprint = "2505.20408",
    archivePrefix = "arXiv",
    primaryClass = "quant-ph",
    reportNumber = "UMD-PP-025-02, IQuS@UW-21-102",
    month = "5",
    year = "2025"
}

@article{Davoudi:2024wyv,
    author = "Davoudi, Zohreh and Hsieh, Chung-Chun and Kadam, Saurabh V.",
    title = "{Scattering wave packets of hadrons in gauge theories: Preparation on a quantum computer}",
    eprint = "2402.00840",
    archivePrefix = "arXiv",
    primaryClass = "quant-ph",
    reportNumber = "UMD-PP-024-02, IQuS@UW-21-071",
    doi = "10.22331/q-2024-11-11-1520",
    journal = "Quantum",
    volume = "8",
    pages = "1520",
    year = "2024"
}

@misc{Surace:2024bht,
    author = "Surace, Federica Maria and others",
    title = "{String-Breaking Dynamics in Quantum Adiabatic and Diabatic Processes}",
    eprint = "2411.10652",
    archivePrefix = "arXiv",
    primaryClass = "quant-ph",
    month = "11",
    year = "2024"
}

@misc{De:2024smi,
    author = "De, Arinjoy and others",
    title = "{Observation of string-breaking dynamics in a quantum simulator}",
    eprint = "2410.13815",
    archivePrefix = "arXiv",
    primaryClass = "quant-ph",
    month = "10",
    year = "2024"
}

@article{Surace:2019dtp,
    author = "Surace, Federica M. and Mazza, Paolo P. and Giudici, Giuliano and Lerose, Alessio and Gambassi, Andrea and Dalmonte, Marcello",
    title = "{Lattice gauge theories and string dynamics in Rydberg atom quantum simulators}",
    eprint = "1902.09551",
    archivePrefix = "arXiv",
    primaryClass = "cond-mat.quant-gas",
    doi = "10.1103/PhysRevX.10.021041",
    journal = "Phys. Rev. X",
    volume = "10",
    number = "2",
    pages = "021041",
    year = "2020"
}

@misc{Borla:2025gfs,
    author = "Borla, Umberto and Osborne, Jesse J. and Moroz, Sergej and Halimeh, Jad C.",
    title = "{String Breaking in a $2+1$D $\mathbb{Z}_2$ Lattice Gauge Theory}",
    eprint = "2501.17929",
    archivePrefix = "arXiv",
    primaryClass = "quant-ph",
    month = "1",
    year = "2025"
}

@misc{Cataldi:2025cyo,
    author = "Cataldi, Giovanni and Orlando, Simone and Halimeh, Jad C.",
    title = "{Real-Time String Dynamics in a $2+1$D Non-Abelian Lattice Gauge Theory: String Breaking, Glueball Formation, Baryon Blockade, and Tension Reduction}",
    eprint = "2509.08868",
    archivePrefix = "arXiv",
    primaryClass = "hep-lat",
    month = "9",
    year = "2025"
}

@misc{Tian:2025mbv,
    author = "Tian, Yizhuo and Srivatsa, N. S. and Xu, Kaidi and Osborne, Jesse J. and Borla, Umberto and Halimeh, Jad C.",
    title = "{Role of Plaquette Term in Genuine $2+1$D String Dynamics on Quantum Simulators}",
    eprint = "2508.05736",
    archivePrefix = "arXiv",
    primaryClass = "quant-ph",
    month = "8",
    year = "2025"
}

@misc{Xu:2025abo,
    author = "Xu, Kaidi and Borla, Umberto and Moroz, Sergej and Halimeh, Jad C.",
    title = "{String Breaking Dynamics and Glueball Formation in a $2+1$D Lattice Gauge Theory}",
    eprint = "2507.01950",
    archivePrefix = "arXiv",
    primaryClass = "hep-lat",
    month = "7",
    year = "2025"
}

@misc{grieninger2025st,
    author = "Grieninger, Sebastian and Kharzeev, Dmitri E. and Marroquin, Eliana",
    title = "{Thermal nature of confining strings}",
    eprint = "2510.23919",
    archivePrefix = "arXiv",
    primaryClass = "hep-ph",
    reportNumber = "IQuS@UW-21-112, NT@UW-25-15",
    month = "10",
    year = "2025"
}

@article{Davoudi2024a,
   author = "Davoudi, Zohreh and Jarzynski, Christopher and Mueller, Niklas and Oruganti, Greeshma and Powers, Connor and Halpern, Nicole Yunger",
    title = "{Quantum Thermodynamics of Nonequilibrium Processes in Lattice Gauge Theories}",
    eprint = "2404.02965",
    archivePrefix = "arXiv",
    primaryClass = "quant-ph",
    reportNumber = "UMD-PP-024-05, IQuS@UW-21-075",
    doi = "10.1103/PhysRevLett.133.250402",
    journal = "Phys. Rev. Lett.",
    volume = "133",
    number = "25",
    pages = "250402",
    year = "2024"
}

@article{Meth:2023wzd,
   author = "Meth, Michael and others",
    title = "{Simulating two-dimensional lattice gauge theories on a qudit quantum computer}",
    eprint = "2310.12110",
    archivePrefix = "arXiv",
    primaryClass = "quant-ph",
    doi = "10.1038/s41567-025-02797-w",
    journal = "Nat. Phys.",
    volume = "21",
    number = "4",
    pages = "570--576",
    year = "2025"
}

@misc{Chernyshev:2025lil,
    author = "Chernyshev, Ivan A. and others",
    title = "{Pathfinding Quantum Simulations of Neutrinoless Double-$\beta$ Decay}",
    eprint = "2506.05757",
    archivePrefix = "arXiv",
    primaryClass = "quant-ph",
    reportNumber = "IQuS@UW-21-103, LA-UR-25-23950",
    month = "6",
    year = "2025"
}

@article{Bauer:2022hpo,
    author = "Bauer, Christian W. and others",
    title = "{Quantum Simulation for High-Energy Physics}",
    eprint = "2204.03381",
    archivePrefix = "arXiv",
    primaryClass = "quant-ph",
    reportNumber = "UMD-PP-022-04, LA-UR-22-22100, RIKEN-iTHEMS-Report-22, RIKEN-iTHEMS-Report-22,
  FERMILAB-PUB-22-249-SQMS-T, FERMILAB-PUB-22-249-SQMS-T, IQuS@UW-21-027, MITRE-21-03848-2, FERMILAB-PUB-22-249-SQMS-T",
    doi = "10.1103/PRXQuantum.4.027001",
    journal = "PRX Quantum",
    volume = "4",
    number = "2",
    pages = "027001",
    year = "2023"
}

@article{Bauer:2019qxa,
    author = "Bauer, Christian W. and de Jong, Wibe A. and Nachman, Benjamin and Provasoli, Davide",
    title = "{Quantum Algorithm for High Energy Physics Simulations}",
    eprint = "1904.03196",
    archivePrefix = "arXiv",
    primaryClass = "hep-ph",
    doi = "10.1103/PhysRevLett.126.062001",
    journal = "Phys. Rev. Lett.",
    volume = "126",
    number = "6",
    pages = "062001",
    year = "2021"
}

@article{Bauer:2023qgm,
    author = "Bauer, Christian W. and Davoudi, Zohreh and Klco, Natalie and Savage, Martin J.",
    title = "{Quantum simulation of fundamental particles and forces}",
    eprint = "2404.06298",
    archivePrefix = "arXiv",
    primaryClass = "hep-ph",
    reportNumber = "IQuS@UW-21-052",
    doi = "10.1038/s42254-023-00599-8",
    journal = "Nature Rev. Phys.",
    volume = "5",
    number = "7",
    pages = "420--432",
    year = "2023"
}

@article{Klco:2021lap,
    author = "Klco, Natalie and Roggero, Alessandro and Savage, Martin J.",
    title = "{Standard model physics and the digital quantum revolution: thoughts about the interface}",
    eprint = "2107.04769",
    archivePrefix = "arXiv",
    primaryClass = "quant-ph",
    reportNumber = "IQuS@UW-21-007",
    doi = "10.1088/1361-6633/ac58a4",
    journal = "Rept. Prog. Phys.",
    volume = "85",
    number = "6",
    pages = "064301",
    year = "2022"
}

@article{DiMeglio:2023nsa,
    author = "Di Meglio, Alberto and others",
    title = "{Quantum Computing for High-Energy Physics: State of the Art and Challenges}",
    eprint = "2307.03236",
    archivePrefix = "arXiv",
    primaryClass = "quant-ph",
    reportNumber = "FERMILAB-PUB-23-468-ETD",
    doi = "10.1103/PRXQuantum.5.037001",
    journal = "PRX Quantum",
    volume = "5",
    number = "3",
    pages = "037001",
    year = "2024"
}

@article{basaglia2024geant4gamechangerhigh,
      author = "Basaglia, Tullio and Bell, Zane W. and D'Agostino, Daniele and Dressendorfer, Paul V. and Giani, Simone and Pia, Maria Grazia and Saracco, Paolo",
    title = "{Geant4: a Game Changer in High Energy Physics and Related Applicative Fields}",
    eprint = "2405.12159",
    archivePrefix = "arXiv",
    primaryClass = "physics.comp-ph",
    journal = {Future Gener. Comput. Syst.},
    volume = {159},
    pages = {411-422},
    doi = "10.1016/j.future.2024.05.042",
    month = "5",
    year = "2024"
}

@article{Muskinja:2020ots,
    author = "Mu{\v{s}}kinja, Miha and Chapman, John Derek and Gray, Heather",
    editor = "Doglioni, C. and Kim, D. and Stewart, G. A. and Silvestris, L. and Jackson, P. and Kamleh, W.",
    collaboration = "ATLAS",
    title = "{Geant4 performance optimization in the ATLAS experiment}",
    doi = "10.1051/epjconf/202024502036",
    journal = "EPJ Web Conf.",
    volume = "245",
    pages = "02036",
    year = "2020"
}

@article{GEANT4:2002zbu,
    author = "Agostinelli, S. and others",
    collaboration = "GEANT4",
    title = "{GEANT4 - A Simulation Toolkit}",
    reportNumber = "SLAC-PUB-9350, FERMILAB-PUB-03-339, CERN-IT-2002-003",
    doi = "10.1016/S0168-9002(03)01368-8",
    journal = "Nucl. Instrum. Meth. A",
    volume = "506",
    pages = "250--303",
    year = "2003"
}

@article{Gray:2022fou,
    author = "Gray, Heather M. and Terashi, Koji",
    title = "{Quantum Computing Applications in Future Colliders}",
    doi = "10.3389/fphy.2022.864823",
    journal = "Front. in Phys.",
    volume = "10",
    pages = "864823",
    year = "2022"
}

@article{BRAHMS:2004adc,
    author = "Arsene, I. and others",
    collaboration = "BRAHMS",
    title = "{Quark gluon plasma and color glass condensate at RHIC? The Perspective from the BRAHMS experiment}",
    eprint = "nucl-ex/0410020",
    archivePrefix = "arXiv",
    doi = "10.1016/j.nuclphysa.2005.02.130",
    journal = "Nucl. Phys. A",
    volume = "757",
    pages = "1--27",
    year = "2005"
}

@article{He_2023,
   author = "He, Min and van Hees, Hendrik and Rapp, Ralf",
    title = "{Heavy-quark diffusion in the quark{\textendash}gluon plasma}",
    eprint = "2204.09299",
    archivePrefix = "arXiv",
    primaryClass = "hep-ph",
    doi = "10.1016/j.ppnp.2023.104020",
    journal = "Prog. Part. Nucl. Phys.",
    volume = "130",
    pages = "104020",
    year = "2023"
}

@article{Lami:2023naw,
    author = "Lami, Guglielmo and Collura, Mario",
    title = "{Nonstabilizerness via Perfect Pauli Sampling of Matrix Product States}",
    eprint = "2303.05536",
    archivePrefix = "arXiv",
    primaryClass = "quant-ph",
    doi = "10.1103/PhysRevLett.131.180401",
    journal = "Phys. Rev. Lett.",
    volume = "131",
    number = "18",
    pages = "180401",
    year = "2023"
}

@misc{Hoshino:2025ine,
    author = "Hoshino, Masahiro and Ashida, Yuto",
    title = "{Stabilizer R{\'e}nyi Entropy Encodes Fusion Rules of Topological Defects and Boundaries}",
    eprint = "2507.10656",
    archivePrefix = "arXiv",
    primaryClass = "quant-ph",
    month = "7",
    year = "2025"
}

@article{Farrell:2022vyh,
    author = "Farrell, Roland C. and Chernyshev, Ivan A. and Powell, Sarah J. M. and Zemlevskiy, Nikita A. and Illa, Marc and Savage, Martin J.",
    title = "{Preparations for quantum simulations of quantum chromodynamics in 1+1 dimensions. II. Single-baryon {\ensuremath{\beta}}-decay in real time}",
    eprint = "2209.10781",
    archivePrefix = "arXiv",
    primaryClass = "quant-ph",
    reportNumber = "IQuS@UW-21-030, NT@UW-22-12",
    doi = "10.1103/PhysRevD.107.054513",
    journal = "Phys. Rev. D",
    volume = "107",
    number = "5",
    pages = "054513",
    year = "2023"
}

@article{Corbett:2025flm,
    author = "Corbett, Benjamin and Miyake, Akimasa",
    title = "{Scaling up the transcorrelated density matrix renormalization group}",
    eprint = "2506.07441",
    archivePrefix = "arXiv",
    primaryClass = "cond-mat.str-el",
    doi = "10.1103/nzrt-l2j1",
    journal = "Phys. Rev. B",
    volume = "112",
    number = "16",
    pages = "165120",
    year = "2025"
}

@Article{fishman2022itensor,
	title={{The ITensor Software Library for Tensor Network Calculations}},
	author={Matthew Fishman and Steven R. White and E. Miles Stoudenmire},
	journal={SciPost Phys. Codebases},
	pages={4},
	year={2022},
	publisher={SciPost},
	doi={10.21468/SciPostPhysCodeb.4},
	url={https://scipost.org/10.21468/SciPostPhysCodeb.4},
archivePrefix = {arXiv},
       eprint = {2007.14822},
 primaryClass = {cs.MS},
}

@book{python3,
 author = {Van Rossum, Guido and Drake, Fred L.},
 title = "{Python 3 Reference Manual}",
 year = {2009},
 isbn = {1441412697},
 publisher = {CreateSpace},
 address = {Scotts Valley, CA}
}

@misc{anaconda,
  title="{Anaconda Software Distribution, Vers. 2-2.4.0}",
  url={https://docs.anaconda.com/},
  journal={Anaconda Documentation},
  author={{Anaconda Inc.}},
  year={2020}
}

@Article{Hunter:2007,
  Author    = {Hunter, J. D.},
  Title     = "{Matplotlib: A 2D graphics environment}",
  Journal   = {Comput. Sci. Eng.},
  Volume    = {9},
  Number    = {3},
  Pages     = {90--95},
  doi       = {10.1109/MCSE.2007.55},
  year      = 2007
}

@Article{PER-GRA:2007,
  Author    = {P\'erez, Fernando and Granger, Brian E.},
  Title     = "{{IP}ython: a System for Interactive Scientific Computing}",
  Journal   = {Comput. Sci. Eng.},
  Volume    = {9},
  Number    = {3},
  Pages     = {21--29},
  month     = may,
  year      = 2007,
  url       = "https://ipython.org",
  ISSN      = "1521-9615",
  doi       = {10.1109/MCSE.2007.53},
  publisher = {IEEE Computer Society},
}

@misc{qiskit,
      title={Quantum computing with {Q}iskit},
      author={Javadi-Abhari, Ali and Treinish, Matthew and Krsulich, Kevin and Wood, Christopher J. and Lishman, Jake and Gacon, Julien and Martiel, Simon and Nation, Paul D. and Bishop, Lev S. and Cross, Andrew W. and Johnson, Blake R. and Gambetta, Jay M.},
      year={2024},
      eprint={2405.08810},
      archivePrefix={arXiv},
      primaryClass={quant-ph}
}

@article{Julia-2017,
    title={Julia: A fresh approach to numerical computing},
    author={Bezanson, Jeff and Edelman, Alan and Karpinski, Stefan and Shah, Viral B},
    eprint = "1411.1607",
    archivePrefix = "arXiv",
    primaryClass = "cs.MS",
    journal={SIAM {R}eview},
    volume={59},
    number={1},
    pages={65--98},
    year={2017},
    publisher={SIAM},
    doi={10.1137/141000671},
    url={https://epubs.siam.org/doi/10.1137/141000671}
}

@article{Florio:2024aix,
    author = "Florio, Adrien and Frenklakh, David and Ikeda, Kazuki and Kharzeev, Dmitri E. and Korepin, Vladimir and Shi, Shuzhe and Yu, Kwangmin",
    title = "{Quantum real-time evolution of entanglement and hadronization in jet production: Lessons from the massive Schwinger model}",
    eprint = "2404.00087",
    archivePrefix = "arXiv",
    primaryClass = "hep-ph",
    doi = "10.1103/PhysRevD.110.094029",
    journal = "Phys. Rev. D",
    volume = "110",
    number = "9",
    pages = "094029",
    year = "2024"
}

@article{Florio:2023dke,
    author = "Florio, Adrien and Frenklakh, David and Ikeda, Kazuki and Kharzeev, Dmitri and Korepin, Vladimir and Shi, Shuzhe and Yu, Kwangmin",
    title = "{Real-Time Nonperturbative Dynamics of Jet Production in Schwinger Model: Quantum Entanglement and Vacuum Modification}",
    eprint = "2301.11991",
    archivePrefix = "arXiv",
    primaryClass = "hep-ph",
    doi = "10.1103/PhysRevLett.131.021902",
    journal = "Phys. Rev. Lett.",
    volume = "131",
    number = "2",
    pages = "021902",
    year = "2023"
}

@article{Tarabunga:2023xmv,
    author = "Tarabunga, Poetri Sonya and Castelnovo, Claudio",
    title = "{Magic in generalized Rokhsar-Kivelson wavefunctions}",
    eprint = "2311.08463",
    archivePrefix = "arXiv",
    primaryClass = "quant-ph",
    doi = "10.22331/q-2024-05-14-1347",
    journal = "Quantum",
    volume = "8",
    pages = "1347",
    year = "2024"
}

@misc{Capecci:2025ull,
    author = "Capecci, Chiara and Santra, Gopal Chandra and Bottarelli, Alberto and Tirrito, Emanuele and Hauke, Philipp",
    title = "{Role of Nonstabilizerness in Quantum Optimization}",
    eprint = "2505.17185",
    archivePrefix = "arXiv",
    primaryClass = "quant-ph",
    month = "5",
    year = "2025"
}

@article{Klco:2020aud,
    author = "Klco, Natalie and Savage, Martin J.",
    title = "{Fixed-point quantum circuits for quantum field theories}",
    eprint = "2002.02018",
    archivePrefix = "arXiv",
    primaryClass = "quant-ph",
    reportNumber = "INT-PUB-20-003",
    doi = "10.1103/PhysRevA.102.052422",
    journal = "Phys. Rev. A",
    volume = "102",
    number = "5",
    pages = "052422",
    year = "2020"
}

@article{Klco:2019yrb,
    author = "Klco, Natalie and Savage, Martin J.",
    title = "{Systematically Localizable Operators for Quantum Simulations of Quantum Field Theories}",
    eprint = "1912.03577",
    archivePrefix = "arXiv",
    primaryClass = "quant-ph",
    reportNumber = "INT-PUB-19-058",
    doi = "10.1103/PhysRevA.102.012619",
    journal = "Phys. Rev. A",
    volume = "102",
    number = "1",
    pages = "012619",
    year = "2020"
}

@article{Verstraete:2008qpa,
    author = "Verstraete, Frank and Cirac, J. Ignacio and Latorre, Jose I.",
    title = "{Quantum circuits for strongly correlated quantum systems}",
    eprint = "0804.1888",
    archivePrefix = "arXiv",
    primaryClass = "quant-ph",
    doi = "10.1103/PhysRevA.79.032316",
    journal = "Phys. Rev. A",
    volume = "79",
    pages = "032316",
    year = "2009"
}

@article{Cervera-Lierta:2018bia,
    author = "Cervera-Lierta, Alba",
    title = "{Exact Ising model simulation on a quantum computer}",
    eprint = "1807.07112",
    archivePrefix = "arXiv",
    doi = "10.22331/q-2018-12-21-114",
    journal = "Quantum",
    volume = "2",
    pages = "114",
    year = "2018"
}

@article{Kuhn:2015zqa,
    author = {K\"uhn, Stefan and Zohar, Erez and Cirac, J. Ignacio and Ba\~nuls, Mari Carmen},
    title = "{Non-Abelian string breaking phenomena with Matrix Product States}",
    eprint = "1505.04441",
    archivePrefix = "arXiv",
    primaryClass = "hep-lat",
    doi = "10.1007/JHEP07(2015)130",
    journal = "JHEP",
    volume = "07",
    pages = "130",
    year = "2015"
}

@article{Lanczos:1950zz,
    author = "Lanczos, Cornelius",
    title = "{An iteration method for the solution of the eigenvalue problem of linear differential and integral operators}",
    doi = "10.6028/jres.045.026",
    journal = "J. Res. Natl. Bur. Stand. B",
    volume = "45",
    pages = "255--282",
    year = "1950"
}

@article{Sala:2018dui,
    author = {Sala, P. and Shi, T. and K\"uhn, S. and Ba\~nuls, M. C. and Demler, E. and Cirac, J. I.},
    title = "{Variational study of U(1) and SU(2) lattice gauge theories with Gaussian states in 1+1 dimensions}",
    eprint = "1805.05190",
    archivePrefix = "arXiv",
    primaryClass = "hep-lat",
    doi = "10.1103/PhysRevD.98.034505",
    journal = "Phys. Rev. D",
    volume = "98",
    number = "3",
    pages = "034505",
    year = "2018"
}

@article{Klco:2019evd,
    author = "Klco, Natalie and Stryker, Jesse R. and Savage, Martin J.",
    title = "{SU(2) non-Abelian gauge field theory in one dimension on digital quantum computers}",
    eprint = "1908.06935",
    archivePrefix = "arXiv",
    primaryClass = "quant-ph",
    reportNumber = "INT-PUB-19-033",
    doi = "10.1103/PhysRevD.101.074512",
    journal = "Phys. Rev. D",
    volume = "101",
    number = "7",
    pages = "074512",
    year = "2020"
}

@article{Farrell:2022wyt,
    author = "Farrell, Roland C. and Chernyshev, Ivan A. and Powell, Sarah J. M. and Zemlevskiy, Nikita A. and Illa, Marc and Savage, Martin J.",
    title = "{Preparations for quantum simulations of quantum chromodynamics in 1+1 dimensions. I. Axial gauge}",
    eprint = "2207.01731",
    archivePrefix = "arXiv",
    primaryClass = "quant-ph",
    reportNumber = "IQuS@UW-21-027, NT@UW-22-05",
    doi = "10.1103/PhysRevD.107.054512",
    journal = "Phys. Rev. D",
    volume = "107",
    number = "5",
    pages = "054512",
    year = "2023"
}

@article{Farrell:2024mgu,
    author = "Farrell, Roland C. and Illa, Marc and Savage, Martin J.",
    title = "{Steps toward quantum simulations of hadronization and energy loss in dense matter}",
    eprint = "2405.06620",
    archivePrefix = "arXiv",
    primaryClass = "quant-ph",
    reportNumber = "IQuS@UW-21-076, NT@UW-24-06",
    doi = "10.1103/PhysRevC.111.015202",
    journal = "Phys. Rev. C",
    volume = "111",
    number = "1",
    pages = "015202",
    year = "2025"
}

@article{Ciavarella:2024lsp,
    author = "Ciavarella, Anthony N.",
    title = "{String breaking in the heavy quark limit with scalable circuits}",
    eprint = "2411.05915",
    archivePrefix = "arXiv",
    primaryClass = "quant-ph",
    doi = "10.1103/PhysRevD.111.054501",
    journal = "Phys. Rev. D",
    volume = "111",
    number = "5",
    pages = "054501",
    year = "2025"
}

@article{Atas:2021ext,
    author = "Atas, Yasar Y. and Zhang, Jinglei and Lewis, Randy and Jahanpour, Amin and Haase, Jan F. and Muschik, Christine A.",
    title = "{SU(2) hadrons on a quantum computer via a variational approach}",
    eprint = "2102.08920",
    archivePrefix = "arXiv",
    primaryClass = "quant-ph",
    doi = "10.1038/s41467-021-26825-4",
    journal = "Nat. Commun.",
    volume = "12",
    number = "1",
    pages = "6499",
    year = "2021"
}

@article{Ciavarella:2021nmj,
    author = "Ciavarella, Anthony and Klco, Natalie and Savage, Martin J.",
    title = "{Trailhead for quantum simulation of SU(3) Yang-Mills lattice gauge theory in the local multiplet basis}",
    eprint = "2101.10227",
    archivePrefix = "arXiv",
    primaryClass = "quant-ph",
    reportNumber = "IQuS@UW-21-001",
    doi = "10.1103/PhysRevD.103.094501",
    journal = "Phys. Rev. D",
    volume = "103",
    number = "9",
    pages = "094501",
    year = "2021"
}

@article{Illa:2024kmf,
    author = "Illa, Marc and Robin, Caroline E. P. and Savage, Martin J.",
    title = "{Qu8its for quantum simulations of lattice quantum chromodynamics}",
    eprint = "2403.14537",
    archivePrefix = "arXiv",
    primaryClass = "quant-ph",
    reportNumber = "IQuS@UW-21-074",
    doi = "10.1103/PhysRevD.110.014507",
    journal = "Phys. Rev. D",
    volume = "110",
    number = "1",
    pages = "014507",
    year = "2024"
}

@misc{Mathematica,
  author = {{Wolfram Research{,} Inc.}},
  title = {Mathematica, {V}ersion 13.0.1},
  url = {https://www.wolfram.com/mathematica},
  note = {{Champaign, IL}},
  year = {2022}
}

@article{Farrell:2024fit,
    author = "Farrell, Roland C. and Illa, Marc and Ciavarella, Anthony N. and Savage, Martin J.",
    title = "{Quantum simulations of hadron dynamics in the Schwinger model using 112 qubits}",
    eprint = "2401.08044",
    archivePrefix = "arXiv",
    primaryClass = "quant-ph",
    reportNumber = "IQuS@UW-21-069, NT@UW-24-1",
    doi = "10.1103/PhysRevD.109.114510",
    journal = "Phys. Rev. D",
    volume = "109",
    number = "11",
    pages = "114510",
    year = "2024"
}

@article{Farrell:2023fgd,
    author = "Farrell, Roland C. and Illa, Marc and Ciavarella, Anthony N. and Savage, Martin J.",
    title = "{Scalable Circuits for Preparing Ground States on Digital Quantum Computers: The Schwinger Model Vacuum on 100 Qubits}",
    eprint = "2308.04481",
    archivePrefix = "arXiv",
    primaryClass = "quant-ph",
    reportNumber = "IQuS@UW-21-060, NT@UW-23-13",
    doi = "10.1103/PRXQuantum.5.020315",
    journal = "PRX Quantum",
    volume = "5",
    number = "2",
    pages = "020315",
    year = "2024"
}

@article{Algaba:2023enr,
    author = "Algaba, Manuel G. and Sriluckshmy, P. V. and Leib, Martin and {\v{S}}imkovic IV, Fedor",
    title = "{Low-depth simulations of fermionic systems on square-grid quantum hardware}",
    eprint = "2302.01862",
    archivePrefix = "arXiv",
    primaryClass = "quant-ph",
    doi = "10.22331/q-2024-04-30-1327",
    journal = "Quantum",
    volume = "8",
    pages = "1327",
    year = "2024"
}

@article{Muller:2023nnk,
    author = {M{\"u}ller, Berndt and Yao, Xiaojun},
    title = "{Simple Hamiltonian for quantum simulation of strongly coupled (2+1)D SU(2) lattice gauge theory on a honeycomb lattice}",
    eprint = "2307.00045",
    archivePrefix = "arXiv",
    primaryClass = "quant-ph",
    reportNumber = "IQuS@UW-21-055",
    doi = "10.1103/PhysRevD.108.094505",
    journal = "Phys. Rev. D",
    volume = "108",
    number = "9",
    pages = "094505",
    year = "2023"
}

@article{Kavaki:2024ijd,
    author = "Kavaki, Ali H. Z. and Lewis, Randy",
    title = "{From square plaquettes to triamond lattices for SU(2) gauge theory}",
    eprint = "2401.14570",
    archivePrefix = "arXiv",
    primaryClass = "hep-lat",
    doi = "10.1038/s42005-024-01697-4",
    journal = "Commun. Phys.",
    volume = "7",
    number = "1",
    pages = "208",
    year = "2024"
}

@article{Turro:2024pxu,
    author = "Turro, Francesco and Ciavarella, Anthony and Yao, Xiaojun",
    title = "{Classical and quantum computing of shear viscosity for (2+1)D SU(2) gauge theory}",
    eprint = "2402.04221",
    archivePrefix = "arXiv",
    primaryClass = "hep-lat",
    reportNumber = "IQuS@UW-21-073",
    doi = "10.1103/PhysRevD.109.114511",
    journal = "Phys. Rev. D",
    volume = "109",
    number = "11",
    pages = "114511",
    year = "2024"
}

@article{Illa:2025dou,
    author = "Illa, Marc and Savage, Martin J. and Yao, Xiaojun",
    title = "{Improved honeycomb and hyperhoneycomb lattice Hamiltonians for quantum simulations of non-Abelian gauge theories}",
    eprint = "2503.09688",
    archivePrefix = "arXiv",
    primaryClass = "hep-lat",
    reportNumber = "IQuS@UW-21-097",
    doi = "10.1103/3rwf-f844",
    journal = "Phys. Rev. D",
    volume = "111",
    number = "11",
    pages = "114520",
    year = "2025"
}

@misc{Illa:2025njz,
    author = "Illa, Marc and Savage, Martin J. and Yao, Xiaojun",
    title = "{Dynamical Local Tadpole-Improvement in Quantum Simulations of Gauge Theories}",
    eprint = "2504.21575",
    archivePrefix = "arXiv",
    primaryClass = "quant-ph",
    reportNumber = "IQuS@UW-21-100",
    month = "4",
    year = "2025"
}

@article{Li:2020uhl,
    author = "Li, Meijian and Zhao, Xingbo and Maris, Pieter and Chen, Guangyao and Li, Yang and Tuchin, Kirill and Vary, James P.",
    title = "{Ultrarelativistic quark-nucleus scattering in a light-front Hamiltonian approach}",
    eprint = "2002.09757",
    archivePrefix = "arXiv",
    primaryClass = "nucl-th",
    doi = "10.1103/PhysRevD.101.076016",
    journal = "Phys. Rev. D",
    volume = "101",
    number = "7",
    pages = "076016",
    year = "2020"
}

@article{Li:2021zaw,
    author = "Li, Meijian and Lappi, Tuomas and Zhao, Xingbo",
    title = "{Scattering and gluon emission in a color field: A light-front Hamiltonian approach}",
    eprint = "2107.02225",
    archivePrefix = "arXiv",
    primaryClass = "hep-ph",
    doi = "10.1103/PhysRevD.104.056014",
    journal = "Phys. Rev. D",
    volume = "104",
    number = "5",
    pages = "056014",
    year = "2021"
}

@article{Barata:2023jgd,
    author = "Barata, Jo\~ao and Gong, Wenjie and Venugopalan, Raju",
    title = "{Realtime dynamics of hyperon spin correlations from string fragmentation in a deformed four-flavor Schwinger model}",
    eprint = "2308.13596",
    archivePrefix = "arXiv",
    primaryClass = "hep-ph",
    doi = "10.1103/PhysRevD.109.116003",
    journal = "Phys. Rev. D",
    volume = "109",
    number = "11",
    pages = "116003",
    year = "2024"
}

@article{Barata:2022wim,
    author = "Barata, Jo\~ao and Du, Xiaojian and Li, Meijian and Qian, Wenyang and Salgado, Carlos A.",
    title = "{Medium induced jet broadening in a quantum computer}",
    eprint = "2208.06750",
    archivePrefix = "arXiv",
    primaryClass = "hep-ph",
    doi = "10.1103/PhysRevD.106.074013",
    journal = "Phys. Rev. D",
    volume = "106",
    number = "7",
    pages = "074013",
    year = "2022"
}

@article{Barata:2021yri,
    author = "Barata, Jo\~ao and Salgado, Carlos A.",
    title = "{A quantum strategy to compute the jet quenching parameter $\hat{q}$}",
    eprint = "2104.04661",
    archivePrefix = "arXiv",
    primaryClass = "hep-ph",
    doi = "10.1140/epjc/s10052-021-09674-9",
    journal = "Eur. Phys. J. C",
    volume = "81",
    number = "10",
    pages = "862",
    year = "2021"
}

@article{Gong:2021bcp,
    author = "Gong, Wenjie and Parida, Ganesh and Tu, Zhoudunming and Venugopalan, Raju",
    title = "{Measurement of Bell-type inequalities and quantum entanglement from \ensuremath{\Lambda}-hyperon spin correlations at high energy colliders}",
    eprint = "2107.13007",
    archivePrefix = "arXiv",
    primaryClass = "hep-ph",
    doi = "10.1103/PhysRevD.106.L031501",
    journal = "Phys. Rev. D",
    volume = "106",
    number = "3",
    pages = "L031501",
    year = "2022"
}

@article{Baker:2017wtt,
    author = "Baker, O. K. and Kharzeev, D. E.",
    title = "{Thermal radiation and entanglement in proton-proton collisions at energies available at the CERN Large Hadron Collider}",
    eprint = "1712.04558",
    archivePrefix = "arXiv",
    primaryClass = "hep-ph",
    doi = "10.1103/PhysRevD.98.054007",
    journal = "Phys. Rev. D",
    volume = "98",
    number = "5",
    pages = "054007",
    year = "2018"
}

@misc{Lovato:2022vgq,
    author = "Lovato, Alessandro and others",
    title = "{Long Range Plan: Dense matter theory for heavy-ion collisions and neutron stars}",
    eprint = "2211.02224",
    archivePrefix = "arXiv",
    primaryClass = "nucl-th",
    reportNumber = "LA-UR-22-31648",
    month = "11",
    year = "2022"
}

@article{Klco:2019xro,
    author = "Klco, Natalie and Savage, Martin J.",
    title = "{Minimally entangled state preparation of localized wave functions on quantum computers}",
    eprint = "1904.10440",
    archivePrefix = "arXiv",
    primaryClass = "quant-ph",
    reportNumber = "INT-PUB-19-013",
    doi = "10.1103/PhysRevA.102.012612",
    journal = "Phys. Rev. A",
    volume = "102",
    number = "1",
    pages = "012612",
    year = "2020"
}

@article{Beane:2010em,
    author = "Beane, S. R. and Detmold, W. and Orginos, K. and Savage, M. J.",
    collaboration = "NPLQCD",
    title = "{Nuclear Physics from Lattice QCD}",
    eprint = "1004.2935",
    archivePrefix = "arXiv",
    primaryClass = "hep-lat",
    reportNumber = "JLAB-THY-10-1166",
    doi = "10.1016/j.ppnp.2010.08.002",
    journal = "Prog. Part. Nucl. Phys.",
    volume = "66",
    pages = "1--40",
    year = "2011"
}

@article{Beane:2021zvo,
    author = "Beane, Silas R. and Farrell, Roland C. and Varma, Mira",
    title = "{Entanglement minimization in hadronic scattering with pions}",
    eprint = "2108.00646",
    archivePrefix = "arXiv",
    primaryClass = "hep-ph",
    reportNumber = "NT@UW-21-05, IQuS@UW-21-008",
    doi = "10.1142/S0217751X21502055",
    journal = "Int. J. Mod. Phys. A",
    volume = "36",
    number = "30",
    pages = "2150205",
    year = "2021"
}

@article{Savage:2016kon,
    author = "Savage, Martin J. and Shanahan, Phiala E. and Tiburzi, Brian C. and Wagman, Michael L. and Winter, Frank and Beane, Silas R. and Chang, Emmanuel and Davoudi, Zohreh and Detmold, William and Orginos, Kostas",
    title = "{Proton-Proton Fusion and Tritium $\beta$ Decay from Lattice Quantum Chromodynamics}",
    eprint = "1610.04545",
    archivePrefix = "arXiv",
    primaryClass = "hep-lat",
    reportNumber = "INT-PUB-16-033, JLAB-THY-16-2362, MIT-CTP-4844, NT@UW-16-12",
    doi = "10.1103/PhysRevLett.119.062002",
    journal = "Phys. Rev. Lett.",
    volume = "119",
    number = "6",
    pages = "062002",
    year = "2017"
}

@article{Chang:2017eiq,
    author = "Chang, Emmanuel and Davoudi, Zohreh and Detmold, William and Gambhir, Arjun S. and Orginos, Kostas and Savage, Martin J. and Shanahan, Phiala E. and Wagman, Michael L. and Winter, Frank",
    collaboration = "NPLQCD",
    title = "{Scalar, Axial, and Tensor Interactions of Light Nuclei from Lattice QCD}",
    eprint = "1712.03221",
    archivePrefix = "arXiv",
    primaryClass = "hep-lat",
    reportNumber = "INT-PUB-17-054, MIT-CTP-4967, JLAB-THY-17-2607, UMD-PP-017-35",
    doi = "10.1103/PhysRevLett.120.152002",
    journal = "Phys. Rev. Lett.",
    volume = "120",
    number = "15",
    pages = "152002",
    year = "2018"
}

@article{Angelides:2023noe,
    author = {Angelides, Takis and Naredi, Pranay and Crippa, Arianna and Jansen, Karl and K{\"u}hn, Stefan and Tavernelli, Ivano and Wang, Derek S.},
    title = "{First-order phase transition of the Schwinger model with a quantum computer}",
    eprint = "2312.12831",
    archivePrefix = "arXiv",
    primaryClass = "hep-lat",
    doi = "10.1038/s41534-024-00950-6",
    journal = "npj Quantum Inf.",
    volume = "11",
    number = "1",
    pages = "6",
    year = "2025"
}

@article{Preskill:2018jim,
    author = "Preskill, John",
    title = "{Quantum Computing in the NISQ era and beyond}",
    eprint = "1801.00862",
    archivePrefix = "arXiv",
    primaryClass = "quant-ph",
    doi = "10.22331/q-2018-08-06-79",
    journal = "Quantum",
    volume = "2",
    pages = "79",
    year = "2018"
}

@article{Asaduzzaman:2022bpi,
    author = "Asaduzzaman, Muhammad and Catterall, Simon and Toga, Goksu Can and Meurice, Yannick and Sakai, Ryo",
    title = "{Quantum simulation of the $N$-flavor Gross-Neveu model}",
    eprint = "2208.05906",
    archivePrefix = "arXiv",
    primaryClass = "hep-lat",
    doi = "10.1103/PhysRevD.106.114515",
    journal = "Phys. Rev. D",
    volume = "106",
    number = "11",
    pages = "114515",
    year = "2022"
}

@article{Schuster:2023klj,
    author = {Schuster, Stephan and K{\"u}hn, Stefan and Funcke, Lena and Hartung, Tobias and Pleinert, Marc-Oliver and von Zanthier, Joachim and Jansen, Karl},
    title = "{Studying the phase diagram of the three-flavor Schwinger model in the presence of a chemical potential with measurement- and gate-based quantum computing}",
    eprint = "2311.14825",
    archivePrefix = "arXiv",
    primaryClass = "hep-lat",
    doi = "10.1103/PhysRevD.109.114508",
    journal = "Phys. Rev. D",
    volume = "109",
    number = "11",
    pages = "114508",
    year = "2024"
}

@article{Borzenkova:2023xaf,
    author = "Borzenkova, O. V. and Struchalin, G. I. and Kondratyev, I. V. and Moiseevskiy, A. D. and Skryabin, N. N. and Dyakonov, I. V. and Straupe, S. S.",
    title = "{Error-mitigated variational algorithm on a photonic processor}",
    eprint = "2311.13985",
    archivePrefix = "arXiv",
    primaryClass = "quant-ph",
    doi = "10.1364/OL.532017",
    journal = "Opt. Lett.",
    volume = "49",
    number = "15",
    pages = "4453--4456",
    year = "2024"
}

@article{Troyer:2004ge,
    author = "Troyer, Matthias and Wiese, Uwe-Jens",
    title = "{Computational complexity and fundamental limitations to fermionic quantum Monte Carlo simulations}",
    eprint = "cond-mat/0408370",
    archivePrefix = "arXiv",
    doi = "10.1103/PhysRevLett.94.170201",
    journal = "Phys. Rev. Lett.",
    volume = "94",
    pages = "170201",
    year = "2005"
}

@article{Schwinger:1962tp,
    author = "Schwinger, Julian S.",
    title = "{Gauge Invariance and Mass. II}",
    doi = "10.1103/PhysRev.128.2425",
    journal = "Phys. Rev.",
    volume = "128",
    pages = "2425--2429",
    year = "1962"
}

@article{Bierlich:2022pfr,
    author = "Bierlich, Christian and others",
    title = "{A comprehensive guide to the physics and usage of PYTHIA 8.3}",
    eprint = "2203.11601",
    archivePrefix = "arXiv",
    primaryClass = "hep-ph",
    reportNumber = "LU-TP 22-16, MCNET-22-04, FERMILAB-PUB-22-227-SCD",
    doi = "10.21468/SciPostPhysCodeb.8",
    journal = "SciPost Phys. Codeb.",
    volume = "2022",
    pages = "8",
    year = "2022"
}

@article{Bauer:2021gup,
    author = "Bauer, Christian W. and Freytsis, Marat and Nachman, Benjamin",
    title = "{Simulating Collider Physics on Quantum Computers Using Effective Field Theories}",
    eprint = "2102.05044",
    archivePrefix = "arXiv",
    primaryClass = "hep-ph",
    doi = "10.1103/PhysRevLett.127.212001",
    journal = "Phys. Rev. Lett.",
    volume = "127",
    number = "21",
    pages = "212001",
    year = "2021"
}

@article{Magnifico:2019kyj,
    author = "Magnifico, Giuseppe and Dalmonte, Marcello and Facchi, Paolo and Pascazio, Saverio and Pepe, Francesco V. and Ercolessi, Elisa",
    title = "{Real Time Dynamics and Confinement in the $\mathbb{Z}_{n}$ Schwinger-Weyl lattice model for 1+1 QED}",
    eprint = "1909.04821",
    archivePrefix = "arXiv",
    primaryClass = "quant-ph",
    reportNumber = "Quantum 4, 281 (2020)",
    doi = "10.22331/q-2020-06-15-281",
    journal = "Quantum",
    volume = "4",
    pages = "281",
    year = "2020"
}

@article{Ferguson:2020qyf,
    author = {Ferguson, Ryan R. and Dellantonio, Luca and Jansen, Karl and Balushi, Abdulrahim Al and D\"ur, Wolfgang and Muschik, Christine A.},
    title = "{Measurement-Based Variational Quantum Eigensolver}",
    eprint = "2010.13940",
    archivePrefix = "arXiv",
    primaryClass = "quant-ph",
    reportNumber = "DESY-20-197",
    doi = "10.1103/PhysRevLett.126.220501",
    journal = "Phys. Rev. Lett.",
    volume = "126",
    number = "22",
    pages = "220501",
    year = "2021"
}

@article{Ciavarella:2023mfc,
    author = "Ciavarella, Anthony N.",
    title = "{Quantum Simulation of Lattice QCD with Improved Hamiltonians}",
    eprint = "2307.05593",
    archivePrefix = "arXiv",
    primaryClass = "hep-lat",
    doi = {10.1103/PhysRevD.108.094513},
    journal = {Phys. Rev. D},
    volume = {108},
    issue = {9},
    pages = {094513},
    numpages = {14},
    year = {2023},
    reportNumber = "IQuS@UW-21-056",
}

@article{Zhang:2023hzr,
    author = "Zhang, Wei-Yong and others",
    title = "{Observation of microscopic confinement dynamics by a tunable topological {\ensuremath{\theta}}-angle}",
    eprint = "2306.11794",
    archivePrefix = "arXiv",
    primaryClass = "cond-mat.quant-gas",
    doi = "10.1038/s41567-024-02702-x",
    journal = "Nat. Phys.",
    volume = "21",
    number = "1",
    pages = "155--160",
    year = "2025"
}

@article{Su:2022glk,
    author = "Su, Guo-Xian and Sun, Hui and Hudomal, Ana and Desaules, Jean-Yves and Zhou, Zhao-Yu and Yang, Bing and Halimeh, Jad C. and Yuan, Zhen-Sheng and Papi{\'c}, Zlatko and Pan, Jian-Wei",
    title = "{Observation of many-body scarring in a Bose-Hubbard quantum simulator}",
    eprint = "2201.00821",
    archivePrefix = "arXiv",
    primaryClass = "cond-mat.quant-gas",
    doi = "10.1103/PhysRevResearch.5.023010",
    journal = "Phys. Rev. Res.",
    volume = "5",
    number = "2",
    pages = "023010",
    year = "2023"
}

@article{Pomarico:2023png,
    author = "Pomarico, Domenico and Cosmai, Leonardo and Facchi, Paolo and Lupo, Cosmo and Pascazio, Saverio and Pepe, Francesco V.",
    title = "{Dynamical Quantum Phase Transitions of the Schwinger Model: Real-Time Dynamics on IBM Quantum}",
    eprint = "2302.01151",
    archivePrefix = "arXiv",
    primaryClass = "quant-ph",
    doi = "10.3390/e25040608",
    journal = "Entropy",
    volume = "25",
    number = "4",
    pages = "608",
    year = "2023"
}

@article{Tran:2020azk,
    author = "Tran, Minh C. and Su, Yuan and Carney, Daniel and Taylor, Jacob M.",
    title = "{Faster Digital Quantum Simulation by Symmetry Protection}",
    eprint = "2006.16248",
    archivePrefix = "arXiv",
    primaryClass = "quant-ph",
    reportNumber = "FERMILAB-PUB-20-240-QIS-T",
    doi = "10.1103/PRXQuantum.2.010323",
    journal = "PRX Quantum",
    volume = "2",
    pages = "010323",
    year = "2021"
}

@article{Charles:2023zbl,
    author = "Charles, Clement and Gustafson, Erik J. and Hardt, Elizabeth and Herren, Florian and Hogan, Norman and Lamm, Henry and Starecheski, Sara and Van de Water, Ruth S. and Wagman, Michael L.",
    title = "{Simulating $\mathbb{Z}_2$ lattice gauge theory on a quantum computer}",
    eprint = "2305.02361",
    archivePrefix = "arXiv",
    primaryClass = "hep-lat",
    reportNumber = "FERMILAB-PUB-23-171-SQMS-T",
    journal = {Phys. Rev. E},
  volume = {109},
  issue = {1},
  pages = {015307},
  numpages = {19},
  year = {2024},
  month = {Jan},
  publisher = {American Physical Society},
  doi = {10.1103/PhysRevE.109.015307},
}

@article{Atas:2022dqm,
    author = "Atas, Yasar Y. and Haase, Jan F. and Zhang, Jinglei and Wei, Victor and Pfaendler, Sieglinde M. -L. and Lewis, Randy and Muschik, Christine A.",
    title = "{Simulating one-dimensional quantum chromodynamics on a quantum computer: Real-time evolutions of tetra- and pentaquarks}",
    eprint = "2207.03473",
    archivePrefix = "arXiv",
    primaryClass = "quant-ph",
    doi = "10.1103/PhysRevResearch.5.033184",
    journal = "Phys. Rev. Res.",
    volume = "5",
    number = "3",
    pages = "033184",
    year = "2023"
}

@article{Riechert:2021ink,
    author = "Riechert, Hannes and Halimeh, Jad C. and Kasper, Valentin and Bretheau, Landry and Zohar, Erez and Hauke, Philipp and Jendrzejewski, Fred",
    title = "{Engineering a U(1) lattice gauge theory in classical electric circuits}",
    eprint = "2108.01086",
    archivePrefix = "arXiv",
    primaryClass = "cond-mat.mes-hall",
    doi = "10.1103/PhysRevB.105.205141",
    journal = "Phys. Rev. B",
    volume = "105",
    number = "20",
    pages = "205141",
    year = "2022"
}

@article{Honda:2021aum,
    author = "Honda, Masazumi and Itou, Etsuko and Kikuchi, Yuta and Nagano, Lento and Okuda, Takuya",
    title = "{Classically emulated digital quantum simulation for screening and confinement in the Schwinger model with a topological term}",
    eprint = "2105.03276",
    archivePrefix = "arXiv",
    primaryClass = "hep-lat",
    reportNumber = "YITP-21-12, UT-Komaba-21-4",
    doi = "10.1103/PhysRevD.105.014504",
    journal = "Phys. Rev. D",
    volume = "105",
    number = "1",
    pages = "014504",
    year = "2022"
}

@article{Mueller:2022xbg,
    author = {Mueller, Niklas and Carolan, Joseph A. and Connelly, Andrew and Davoudi, Zohreh and Dumitrescu, Eugene F. and Yeter-Aydeniz, K\"ubra},
    title = "{Quantum Computation of Dynamical Quantum Phase Transitions and Entanglement Tomography in a Lattice Gauge Theory}",
    eprint = "2210.03089",
    archivePrefix = "arXiv",
    primaryClass = "quant-ph",
    journal = {PRX Quantum},
  volume = {4},
  issue = {3},
  pages = {030323},
  numpages = {32},
  year = {2023},
  month = {Aug},
  doi = {10.1103/PRXQuantum.4.030323},
}

@article{Yang:2020yer,
    author = "Yang, Bing and Sun, Hui and Ott, Robert and Wang, Han-Yi and Zache, Torsten V. and Halimeh, Jad C. and Yuan, Zhen-Sheng and Hauke, Philipp and Pan, Jian-Wei",
    title = "{Observation of gauge invariance in a 71-site Bose\textendash{}Hubbard quantum simulator}",
    eprint = "2003.08945",
    archivePrefix = "arXiv",
    primaryClass = "cond-mat.quant-gas",
    doi = "10.1038/s41586-020-2910-8",
    journal = "Nature",
    volume = "587",
    number = "7834",
    pages = "392--396",
    year = "2020"
}

@article{Klco:2018kyo,
    author = "Klco, N. and Dumitrescu, E. F. and McCaskey, A. J. and Morris, T. D. and Pooser, R. C. and Sanz, M. and Solano, E. and Lougovski, P. and Savage, M. J.",
    title = "{Quantum-classical computation of Schwinger model dynamics using quantum computers}",
    eprint = "1803.03326",
    archivePrefix = "arXiv",
    primaryClass = "quant-ph",
    reportNumber = "INT-PUB-18-013",
    doi = "10.1103/PhysRevA.98.032331",
    journal = "Phys. Rev. A",
    volume = "98",
    number = "3",
    pages = "032331",
    year = "2018"
}

@article{ARahman:2022tkr,
    author = "A Rahman, Sarmed and Lewis, Randy and Mendicelli, Emanuele and Powell, Sarah",
    title = "{Self-mitigating Trotter circuits for SU(2) lattice gauge theory on a quantum computer}",
    eprint = "2205.09247",
    archivePrefix = "arXiv",
    primaryClass = "hep-lat",
    doi = "10.1103/PhysRevD.106.074502",
    journal = "Phys. Rev. D",
    volume = "106",
    number = "7",
    pages = "074502",
    year = "2022"
}

@article{Ciavarella:2021lel,
    author = "Ciavarella, Anthony N. and Chernyshev, Ivan A.",
    title = "{Preparation of the SU(3) lattice Yang-Mills vacuum with variational quantum methods}",
    eprint = "2112.09083",
    archivePrefix = "arXiv",
    primaryClass = "quant-ph",
    reportNumber = "IQuS@UW-21-017",
    doi = "10.1103/PhysRevD.105.074504",
    journal = "Phys. Rev. D",
    volume = "105",
    number = "7",
    pages = "074504",
    year = "2022"
}

@article{ARahman:2021ktn,
    author = "A Rahman, Sarmed and Lewis, Randy and Mendicelli, Emanuele and Powell, Sarah",
    title = "{SU(2) lattice gauge theory on a quantum annealer}",
    eprint = "2103.08661",
    archivePrefix = "arXiv",
    primaryClass = "hep-lat",
    doi = "10.1103/PhysRevD.104.034501",
    journal = "Phys. Rev. D",
    volume = "104",
    number = "3",
    pages = "034501",
    year = "2021"
}

@article{Nguyen:2021hyk,
    author = "Nguyen, Nhung H. and Tran, Minh C. and Zhu, Yingyue and Green, Alaina M. and Alderete, C. Huerta and Davoudi, Zohreh and Linke, Norbert M.",
    title = "{Digital Quantum Simulation of the Schwinger Model and Symmetry Protection with Trapped Ions}",
    eprint = "2112.14262",
    archivePrefix = "arXiv",
    primaryClass = "quant-ph",
    reportNumber = "MIT-CTP/5384, UMD-PP-021-08",
    doi = "10.1103/PRXQuantum.3.020324",
    journal = "PRX Quantum",
    volume = "3",
    number = "2",
    pages = "020324",
    year = "2022"
}

@article{Zhou:2021kdl,
    author = {Zhou, Zhao-Yu and Su, Guo-Xian and Halimeh, Jad C. and Ott, Robert and Sun, Hui and Hauke, Philipp and Yang, Bing and Yuan, Zhen-Sheng and Berges, J\"urgen and Pan, Jian-Wei},
    title = "{Thermalization dynamics of a gauge theory on a quantum simulator}",
    eprint = "2107.13563",
    archivePrefix = "arXiv",
    primaryClass = "cond-mat.quant-gas",
    doi = "10.1126/science.abl6277",
    journal = "Science",
    volume = "377",
    number = "6603",
    pages = "311-314",
    year = "2022"
}

@article{deJong:2021wsd,
    author = "de Jong, Wibe A. and Lee, Kyle and Mulligan, James and P\l{}osko\'n, Mateusz and Ringer, Felix and Yao, Xiaojun",
    title = "{Quantum simulation of nonequilibrium dynamics and thermalization in the Schwinger model}",
    eprint = "2106.08394",
    archivePrefix = "arXiv",
    primaryClass = "quant-ph",
    reportNumber = "MIT-CTP/5308",
    doi = "10.1103/PhysRevD.106.054508",
    journal = "Phys. Rev. D",
    volume = "106",
    number = "5",
    pages = "054508",
    year = "2022"
}

@article{Mil:2019pbt,
    author = {Mil, Alexander and Zache, Torsten V. and Hegde, Apoorva and Xia, Andy and Bhatt, Rohit P. and Oberthaler, Markus K. and Hauke, Philipp and Berges, J\"urgen and Jendrzejewski, Fred},
    title = "{A scalable realization of local U(1) gauge invariance in cold atomic mixtures}",
    eprint = "1909.07641",
    archivePrefix = "arXiv",
    primaryClass = "cond-mat.quant-gas",
    doi = "10.1126/science.aaz5312",
    journal = "Science",
    volume = "367",
    number = "6482",
    pages = "1128--1130",
    year = "2020"
}

@article{Lu:2018pjk,
    author = {Lu, Hsuan-Hao and Klco, Natalie and Lukens, Joseph M. and Morris, Titus D. and Bansal, Aaina and Ekstr\"om, Andreas and Hagen, Gaute and Papenbrock, Thomas and Weiner, Andrew M. and Savage, Martin J. and Lougovski, Pavel},
    title = "{Simulations of Subatomic Many-Body Physics on a Quantum Frequency Processor}",
    eprint = "1810.03959",
    archivePrefix = "arXiv",
    primaryClass = "quant-ph",
    doi = "10.1103/PhysRevA.100.012320",
    journal = "Phys. Rev. A",
    volume = "100",
    number = "1",
    pages = "012320",
    year = "2019"
}

@article{Kokail:2018eiw,
    author = {{Kokail}, C. and {Maier}, C. and {van Bijnen}, R. and {Brydges}, T. and {Joshi}, M.~K. and {Jurcevic}, P. and {Muschik}, C.~A. and {Silvi}, P. and {Blatt}, R. and {Roos}, C.~F. and {Zoller}, P.},
    title = "{Self-verifying variational quantum simulation of lattice models}",
    eprint = "1810.03421",
    archivePrefix = "arXiv",
    primaryClass = "quant-ph",
    doi = "10.1038/s41586-019-1177-4",
    journal = "Nature",
    volume = "569",
    number = "7756",
    pages = "355--360",
    year = "2019"
}

@article{Martinez:2016yna,
    author = "{Martinez}, Esteban A. and {Muschik}, Christine A. and {Schindler}, Philipp and {Nigg}, Daniel and {Erhard}, Alexander and {Heyl}, Markus and {Hauke}, Philipp and {Dalmonte}, Marcello and {Monz}, Thomas and {Zoller}, Peter and {Blatt}, Rainer",
    title = "{Real-time dynamics of lattice gauge theories with a few-qubit quantum computer}",
    eprint = "1605.04570",
    archivePrefix = "arXiv",
    primaryClass = "quant-ph",
    doi = "10.1038/nature18318",
    journal = "Nature",
    volume = "534",
    pages = "516--519",
    year = "2016"
}

@article{Banuls:2019bmf,
    author = "Ba\~nuls, M. C. and others",
    title = "{Simulating Lattice Gauge Theories within Quantum Technologies}",
    eprint = "1911.00003",
    archivePrefix = "arXiv",
    primaryClass = "quant-ph",
    doi = "10.1140/epjd/e2020-100571-8",
    journal = "Eur. Phys. J. D",
    volume = "74",
    number = "8",
    pages = "165",
    year = "2020"
}

@article{Weinberg:1967tq,
    author = "Weinberg, Steven",
    title = "{A Model of Leptons}",
    doi = "10.1103/PhysRevLett.19.1264",
    journal = "Phys. Rev. Lett.",
    volume = "19",
    pages = "1264--1266",
    year = "1967"
}

@article{Glashow:1961tr,
    author = "Glashow, S.L.",
    title = "{Partial Symmetries of Weak Interactions}",
    doi = "10.1016/0029-5582(61)90469-2",
    journal = "Nucl. Phys.",
    volume = "22",
    pages = "579--588",
    year = "1961"
}

@article{Salam:1968rm,
    author = "Salam, Abdus",
    title = "{Weak and Electromagnetic Interactions}",
    doi = "10.1142/9789812795915\_0034",
    journal = "Conf. Proc. C",
    volume = "680519",
    pages = "367--377",
    year = "1968"
}

@article{Politzer:1973fx,
    author = "Politzer, H.David",
    editor = "Taylor, J.C.",
    title = "{Reliable Perturbative Results for Strong Interactions?}",
    doi = "10.1103/PhysRevLett.30.1346",
    journal = "Phys. Rev. Lett.",
    volume = "30",
    pages = "1346--1349",
    year = "1973"
}

@article{Wilson:1974sk,
    author = "Wilson, Kenneth G.",
    editor = "Taylor, J. C.",
    title = "{Confinement of Quarks}",
    reportNumber = "CLNS-262",
    doi = "10.1103/PhysRevD.10.2445",
    journal = "Phys. Rev. D",
    volume = "10",
    pages = "2445--2459",
    year = "1974"
}

@article{BMW:2008jgk,
    author = "Durr, S. and others",
    collaboration = "BMW",
    title = "{Ab-Initio Determination of Light Hadron Masses}",
    eprint = "0906.3599",
    archivePrefix = "arXiv",
    primaryClass = "hep-lat",
    doi = "10.1126/science.1163233",
    journal = "Science",
    volume = "322",
    pages = "1224--1227",
    year = "2008"
}

@article{Borsanyi:2013bia,
    author = "Borsanyi, Szabocls and Fodor, Zoltan and Hoelbling, Christian and Katz, Sandor D. and Krieg, Stefan and Szabo, Kalman K.",
    title = "{Full result for the QCD equation of state with 2+1 flavors}",
    eprint = "1309.5258",
    archivePrefix = "arXiv",
    primaryClass = "hep-lat",
    doi = "10.1016/j.physletb.2014.01.007",
    journal = "Phys. Lett. B",
    volume = "730",
    pages = "99--104",
    year = "2014"
}

@article{HotQCD:2014kol,
    author = "Bazavov, A. and others",
    collaboration = "HotQCD",
    title = "{Equation of state in ( 2+1 )-flavor QCD}",
    eprint = "1407.6387",
    archivePrefix = "arXiv",
    primaryClass = "hep-lat",
    reportNumber = "BNL-105928-2014-JA",
    doi = "10.1103/PhysRevD.90.094503",
    journal = "Phys. Rev. D",
    volume = "90",
    pages = "094503",
    year = "2014"
}

@article{NPLQCD:2012mex,
    author = "Beane, S. R. and Chang, E. and Cohen, S. D. and Detmold, William and Lin, H. W. and Luu, T. C. and Orginos, K. and Parreno, A. and Savage, M. J. and Walker-Loud, A.",
    collaboration = "NPLQCD",
    title = "{Light Nuclei and Hypernuclei from Quantum Chromodynamics in the Limit of SU(3) Flavor Symmetry}",
    eprint = "1206.5219",
    archivePrefix = "arXiv",
    primaryClass = "hep-lat",
    reportNumber = "ICCUB-12-309, JLAB-THY-12-1581, NT-UW-12-09, NT-LBNL-12-012, UCB-NPAT-12-011, UNH-12-04",
    doi = "10.1103/PhysRevD.87.034506",
    journal = "Phys. Rev. D",
    volume = "87",
    number = "3",
    pages = "034506",
    year = "2013"
}

@article{FlavourLatticeAveragingGroupFLAG:2024oxs,
    author = "Aoki, Y. and others",
    collaboration = "Flavour Lattice Averaging Group (FLAG)",
    title = "{FLAG review 2024}",
    eprint = "2411.04268",
    archivePrefix = "arXiv",
    primaryClass = "hep-lat",
    reportNumber = "CERN-TH-2024-192, FERMILAB-PUB-24-0785-T",
    doi = "10.1103/nfzp-p5dn",
    journal = "Phys. Rev. D",
    volume = "113",
    number = "1",
    pages = "014508",
    year = "2026"
}

@article{Gross:1973id,
    author = "Gross, David J. and Wilczek, Frank",
    editor = "Taylor, J.C.",
    title = "{Ultraviolet Behavior of Nonabelian Gauge Theories}",
    doi = "10.1103/PhysRevLett.30.1343",
    journal = "Phys. Rev. Lett.",
    volume = "30",
    pages = "1343--1346",
    year = "1973"
}

@article{Higgs:1964pj,
    author = "Higgs, Peter W.",
    editor = "Taylor, J. C.",
    title = "{Broken Symmetries and the Masses of Gauge Bosons}",
    doi = "10.1103/PhysRevLett.13.508",
    journal = "Phys. Rev. Lett.",
    volume = "13",
    pages = "508--509",
    year = "1964",
    url = "https://link.aps.org/doi/10.1103/PhysRevLett.13.508"
}

@article{Grimsley_2019,
	author = {{Grimsley}, Harper R. and {Economou}, Sophia E. and {Barnes}, Edwin and {Mayhall}, Nicholas J.},
        title = "{An adaptive variational algorithm for exact molecular simulations on a quantum computer}",
      journal = {Nat. Commun.},
         year = 2019,
        month = jul,
       volume = {10},
          eid = {3007},
        pages = {3007},
          doi = {10.1038/s41467-019-10988-2},
archivePrefix = {arXiv},
       eprint = {1812.11173},
 primaryClass = {quant-ph},
}

@article{Kogut:1974ag,
    author = "Kogut, John B. and Susskind, Leonard",
    title = "{Hamiltonian Formulation of Wilson's Lattice Gauge Theories}",
    reportNumber = "Print-74-1186 (CORNELL)",
    doi = "10.1103/PhysRevD.11.395",
    journal = "Phys. Rev. D",
    volume = "11",
    pages = "395--408",
    year = "1975"
}

@article{Belyansky:2023rgh,
    author = "Belyansky, Ron and Whitsitt, Seth and Mueller, Niklas and Fahimniya, Ali and Bennewitz, Elizabeth R. and Davoudi, Zohreh and Gorshkov, Alexey V.",
    title = "{High-Energy Collision of Quarks and Mesons in the Schwinger Model: From Tensor Networks to Circuit QED}",
    eprint = "2307.02522",
    archivePrefix = "arXiv",
    primaryClass = "quant-ph",
    reportNumber = "UMD-PP-023-02, IQuS@UW-21-050",
    doi = "10.1103/PhysRevLett.132.091903",
    journal = "Phys. Rev. Lett.",
    volume = "132",
    number = "9",
    pages = "091903",
    year = "2024"
}

@article{Ohliger:2013efs,
       author = {{Ohliger}, M. and {Nesme}, V. and {Eisert}, J.},
        title = "{Efficient and feasible state tomography of quantum many-body systems}",
      journal = {New J. Phys.},
         year = 2013,
        month = jan,
       volume = {15},
       number = {1},
          eid = {015024},
        pages = {015024},
          doi = {10.1088/1367-2630/15/1/015024},
archivePrefix = {arXiv},
       eprint = {1204.5735},
 primaryClass = {quant-ph},
}

@article{Nation:2021kye,
    author = "Nation, Paul D. and Kang, Hwajung and Sundaresan, Neereja and Gambetta, Jay M.",
    title = "{Scalable mitigation of measurement errors on quantum computers}",
    eprint = "2108.12518",
    archivePrefix = "arXiv",
    primaryClass = "quant-ph",
    doi = "10.1103/PRXQuantum.2.040326",
    journal = "PRX Quantum",
    volume = "2",
    pages = "040326",
    year = "2021"
}

@article{Yu:2022ivm,
    author = "Yu, Hongye and Zhao, Yusheng and Wei, Tzu-Chieh",
    title = "{Simulating large-size quantum spin chains on cloud-based superconducting quantum computers}",
    eprint = "2207.09994",
    archivePrefix = "arXiv",
    primaryClass = "quant-ph",
    doi = "10.1103/PhysRevResearch.5.013183",
    journal = "Phys. Rev. Res.",
    volume = "5",
    number = "1",
    pages = "013183",
    year = "2023"
}

@misc{Shtanko:2023tjn,
    author = "Shtanko, Oles and Wang, Derek S. and Zhang, Haimeng and Harle, Nikhil and Seif, Alireza and Movassagh, Ramis and Minev, Zlatko",
    title = "{Uncovering Local Integrability in Quantum Many-Body Dynamics}",
    eprint = "2307.07552",
    archivePrefix = "arXiv",
    primaryClass = "quant-ph",
    month = "7",
    year = "2023"
}

@misc{Chen:2023tfg,
    author = "Chen, Edward H. and others",
    title = "{Realizing the Nishimori transition across the error threshold for constant-depth quantum circuits}",
    eprint = "2309.02863",
    archivePrefix = "arXiv",
    primaryClass = "quant-ph",
    month = "9",
    year = "2023"
}

@misc{Liao:2023eug,
    author = "Liao, Haoran and Wang, Derek S. and Sitdikov, Iskandar and Salcedo, Ciro and Seif, Alireza and Minev, Zlatko K.",
    title = "{Machine Learning for Practical Quantum Error Mitigation}",
    eprint = "2309.17368",
    archivePrefix = "arXiv",
    primaryClass = "quant-ph",
    month = "9",
    year = "2023"
}

@article{Kim:2023bwr,
    author = {{Kim}, Youngseok and {Eddins}, Andrew and {Anand}, Sajant and {Wei}, Ken Xuan and {van den Berg}, Ewout and {Rosenblatt}, Sami and {Nayfeh}, Hasan and {Wu}, Yantao and {Zaletel}, Michael and {Temme}, Kristan and {Kandala}, Abhinav},
    title = "{Evidence for the utility of quantum computing before fault tolerance}",
    doi = "10.1038/s41586-023-06096-3",
    journal = "Nature",
    volume = "618",
    number = "7965",
    pages = "500--505",
    year = "2023"
}

@misc{quantinuumweb,
  url = {https://www.quantinuum.com/hardware},
  title = {{Quantinuum hardware}},
  publisher = {Quantinuum},
  year = {2024}
}

@article{Zohar:2012ay,
    author = "Zohar, Erez and Cirac, J. Ignacio and Reznik, Benni",
    title = "{Simulating Compact Quantum Electrodynamics with ultracold atoms: Probing confinement and nonperturbative effects}",
    eprint = "1204.6574",
    archivePrefix = "arXiv",
    primaryClass = "quant-ph",
    doi = "10.1103/PhysRevLett.109.125302",
    journal = "Phys. Rev. Lett.",
    volume = "109",
    pages = "125302",
    year = "2012"
}

@article{Bluvstein:2023zmt,
    author = "Bluvstein, Dolev and others",
    title = "{Logical quantum processor based on reconfigurable atom arrays}",
    eprint = "2312.03982",
    archivePrefix = "arXiv",
    primaryClass = "quant-ph",
    month = "12",
    year = "2023",
   DOI={10.1038/s41586-023-06927-3},
   journal={Nature},
}

@article{Mildenberger:2022jqr,
    author = "Mildenberger, Julius and Mruczkiewicz, Wojciech and Halimeh, Jad C. and Jiang, Zhang and Hauke, Philipp",
    title = "{Confinement in a ${{\mathbb{Z}}}_{2}$ lattice gauge theory on a quantum computer}",
    eprint = "2203.08905",
    archivePrefix = "arXiv",
    primaryClass = "quant-ph",
    doi = "10.1038/s41567-024-02723-6",
    journal = "Nat. Phys.",
    volume = "21",
    number = "2",
    pages = "312--317",
    year = "2025"
}

@article{Lee:2023urk,
    author = "Lee, Kyle and Mulligan, James and Ringer, Felix and Yao, Xiaojun",
    title = "{Liouvillian dynamics of the open Schwinger model: String breaking and kinetic dissipation in a thermal medium}",
    eprint = "2308.03878",
    archivePrefix = "arXiv",
    primaryClass = "quant-ph",
    reportNumber = "JLAB-THY-23-3894, MIT-CTP-5592, YITP-SB-2023-23, IQuS@UW-21-061",
    doi = "10.1103/PhysRevD.108.094518",
    journal = "Phys. Rev. D",
    volume = "108",
    number = "9",
    pages = "094518",
    year = "2023"
}

@article{Buyens:2015tea,
    author = "Buyens, Boye and Haegeman, Jutho and Verschelde, Henri and Verstraete, Frank and Van Acoleyen, Karel",
    title = "{Confinement and string breaking for QED$_2$ in the Hamiltonian picture}",
    eprint = "1509.00246",
    archivePrefix = "arXiv",
    primaryClass = "hep-lat",
    doi = "10.1103/PhysRevX.6.041040",
    journal = "Phys. Rev. X",
    volume = "6",
    number = "4",
    pages = "041040",
    year = "2016"
}

@article{Papaefstathiou:2024zsu,
    author = "Papaefstathiou, Irene and Knolle, Johannes and Ba{\~n}uls, Mari Carmen",
    title = "{Real-time scattering in the lattice Schwinger model}",
    eprint = "2402.18429",
    archivePrefix = "arXiv",
    primaryClass = "hep-lat",
    doi = "10.1103/PhysRevD.111.014504",
    journal = "Phys. Rev. D",
    volume = "111",
    number = "1",
    pages = "014504",
    year = "2025"
}

@article{PHOBOS:2004zne,
    author = "Back, B. B. and others",
    collaboration = "PHOBOS",
    title = "{The PHOBOS perspective on discoveries at RHIC}",
    eprint = "nucl-ex/0410022",
    archivePrefix = "arXiv",
    doi = "10.1016/j.nuclphysa.2005.03.084",
    journal = "Nucl. Phys. A",
    volume = "757",
    pages = "28--101",
    year = "2005"
}

@article{STAR:2005gfr,
    author = "Adams, John and others",
    collaboration = "STAR",
    title = "{Experimental and theoretical challenges in the search for the quark gluon plasma: The STAR Collaboration's critical assessment of the evidence from RHIC collisions}",
    eprint = "nucl-ex/0501009",
    archivePrefix = "arXiv",
    doi = "10.1016/j.nuclphysa.2005.03.085",
    journal = "Nucl. Phys. A",
    volume = "757",
    pages = "102--183",
    year = "2005"
}

@article{PHENIX:2004vcz,
    author = "Adcox, K. and others",
    collaboration = "PHENIX",
    title = "{Formation of dense partonic matter in relativistic nucleus-nucleus collisions at RHIC: Experimental evaluation by the PHENIX collaboration}",
    eprint = "nucl-ex/0410003",
    archivePrefix = "arXiv",
    doi = "10.1016/j.nuclphysa.2005.03.086",
    journal = "Nucl. Phys. A",
    volume = "757",
    pages = "184--283",
    year = "2005"
}

@article{Illa:2022jqb,
    author = "Illa, Marc and Savage, Martin J.",
    title = "{Basic elements for simulations of standard-model physics with quantum annealers: Multigrid and clock states}",
    eprint = "2202.12340",
    archivePrefix = "arXiv",
    primaryClass = "quant-ph",
    reportNumber = "IQuS@UW-21-020",
    doi = "10.1103/PhysRevA.106.052605",
    journal = "Phys. Rev. A",
    volume = "106",
    number = "5",
    pages = "052605",
    year = "2022"
}

@article{Mazzola:2021hma,
    author = "Mazzola, Giulia and Mathis, Simon V. and Mazzola, Guglielmo and Tavernelli, Ivano",
    title = "{Gauge-invariant quantum circuits for U(1) and Yang-Mills lattice gauge theories}",
    eprint = "2105.05870",
    archivePrefix = "arXiv",
    primaryClass = "quant-ph",
    doi = "10.1103/PhysRevResearch.3.043209",
    journal = "Phys. Rev. Res.",
    volume = "3",
    number = "4",
    pages = "043209",
    year = "2021"
}

@article{Gustafson:2023aai,
    author = "Gustafson, Erik and Van de Water, Ruth",
    title = "{Improved Fermion Hamiltonians for Quantum Simulation}",
    eprint = "2402.04317",
    archivePrefix = "arXiv",
    primaryClass = "hep-lat",
    reportNumber = "FERMILAB-CONF-23-0877-T",
    doi = "10.22323/1.453.0215",
    journal = "PoS",
    volume = "LATTICE2023",
    pages = "215",
    year = "2024"
}

@article{Barata:2023clv,
    author = "Barata, Jo\~ao and Du, Xiaojian and Li, Meijian and Qian, Wenyang and Salgado, Carlos A.",
    title = "{Quantum simulation of in-medium QCD jets: Momentum broadening, gluon production, and entropy growth}",
    eprint = "2307.01792",
    archivePrefix = "arXiv",
    primaryClass = "hep-ph",
    doi = "10.1103/PhysRevD.108.056023",
    journal = "Phys. Rev. D",
    volume = "108",
    number = "5",
    pages = "056023",
    year = "2023"
}

@article{Hebenstreit:2013baa,
    author = {Hebenstreit, Florian and Berges, J\"urgen and Gelfand, Daniil},
    title = "{Real-time dynamics of string breaking}",
    eprint = "1307.4619",
    archivePrefix = "arXiv",
    primaryClass = "hep-ph",
    doi = "10.1103/PhysRevLett.111.201601",
    journal = "Phys. Rev. Lett.",
    volume = "111",
    pages = "201601",
    year = "2013"
}

@article{Kasper:2015cca,
    author = "Kasper, V. and Hebenstreit, F. and Oberthaler, M. and Berges, J.",
    title = "{Schwinger pair production with ultracold atoms}",
    eprint = "1506.01238",
    archivePrefix = "arXiv",
    primaryClass = "cond-mat.quant-gas",
    doi = "10.1016/j.physletb.2016.07.036",
    journal = "Phys. Lett. B",
    volume = "760",
    pages = "742--746",
    year = "2016"
}

@misc{Yao:2022eqm,
    author = "Yao, Xiaojun",
    title = "{Quantum Simulation of Light-Front QCD for Jet Quenching in Nuclear Environments}",
    eprint = "2205.07902",
    archivePrefix = "arXiv",
    primaryClass = "hep-ph",
    reportNumber = "MIT-CTP 5435, IQuS@UW-21-037",
    month = "5",
    year = "2022"
}

@article{Leone:2021rzd,
    author = "Leone, Lorenzo and Oliviero, Salvatore F. E. and Hamma, Alioscia",
    title = "{Stabilizer R\'enyi Entropy}",
    eprint = "2106.12587",
    archivePrefix = "arXiv",
    primaryClass = "quant-ph",
    doi = "10.1103/PhysRevLett.128.050402",
    journal = "Phys. Rev. Lett.",
    volume = "128",
    number = "5",
    pages = "050402",
    year = "2022"
}

@article{Illa:2022zgu,
    author = "Illa, Marc and Savage, Martin J.",
    title = "{Multi-Neutrino Entanglement and Correlations in Dense Neutrino Systems}",
    eprint = "2210.08656",
    archivePrefix = "arXiv",
    primaryClass = "nucl-th",
    reportNumber = "IQuS@UW-21-034",
    doi = "10.1103/PhysRevLett.130.221003",
    journal = "Phys. Rev. Lett.",
    volume = "130",
    number = "22",
    pages = "221003",
    year = "2023"
}

@article{Zemlevskiy:2023eyw,
    author = "Zemlevskiy, Nikita A. and Froland, Henry F. and Caspar, Stephan",
    title = "{Optimization of algorithmic errors in analog quantum simulations}",
    eprint = "2308.02642",
    archivePrefix = "arXiv",
    primaryClass = "quant-ph",
    reportNumber = "IQuS@UW-21-059",
    doi = "10.1103/PhysRevA.109.052425",
    journal = "Phys. Rev. A",
    volume = "109",
    number = "5",
    pages = "052425",
    year = "2024"
}

@article{Cai:2022rnq,
    author = "Cai, Zhenyu and Babbush, Ryan and Benjamin, Simon C. and Endo, Suguru and Huggins, William J. and Li, Ying and McClean, Jarrod R. and O'Brien, Thomas E.",
    title = "{Quantum error mitigation}",
    eprint = "2210.00921",
    archivePrefix = "arXiv",
    primaryClass = "quant-ph",
    doi = "10.1103/RevModPhys.95.045005",
    journal = "Rev. Mod. Phys.",
    volume = "95",
    number = "4",
    pages = "045005",
    year = "2023"
}

@article{Georgopoulos:2021fyi,
    author = "Georgopoulos, Konstantinos and Emary, Clive and Zuliani, Paolo",
    title = "{Modeling and simulating the noisy behavior of near-term quantum computers}",
    eprint = "2101.02109",
    archivePrefix = "arXiv",
    primaryClass = "quant-ph",
    doi = "10.1103/PhysRevA.104.062432",
    journal = "Phys. Rev. A",
    volume = "104",
    number = "6",
    pages = "062432",
    year = "2021"
}

@article{Childs:2019rty,
    author = "Childs, Andrew M. and Su, Yuan",
    title = "{Nearly Optimal Lattice Simulation by Product Formulas}",
    doi = "10.1103/PhysRevLett.123.050503",
    journal = "Phys. Rev. Lett.",
    volume = "123",
    number = "5",
    pages = "050503",
    year = "2019"
}

@article{Childs:2019hts,
    author = "Childs, Andrew M. and Su, Yuan and Tran, Minh C. and Wiebe, Nathan and Zhu, Shuchen",
    title = "{Theory of Trotter Error with Commutator Scaling}",
    eprint = "1912.08854",
    archivePrefix = "arXiv",
    primaryClass = "quant-ph",
    doi = "10.1103/PhysRevX.11.011020",
    journal = "Phys. Rev. X",
    volume = "11",
    number = "1",
    pages = "011020",
    year = "2021"
}

@article{Keever:2023gqo,
    author = "Keever, Conor Mc and Lubasch, Michael",
    title = "{Towards Adiabatic Quantum Computing Using Compressed Quantum Circuits}",
    eprint = "2311.05544",
    archivePrefix = "arXiv",
    primaryClass = "quant-ph",
    doi = "10.1103/PRXQuantum.5.020362",
    journal = "PRX Quantum",
    volume = "5",
    number = "2",
    pages = "020362",
    year = "2024"
}

@article{Flannigan:2022bwi,
    author = "Flannigan, S. and Pearson, N. and Low, G. H. and Buyskikh, A. and Bloch, I. and Zoller, P. and Troyer, M. and Daley, A. J.",
    title = "{Propagation of errors and quantitative quantum simulation with quantum advantage}",
    eprint = "2204.13644",
    archivePrefix = "arXiv",
    primaryClass = "quant-ph",
    doi = "10.1088/2058-9565/ac88f5",
    journal = "Quantum Sci. Technol.",
    volume = "7",
    number = "4",
    pages = "045025",
    year = "2022"
}

@article{Shukla:2023oub,
    author = "Shukla, Alok and Vedula, Prakash",
    title = "{A quantum approach for digital signal processing}",
    eprint = "2309.06570",
    archivePrefix = "arXiv",
    primaryClass = "quant-ph",
    doi = "10.1140/epjp/s13360-023-04730-7",
    journal = "Eur. Phys. J. Plus",
    volume = "138",
    number = "12",
    pages = "1121",
    year = "2023"
}

@article{Amitrano:2022yyn,
    author = "Amitrano, Valentina and Roggero, Alessandro and Luchi, Piero and Turro, Francesco and Vespucci, Luca and Pederiva, Francesco",
    title = "{Trapped-ion quantum simulation of collective neutrino oscillations}",
    eprint = "2207.03189",
    archivePrefix = "arXiv",
    primaryClass = "quant-ph",
    doi = "10.1103/PhysRevD.107.023007",
    journal = "Phys. Rev. D",
    volume = "107",
    number = "2",
    pages = "023007",
    year = "2023"
}

@article{Yeter-Aydeniz:2021olz,
    author = {Yeter-Aydeniz, K\"ubra and Bangar, Shikha and Siopsis, George and Pooser, Raphael C.},
    title = "{Collective neutrino oscillations on a quantum computer}",
    eprint = "2104.03273",
    archivePrefix = "arXiv",
    primaryClass = "quant-ph",
    doi = "10.1007/s11128-021-03348-x",
    journal = "Quant. Inf. Proc.",
    volume = "21",
    number = "3",
    pages = "84",
    year = "2022"
}

@article{Li:2024uzc,
    author = "Li, Junxu and Gao, Xingyu",
    title = "{Quantum circuit for high order perturbation theory corrections}",
    eprint = "2404.05162",
    archivePrefix = "arXiv",
    primaryClass = "quant-ph",
    doi = "10.1038/s41598-024-64854-3",
    journal = "Sci. Rep.",
    volume = "14",
    number = "1",
    pages = "13963",
    year = "2024"
}

@article{Li:2022zbj,
    author = "Li, Junxu and Jones, Barbara A. and Kais, Sabre",
    title = "{Toward perturbation theory methods on a quantum computer}",
    eprint = "2206.14955",
    archivePrefix = "arXiv",
    primaryClass = "quant-ph",
    doi = "10.1126/sciadv.adg4576",
    journal = "Sci. Adv.",
    volume = "9",
    number = "19",
    pages = "adg4576",
    year = "2023"
}

@article{Mitarai:2022via,
    author = "Mitarai, Kosuke and Toyoizumi, Kiichiro and Mizukami, Wataru",
    title = "{Perturbation theory with quantum signal processing}",
    eprint = "2210.00718",
    archivePrefix = "arXiv",
    primaryClass = "quant-ph",
    doi = "10.22331/q-2023-05-12-1000",
    journal = "Quantum",
    volume = "7",
    pages = "1000",
    year = "2023"
}

@article{Sun:2021mdd,
    author = "Sun, Jinzhao and Endo, Suguru and Lin, Huiping and Hayden, Patrick and Vedral, Vlatko and Yuan, Xiao",
    title = "{Perturbative Quantum Simulation}",
    eprint = "2106.05938",
    archivePrefix = "arXiv",
    primaryClass = "quant-ph",
    doi = "10.1103/PhysRevLett.129.120505",
    journal = "Phys. Rev. Lett.",
    volume = "129",
    number = "12",
    pages = "120505",
    year = "2022"
}

@article{Siwach:2022xhx,
    author = "Siwach, Pooja and Suliga, Anna M. and Balantekin, A. Baha",
    title = "{Entanglement in three-flavor collective neutrino oscillations}",
    eprint = "2211.07678",
    archivePrefix = "arXiv",
    primaryClass = "hep-ph",
    reportNumber = "N3AS-22-023, INT-PUB-22-051",
    doi = "10.1103/PhysRevD.107.023019",
    journal = "Phys. Rev. D",
    volume = "107",
    number = "2",
    pages = "023019",
    year = "2023"
}

@inbook{Patwardhan:2022mxg,
    author = "Patwardhan, Amol V. and Cervia, Michael J. and Rrapaj, Ermal and Siwach, Pooja and Balantekin, A. B.",
    editor = "Tanihata, Isao and Toki, Hiroshi and Kajino, Toshitaka",
    title = "{Many-Body Collective Neutrino Oscillations: Recent Developments}",
    booktitle = "{Handbook of Nuclear Physics}",
    publisher={Springer Nature Singapore},
    eprint = "2301.00342",
    archivePrefix = "arXiv",
    primaryClass = "hep-ph",
    doi = "10.1007/978-981-15-8818-1_126-1",
    pages = "1--16",
    year = "2023"
}

@article{Balantekin:2023qvm,
    author = "Balantekin, A. B. and Cervia, Michael J. and Patwardhan, Amol V. and Rrapaj, Ermal and Siwach, Pooja",
    title = "{Quantum information and quantum simulation of neutrino physics}",
    eprint = "2305.01150",
    archivePrefix = "arXiv",
    primaryClass = "nucl-th",
    doi = "10.1140/epja/s10050-023-01092-7",
    journal = "Eur. Phys. J. A",
    volume = "59",
    number = "8",
    pages = "186",
    year = "2023"
}

@article{Siwach:2023wzy,
    author = "Siwach, Pooja and Harrison, Kaytlin and Balantekin, A. Baha",
    title = "{Collective neutrino oscillations on a quantum computer with hybrid quantum-classical algorithm}",
    eprint = "2308.09123",
    archivePrefix = "arXiv",
    primaryClass = "quant-ph",
    reportNumber = "LLNL-JRNL-853275-DRAFT",
    doi = "10.1103/PhysRevD.108.083039",
    journal = "Phys. Rev. D",
    volume = "108",
    number = "8",
    pages = "083039",
    year = "2023"
}

@article{Balantekin:2023ayx,
    author = "Balantekin, A. Baha and Cervia, Michael J. and Patwardhan, Amol V. and Surman, Rebecca and Wang, Xilu",
    title = "{Collective Neutrino Oscillations and Heavy-element Nucleosynthesis in Supernovae: Exploring Potential Effects of Many-body Neutrino Correlations}",
    eprint = "2311.02562",
    archivePrefix = "arXiv",
    primaryClass = "astro-ph.HE",
    reportNumber = "N3AS-24-004",
    doi = "10.3847/1538-4357/ad393d",
    journal = "Astrophys. J.",
    volume = "967",
    number = "2",
    pages = "146",
    year = "2024"
}

@article{Loreti:1994ry,
    author = "Loreti, F. N. and Balantekin, A. B.",
    title = "{Neutrino oscillations in noisy media}",
    eprint = "nucl-th/9406003",
    archivePrefix = "arXiv",
    reportNumber = "MAD-NT-94-07",
    doi = "10.1103/PhysRevD.50.4762",
    journal = "Phys. Rev. D",
    volume = "50",
    pages = "4762--4770",
    year = "1994"
}

@article{Balantekin:2004ug,
    author = "Balantekin, A. B. and Yuksel, H.",
    title = "{Neutrino mixing and nucleosynthesis in core-collapse supernovae}",
    eprint = "astro-ph/0411159",
    archivePrefix = "arXiv",
    doi = "10.1088/1367-2630/7/1/051",
    journal = "New J. Phys.",
    volume = "7",
    pages = "51",
    year = "2005"
}

@article{Pehlivan:2011hp,
    author = "Pehlivan, Y. and Balantekin, A. B. and Kajino, Toshitaka and Yoshida, Takashi",
    title = "{Invariants of Collective Neutrino Oscillations}",
    eprint = "1105.1182",
    archivePrefix = "arXiv",
    primaryClass = "astro-ph.CO",
    doi = "10.1103/PhysRevD.84.065008",
    journal = "Phys. Rev. D",
    volume = "84",
    pages = "065008",
    year = "2011"
}

@article{Balantekin:2006tg,
    author = "Balantekin, A. B. and Pehlivan, Y.",
    title = "{Neutrino-Neutrino Interactions and Flavor Mixing in Dense Matter}",
    eprint = "astro-ph/0607527",
    archivePrefix = "arXiv",
    doi = "10.1088/0954-3899/34/1/004",
    journal = "J. Phys. G",
    volume = "34",
    pages = "47--66",
    year = "2007"
}

@article{Izaguirre:2016gsx,
    author = "Izaguirre, Ignacio and Raffelt, Georg and Tamborra, Irene",
    title = "{Fast Pairwise Conversion of Supernova Neutrinos: A Dispersion-Relation Approach}",
    eprint = "1610.01612",
    archivePrefix = "arXiv",
    primaryClass = "hep-ph",
    reportNumber = "INT-PUB-16-023, MPP-2016-266",
    doi = "10.1103/PhysRevLett.118.021101",
    journal = "Phys. Rev. Lett.",
    volume = "118",
    number = "2",
    pages = "021101",
    year = "2017"
}

\end{document}